# The Earth-Moon CR3BP: a Full Atlas of Low-Energy Fast Periodic Transfer Orbits

A. M. Leiva

Observatorio Astronómico de Córdoba,
Universidad Nacional de Córdoba
Laprida 854, 5000 Córdoba, Argentina
e-mail: mleiva@oac.uncor.edu

C. B. Briozzo

Facultad de Matemática, Astronomía y Física,
Universidad Nacional de Córdoba
Medina Allende S/N, Ciudad Universitaria, 5000 Córdoba, Argentina
e-mail: briozzo@famaf.unc.edu.ar

December 15, 2006

## Abstract

In the framework of the planar CR3BP for mass parameter μ = 0.0121505, corresponding to the Earth-Moon system, we identify and describe 80 families of periodic orbits encircling both the Earth and the Moon ("transfer" orbits). All the orbits in these families have very low energies, most of them corresponding to values of the Jacobi constant $C$ for which the Hill surface is closed at the Lagrangian point L2. All of these orbits have also short period $T$, generally under six months. Most of the families are composed of orbits, which are asymmetric with respect to the Earth-Moon axis.

The main results presented for each family are: (i) the characteristic curves $T(h)$, $y(h)$, $v_y(h)$, and $v_x(h)$ on the Poincaré Section $\Sigma_1 = \{ x=0{,}836915310 \ , \ y \ , \ v_x > 0 \ , \ v_y \}$ normal to the Earth-Moon axis at the Lagrangian point L1, parameterized by their Jacobi constant $h = -C/2$ in the synodic coordinate system; (ii) the stability parameter along each family; (iii) the intersections $x_i(h)$ of the orbits with the Earth-Moon axis, on the Poincaré section $\Sigma_2 = \{ x \ , y=0 \ , \ v_x \ , \ v_y > 0 \}$; (iv) plots of some selected orbits and details of their circumlunar region; and (v) numerical data for the intersection of an orbit with $\Sigma_1$ at a reference value of $h$. Some possible extensions and applications of this work are also discussed.

# 1. Introduction

A decade ago, chaos used to de considered as undesirable when designing a space mission: the extreme sensitivity to slight changes in initial conditions made hard to predict future trajectories, and required frequent and wasteful control impulses to keep a probe on the desired path. However, works by Bollt and Meiss (1995), Schroer and Ott (1997), and more recently Ross *et al.* (2003), showed that chaos can be advantageous in providing low-energy transfer orbits between different astronomical bodies. The disadvantages of their approach are that it requires considering large quantities of orbit arcs, choosing very carefully which ones to patch together, that it requires several control impulses to achieve the patching, and that it results generally in a slow transfer. Moreover, these works, as well as the recent ones by Yagasaki (2004 a, b), aim to find a single, one-time transfer orbit from the Earth to the Moon which minimizes fuel consumption or achieves some compromise between consumption and transfer time.

On the other hand, since the pioneering work by Ott *et al.* (1990), the possibility of controlling chaos has opened new opportunities in this field. The control methods are based on two features of chaotic systems. First, chaos in a Hamiltonian system usually involves the presence of a dense set of unstable periodic orbits (UPOs) (Otani and Jones 1997), and results from the orbit being pulled towards each of them along their stable manifolds, only to be pushed away along the unstable ones. Second, the very same sensitivity to initial conditions allows any one of these UPOs to be easily stabilized, by applying carefully chosen but very small control impulses carrying the actual orbit onto stable manifold of the desired one. Thus, to implement control of chaos in a mission design, we need first a survey of the UPOs in the region of interest, and then a suitable control algorithm allowing to stabilize the chosen one. This approach has two additional advantages: it leads to periodic orbits, with the possibility of repeatedly passing near the astronomical bodies of interest, and it allows to switch to another UPO by suitable control impulse, adding flexibility to the mission design.

In a recent work (Leiva and Briozzo 2006a) we implemented this program for periodic transfer orbits in the Earth-Moon planar Circular Restricted Three Body Problem (CR3BP). By a numerical survey we found 287 orbits that were fast and of low energy, performed periodic transfers between the Earth and the Moon, and were successfully stabilized by a simple control algorithm requiring modest amounts of fuel. The algorithm was designed to push the perturbed orbit to the stable manifold of the desired UPO, which is much easier than trying to re-target the UPO itself.

At a difference with CR3BP models for other systems like the Sun-Jupiter one, where the influence of the remaining Solar System bodies is small an can be accounted for perturbatively, in the case of the Earth-Moon system the influence of the Sun on this kind of orbits is large, and the control method introduced in (Leiva and Briozzo 2006a) is unable to cope with it. In a related work (Leiva and Briozzo 2005) we addressed the problem of accounting for the Sun's influence, showing how under suitable conditions some UPOs could be extended from the Earth-Moon CR3BP to a more realistic model like the Sun-Earth-Moon Quasi-Bicircular Problem (QBCP) (Andreu 1998, 2002, 2003). In this model an approximate closed solution is obtained for the motion the primaries under their mutual gravitational interactions. At a difference with the simpler Bicircular Problem, in the QBCP the motion of the primaries is both energy-conserving (with the same approximation of the primaries solution) and periodic, allowing the use of standard techniques for periodic Hamiltonian systems while remaining a fairly realistic approximation to the real Sun-Earth-Moon system.

Together, the previous two works suggested the goal of finding as many UPOs of low energies and short periods in the CR3BP as we could, and extend the suitable ones to the QBCP. For, given a large enough number of transfer UPO families in the CR3BP, we could then find a periodic orbit in the QBCP suited to almost any mission goals requiring repeated passages of a probe near both the Earth and the Moon, and having the desirable properties of a short period and low energy.

In the present work we report the results of this survey in the form of a full Atlas of short-period, low-energy periodic orbits (POs) encircling both the Earth and the Moon, in the framework of the Earth-Moon CR3BP. We must emphasize that the term "transfer orbit" is used here only in the sense that the orbits encircle both primaries, and is not intended to imply that they pass particularly close to any of them, though many of the orbits found pass close to the Moon or even collide with its surface.

Finally, we want to remark that the PO families presented here include both symmetric and asymmetric orbits (in fact, most of the UPOs are composed of asymmetric orbits) for the Earth-Moon value of the mass parameter $\mu$, while a recent similar work by Bruno and Varin (2006) concerns only symmetric orbits in the CR3BP (though for $0 \le \mu \le 1/2$ ). We think the intersection between both works provides a desirable cross-check of the results.

In Section 2 we introduce the coordinate system, parameters and equations of motion for the Earth-Moon CR3BP as used in this work, and review some properties of periodic orbits which are useful to find and characterize them. In Section 3 we describe in detail the numerical algorithms employed in the survey for the PO families. In Section 4 we describe the structure of the Atlas. In Section 5, we show complete data sets for the 80 families and give the numerical data for a reference PO on each one of the families we found, allowing the reader to reproduce our results if wanted. Finally, in Section 6 we discuss to some length possible applications and extensions of the present work.

## 2. The CR3BP

The CR3BP is a special case of the general Three Body Problem, where we assume two primaries of masses $m_E$ and $m_M$ in plane circular orbits around their center of mass, and an infinitesimal third mass $m$ moving in the orbital plane of the primaries without disturbing them (Szebehely 1967). The orbits of the primaries are taken to lie in the $(x, y)$ plane, with their center of mass at the origin. Adimensional units are used, with $m_E = 1 - \mu$ and $m_M = \mu$, distance between the primaries unity, and orbital period $2\pi$ (giving unit angular frequency). For the Earth-Moon system this gives $\mu \approx 0.0121505$, times units of ~104h (one sidereal month /$2\pi$), length units of ~384400km, and velocity units of ~1024m/s. The values for all Solar System parameters are taken from the Jet Propulsion Laboratory (JPL) ephemeris, available on-line at NASA's JPL site (`http://ssd.nasa.gov/?ephemerides#planets`). We will work in a coordinate system rotating with angular velocity $n = 1$ (synodic system), with the primaries lying on the $x$ axis at $(x_E; y_E) = (-\mu;0)$, $(x_M; y_M) = (1 - \mu,0)$ (see Fig. 1).

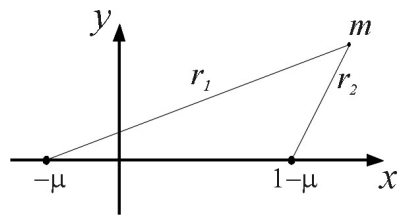

Figure 1. Synodic System (CR3BP). Earth is at left, Moon is at right.

The Hamiltonian for the motion of the infinitesimal mass $m$ is then (Arnold 1993)

$$H_{CR3BP} = \frac{1}{2}\left(p_x^2 + p_y^2\right) + yp_x - xp_y - \frac{1-\mu}{r_1} - \frac{\mu}{r_2}, \qquad (1)$$

where $p_x = \dot{x} - y$, $p_y = \dot{y} + x$, $r_1^2 = \left(x - x_E\right)^2 + y^2$ and $r_2^2 = \left(x - x_M\right)^2 + y^2$. The equations of motion for the mass $m$ have a well known integral of motion, the Jacobi integral $C$; however, in this work prefer to use instead the energy $H_{CR3BP} = h = -C/2$ in the synodic System. The name "energy" is used here in the sense of it being the canonically conjugate variable to the time $t$ (Arnold 1993), not in the sense of being the sum of the kinetic and potential energies. The equations of motion for the infinitesimal mass $m$ are

$$\ddot{x} = 2\dot{y} + x - \left(1 - \mu\right)\frac{\left(x - x_E\right)}{r_1^3} - \mu\frac{\left(x - x_M\right)}{r_2^3} \qquad (2)$$

$$\ddot{y} = -2\dot{x} + y - \left(1 - \mu\right)\frac{y}{r_1^3} - \mu\frac{y}{r_2^3} \qquad (3)$$

The issue of the existence and properties of periodic orbits (POs) in autonomous Hamiltonian systems has a long history (Hénon 1997; Szebehely 1967), and a detailed account is far beyond the scope of this work. Here we briefly review some properties de POs in the CR3BP which are helpful for the task of finding them, and discuss how to characterize their stability.

A periodic orbit with initial condition $\vec{x}(t = 0) = \vec{x}_0$ can be represented as a closed curve $\vec{\xi}_t(\vec{x}_0)$ in phase-space. The problem of finding a $T$-periodic orbit is equivalent to that of finding a root of the function $\vec{G}(\vec{x}_0) = \vec{\xi}_T(\vec{x}_0) - \vec{x}_0$, but since the CR3BP Hamiltonian (1) is independent of time, if $\vec{x}_0$ is a solution then $\vec{\xi}_T(\vec{x}_0)$ is also a solution for all $t$. Thus it is better not to select the period $T$ beforehand, but to consider a Poincaré section $\Sigma$ transverse to the flux and restate the problem as that of finding fixed points of the corresponding Poincaré map.

Also due to the time invariance of the Hamiltonian, in the CR3BP and for a given value of $\mu$ each PO is a member of some monoparametric family of POs parameterized by $h$ (Hénon 1997). Each family can be completely characterized by a function $\vec{\phi}(t,h)$ and an analytic function $T^*(h)$, its *period-in-family*, which varies smoothly along the family.

The stability of a PO can be analyzed directly on the surface section by linearization of the corresponding Poincaré map around the fixed point associated with the PO. In the CR3BP the section surface $\Sigma$ is two-dimensional (Hénon 1997), and the linearized map is given by a symplectic $2 \times 2$ matrix. Hence if $\eta$ is an eingenvalue the other one is $1/\eta$, and the characteristic polynomial can be expressed as

$$p(\eta) = (\eta^2 - s\eta + 1) \qquad (4)$$

where $s = \eta + 1/\eta$ is usually called the stability parameter. In what follows we will denote the eingenvalues by $\lambda_u$ and $\lambda_s$. According to their value we have the following four cases (see Fig. 2):

1. $\lambda_u$ and $\lambda_s$ real, with $\lambda_u < -1 < \lambda_s < 0$. The fixed point is a *reflection hyperbolic point*. The eigenvector of the linearized map belonging to $\lambda_u$ (re. $\lambda_s$) is tangent to the unstable (re. stable) manifold of the fixed point. The PO is *linearly unstable*.

2. $\lambda_u$ and $\lambda_s$ real, with $0 < \lambda_s < 1 < \lambda_u$. The fixed point is a *ordinary hyperbolic point*. The eingenvector of the linearized map belonging to $\lambda_u$ (re. $\lambda_s$) is tangent to the unstable (re. stable) manifold of the fixed point. The PO is *linearly unstable*.

3. $\lambda_u$ and $\lambda_s$ complex, with $\lambda_u = e^{i\phi}$, $\lambda_s = e^{-i\phi}$, $0 < \phi < \pi$. The fixed point is an *elliptic point*. The eigenvectors of the linearized map are complex unit vectors, conjugate to each other, and the linearized map itself is represented by a $2 \times 2$ rotation matrix. The PO is *orbitally stable* (Verhulst 1990).

4. $\lambda_u = \lambda_s = 1$, or $\lambda_u = \lambda_s = -1$. The stability of the fixed point cannot be characterized by the linear analysis, as it lies on the central manifold of the map. The stability of the PO can neither be characterized by linear analysis. The former (re. latter) case corresponds to *intersection* (re. *bifurcation*) *orbits* (Hénon 1997).

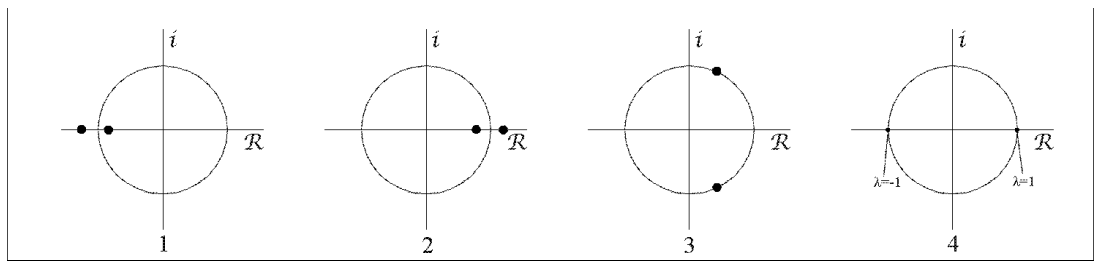

Figure 2. Possible values for the eingenvalues of the fixed point on the complex plane (schematic).

# 3. Numerical Survey for PO families

## 3.1 Initial search for POs

The survey for short-period, low energy PO families started with a numerical search in the adequate region of parameter space. This search and its results have been reported in a previous work (Leiva and Briozzo 2006a). Here will briefly review its main features.

The level sets of energy $h$ (zero-velocity curves) divide the $(x, y)$ plane into allowed and forbidden regions of the motion. For $h < h_1 \approx -1.59407$ the allowed regions around the Earth and the Moon are disconnected, and no transfer orbits are possible (see Fig. 3). For $h > h_2 \approx -1.58617$ the inner and outer regions are connected and transfer orbits can escape to infinity. Thus we limited the initial search to the range $h_1 < h < h_2$.

We select a Poincaré $\Sigma_1 = \{ x=x_{L1}, \, y \, , \, v_x > 0 \, , \, v_y \}$, where $x_{L1} = 0.836915310$ is the $x$-coordinate of the collinear Lagrangian point L1, since for $h_1 < h < h_2$ any transfer orbit necessarily intersect this section. On $\Sigma_1$ and for a given value of $h$, the condition $H_{CR3BP} = h$ defines a convex allowed region in the $(y, \dot{y})$ plane, inscribe in a rectangle $[- y_{max} , y_{max}] \times [- \dot{y}_{max} , \dot{y}_{max}]$. A grid the initial conditions was generated by varying $h$ from $h_1$ to $h_2$ in steps $\Delta h = 10^{-4}$, $y$ from $- y_{max}$ to $y_{max}$ in steps $\Delta y = 10^{-2} y_{max}$, and $\dot{y}$ from $- \dot{y}_{max}$ to $\dot{y}_{max}$ in steps $\Delta y = 10^{-2} \dot{y}_{max}$, and determining the corresponding value of $\dot{x}$ from $H_{CR3BP} = h$, discarding those lying outside the allowed region.

For each initial condition $(x_0, y_0, \dot{x}_0, \dot{y}_0)$ thus generated, the equations of motion (2,3) were integrated by a variable-step Burlisch-Stoer algorithm (Press *et. al.* 1992) with relative precision $10^{-14}$. From these integrations we only retained orbits returning to $\Sigma_1$ at $T < 40$, with $\left| (y(T) - y_0)^2 + (\dot{y}(T) - \dot{y}_0)^2 \right| < 2 \times 10^{-3}$, and having passed behind both the Moon and the Earth. Their evolution was then considered as a Poincaré map

$$M : (y, \dot{y}) \rightarrow (y, \dot{y}) \qquad (5)$$

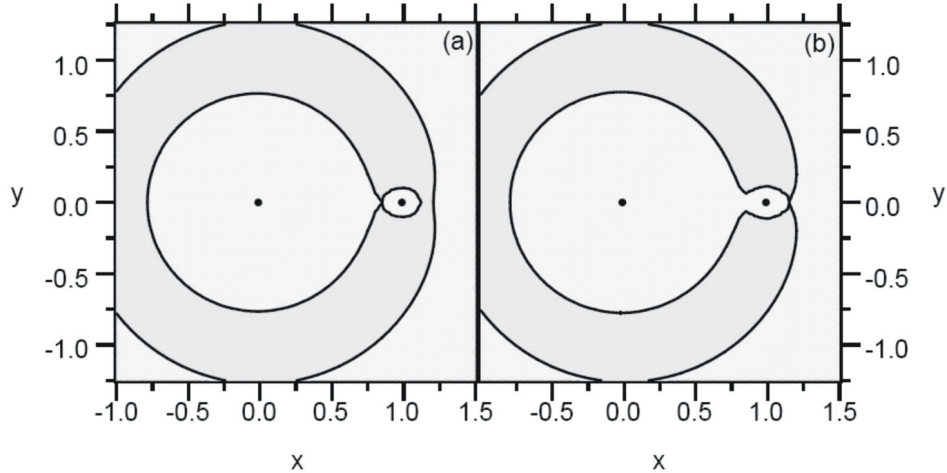

Figure 3. Zero-Velocity curves for the Earth-Moon system. In each plot Earth is the dot at the left, Moon the dot at the right. The allowed regions are lightly shaded, the forbidden regions are darker. (a) $h \approx -1.5941$. (b) $h \approx -1.5860$.

of the $(y, \dot{y})$ plane onto itself, for which the initial condition $(y_0, \dot{y}_0)$ was closed to a fixed point $(y^*, \dot{y}^*)$ satisfying $M(y^*, \dot{y}^*) = (y^*, \dot{y}^*)$. The initial condition was then used to seed a Newton-Raphson algorithm (Press *et. al.* 1992) locating the fixed point to a relative precision of $10^{-12}$. Each of these fixed points corresponds to a periodic orbit of the dynamics described by Eqs. (2,3). After pruning of repeated orbits, we were left with a set of 287 initial conditions giving transfer orbits which were periodic with a relative precision of $10^{-12}$. These initial conditions are plotted in the period-energy (*T-h*) plane in Fig. 4, where their ordering on families is already apparent.

## 3.2 Family reconstruction: analytical continuation

Once the initial search for POs was completed, we proceeded to reconstruct the monoparametric family to which each of them belonged. To this end we proceeded to analytically continue each of these POs in the parameter $h$, in both the two possible directions.

For each PO we started by varying $h$ in steps $\Delta h = 10^{-8}$, using the previous fixed point on $\Sigma_1$ to seed a Newton-Raphson algorithm which reconverged it to the new fixed point with relative precision $10^{-9}$. At each step of this algorithm, the necessary integrations of Eqs. (2,3) were carried out by a variable-step Bulirsch-Stoer algorithm with relative precision $10^{-14}$, as before.

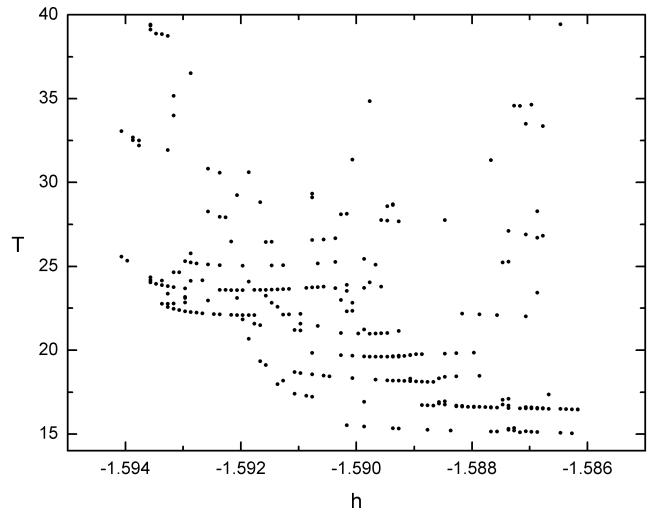

Figure 4. Period-energy ($T$ - $h$) plot of the 287 periodic orbits found in the initial numerical search.

After having accumulated a few points along a branch of a family (usually six), we accelerated the continuation by implementing a polynomial extrapolation (Press *et. al.* 1992) using the last few points computed to the obtain a seed for the next one. This allowed us to take much larger increments in $h$, typically as large as $10^{-5}$.

In all cases the continuation was ended when, even taking a very small $\Delta h$, the Newton-Raphson algorithm failed to converge in less than 200 steps. This happened either when the POs neared collision with a primary, or when the unstable eigenvalue of the linearized map became too large. In the first case, no attempt was made to regularize the orbit past the collision. In the second case, the high instability of the orbits led to a large accumulation of numerical errors, either precluding their integration to the next intersection with $\Sigma_1$, or leading to an intersection so far from the fixed point that the Newton-Raphson algorithm failed to converge from it.

In several cases the continuation reached a branching point of the family corresponding to a bifurcation orbit (Hénon 1997). In these cases, standard techniques (Parker and Chua 1989) were implemented to find all of the emanating branches. Each new branch found was then continued away from the bifurcation by the same methods described above. In the cases where the branching was simply a return point in $h$ (the family became singular in $h$) we just switched temporarily to analytical continuation in $y$ to allow continuation into the next branch.

It is evident from Fig. 4 that in many cases several of the POs found in the initial search belong to the same family. Whenever this happened we arbitrarily selected one of them as the PO from which the continuation was started.

## 3.3 Atlas completion: use of symmetries

The CR3BP equations of motion (2,3) are invariant under the discrete symmetry

$$S : (x, y, \dot{x}, \dot{y}, t) \rightarrow (x, -y, -\dot{x}, \dot{y}, t) \qquad (6)$$

So applying $S$ to a given PO will give another (or possibly the same) PO. Following Hénon (1997), when referring to POs two meanings of symmetry will be used, which must be carefully distinguished. On one hand, we will call a PO *symmetric* if applying $S$ to it gives the same PO (apart of a time translation), and *asymmetric* otherwise. On the other hand, the PO obtained by applying $S$ to a given one will be called its *symmetrical*. Thus two asymmetric POs can be the symmetricals of each other, but a symmetric PO is always its own symmetrical.

Also following Hénon (1997), we will call a family *symmetric* if it is invariant under $S$, that is, if applying $S$ to any PO of the family gives a PO belonging to the same family. Otherwise, the family is called *asymmetric*. Thus we have three possible cases:

- a symmetric family of symmetric orbits (abbreviated Ss);
- a symmetric family of asymmetric orbits (abbreviated Sa);
- an asymmetric family of asymmetric orbits (abbreviated Aa).

The function $\vec{\phi}(t, h)$ characterizing a family of POs is a monoparametric family of curves $\vec{\phi}(t)$ in the phase-space, parameterized by $h$. In general the parameter $h$ will not vary smoothly over entire family (Hénon 1997), and two or more POs will then correspond to the same value of $h$. These POs are said to lie on different *family segments*, along each of which $h$ varies smoothly.

In completing the Atlas presented here, the symmetry properties of POs and families have been used in two ways:

- for an asymmetric family of asymmetric orbits, applying $S$ to the whole family leads to a different one, which is also an asymmetric family of asymmetric orbits;
- for a symmetric family of asymmetric orbits, applying $S$ to one family segment gives another segment of the same family (see Fig. 5).

An example of first case is the family 027 in the Atlas, obtained by applying $S$ to the family 026. Though the number assigned to these families shows that POs belonging to both were found in the initial survey, the family 027 had not been amenable to reconstruction by analytical continuation, since the only orbits we had pertaining to it were too unstable for this method to converge.

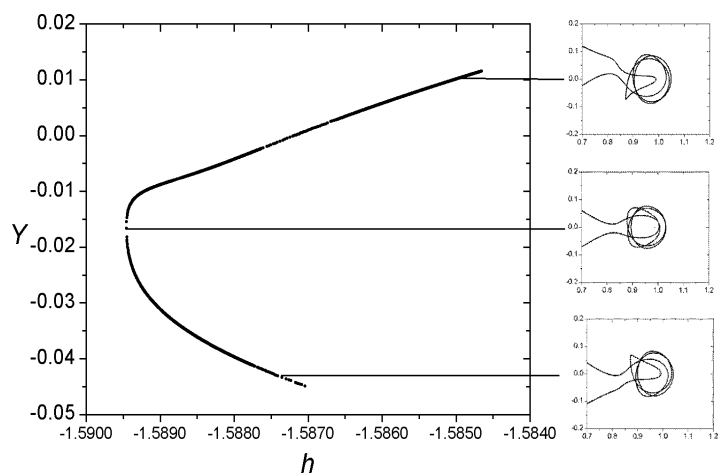

Figure 5. *Left:* The $y(h)$ characteristic for a symmetric family of asymmetric orbits (Sa) comprising two segments of mutually symmetrical orbits. *Right:* Circumlunar regions of one orbit on each segment (top and bottom) and the symmetric orbit where they join (center).

## 3.4 Atlas completion: period doubling

Due to the chosen limitation to periods $T \leq 40$, most families presented in this work are 1-periodic on the section $\Sigma_1$. Exceptions are the families 026, 027, and 300 A-B-C-D, which are 2-periodic on $\Sigma_1$. Although the families 026 and 027 were found in the initial search, the family 300 A arose from an intersection with family 197 A through a period doubling: at the intersection, the orbit of family 300 A is the same as that of family 197 A repeated twice.

Additional candidates to generate new families with $T \leq 40$ through period doubling are families 357, 037, 043, 056, 053, 084, 077, 180 A-B, 197 B-C-D, and 146 A-B-C. This possibility has not been yet exploited, but is the subject of current research; the results will be added to this Atlas as soon as they become available.

## 3.5 Atlas completion: continuation in $\mu$

If instead of continuating a given PO in $h$ we perform analytical continuation in the mass parameter $\mu$, we obtain a monoparametric family of POs parameterized by $\mu$ (Hénon 1997, Bruno and Varin 2006). Analytical continuation of a whole PO family parameterized by $h$ thus leads to a *biparametric* family of POs parameterized by both $h$ and $\mu$.

As part of research currently under way, we have recently performed such a continuation in $\mu$ for the Atlas families 037, 043, and 056, from the value of $\mu$ corresponding to the Earth-Moon system towards zero. We found that these three families are parts of a single biparametric family, having a very complicated structure, which shows several bifurcations, intersections and reconnections between families parameterized by $h$ as $\mu$ is varied. Somewhat unexpectedly, we discovered a new family in $h$ belonging to this same biparametric family; we then continuated it in $\mu$ up to the Earth-Moon value, and added it to the Atlas as family 357.

Families 037 and 043 were initially selected for continuation in $\mu$ because, though they are both asymmetric families of asymmetric orbits, they are symmetrical to each other, suggesting the possibility of a connection like the one we found. Their connection to family 056, which is a symmetric family of symmetric orbits, was discovered through the continuation in $\mu$, though in hindsight the similarity in period, energy and morphology between the POs of family 056 and those of families 037 and 043 makes it somewhat to be expected.

This raises the possibility of finding an equivalent connection between other pairs of asymmetric families of asymmetric orbits in the Atlas which are symmetrical to each other, probably including a symmetric family of symmetric orbits similar to them in period, energy and morphology. The possibility of discovering some additional families in $h$ currently overlooked in the Atlas seems also very attractive. Some good candidates for this search are the families 077 and 084, and families 146 A and B. This is the subject of current research, and the results will be added to this Atlas as soon as they become available.

# 4. Structure of the Atlas

The data for each family (or branch in the case of bifurcations) in the Atlas, are structured as follows:

**Family number:** To each family we assign a three-digit number taken from the PO found in the initial search (Sec. 3.1) which we used as starting point for the analytical continuation (Sec. 3.2). This number simply denotes the order in which the POs were found in the initial search, and can be considered arbitrary. Families having numbers larger than 287, which is the number of POs found in the initial search, have been found by the techniques described in Secs. 3.3, 3.4, or 3.5; some families with numbers below 287 were also reconstructed with the help of these techniques (*e.g.* family 027, see Sec. 3.3).

**Branches:** When a family has bifurcations, all branches found have been assigned the same family number, and are distinguished by appended capital letters (A, B, etc.). In general, the reference branch A consists of symmetric orbits, and the remaining ones of asymmetric orbits.

**Symmetry properties:** For each family, and for each branch when bifurcations are present, the symmetry of both the family or branch, and of the orbits, is stated next to its number.

**Energy and period ranges:** For each family (or branch) we give the lowest and highest energies found along it, denoted respectively by $h_{min}$ and $h_{max}$. We also give the shortest and longest periods found along it, denoted respectively by $T_{min}$ and $T_{max}$. It must be kept in mind that in many cases the continuation has not been carried to a natural termination of the family (see Sec. 3.2), so these numbers should be considered just indicative.

**Branching points:** For families having bifurcations, the branching points are designated by $P_1$, $P_2$, etc. For each branching point we give the energy $h$, the period $T$, and the initial conditions $y$, $\dot{x}$, and $\dot{y}$ for the corresponding bifurcation orbit. The branching points are always taken on the Poincaré section $\Sigma_1$, so $x = 0.836915310$ for all of them.

**Characteristic curves:** For each family, and for each branch when bifurcations are present, we show the following graphs (see Fig.6 for an example):

- the period vs. energy curve $T(h)$;
- the characteristics $y(h)$, $\dot{y}(h)$, and $\dot{x}(h)$ on $\Sigma_1$;
- the stability index $sign(\lambda_u)\log\|\lambda_u\|$ vs. $h$;
- characteristics $h(x_i)$ on $\Sigma_2$.

The stability index is defined from the unstable eigenvalue $\lambda_u$ of the linearized Poincaré map on $\Sigma_1$. It allows to distinguish between elliptic (zero), ordinary hyperbolic (positive) and reflection hyperbolic (negative) points. For orbits showing large excursions in the value of $\lambda_u$, it also allows to show details of the stability behaviour, which would be lost in a linear scale graph. Note that the indeterminacy of the sign function for complex eigenvalues is inconsequential, since for these cases $\log\|\lambda_u\|$ vanishes.

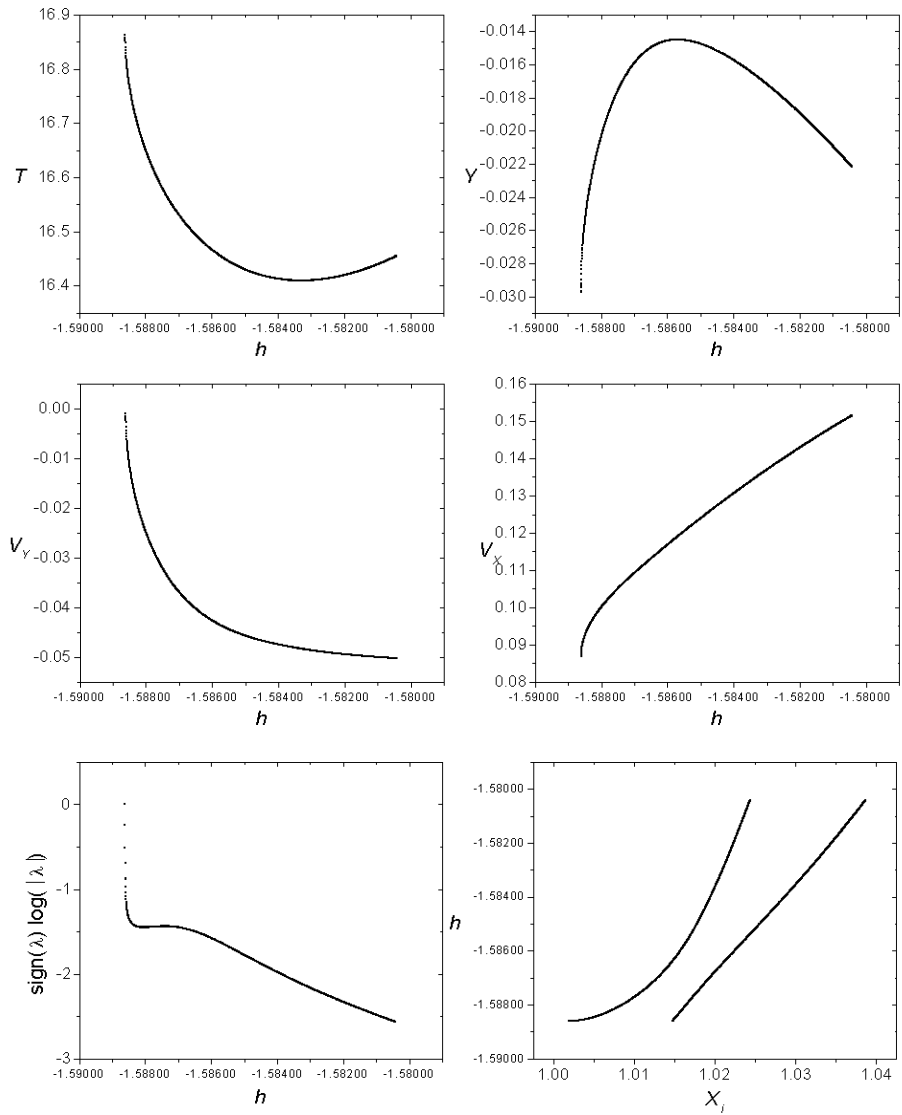

Figure 6. Example of characteristic curves $T(h)$, $y(h)$, $v_y(h)$, and $v_x(h)$ on $\Sigma_1$, a stability index curve, and a characteristic curve $x_i(h)$ on $\Sigma_2$.

The characteristics $h(x_i)$ are defined on a new Poincaré section $\Sigma_2 = \{ x, y = 0, v_x, v_y > 0\}$ and show only the (possibly several) intersections $x_i$ with the orbit for $x_i > 0.7$. These characteristics are similar to those in Szebehely (1967), and are useful to follow the evolution in the orbit shape in the circumlunar region.

For families having bifurcations, we additionally show graphs of $T(h)$, $y(h)$, and $h(x_i)$ for all branches together, identifying the branching points and showing the bifurcation structure (see Fig. 7 for an example).

We also show graphs of $T(h)$, $y(h)$, $\dot{y}(h)$, $\dot{x}(h)$, $\dot{y}(y)$ and $\dot{y}(\dot{x})$ grouping the curves of all families in the Atlas.

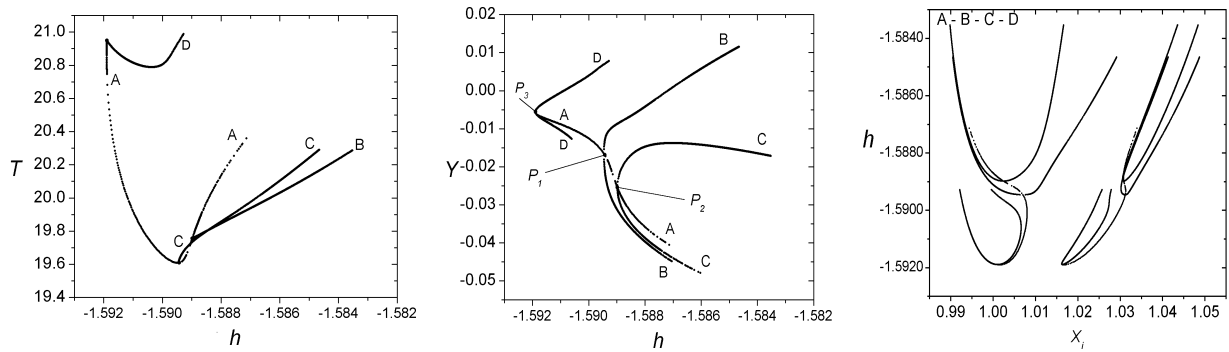

Figure 7. Example graphs of $T(h)$, $y(h)$, and $h(x_i)$ for all branches of a bifurcating family.

**Orbit geometry:** For all families we show graphs of the geometrical shape of one or more full orbits in the $(x, y)$ plane, at some selected values of $h$. We also show details of the circumlunar regions, in this case for all branches if there is more than one, for this is the region where the orbit shape varies widely with $h$ and between branches.

**Reference points:** For each family, and for each branch when bifurcations are present, we give full numerical data (initial conditions) for a reference PO in that family or branch.

All the data mentioned are given in ascending order of the minimal period $T_{\min}$.

# 5. Description of all families

We begin by listing the symmetries and minimal period (Table 1) and the initial conditions for a reference orbit (Table 2) for each branch of each family in the Atlas.

Table 1. Symmetries and minimal period $T_{min}$ for each family in the Atlas.

| Family | Symmetry | $T_{min}$ | Family | Symmetry | $T_{min}$ |
|---|---|---|---|---|---|
| 357 | Ss | 14.487454 | 286 B | Sa | 26.022460 |
| 037 | Aa | 14.786812 | 171 | Ss | 26.459669 |
| 043 | Aa | 14.786812 | 020 | Aa | 26.483357 |
| 056 | Ss | 14.975518 | 021 | Aa | 26.483355 |
| 053 | Ss | 16.220757 | 013 | Ss | 26.574839 |
| 084 | Aa | 16.409829 | 222 A | Ss | 27.850288 |
| 077 | Aa | 16.409829 | 222 B | Sa | 28.271856 |
| 180 A | Ss | 16.459140 | 081 A | Ss | 27.684763 |
| 180 B | Aa | 17.915057 | 081 B | Sa | 28.274740 |
| 197 A | Ss | 18.122697 | 081 C | Sa | 28.637522 |
| 197 B | Sa | 18.450933 | 200 | Ss | 28.827241 |
| 197 C | Sa | 18.148249 | 172 | Ss | 29.220943 |
| 197 D | Sa | 19.401657 | 209 | Aa | 30.572004 |
| 146 A | Ss | 19.607828 | 232 | Aa | 30.572004 |
| 146 B | Sa | 19.608057 | 058 A | Ss | 30.966292 |
| 146 C | Sa | 19.755857 | 058 B | Sa | 31.347506 |
| 146 D | Sa | 20.788368 | 263 A | Ss | 31.933083 |
| 178 | Ss | 20. 992907 | 263 B | Sa | 32.489625 |
| 136 A | Ss | 21.010742 | 287 | Ss | 32.562769 |
| 136 B | Sa | 21.115366 | 032 A | Ss | 33.370152 |
| 144 A | Ss | 21.614173 | 032 B | Sa | 33.891167 |
| 144 B | Sa | 23.636883 | 254 | Aa | 33.984921 |
| 187 A | Ss | 22.099407 | 255 | Aa | 33.984922 |
| 187 B | Sa | 22.743109 | 026 | Aa | 34.017748 |
| 147 | Ss | 22.350225 | 027 | Aa | 34.017748 |
| 194 | Aa | 22.812098 | 044 | Aa | 34.285575 |
| 244 | Aa | 22.812096 | 045 | Aa | 34.289797 |
| 305 | Ss | 23.268005 | 132 | Aa | 34.758144 |
| 018 A | Ss | 23.425287 | 133 | Aa | 34.758144 |
| 018 B | Sa | 25.007851 | 256 A | Ss | 35.055198 |
| 188 A | Ss | 23.585381 | 256 B | Sa | 35.263374 |
| 188 B | Sa | 24.137851 | 300 A | Ss | 36.245398 |
| 157A | Ss | 23.706160 | 300 B | Sa | 36.506826 |
| 157 B | Sa | 24.653557 | 300 C | Ss | 36.260612 |
| 250 | Aa | 24.659896 | 300 D | Ss | 36.625264 |
| 251 | Aa | 24.659896 | 238 | Ss | 36.403708 |
| 302 | Aa | 24.955219 | 262 A | Ss | 38.707003 |
| 301 | Aa | 25.033123 | 262 B | Sa | 39.357787 |
| 192 | Ss | 25.039939 | 262 C | Sa | 38.815998 |
| 286 A | Ss | 25.304381 | 008 | Ss | 39.423500 |

Table 2. Initial conditions for a reference orbit on each segment of each family in the Atlas.

| Family | h $/$ T | x $/$ y | vx $/$ vy |
|---|---|---|---|
| 357 | -0.1553849931959387E+01 | 0.8369153095696800E+00 | 0.1882861991773726E+00 |
|  | 0.1535213364809199E+02 | -0.1171235689440371E+00 | -0.5721969437090824E-01 |
|  | -0.1580149826694191E+01 | 0.8369153095696800E+00 | 0.7032232204724576E-01 |
|  | 0.1495200383953257E+02 | -0.7523099816467904E-01 | 0.5260578355402396E-01 |
| 037 | -0.1557940386649792E+01 | 0.8369153095696800E+00 | 0.1931044766332113E+00 |
|  | 0.1550020353331435E+02 | -0.9526883178836698E-01 | -0.6987584622079623E-01 |
|  | -0.1586516528824736E+01 | 0.8369153095696800E+00 | 0.6685301616472657E-01 |
|  | 0.1550080549180712E+02 | -0.4943421181923025E-01 | 0.3675197350381332E-01 |
| 043 | -0.1557943386649809E+01 | 0.8369153095696800E+00 | 0.1747791691425595E+00 |
|  | 0.1550004613291211E+02 | 0.7563171121319101E-01 | -0.1462492872874179E+00 |
|  | -0.1587200386649491E+01 | 0.8369153095696800E+00 | 0.7120385411381607E-01 |
|  | 0.1539115344522493E+02 | 0.1475375273725499E-01 | -0.8929019801282878E-01 |
| 056 | -0.1588349960008820E+01 | 0.8369153095696800E+00 | 0.6066256122650340E-01 |
|  | 0.1620482478505905E+02 | 0.6242157762308409E-02 | -0.8831464679343704E-01 |
|  | -0.1589645386649505E+01 | 0.8369153095696800E+00 | 0.6968630586685100E-01 |
|  | 0.1540240149361518E+02 | -0.1334575082907463E-03 | -0.6475912987563248E-01 |
| 053 | -0.1587528386649445E+01 | 0.8369153095696800E+00 | 0.9693928261096405E-01 |
|  | 0.1680918745218759E+02 | -0.3099238100450918E-01 | 0.2853422136741557E-02 |
|  | -0.1587160386649504E+01 | 0.8369153095696800E+00 | 0.8253664490065066E-01 |
|  | 0.1720004312056572E+02 | -0.3972985744555868E-01 | 0.3064538598273736E-01 |
| 084 | -0.1587650386649494E+01 | 0.8369153095696800E+00 | 0.1037082790039896E+00 |
|  | 0.1660054767367177E+02 | -0.1828051285584564E-01 | -0.3019245820910044E-01 |
|  | -0.1588053364091826E+01 | 0.8369153095696800E+00 | 0.7355944416921466E-01 |
|  | 0.1703894171696143E+02 | -0.3948519252295191E-01 | 0.2506517857797677E-01 |
| 077 | -0.1587690386649495E+01 | 0.8369153095696800E+00 | 0.9635113669137989E-01 |
|  | 0.1660578023423538E+02 | 0.6675512817859719E-03 | -0.6061856325980430E-01 |
|  | -0.1572663481926473E+01 | 0.8369153095696800E+00 | 0.1435130650304809E+00 |
|  | 0.1690770453287855E+02 | 0.4249424271921498E-01 | -0.1236512305582949E+00 |
| 180 A | -0.1590871315343652E+01 | 0.8369153095696800E+00 | 0.4899890348953655E-01 |
|  | 0.1848613856454854E+02 | -0.1765821750574048E-02 | -0.6468630844164473E-01 |
|  | -0.1591217886649474E+01 | 0.8369153095696800E+00 | 0.5807507130564620E-01 |
|  | 0.1753808067116183E+02 | -0.4584351507720186E-02 | -0.4944867242757869E-01 |
| 180 B | -0.1591320049933672E+01 | 0.8369153095696800E+00 | 0.4877881292126878E-01 |
|  | 0.1803179957207295E+02 | -0.3216750199024787E-02 | -0.5725728187576357E-01 |
|  | -0.1591356179789391E+01 | 0.8369153095696800E+00 | 0.5028623748055229E-01 |
|  | 0.1803444114349784E+02 | -0.3709002918987459E-02 | -0.5516043108275071E-01 |
| 197 A | -0.1591534233746611E+01 | 0.8369153095696800E+00 | 0.4677586935493131E-01 |
|  | 0.1990409728772504E+02 | -0.4033896512295243E-02 | -0.5492598760984180E-01 |
|  | -0.1590804386649470E+01 | 0.8369153095696800E+00 | 0.7060539590575120E-01 |
|  | 0.1857121166664419E+02 | -0.7488597361791083E-02 | -0.3891857155846878E-01 |
| 197 B | -0.1588158386649521E+01 | 0.8369153095696800E+00 | 0.7902576165227528E-01 |
|  | 0.1846291802803064E+02 | -0.3582306666752050E-01 | 0.2534035476950286E-01 |
|  | -0.1588310386649478E+01 | 0.8369153095696800E+00 | 0.9446825844325772E-01 |
|  | 0.1845614355380422E+02 | -0.2621319483400342E-01 | 0.4260564831710667E-03 |
| 197 C | -0.1589075386649491E+01 | 0.8369153095696800E+00 | 0.9032508725967645E-01 |
|  | 0.1831088254713460E+02 | -0.4367194730625190E-02 | -0.4418503137018659E-01 |
|  | -0.1589073386649482E+01 | 0.8369153095696800E+00 | 0.9196108281116112E-01 |
|  | 0.1816602136636390E+02 | -0.9549369378169526E-02 | -0.3687704582003382E-01 |
| 197 D | -0.1590725942073130E+01 | 0.8369153095696800E+00 | 0.4800846082984926E-01 |
|  | 0.1946665328443075E+02 | 0.1378305921230418E-03 | -0.6770522906295484E-01 |
|  | -0.1591028442065470E+01 | 0.8369153095696800E+00 | 0.4676325477267651E-01 |
|  | 0.1949867690962299E+02 | -0.1109138777605878E-02 | -0.6396862533228709E-01 |
| 146 A | -0.1591893392692515E+01 | 0.8369153095696800E+00 | 0.4815159550464603E-01 |
|  | 0.2092593115377287E+02 | -0.5618439132113275E-02 | -0.4587594951423138E-01 |
|  | -0.1591890747692515E+01 | 0.8369153095696800E+00 | 0.4790090879221719E-01 |
|  | 0.2094883850543574E+02 | -0.5568775499890479E-02 | -0.4621982396706772E-01 |
| 146 B | -0.1589213386649455E+01 | 0.8369153095696800E+00 | 0.9225793685920553E-01 |
|  | 0.1969192689242335E+02 | -0.1000036488298939E-01 | -0.3144613798866303E-01 |
|  | -0.1589075186649445E+01 | 0.8369153095696800E+00 | 0.8003401269921176E-01 |
|  | 0.1971851609173958E+02 | -0.3012325404188321E-01 | 0.1074650703682067E-01 |

| Family | $h$ $T$ | $x$ $y$ | $vx$ $vy$ |
|---|---|---|---|
| *146 C* | -0.1588860386649483E+01 0.1976641840116849E+02 | 0.8369153095696800E+00 -0.2096590403532161E-01 | 0.9350465317792127E-01 -0.8727745756317518E-02 |
| | -0.1588679386649491E+01 0.1978309021887632E+02 | 0.8369153095696800E+00 -0.3288082164004175E-01 | 0.7905407945234576E-01 0.1952100717088873E-01 |
| *146 D* | -0.1590894168192515E+01 0.2080771022440213E+02 | 0.8369153095696800E+00 -0.1097469453844724E-01 | 0.7042057008683775E-01 -0.3309882881620752E-01 |
| | -0.1590851667992515E+01 0.2080494339742744E+02 | 0.8369153095696800E+00 0.8907003238434186E-04 | 0.4701572233939809E-01 -0.6653515057514348E-01 |
| *178* | -0.1591978386649499E+01 0.2185896545716157E+02 | 0.8369153095696800E+00 -0.6784203181706472E-02 | 0.5179565710415293E-01 -0.3886598514191153E-01 |
| | -0.1592037742267489E+01 0.2220079038884092E+02 | 0.8369153095696800E+00 -0.6025971335922864E-02 | 0.4686797409451426E-01 -0.4379758446277871E-01 |
| *136 A* | -0.1591051386649485E+01 0.2164313022727671E+02 | 0.8369153095696800E+00 0.1490013646822186E-01 | 0.5166358185363410E-01 0.5151525155512848E-01 |
| | -0.1591271658727424E+01 0.2239865141919652E+02 | 0.8369153095696800E+00 0.2322732638920666E-01 | 0.4529973545830987E-01 0.3926096309974635E-01 |
| *136 B* | -0.1589547564949489E+01 0.2140407128696851E+02 | 0.8369153095696800E+00 -0.1421170068091016E-01 | 0.6159039744724075E-01 0.6796577273452353E-01 |
| | -0.1589525935943242E+01 0.2171629826333433E+02 | 0.8369153095696800E+00 -0.2092941350001154E-01 | 0.6361852654138159E-01 0.5871209453514303E-01 |
| *144 A* | -0.1590282386649499E+01 0.2300017727749036E+02 | 0.8369153095696800E+00 0.3447487101916202E-01 | 0.5124877789446881E-01 -0.1949228778295975E-01 |
| | -0.1588712153442192E+01 0.2400131994738804E+02 | 0.8369153095696800E+00 0.4596725402882385E-01 | 0.4973190886169957E-01 0.1300537675507485E-01 |
| *144 B* | -0.1589457264791053E+01 0.2364662444809076E+02 | 0.8369153095696800E+00 0.4277812138995607E-01 | 0.4565604206386054E-01 0.1107420627403144E-01 |
| | -0.1589203869779061E+01 0.2364096509435769E+02 | 0.8369153095696800E+00 0.4111377880355605E-01 | 0.5584903670330011E-01 -0.1099800548212215E-01 |
| *187 A* | -0.1592450386649498E+01 0.2215284432733032E+02 | 0.8369153095696800E+00 -0.1007327286455600E-01 | 0.4937089063761972E-01 0.2414262114340932E-01 |
| | -0.1593361584896506E+01 0.2275405307944939E+02 | 0.8369153095696800E+00 0.1057949157989069E-02 | 0.3452183203054967E-01 0.2034701552463669E-01 |
| *187 B* | -0.1592484386649499E+01 0.2299124534981404E+02 | 0.8369153095696800E+00 -0.9998054501750418E-02 | 0.5328626757033673E-01 0.1091561940583316E-01 |
| | -0.1590234285147764E+01 0.2369502584524777E+02 | 0.8369153095696800E+00 -0.2087964878169224E-01 | 0.7582926625656319E-01 0.1833098757504903E-01 |
| *147* | -0.1592107386649489E+01 0.2320016112079927E+02 | 0.8369153095696800E+00 -0.6835457934849027E-02 | 0.4945687067354360E-01 -0.3855494689967946E-01 |
| | -0.1592136265908004E+01 0.2336835911930623E+02 | 0.8369153095696800E+00 -0.6454209393668364E-02 | 0.4693487895964341E-01 -0.4114349969801023E-01 |
| *194* | -0.1592878998464598E+01 0.2301221993318838E+02 | 0.8369153095696800E+00 -0.1973400183318042E-02 | 0.3862452328281195E-01 0.3278372086883730E-01 |
| | -0.1592809604911915E+01 0.2335479213615499E+02 | 0.8369153095696800E+00 0.1013137622084948E-01 | 0.3583157390494823E-01 0.3183035986136762E-01 |
| *244* | -0.1592938917782559E+01 0.2324096048277420E+02 | 0.8369153095696800E+00 -0.4541217497065819E-02 | 0.2472542930143639E-01 -0.4202507162548586E-01 |
| | -0.1592925527275155E+01 0.2304977905784814E+02 | 0.8369153095696800E+00 -0.4815814927688233E-02 | 0.2416324216252356E-01 -0.4254090520800313E-01 |
| *305* | -0.1590460875111452E+01 0.2423902278038772E+02 | 0.8369153095696800E+00 0.3689276031620281E-01 | 0.4196341010629214E-01 -0.1487725277489406E-01 |
| | -0.1590451647388306E+01 0.2390481675359980E+02 | 0.8369153095696800E+00 0.3507482594607048E-01 | 0.4410521203381093E-01 -0.2367925237553692E-01 |
| *018 A* | -0.1589904073050286E+01 0.2500023632541410E+02 | 0.8369153095696800E+00 0.3893206713224617E-01 | 0.4998751021796083E-01 0.2111916205189246E-02 |
| | -0.1589600386649266E+01 0.2400103445230148E+02 | 0.8369153095696800E+00 0.3394098505436549E-01 | 0.6266530658628690E-01 -0.2419618807560653E-01 |
| *018 B* | -0.1584599950525309E+01 0.2539185874675518E+02 | 0.8369153095696800E+00 0.5667115856948612E-01 | 0.8253459621215227E-01 -0.1000491735351261E-01 |
| | -0.1589646544462673E+01 0.2501745799360442E+02 | 0.8369153095696800E+00 0.4088568665056691E-01 | 0.4822238156595812E-01 0.1000491491150513E-01 |
| *188 A* | -0.1593543337488764E+01 0.2435912802205522E+02 | 0.8369153095696800E+00 -0.8454724863081436E-02 | 0.2359307653638891E-01 -0.2003962929636598E-01 |
| | -0.1591730386649494E+01 0.2359668923213236E+02 | 0.8369153095696800E+00 -0.1514767044669224E-01 | 0.6272107545685496E-01 0.7455488248810926E-03 |

| Family | $h$ $T$ | $x$ $y$ | $vx$ $vy$ |
|---|---|---|---|
| 188 B | -0.1592745886662465E+01 0.2415614839483149E+02 | 0.8369153095696800E+00 -0.5483103513041265E-02 | 0.3063249399388768E-01 -0.4226182486620905E-01 |
| | -0.1593224386649501E+01 0.2414056612238386E+02 | 0.8369153095696800E+00 -0.1001075508652482E-01 | 0.3827277559173141E-01 -0.3567026344702606E-02 |
| 157A | -0.1592191964353470E+01 0.2465847307455832E+02 | 0.8369153095696800E+00 -0.6430848378560066E-02 | 0.4556676874154115E-01 -0.4134233474790096E-01 |
| 157 B | -0.1592060214353470E+01 0.2465517039100745E+02 | 0.8369153095696800E+00 -0.5000523068913410E-02 | 0.4274824297693070E-01 -0.4784657264146652E-01 |
| | -0.1592058214553470E+01 0.2465513410207921E+02 | 0.8369153095696800E+00 -0.8033158976711974E-02 | 0.5198325358494582E-01 -0.3542491854516428E-01 |
| 250 | -0.1593246274154317E+01 0.2497331486356555E+02 | 0.8369153095696800E+00 -0.5003518489752779E-02 | 0.2810857346789752E-01 -0.3089289846718568E-01 |
| | -0.1593285928010960E+01 0.2466761597220892E+02 | 0.8369153095696800E+00 -0.6003726901685158E-02 | 0.3861127072304951E-01 -0.1134642052163021E-01 |
| 251 | -0.1593403856664003E+01 0.2490038203735014E+02 | 0.8369153095696800E+00 0.1276841068222412E-01 | 0.2498820522384168E-01 0.1534849683566719E-01 |
| | -0.1593564006664003E+01 0.2477498087362493E+02 | 0.8369153095696800E+00 0.1092522300066126E-01 | 0.2332803442468993E-01 0.1323548863399246E-01 |
| 302 | -0.1590330819854882E+01 0.2558158666970328E+02 | 0.8369153095696800E+00 0.3402421599477634E-01 | 0.5179559494248327E-01 -0.1862081902402224E-01 |
| 301 | -0.1590401296616310E+01 0.2567052447448607E+02 | 0.8369153095696800E+00 0.3551658888694931E-01 | 0.4401105657723107E-01 -0.2342974251052965E-01 |
| 192 | -0.1592388896568739E+01 0.2606872695654292E+02 | 0.8369153095696800E+00 -0.6479046670607502E-02 | 0.3556858138560868E-01 -0.4608639682378168E-01 |
| | -0.1592170386649502E+01 0.2505236653037481E+02 | 0.8369153095696800E+00 -0.1128091847639889E-01 | 0.5718451492239828E-01 -0.1428337208840003E-01 |
| 286 A | -0.1594114797979418E+01 0.2605582221146800E+02 | 0.8369153095696800E+00 0.2600411699121463E-02 | 0.9116567315471112E-02 -0.1559671296163800E-03 |
| | -0.1594120310970987E+01 0.2639626815475656E+02 | 0.8369153095696800E+00 0.3035282692414990E-02 | 0.7782215519529092E-02 -0.1177693047825480E-02 |
| 286 B | -0.1594090980356078E+01 0.2611118837118277E+02 | 0.8369153095696800E+00 0.1891857299902256E-02 | 0.1194012206700557E-01 0.1184412500323528E-02 |
| | -0.1594035872356078E+01 0.2603994436465331E+02 | 0.8369153095696800E+00 0.6683084095436609E-02 | 0.8861025727436035E-02 -0.2345070984638653E-02 |
| 171 | -0.1592513386649478E+01 0.2658399509189787E+02 | 0.8369153095696800E+00 -0.8956787716138837E-02 | 0.4915329117804571E-01 -0.2379022087884103E-01 |
| | -0.1589594386649519E+01 0.2700420521305784E+02 | 0.8369153095696800E+00 -0.2515798814606926E-01 | 0.8054982942395600E-01 0.9210655841671134E-02 |
| 020 | -0.1586420386649488E+01 0.2656049367597927E+02 | 0.8369153095696800E+00 0.2754316985731348E-01 | 0.1079604793100152E+00 0.2764373045552931E-01 |
| | -0.1586844750431270E+01 0.2750128280470764E+02 | 0.8369153095696800E+00 0.4021829245397690E-01 | 0.9068800368423402E-01 0.3160058334041871E-02 |
| 021 | -0.1586075386629491E+01 0.2650410724308010E+02 | 0.8369153095696800E+00 0.4258664566081234E-01 | 0.7272690130733689E-01 0.6118059374460039E-01 |
| | -0.1586822770431270E+01 0.2752140194111188E+02 | 0.8369153095696800E+00 0.4733357023119811E-01 | 0.6185873362203014E-01 0.4609221610984804E-01 |
| 013 | -0.1587262386649484E+01 0.2702272086527108E+02 | 0.8369153095696800E+00 0.3898552763201312E-01 | 0.7444367877202124E-01 0.4721042874942594E-01 |
| | -0.1587041766976160E+01 0.2807470564823058E+02 | 0.8369153095696800E+00 0.4648159348972492E-01 | 0.6560150705268036E-01 0.3877992156563437E-01 |
| 222 A | -0.1592548966629492E+01 0.2843779399125070E+02 | 0.8369153095696800E+00 -0.7143864895502006E-02 | 0.4001932030199090E-01 -0.3781440526517876E-01 |
| | -0.1592311386649489E+01 0.2794133175124106E+02 | 0.8369153095696800E+00 -0.8844517184435796E-02 | 0.5202192050927629E-01 -0.2622960996204105E-01 |
| 222 B | -0.1592119386449492E+01 0.2841905600839925E+02 | 0.8369153095696800E+00 -0.1098389999142135E-01 | 0.5520631374785395E-01 -0.2357080595390426E-01 |
| | -0.1592281386649492E+01 0.2836768900823477E+02 | 0.8369153095696800E+00 -0.5124959324225742E-02 | 0.3915370362386649E-01 -0.4621830192043142E-01 |
| 081 A | -0.1586358386660272E+01 0.2895462990143625E+02 | 0.8369153095696800E+00 0.3608658400881026E-01 | 0.9641351845778733E-01 -0.3344317948835496E-01 |
| | -0.1588640386649493E+01 0.2773086555469934E+02 | 0.8369153095696800E+00 0.3024358440015540E-01 | 0.8493372391447379E-01 -0.1217183723961357E-01 |

| Family | $h$ $T$ | $x$ $y$ | $vx$ $vy$ |
|---|---|---|---|
| 081 B | -0.1586748386649500E+01 0.2830552151358616E+02 | 0.8369153095696800E+00 0.3622436265996543E-01 | 0.9247993276871216E-01 -0.3229586692271269E-01 |
| | -0.1586982474751493E+01 0.2828669733375939E+02 | 0.8369153095696800E+00 0.3354310103126527E-01 | 0.9548052372690739E-01 -0.2713411359213236E-01 |
| 081 C | -0.1589060510443014E+01 0.2881899026175200E+02 | 0.8369153095696800E+00 0.3968026547503412E-01 | 0.5765352118801918E-01 0.2534255838097366E-01 |
| | -0.1587816055440224E+01 0.2945978666621907E+02 | 0.8369153095696800E+00 0.4141151570525281E-01 | 0.7686656491445916E-01 -0.3789314984814496E-02 |
| 200 | -0.1592034386649461E+01 0.2899641924794470E+02 | 0.8369153095696800E+00 0.8466493311599654E-03 | 0.4858642589005642E-01 0.4368508009089278E-01 |
| 172 | -0.1592284480395803E+01 0.2932901103502085E+02 | 0.8369153095696800E+00 -0.8426368989158384E-02 | 0.5147040033914457E-01 -0.2878731353353692E-01 |
| | -0.1589840386649566E+01 0.2968513977815582E+02 | 0.8369153095696800E+00 -0.2098881590120527E-01 | 0.8272416003864516E-01 -0.3481042350884422E-02 |
| 209 | -0.1591958406649484E+01 0.3059995715851163E+02 | 0.8369153095696800E+00 -0.8453365304147198E-02 | 0.4950502452393419E-01 0.4095643031126972E-01 |
| | -0.1592457472609269E+01 0.3103011237357483E+02 | 0.8369153095696800E+00 0.1222999266425262E-01 | 0.3986090191299404E-01 0.3491698556918177E-01 |
| 232 | -0.1592479386650757E+01 0.3100101958282032E+02 | 0.8369153095696800E+00 -0.1002462670558041E-01 | 0.5420348788972473E-01 0.5325857844591306E-02 |
| | -0.1590865366649492E+01 0.3101082159053041E+02 | 0.8369153095696800E+00 -0.1896472175192954E-01 | 0.7051298743485014E-01 0.1270051910119174E-01 |
| 058 A | -0.1590010961650990E+01 0.3150197166154840E+02 | 0.8369153095696800E+00 0.3532842573406780E-01 | 0.5752422287891644E-01 0.3025337042126708E-02 |
| | -0.1588336386649453E+01 0.3112836698097849E+02 | 0.8369153095696800E+00 0.338572209468157E-01 | 0.7989485238904955E-01 -0.2606363802666500E-01 |
| 058 B | -0.1586871632976964E+01 0.3155416214079344E+02 | 0.8369153095696800E+00 0.3646706459572613E-01 | 0.8967746005340747E-01 -0.3519916040752946E-01 |
| | -0.1587140386449497E+01 0.3147856162054910E+02 | 0.8369153095696800E+00 0.3661822098614209E-01 | 0.8647519257465378E-01 -0.3497125463967166E-01 |
| 263 A | -0.1591560606663358E+01 0.3241644731706219E+02 | 0.8369153095696800E+00 -0.1670261630835324E-01 | 0.6327567084906749E-01 0.8227896765580445E-02 |
| | -0.1593739436643733E+01 0.3279126325968858E+02 | 0.8369153095696800E+00 0.477994600475619E-02 | 0.2494281475661245E-01 0.1204382967503892E-01 |
| 263 B | -0.1593663392104221E+01 0.3254249197851610E+02 | 0.8369153095696800E+00 0.6759730097794271E-02 | 0.2504928317821077E-01 0.1404452165415319E-01 |
| | -0.1593524930249493E+01 0.3259048022680659E+02 | 0.8369153095696800E+00 -0.4789090336521973E-02 | 0.3439065709748288E-01 0.3623655858103092E-02 |
| 287 | -0.1594064399149644E+01 0.3302221229709885E+02 | 0.8369153095696800E+00 -0.4914971462564522E-02 | 0.1021528996798942E-01 -0.2738335578989837E-02 |
| | -0.1594133562924557E+01 0.3441644056618387E+02 | 0.8369153095696800E+00 -0.3283097208401919E-02 | 0.2954296163168296E-02 0.4497500880290772E-02 |
| 032 A | -0.1586950386649490E+01 0.3341491029341199E+02 | 0.8369153095696800E+00 0.1626893614484964E-01 | 0.1142674781818364E+00 0.1712761021773979E-01 |
| | -0.1587123293979803E+01 0.3388226279581497E+02 | 0.8369153095696800E+00 0.2387235129579635E-01 | 0.1084740858334866E+00 -0.1202975034482634E-02 |
| 032 B | -0.1586922168779803E+01 0.3404352421248977E+02 | 0.8369153095696800E+00 0.3002054598240433E-01 | 0.1036930323298946E+00 -0.9954157557323234E-02 |
| | -0.1587059168979803E+01 0.3393530809533266E+02 | 0.8369153095696800E+00 0.2002654273954896E-01 | 0.1120856054091730E+00 0.3810674395795743E-02 |
| 254 | -0.1593109790255933E+01 0.3412035135263290E+02 | 0.8369153095696800E+00 -0.8542926285183938E-02 | 0.3205285778514753E-01 -0.2813780541796263E-01 |
| | -0.1592897904809534E+01 0.3401990875740250E+02 | 0.8369153095696800E+00 -0.1050165280298288E-01 | 0.4310102837646588E-01 -0.1520784759868654E-01 |
| 255 | -0.1593174917899376E+01 0.3400454484594235E+02 | 0.8369153095696800E+00 -0.8021869369981616E-02 | 0.2460158073291462E-01 -0.3345576281175600E-01 |
| | -0.1593130071705673E+01 0.3410050385013402E+02 | 0.8369153095696800E+00 -0.8191911130492002E-02 | 0.2503730826041883E-01 -0.3429225928925916E-01 |
| 026 | -0.1587186829790604E+01 0.3450813324725891E+02 | 0.8369153095696800E+00 -0.1167179513351546E-01 | 0.7567269040899081E-01 -0.8762368976487750E-01 |
| | -0.1586341511646894E+01 0.3474867198144949E+02 | 0.8369153095696800E+00 -0.7151597093205453E-02 | 0.7360392616494185E-01 -0.1001418498877573E+00 |

| Family | $h$<br>$T$ | $x$<br>$y$ | $vx$<br>$vy$ |
|---|---|---|---|
| 027 | -0.1586707936444758E+01<br>0.3470055676091164E+02 | 0.8369153095696800E+00<br>-0.4964189804856245E-01 | 0.6805920636129564E-01<br>0.2702465111973015E-01 |
| | -0.1587185471368537E+01<br>0.3450692281378370E+02 | 0.8369153095696800E+00<br>-0.4207424722717690E-01 | 0.8340136758100515E-01<br>0.4417483935640121E-02 |
| 044 | -0.1587378059810589E+01<br>0.3490537368636274E+02 | 0.8369153095696800E+00<br>0.4066794607240354E-02 | 0.7015303609088214E-01<br>-0.9270728761388103E-01 |
| | -0.1587367386649483E+01<br>0.3460955930235731E+02 | 0.8369153095696800E+00<br>0.2238924503231488E-02 | 0.7796263304113228E-01<br>-0.8664317532249299E-01 |
| 045 | -0.1587743636853128E+01<br>0.3469818792127938E+02 | 0.8369153095696800E+00<br>-0.3746745659702810E-01 | 0.8481342693333097E-01<br>-0.7690533516608866E-02 |
| | -0.1587333139980178E+01<br>0.3491234432807354E+02 | 0.8369153095696800E+00<br>-0.4729536144267144E-01 | 0.6764351657185290E-01<br>0.1918082941071386E-01 |
| 132 | -0.1590016994896724E+01<br>0.3510856069697225E+02 | 0.8369153095696800E+00<br>-0.3138167685399651E-02 | 0.5561472772748228E-01<br>-0.7192323066634110E-01 |
| | -0.1589935386649493E+01<br>0.3492140348141709E+02 | 0.8369153095696800E+00<br>-0.3452185020077714E-02 | 0.5919340882176674E-01<br>-0.7011938267785789E-01 |
| 133 | -0.1590042136749493E+01<br>0.3501387755626509E+02 | 0.8369153095696800E+00<br>-0.2241391306058183E-01 | 0.5697633389365831E-01<br>-0.5436700512278136E-01 |
| | -0.1590049424463438E+01<br>0.3504803246725509E+02 | 0.8369153095696800E+00<br>-0.2185271567270775E-01 | 0.5600129770396809E-01<br>-0.5613902650543423E-01 |
| 256 A | -0.1593178276216191E+01<br>0.3526266107906743E+02 | 0.8369153095696800E+00<br>-0.8309136663657846E-02 | 0.3384624983642573E-01<br>-0.2351289640137821E-01 |
| | -0.1592677144669925E+01<br>0.3507005980156986E+02 | 0.8369153095696800E+00<br>-0.1001119922384675E-01 | 0.4885412461044771E-01<br>-0.1361068781865038E-01 |
| 256 B | -0.1593132011214191E+01<br>0.3528378444614044E+02 | 0.8369153095696800E+00<br>-0.7796840214774115E-02 | 0.3223303085209571E-01<br>-0.2803716470412443E-01 |
| | -0.1593067761014191E+01<br>0.3531320582618160E+02 | 0.8369153095696800E+00<br>-0.9303914387642442E-02 | 0.3888427599667085E-01<br>-0.1830624307899498E-01 |
| 300 A | -0.1588477709895878E+01<br>0.3683445905374722E+02 | 0.8369153095696800E+00<br>-0.2893419472153624E-01 | 0.8912479300170050E-01<br>0.7078584256490217E-02 |
| | -0.1589236754795546E+01<br>0.3635165743905721E+02 | 0.8369153095696800E+00<br>-0.1165050308557625E-01 | 0.9056177929578309E-01<br>-0.3324182225271954E-01 |
| 300 B | -0.1588680637500227E+01<br>0.3651070805366049E+02 | 0.8369153095696800E+00<br>-0.3366585614677931E-01 | 0.7866390131023905E-01<br>0.1537234649936536E-01 |
| | -0.1588580418170229E+01<br>0.3670619466106999E+02 | 0.8369153095696800E+00<br>-0.3469041060695943E-01 | 0.7779012661720475E-01<br>0.1735337585608576E-01 |
| 300 C | -0.1588950230998127E+01<br>0.3650988061886180E+02 | 0.8369153095696800E+00<br>-0.2647735496429959E-01 | 0.8706439017117680E-01<br>-0.3000823032380140E-02 |
| | -0.1588966475602281E+01<br>0.3640003431956882E+02 | 0.8369153095696800E+00<br>-0.224187567603451oE-01 | 0.9056039400082534E-01<br>-0.1230185356180868E-01 |
| 300 D | -0.1588495345064487E+01<br>0.3663438378866425E+02 | 0.8369153095696800E+00<br>-0.2157314103491181E-01 | 0.9626732193872402E-01<br>-0.1332421612519141E-01 |
| | -0.1588454491964113E+01<br>0.3663914702411404E+02 | 0.8369153095696800E+00<br>-0.3034983173071382E-01 | 0.8698838273190614E-01<br>0.1188817052383229E-01 |
| 238 | -0.1593206386649482E+01<br>0.3675227638874794E+02 | 0.8369153095696800E+00<br>-0.4898871365064395E-03 | 0.3926506190167107E-01<br>0.1962802950496534E-01 |
| | -0.1593173446125017E+01<br>0.3761473913044777E+02 | 0.8369153095696800E+00<br>0.1132948224467567E-01 | 0.3253451313459106E-01<br>0.2012333513799902E-01 |
| 262 A | -0.1593567386649491E+01<br>0.3944546633584496E+02 | 0.8369153095696800E+00<br>0.3357299214377829E-02 | 0.2980070885502215E-01<br>0.1646764453950930E-01 |
| | -0.1593537386649481E+01<br>0.3900631346781820E+02 | 0.8369153095696800E+00<br>-0.2077048713518326E-02 | 0.3254291778614669E-01<br>0.1375017950102470E-01 |
| 262 B | -0.1593578523168779E+01<br>0.3935857289039138E+02 | 0.8369153095696800E+00<br>0.2821527954439981E-02 | 0.2978164570341811E-01<br>0.1624096505161523E-01 |
| | -0.1593554368873907E+01<br>0.3937788358976672E+02 | 0.8369153095696800E+00<br>0.9378879806565266E-04 | 0.3172659732870578E-01<br>0.1501406305726405E-01 |
| 262 C | -0.1593270386649493E+01<br>0.3899621389914385E+02 | 0.8369153095696800E+00<br>-0.6188096823394496E-02 | 0.3838437521962948E-01<br>0.1296112119202922E-01 |
| | -0.1593363386649492E+01<br>0.3886406962619712E+02 | 0.8369153095696800E+00<br>-0.5739802309933956E-02 | 0.3704393089958365E-01<br>0.1025823071430699E-01 |
| 008 | -0.1587079670517290E+01<br>0.3960000287872858E+02 | 0.8369153095696800E+00<br>-0.2670357135682622E-01 | 0.1040248517420310E+00<br>0.2147558120768167E-01 |
| | -0.1586985207610502E+01<br>0.3988480321556160E+02 | 0.8369153095696800E+00<br>-0.3181792715666276E-01 | 0.9415325605492016E-01<br>0.3772643946916077E-01 |

## 5.1 Characteristic curves for all families in the Atlas

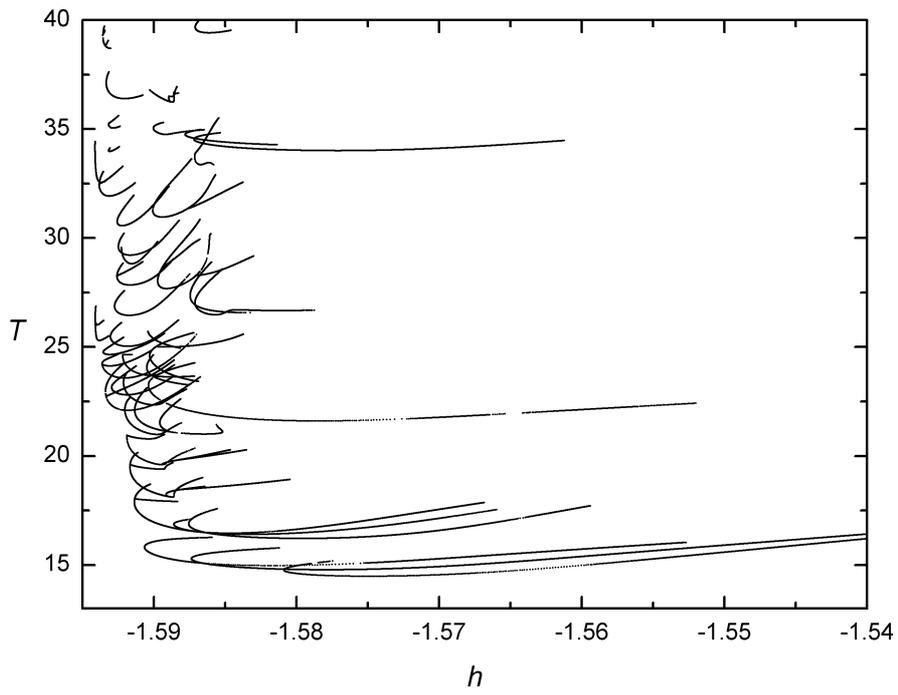

Characteristic curves $T(h)$ for all branches of all families in the Atlas.

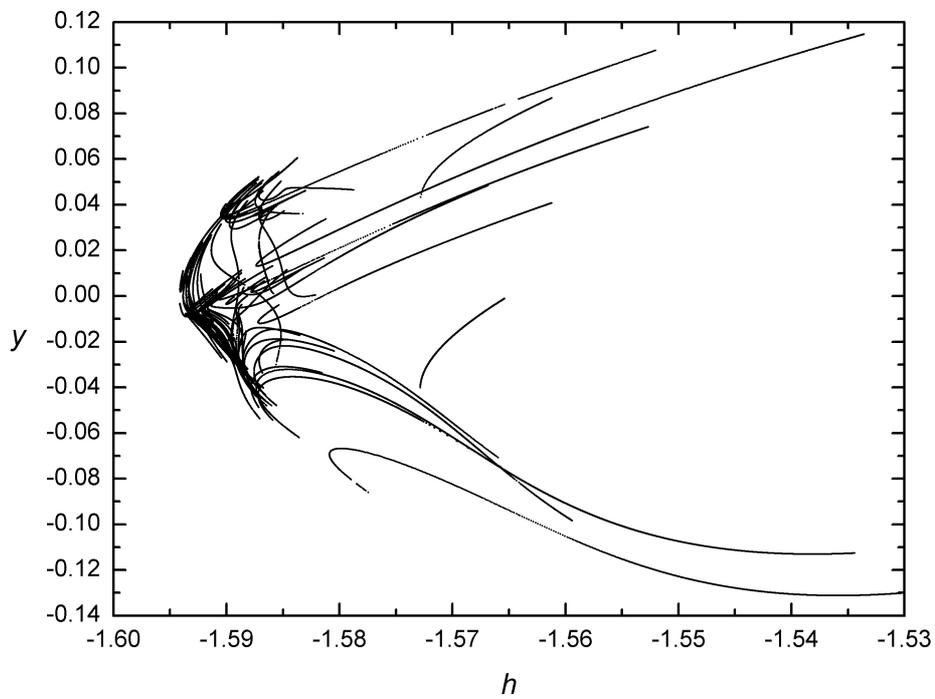

Characteristic curves $y(h)$ on $\Sigma_1$ for all branches of all families in the Atlas.

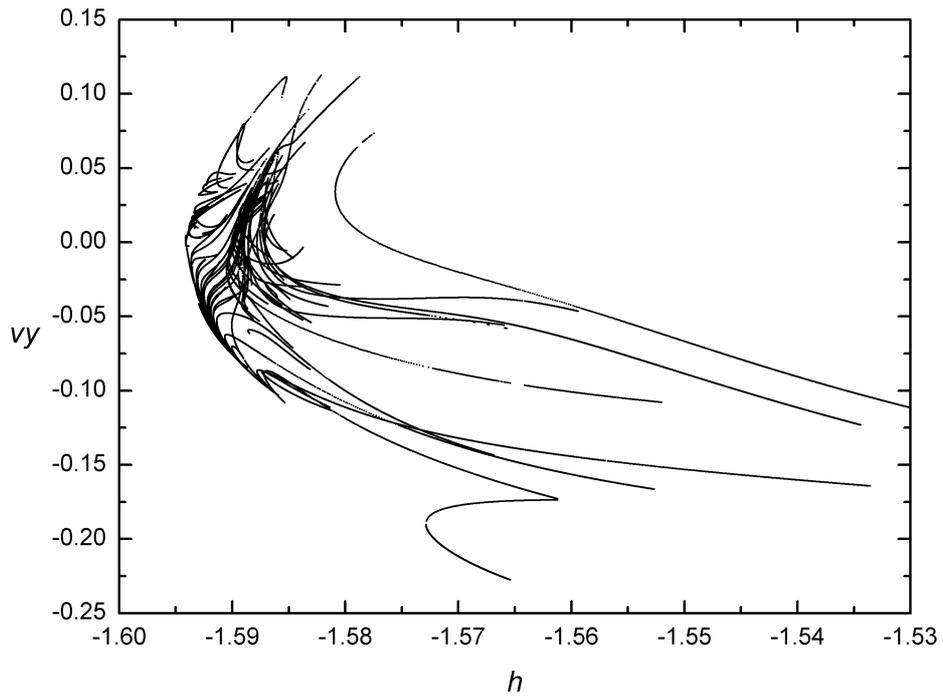

Characteristic curves $v_y(h)$ on $\Sigma_1$ for all branches of all families in the Atlas.

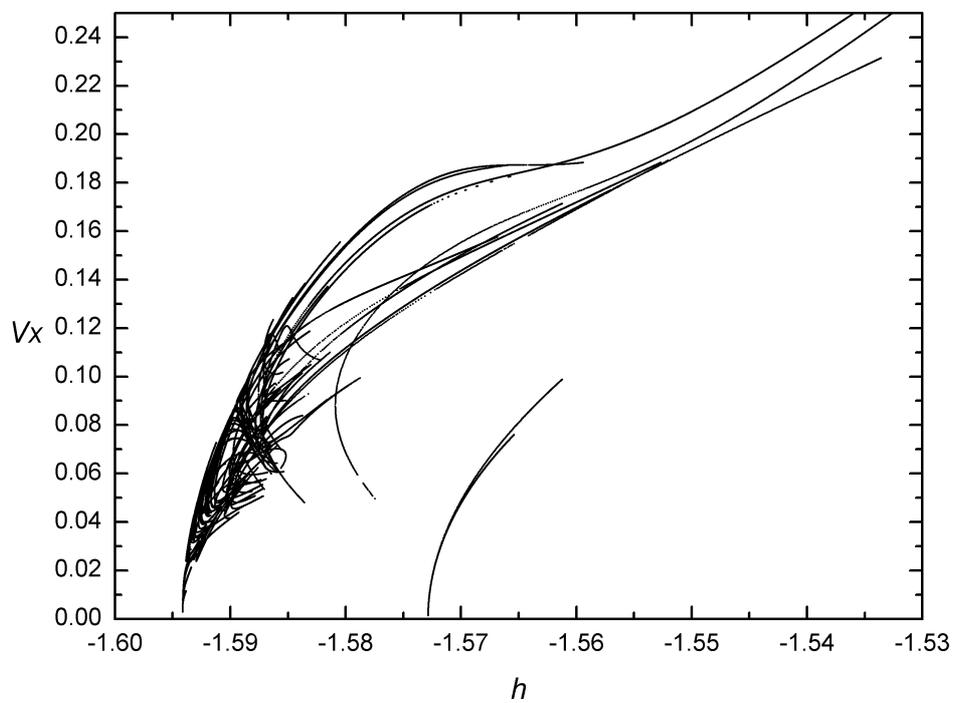

Characteristic curves $v_x(h)$ on $\Sigma_1$ for all branches of all families in the Atlas.

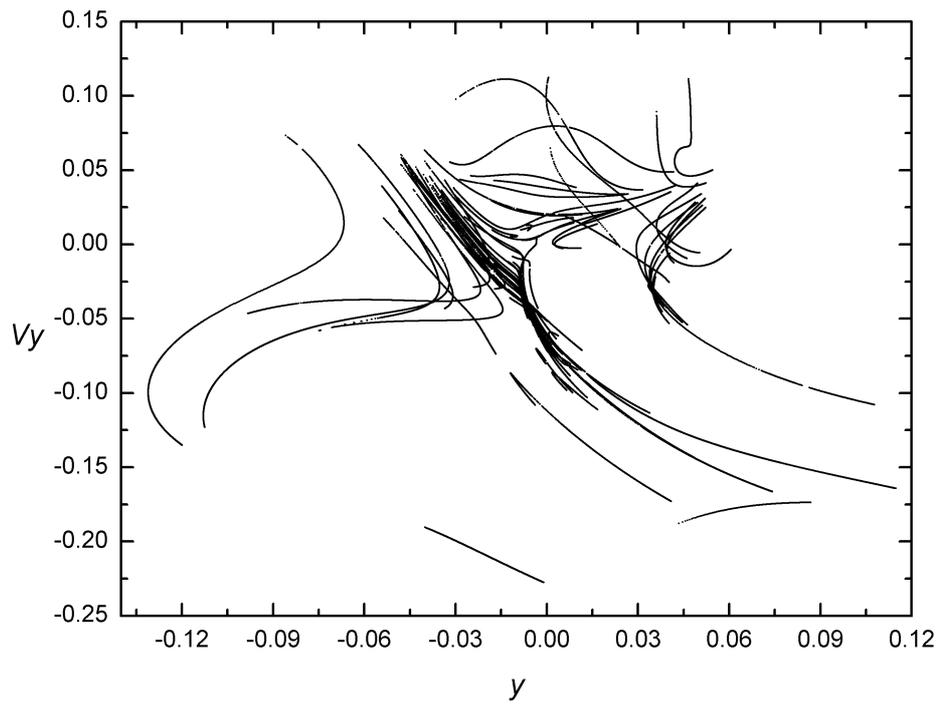

Curves of initial conditions $(y, v_y)$ on $\Sigma_1$ for all branches of all families in the Atlas.

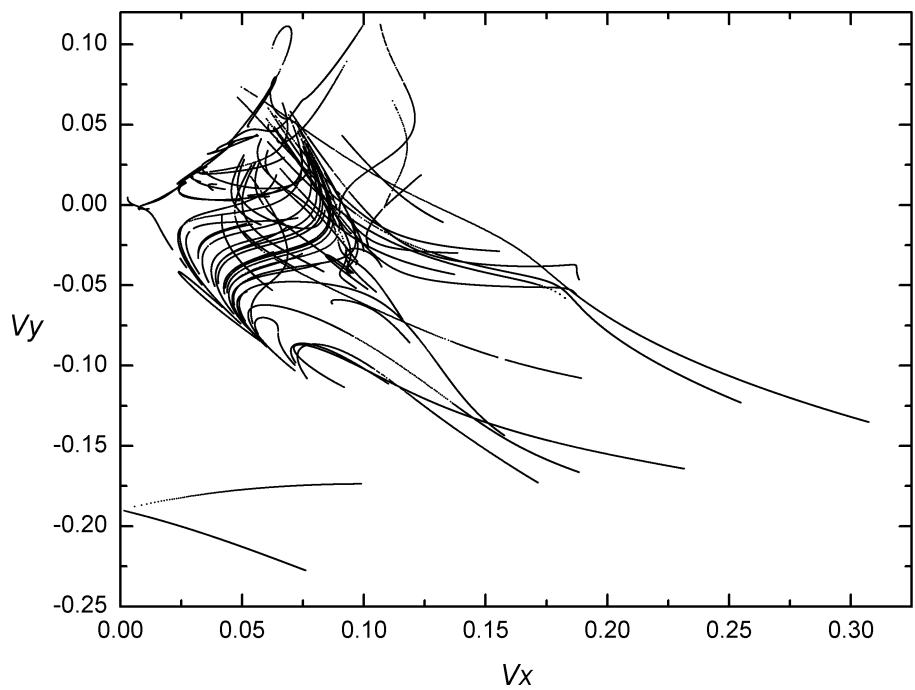

Curves of initial velocity $(v_x, v_y)$ for all branches of all families in the Atlas.

## 5.2 Families of low-energy fast periodic transfer orbits

### Family 357 - *Symmetric family of symmetric POs*

$h_{min} = -1.580898, \quad h_{max} = -1.516112, \quad T_{min} = 14.487454, \quad T_{max} = 17.812637$

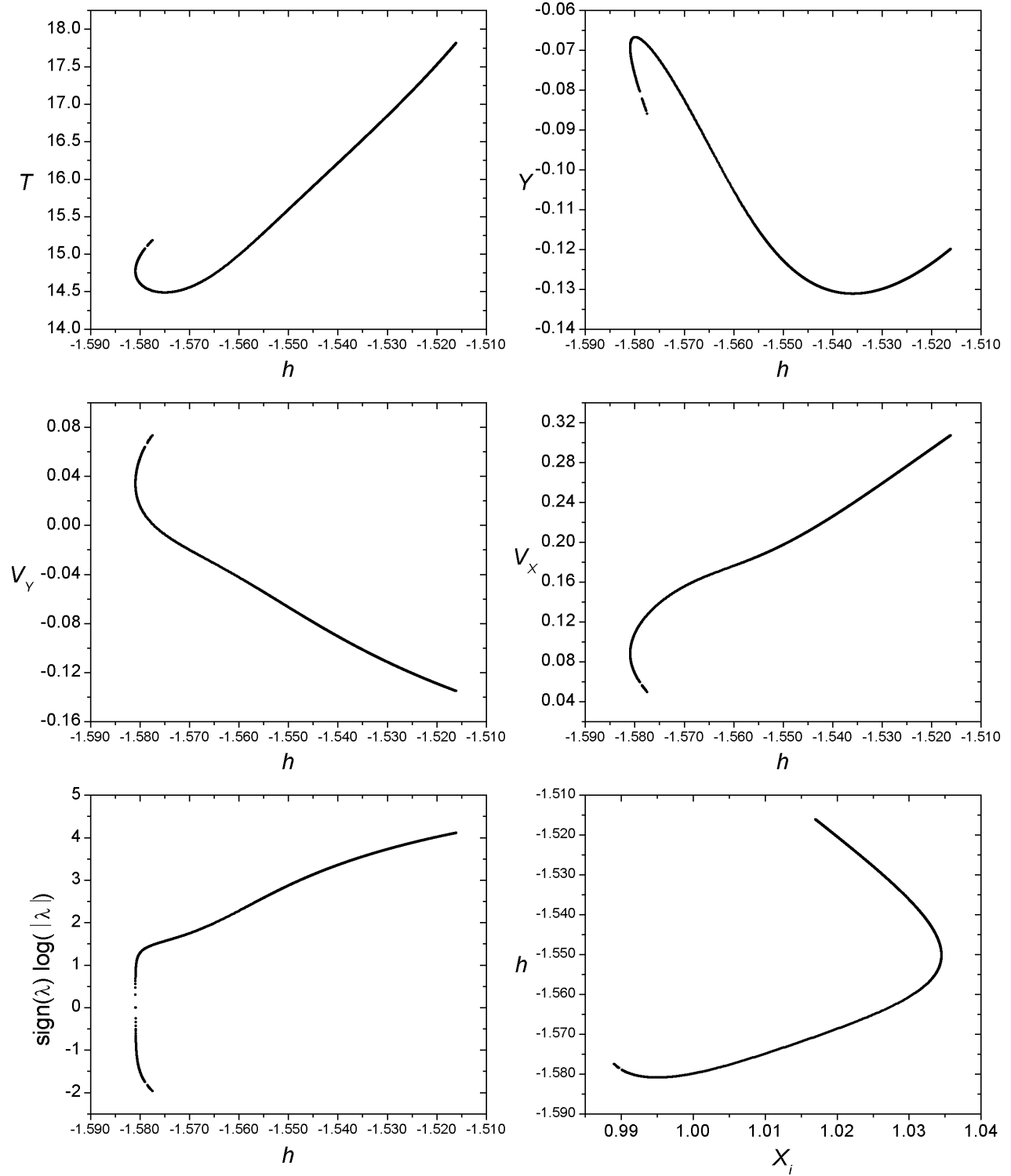

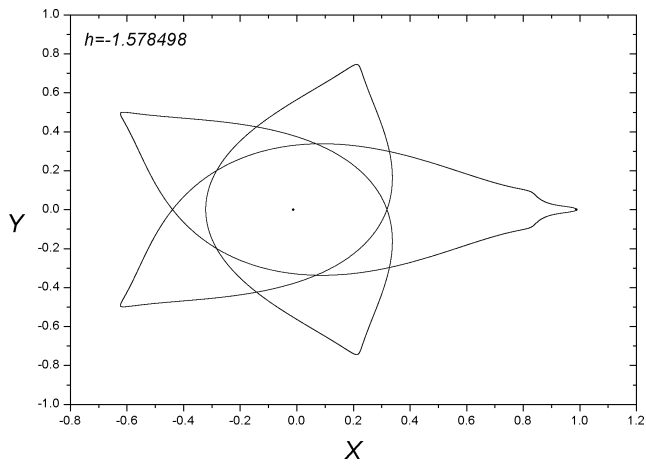
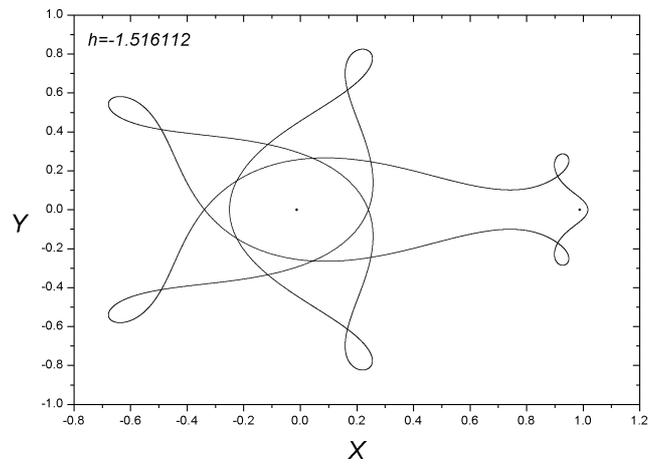

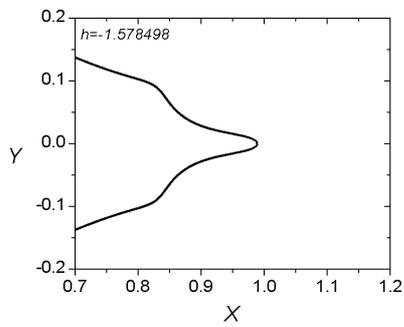
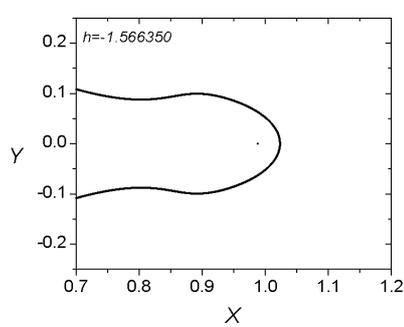
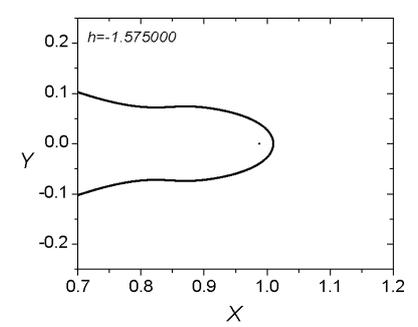

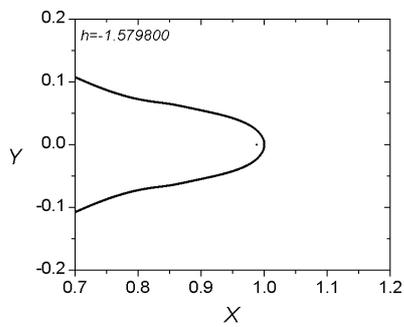
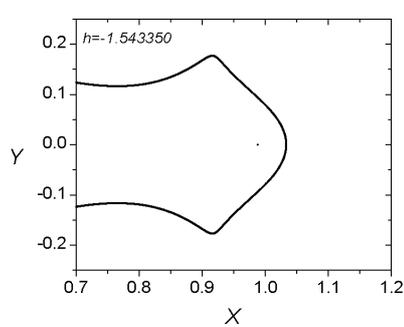
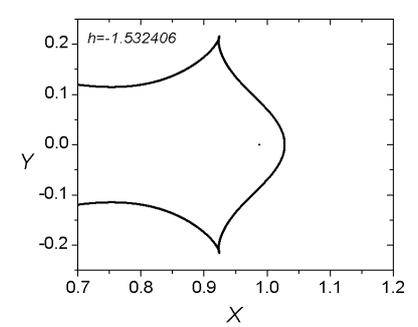

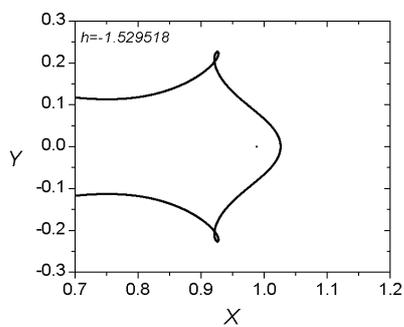
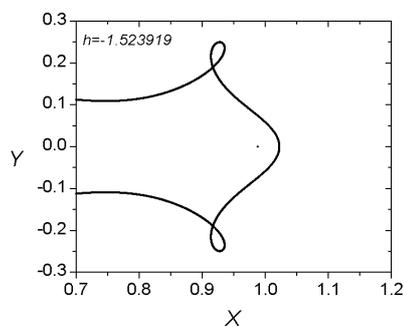
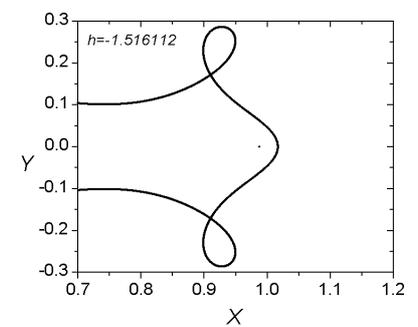

### *Family 037 - Asymmetric family of asymmetric POs*

$h_{min} = -1.587381$, $h_{max} = -1.534414$, $T_{min} = 14.786812$, $T_{max} = 16.716217$

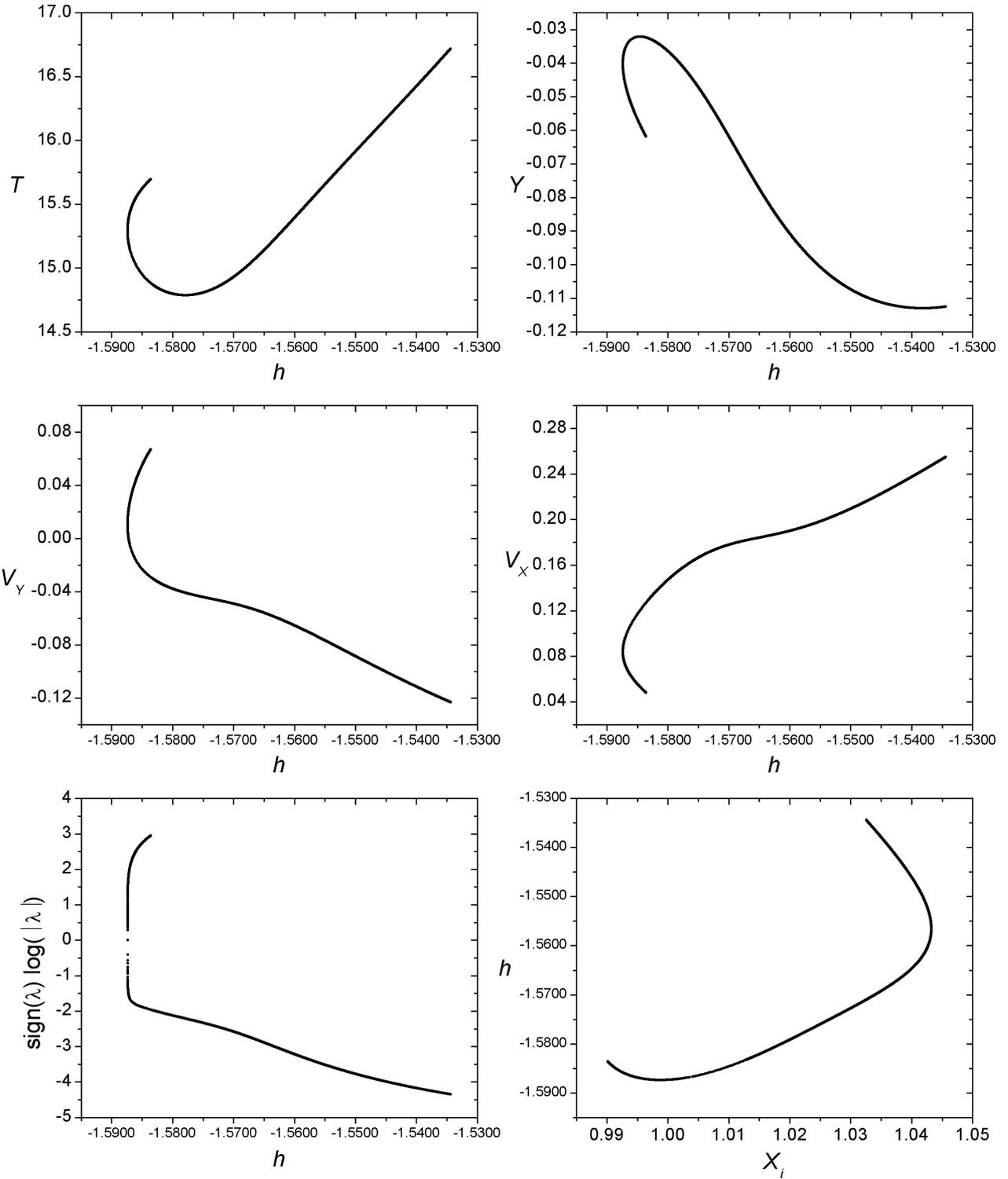

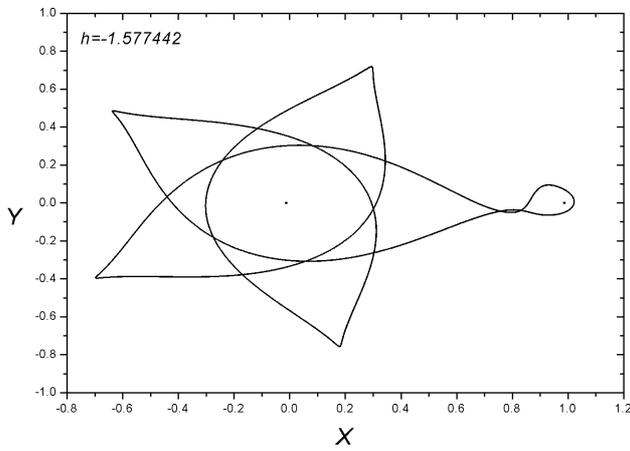

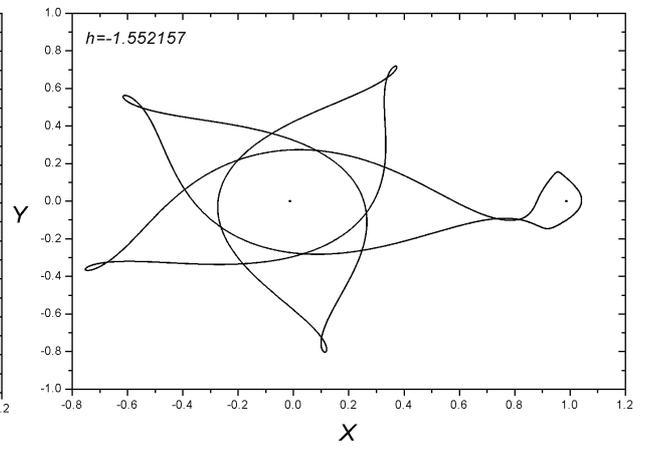

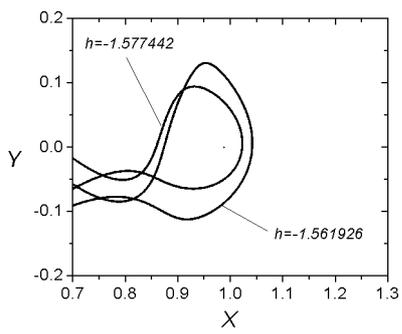

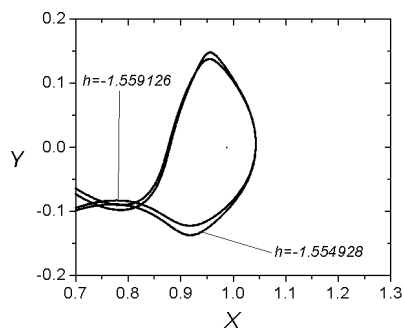

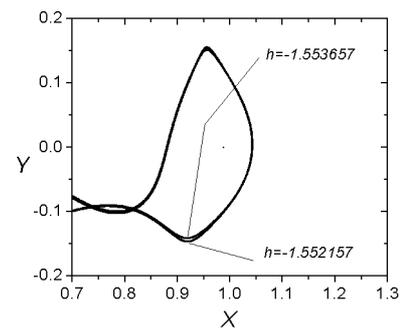

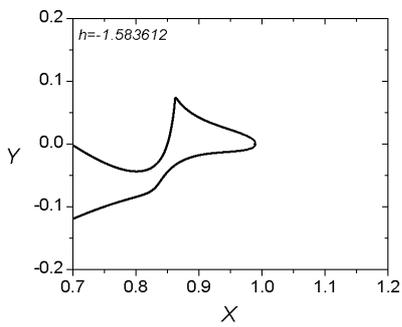

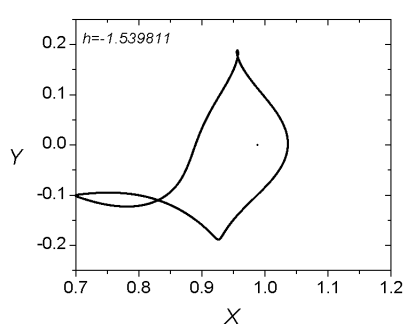

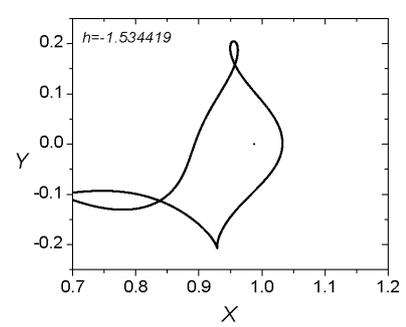

## Family 043 - Asymmetric family of asymmetric POs

$h_{min} = -1.587380$,  $h_{max} = -1.533589$,  $T_{min} = 14.786812$,  $T_{max} = 16.760541$

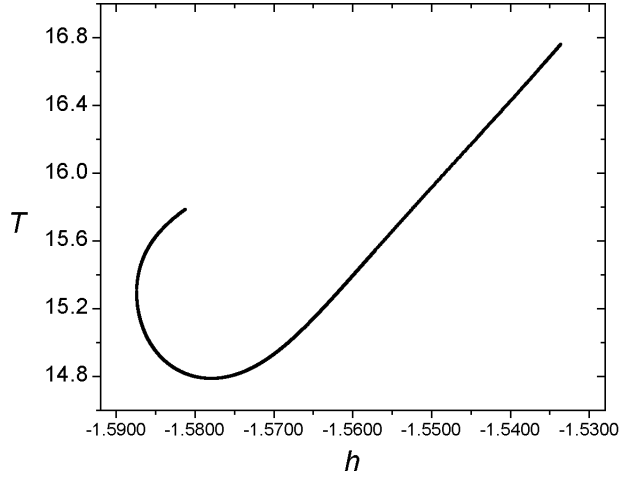

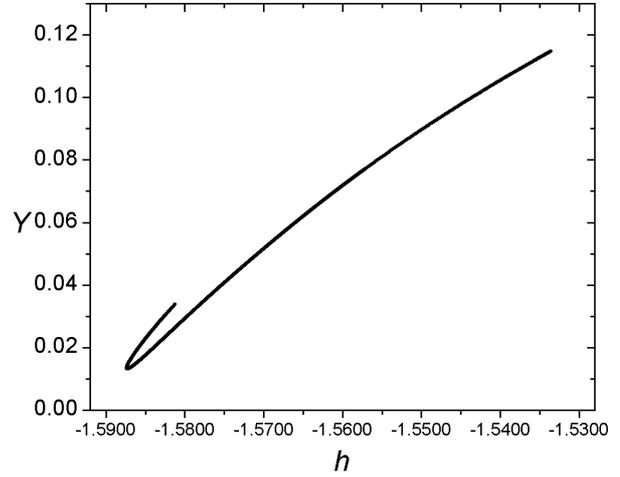

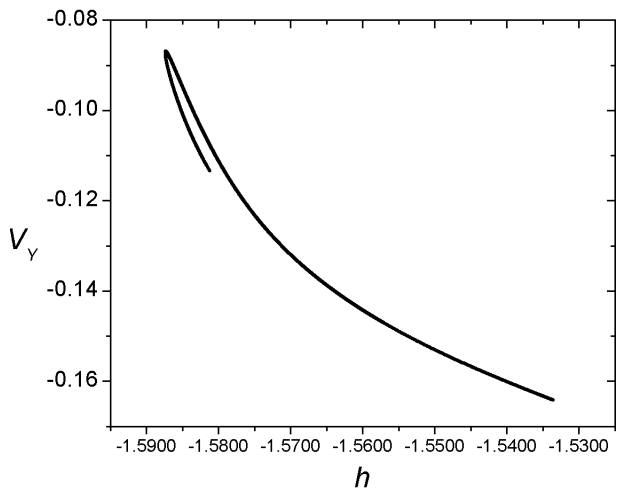

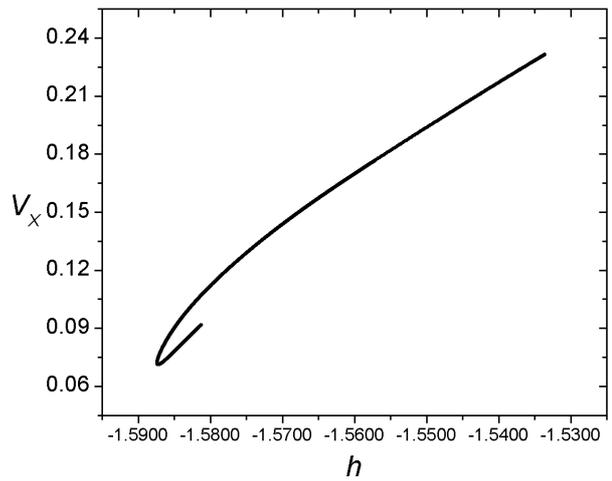

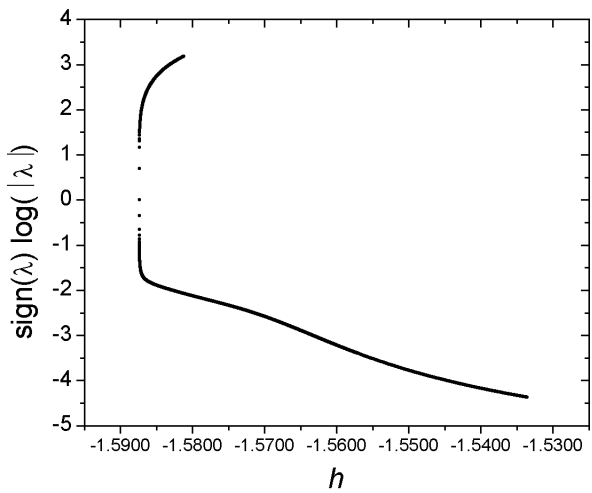

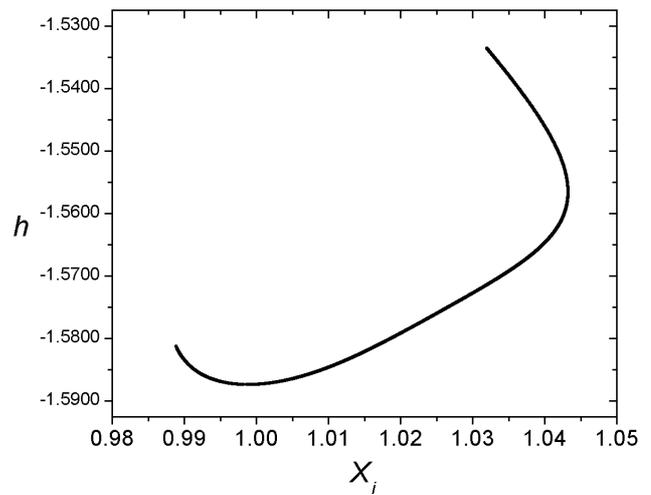

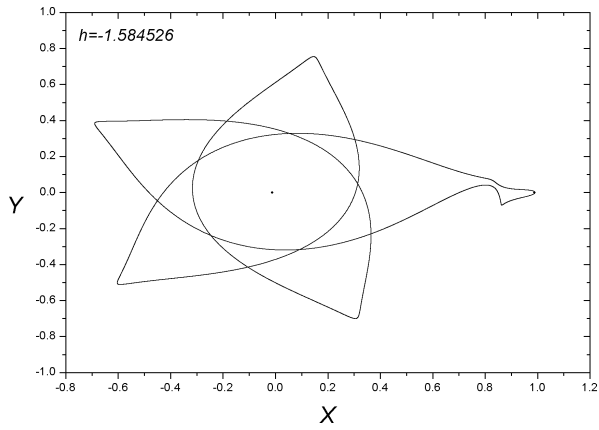
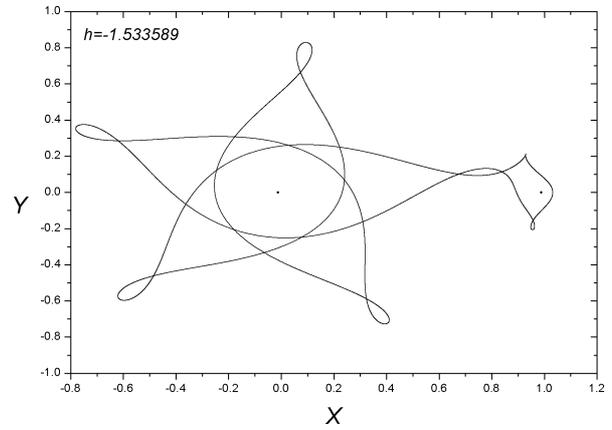

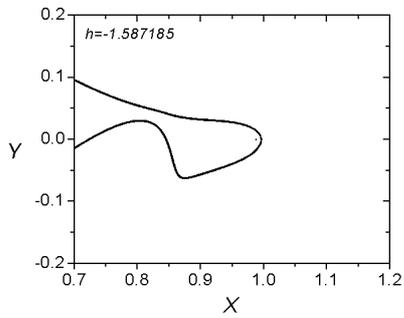
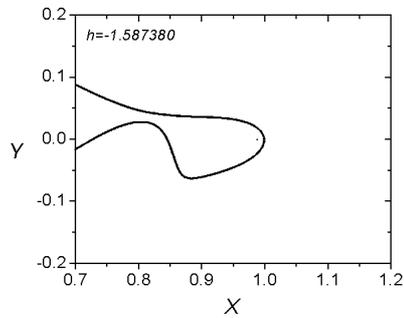
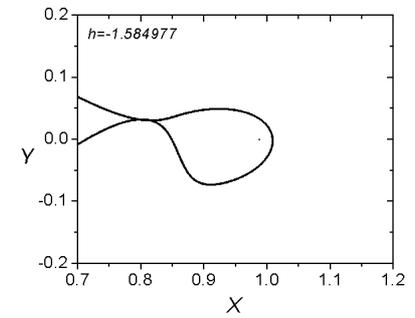

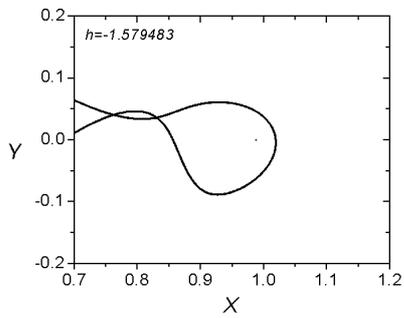
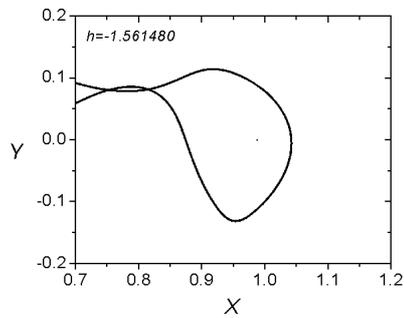
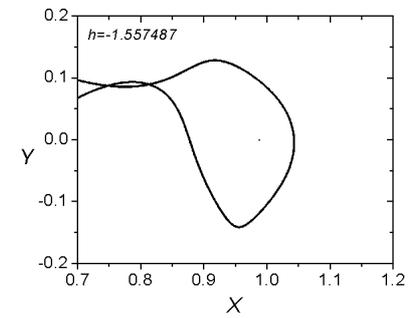

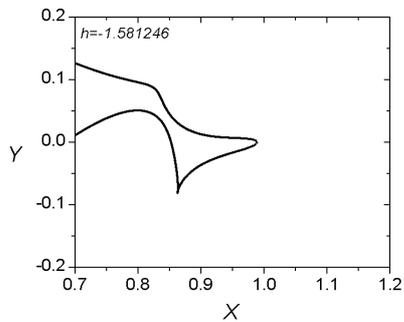
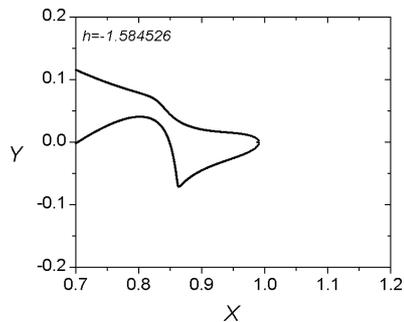
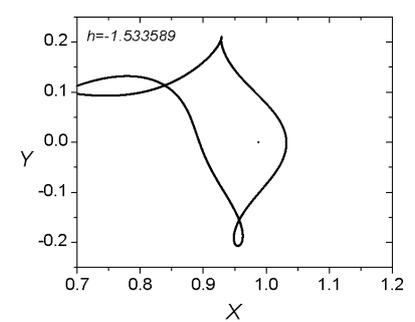

## *Family 056 – Symmetric family of symmetric POs*

$h_{min} = -1.590666, \quad h_{max} = -1.552698, \quad T_{min} = 14.975518, \quad T_{max} = 16.268286$

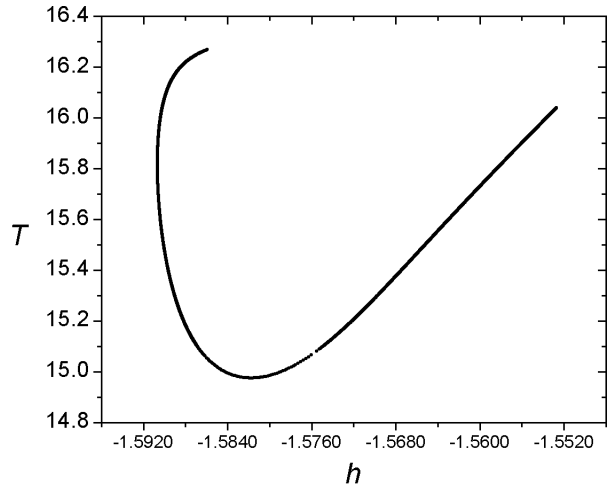
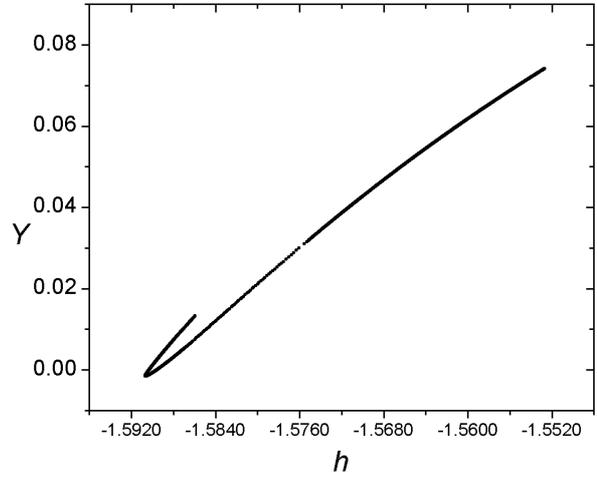

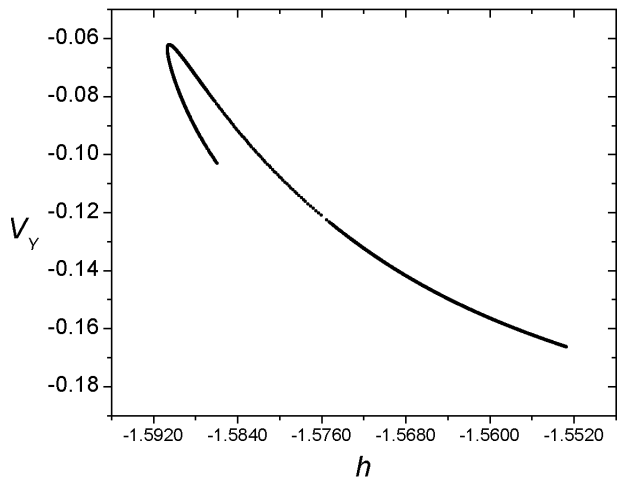
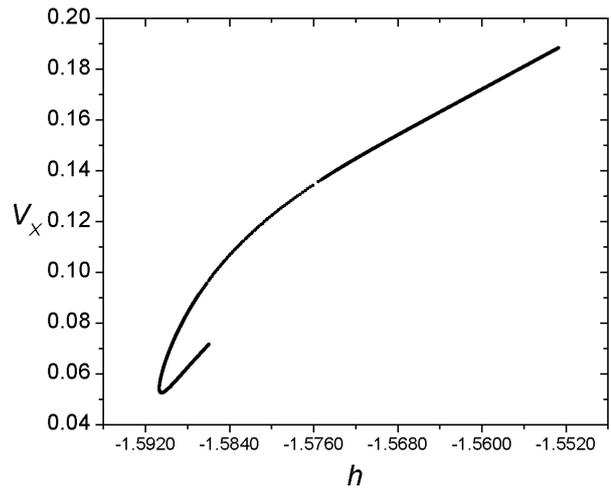

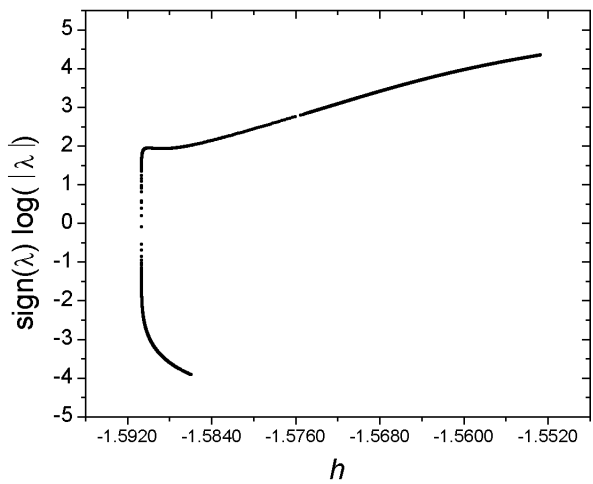
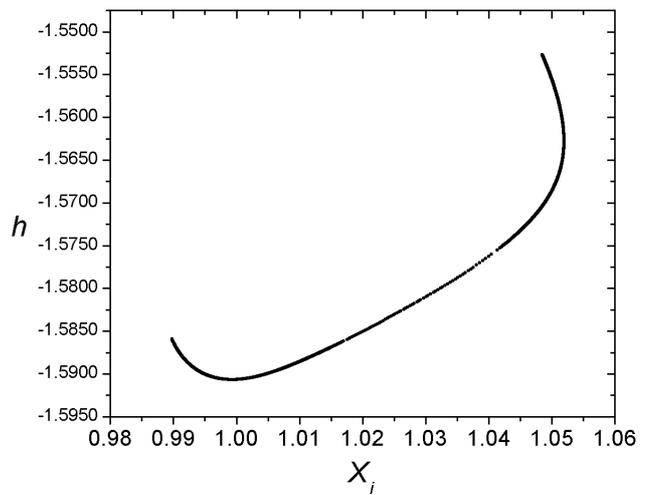

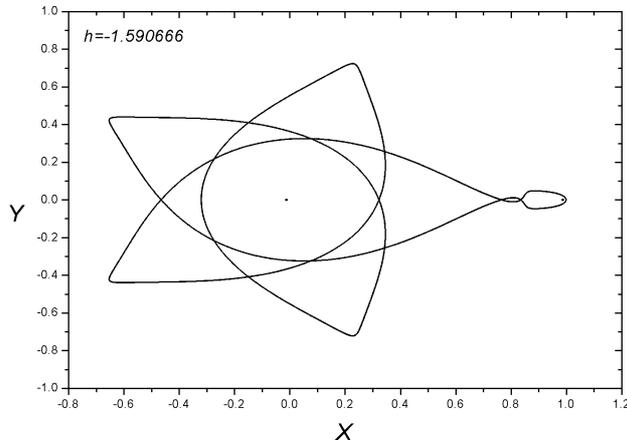

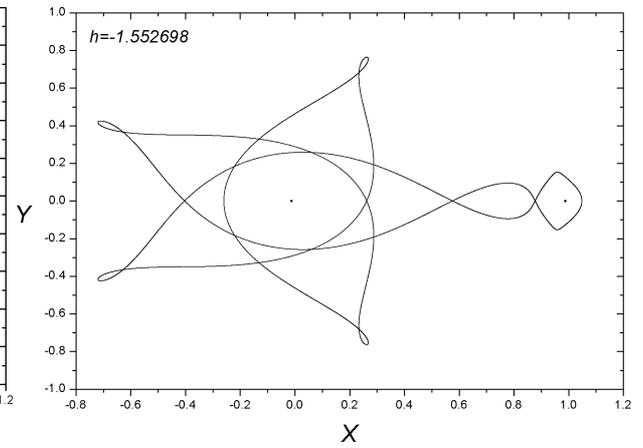

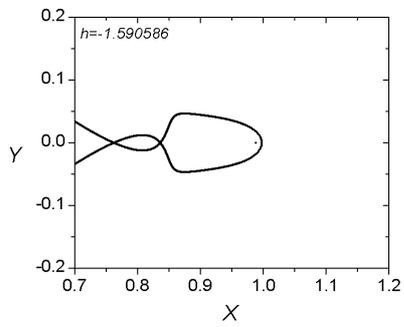

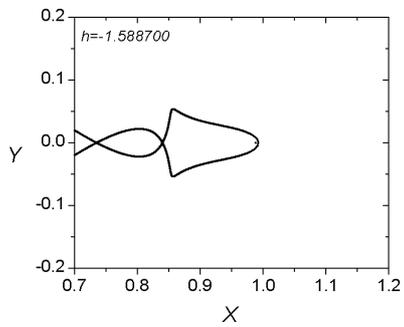

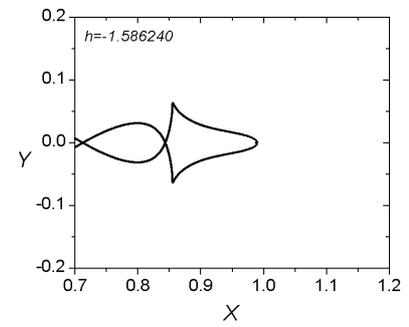

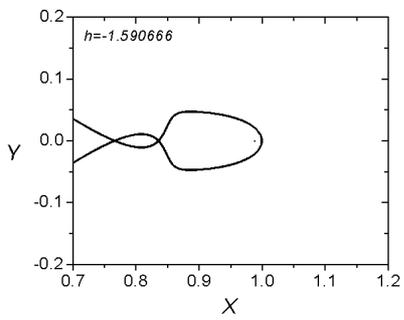

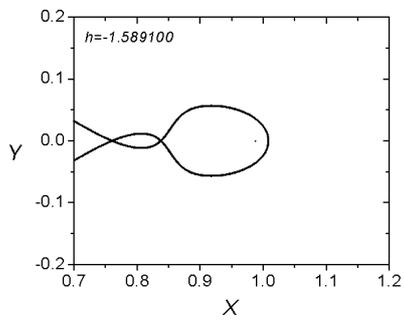

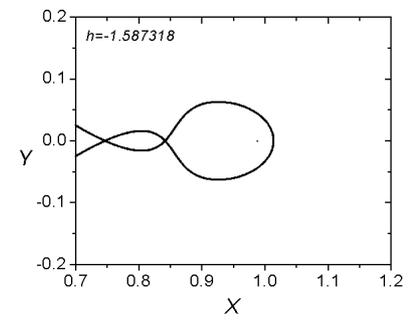

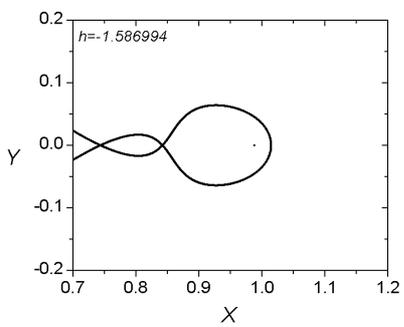

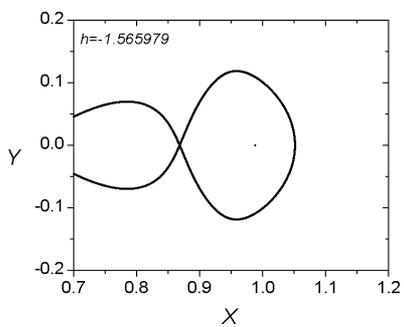

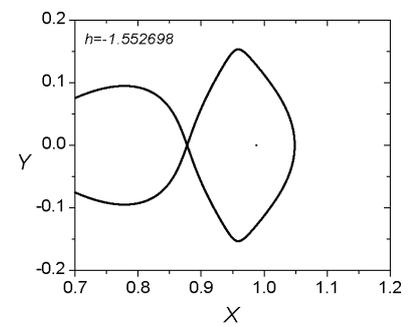

## *Family 053 - Symmetric family of symmetric POs*

$h_{min} = -1.587566, \; h_{max} = -1.559423, \; T_{min} = 16.220757, \; T_{max} = 17.718496$

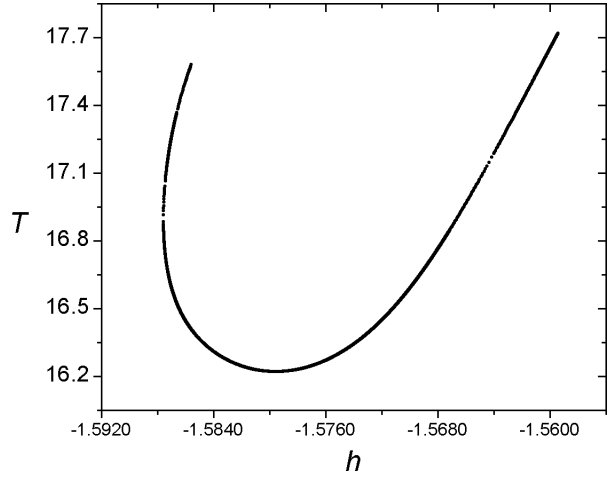
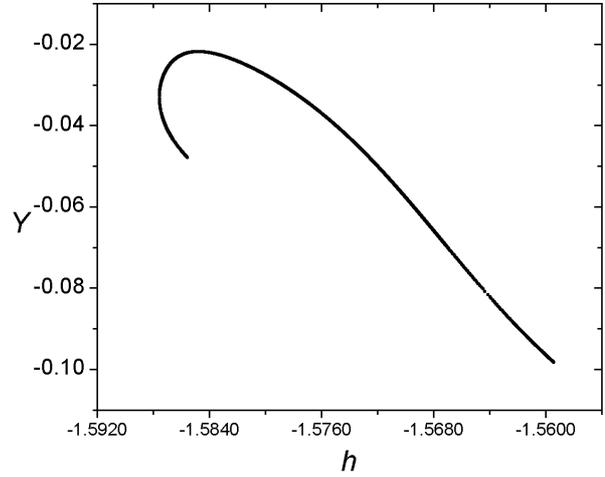
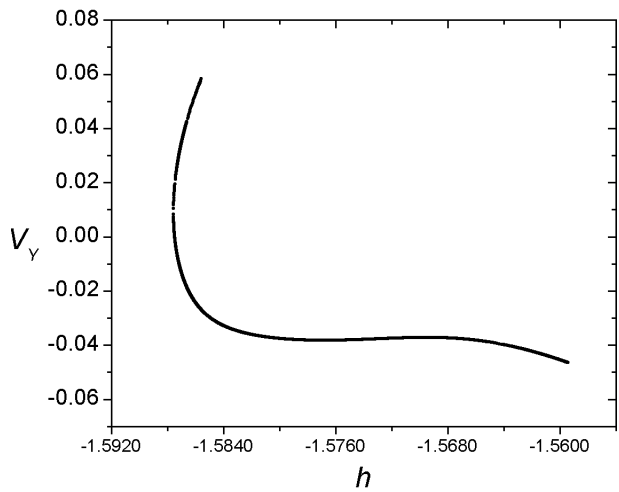
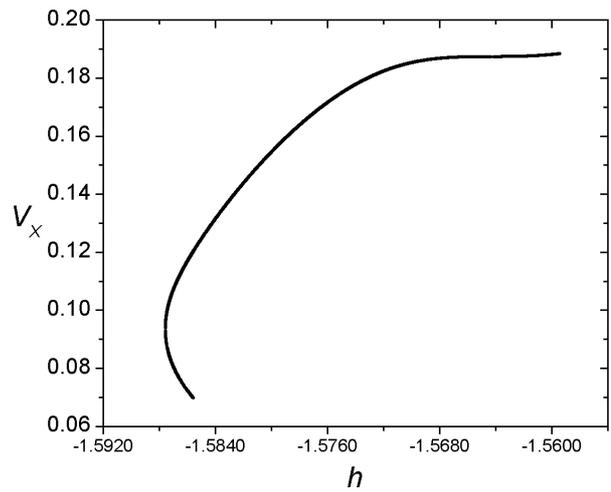
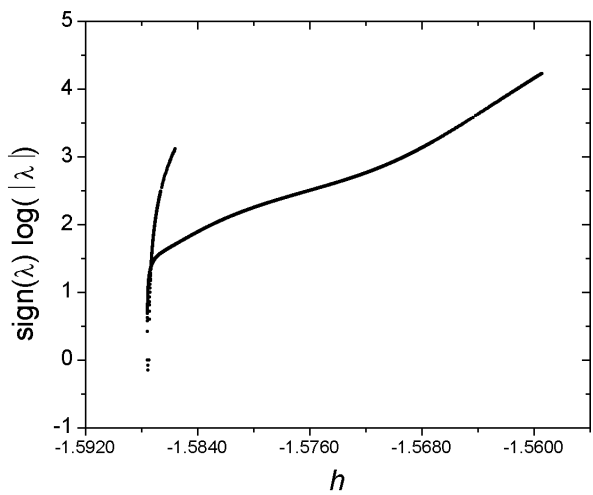
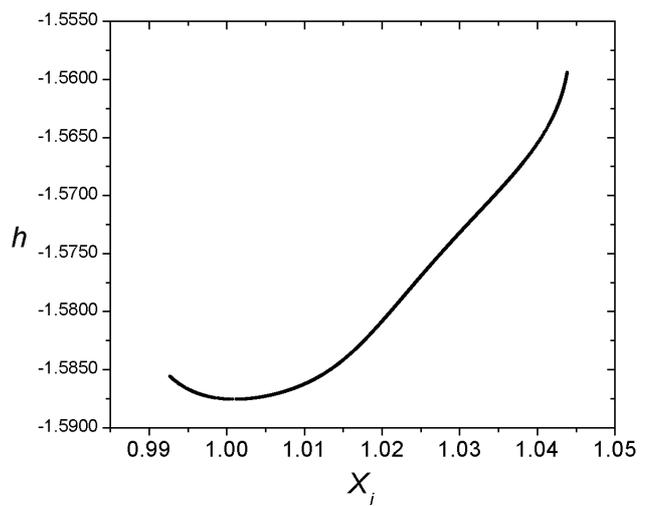

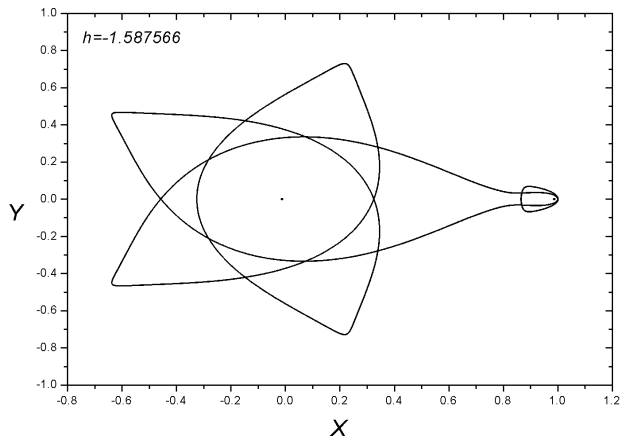
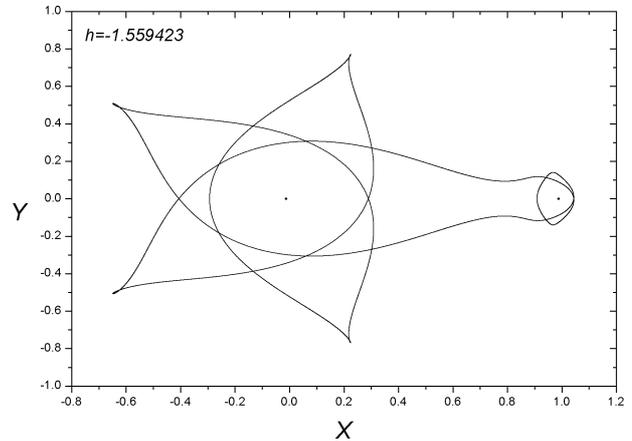

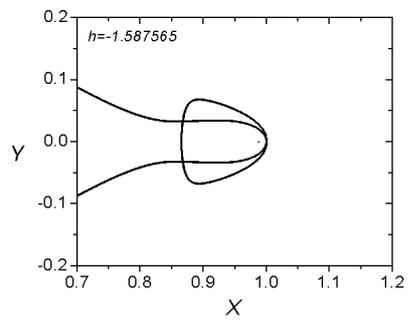
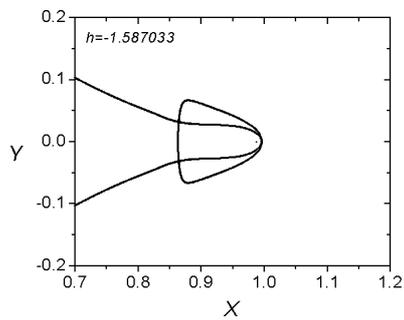
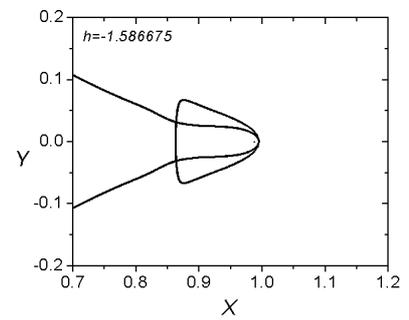

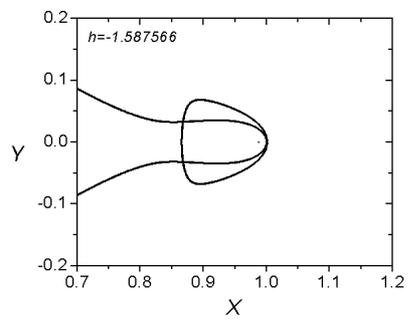
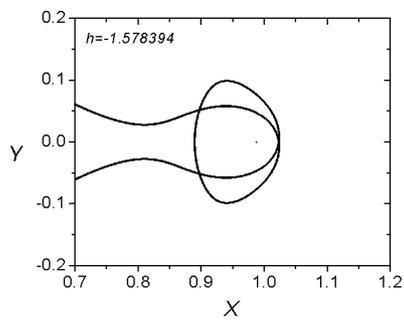
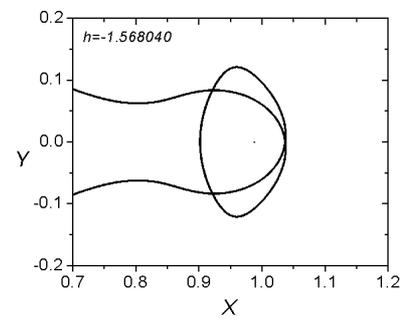

## Family 084 - Asymmetric family of asymmetric POs

$h_{min} = -1.588610, \quad h_{max} = -1.565978, \quad T_{min} = 16.409829, \quad T_{max} = 17.543807$

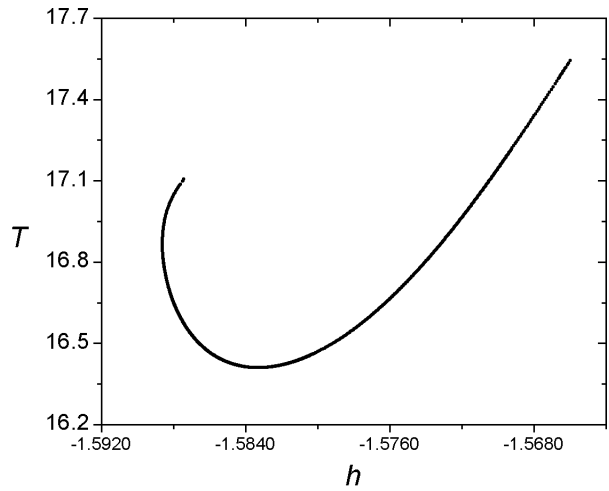
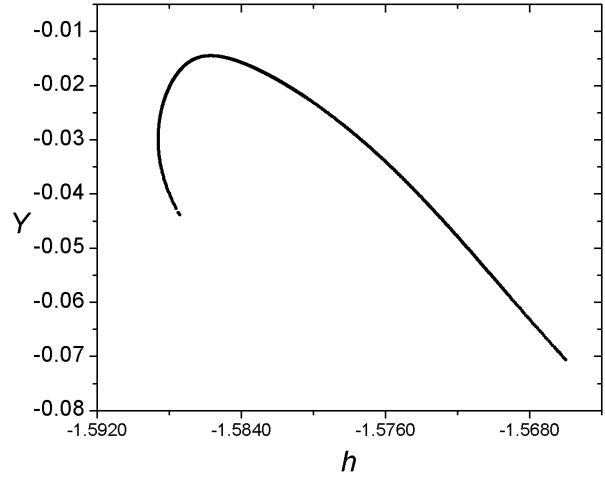

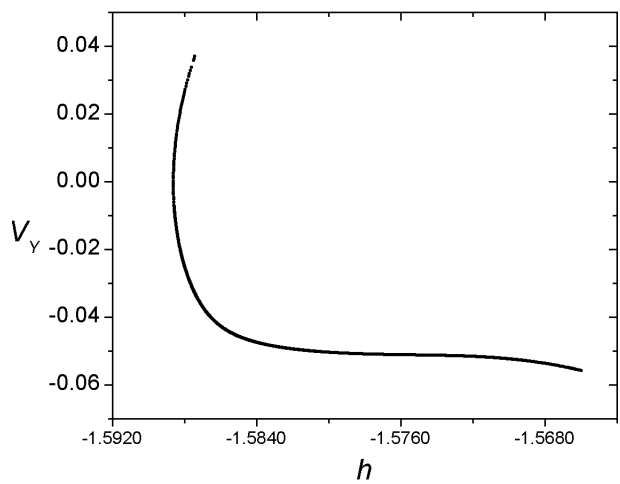
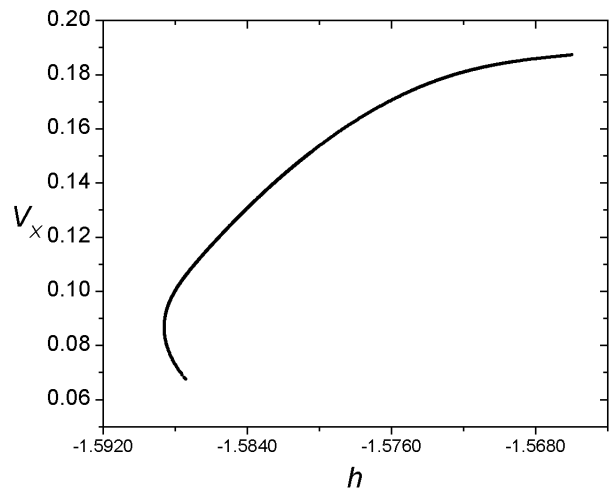

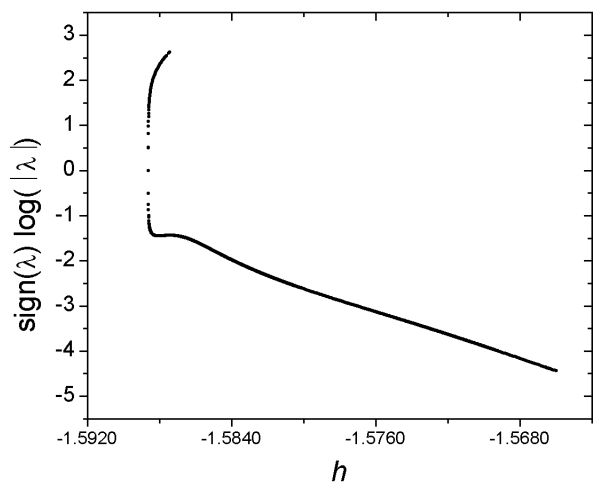
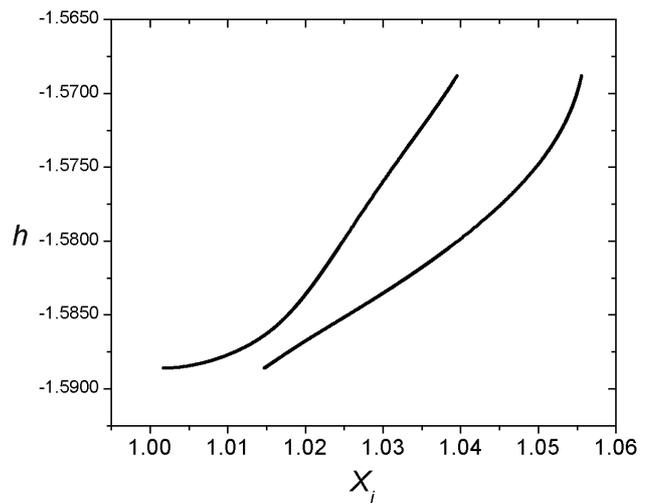

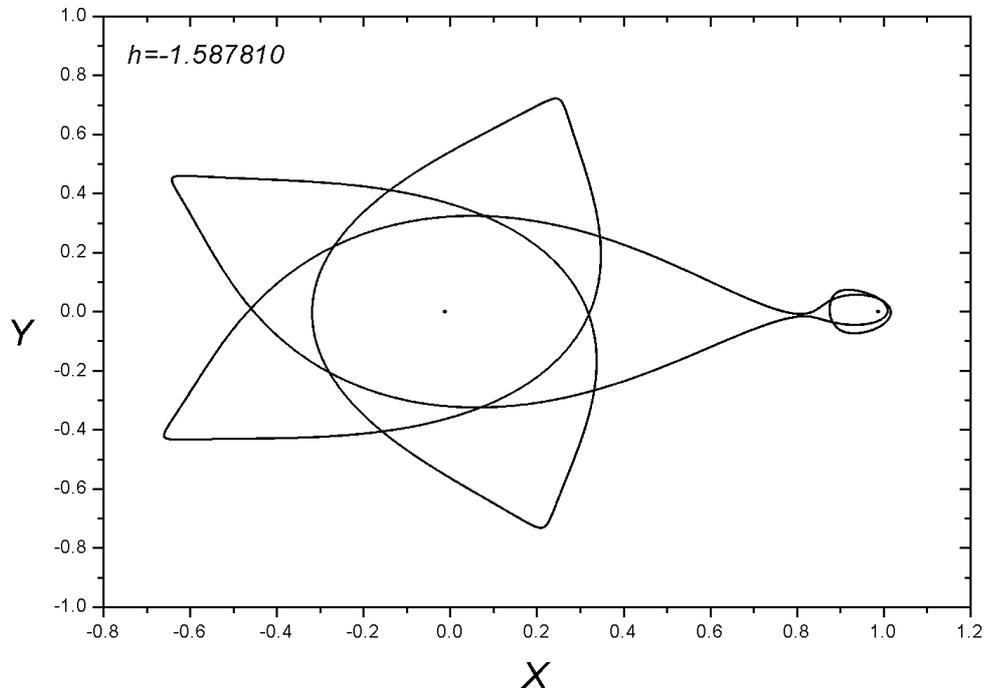

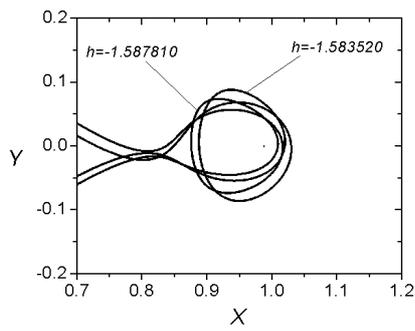
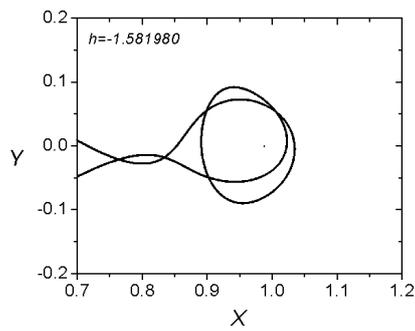
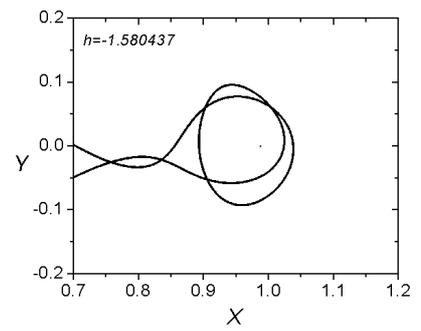

## Family 077 - Asymmetric family of asymmetric POs

$h_{min} = -1.588610, \quad h_{max} = -1.563229, \quad T_{min} = 16.409829, \quad T_{max} = 17.828708$

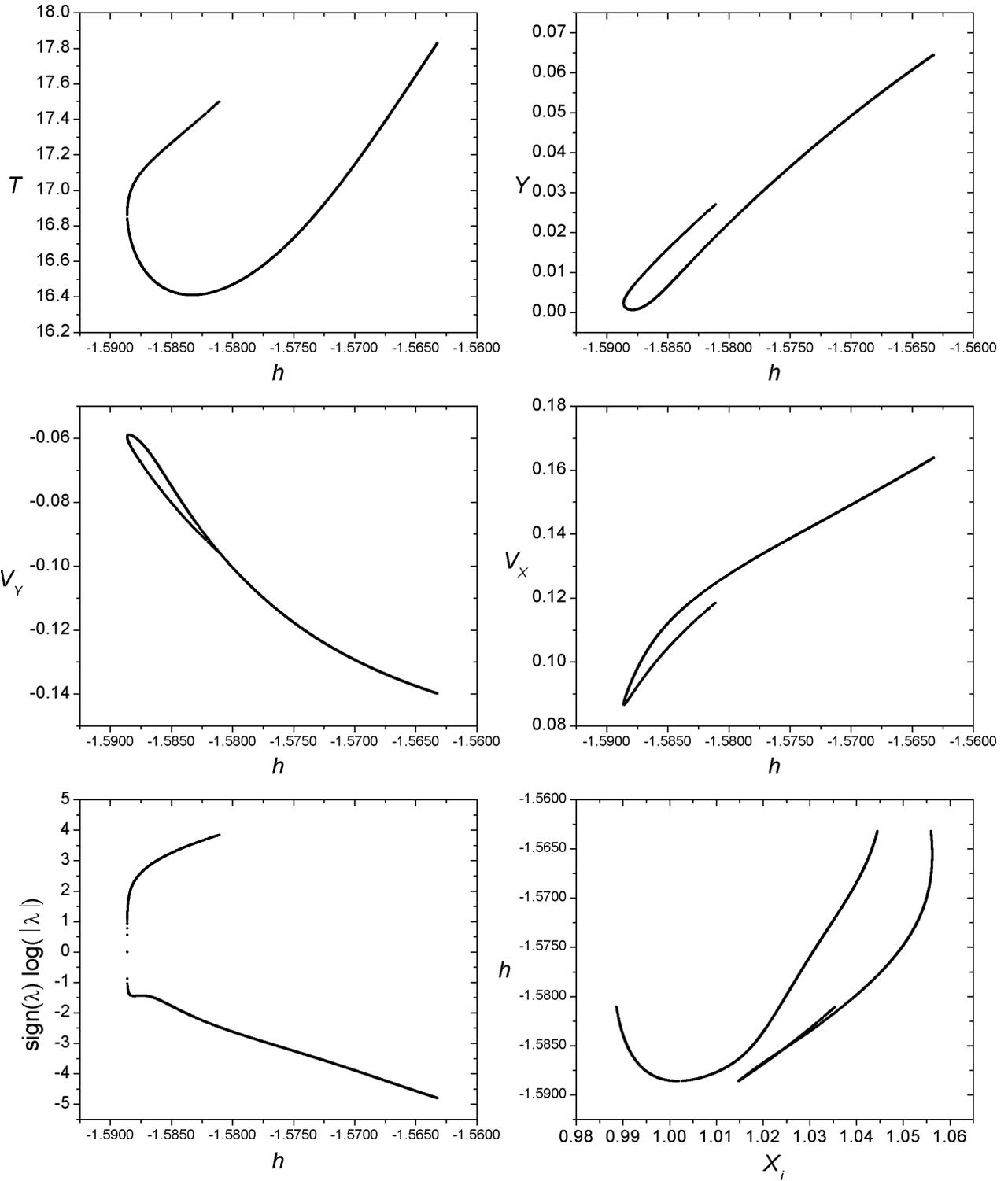

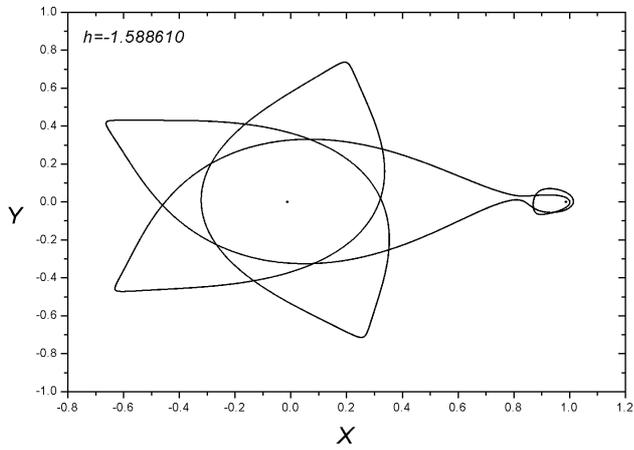
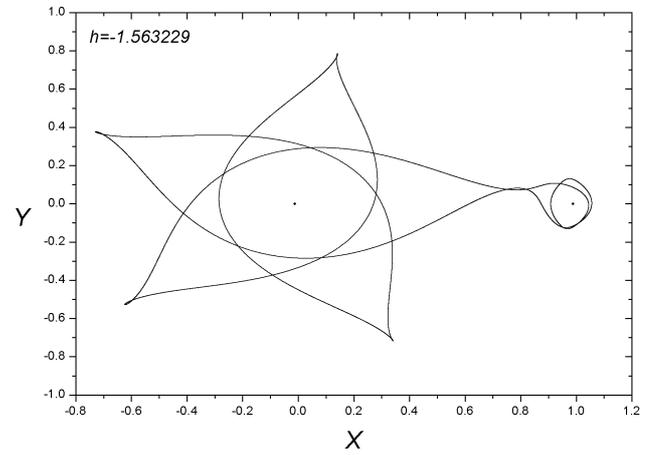

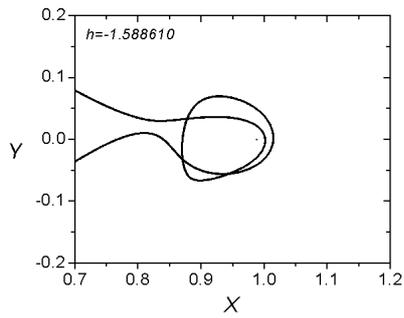
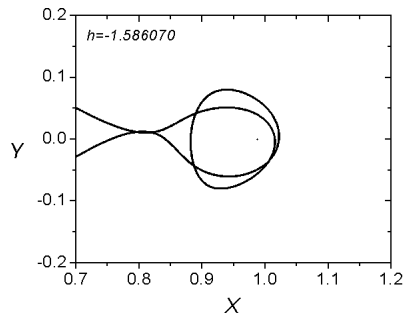
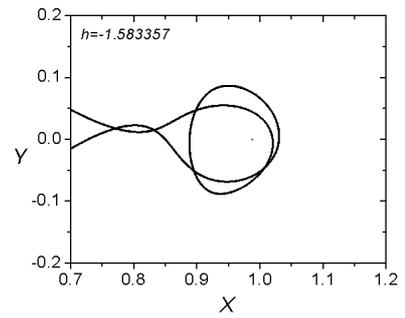

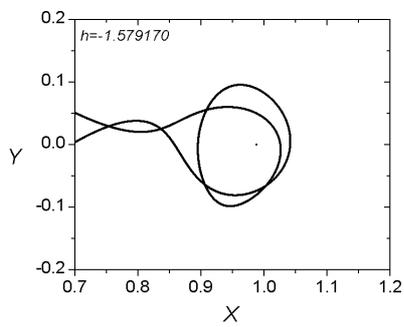
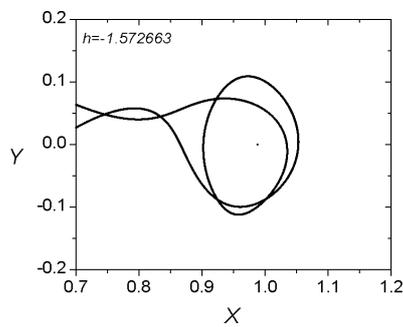
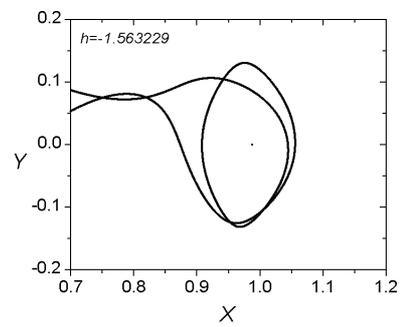

## Families 180A - 180 B

*Bifurcation Point*

|     | $h$       | $T$       | $y$       | $v_y$     | $v_x$    |
|-----|-----------|-----------|-----------|-----------|----------|
| $P_1$ | -1.591355 | 18.034195 | -0.003631 | -0.055498 | 0.049959 |

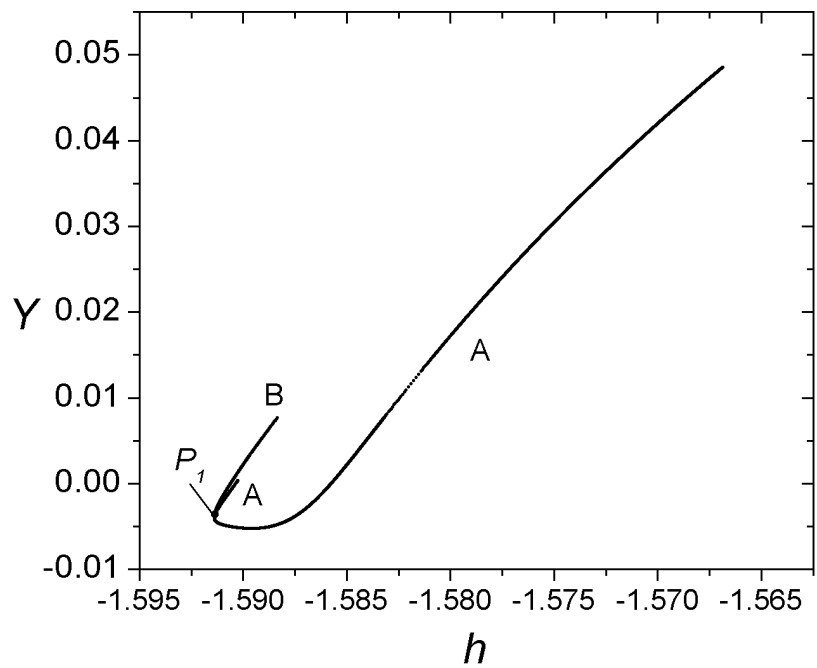

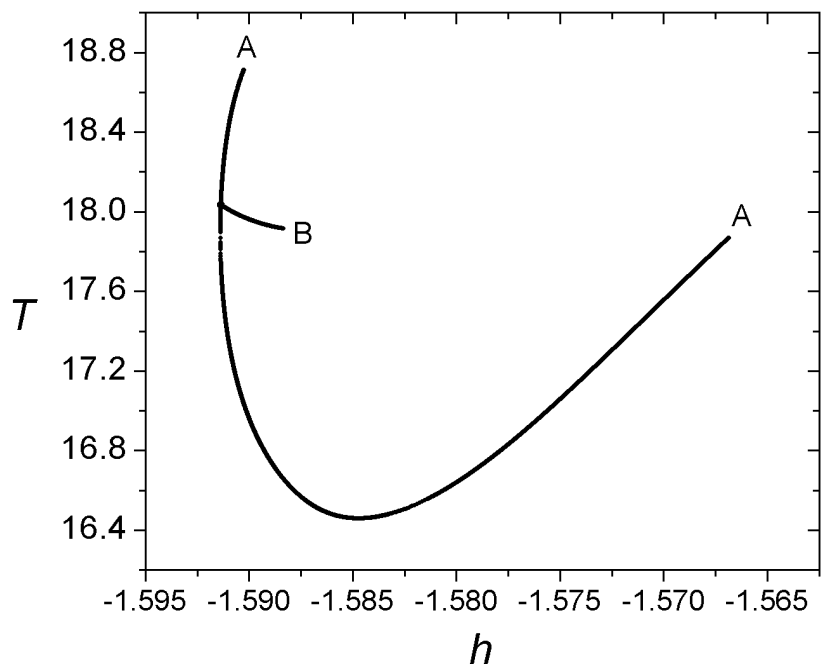

## Family 180 A - Symmetric family of symmetric POs

$h_{min} = -1.591381, \quad h_{max} = -1.566854, \quad T_{min} = 16.459140, \quad T_{max} = 18.712675$

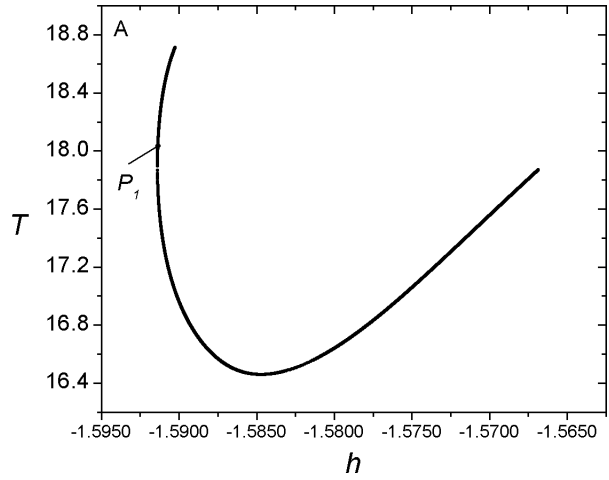
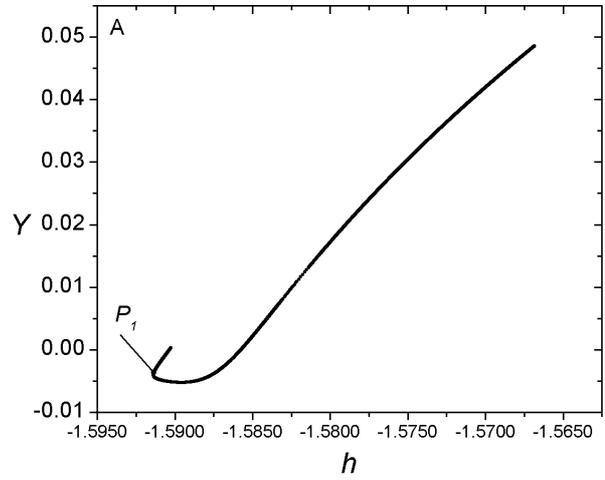

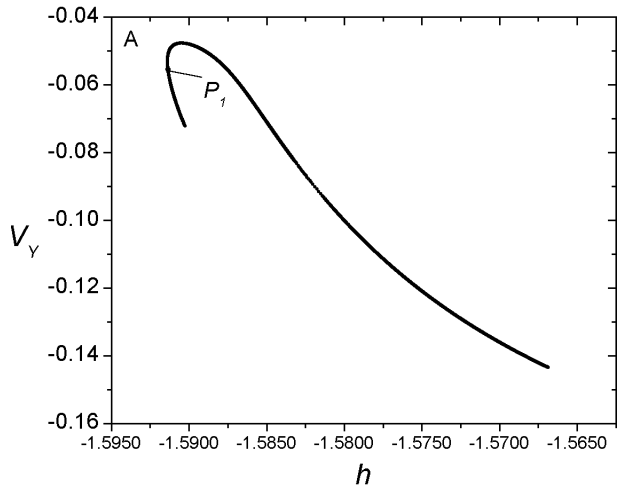
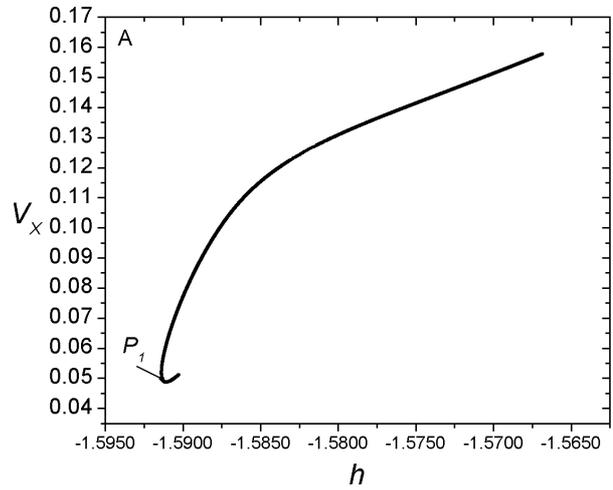

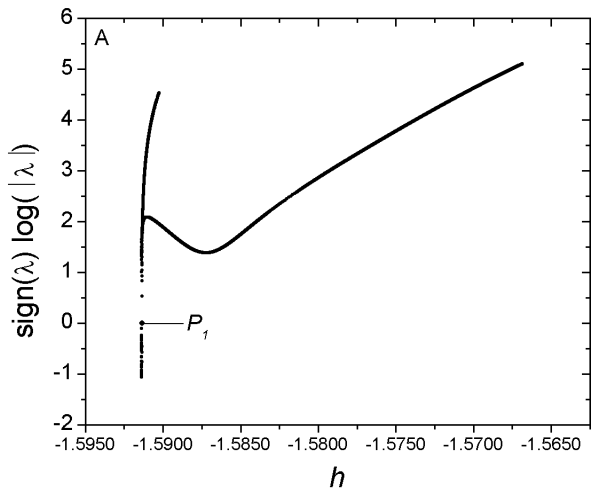
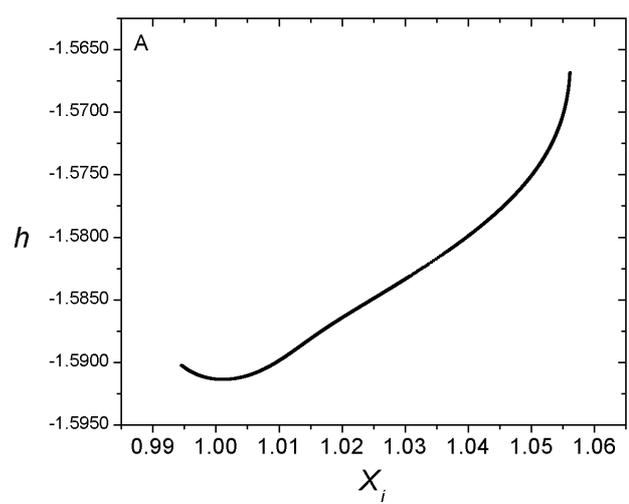

## Family 180 B - Asymmetric family of asymmetric POs

$h_{min} = -1.591356$,  $h_{max} = -1.588341$,  $T_{min} = 17.915057$,  $T_{max} = 18.034441$

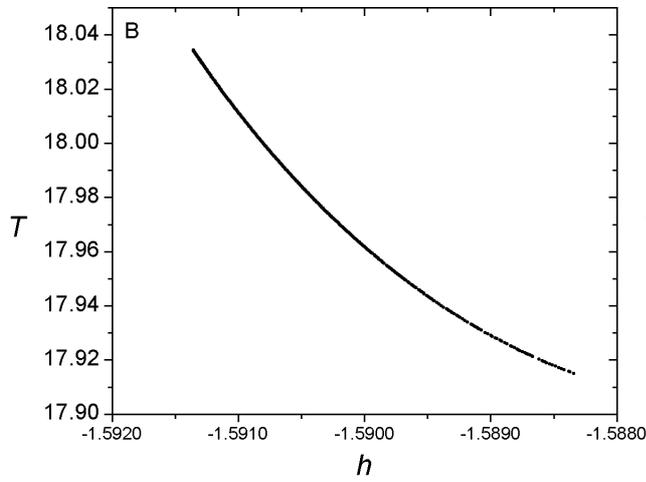

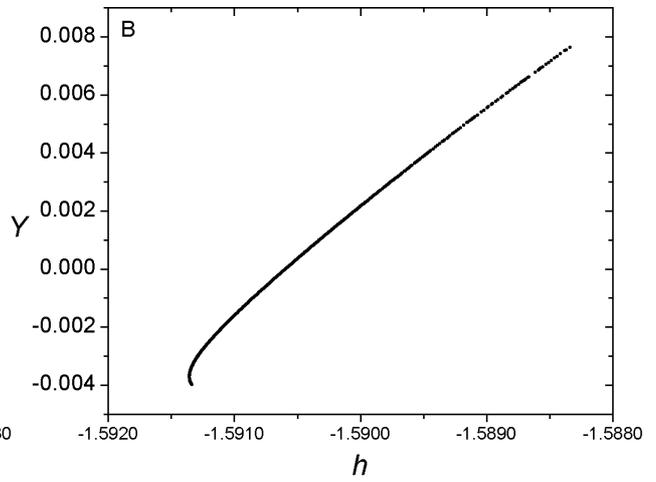

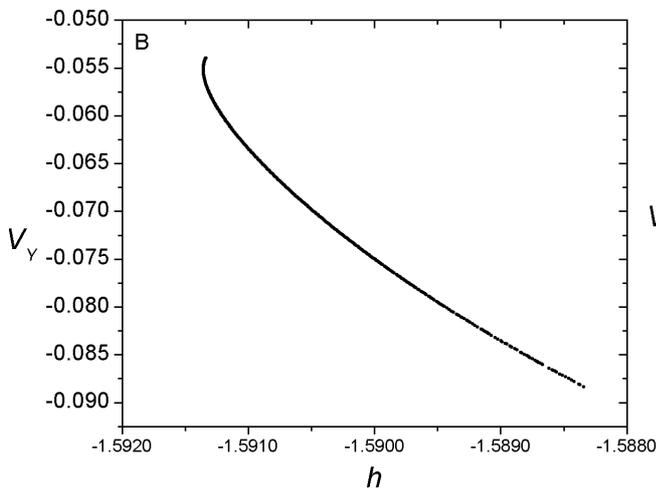

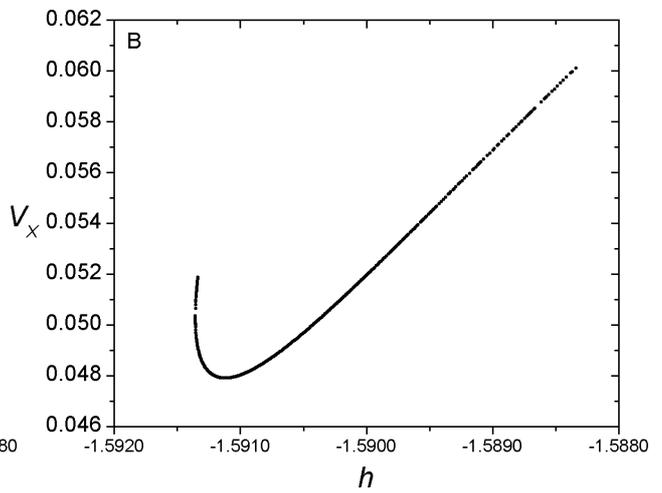

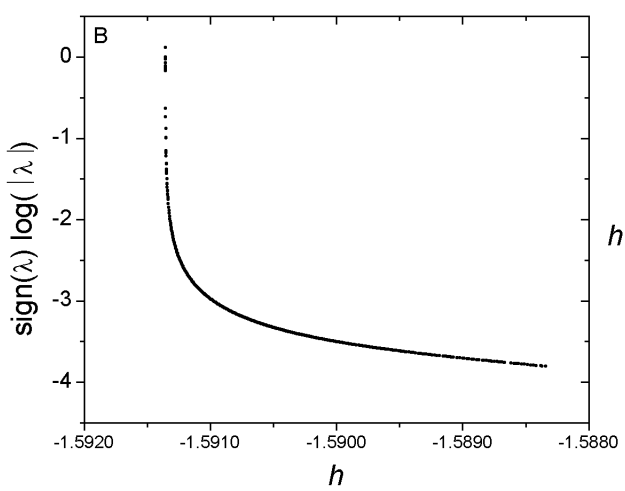

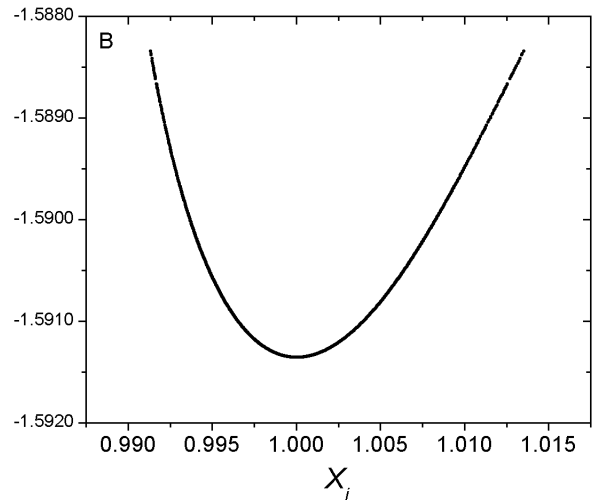

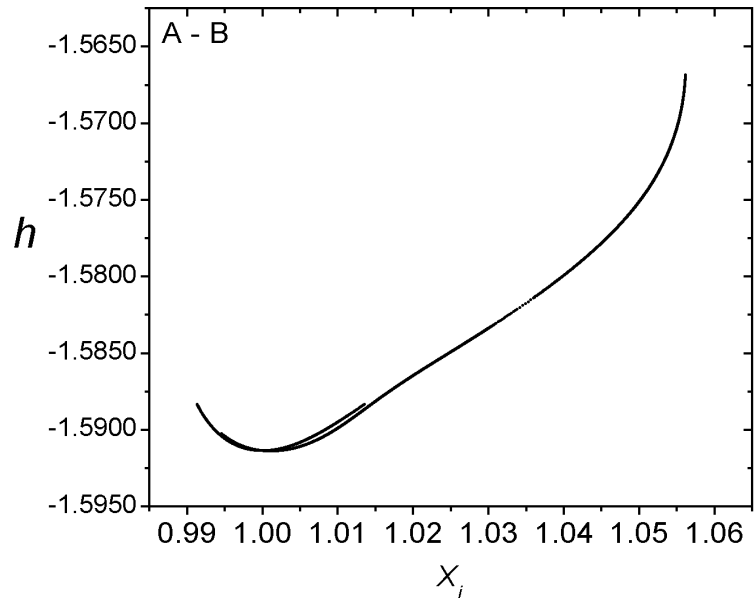

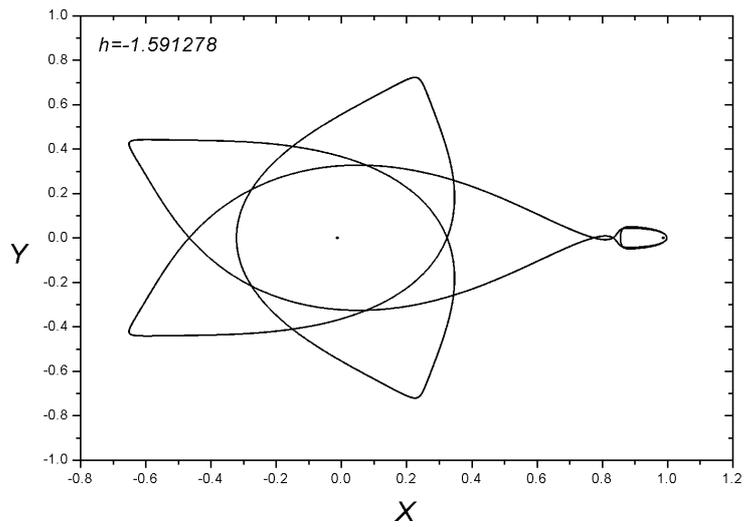

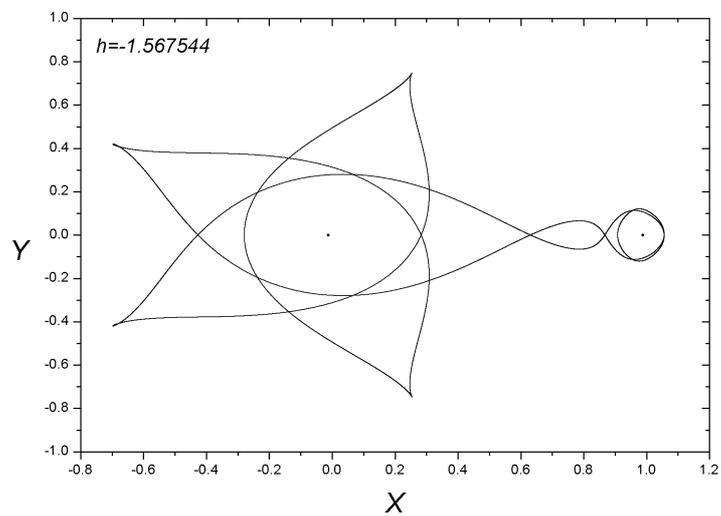

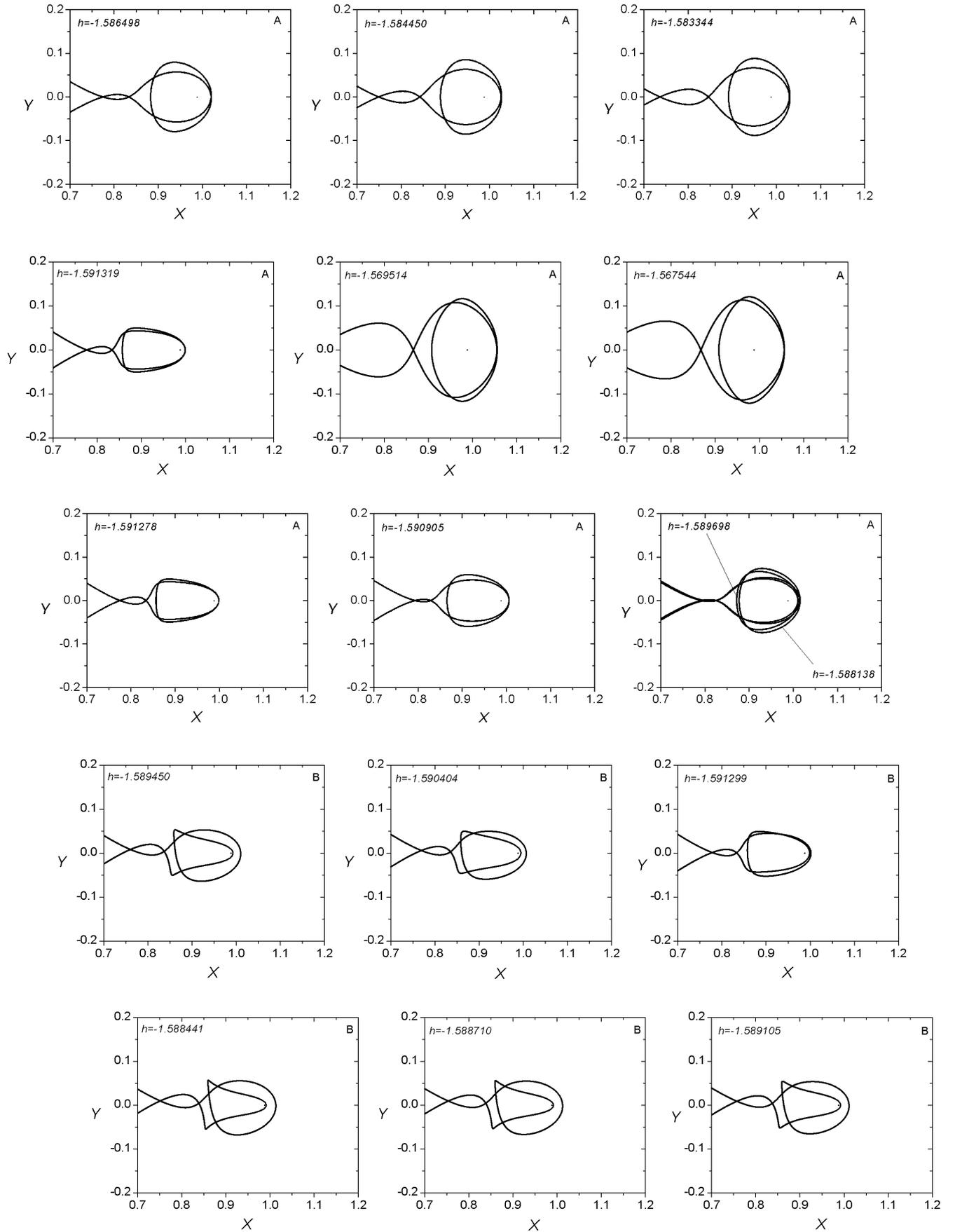

## Families 197A - 197 B - 197 C - 197 D

_Bifurcation Points_

|       | h         | T         | y         | $v_y$     | $v_x$    |
|-------|-----------|-----------|-----------|-----------|----------|
| $P_1$ | -1.588447 | 18.440386 | -0.029588 | 0.008950  | 0.088446 |
| $P_2$ | -1.589032 | 18.148169 | -0.012628 | -0.031732 | 0.092794 |
| $P_3$ | -1.591680 | 19.608840 | -0.004857 | -0.049800 | 0.049017 |

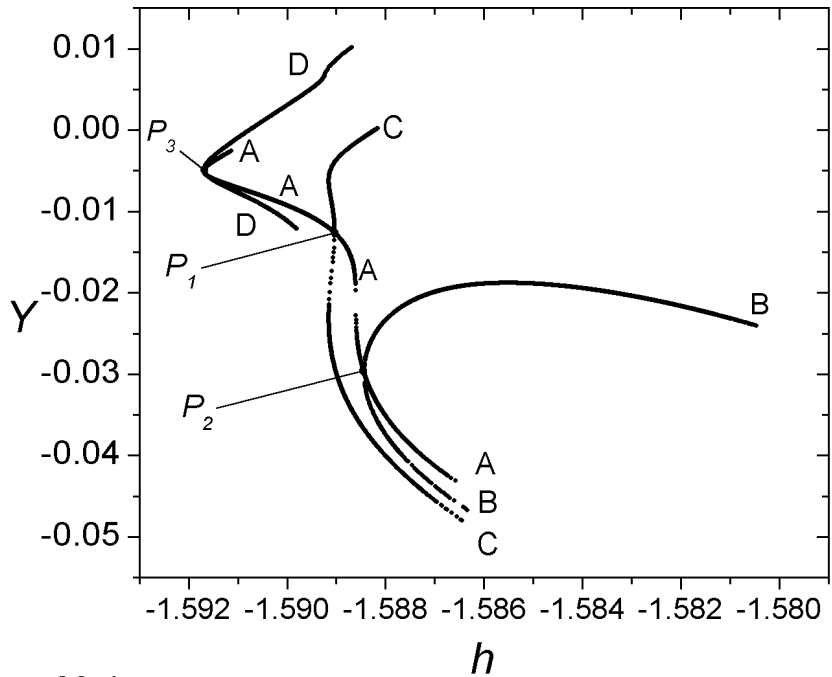

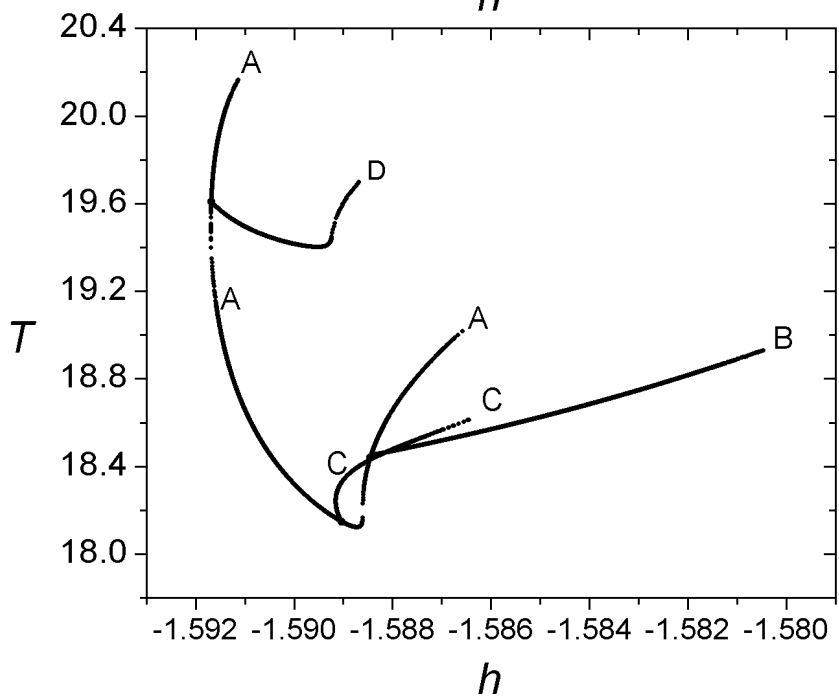

# Family 197 A - Symmetric family of symmetric POs

$h_{min} = -1.591688$, $h_{max} = -1.586582$, $T_{min} = 18.122697$, $T_{max} = 20.165072$

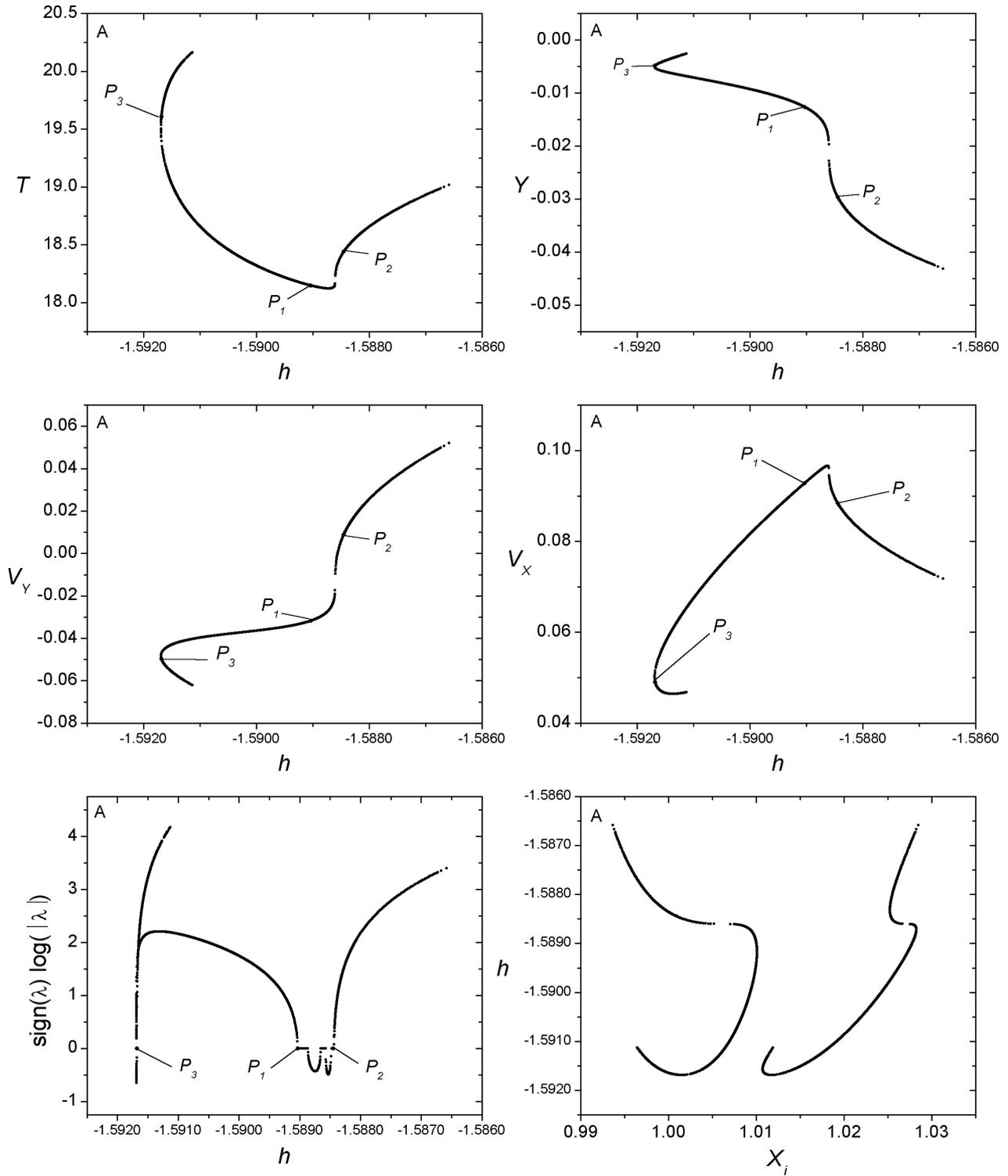

# Family 197 B - Symmetric family of asymmetric POs

$h_{min} = -1.588429, \ h_{max} = -1.580468, \ T_{min} = 18.450933, \ T_{max} = 18.929431$

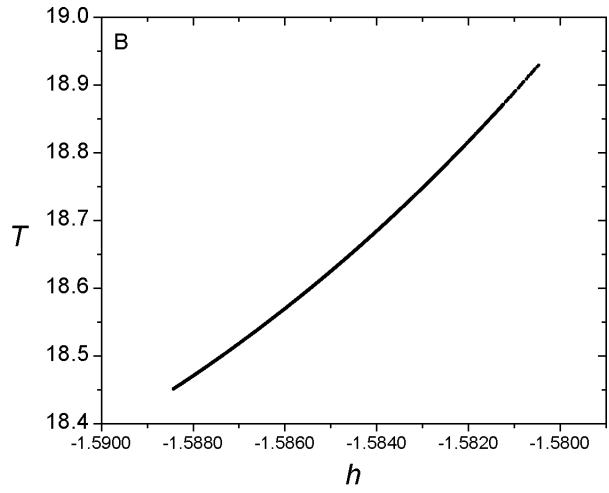
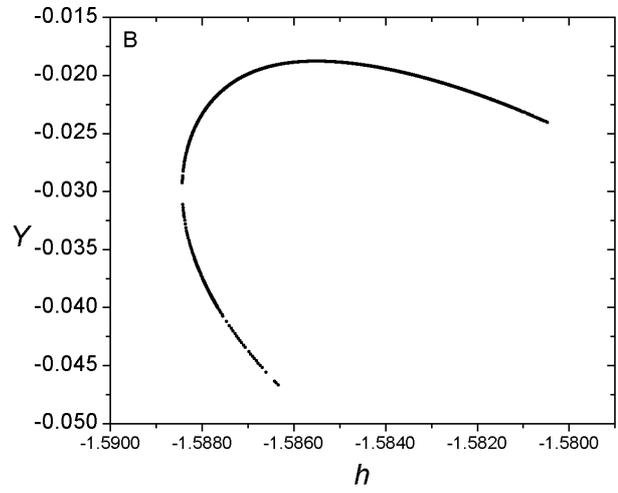

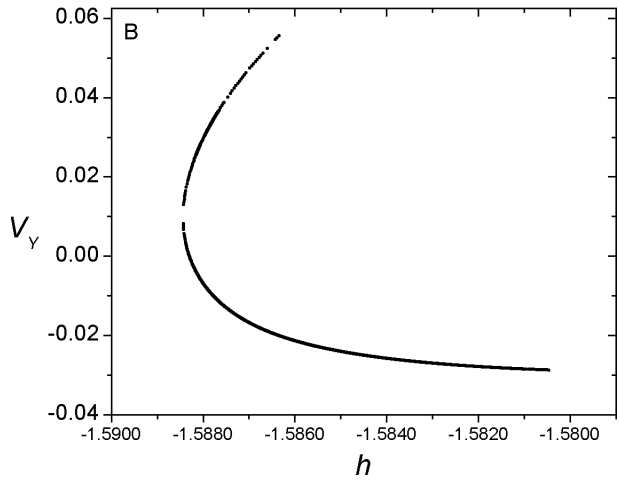
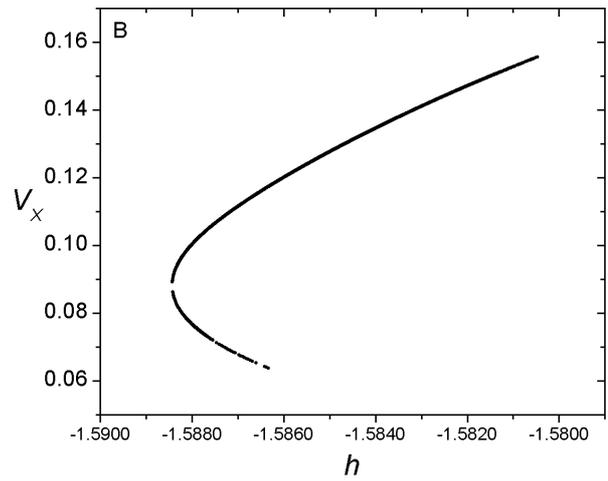

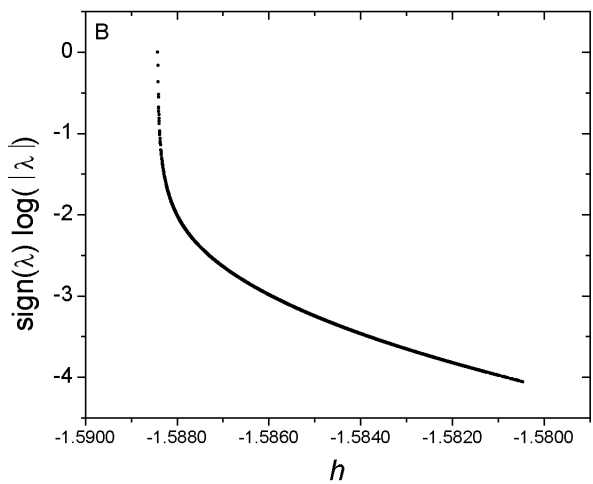
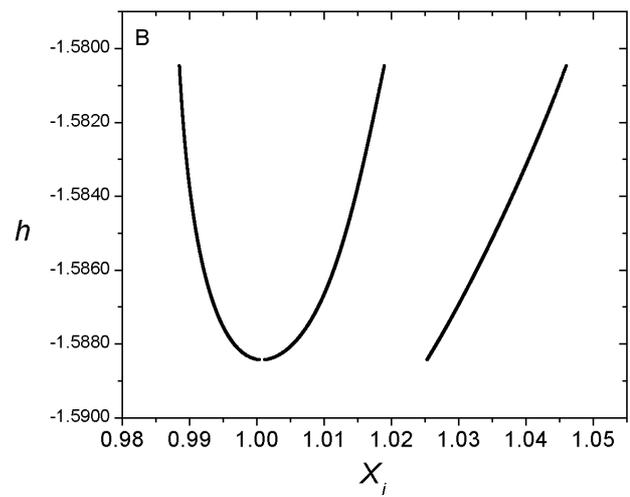

# Family 197 C - Symmetric family of asymmetric POs

$h_{min} = -1.589152, \quad h_{max} = -1.586448, \quad T_{min} = 18.148249, \quad T_{max} = 18.612661$

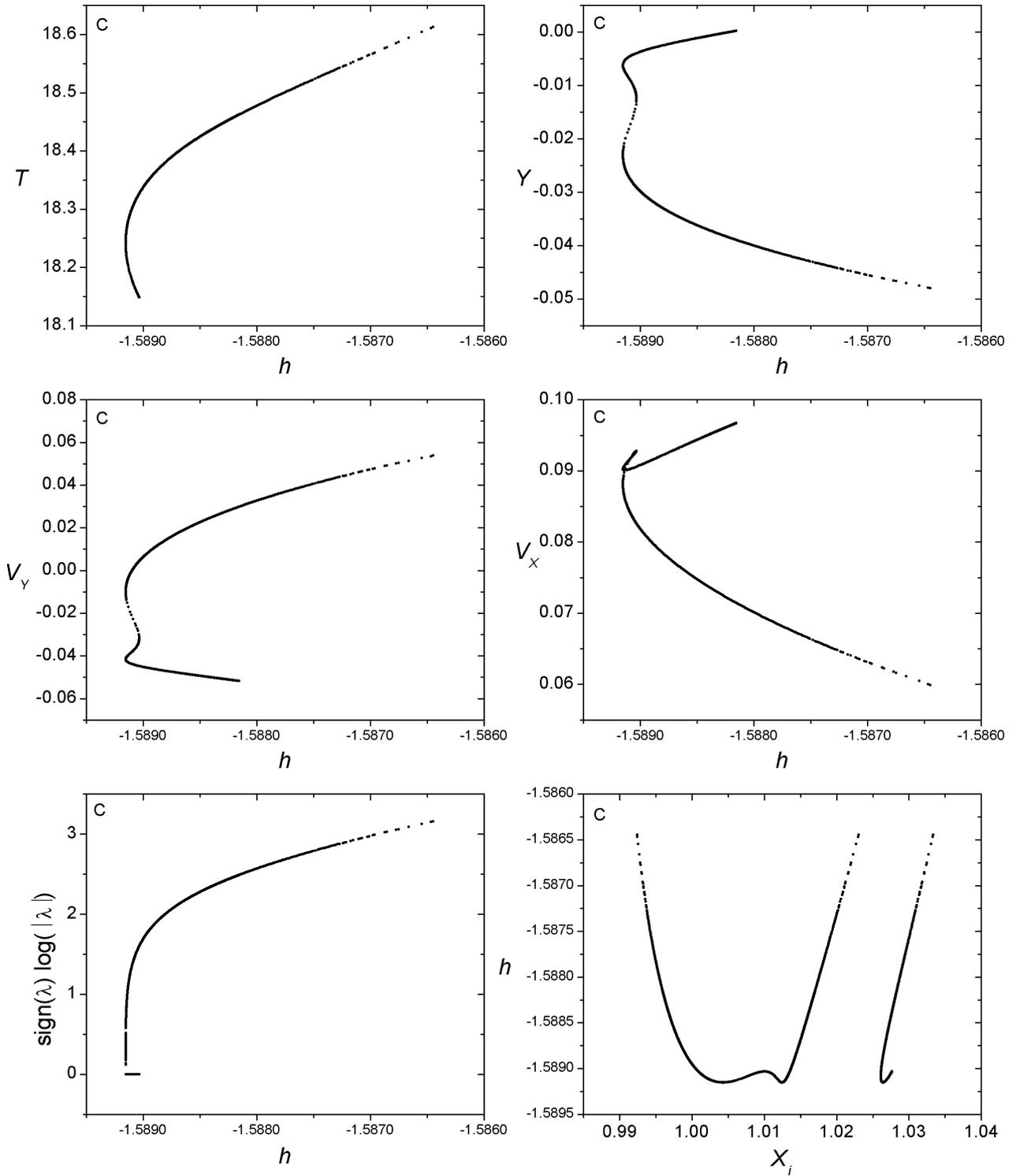

# Family 197 D - Symmetric family of asymmetric POs

$h_{min} = -1.5916802$, $h_{max} = -1.588681$, $T_{min} = 19.401657$, $T_{max} = 19.696562$

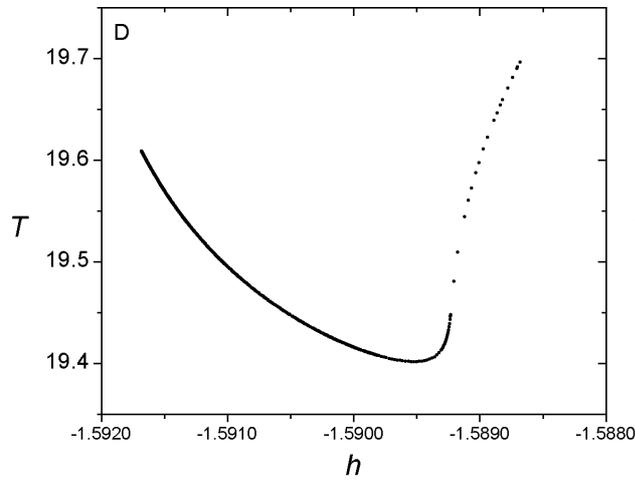
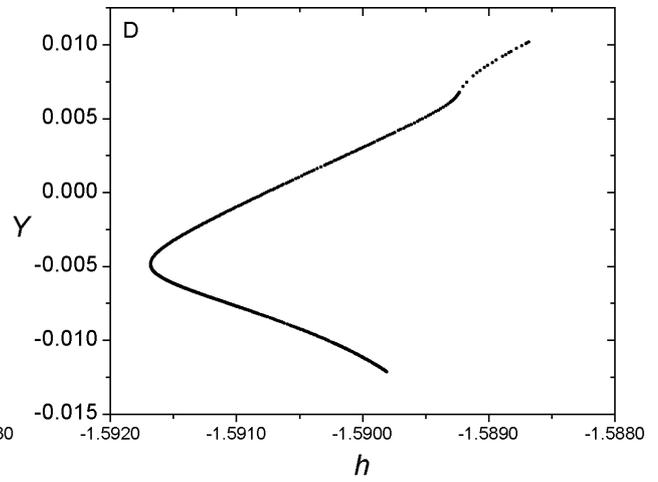

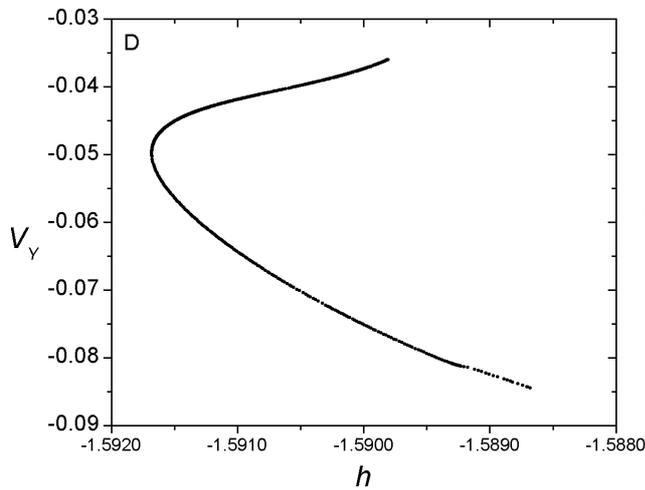
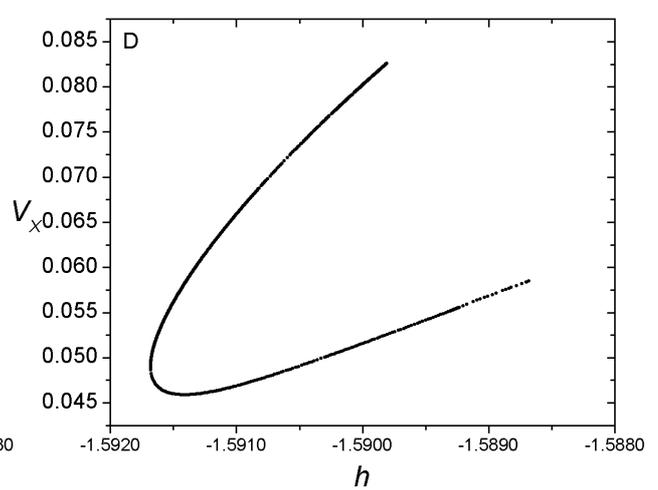

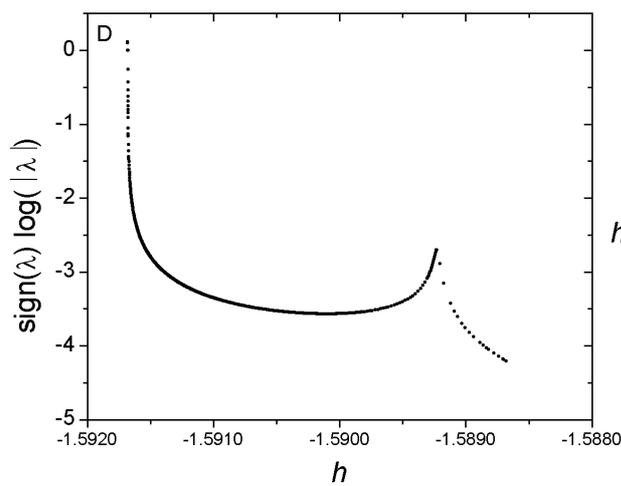
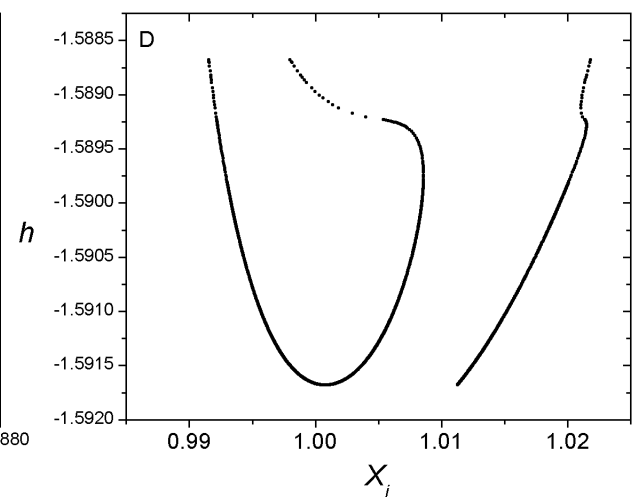

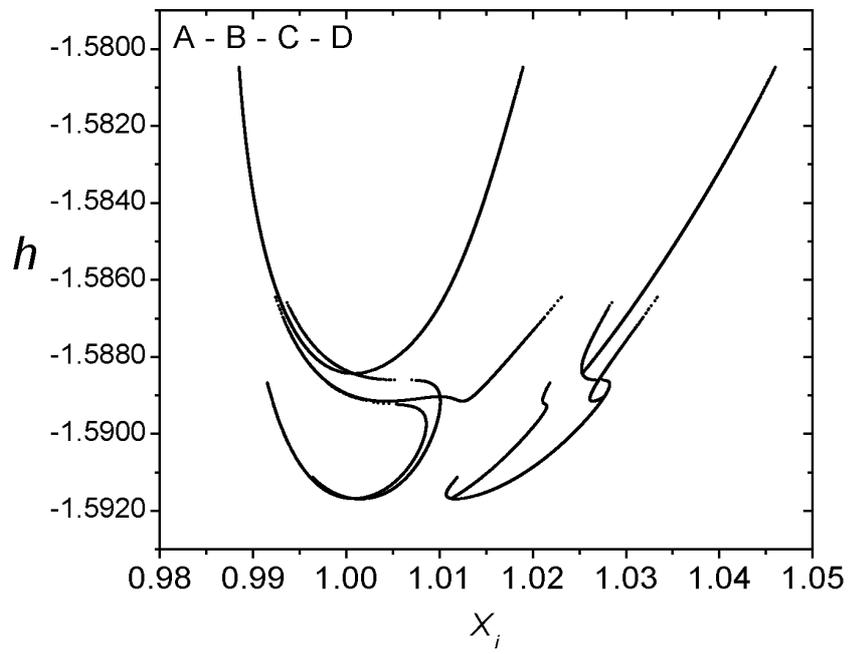

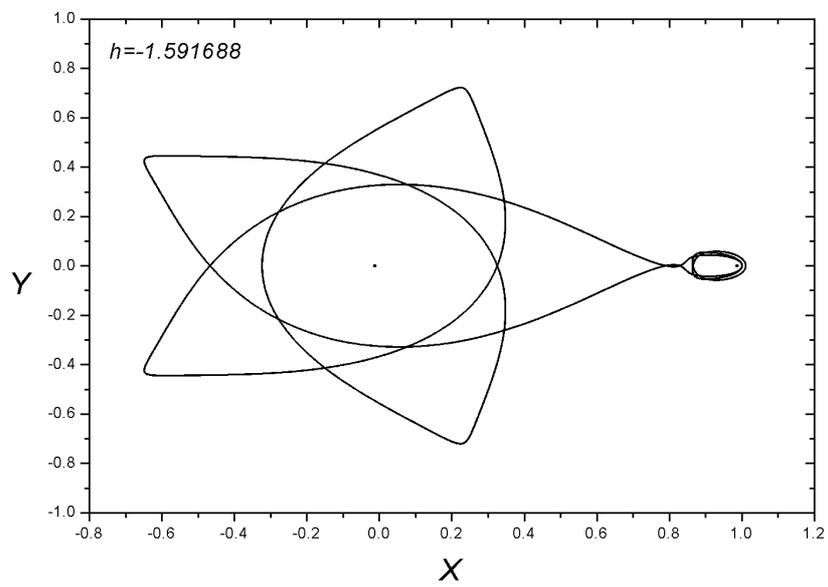

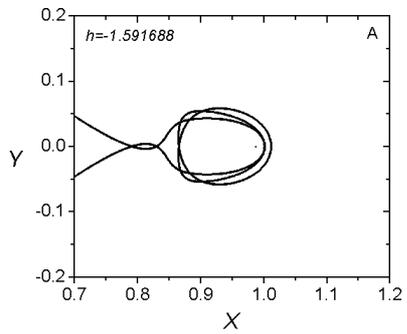
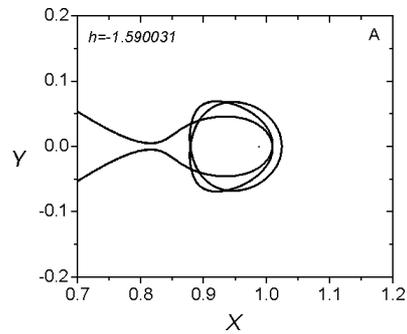
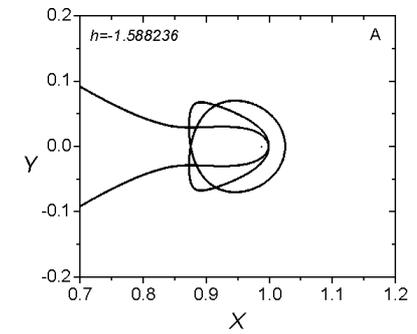

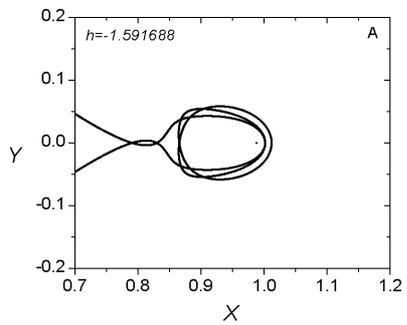
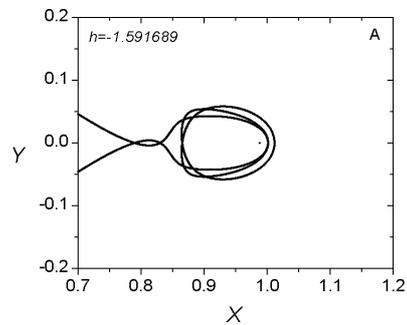
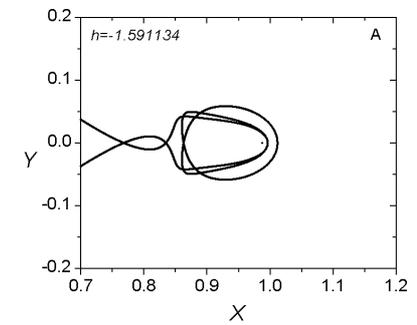

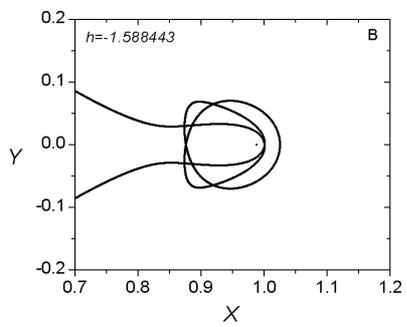
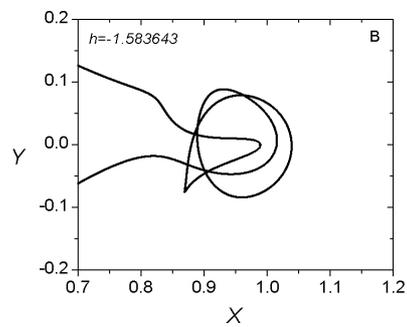
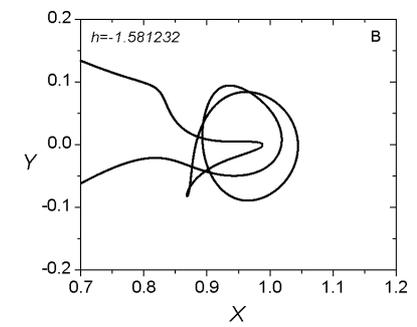

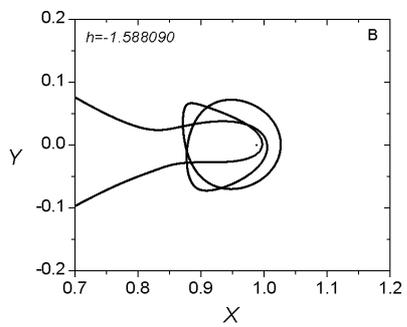
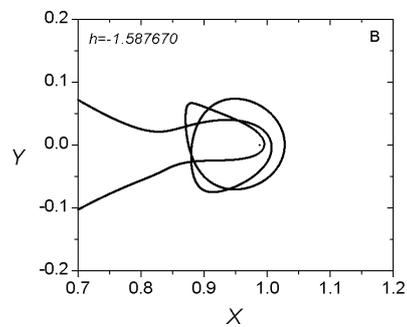
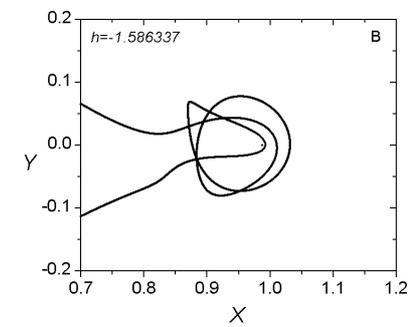

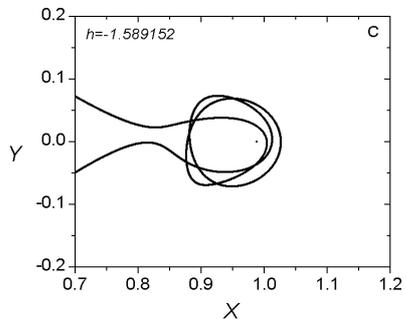
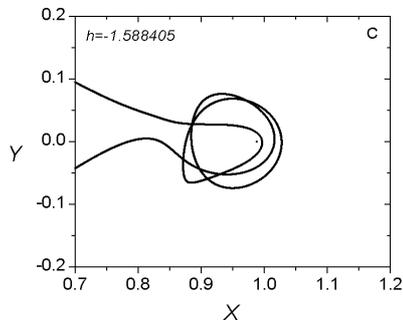
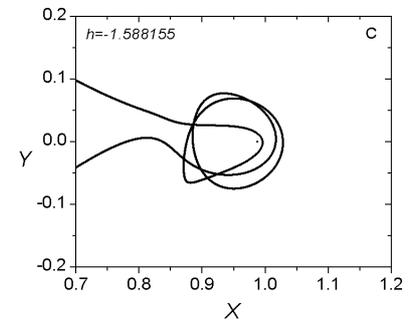

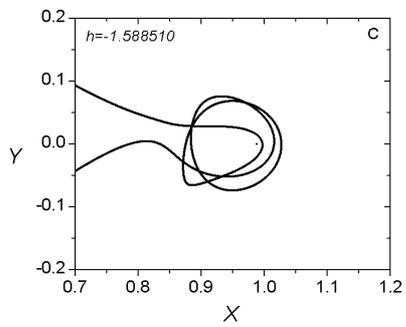
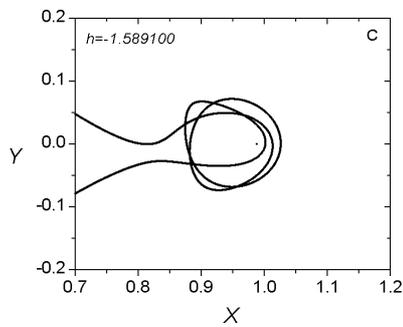
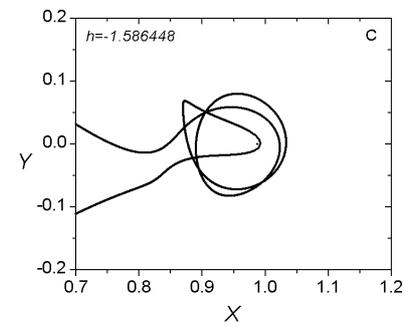

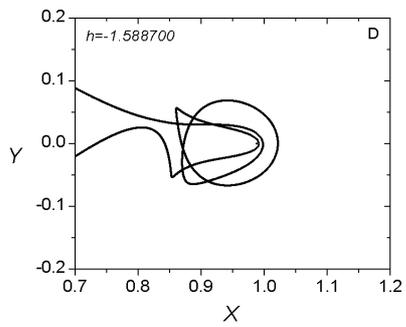
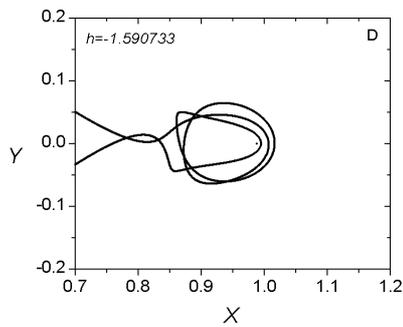
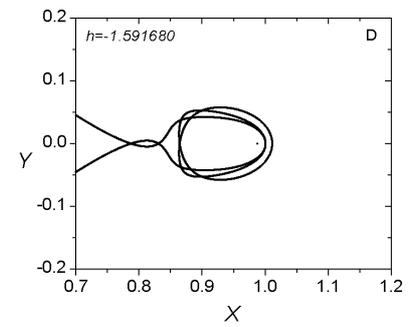

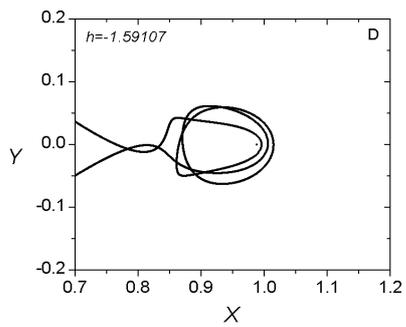
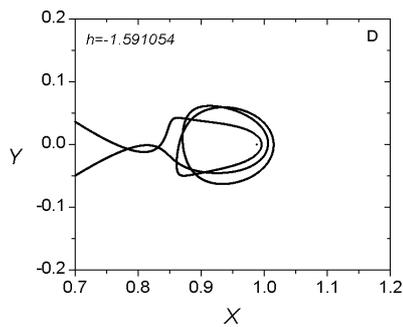
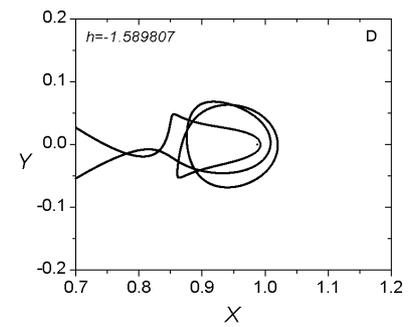

## *Families 146 A - 146 B - 146 C - 146 D*

*Bifurcation Points*

|       | h         | T         | y         | $v_y$      | $v_x$     |
|-------|-----------|-----------|-----------|-----------|-----------|
| $P_1$ | -1.589430 | 19.608724 | -0.016899 | -0.018996 | 0.089130  |
| $P_2$ | -1.588979 | 19.754793 | -0.025445 | 0.001567  | 0.087992  |
| $P_3$ | -1.591891 | 20.948216 | -0.005529 | -0.046382 | 0.047764  |

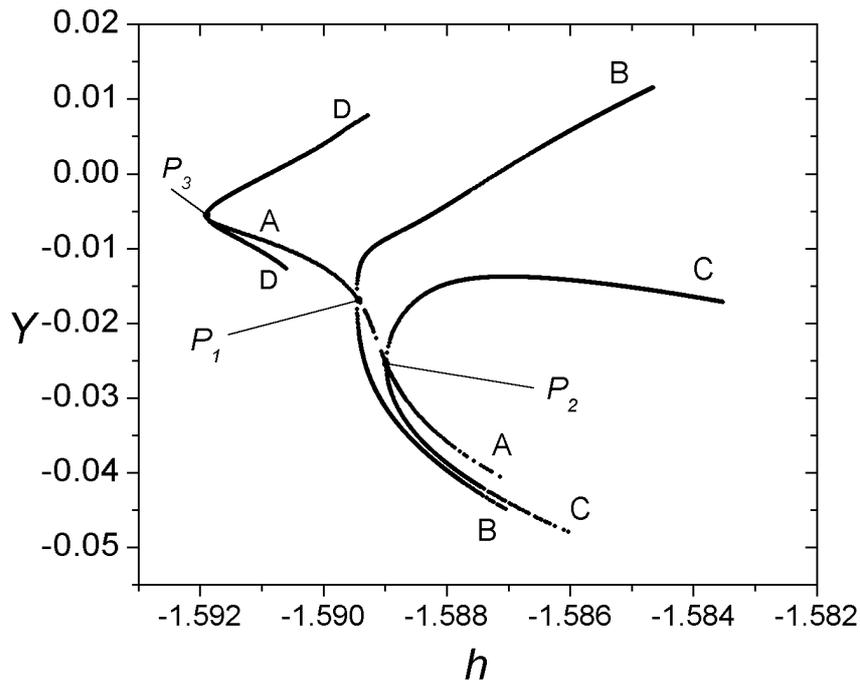

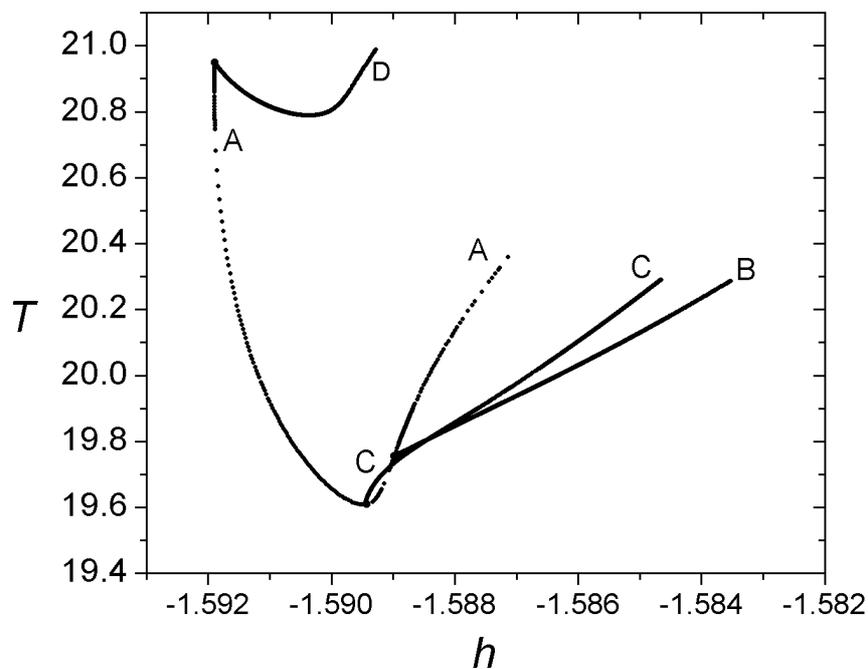

### *Family 146 A - Symmetric family of symmetric POs*

$h_{min} = -1.591896$, $h_{max} = -1.587133$, $T_{min} = 19.607828$, $T_{max} = 20.949069$

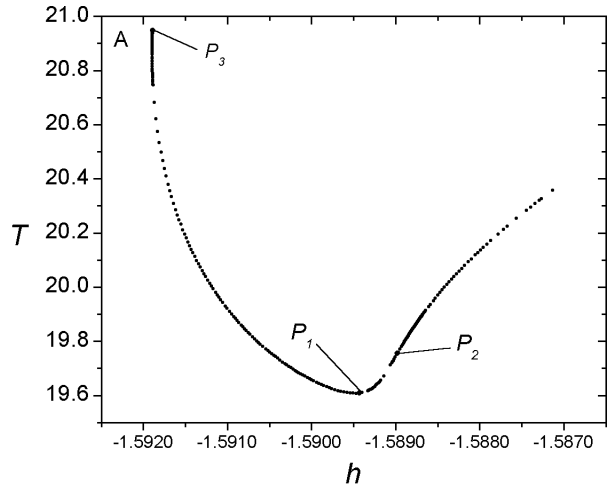
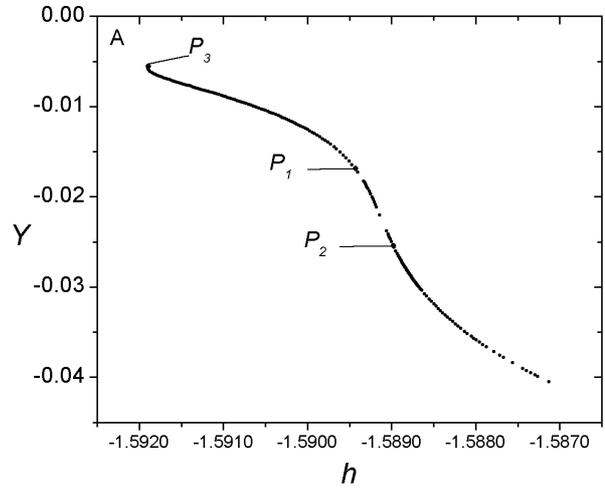

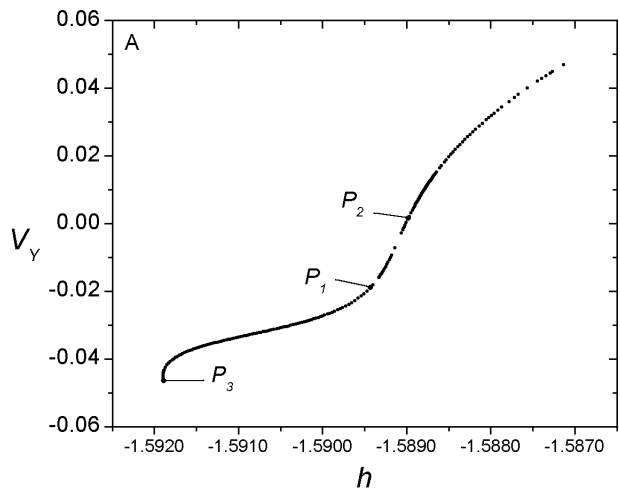
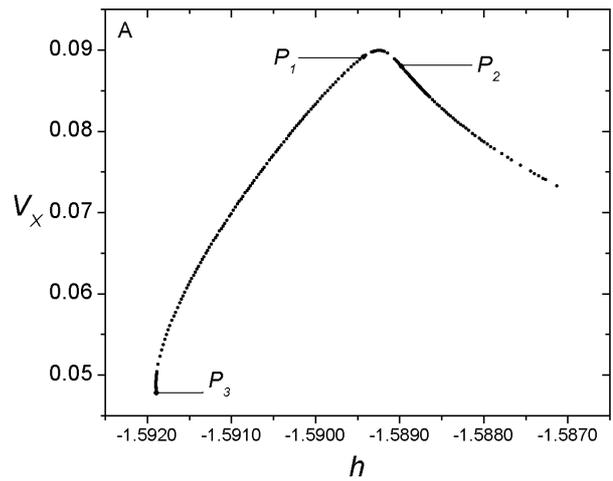

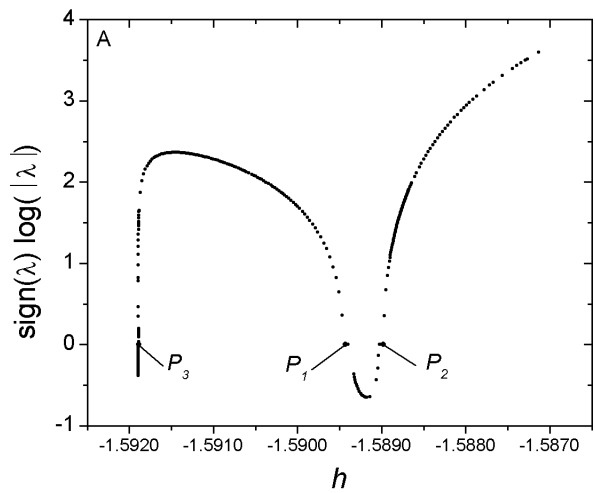
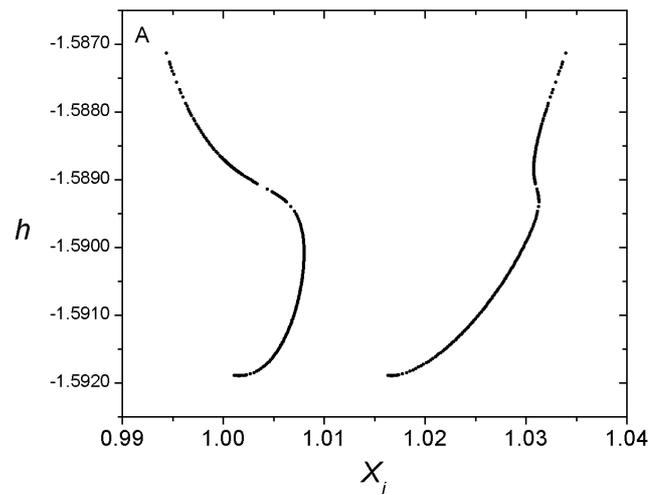

# Family 146 B - *Symmetric family of asymmetric POs*

$h_{min} = -1.589455, \ h_{max} = -1.584655, \ T_{min} = 19.608057, \ T_{max} = 20.290378$

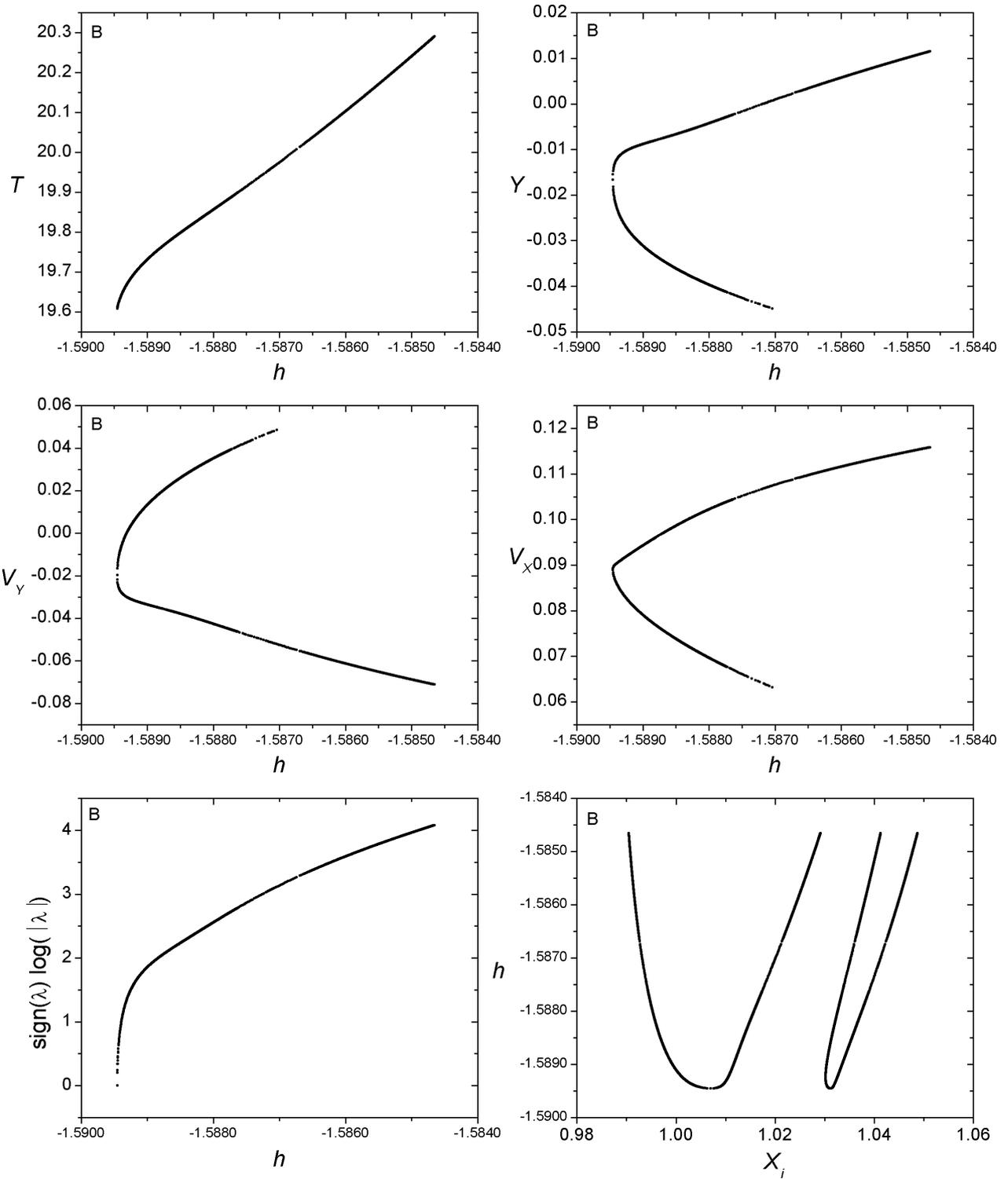

## Family 146 C - *Symmetric family of asymmetric POs*

$h_{min} = -1.588976, \quad h_{max} = -1.583528, \quad T_{min} = 19.755857, \quad T_{max} = 20.286207$

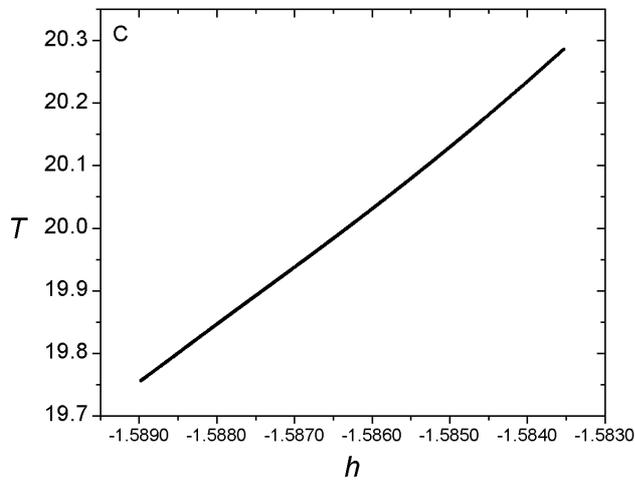
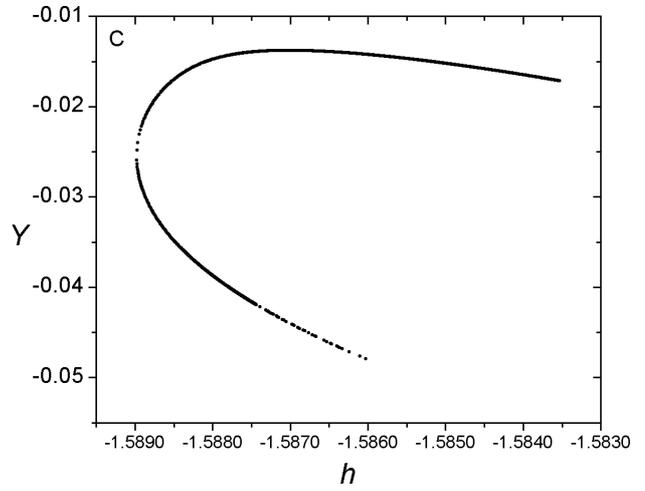
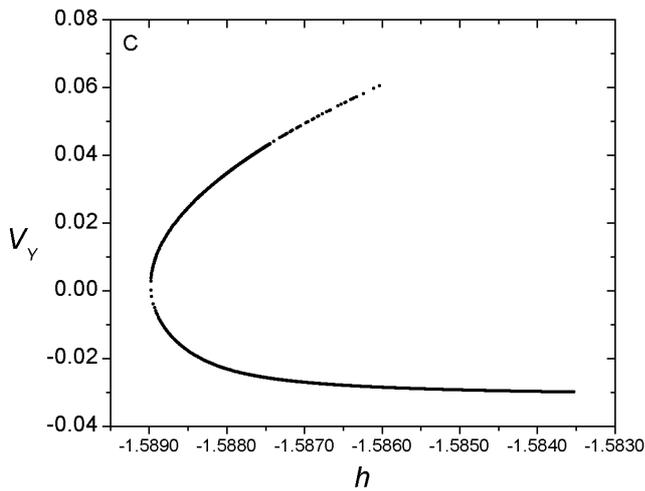
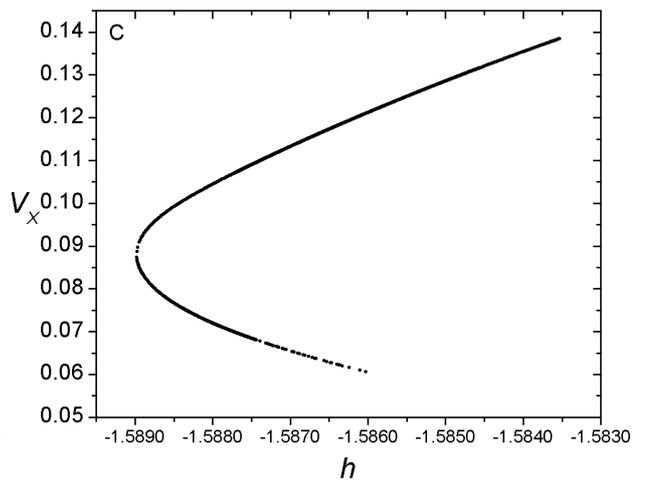
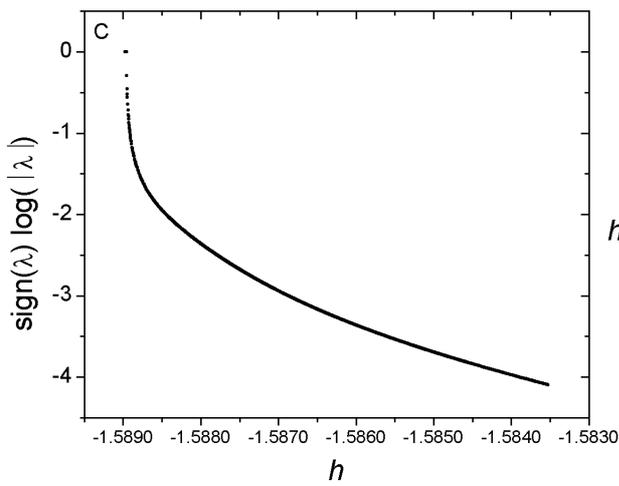
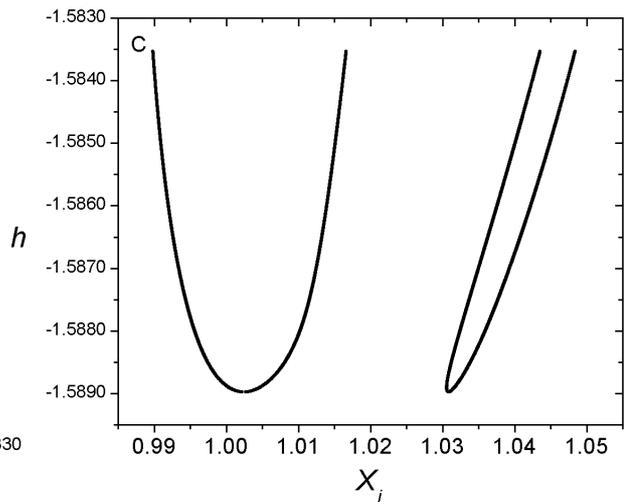

## Family 146 D - Symmetric family of asymmetric POs

$h_{min} = -1.591891, \quad h_{max} = -1.589277, \quad T_{min} = 20.788368, \quad T_{max} = 20.988249$

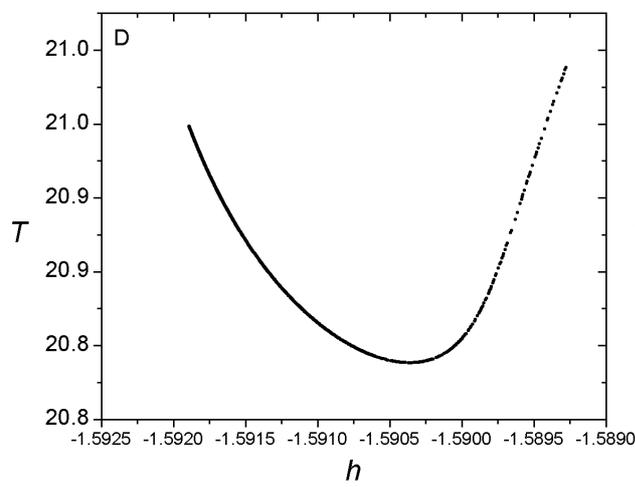
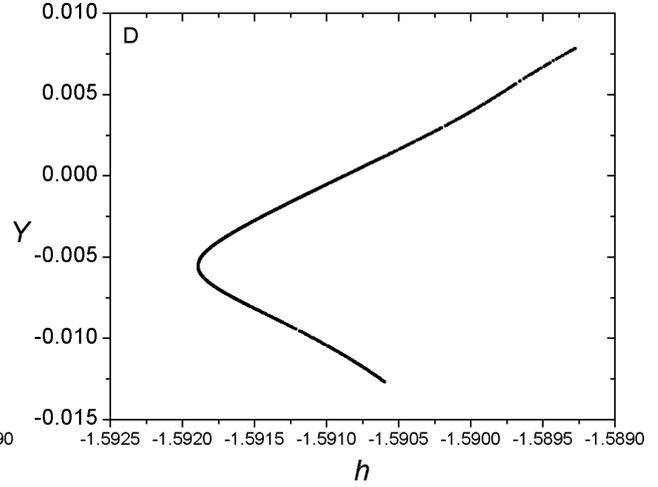

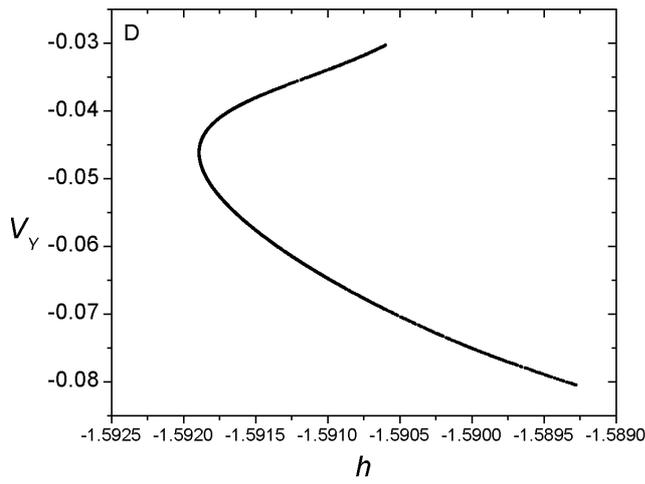
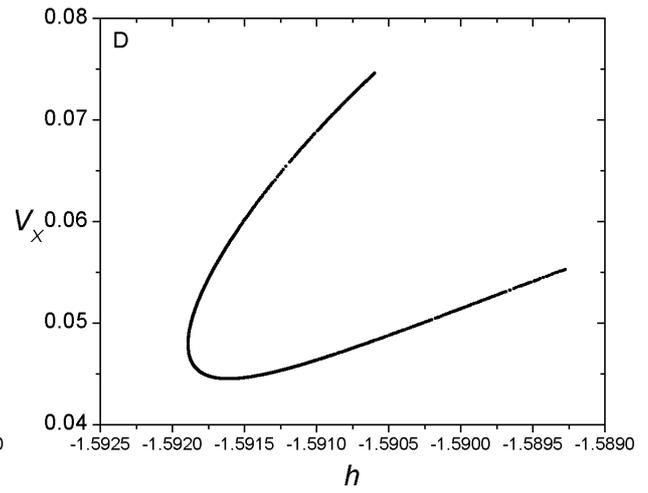

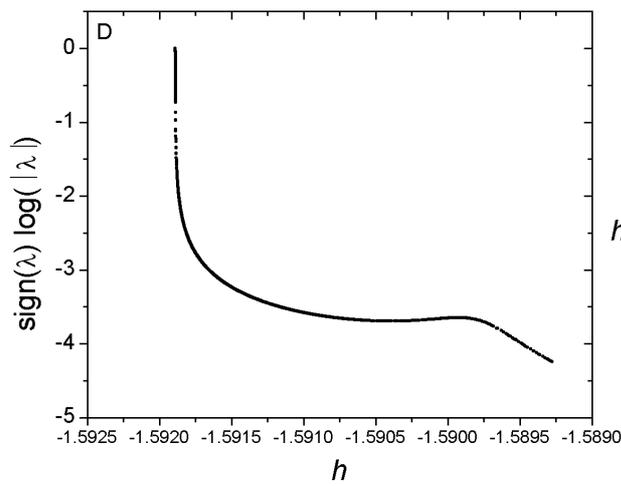
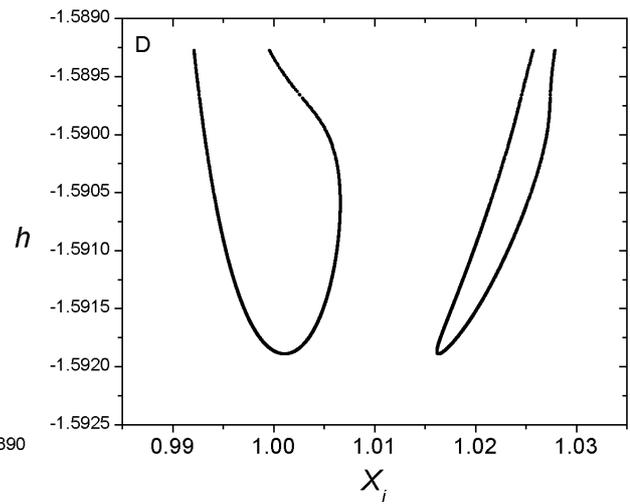

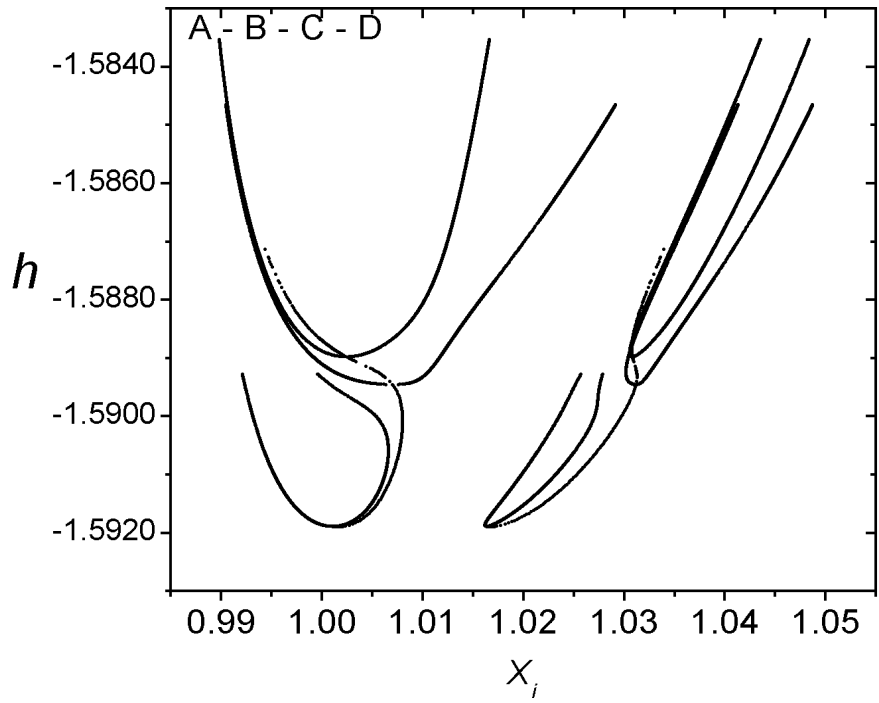

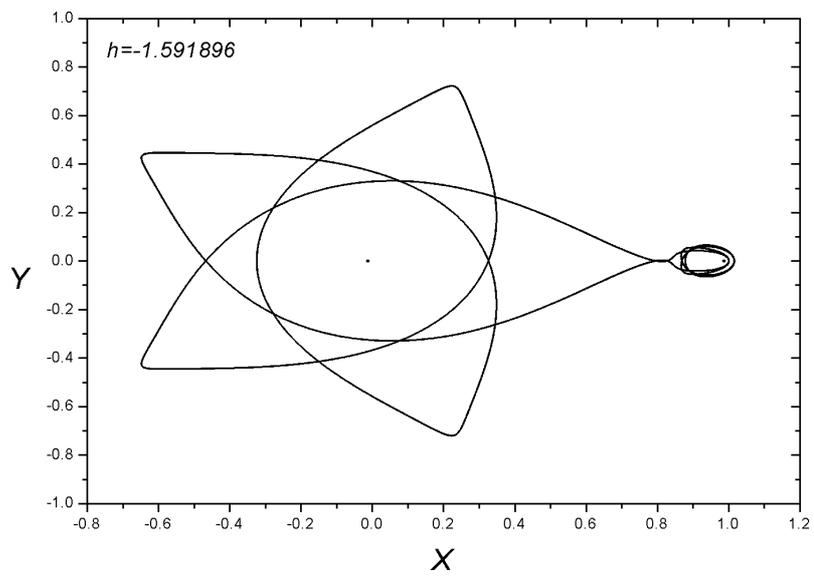

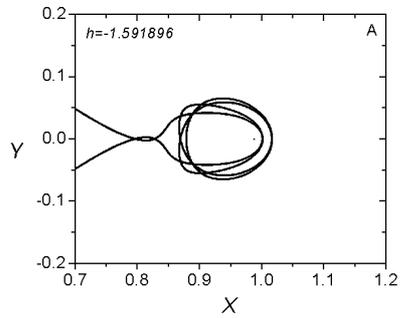
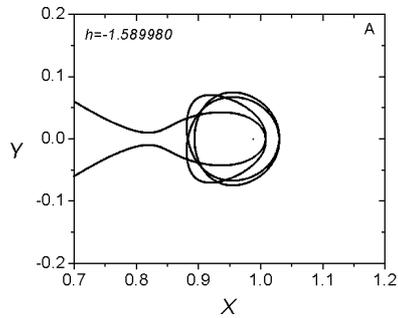
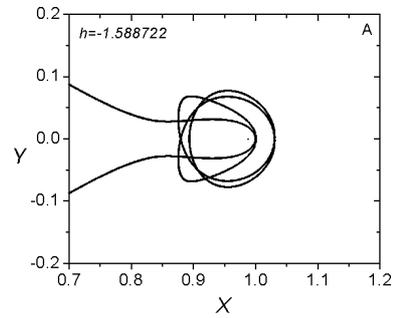

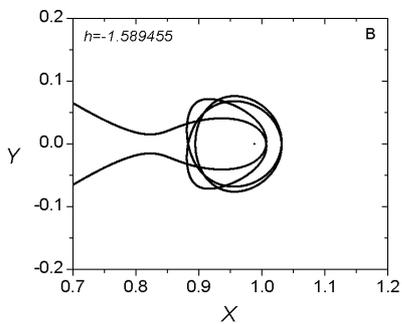
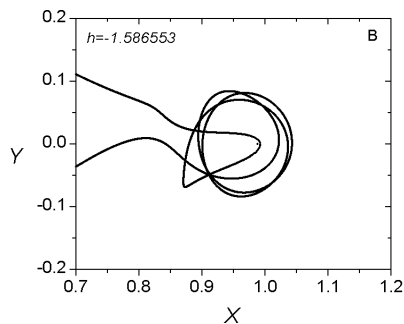
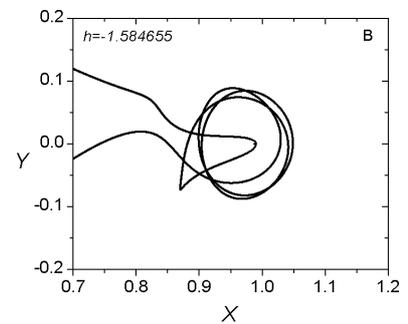

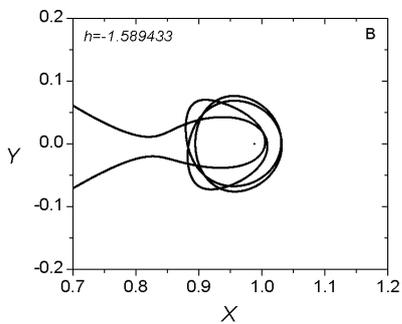
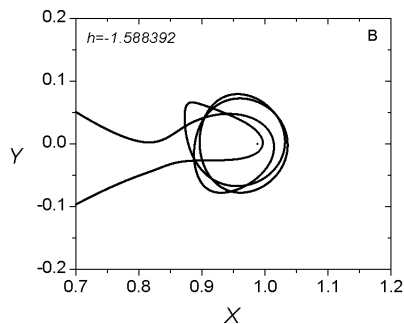
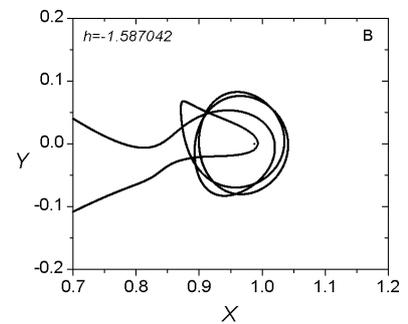

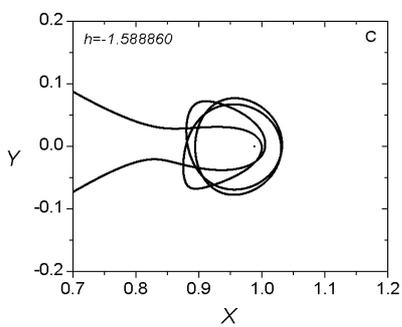
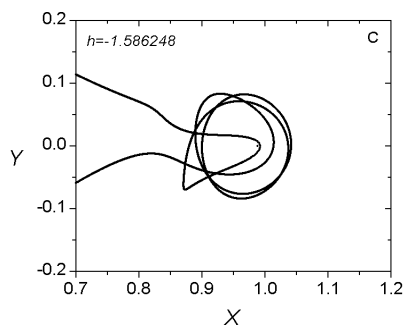
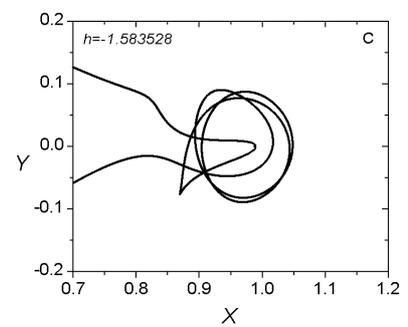

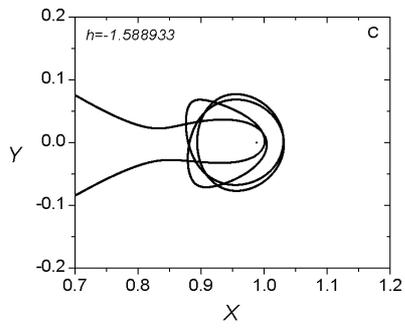

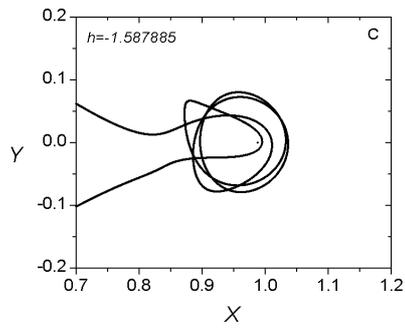

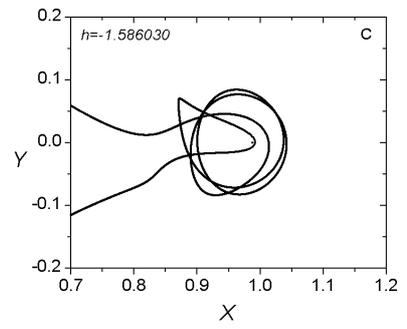

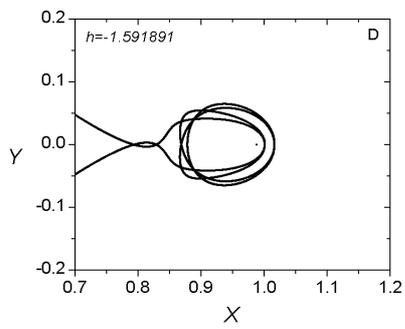

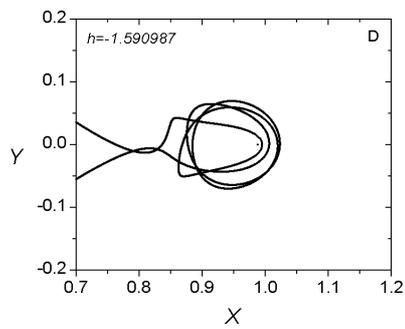

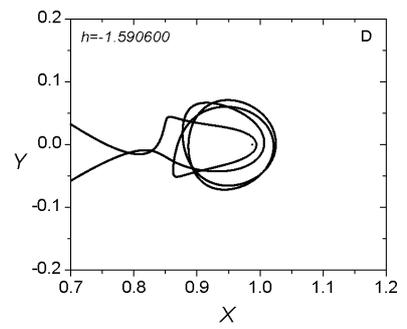

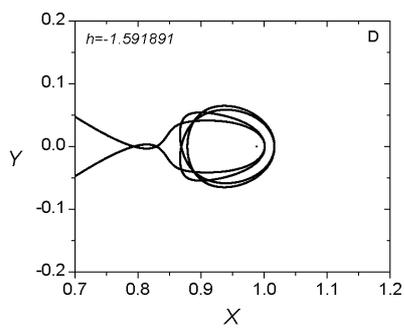

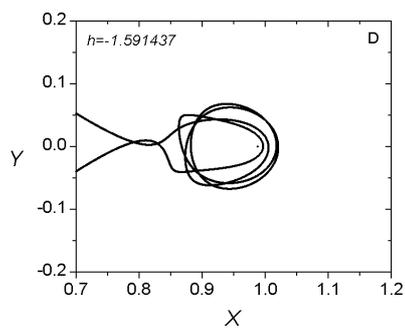

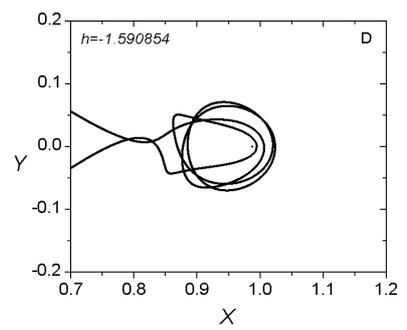

## Family 178 - Symmetric family of symmetric POs

$h_{min} = -1.592042, \quad h_{max} = -1.588089, \quad T_{min} = 20.992907, \quad T_{max} = 22.679572$

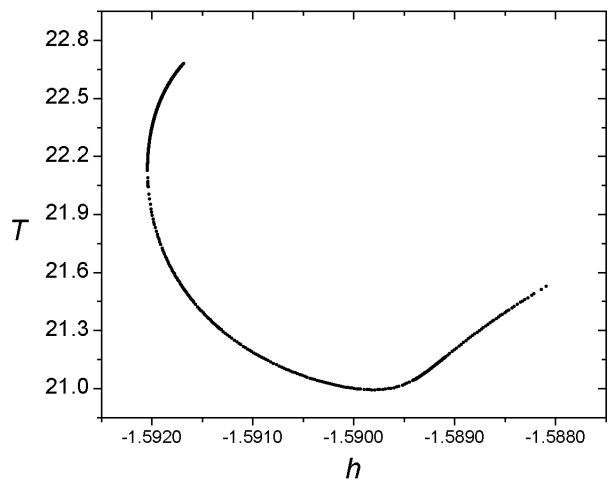
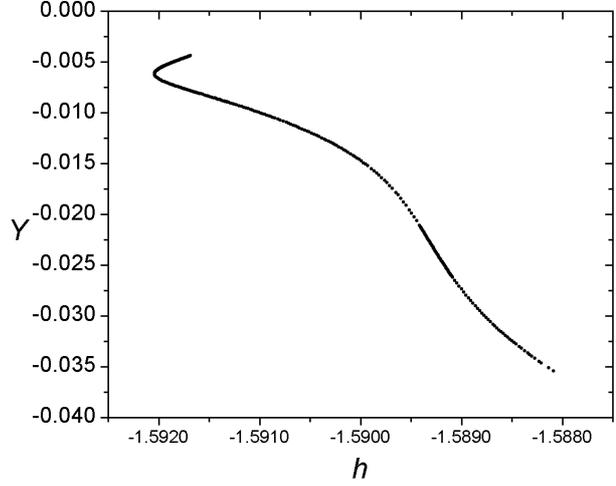

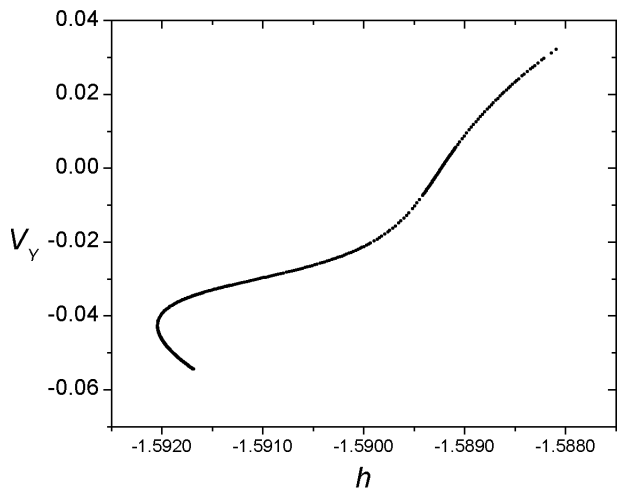
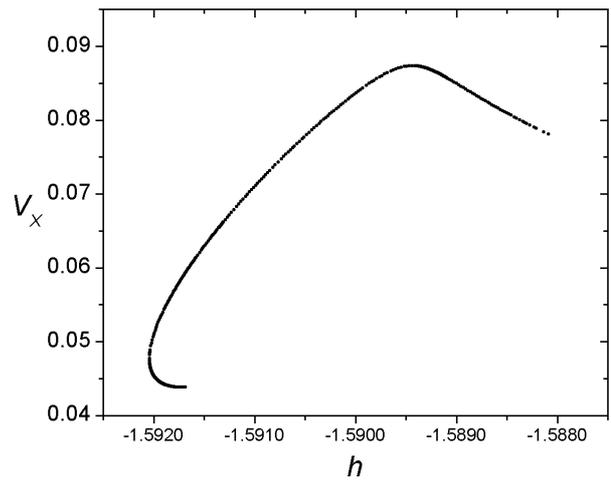

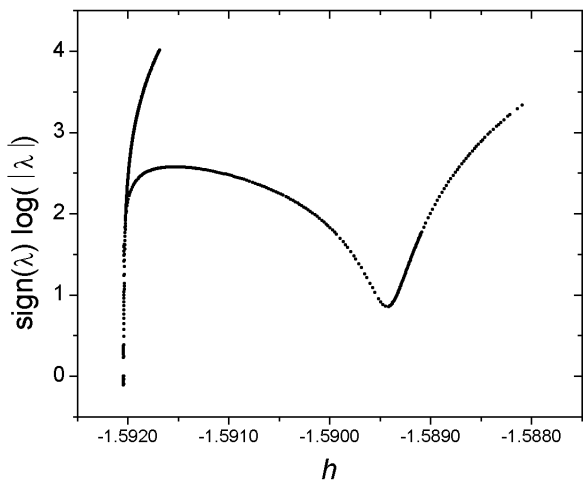
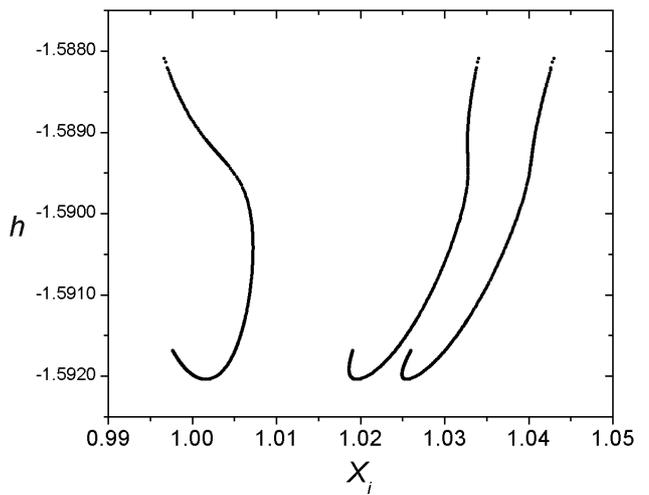

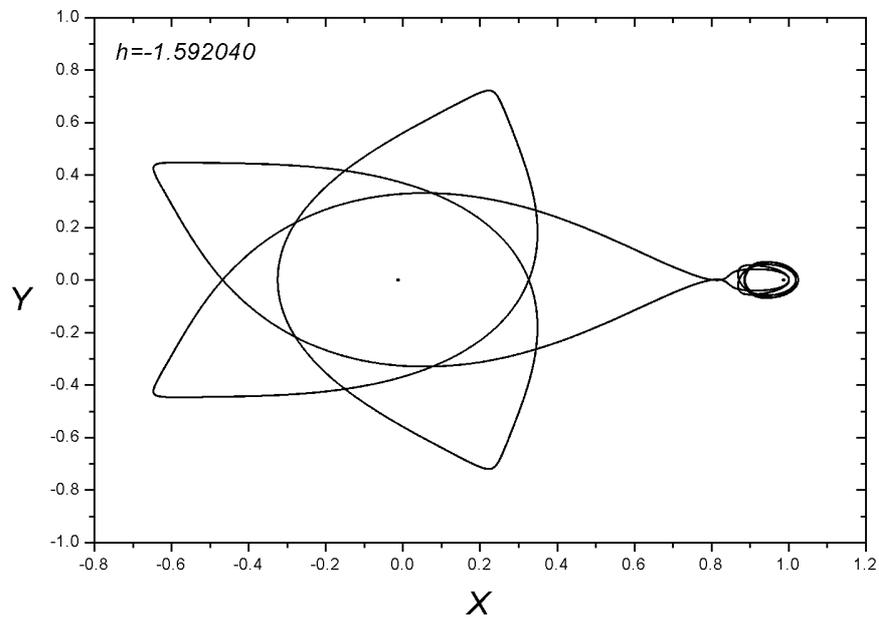

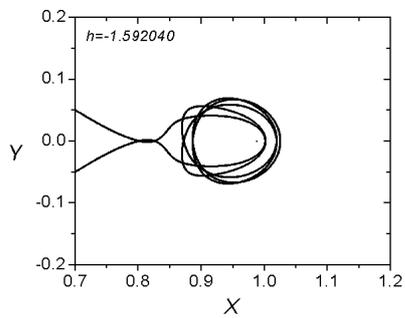
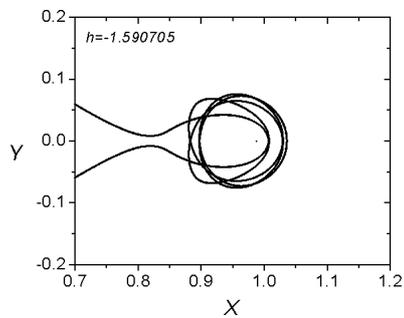
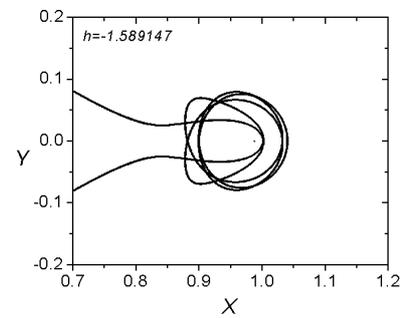

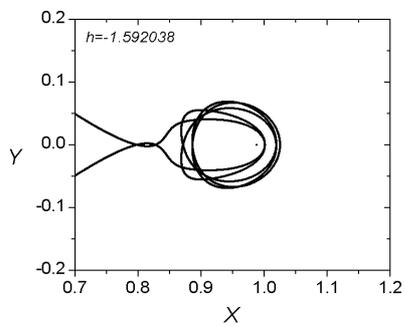
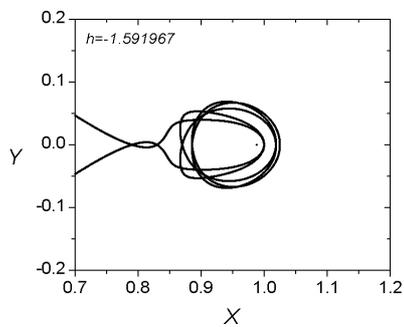
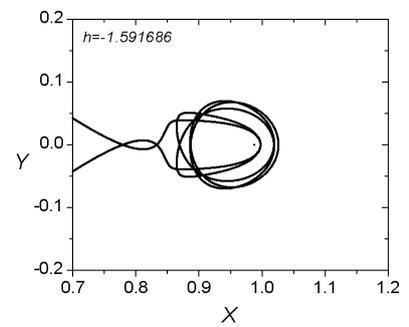

## Families 136 A - 136 B

*Bifurcation Point*

|       | $h$       | $T$        | $y$        | $v_y$      | $v_x$      |
|-------|-----------|------------|------------|------------|------------|
| $P_1$ | -1.588887 | 21.113038  | 0.0063551  | 0.079397   | 0.063989   |

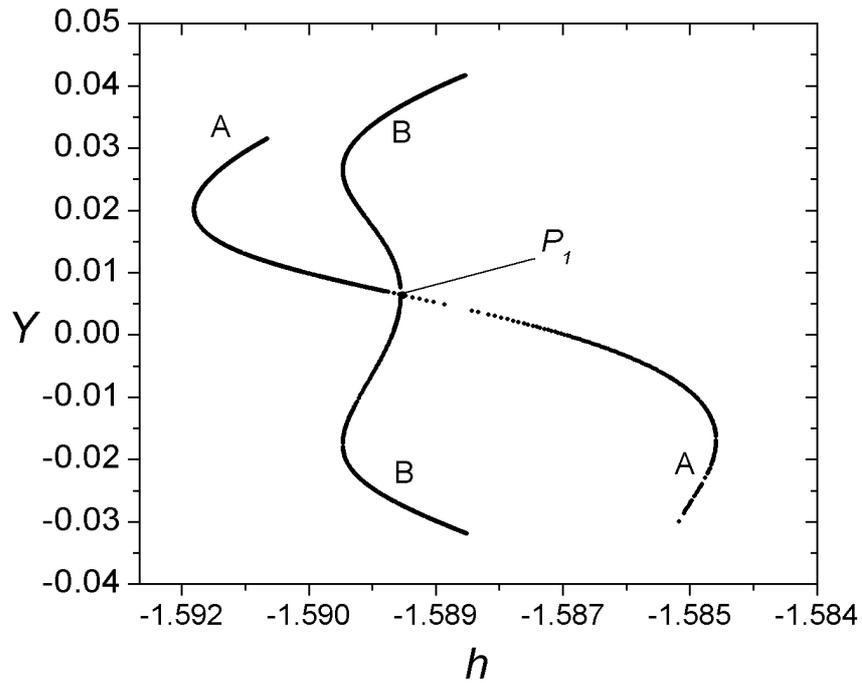

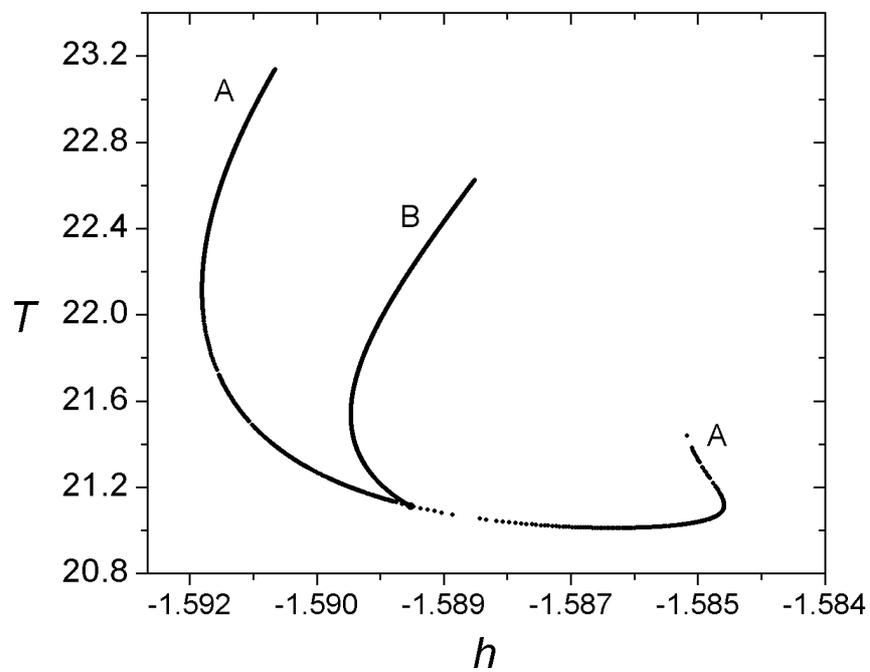

## Family 136 A - Symmetric family of symmetric POs

$h_{min} = -1.591352, \quad h_{max} = -1.585188, \quad T_{min} = 21.010742, \quad T_{max} = 23.139351$

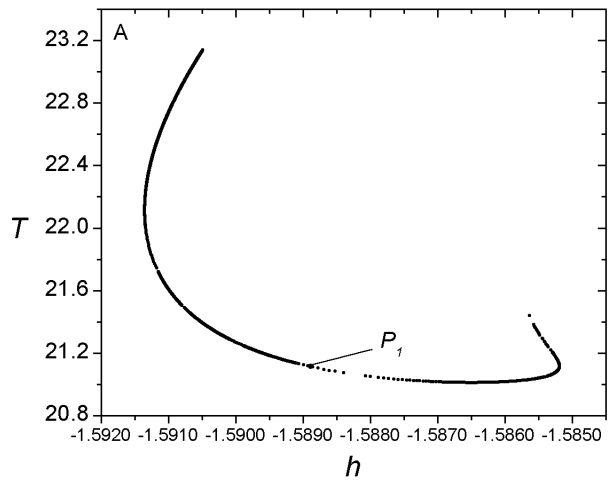

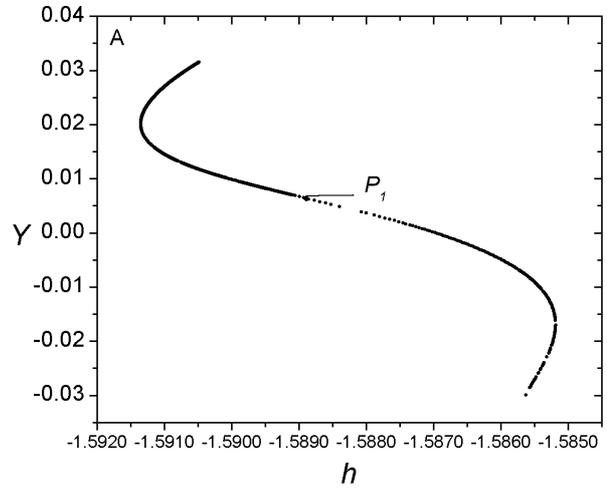

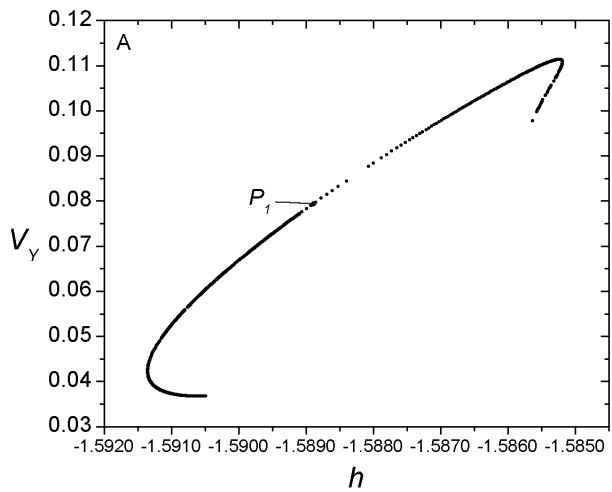

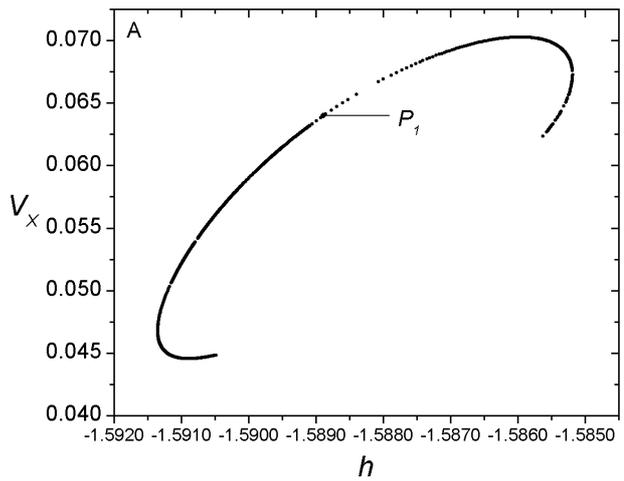

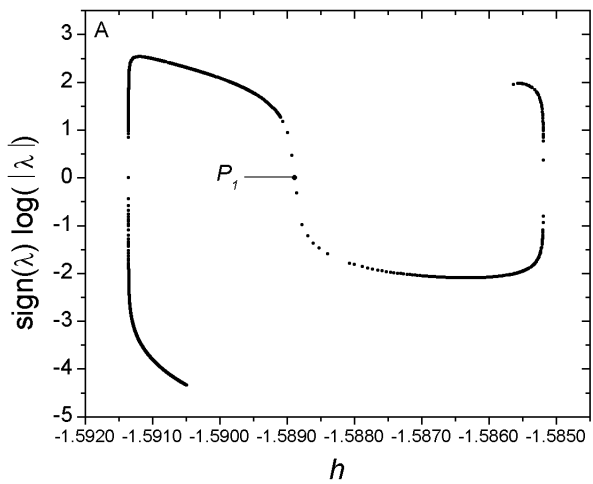

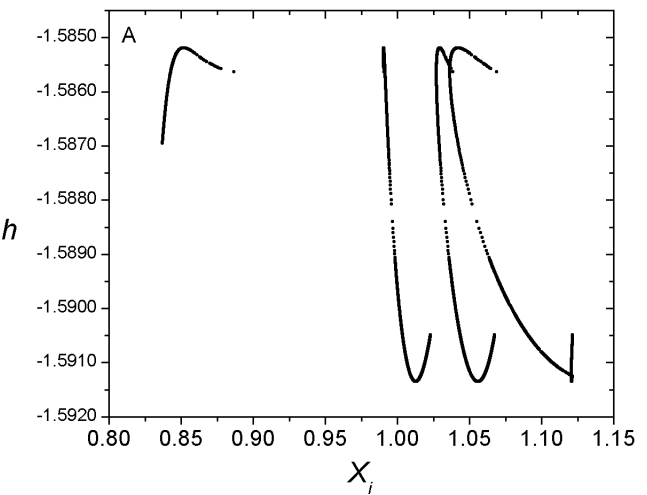

# Family 136 B - Symmetric family of asymmetric POs

$h_{min} = -1.589596$, $h_{max} = -1.588134$, $T_{min} = 21.115366$, $T_{max} = 22.624673$

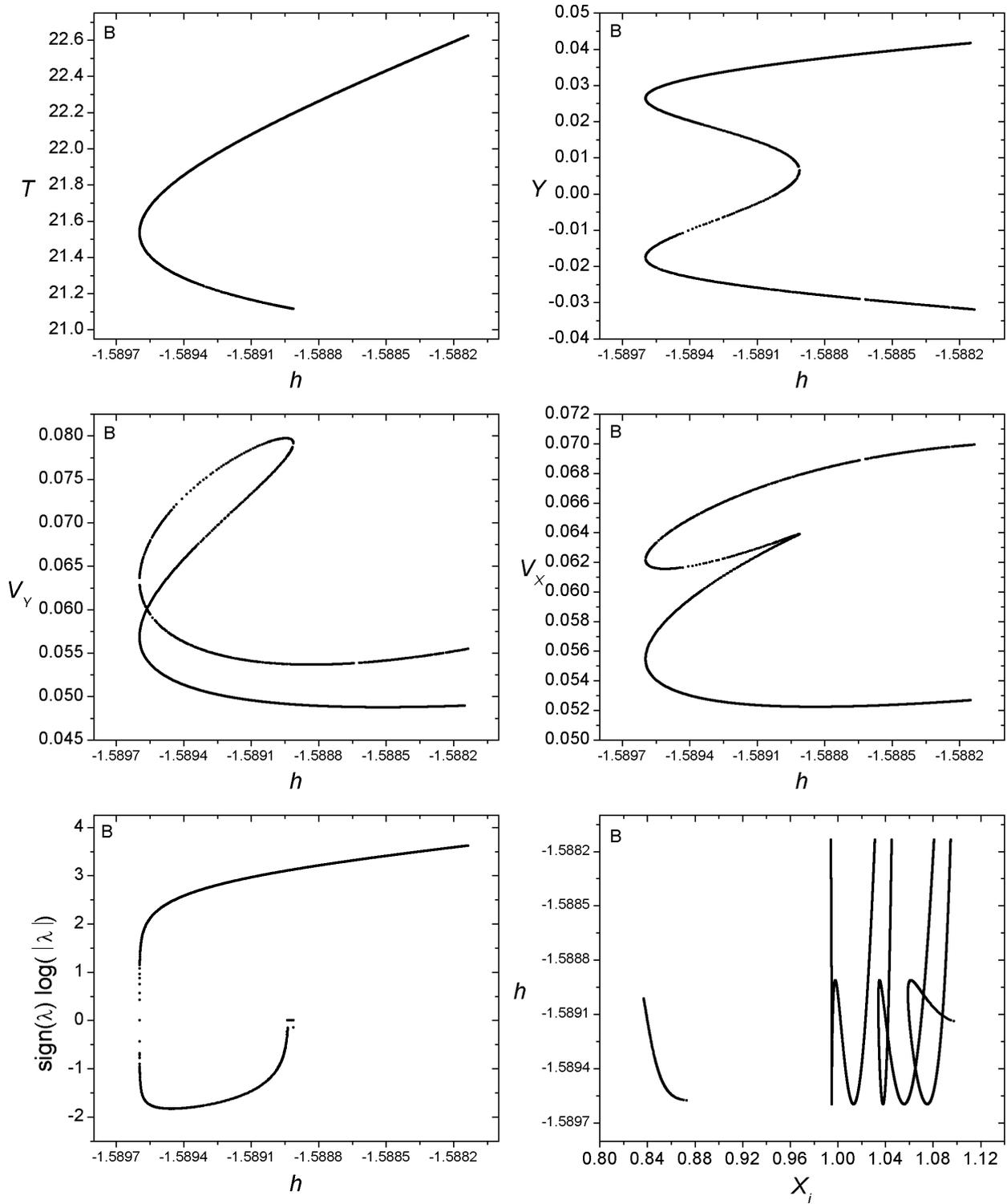

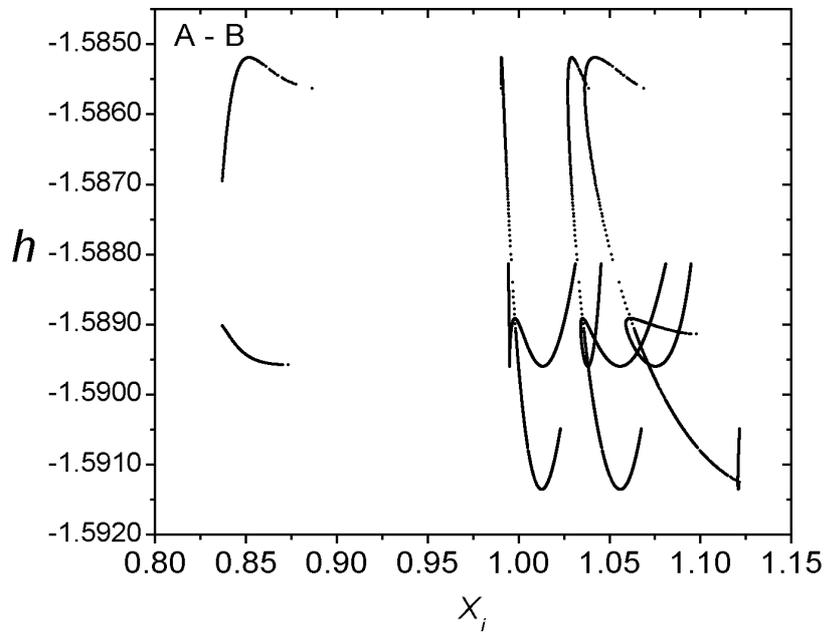

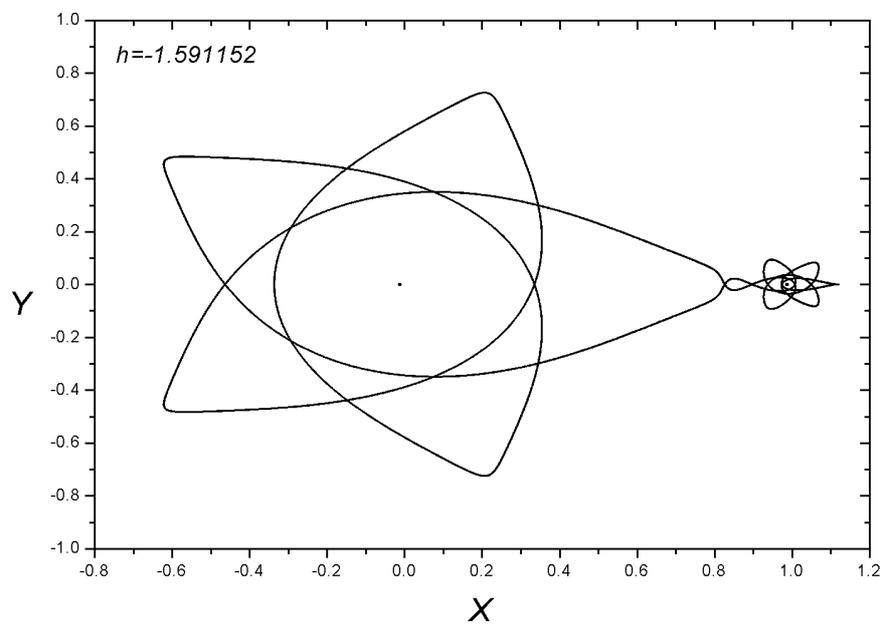

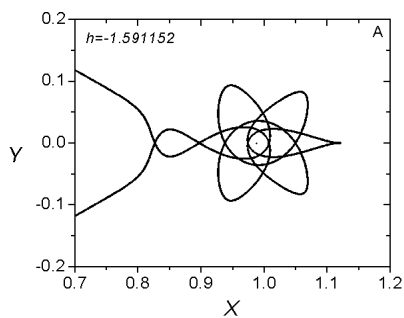
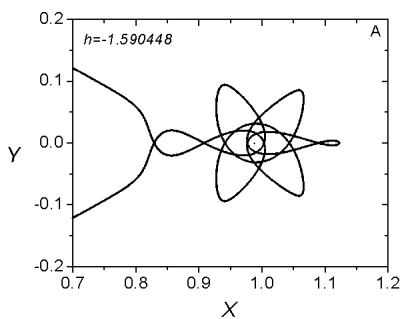
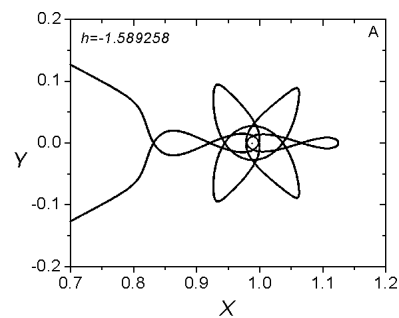

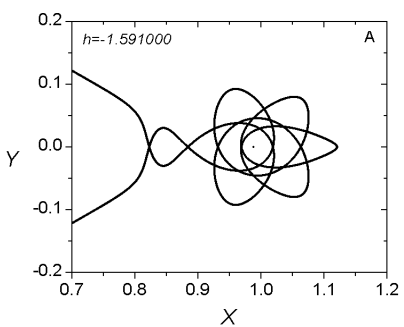
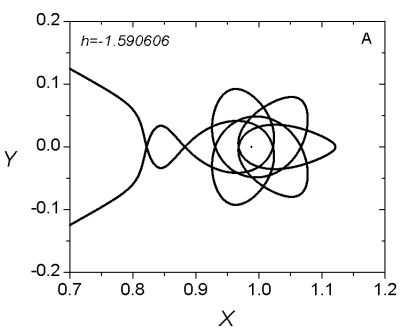
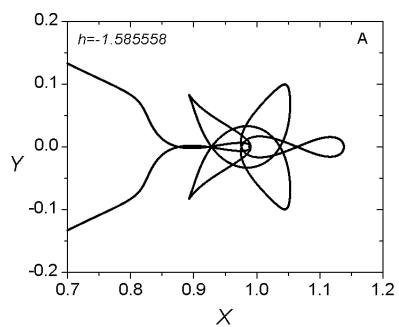

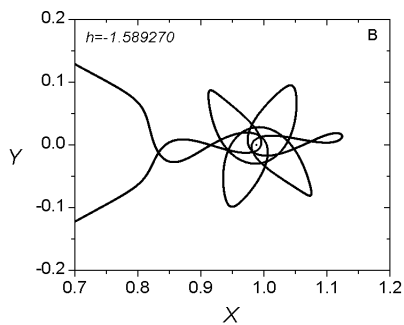
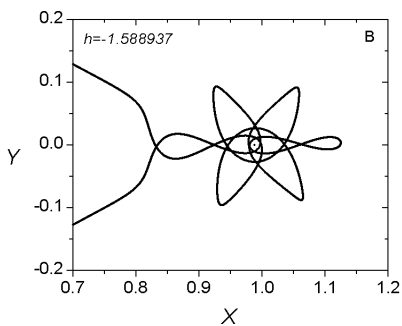
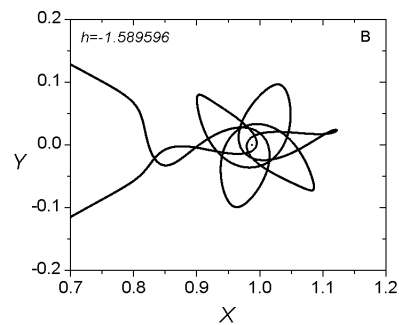

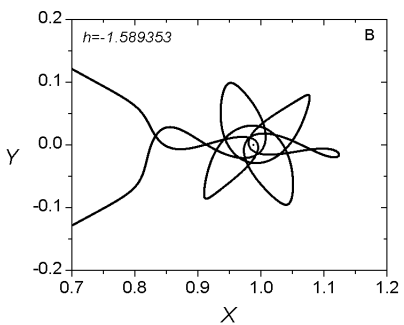
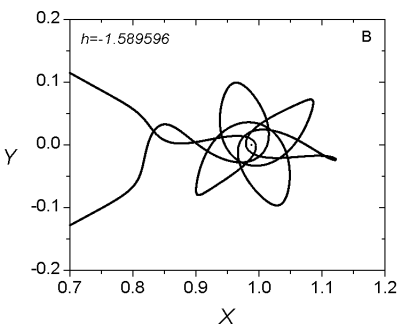
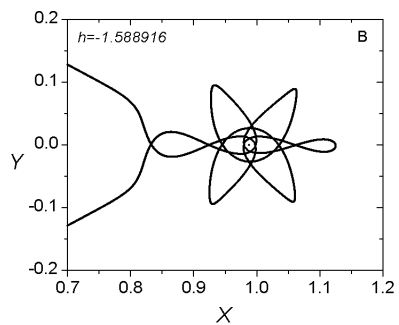

## Families 144 A - 144 B

*Bifurcation Point*

|        | h         | T         | y        | $v_y$     | $v_x$    |
|--------|-----------|-----------|----------|-----------|----------|
| $P_1$  | -1. 589922 | 23.664873 | 0.039763 | -0.000146 | 0.047097 |

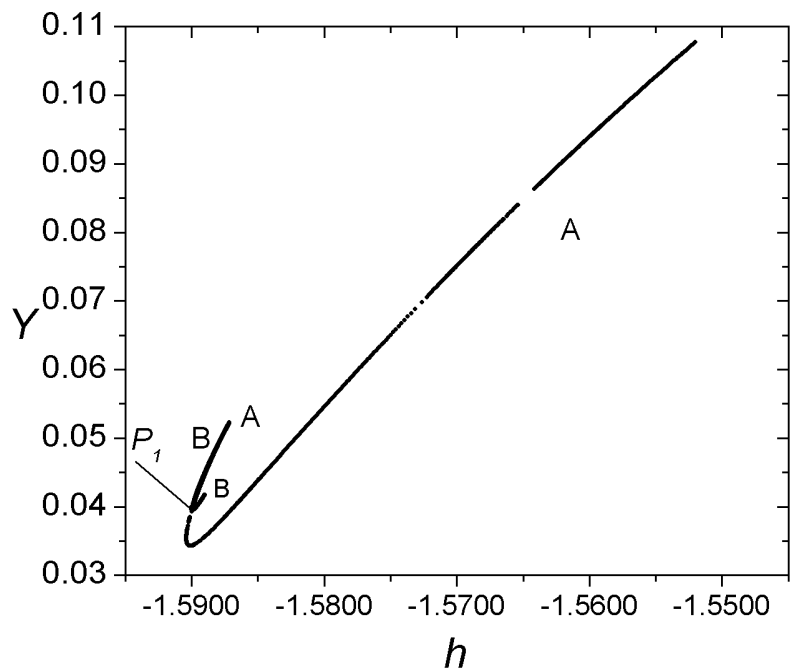

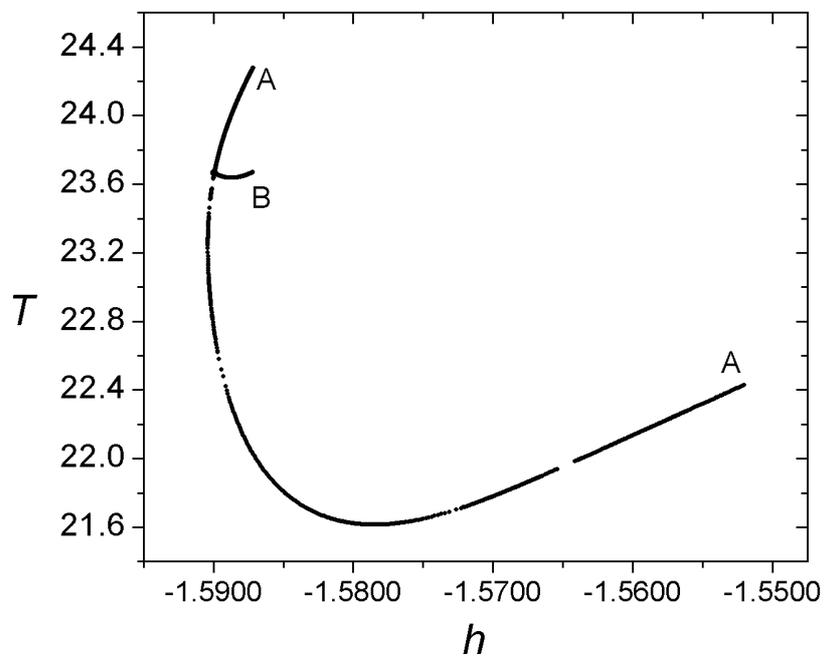

## *Family 144 A*  *- Symmetric family of symmetric POs*

$h_{min} = -1.590391$, $h_{max} = -1.552026$, $T_{min} = 21.614173$, $T_{max} = 24.278568$

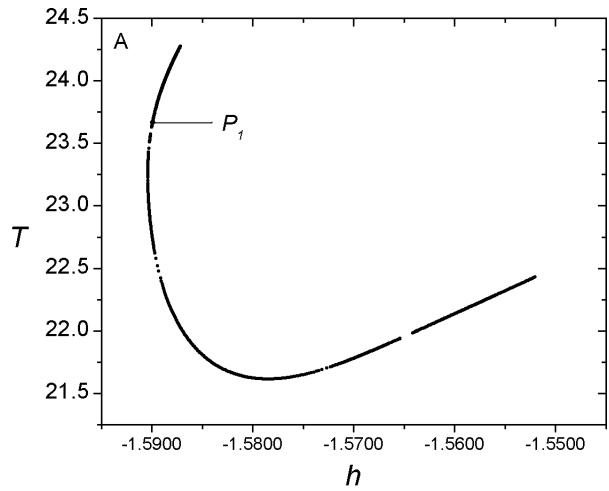

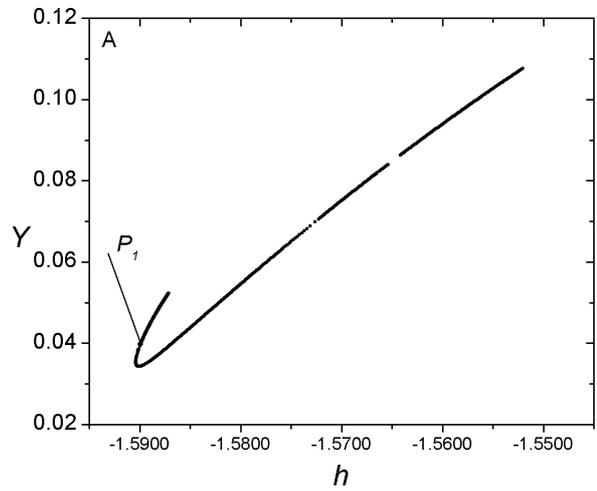

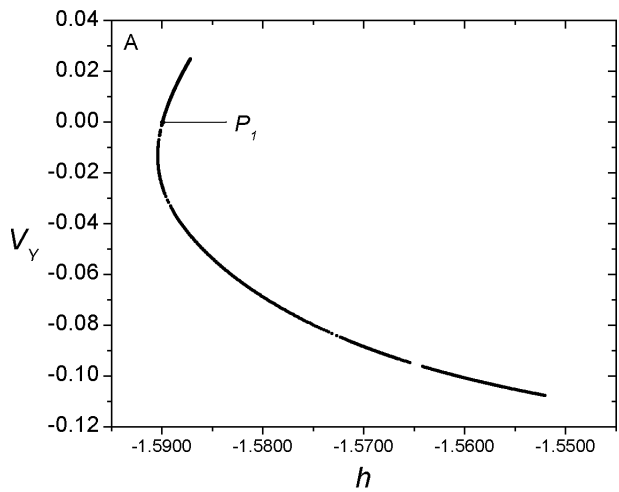

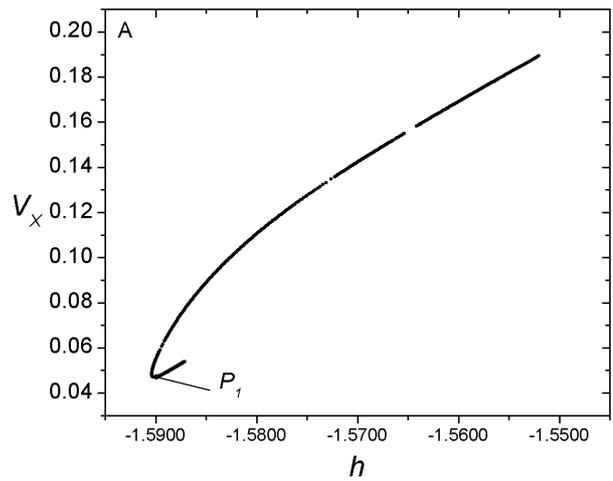

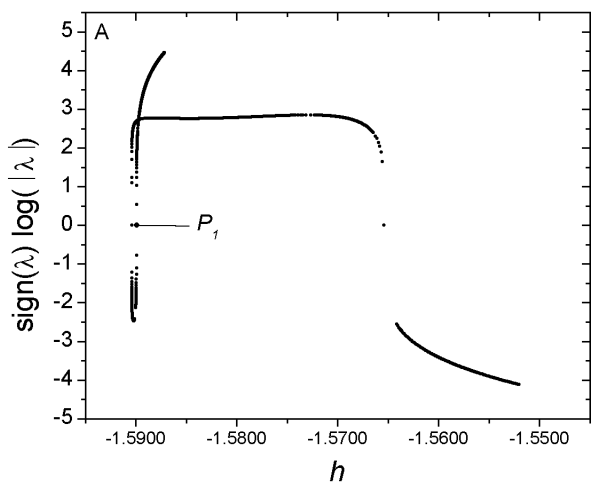

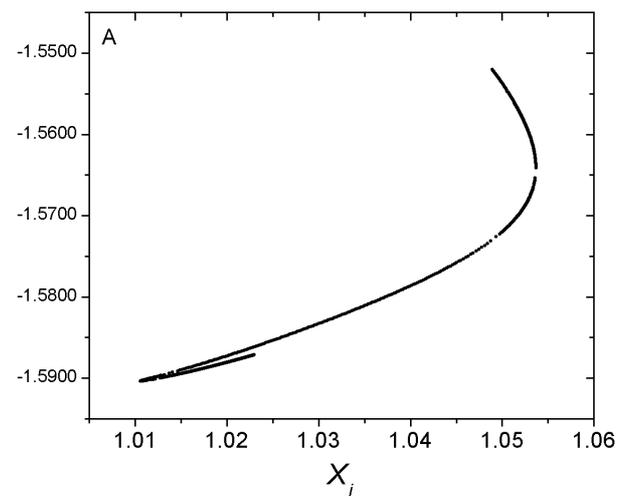

# Family 144 B - Symmetric family of asymmetric POs

$h_{min} = -1.589921$, $h_{max} = -1.587170$, $T_{min} = 23.636883$, $T_{max} = 23.670543$

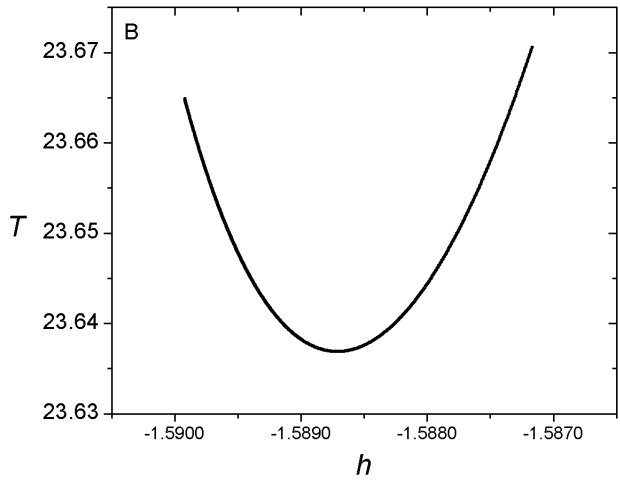
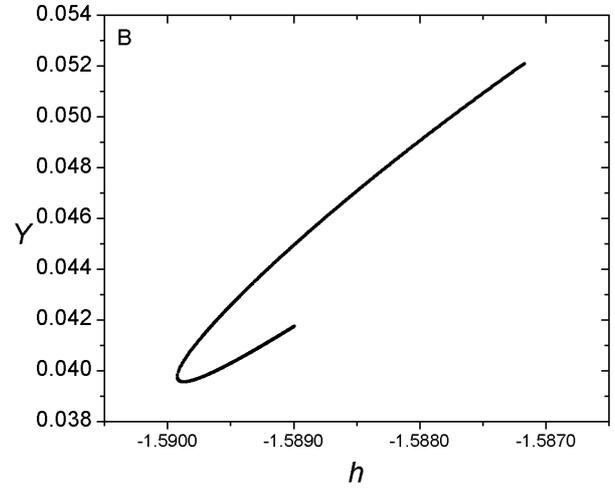

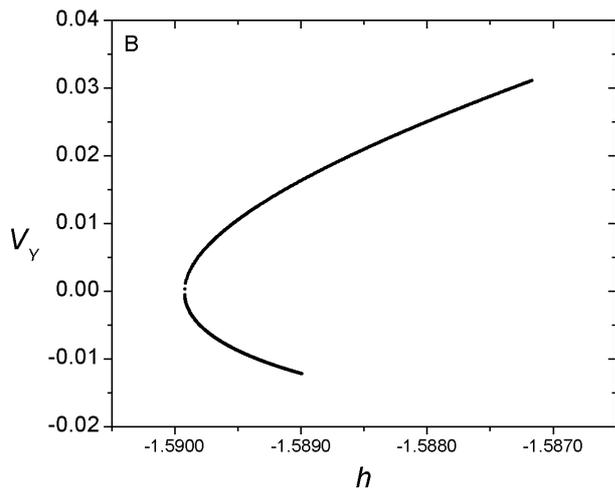
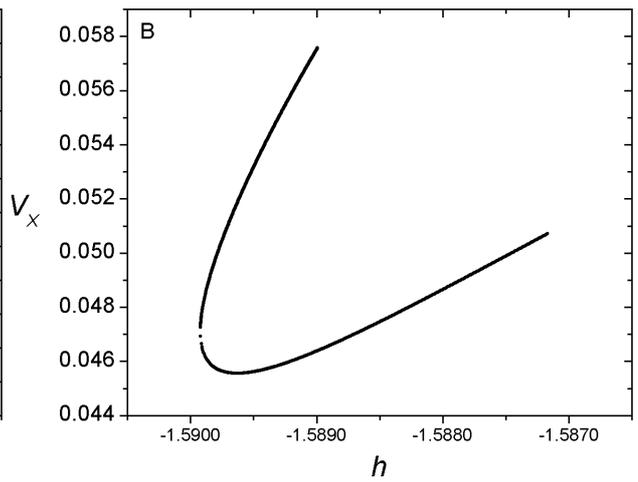

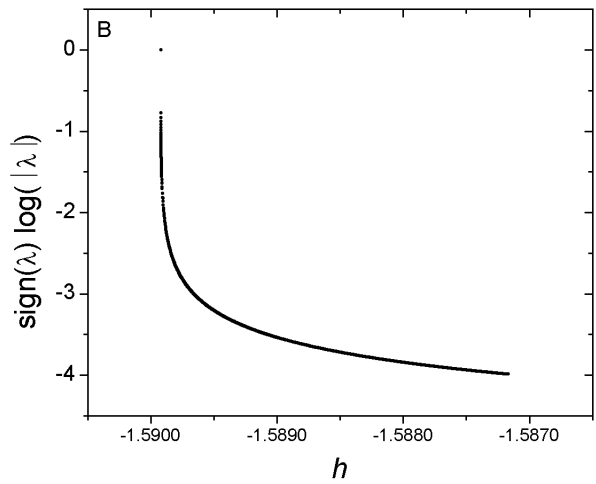
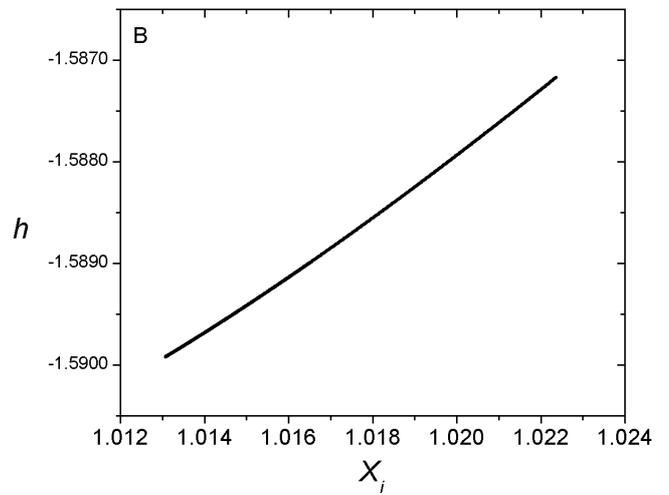

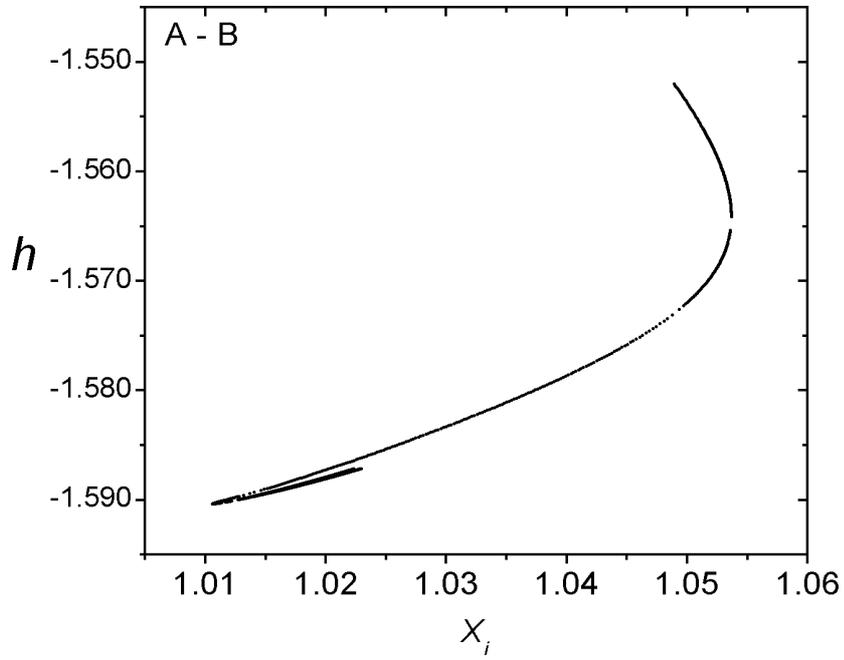

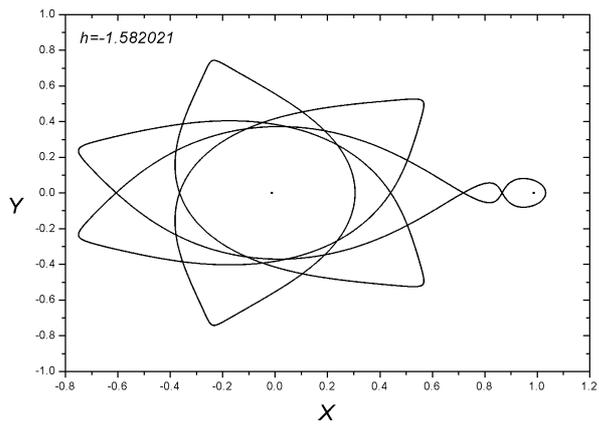

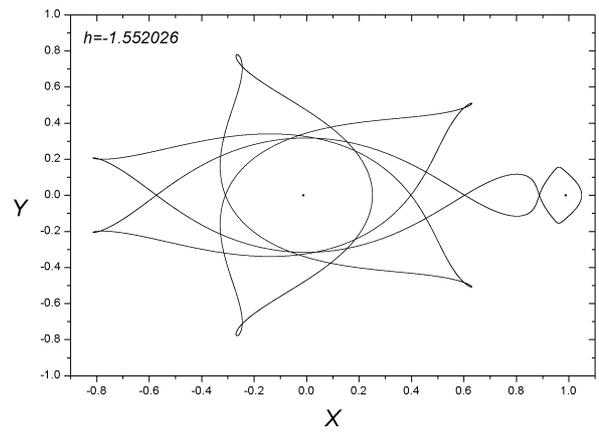

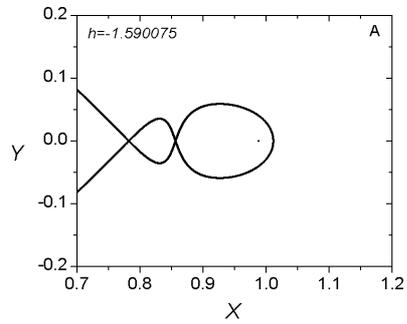
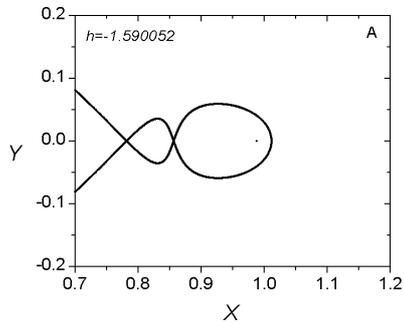
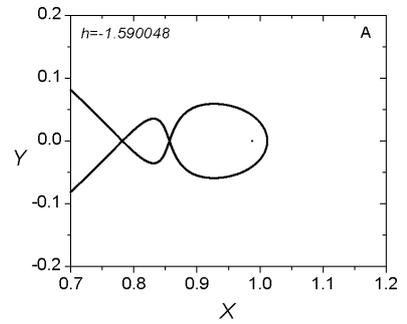

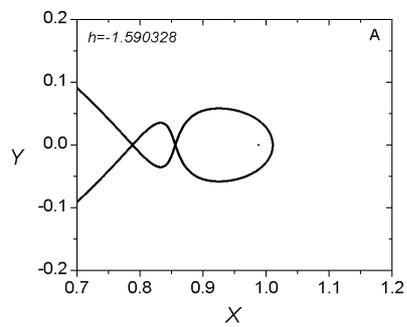
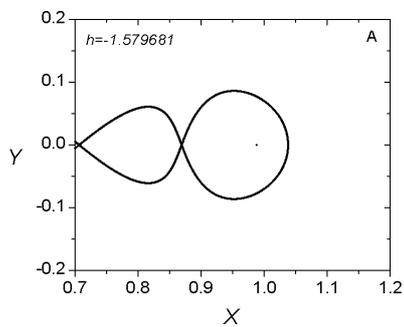
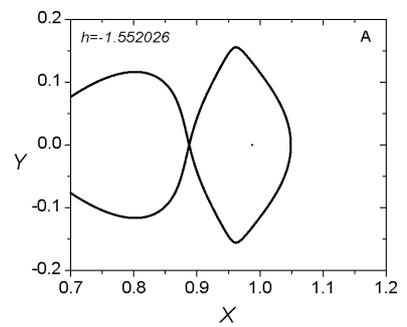

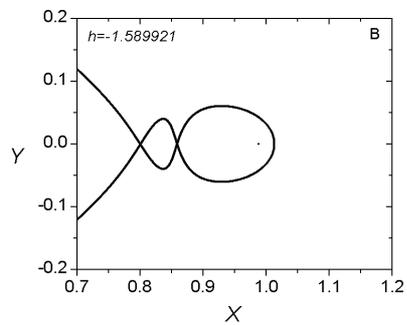
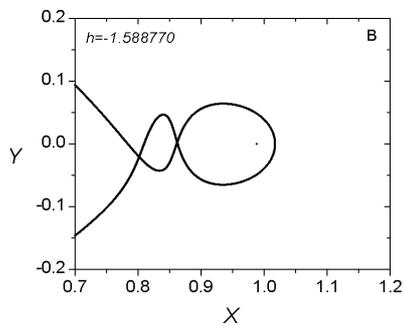
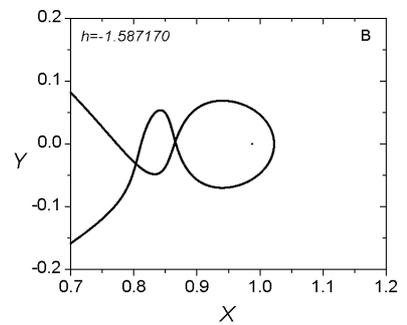

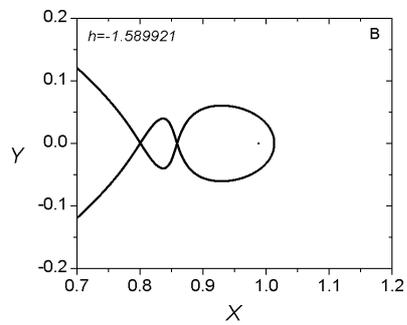
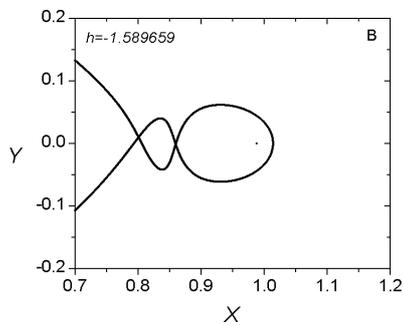
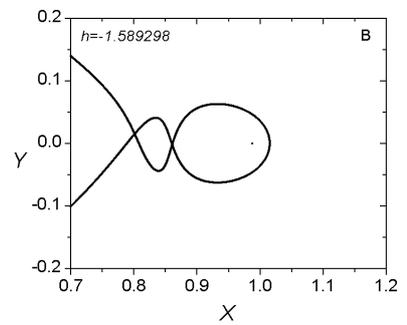

## Families 187 A - 187 B

*Bifurcation Point*

|        | $h$        | $T$        | $y$       | $v_y$     | $v_x$     |
|--------|------------|------------|-----------|-----------|-----------|
| $P_1$  | −1.593358  | 22.736176  | 0.000813  | 0.020375  | 0.034729  |

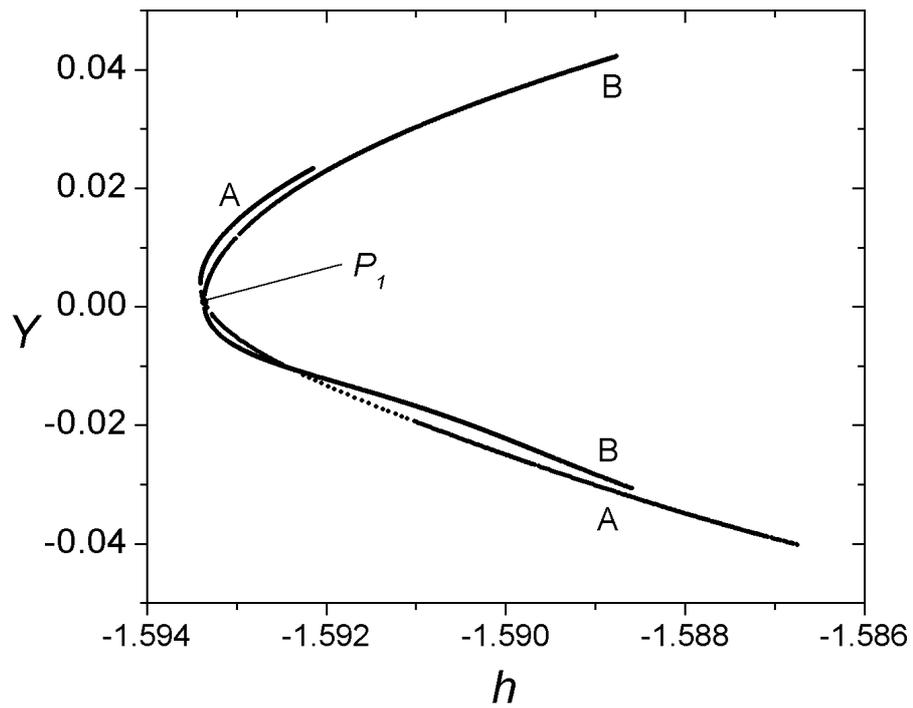

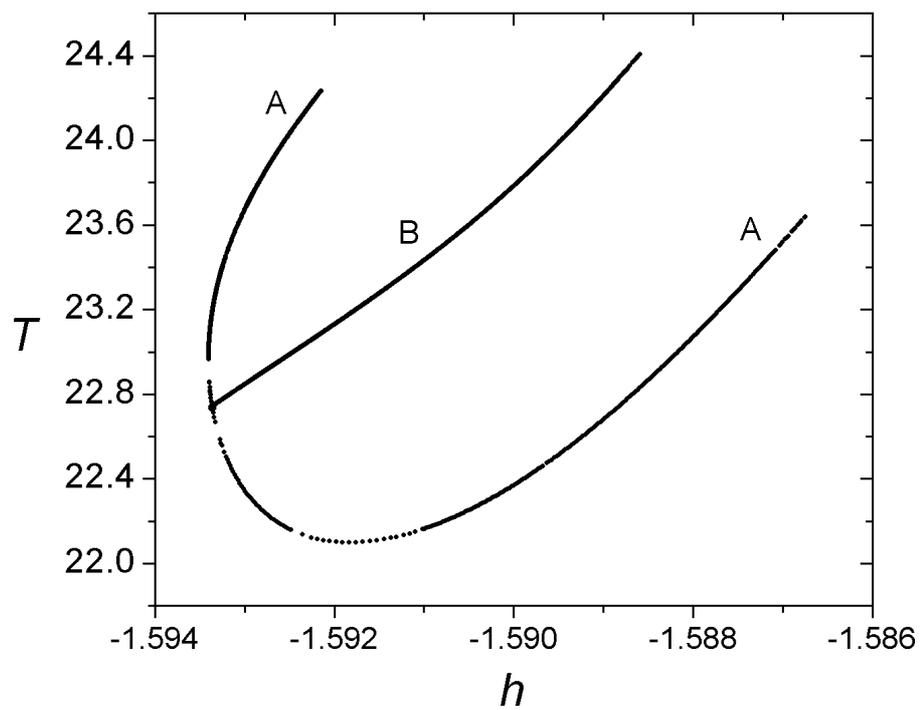

### Family 187 A - *Symmetric family of symmetric POs*

$h_{min} = -1.593401, \quad h_{max} = -1.586745, \quad T_{min} = 22.099407, \quad T_{max} = 24.234116$

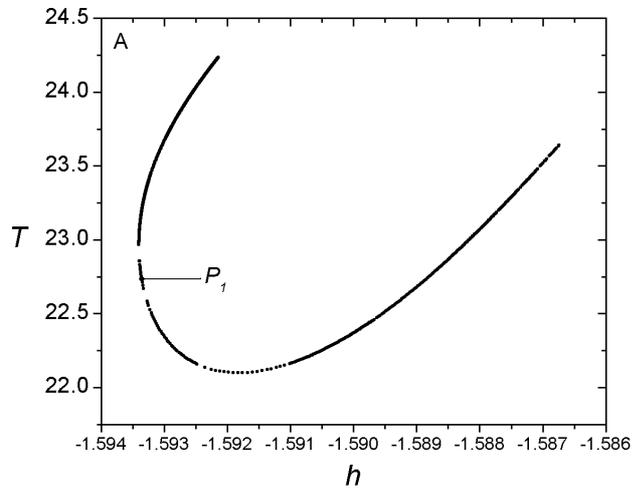

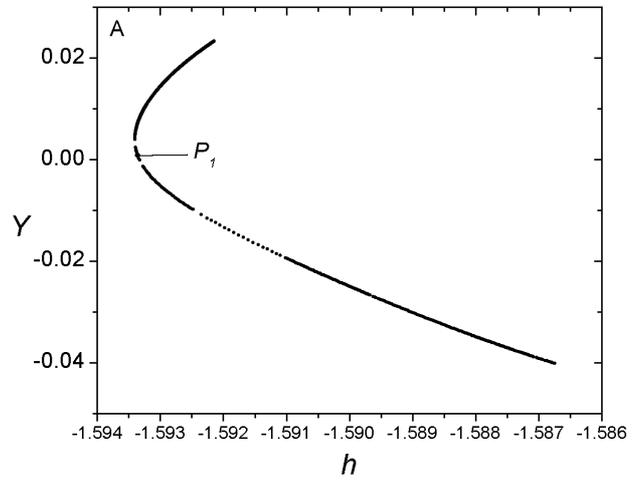

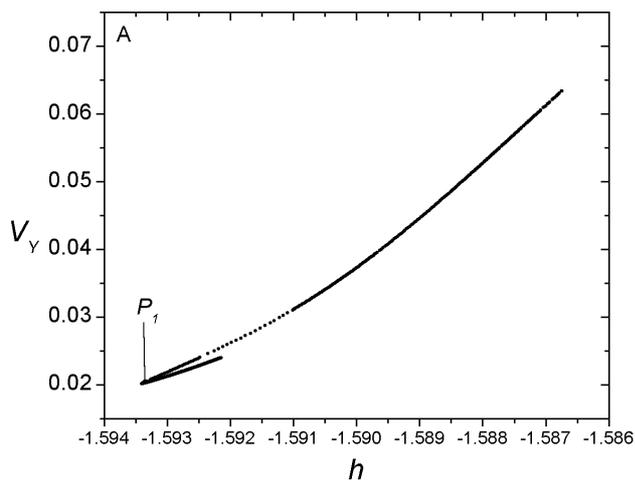

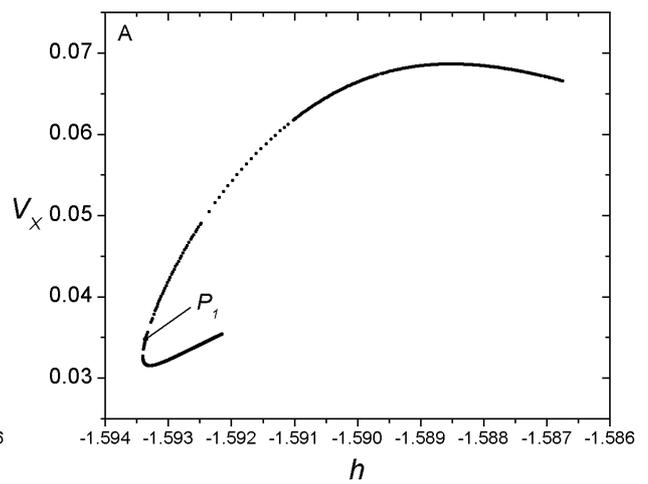

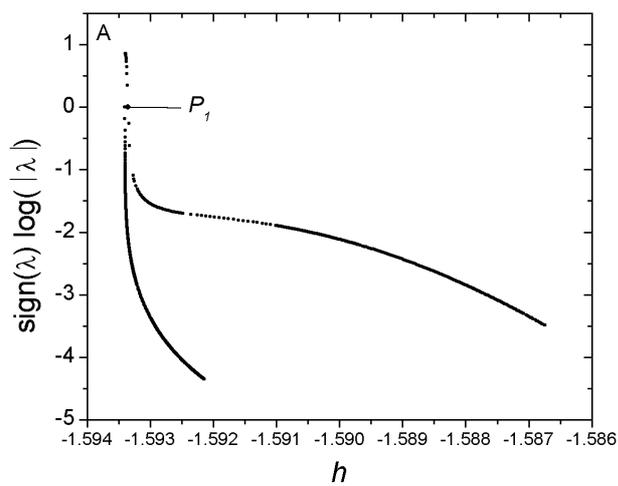

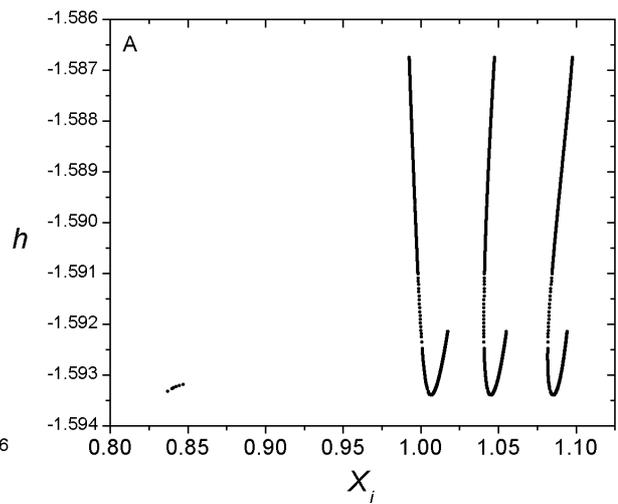

## *Family 187 B - Symmetric family of asymmetric POs*

$h_{min} = -1.593360$,   $h_{max} = -1.588594$,   $T_{min} = 22.743109$,   $T_{max} = 24.405745$

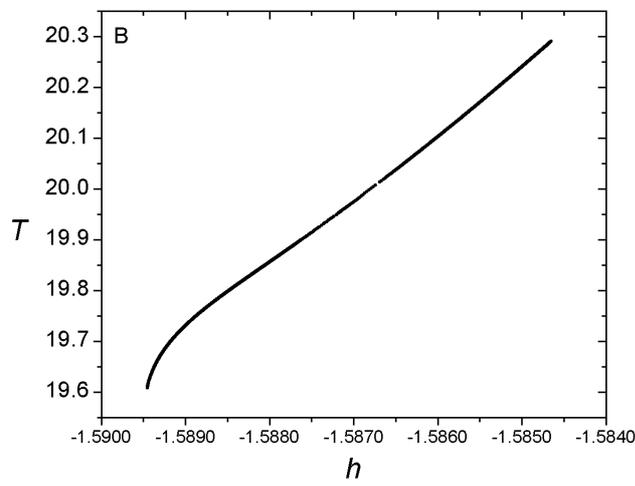

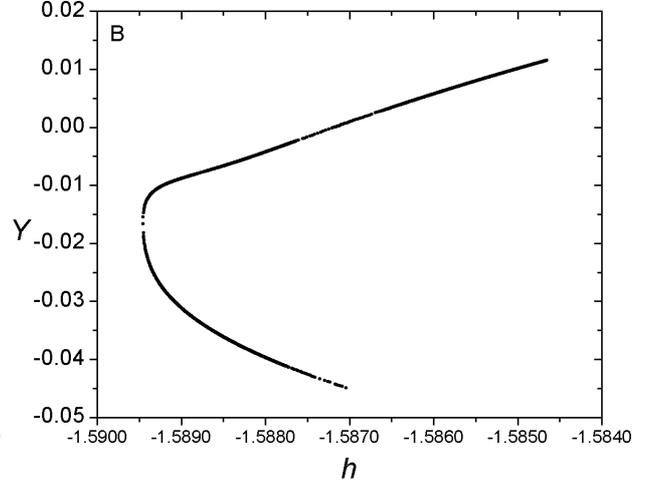

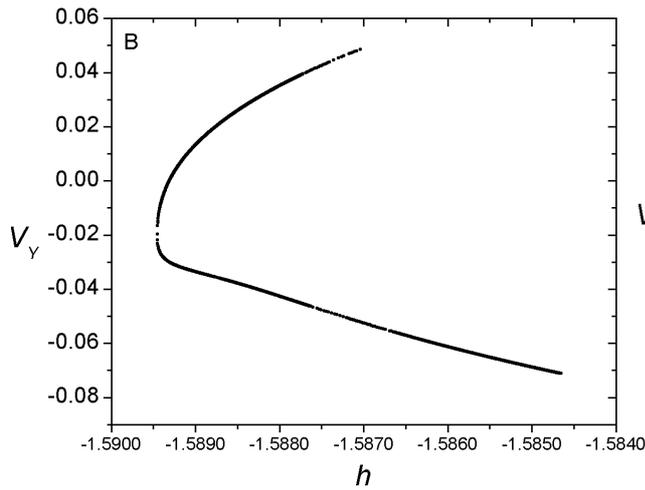

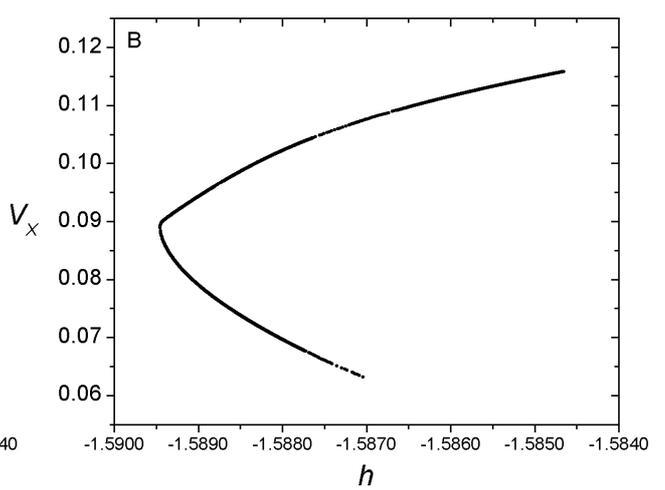

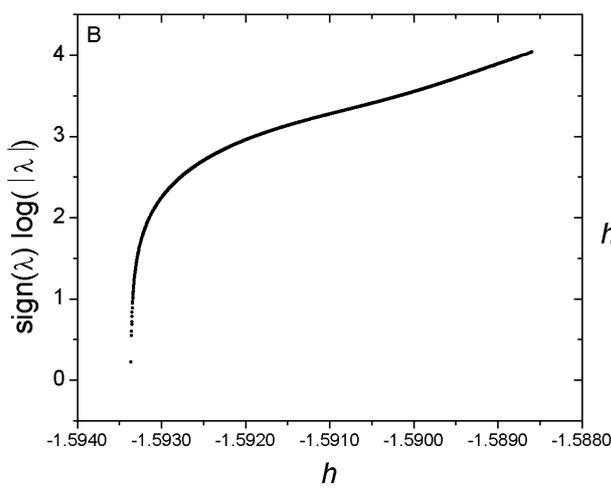

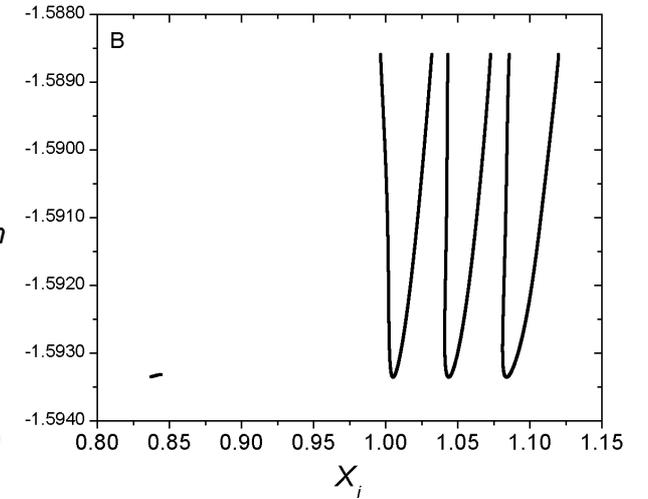

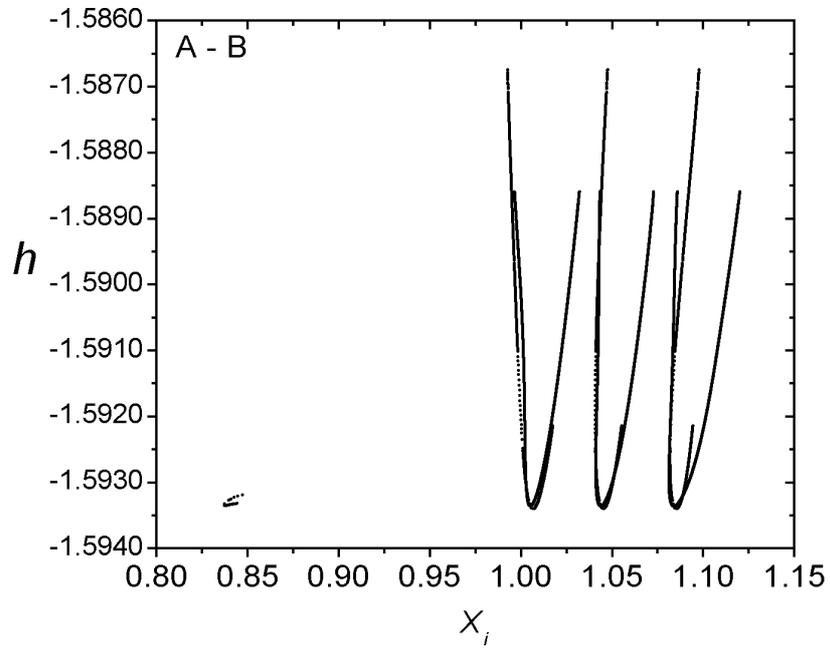

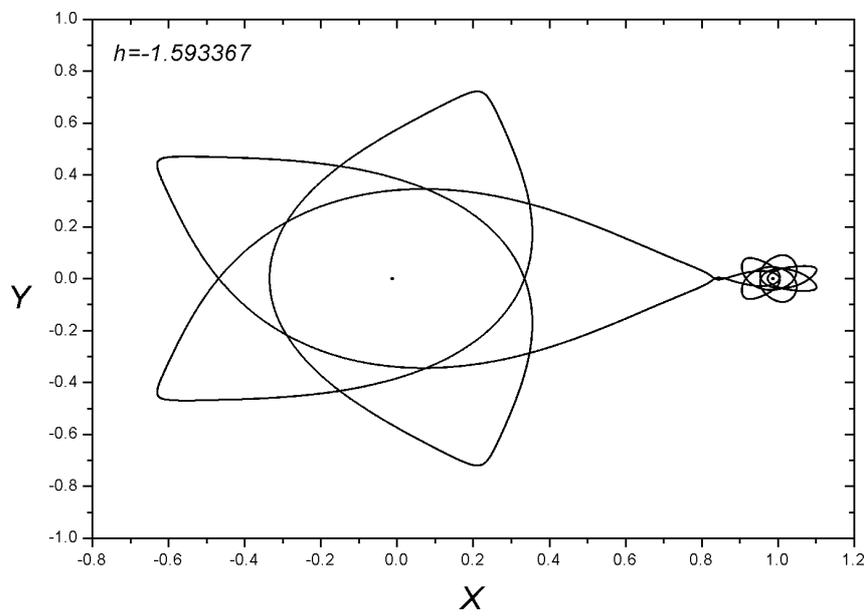

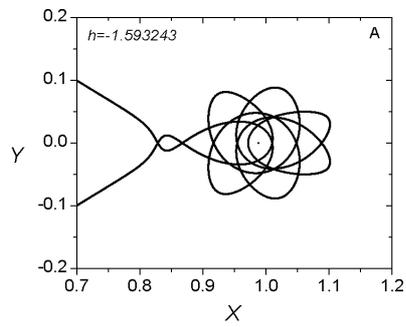
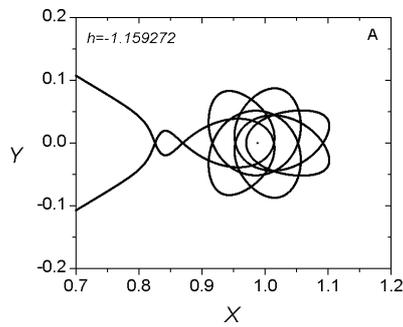
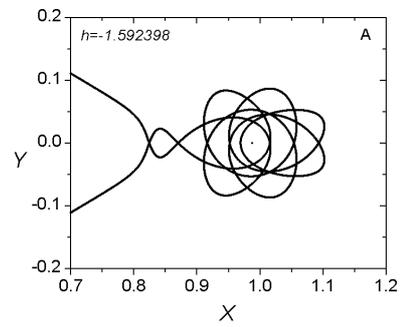

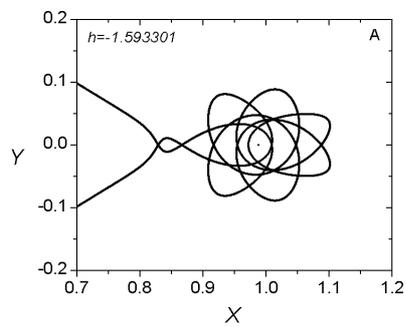
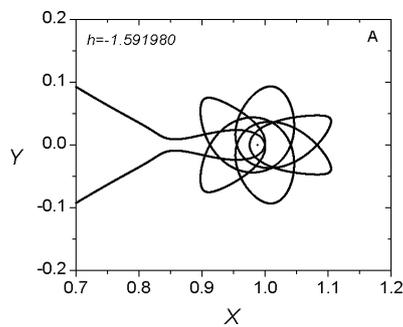
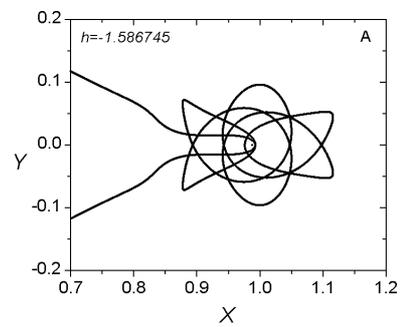

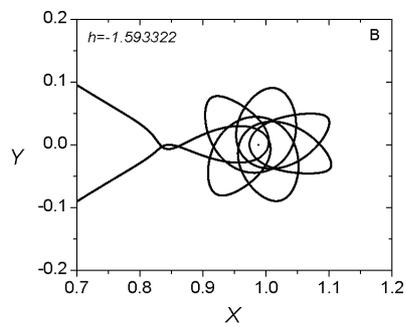
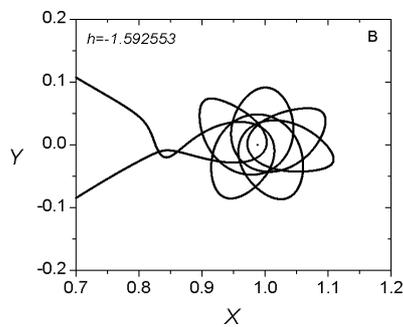
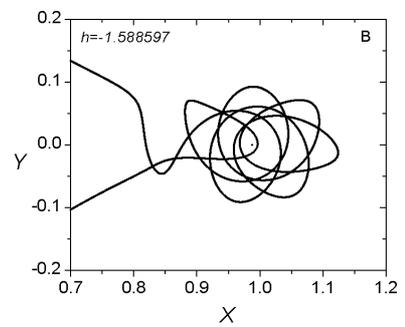

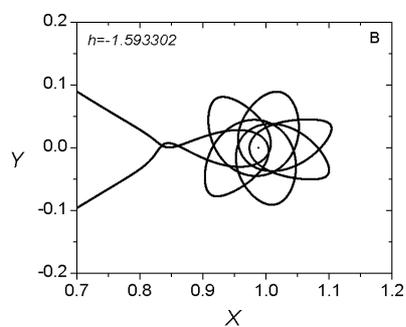
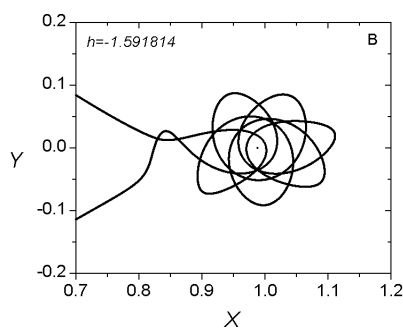
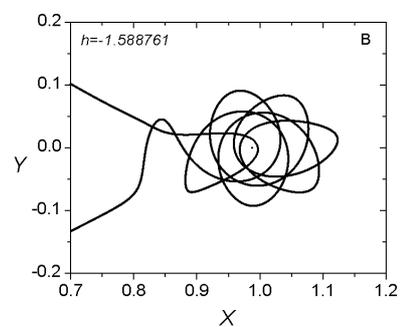

## *Family 147 - Symmetric family of symmetric POs*

$h_{min} = -1.592136, \quad h_{max} = -1.587726, \quad T_{min} = 22.350225, \quad T_{max} = 23.857624$

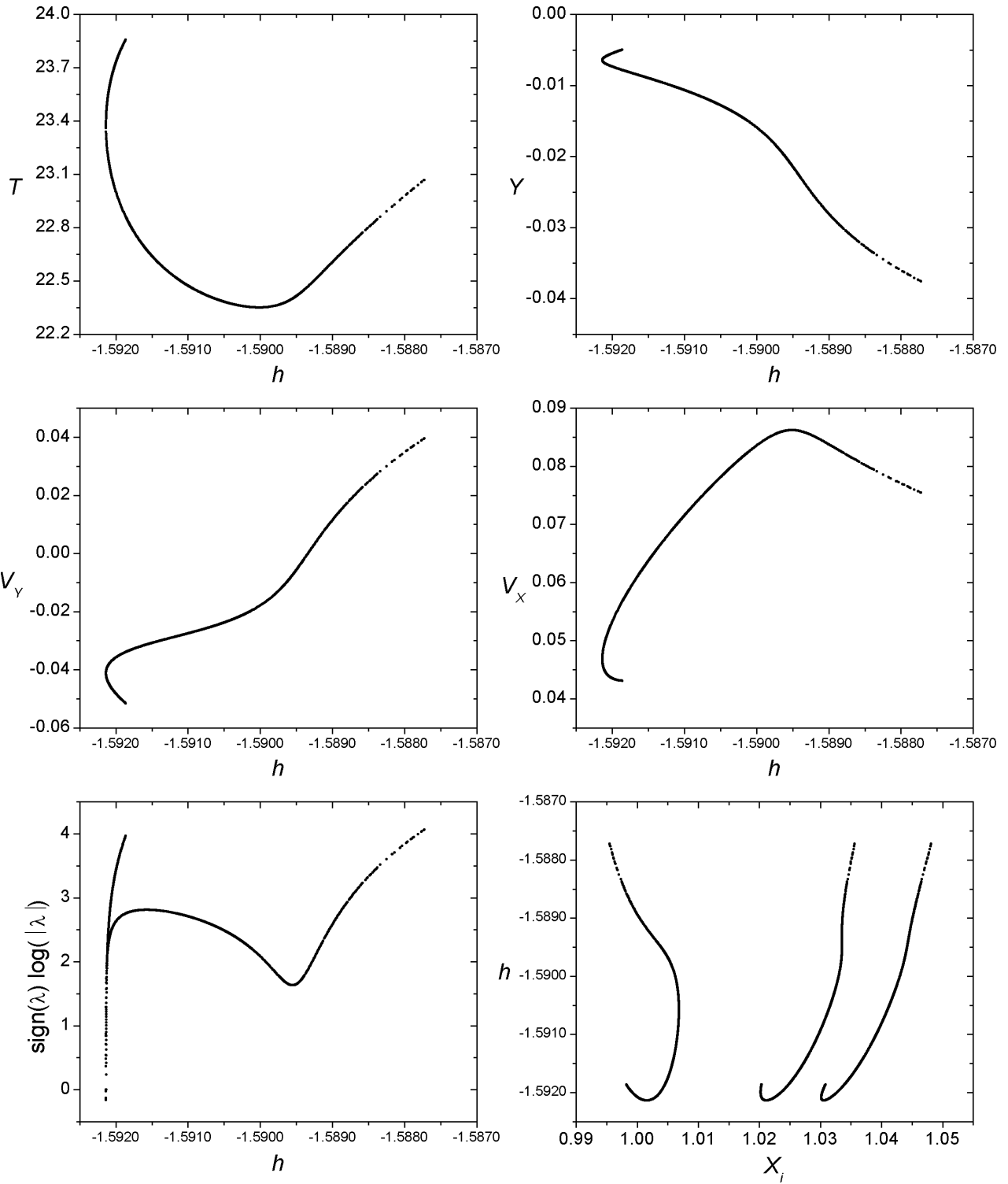

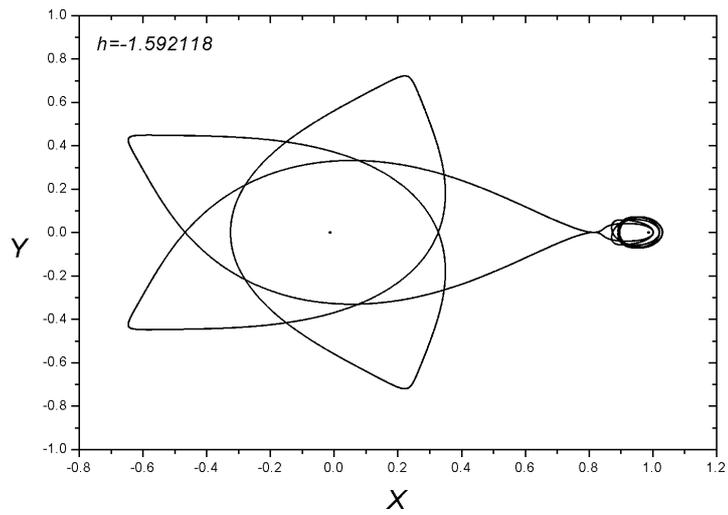

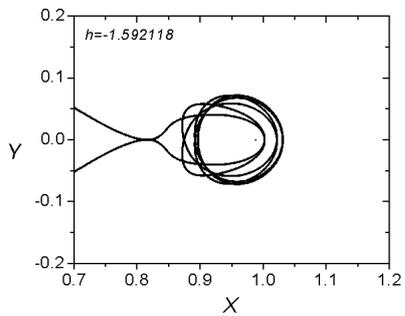
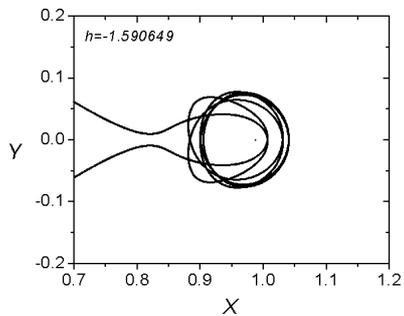
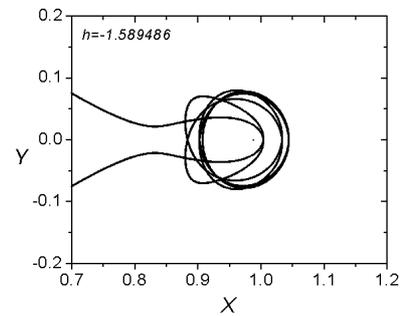

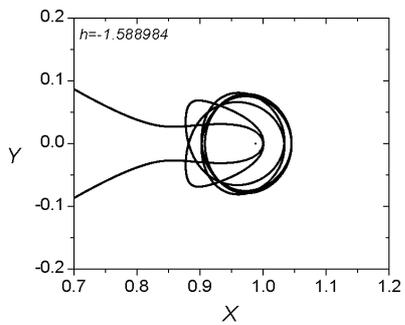
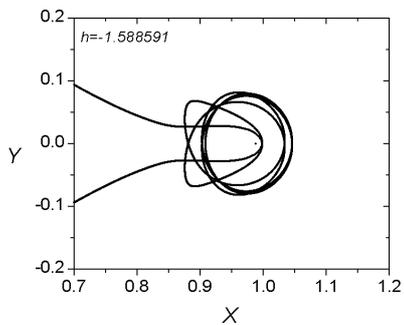
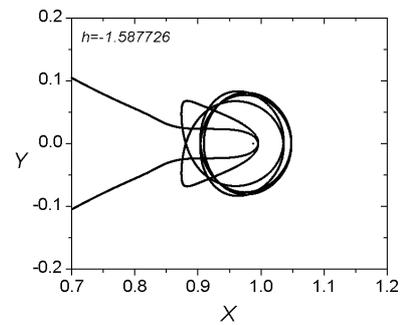

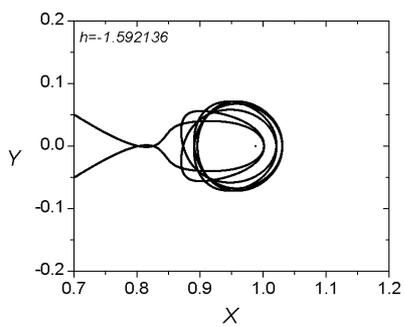
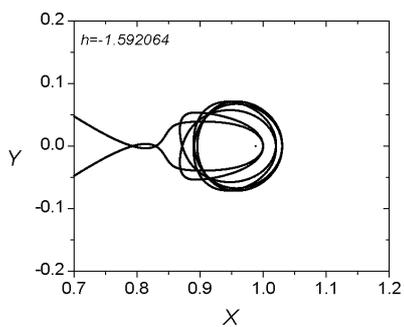
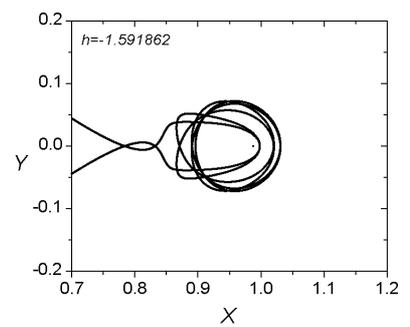

## Family 194 - *Asymmetric family of asymmetric POs*

$h_{min} = -1.592976$, $h_{max} = -1.589971$, $T_{min} = 22.812098$, $T_{max} = 24.091209$

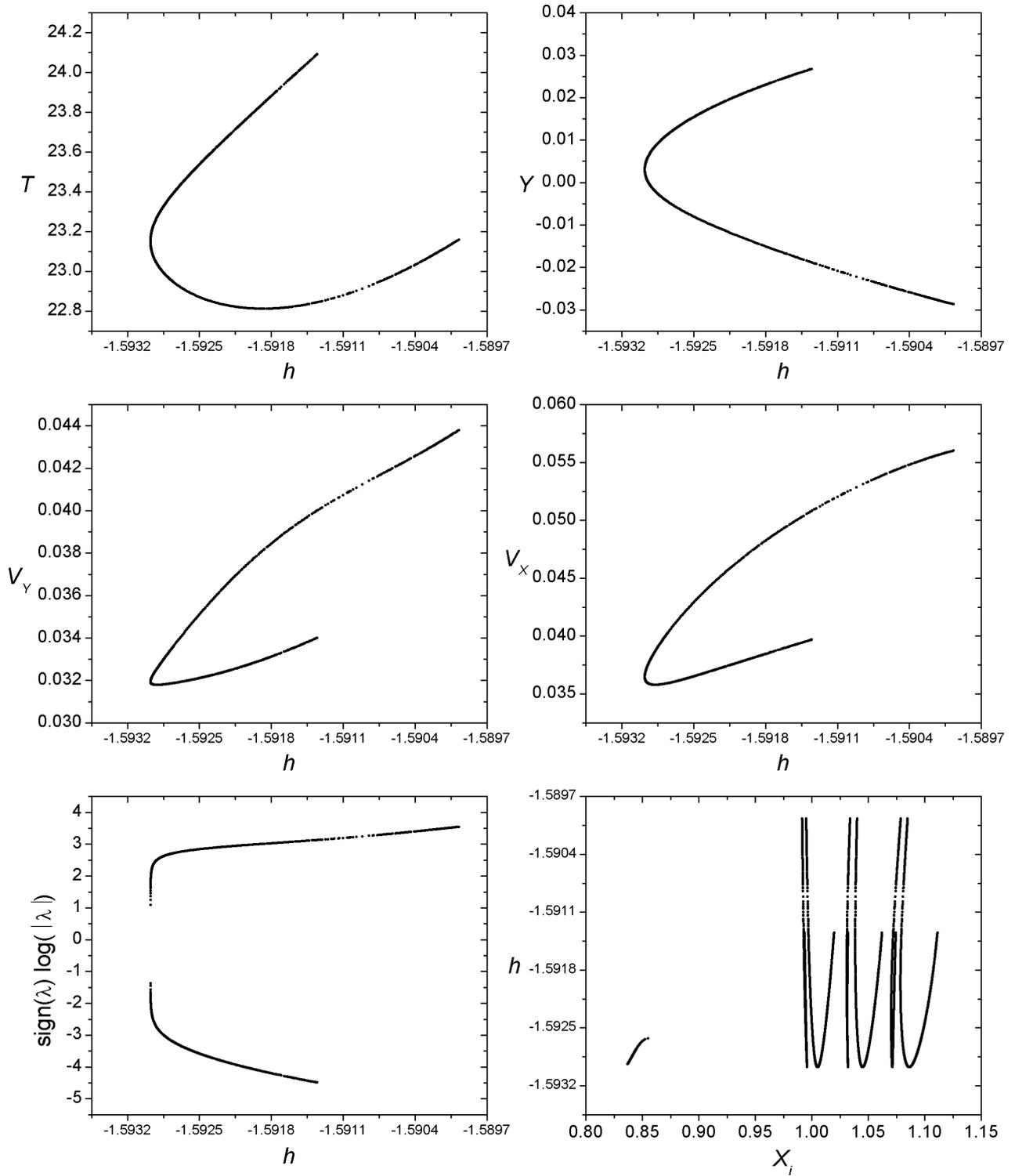

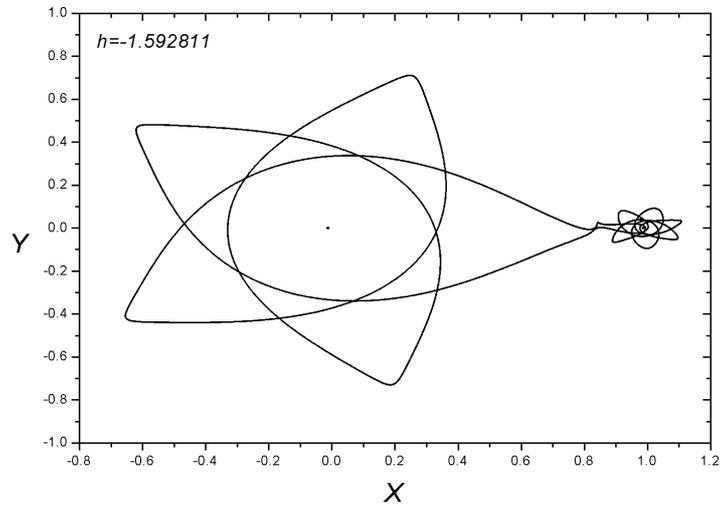

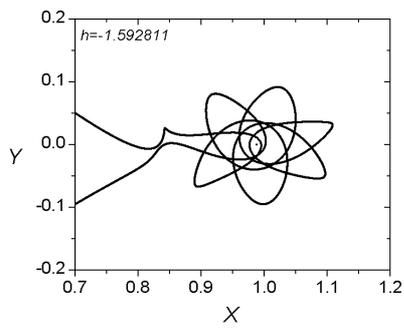
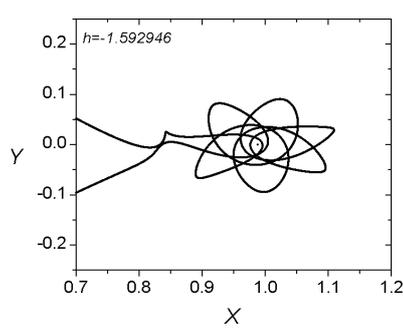
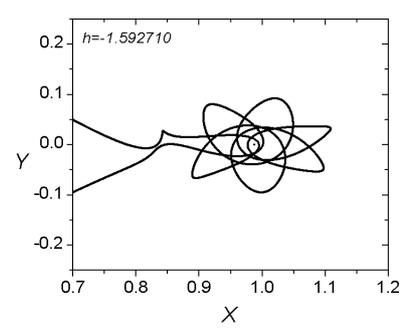

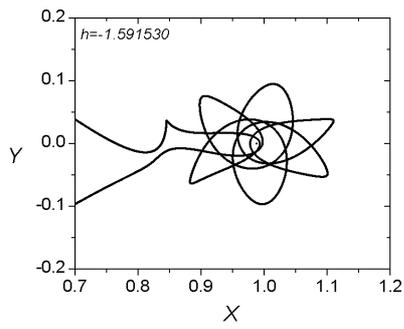
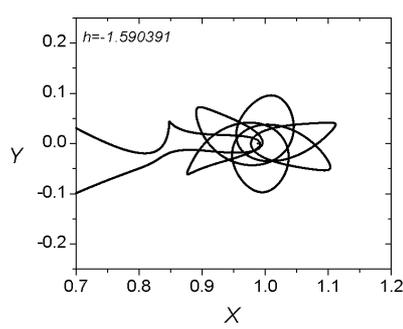
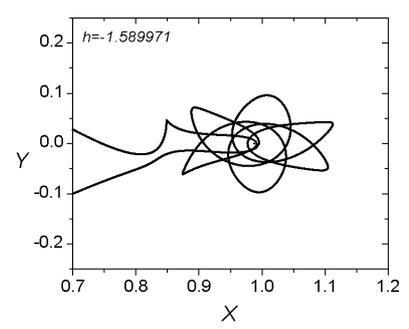

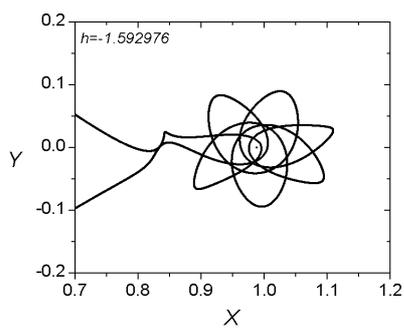
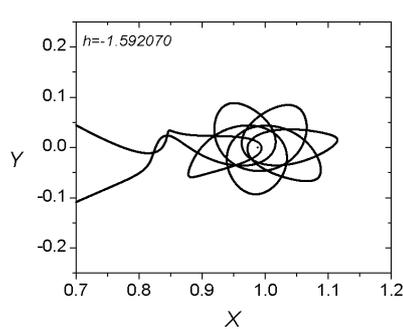
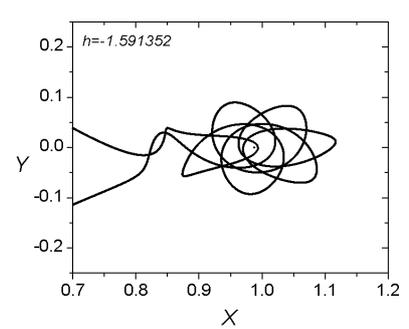

## Family 244  - *Asymmetric family of asymmetric POs*

$h_{min} = -1.592975, \ h_{max} = -1.588677, \ T_{min} = 22.812096, \ T_{max} = 24.1358292$

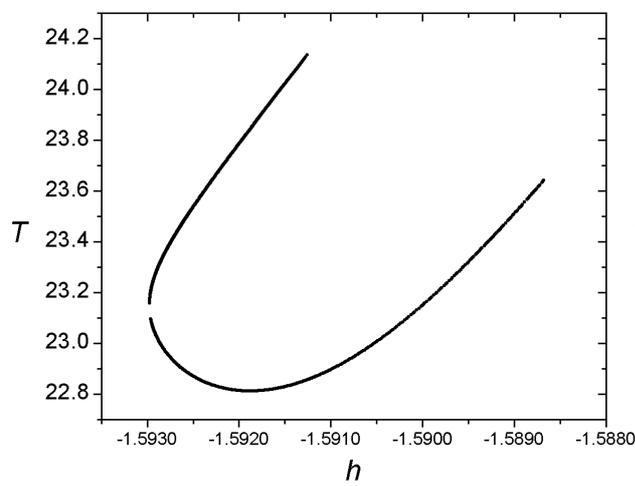

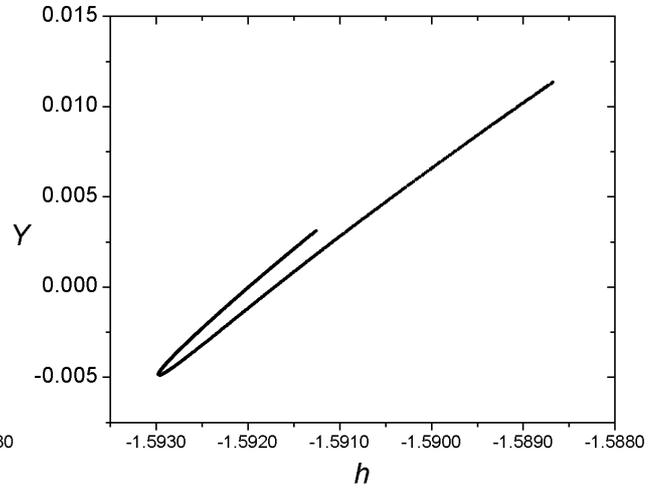

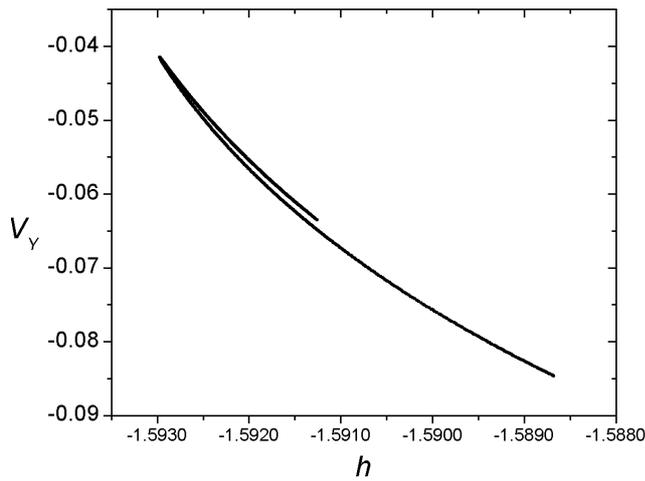

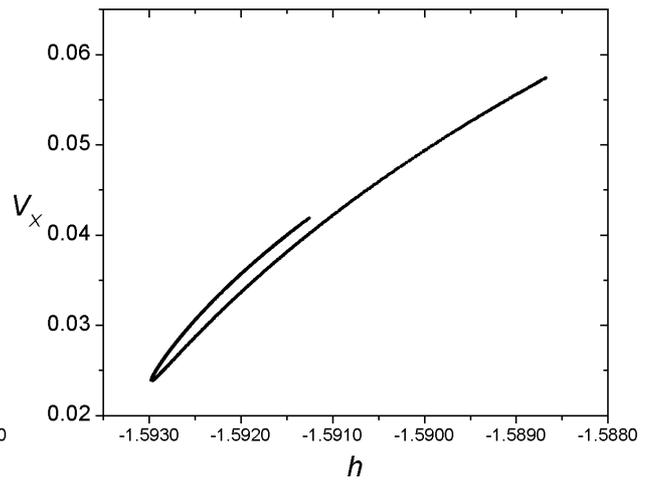

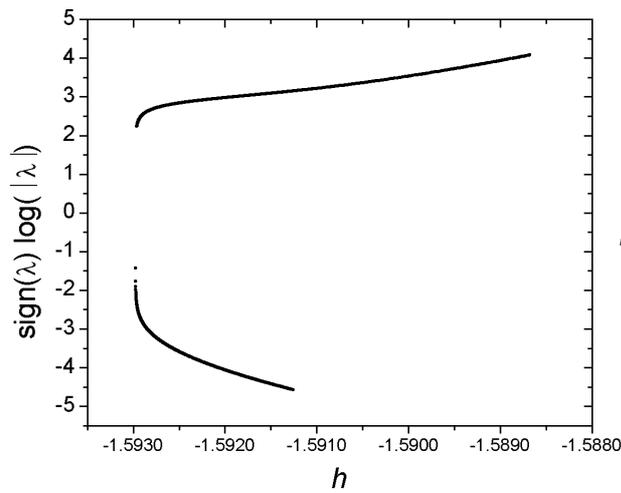

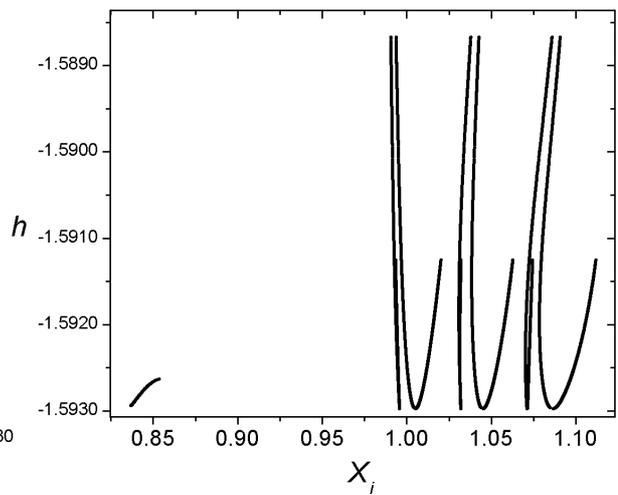

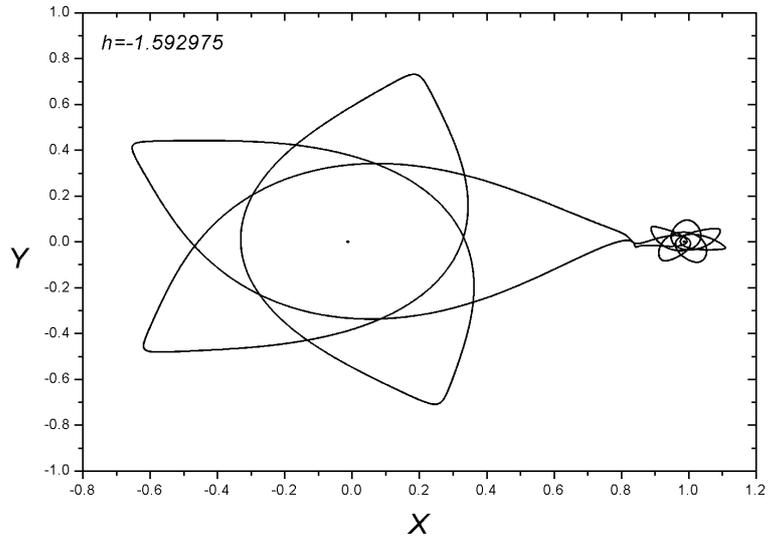

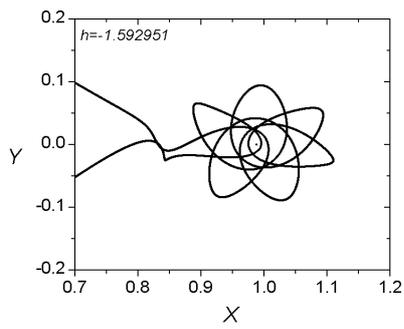
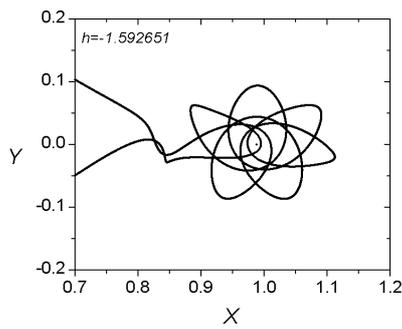
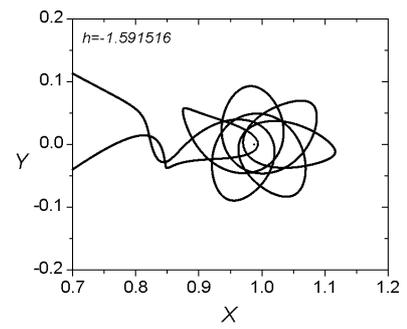

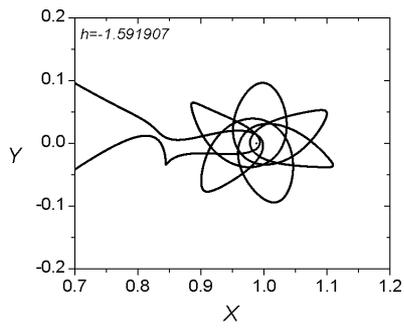
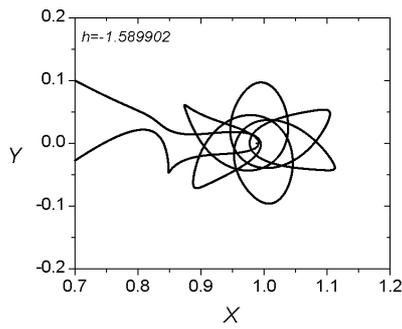
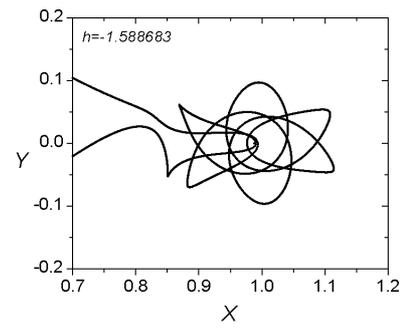

### Family 305 - *Symmetric family of symmetric POs*

$h_{min} = -1.590507$, $h_{max} = -1.587612$, $T_{min} = 23.268005$, $T_{max} = 24.639195$

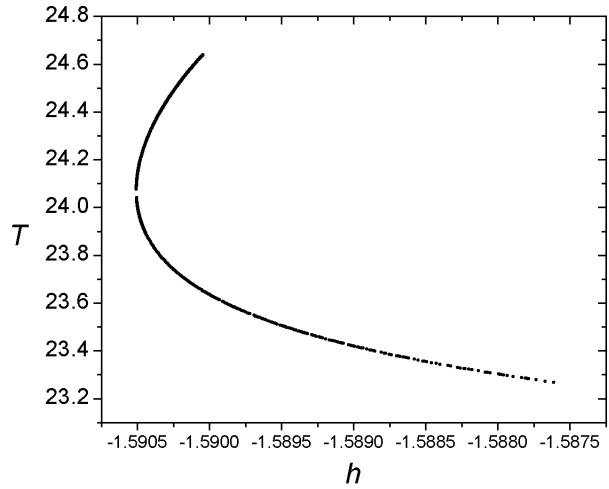
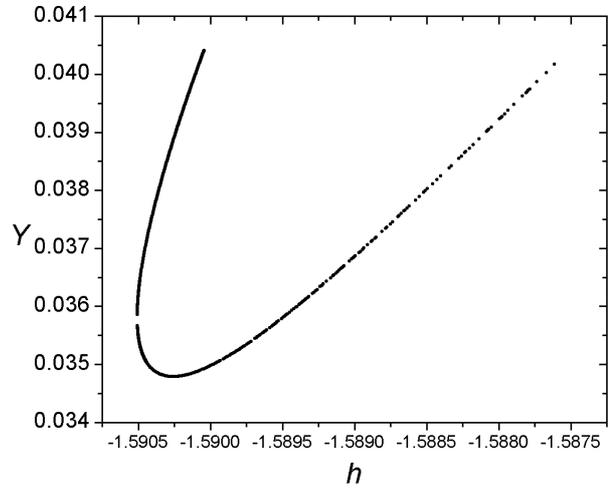

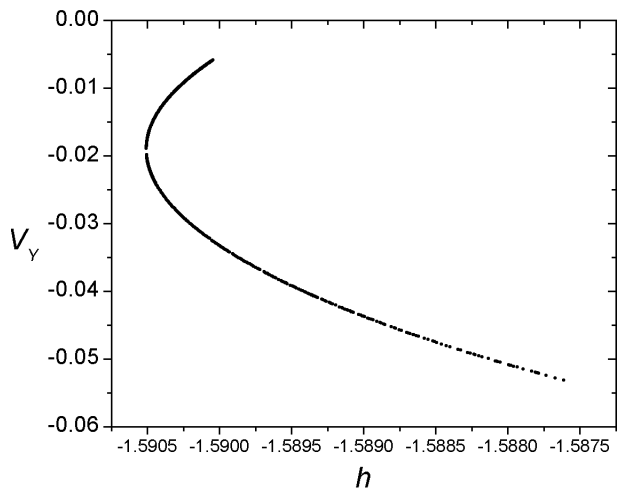
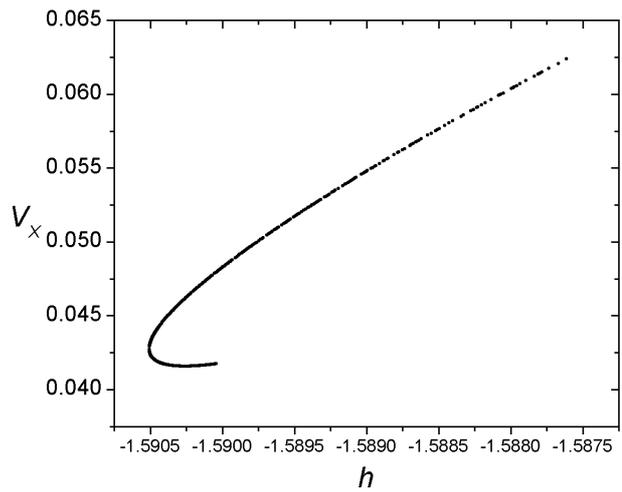

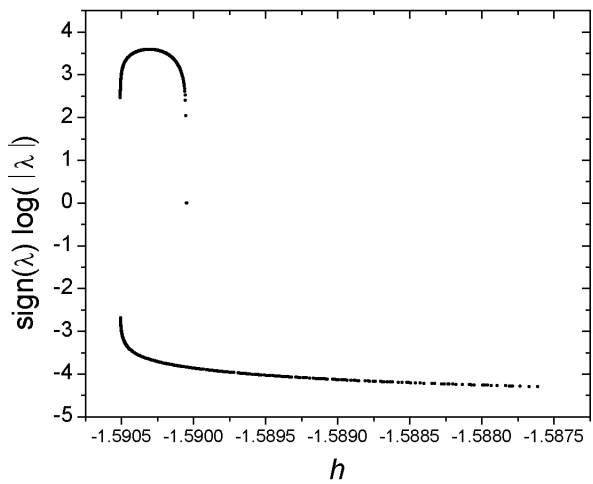
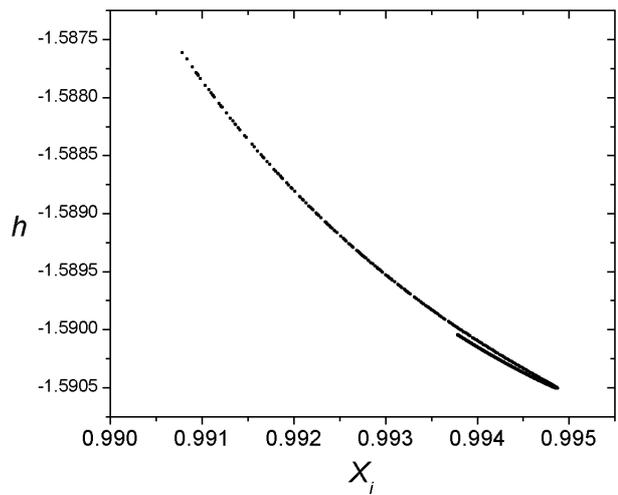

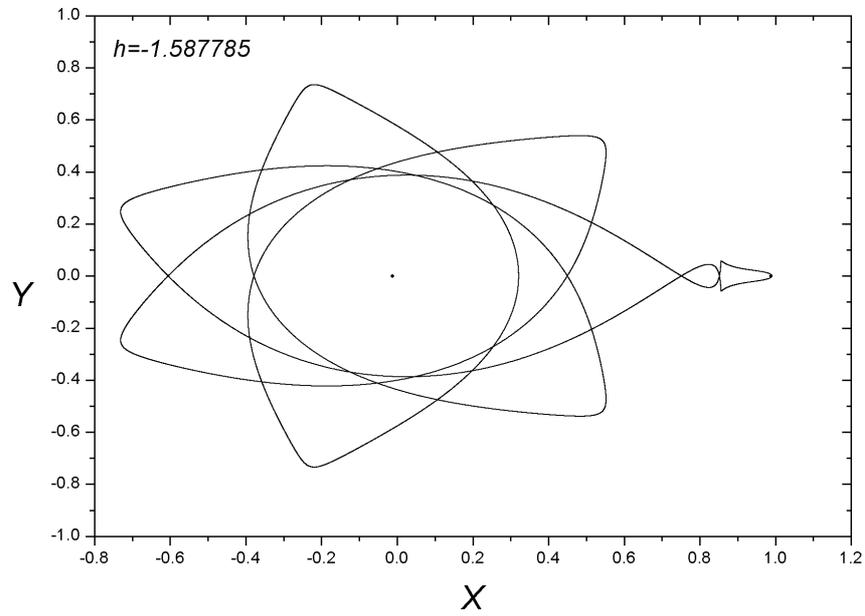

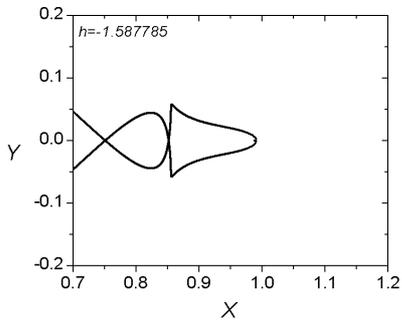
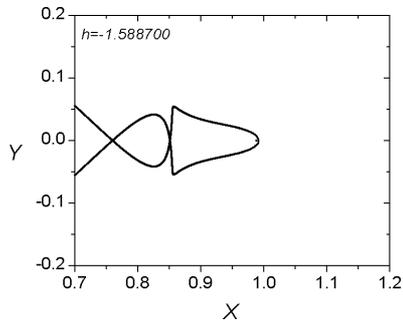
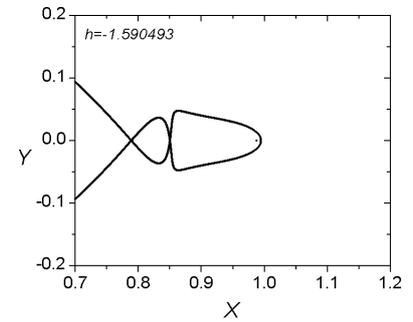

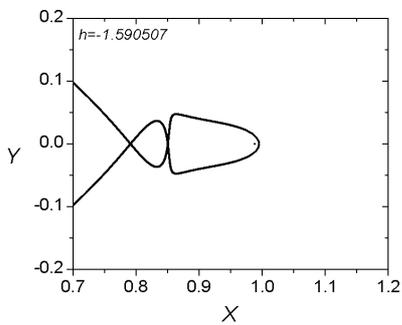
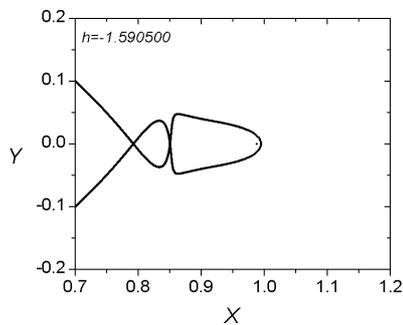
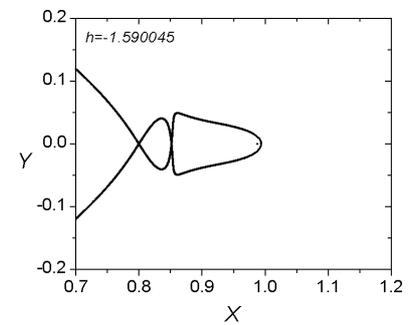

## Families 018 A - 018 B

*Bifurcation Point*

|       | h         | T         | y        | v<sub>y</sub> | v<sub>x</sub> |
|-------|-----------|-----------|----------|----------|----------|
| $P_1$ | -1.589846 | 25.025445 | 0.039325 | 0.003202 | 0.049917 |

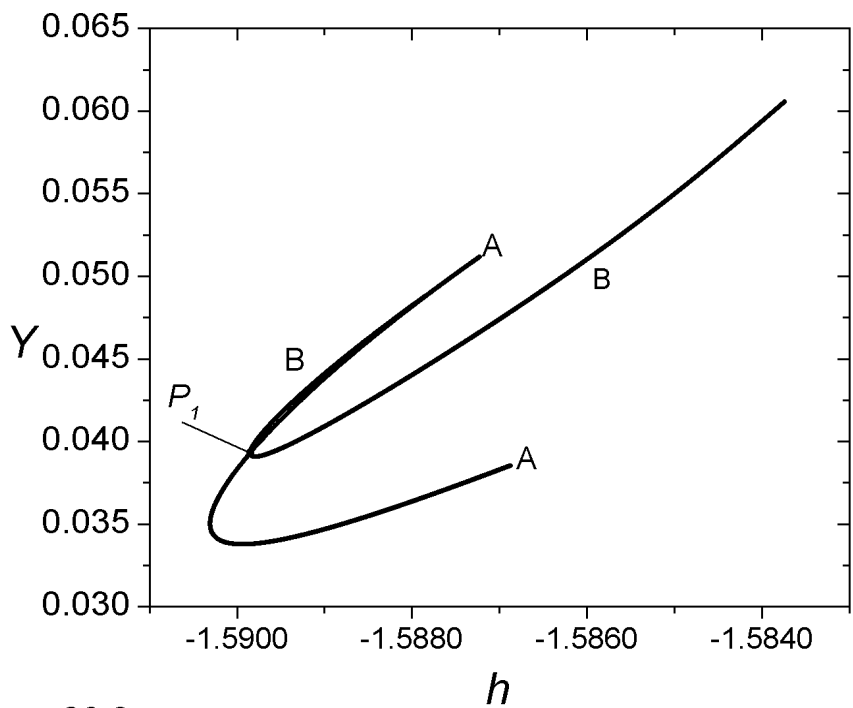

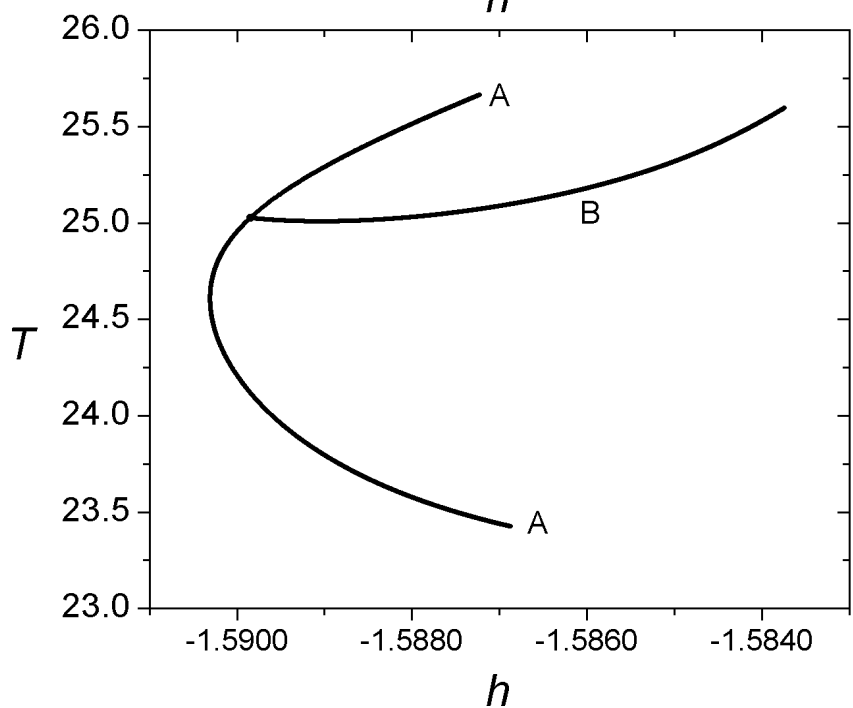

### Family 018 A - Symmetric family of symmetric POs

$h_{min} = -1.590306, \ \ h_{max} = -1.586872, \ \ T_{min} = 23.425287, \ T_{max} = 25.663849$

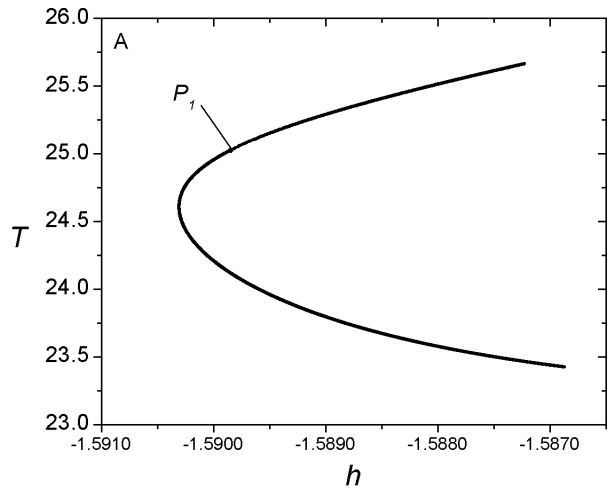
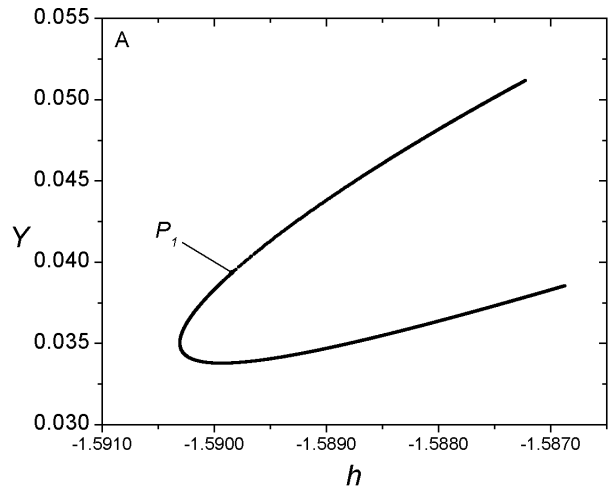

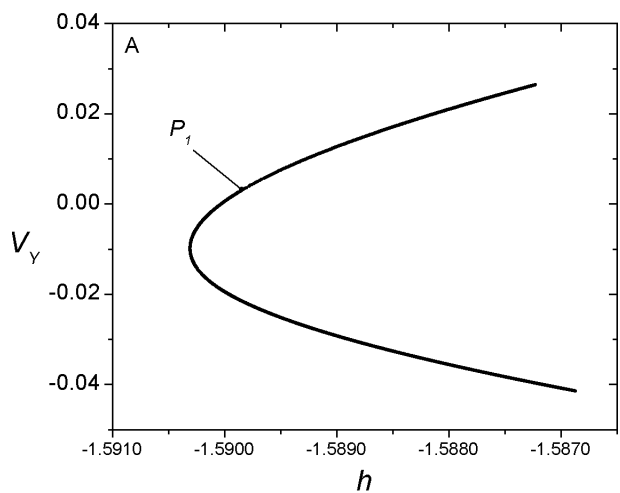
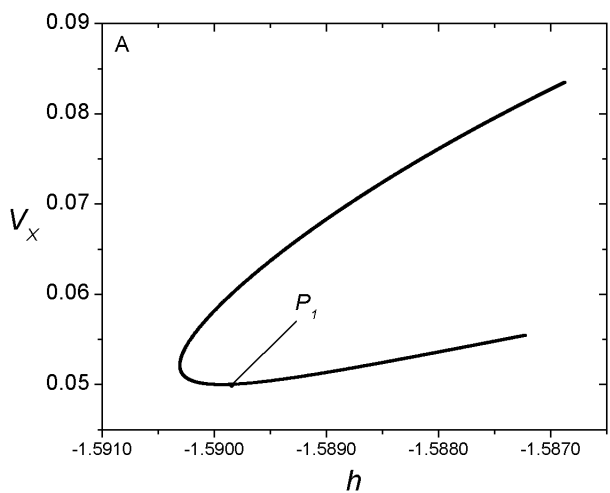

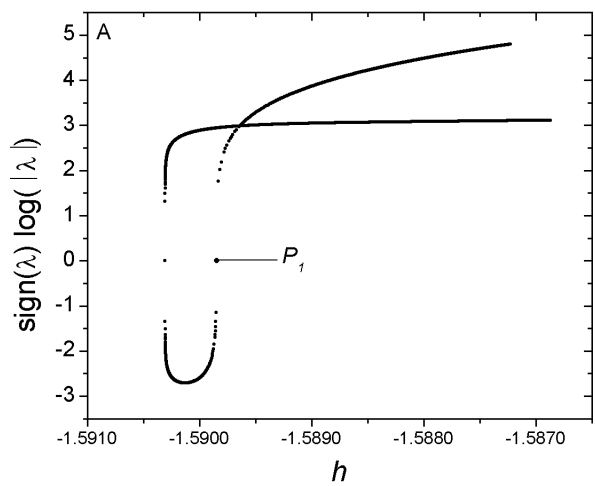
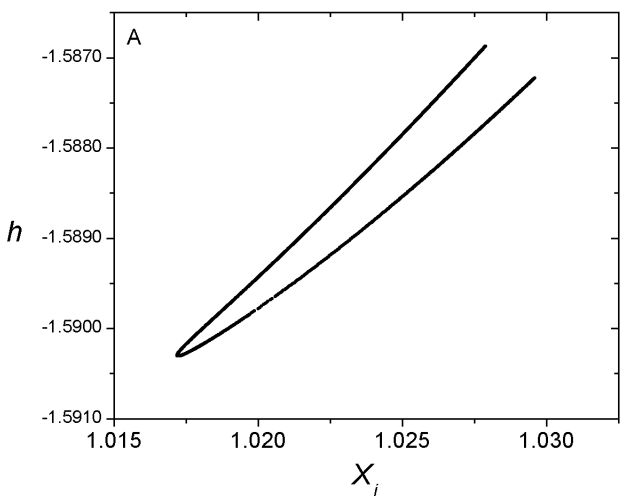

## *Family 018 B - Symmetric family of asymmetric POs*

$h_{min} = -1.589846, \quad h_{max} = -1.583736, \quad T_{min} = 25.007851, \quad T_{max} = 25.596797$

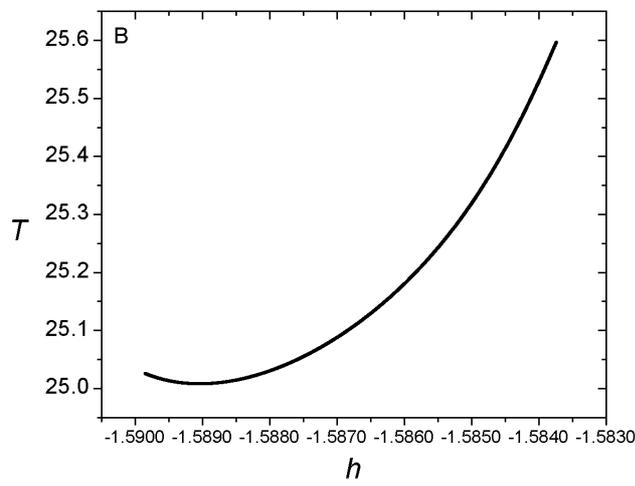

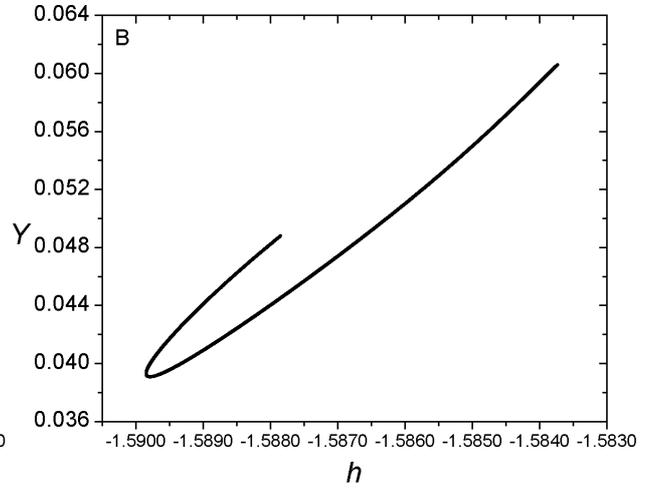

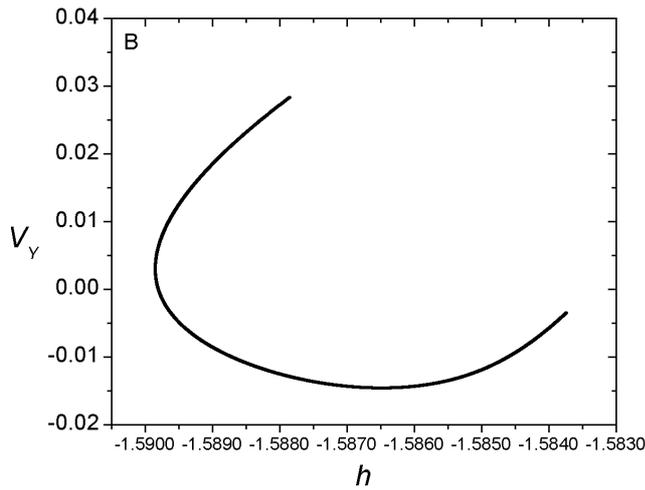

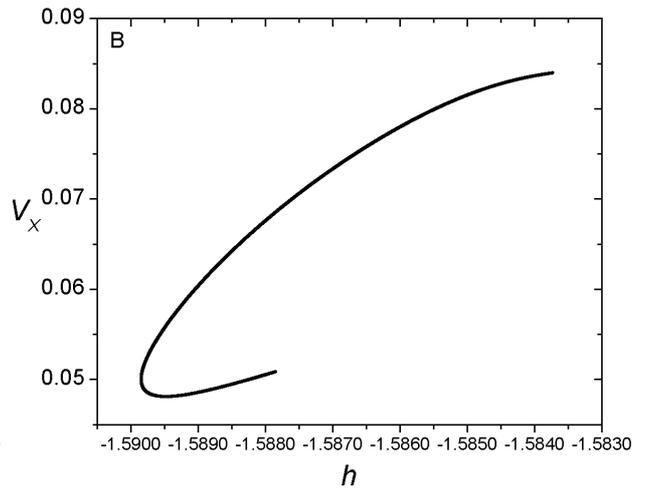

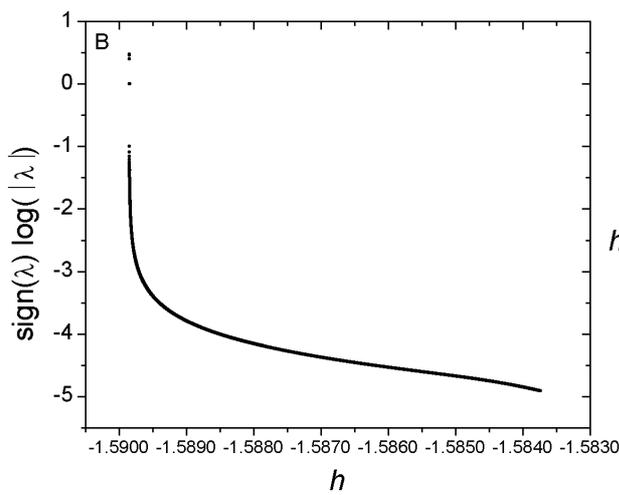

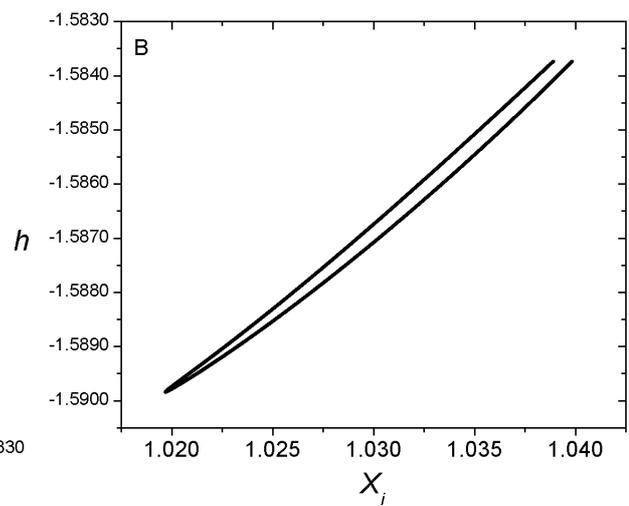

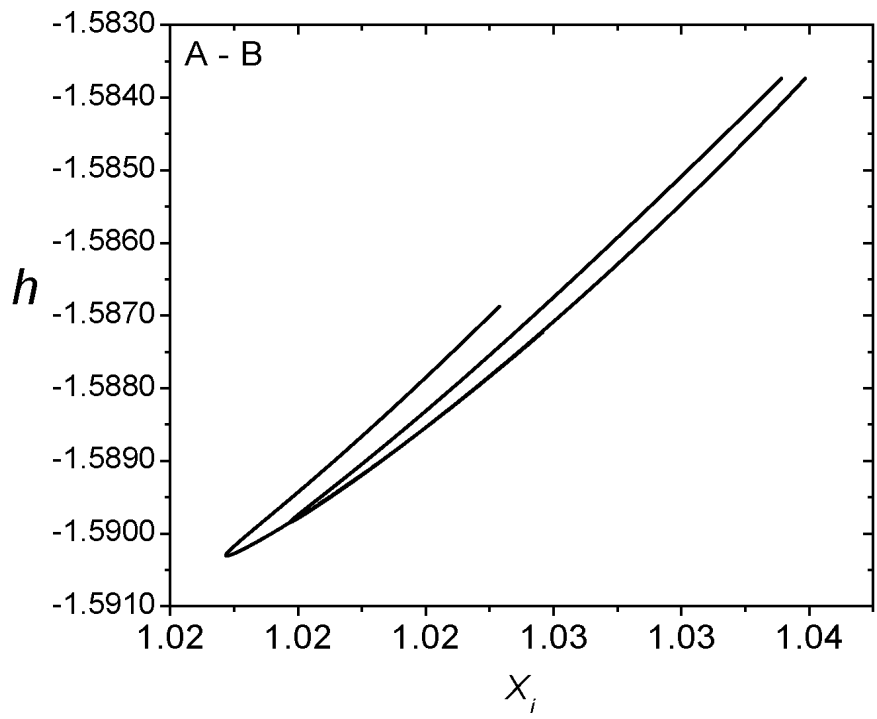

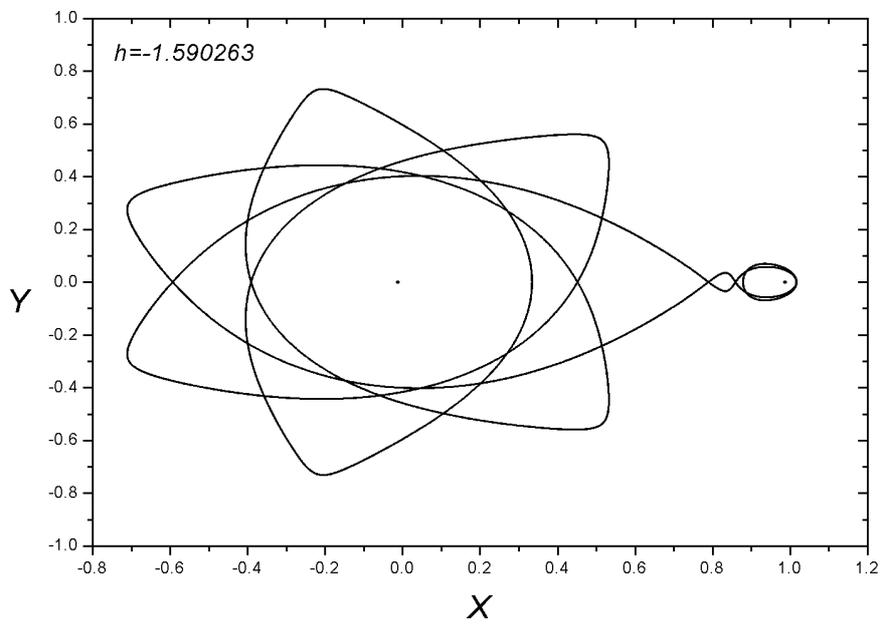

## Families 188 A - 188 B

*Bifurcation Point*

|       | $h$       | $T$       | $y$        | $v_y$      | $v_x$    |
|-------|-----------|-----------|------------|------------|----------|
| $P_1$ | −1.593636 | 24.201099 | −0.00 7716 | −0.012354  | 0.025863 |

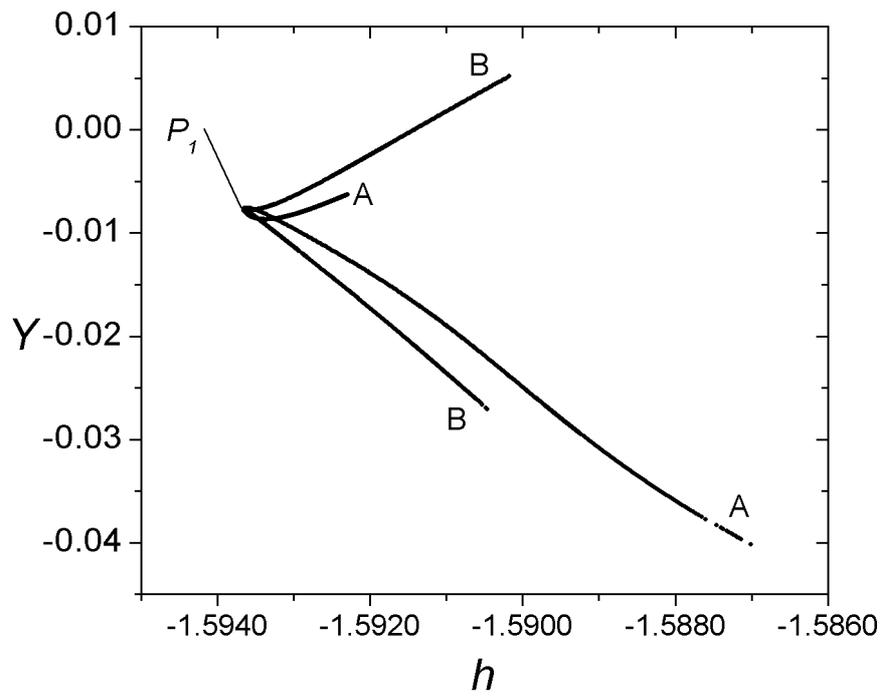

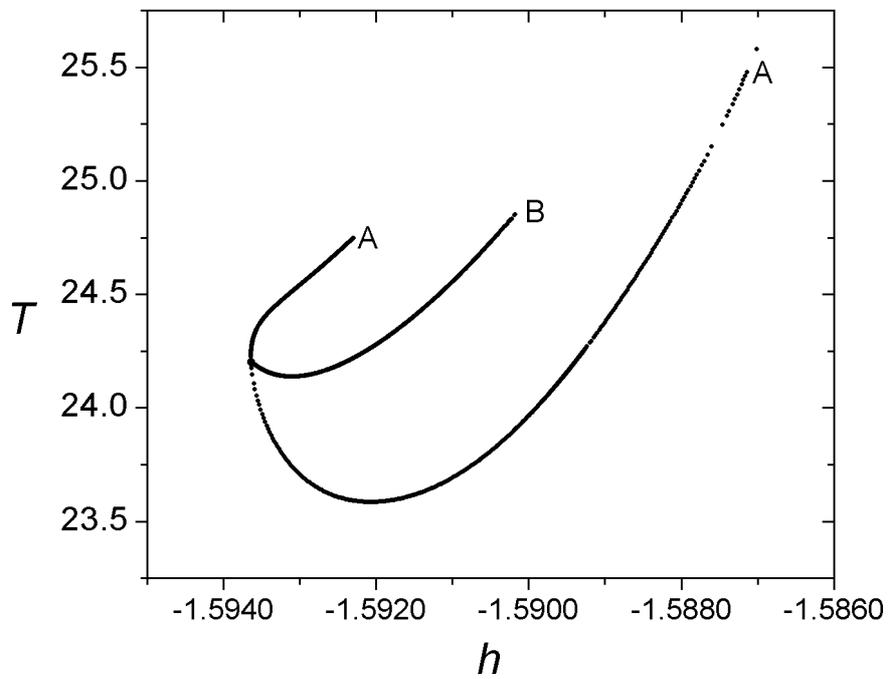

## Family 188 A - Symmetric family of symmetric POs

$h_{min} = -1.593638,\ h_{max} = -1.587012,\ T_{min} = 23.585381,\ T_{max} = 25.580361$

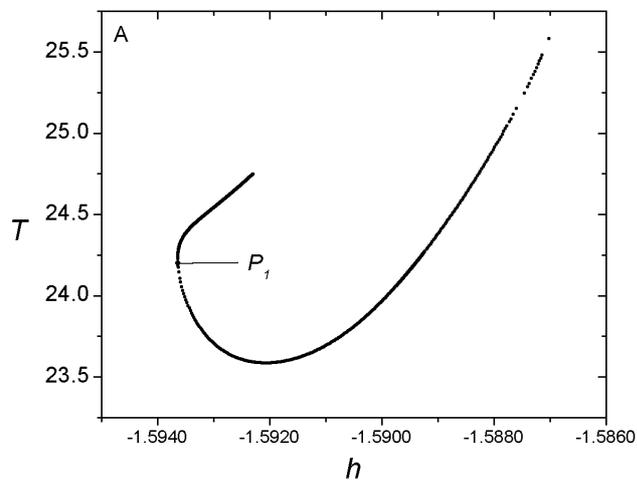

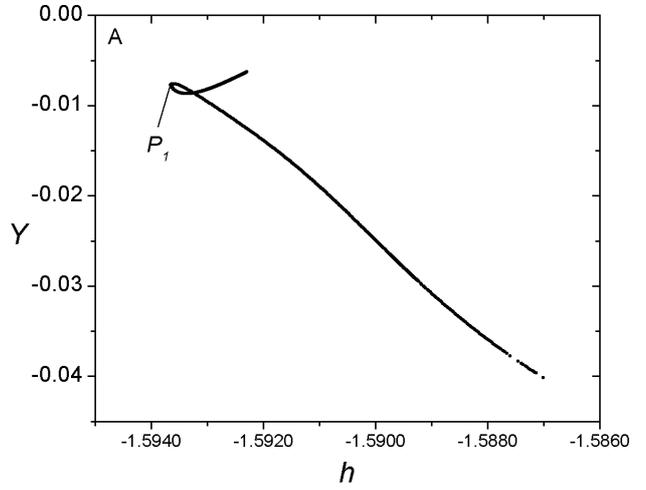

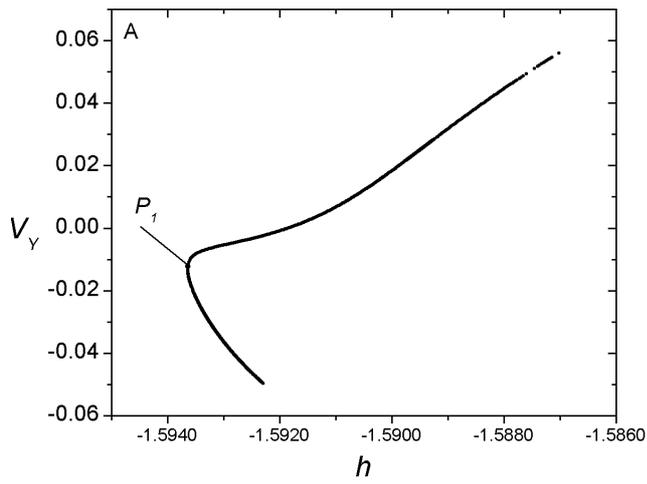

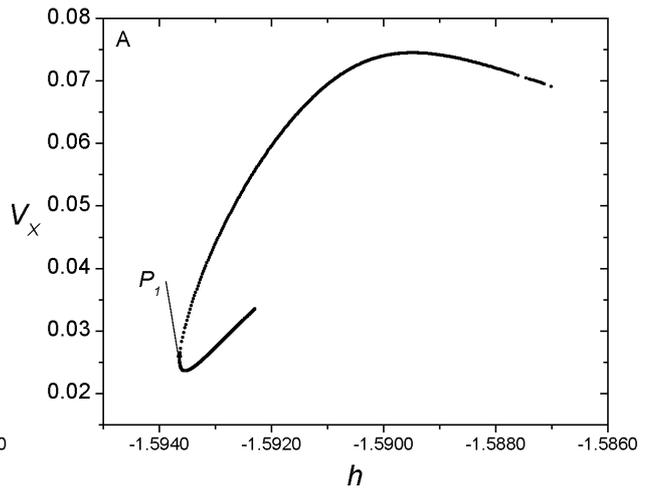

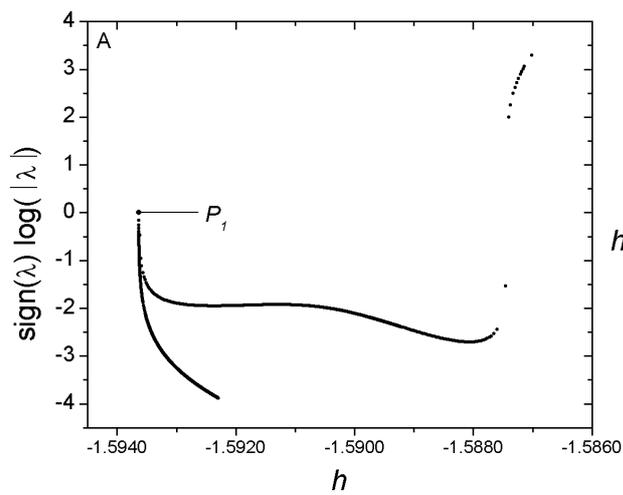

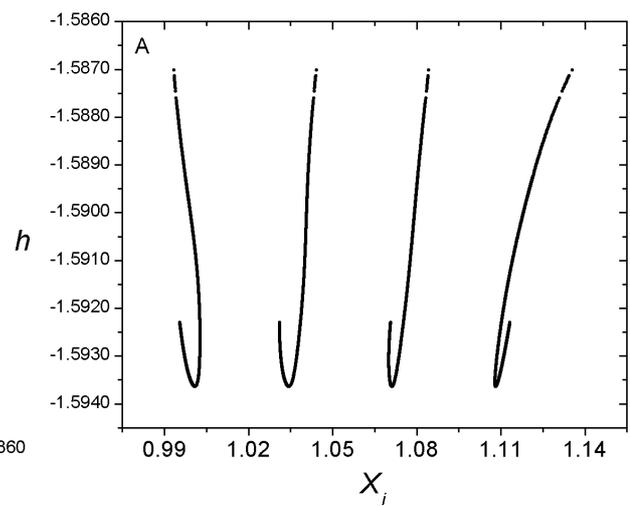

## *Family 188 B - Symmetric family of asymmetric POs*

$h_{min} = -1.593637, \quad h_{max} = -1.590172, \quad T_{min} = 24.137851, \quad T_{max} = 24.853180$

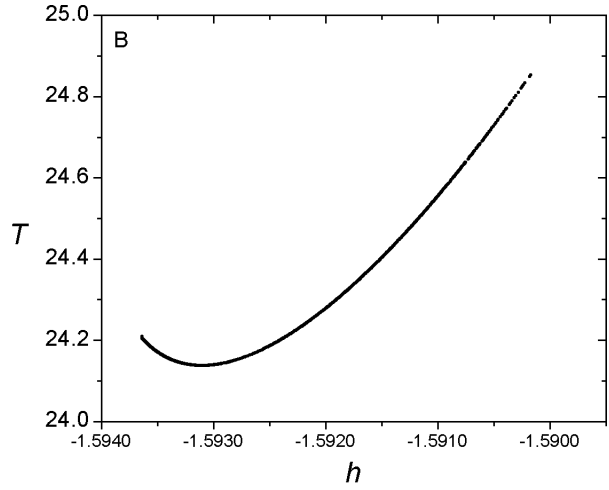
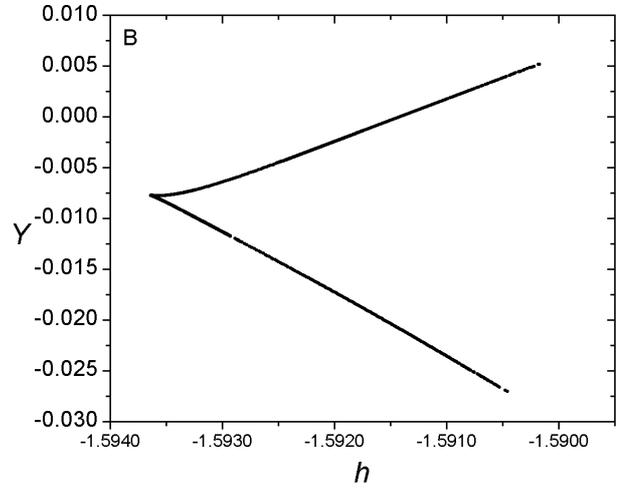

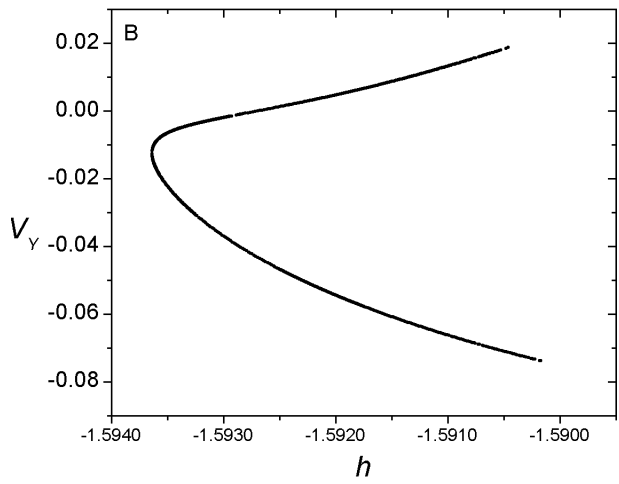
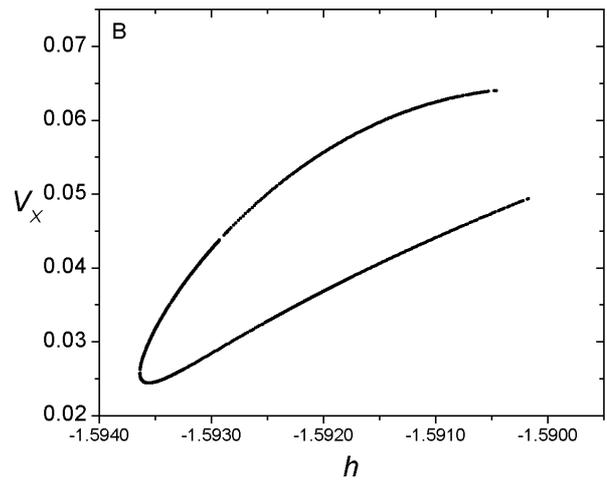

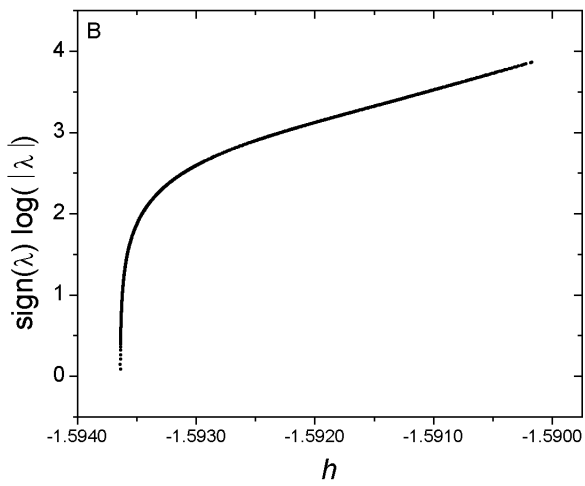
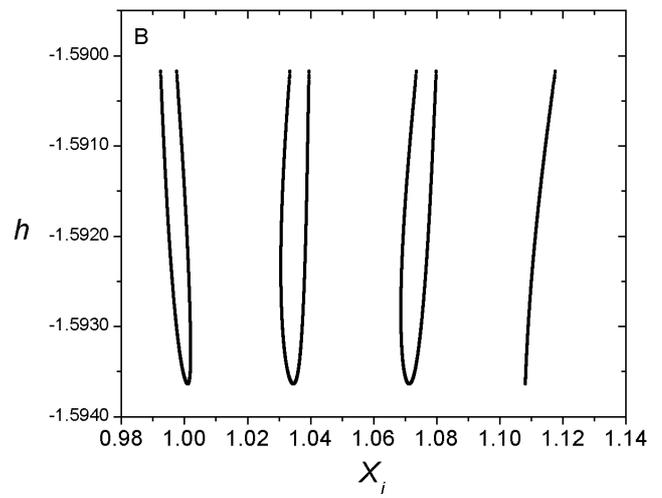

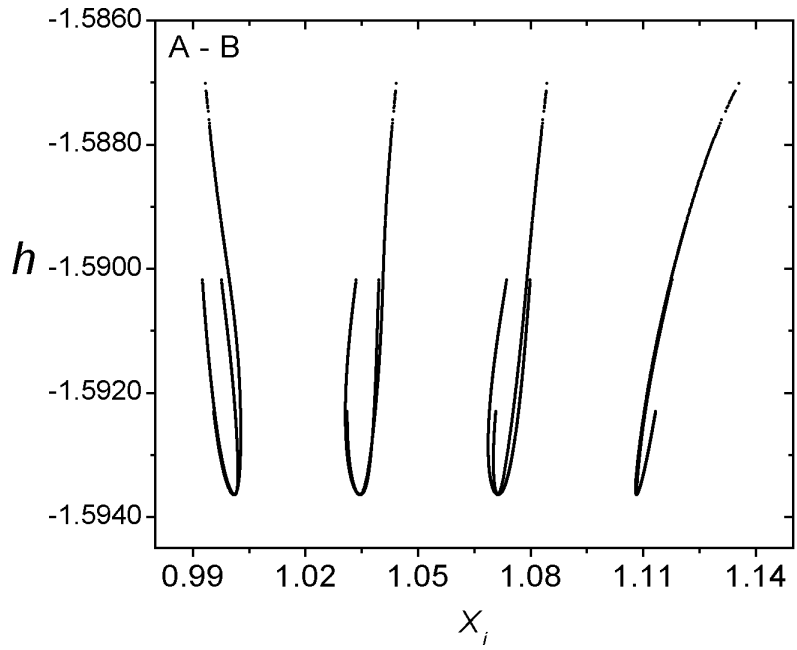

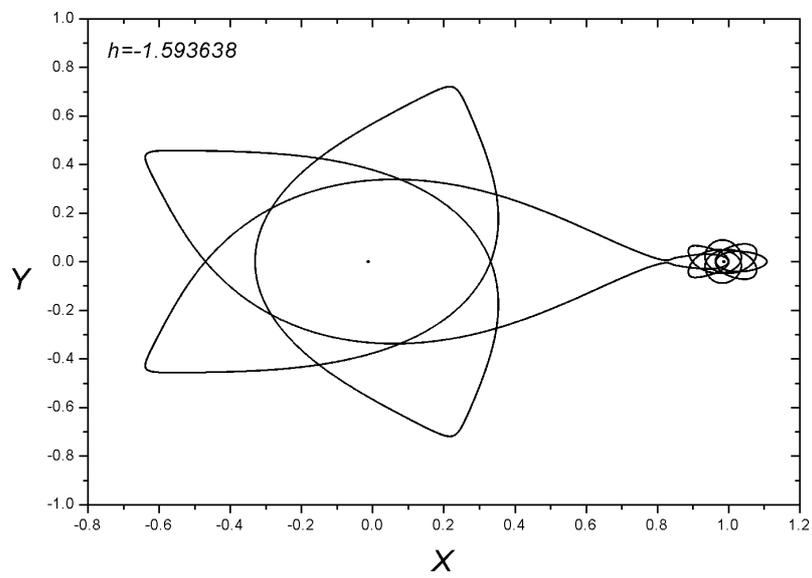

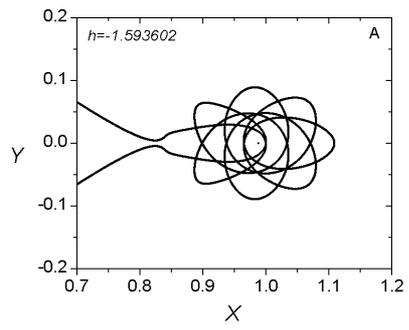
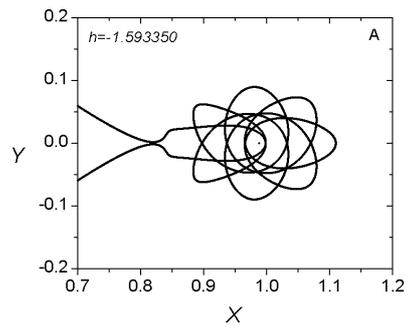
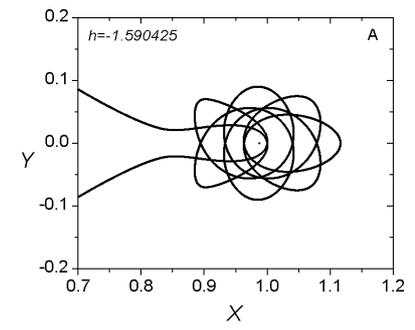

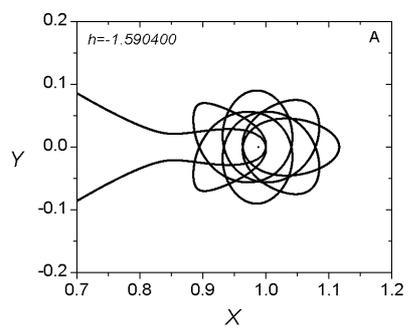
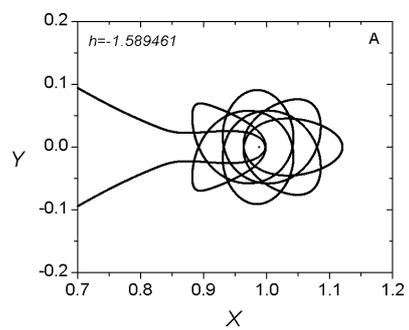
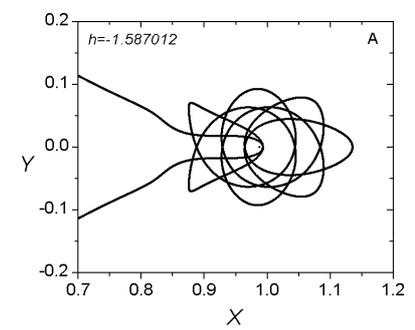

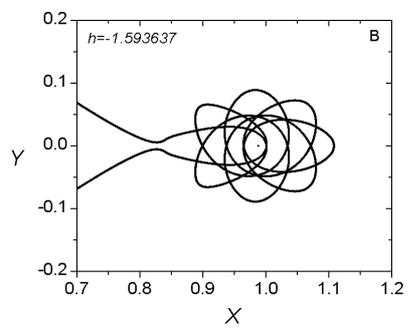
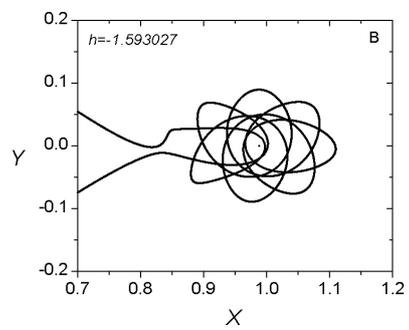
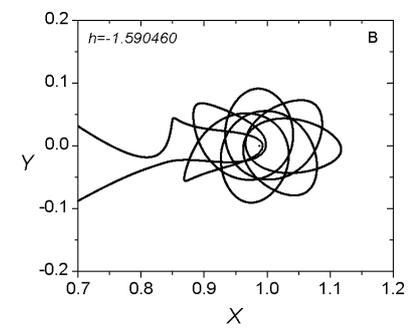

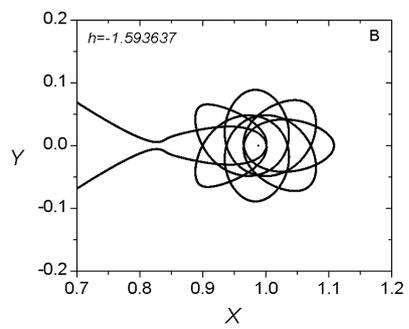
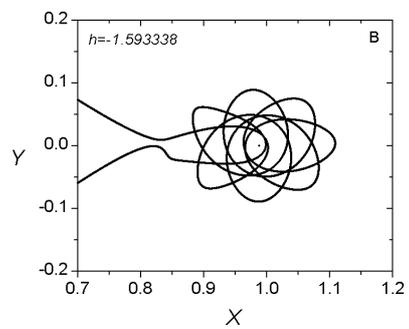
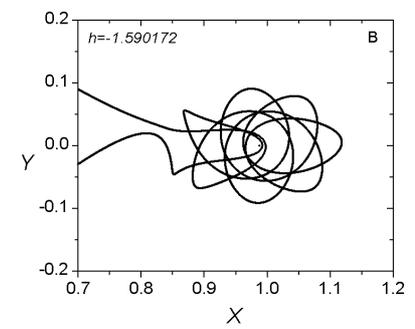

## Families 157 A - 157 B

*Bifurcation Point*

|       | $h$       | $T$       | $y$       | $v_y$     | $v_x$    |
|-------|-----------|-----------|-----------|-----------|----------|
| $P_1$ | −1.592192 | 24.658483 | −0.006472 | −0.041161 | 0.045700 |

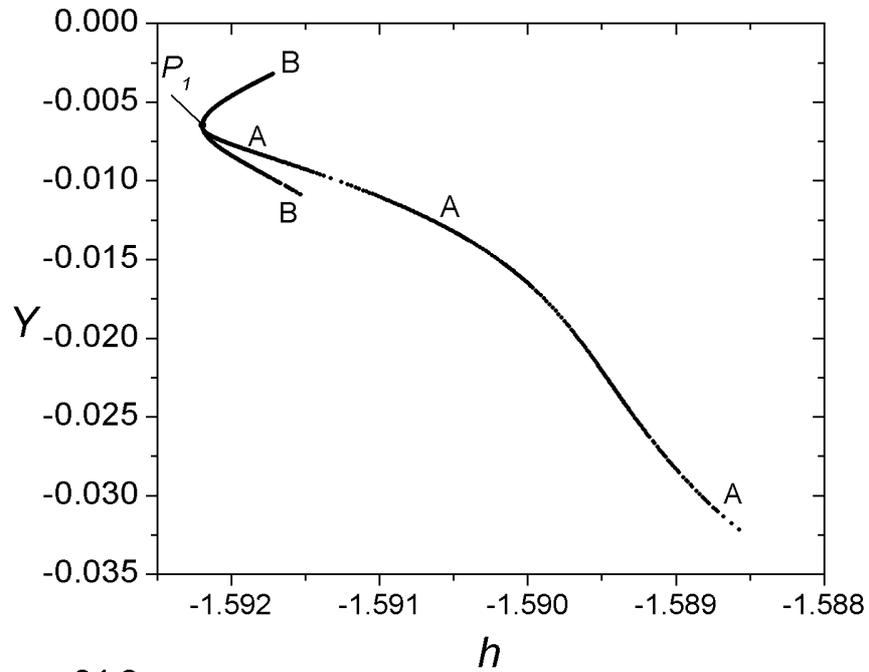

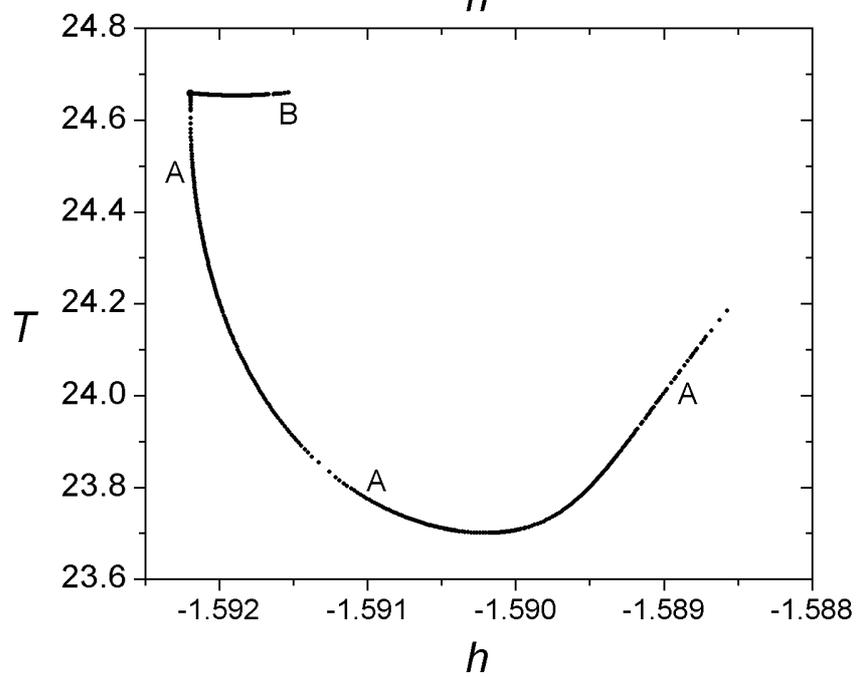

### Family 157 A - Symmetric family of symmetric POs

$h_{min} = -1.592194, \; h_{max} = -1.588574, \; T_{min} = 23.700616, \; T_{max} = 24.658482$

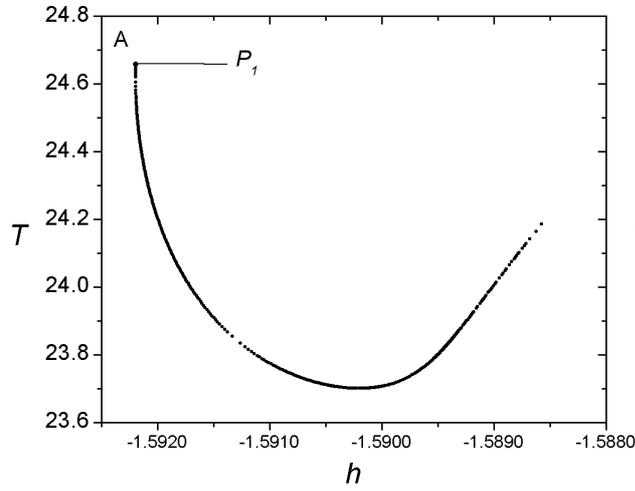

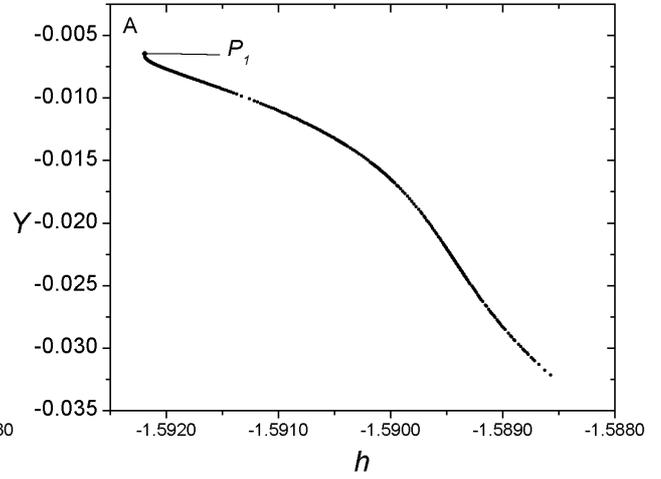

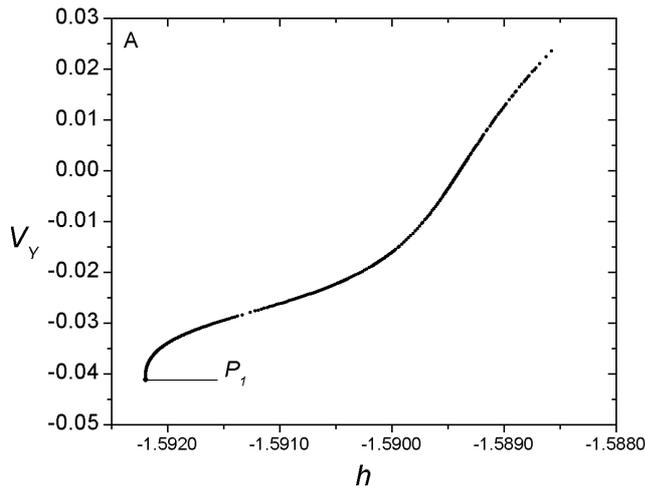

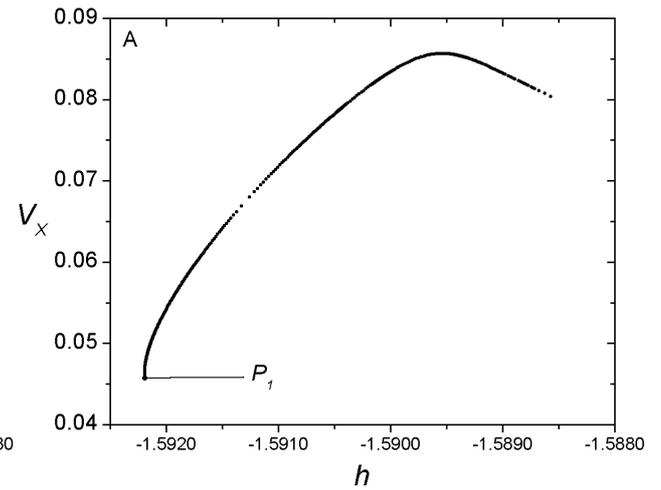

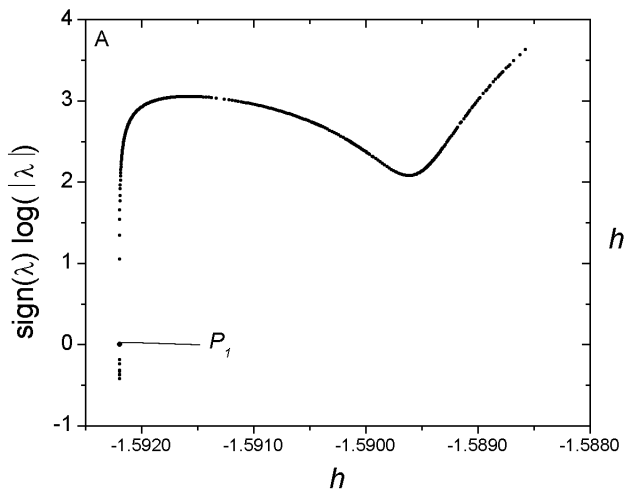

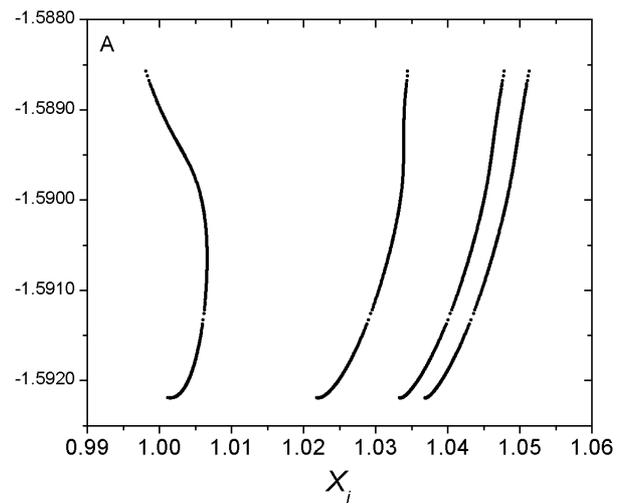

## Family 157 B - Symmetric family of asymmetric POs

*$h_{min} = -1.592192$, $h_{max} = -1.591531$, $T_{min} = 24.653557$, $T_{max} = 24.659776$*

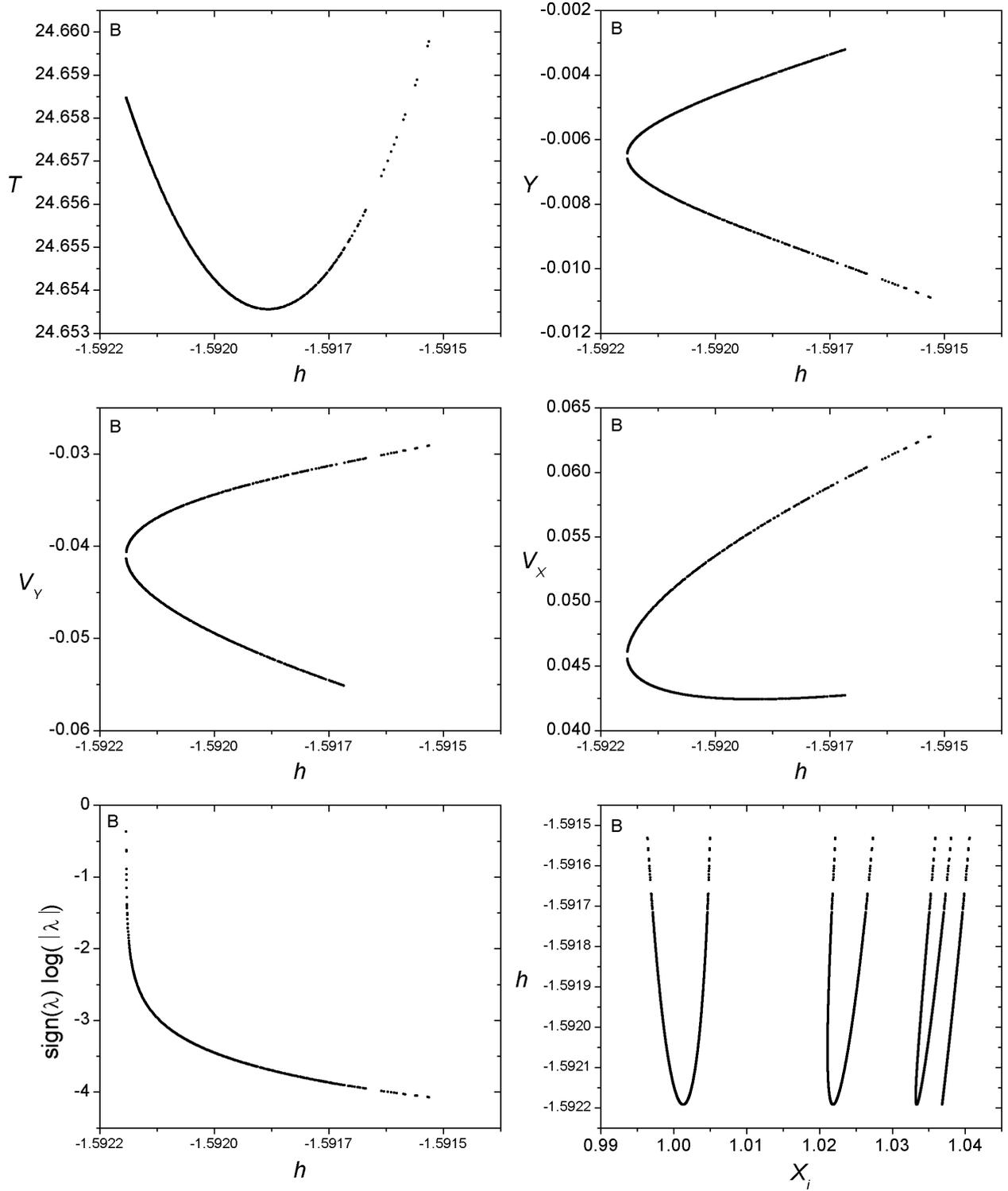

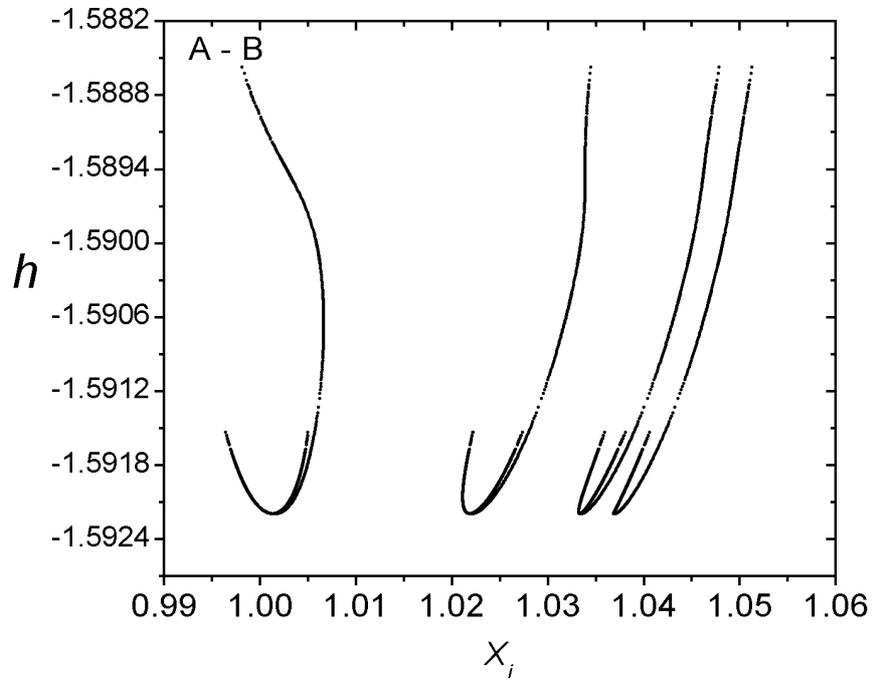

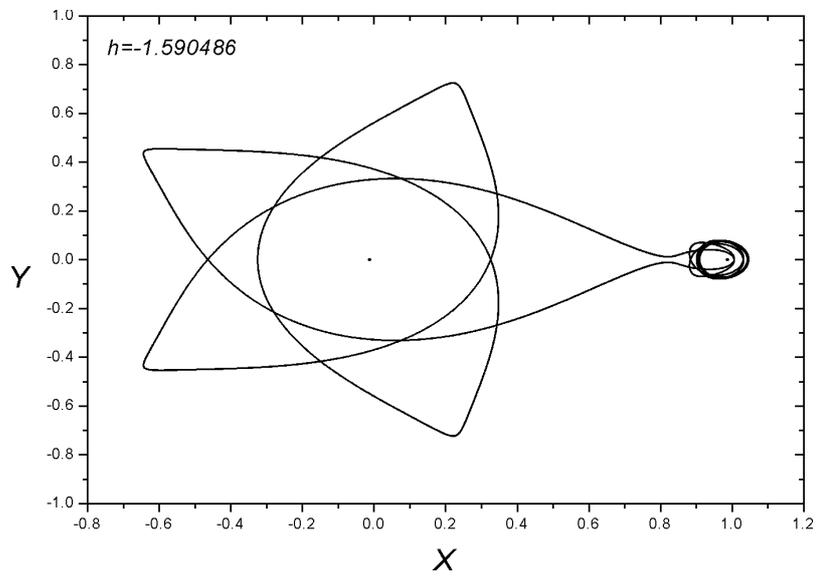

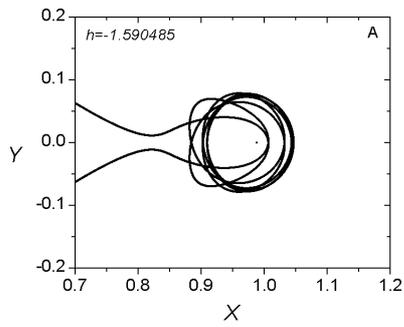
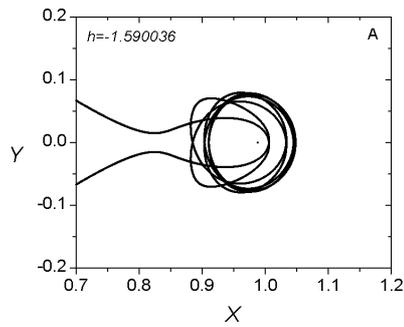
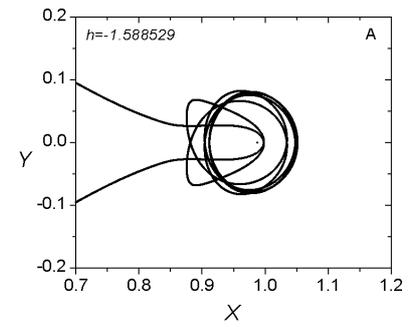

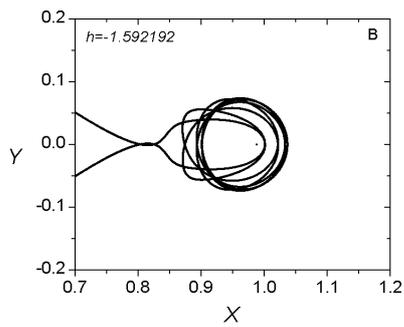
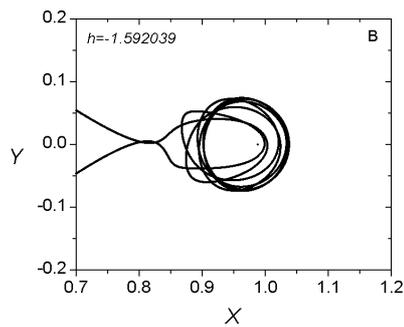
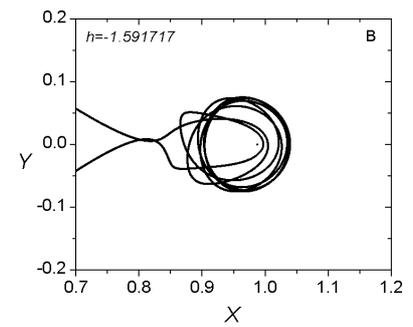

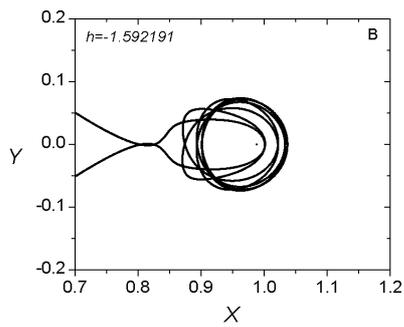
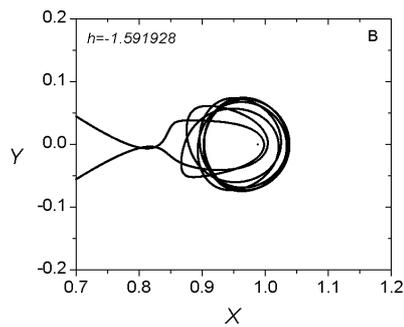
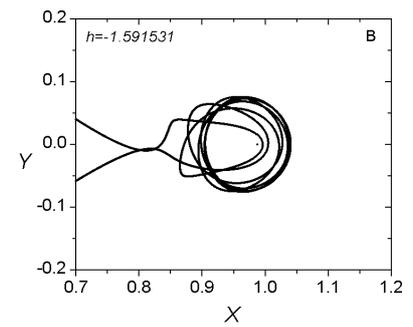

## *Family 250  - Asymmetric family of asymmetric POs*

$h_{min} = -1.593564, \ \ h_{max} = -1.591225, \ \ T_{min} = 24.659896, \ T_{max} = 25.202439$

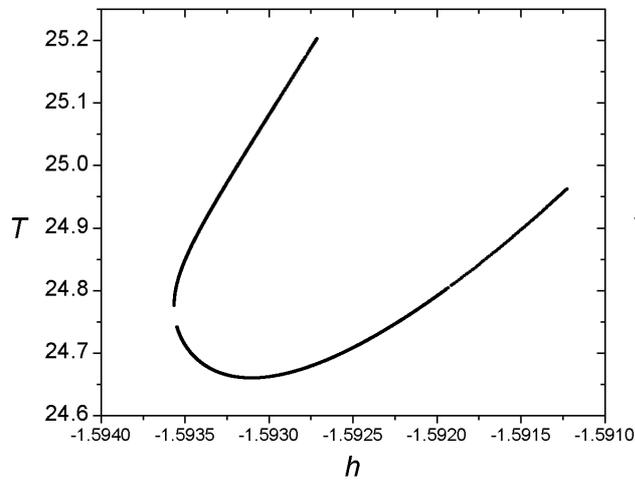
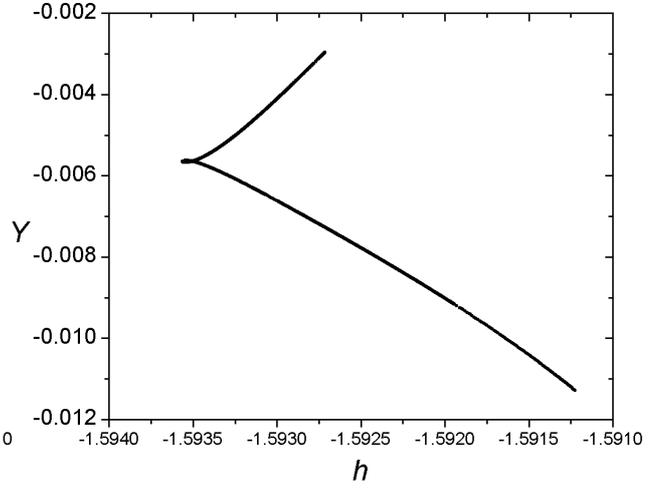

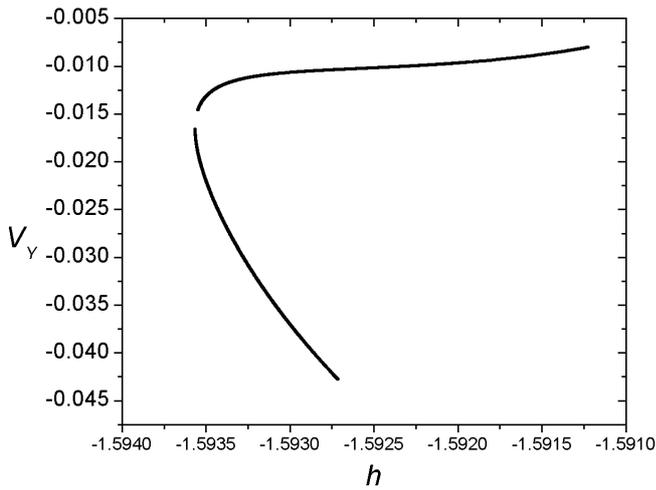
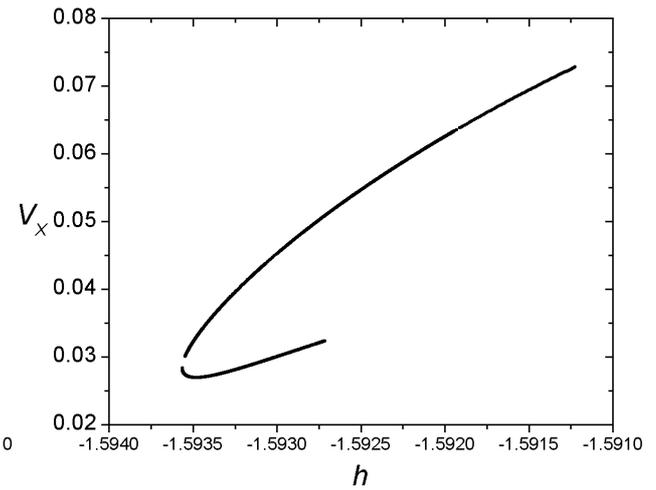

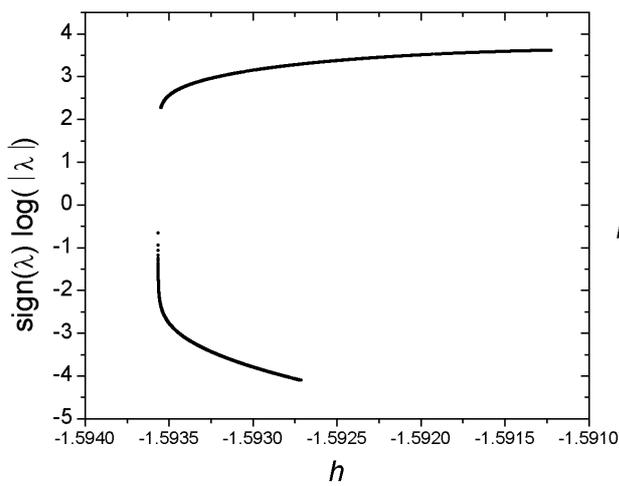
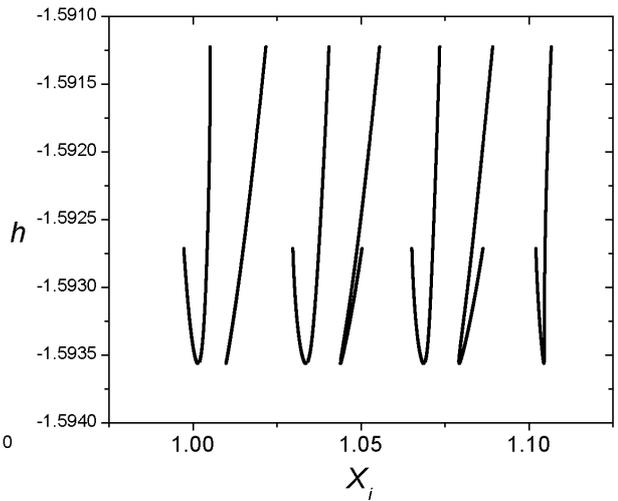

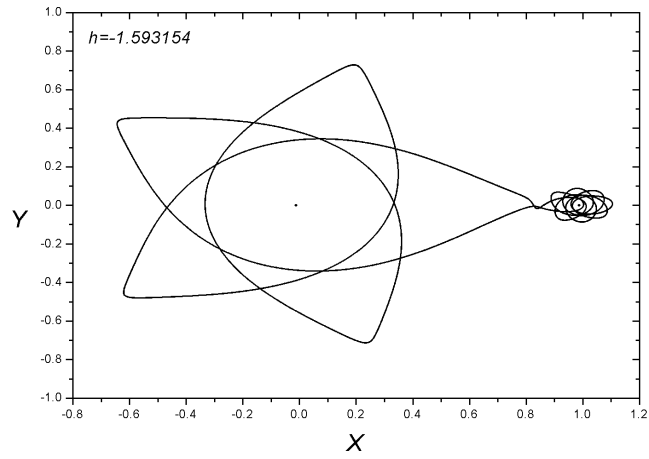

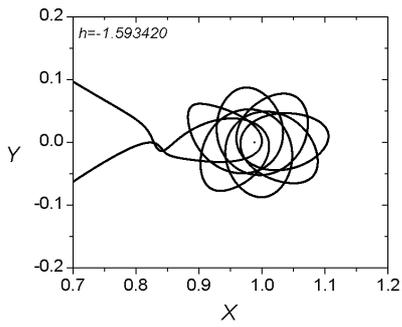
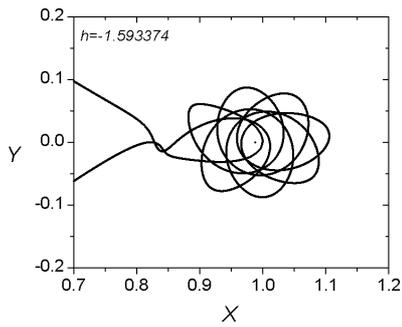
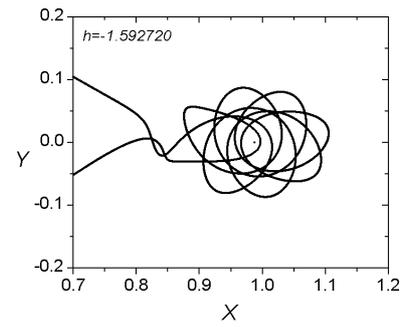

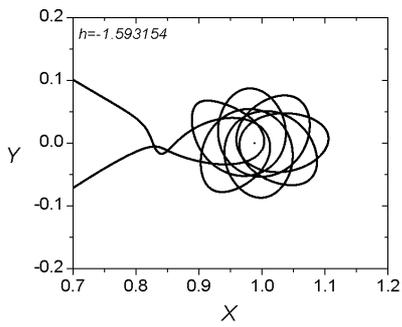
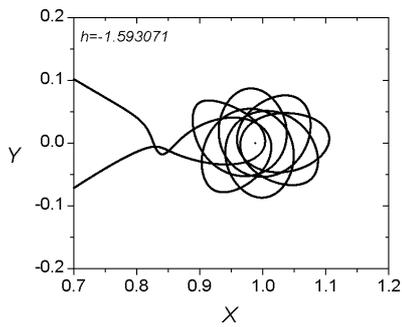
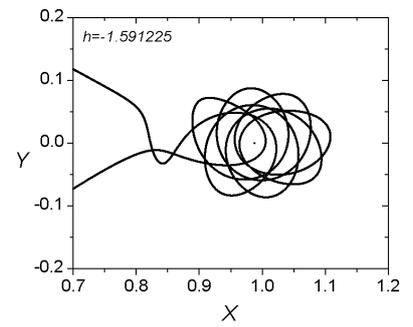

# *Family 251 - Asymmetric family of asymmetric POs*

$h_{min} = -1.593564, \quad h_{max} = -1.589273, \quad T_{min} = 24.659896, \quad T_{max} = 25.633381$

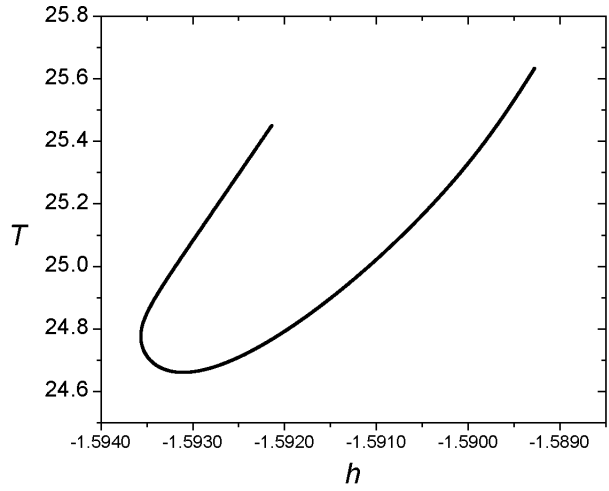

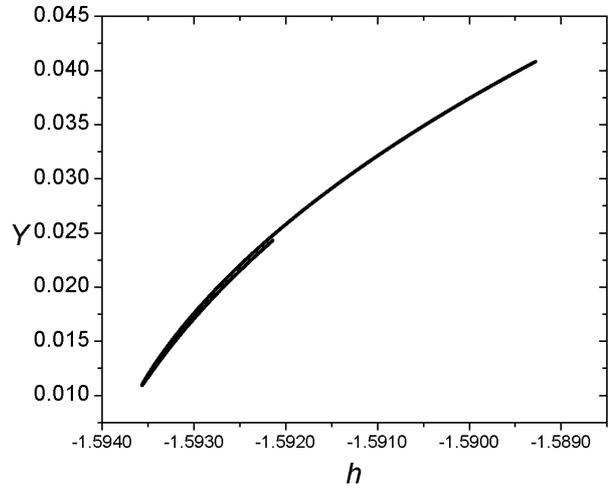

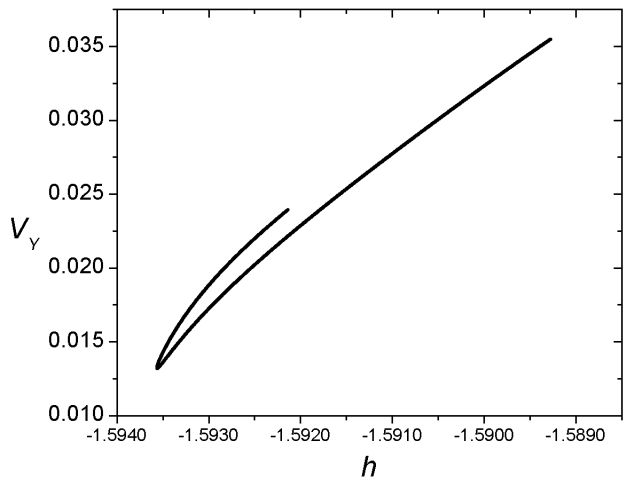

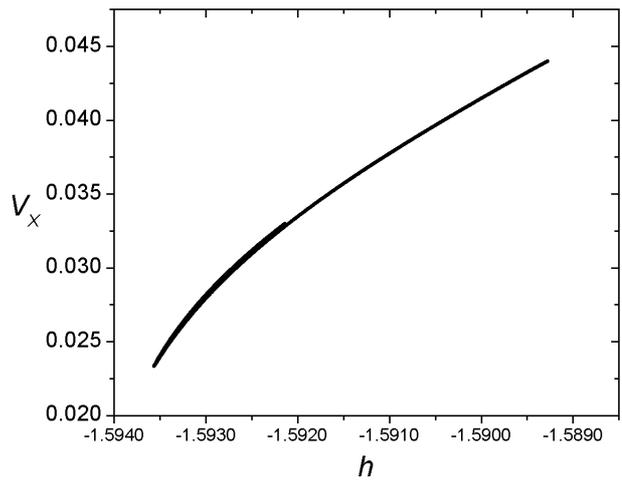

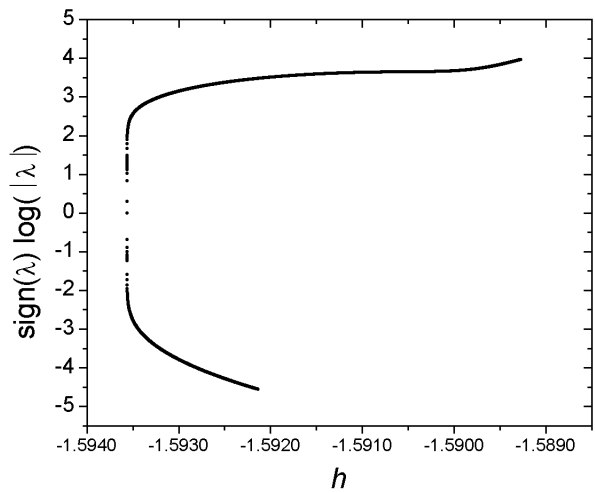

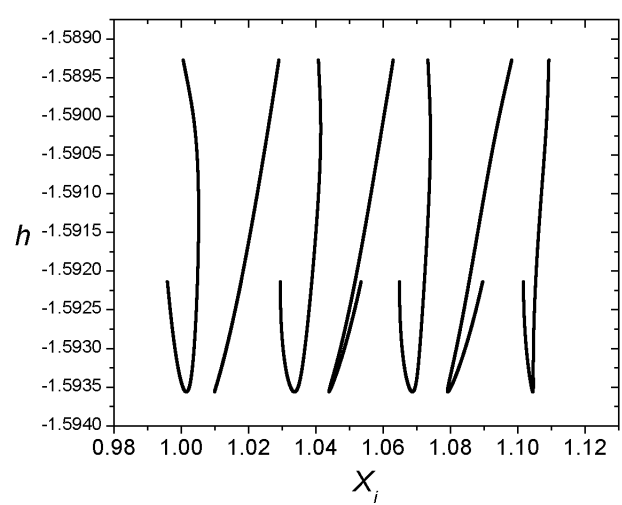

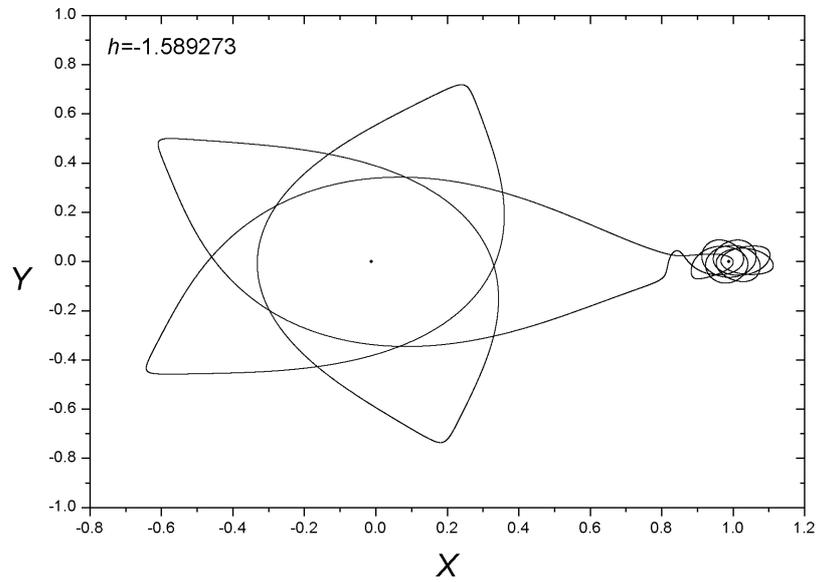

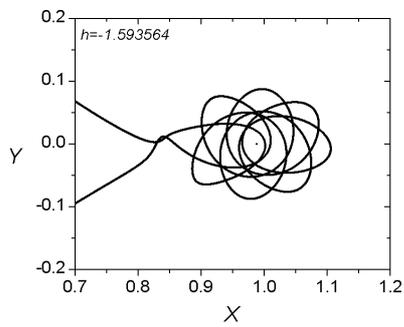
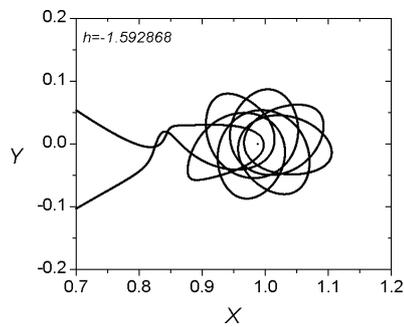
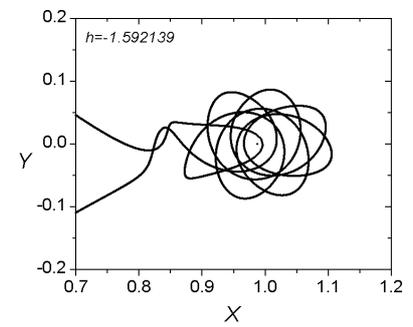

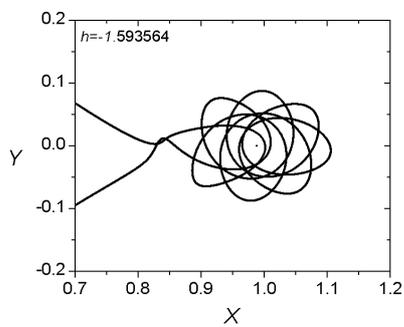
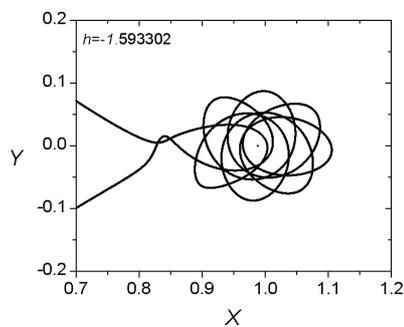
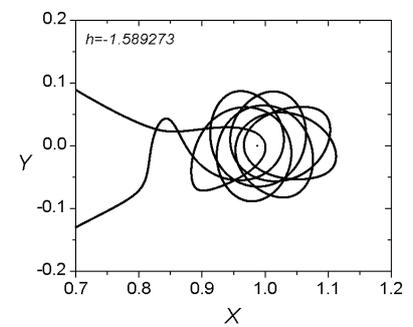

## Family 302  - *Asymmetric family of asymmetric POs*

$h_{min} = -1.590399, \ h_{max} = -1.588149, \ T_{min} = 24.955219, \ T_{max} = 25.666963$

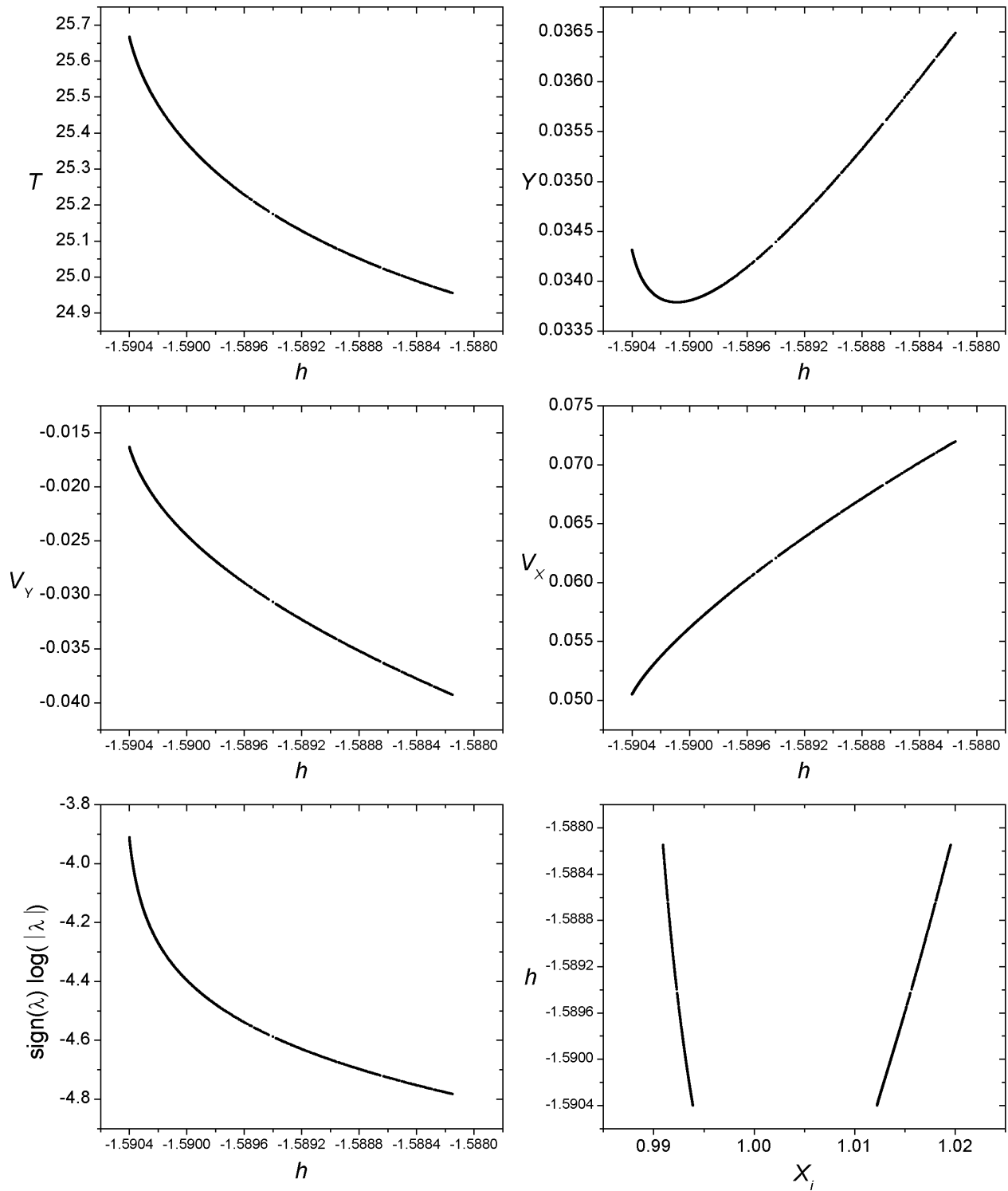

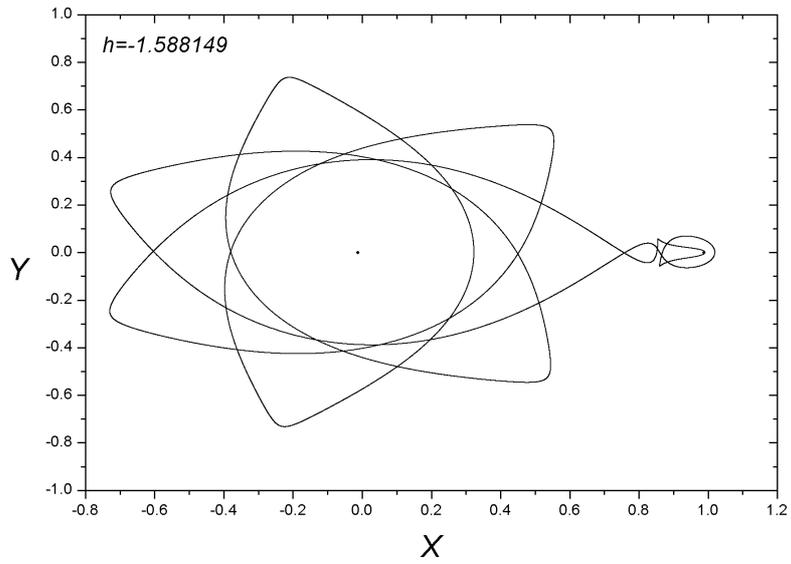

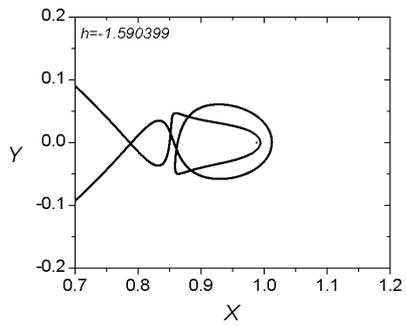

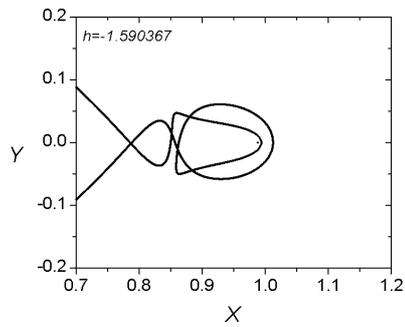

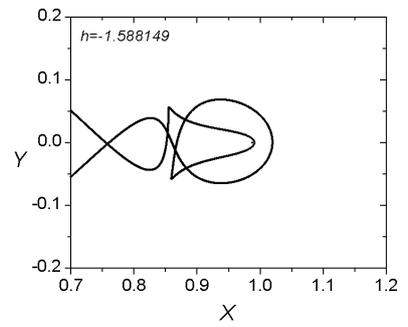

## *Family 301  - Asymmetric family of asymmetric POs*

*$h_{min} = -1.590428$,  $h_{max} = -1.588695$,  $T_{min} = 25.033123$, $T_{max} = 25.727648$*

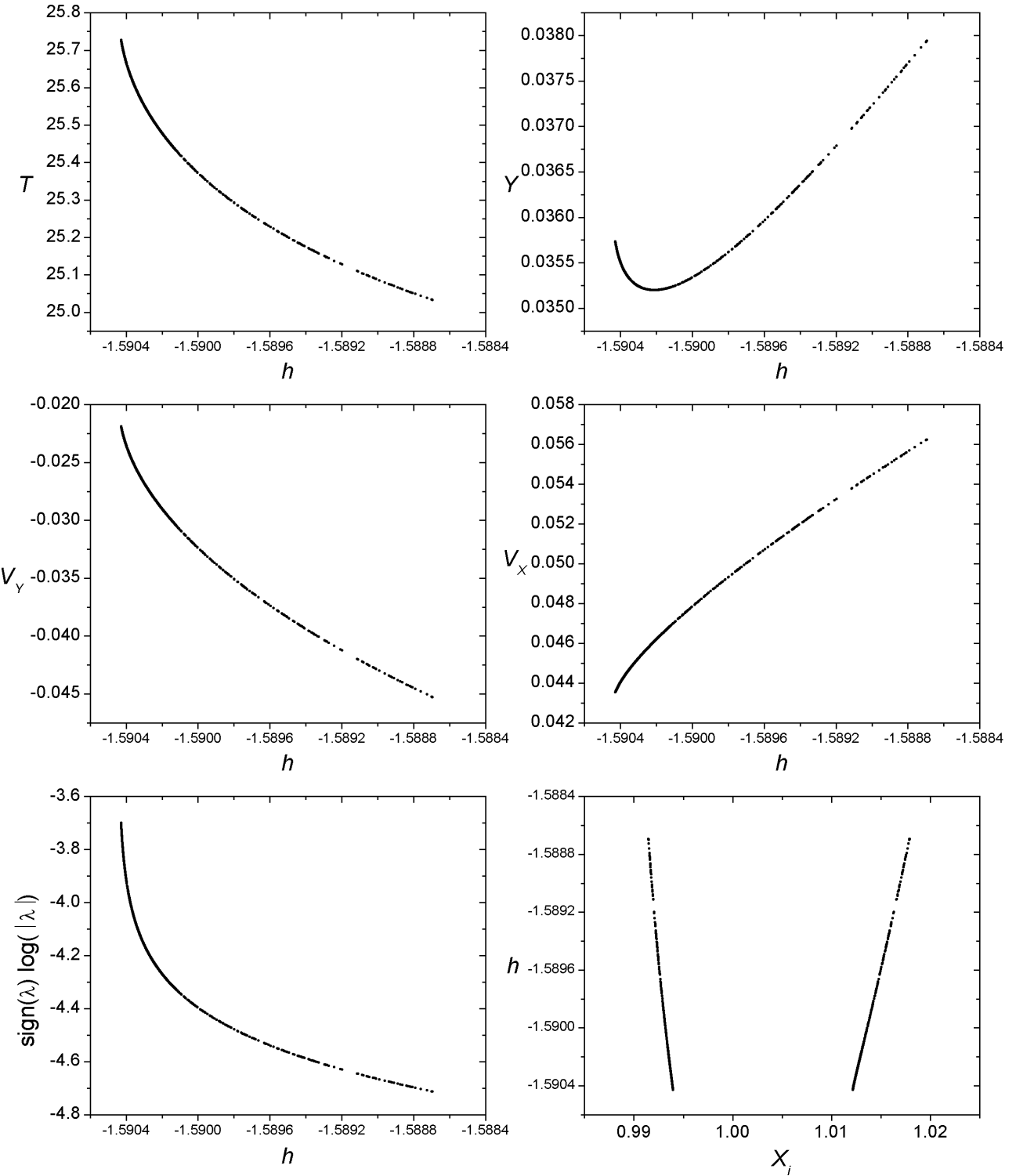

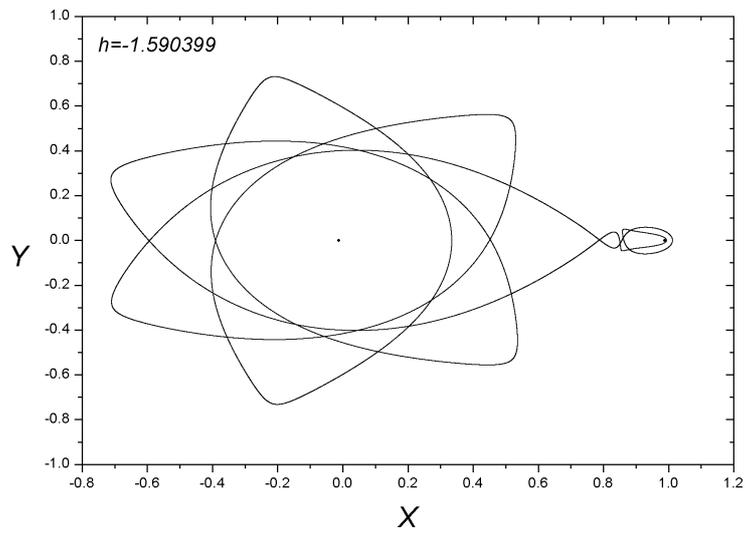

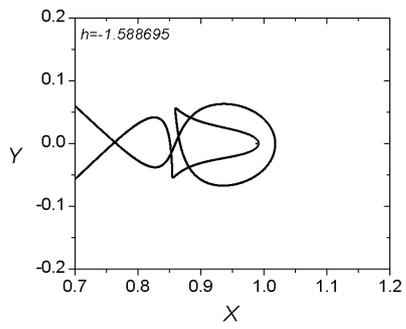
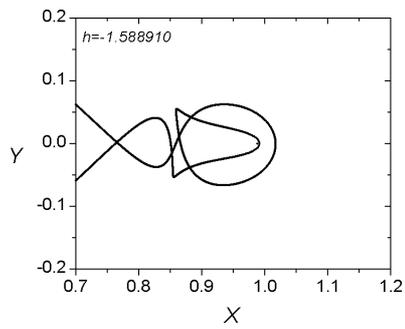
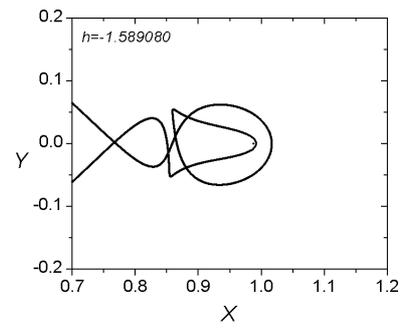

## Family 192 - *Symmetric family of symmetric POs*

$h_{min} = -1.593056$, $h_{max} = -1.588285$, $T_{min} = 25.039939$, $T_{max} = 26.244539$

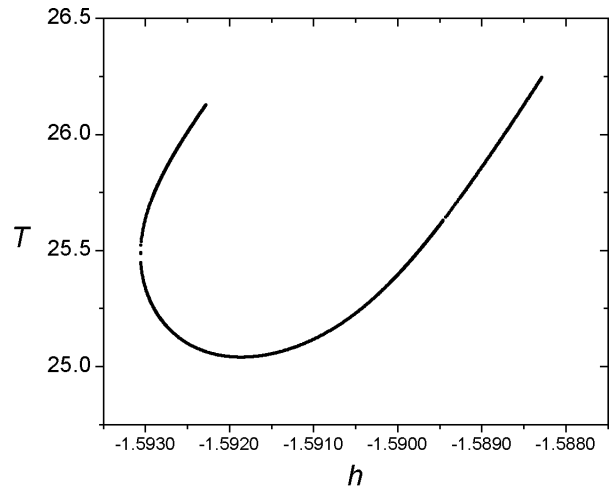
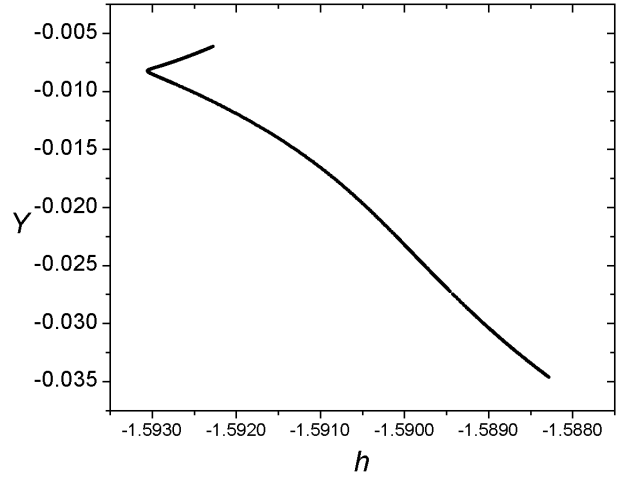

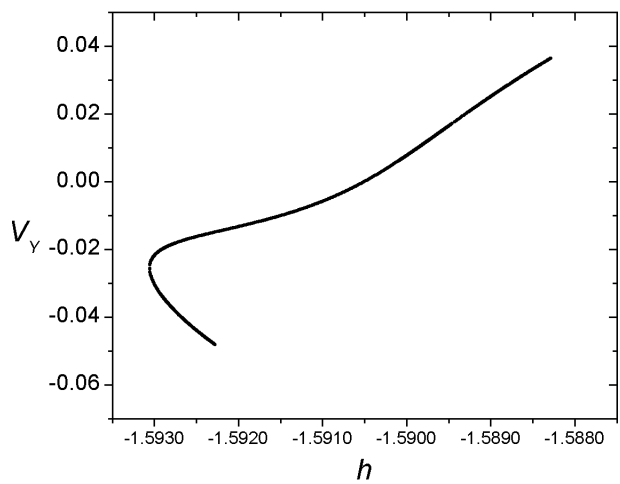
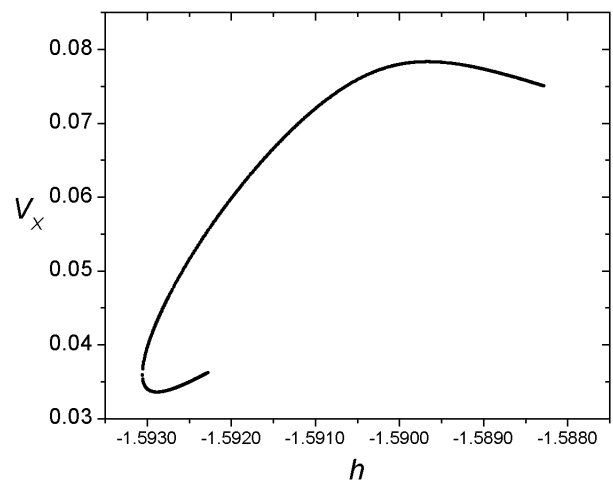

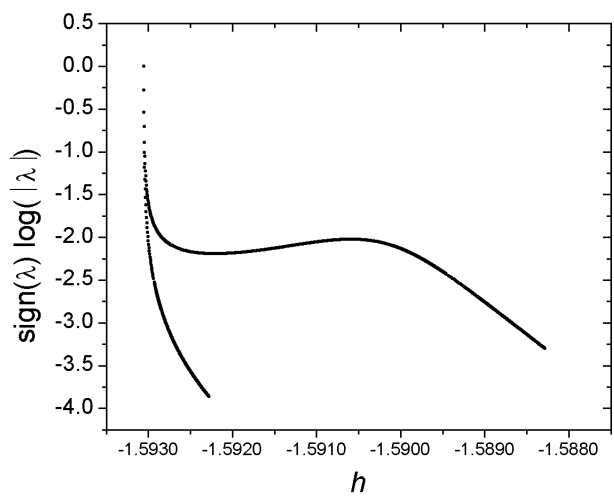
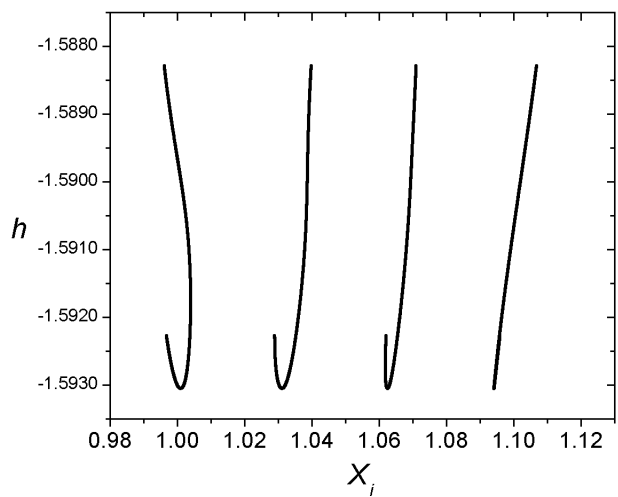

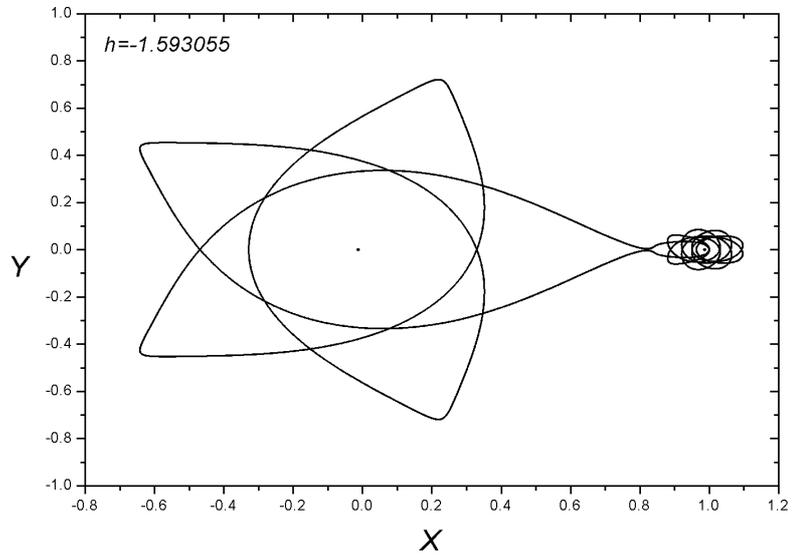

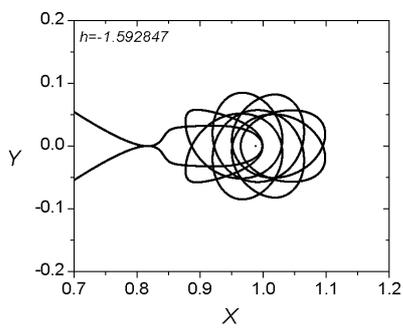
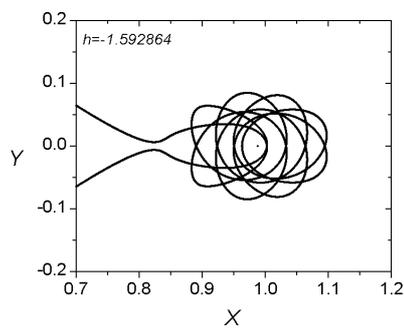
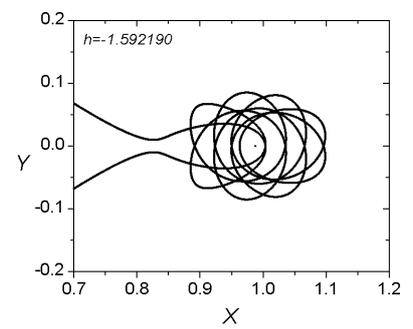

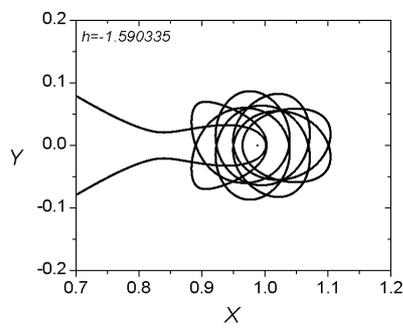
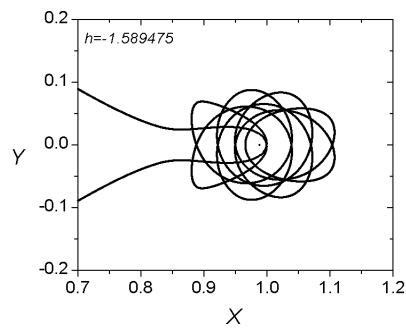
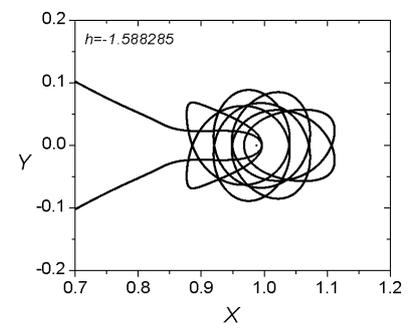

## Families 286 A - 286 B

*Bifurcation Point*

|       | $h$       | $T$       | $y$      | $v_y$     | $v_x$    |
|-------|-----------|-----------|----------|-----------|----------|
| $P_1$ | −1.594119 | 26.183033 | 0.002700 | −0.000584 | 0.008517 |

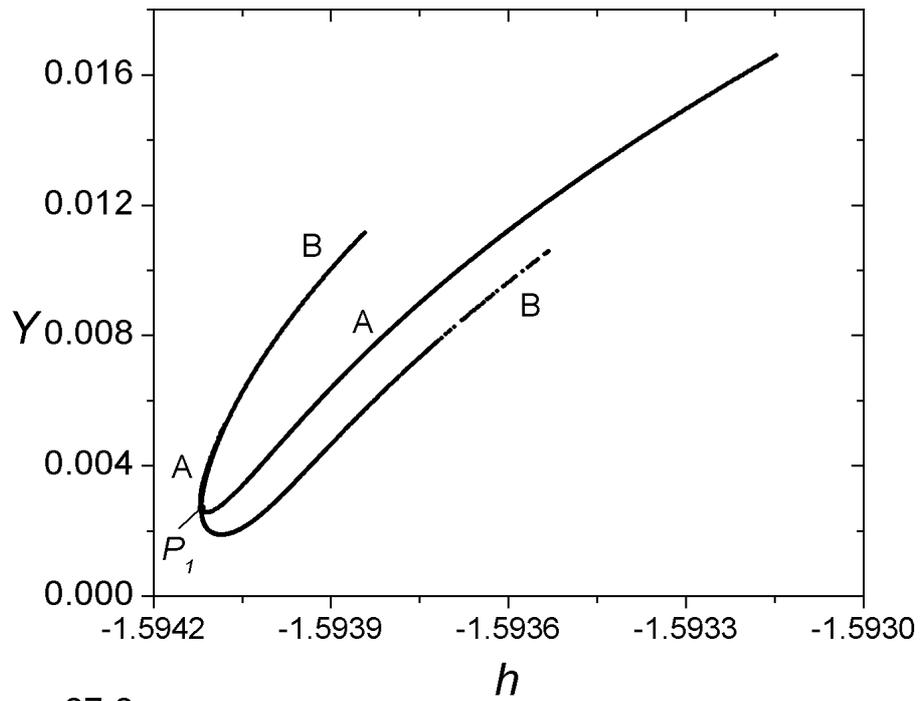

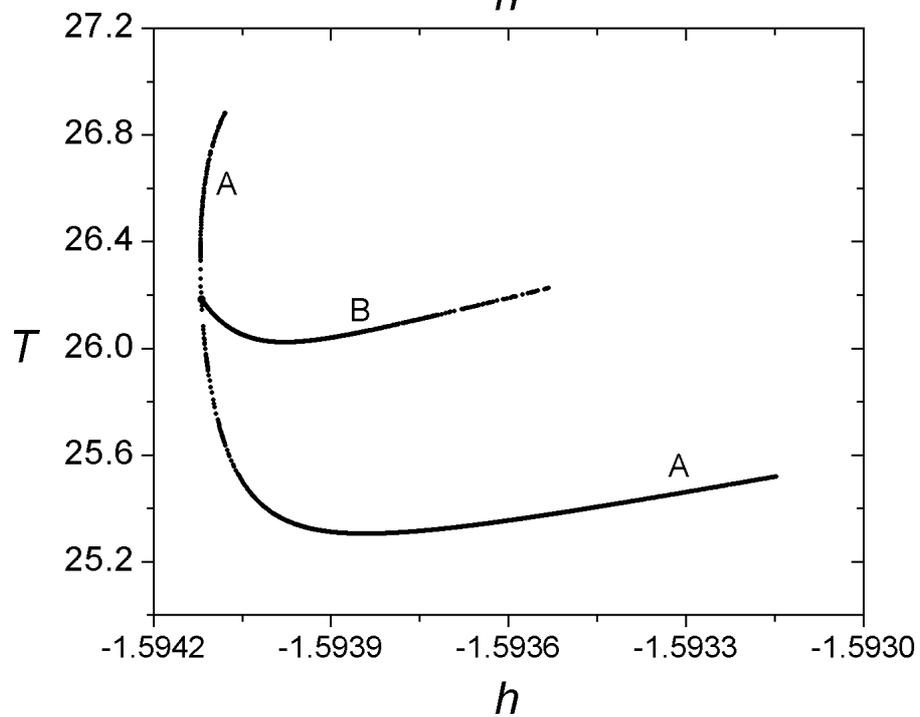

## Family 286 A - Symmetric family of symmetric POs

$h_{min} = -1.594121$,  $h_{max} = -1.593147$,  $T_{min} = 25.304381$, $T_{max} = 26.881957$

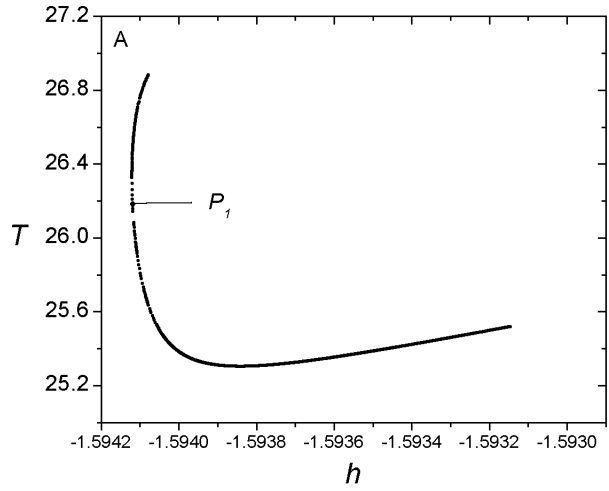
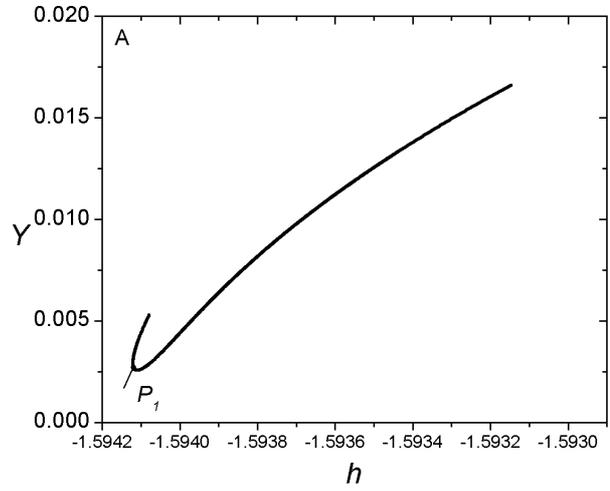

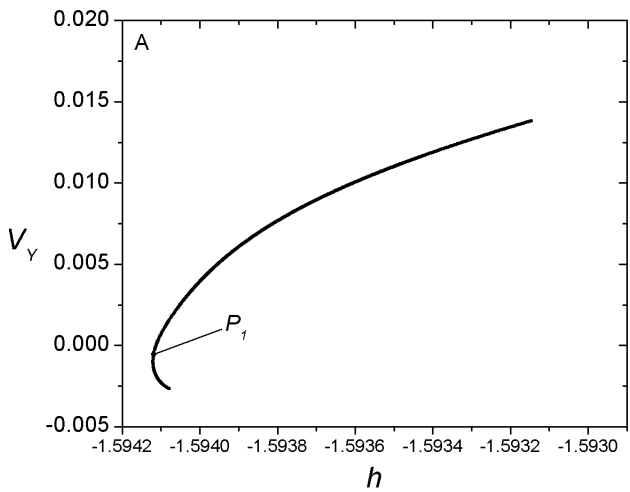
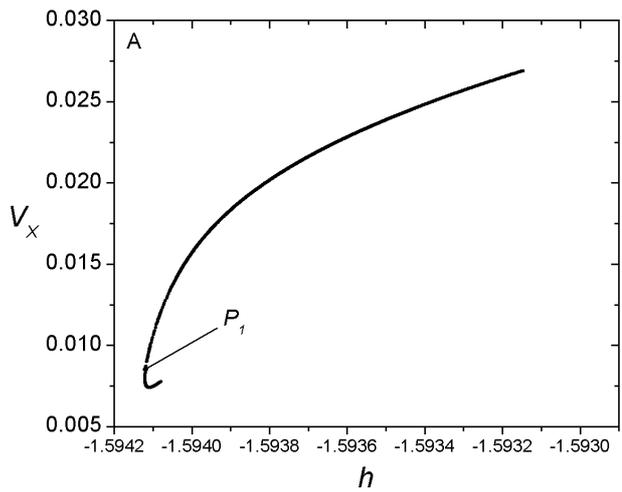

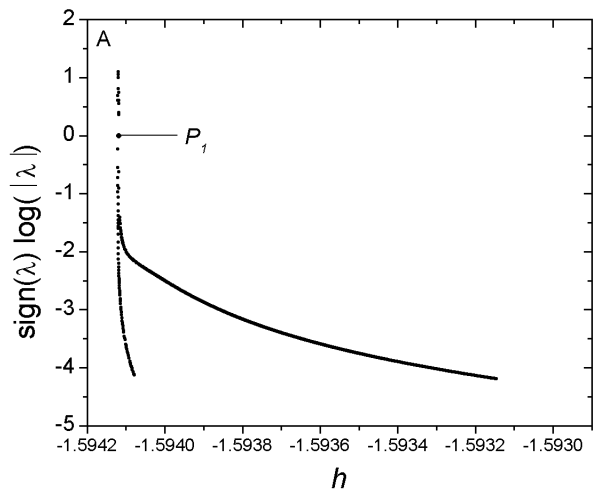
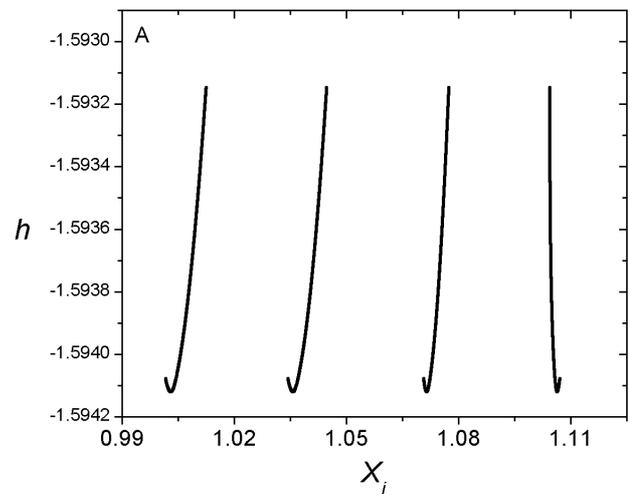

## Family 286 B - Symmetric family of asymmetric POs

$h_{min} = -1.594119$,  $h_{max} = -1.593532$,  $T_{min} = 26.022460$,  $T_{max} = 26.224898$

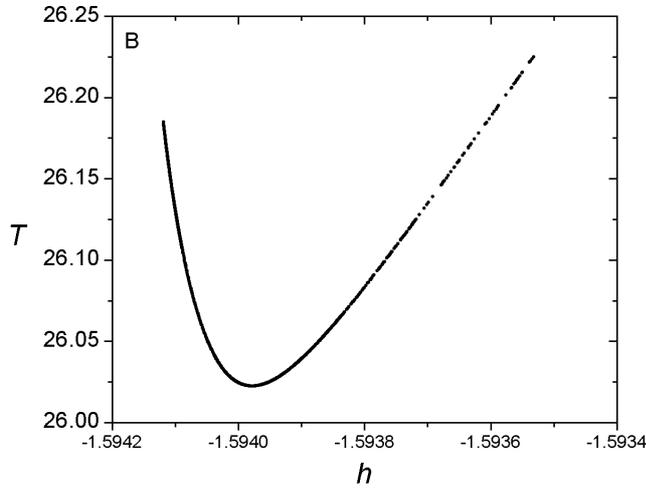

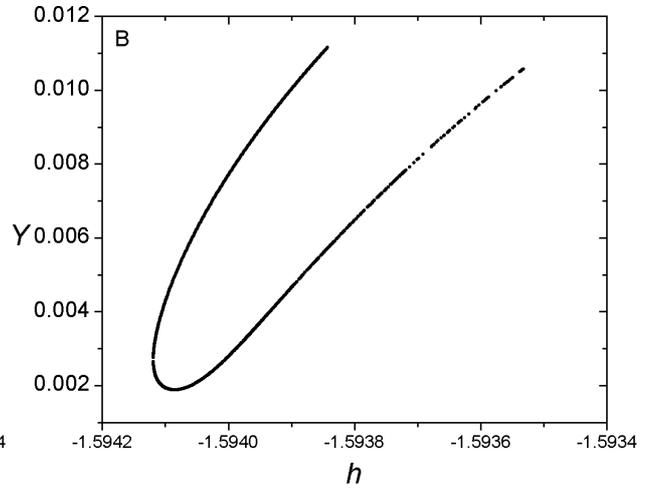

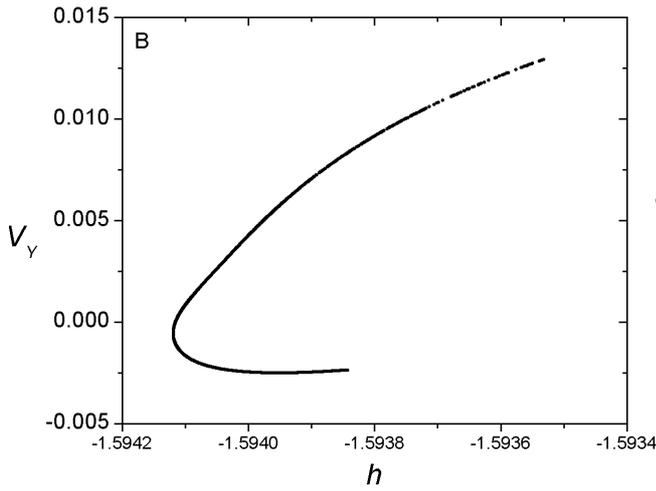

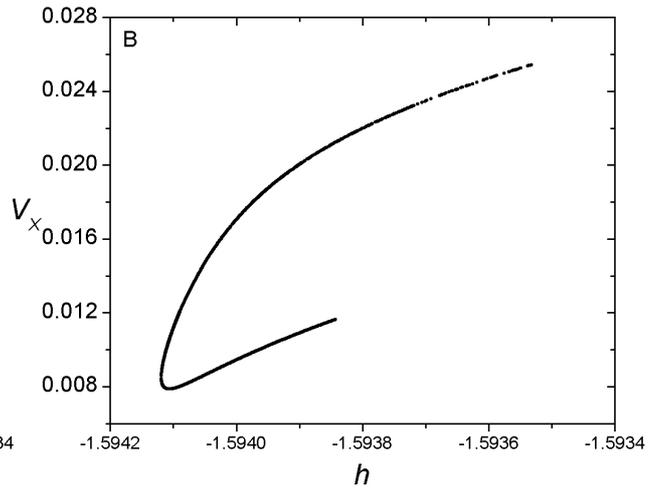

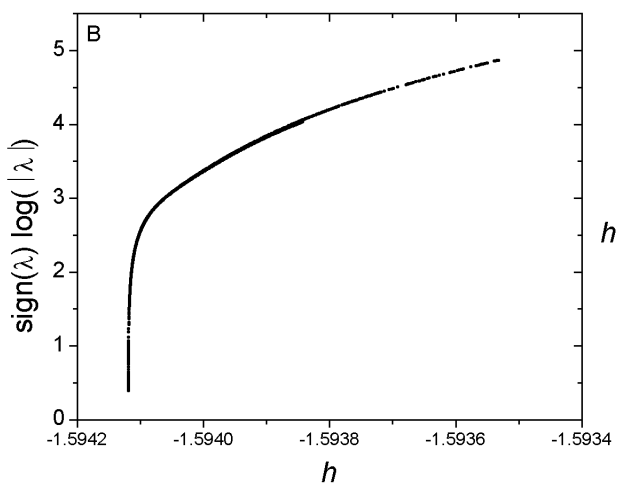

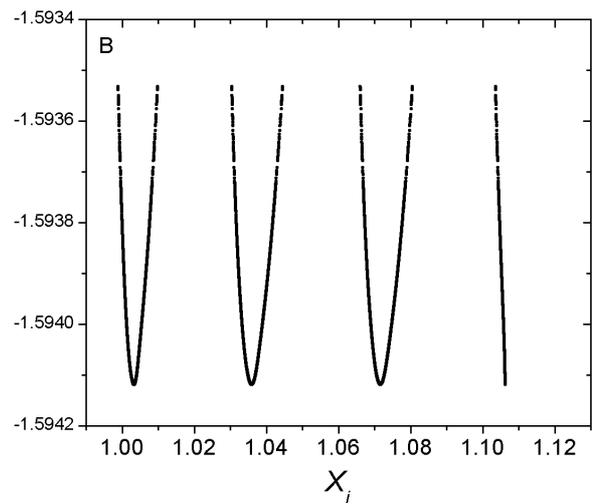

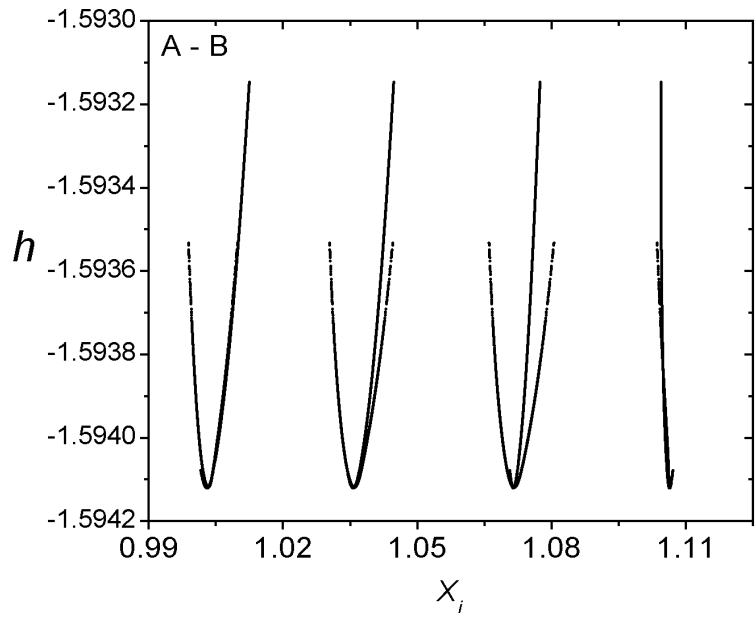

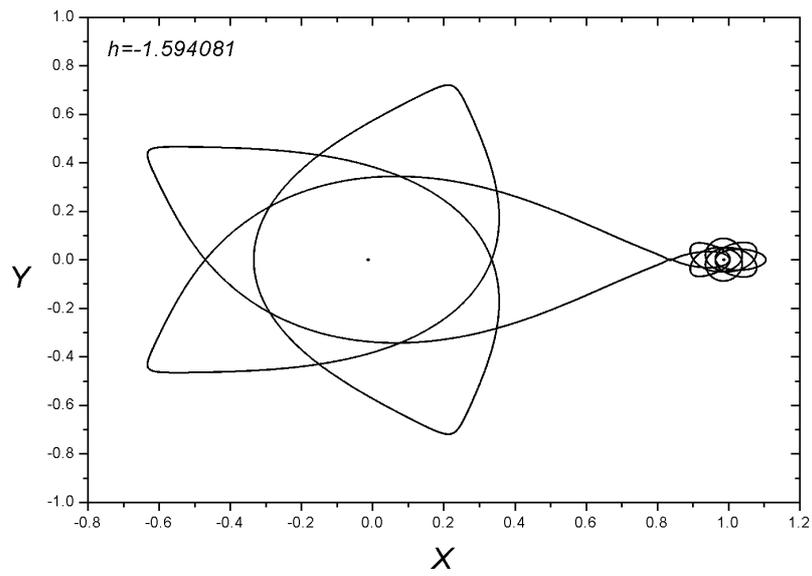

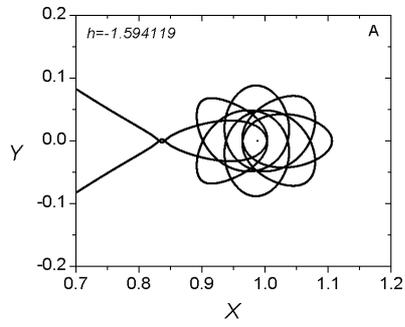
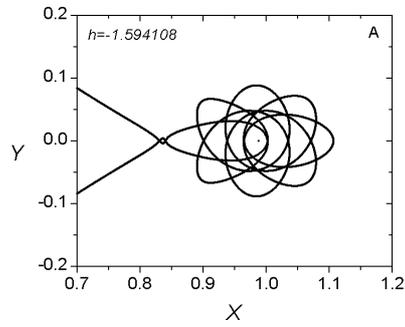
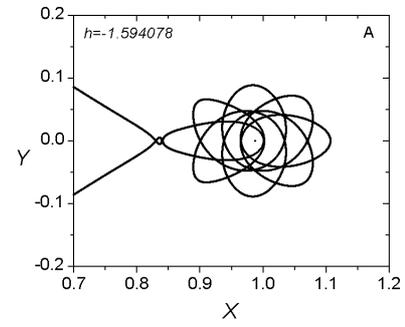

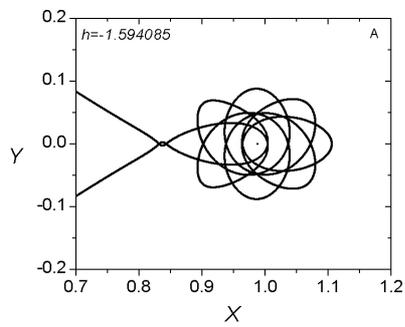
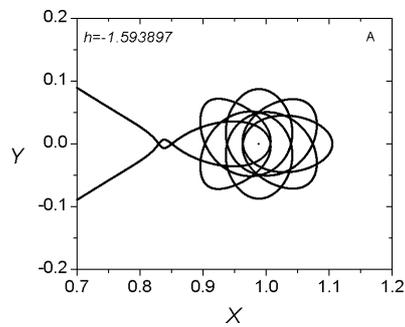
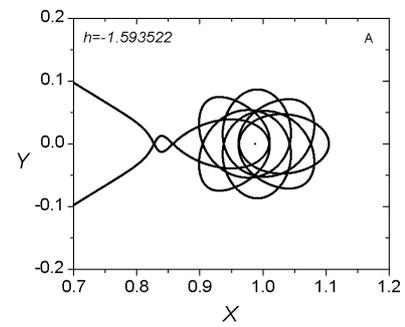

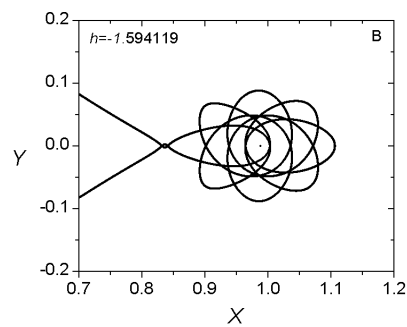
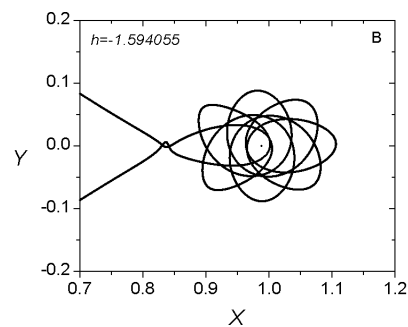
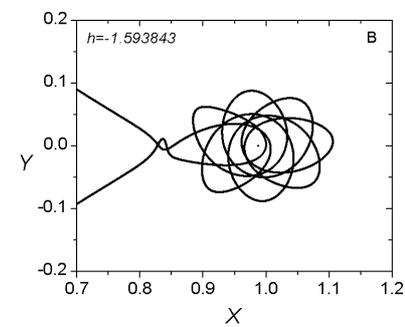

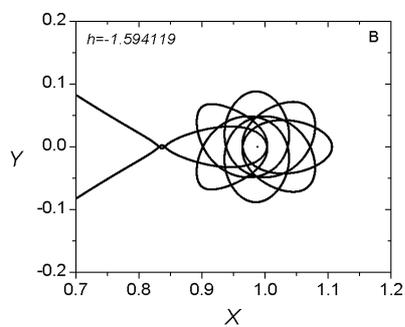
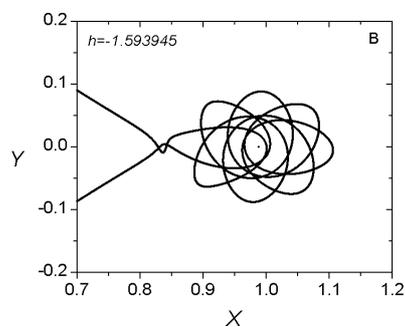
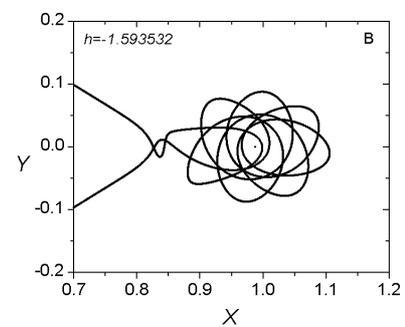

## Family 171 - Symmetric family of symmetric POs

$h_{min} = -1.592765$,  $h_{max} = -1.587517$,  $T_{min} = 26.459669$,  $T_{max} = 28.347150$

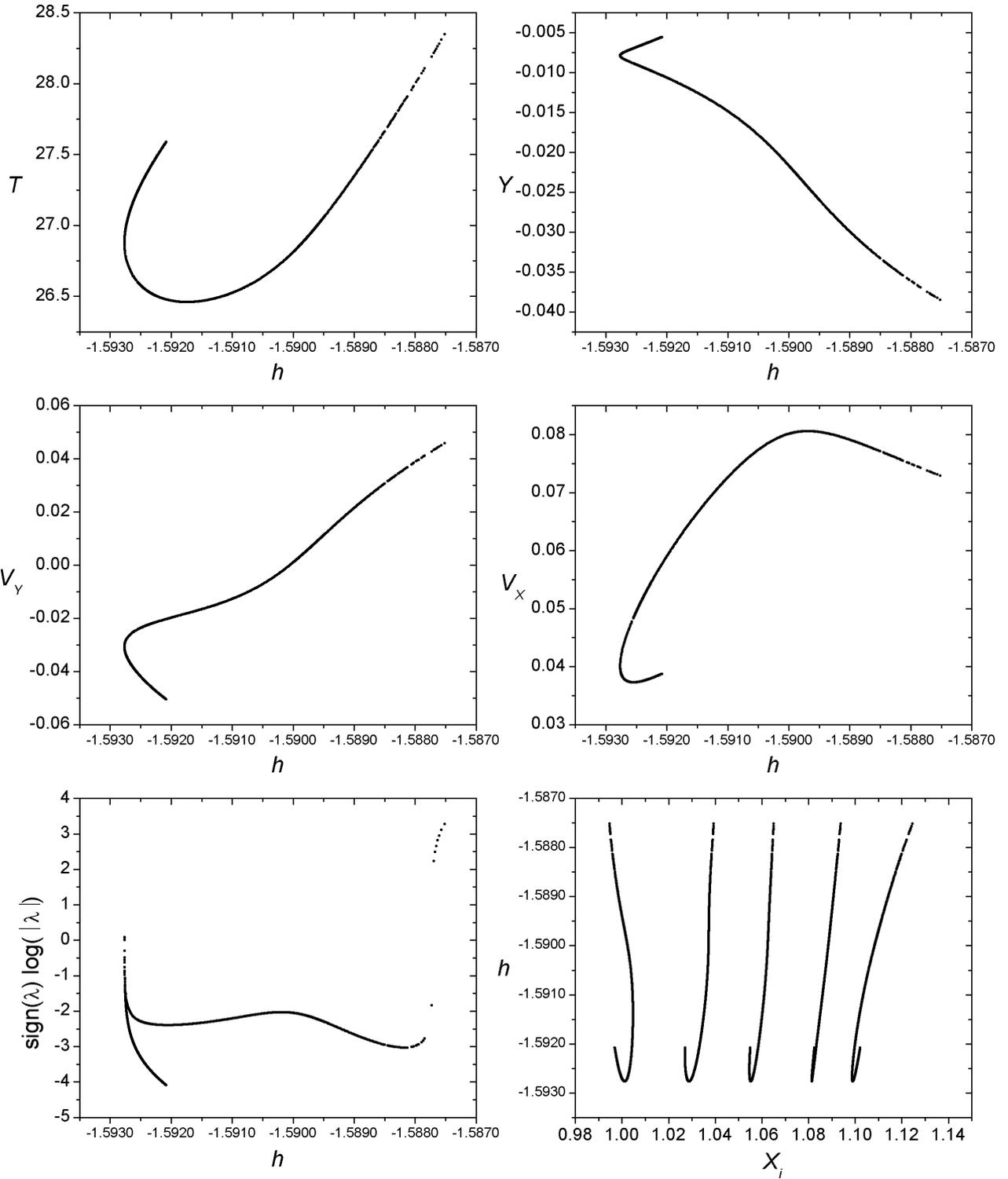

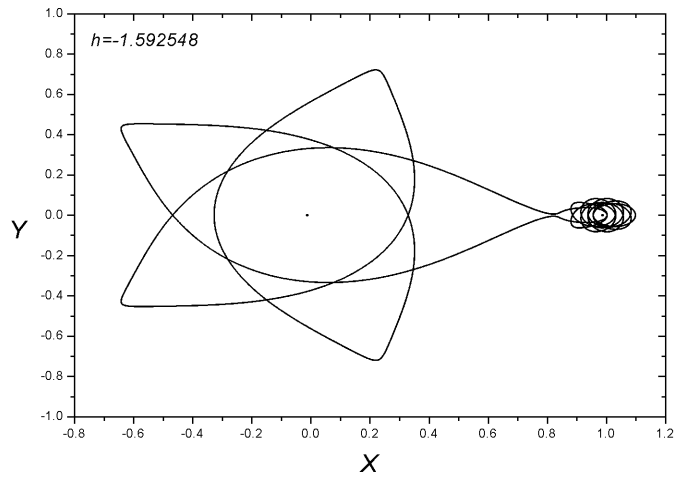

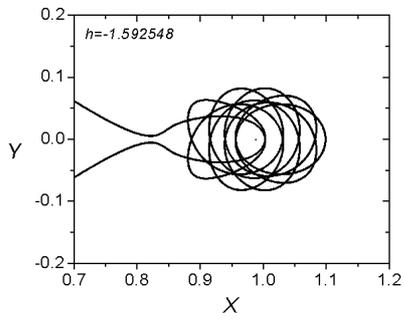
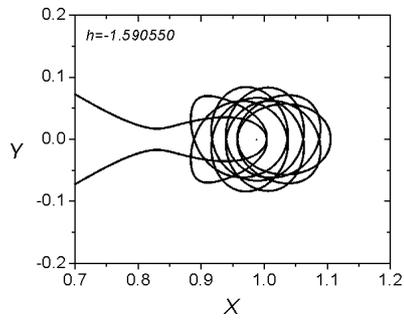
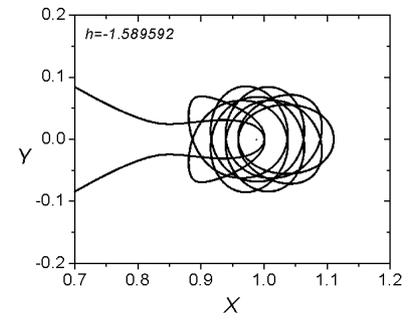

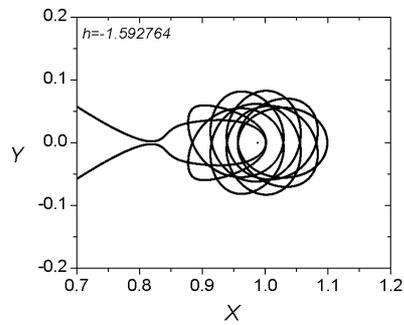
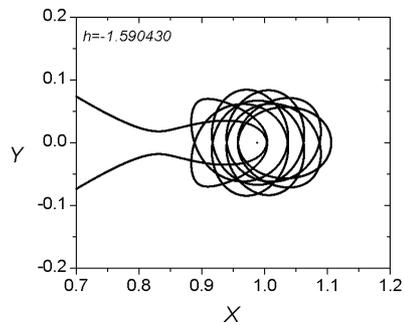
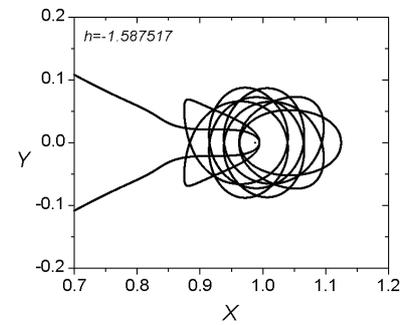

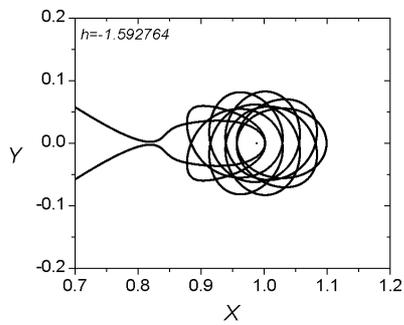
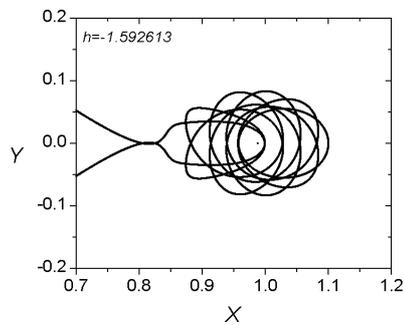
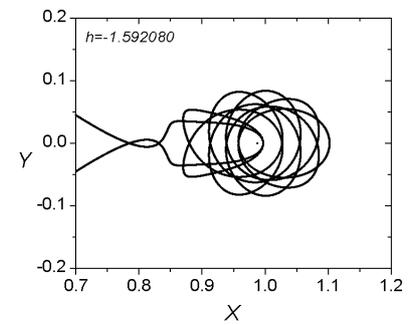

## Family 020 - *Asymmetric family of asymmetric POs*

$h_{min} = -1.587128$, $h_{max} = -1.582154$, $T_{min} = 26.483357$, $T_{max} = 28.588995$

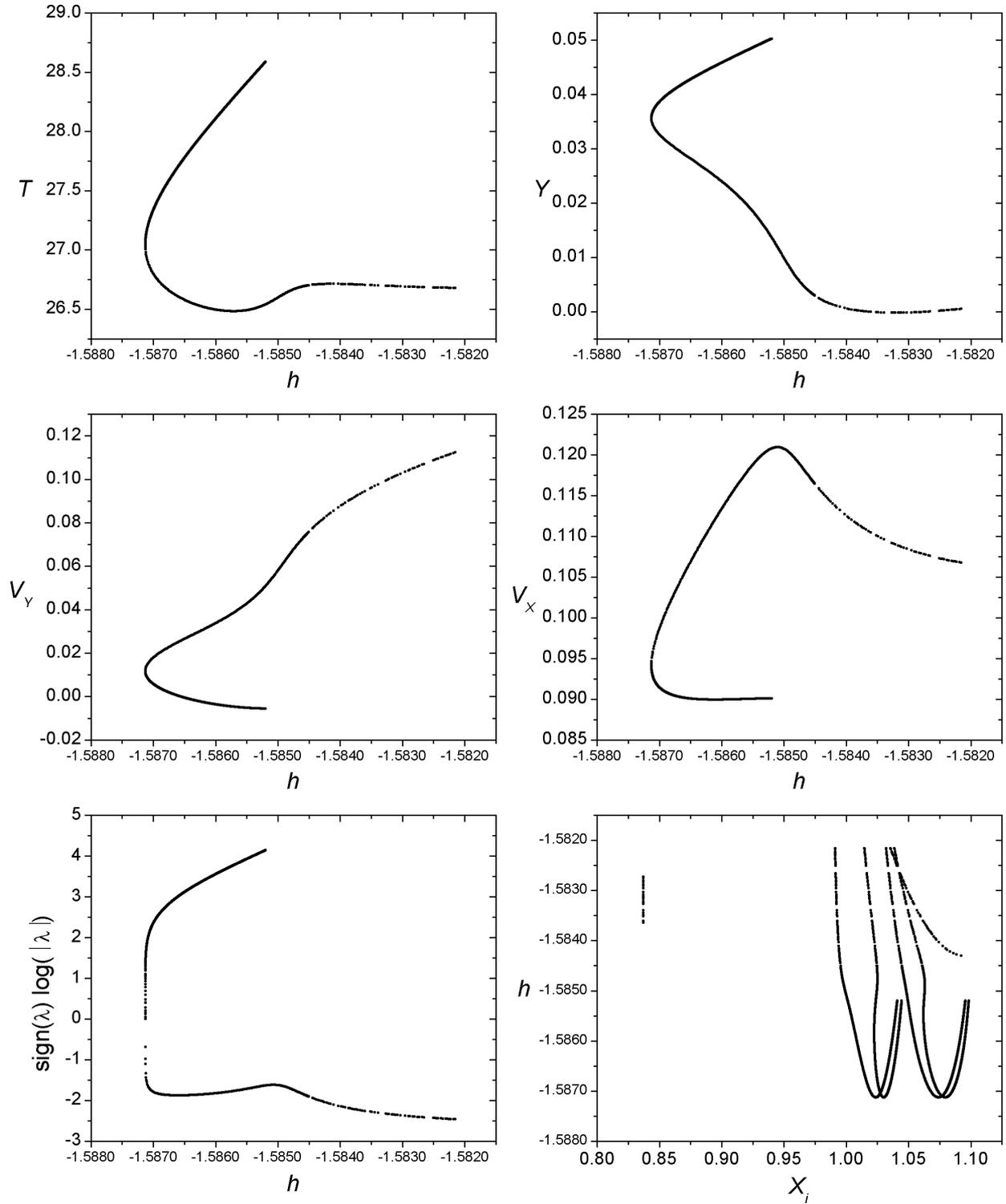

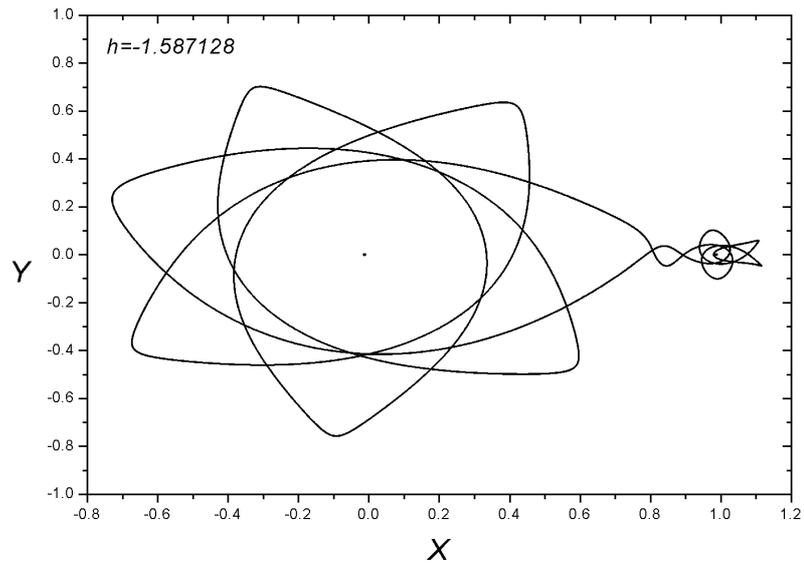

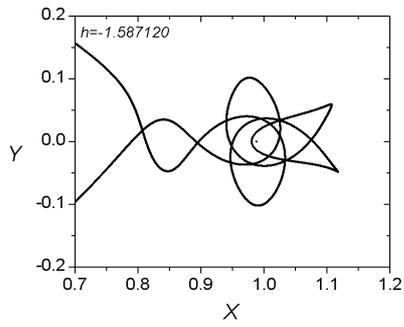
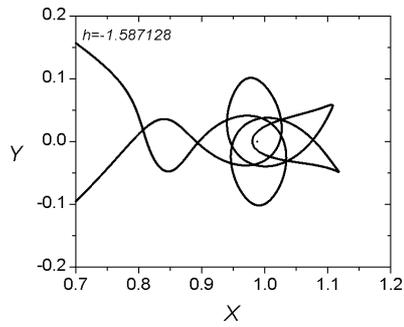
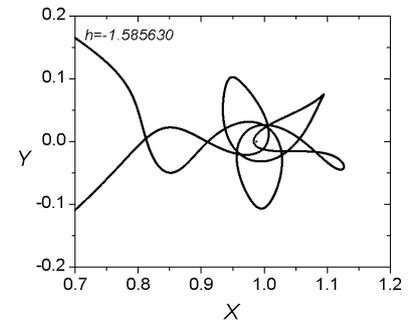

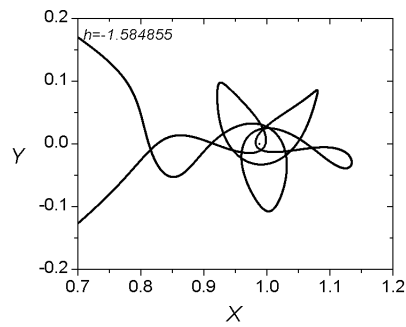
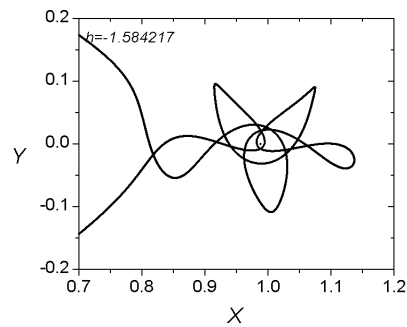
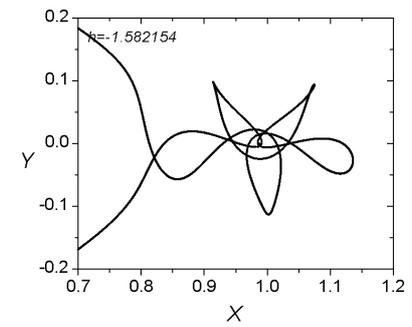

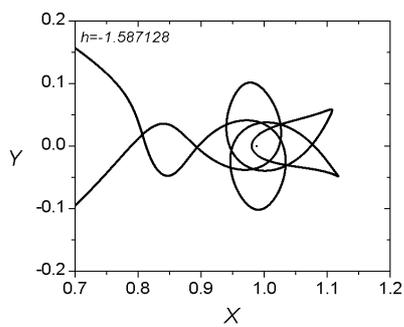
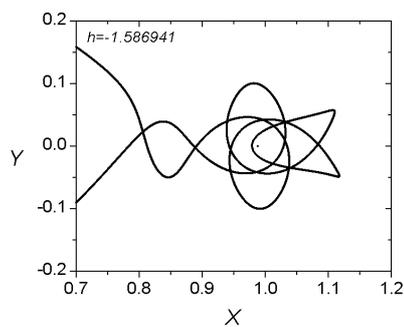
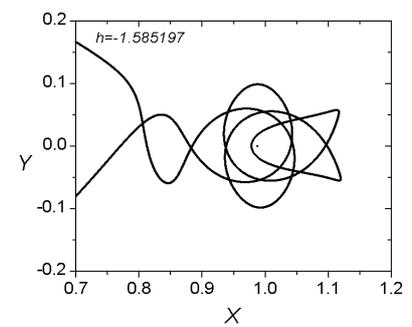

## Family 021 - *Asymmetric family of asymmetric POs*

$h_{min} = -1.587128, \ h_{max} = -1.578773, \ T_{min} = 26.483355, \ T_{max} = 28.485516$

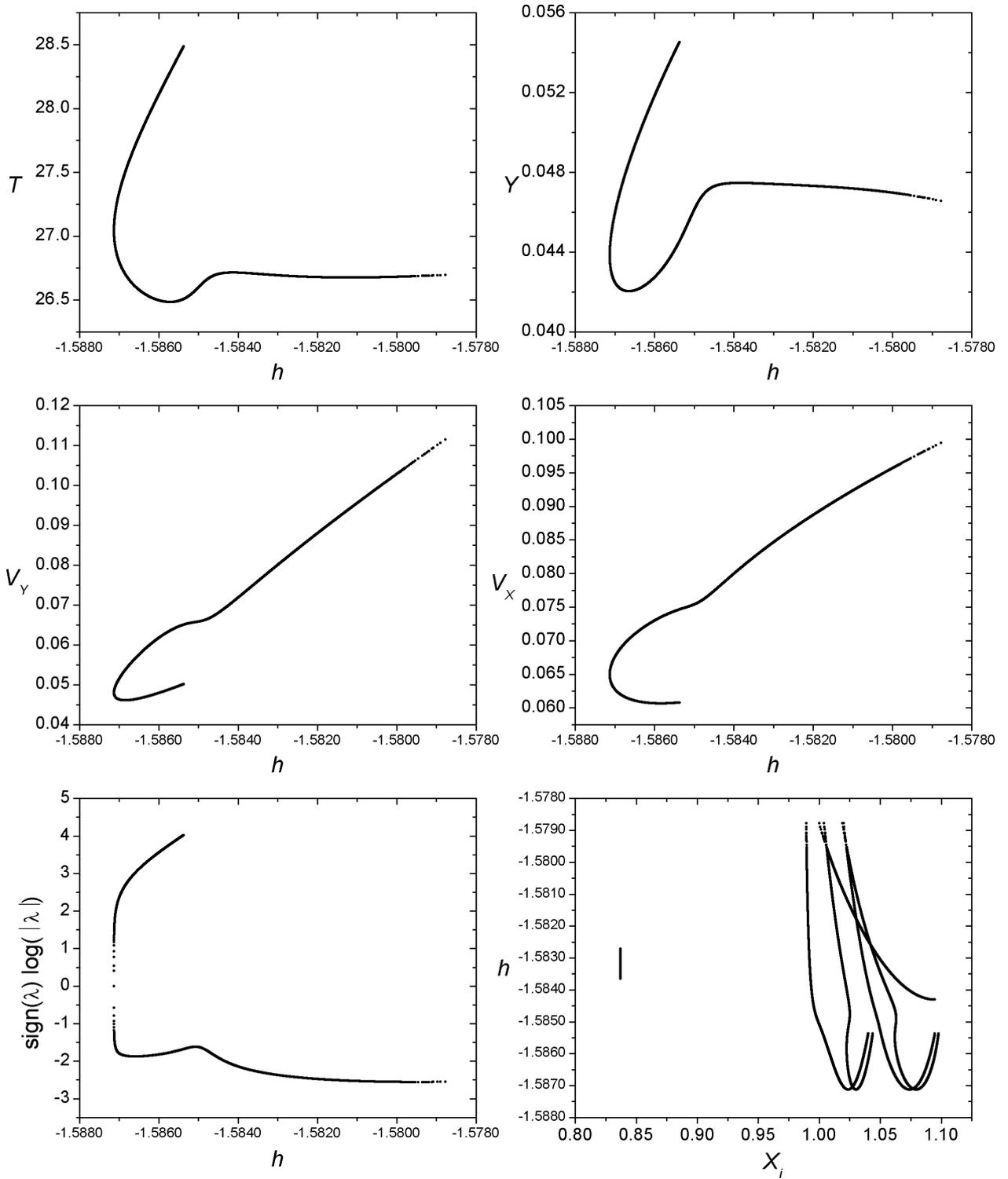

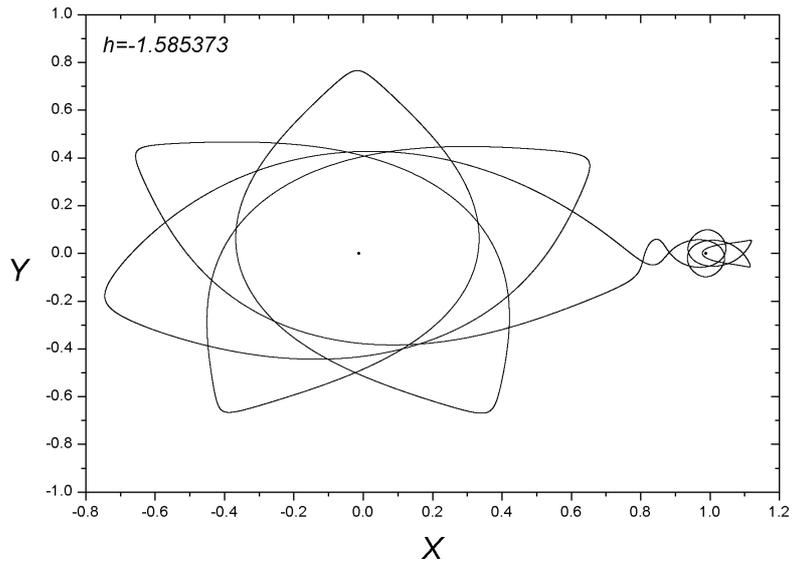

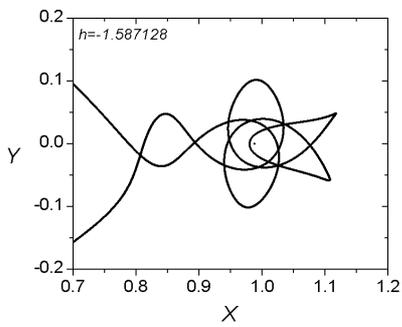
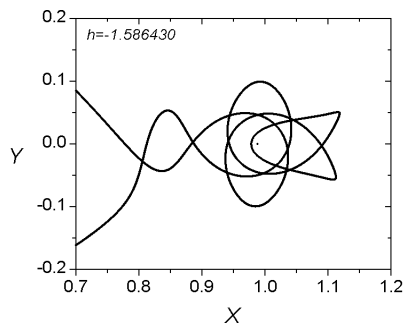
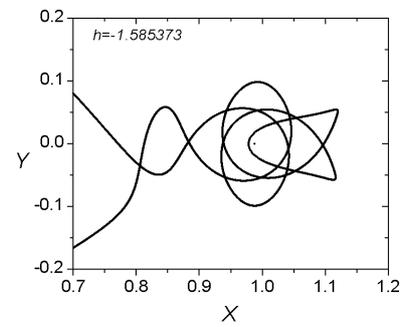

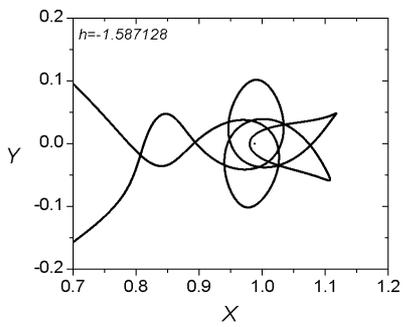
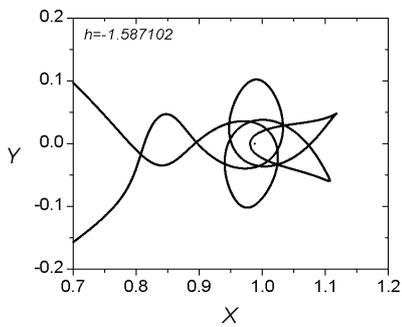
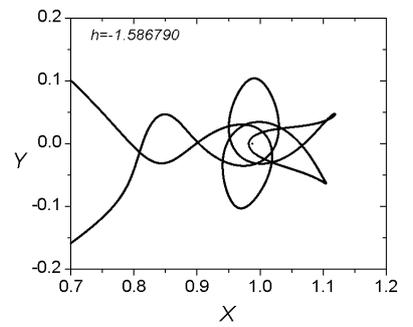

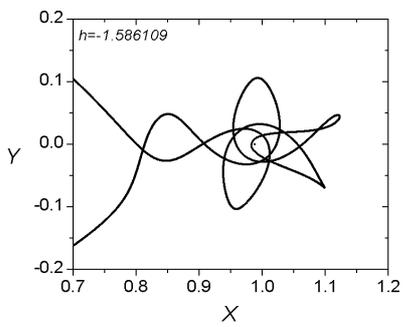
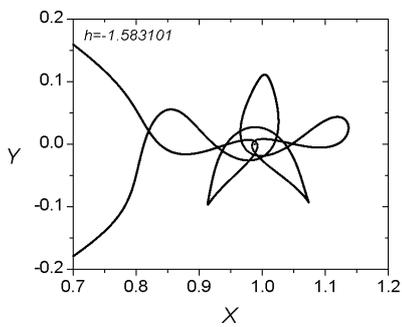
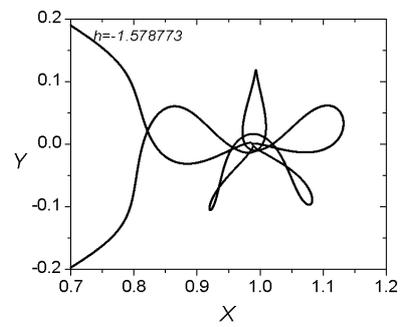

## Family 013 - *Symmetric family of symmetric POs*

$h_{min} = -1.587468$, $h_{max} = -1.583251$, $T_{min} = 26.574839$, $T_{max} = 28.904140$

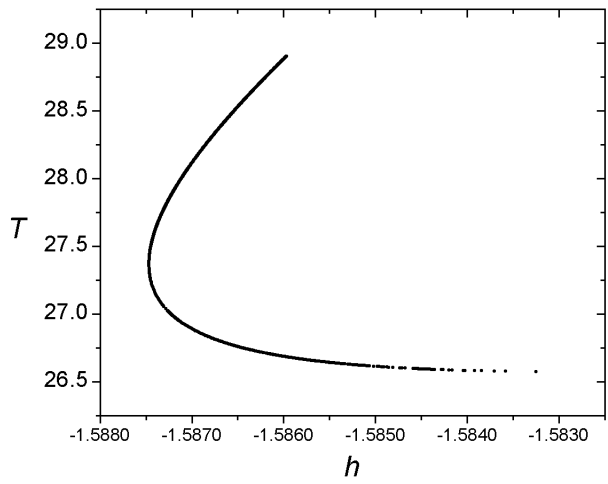
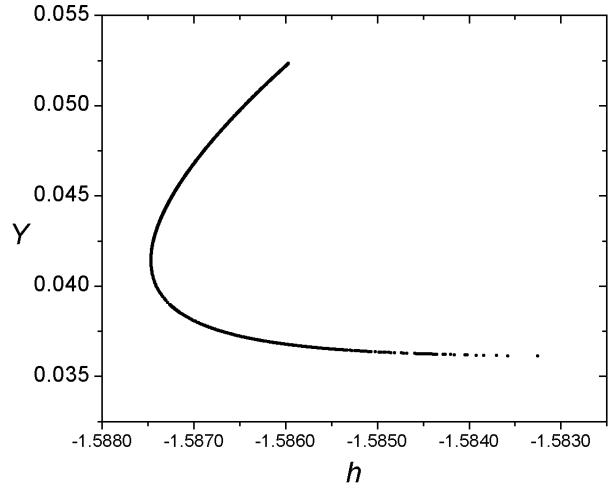

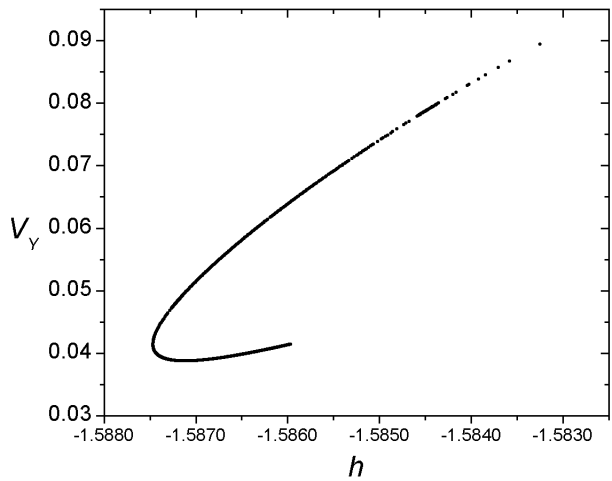
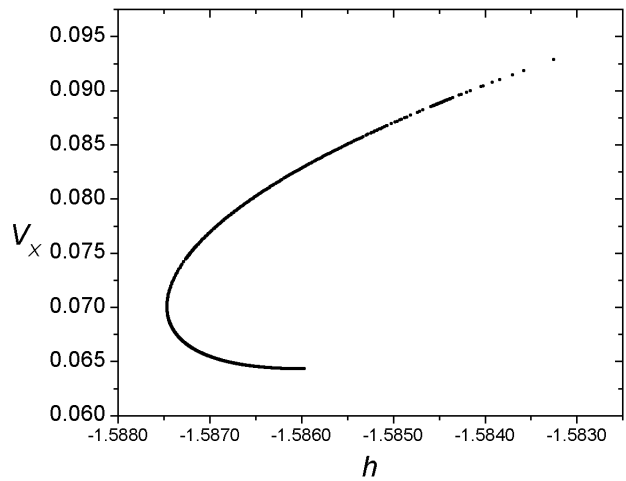

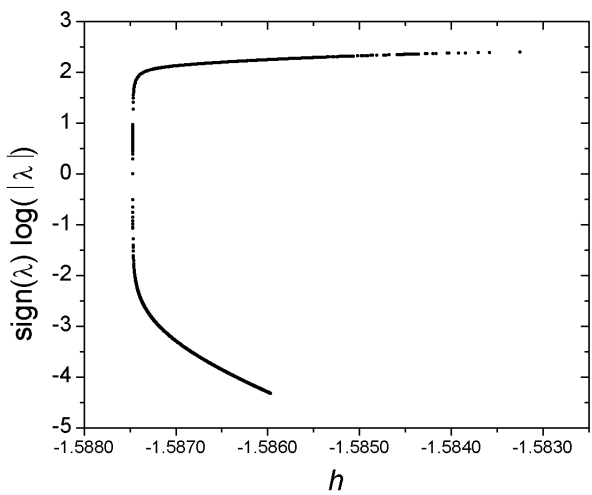
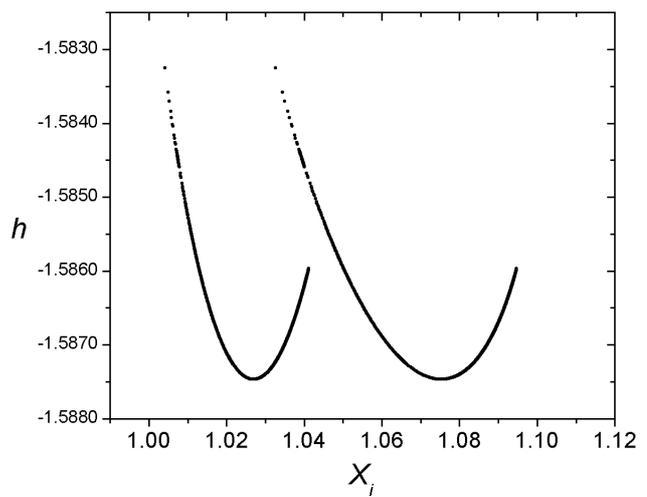

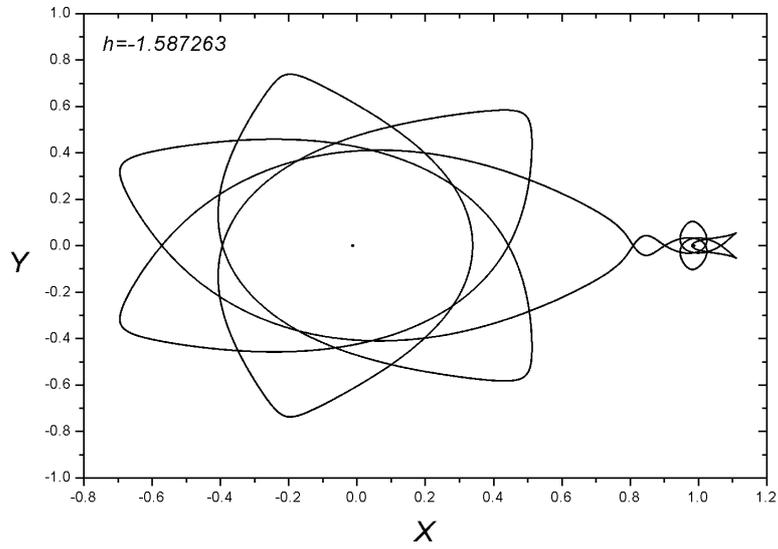

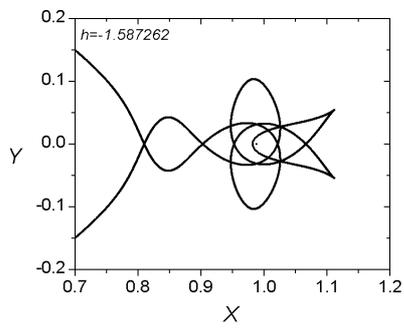
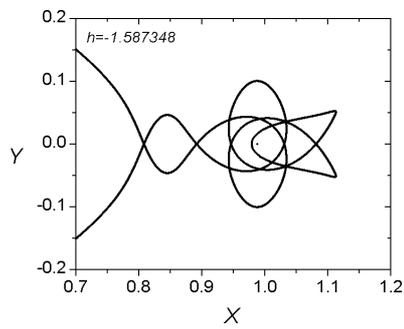
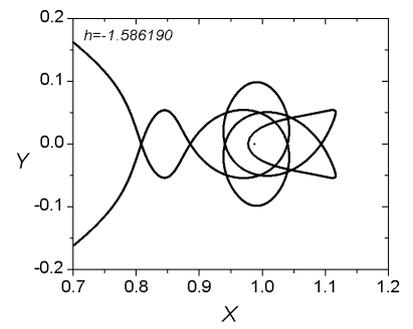

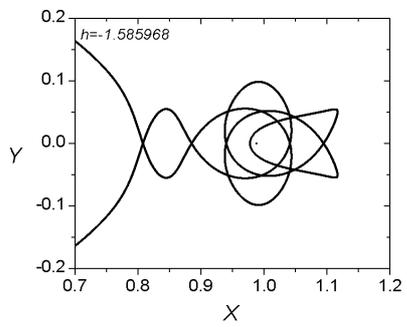
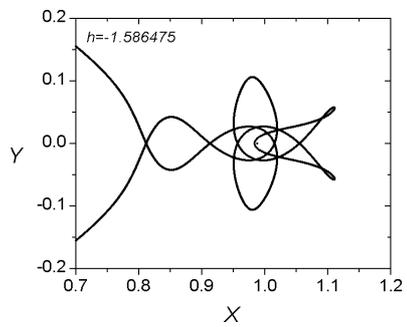
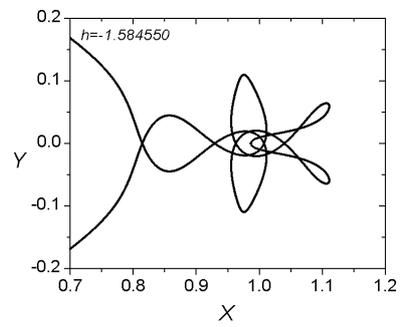

## Families 222 A - 222 B

Bifurcation Point

|       | h         | T         | y         | $v_y$     | $v_x$    |
|-------|-----------|-----------|-----------|-----------|----------|
| $P_1$ | −1.592598 | 28.271856 | -0.007523 | -0.033981 | 0.041892 |

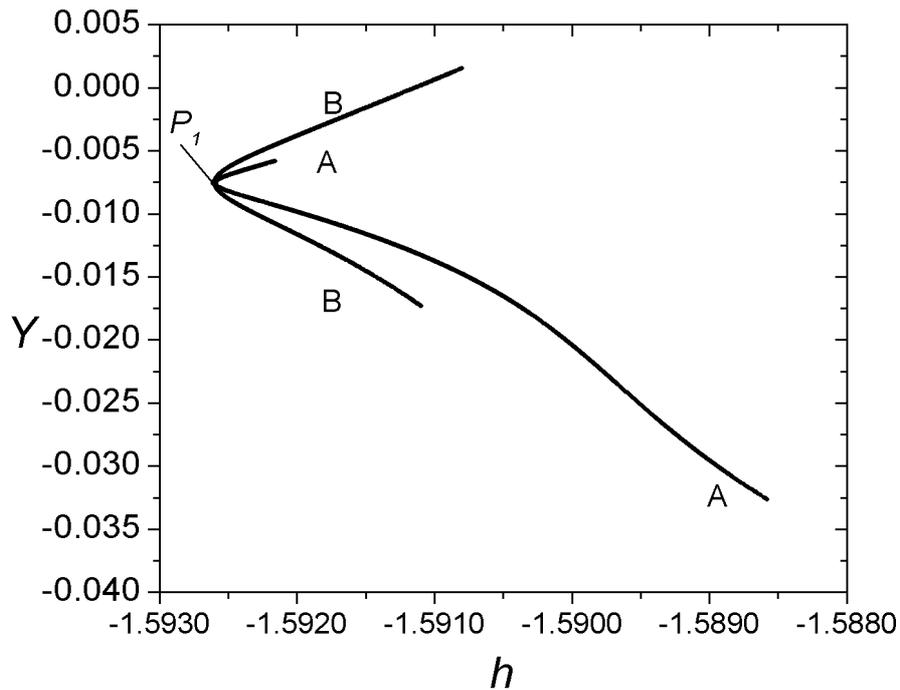

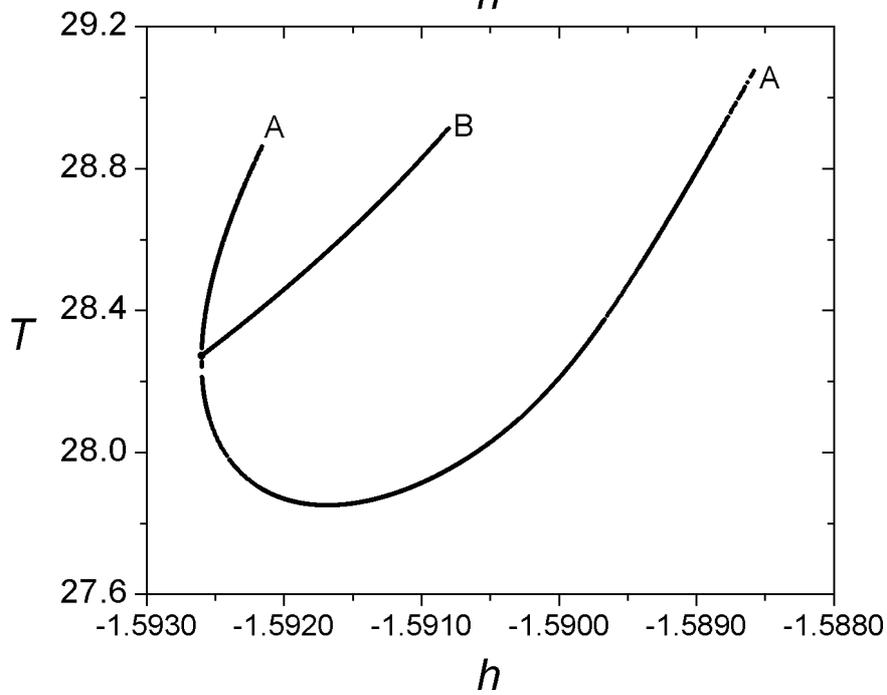

### Family 222 A - Symmetric family of symmetric POs

$h_{min} = -1.592599, \ h_{max} = -1.588578, \ T_{min} = 27.850288, \ T_{max} = 29.075278$

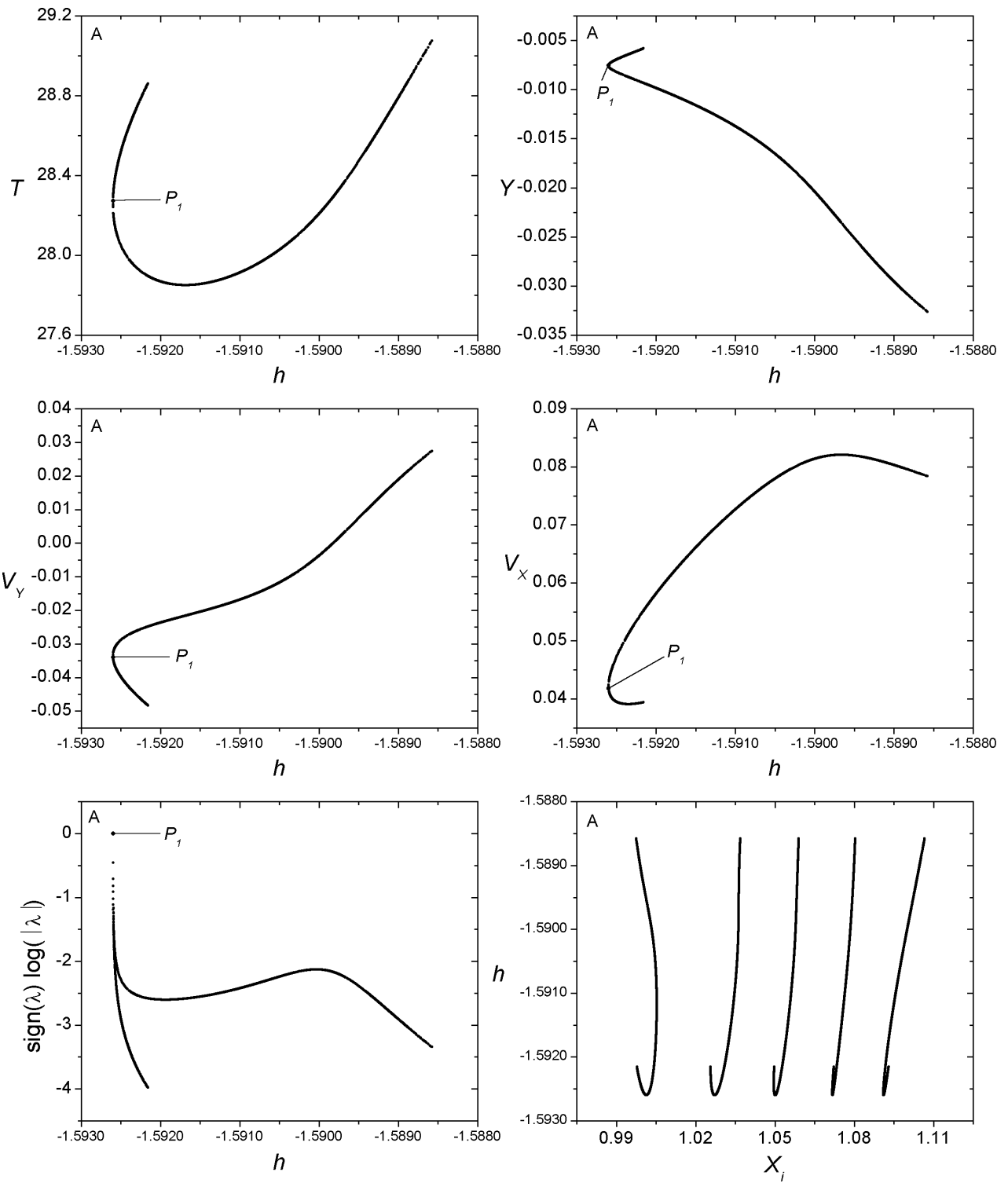

# Family 222 B - Symmetric family of asymmetric POs

$h_{min} = -1.592598$, $h_{max} = -1.590801$, $T_{min} = 28.271856$, $T_{max} = 28.913276$

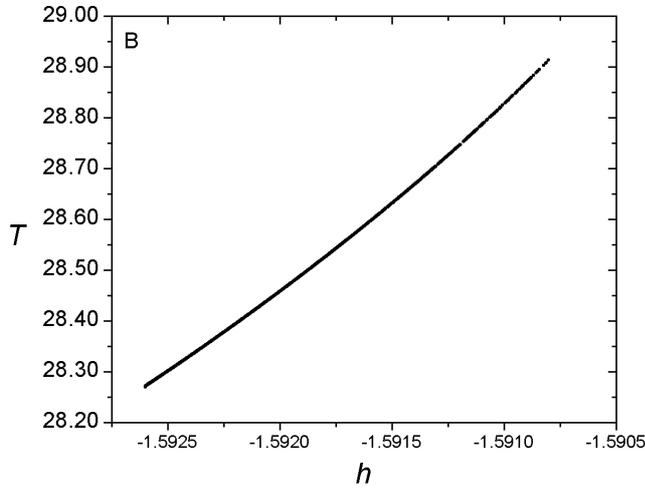
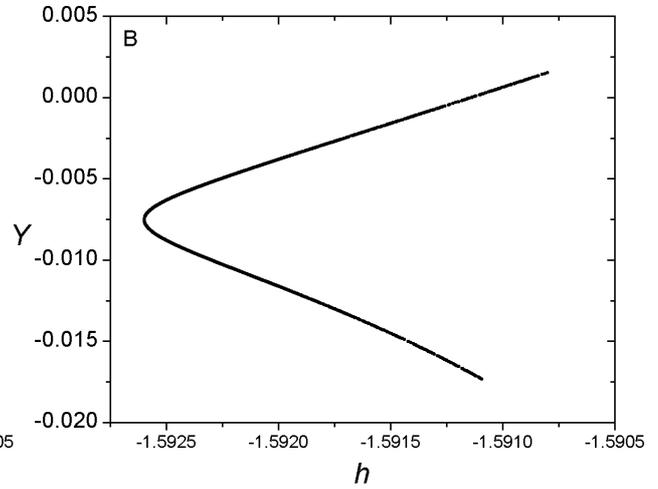

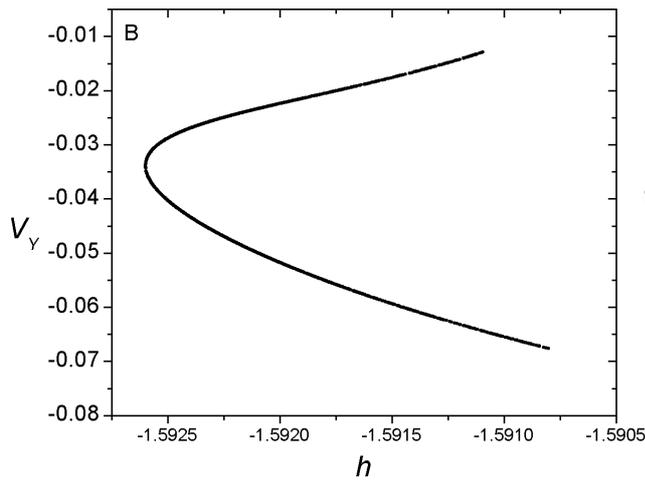
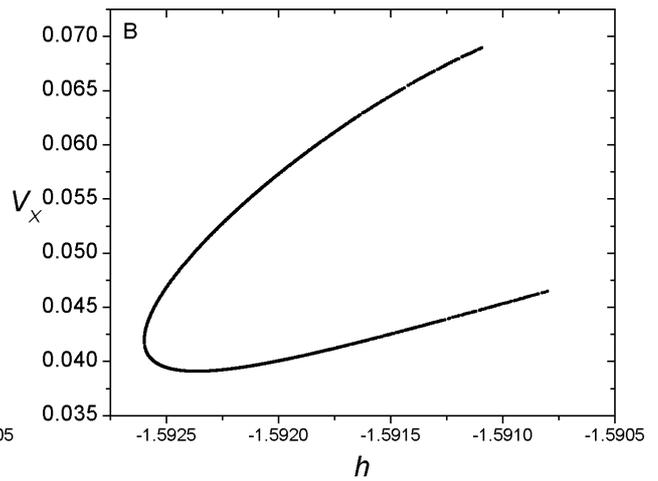

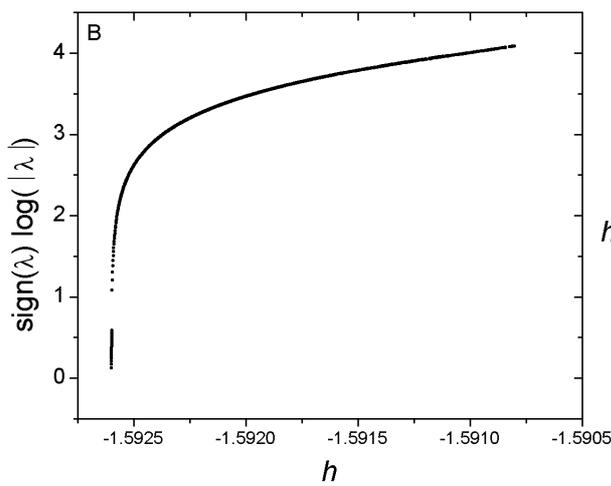
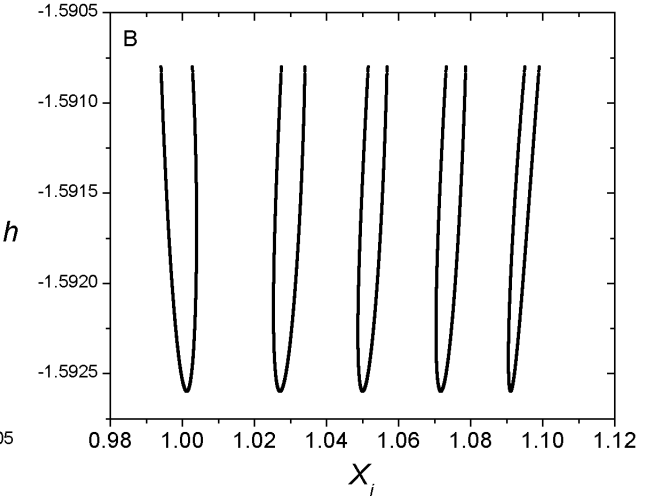

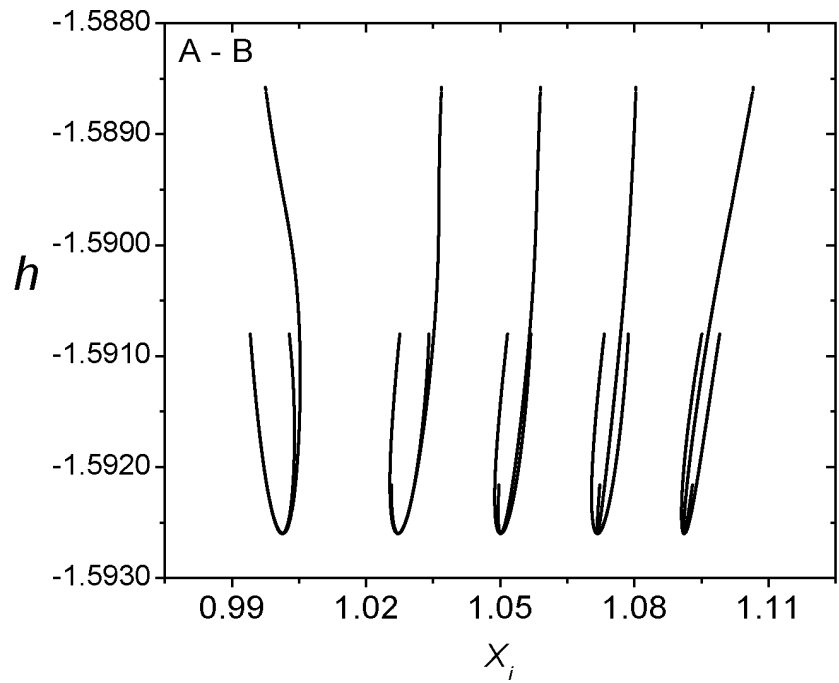

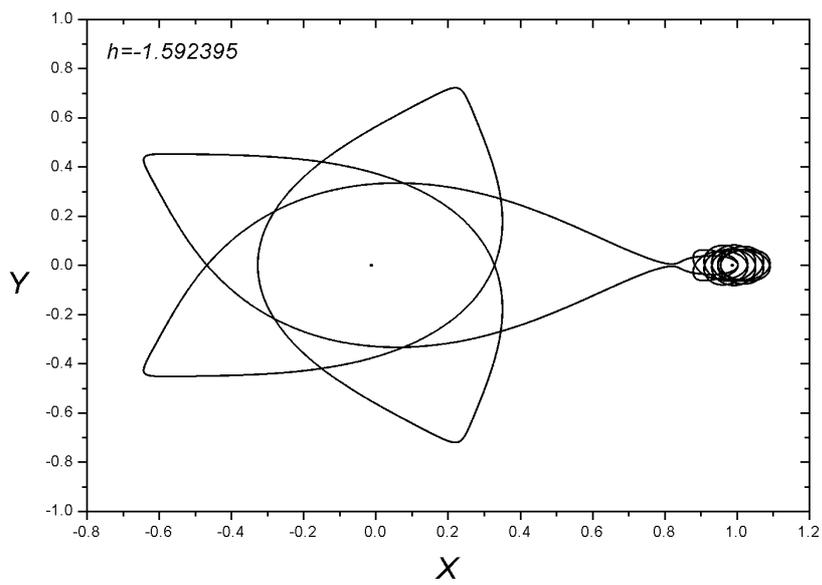

## Families 081 A - 081 B - 081 C

*Bifurcation Points*

|       | $h$       | $T$        | $y$       | $v_y$      | $v_x$     |
|-------|-----------|------------|-----------|------------|-----------|
| $P_1$ | −1.587073 | 28.274740  | 0.034129  | -0.027860  | 0.093490  |
| $P_2$ | -1.589428 | 28.643937  | 0.036479  | 0.016869   | 0.062296  |

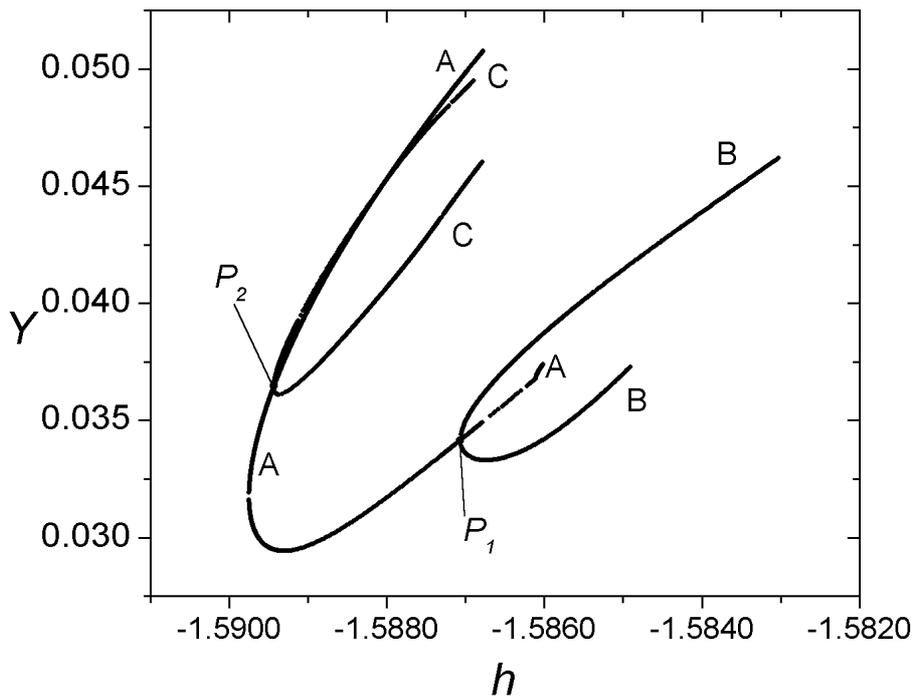

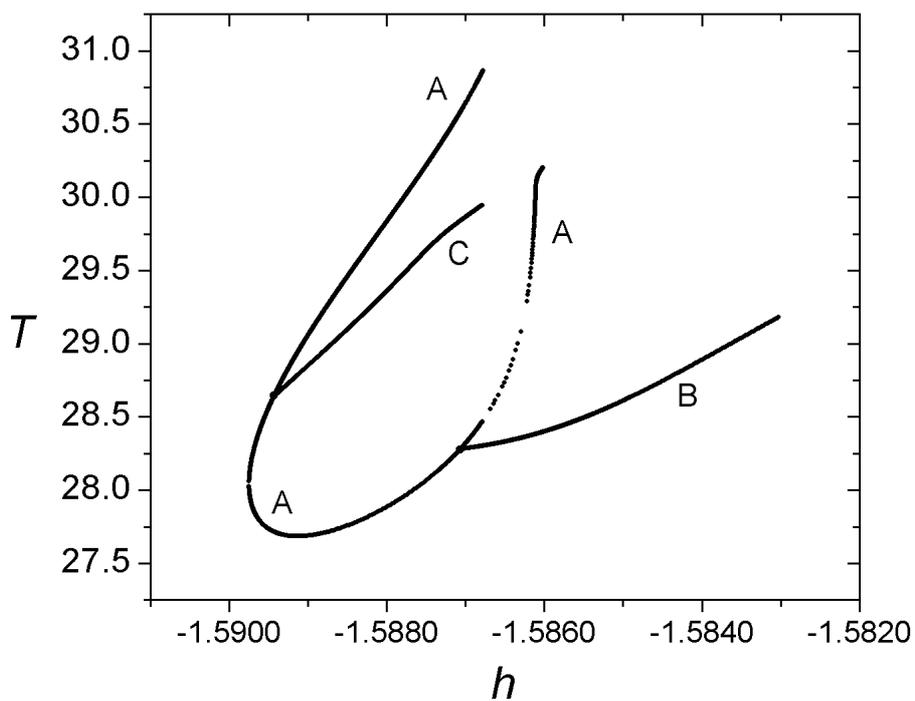

## *Family 081 A - Symmetric family of symmetric POs*

$h_{min} = -1.589745, \ h_{max} = -1.586016, \ T_{min} = 27.684763, \ T_{max} = 30.866992$

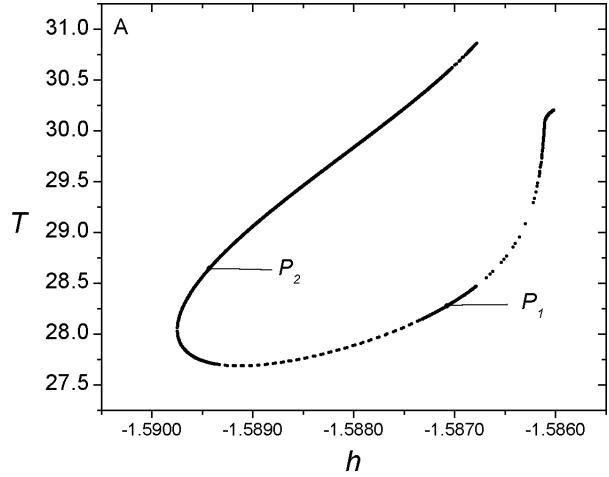

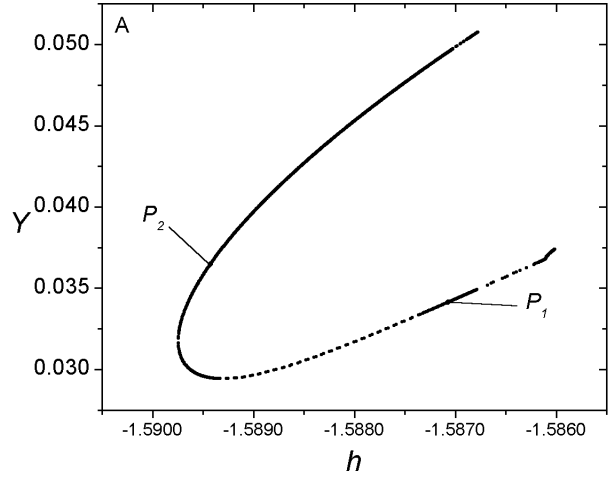

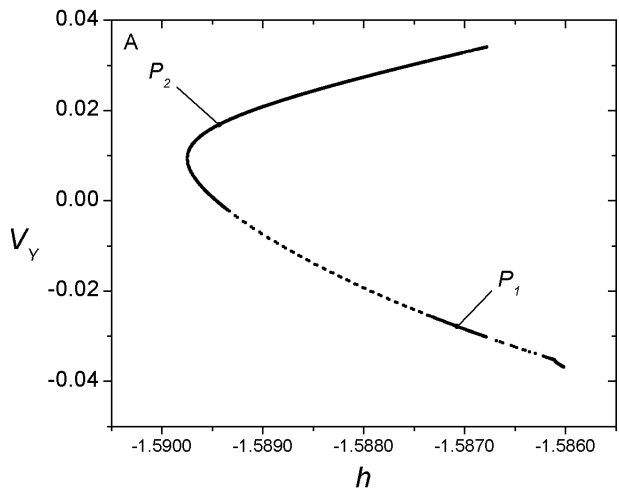

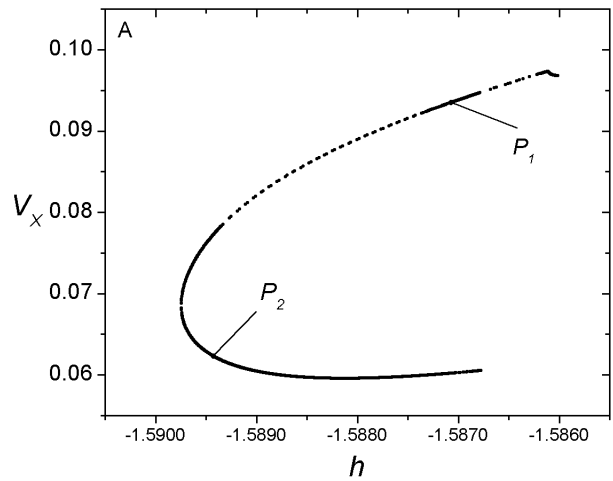

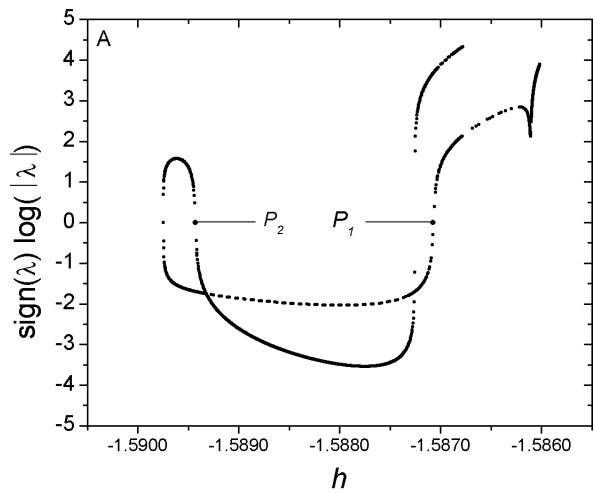

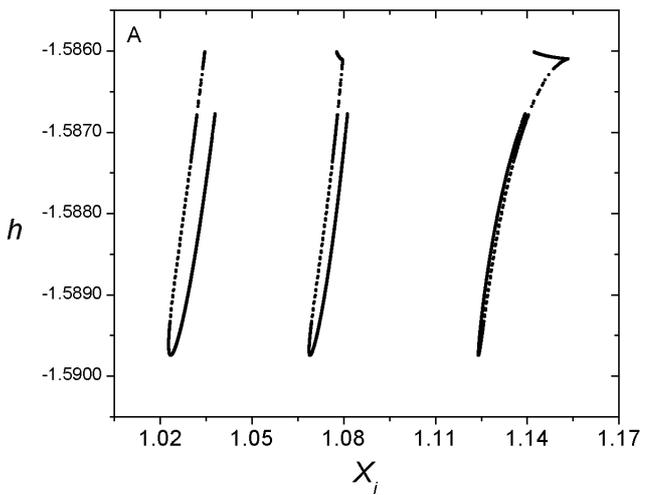

## Family 081 B - Symmetric family of asymmetric POs

$h_{min} = -1.587073, \ h_{max} = -1.583023, \ T_{min} = 28.274740, \ T_{max} = 29.180289$

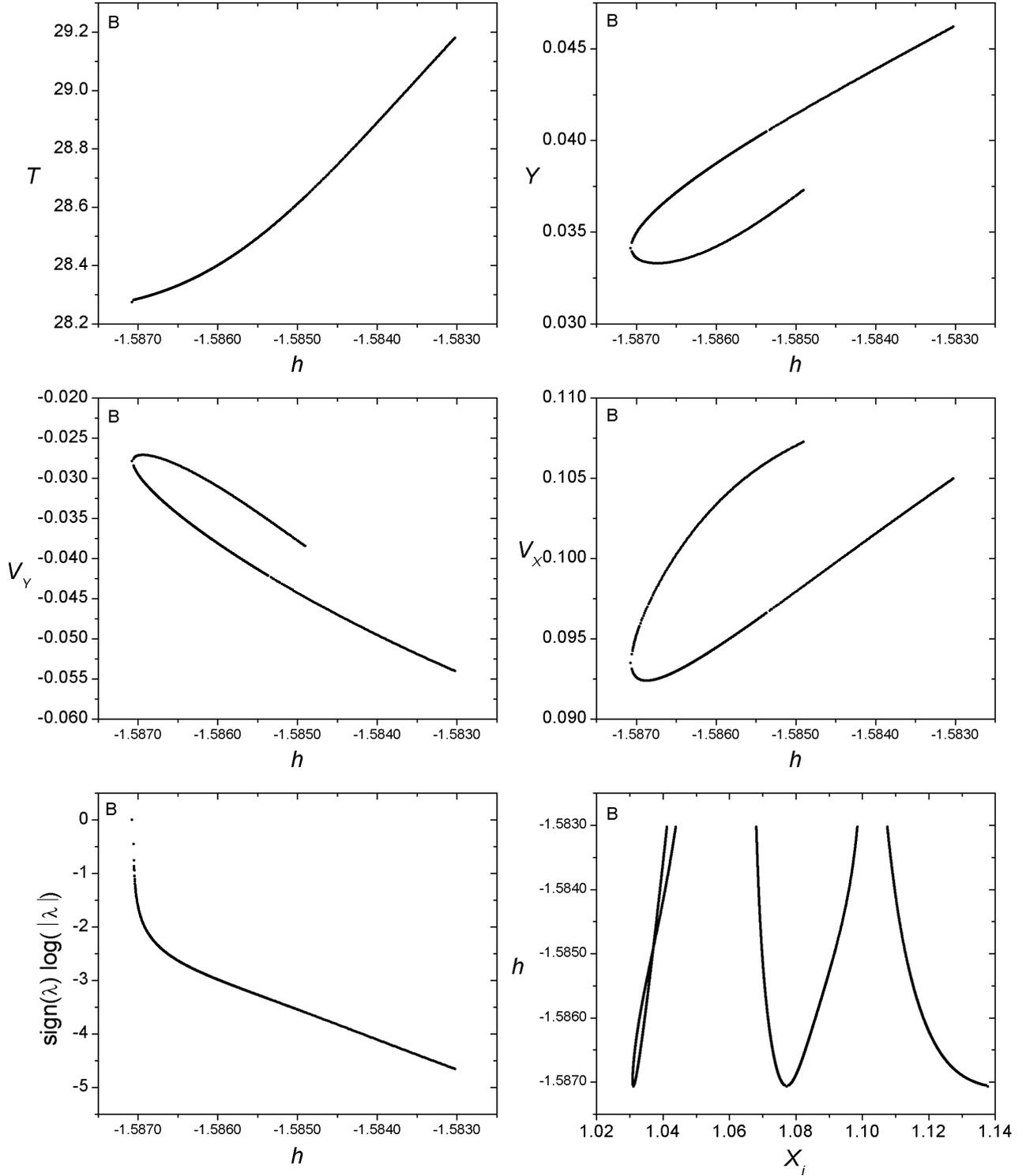

### *Family 081 C - Symmetric family of asymmetric POs*

$h_{min} = -1.589433, \quad h_{max} = -1.586790, \quad T_{min} = 28.637522, \quad T_{max} = 29.945159$

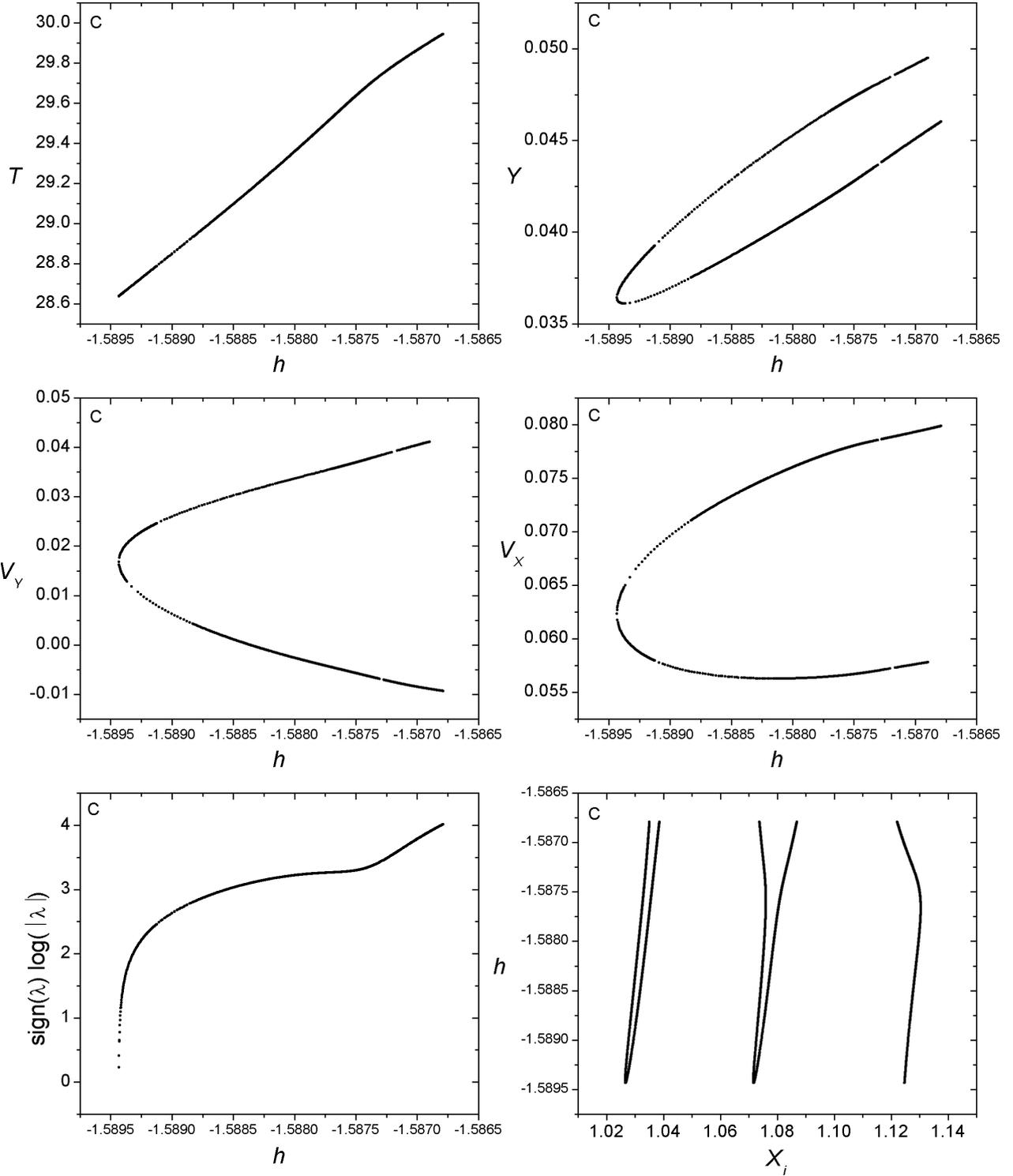

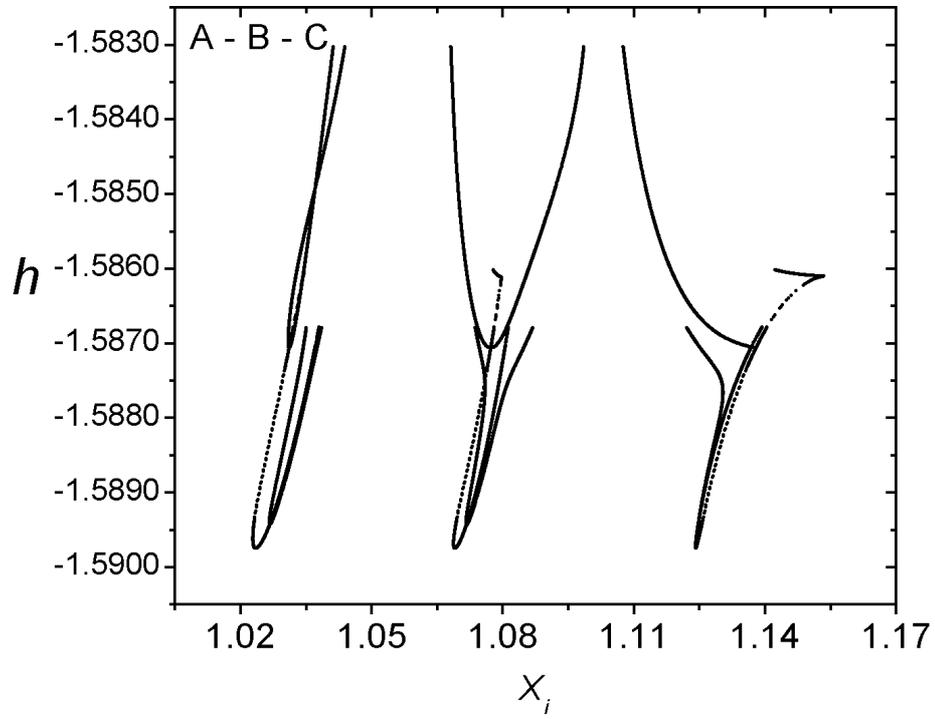

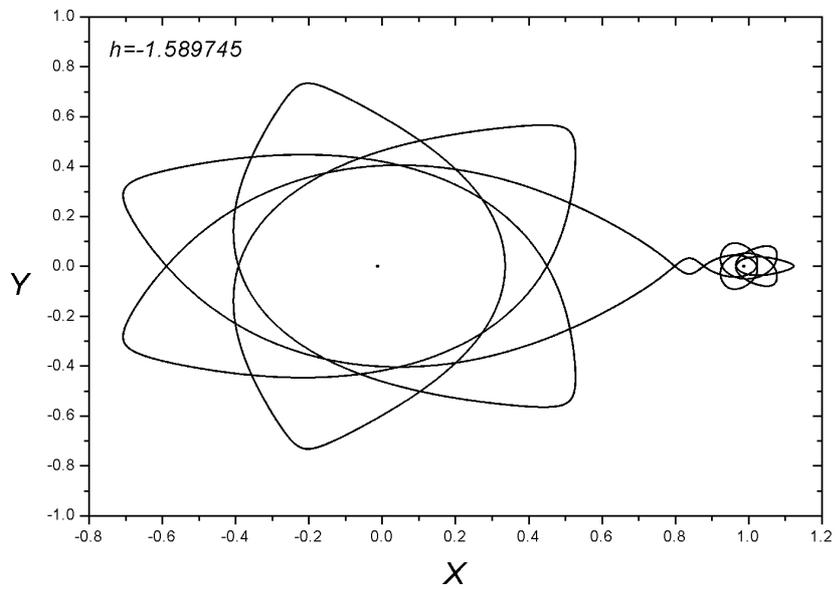

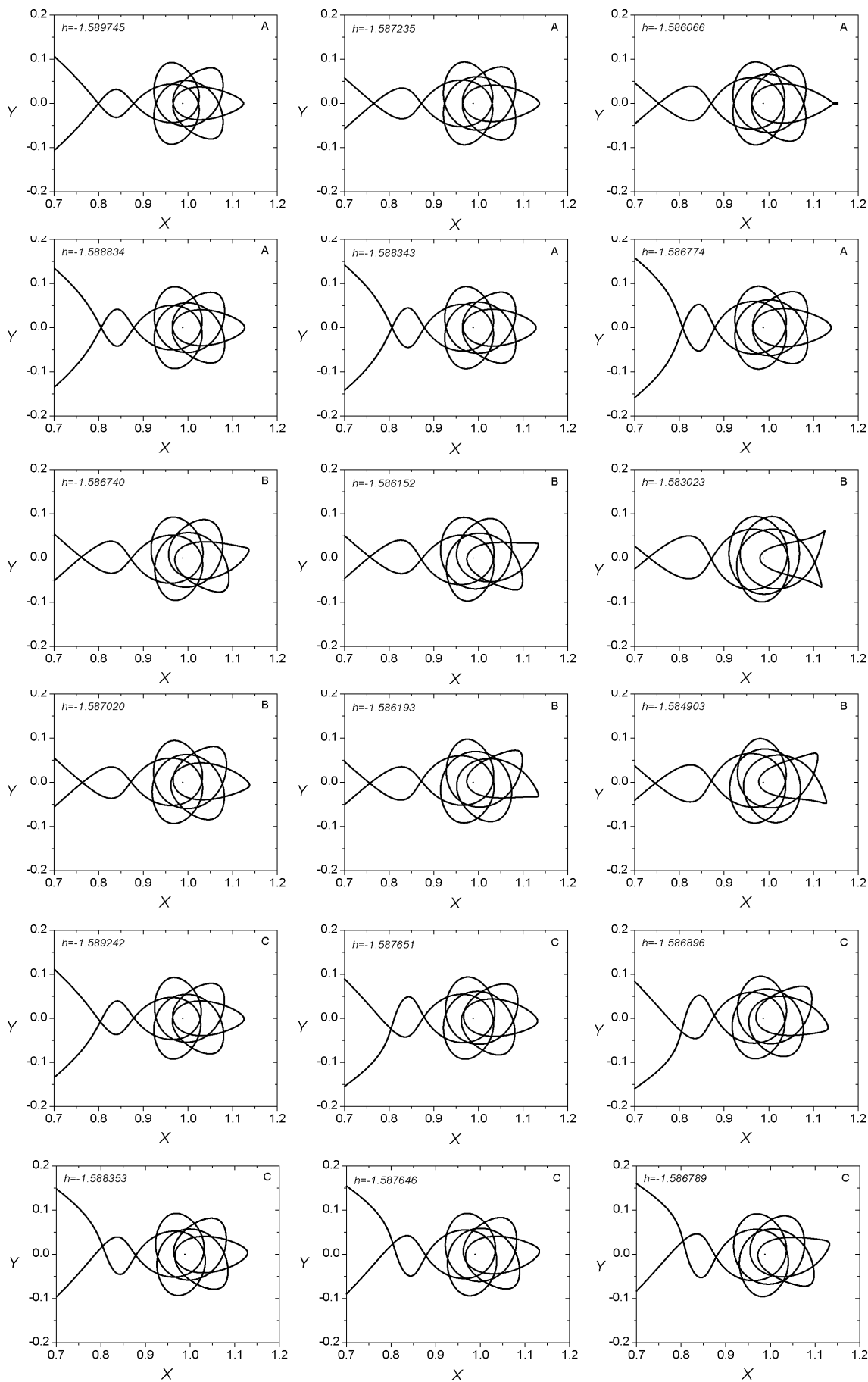

## Family 200 - *Symmetric family of symmetric POs*

$h_{min} = -1.592281$, $h_{max} = -1.589747$, $T_{min} = 28.827241$, $T_{max} = 29.843255$

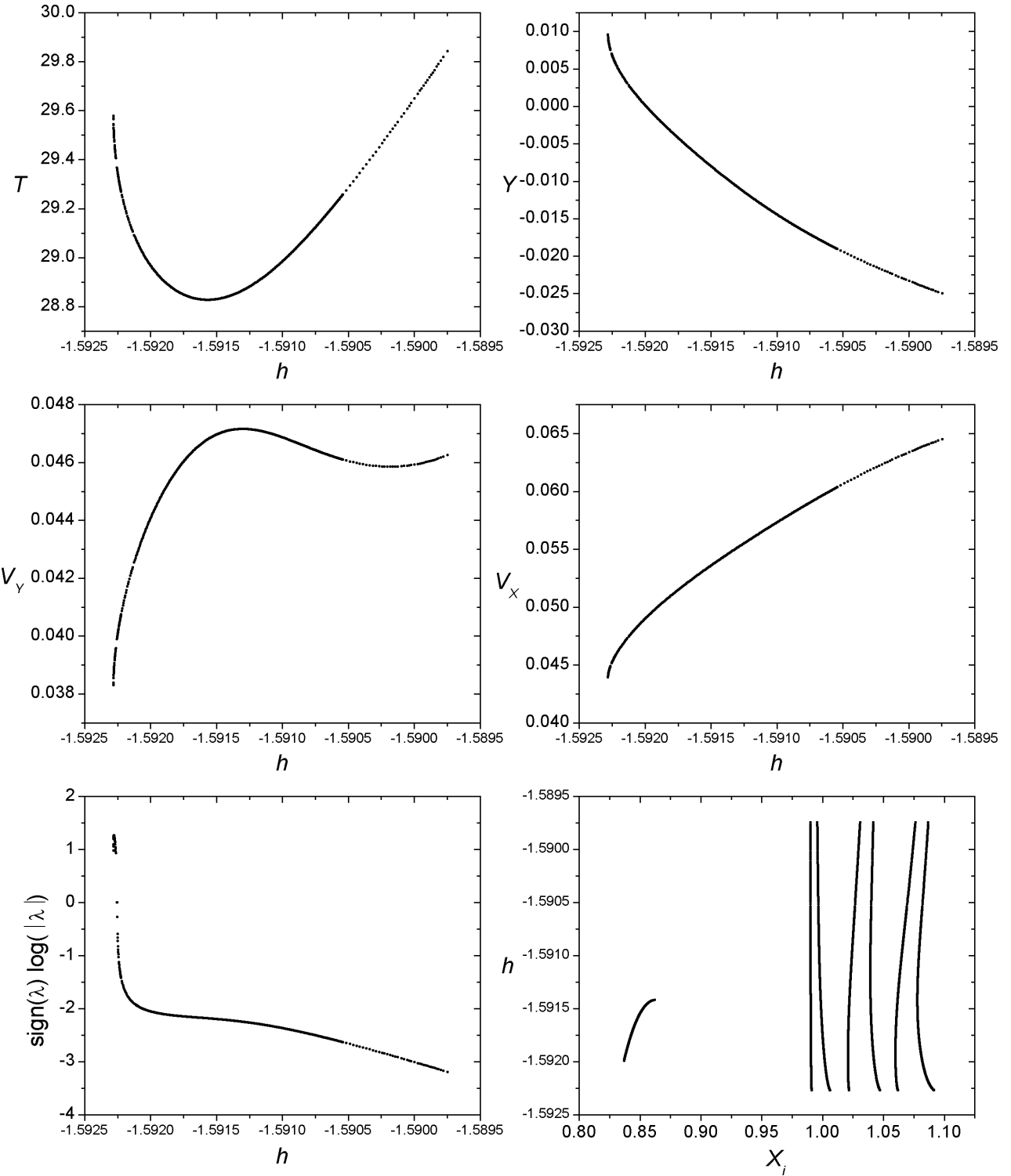

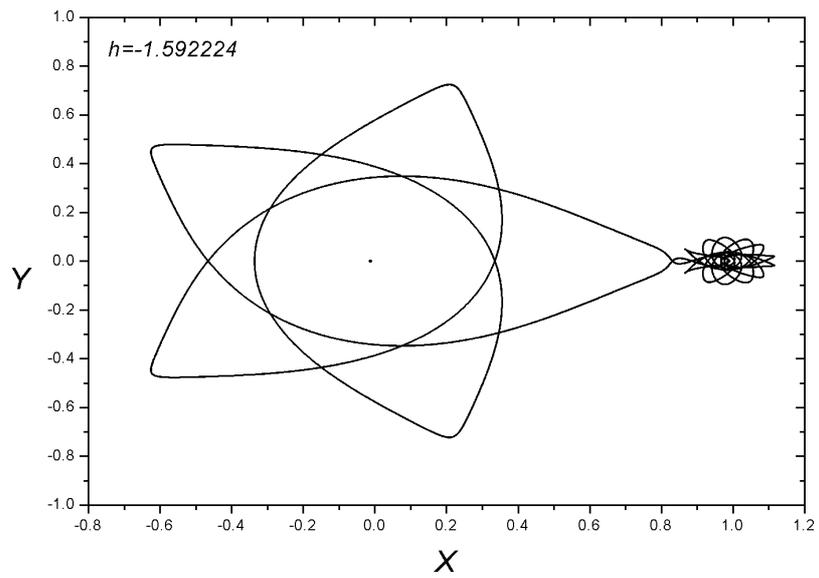

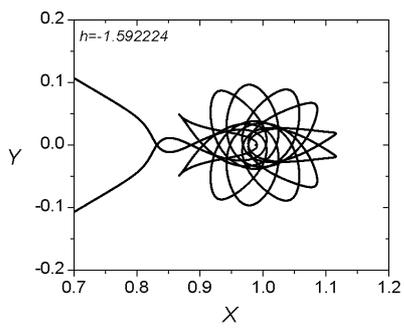
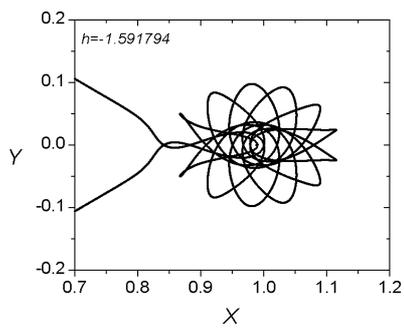
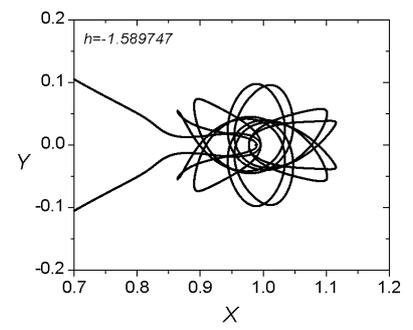

## Family 172 - *Symmetric family of symmetric POs*

$h_{min} = -1.592498, \quad h_{max} = -1.588235, \quad T_{min} = 29.220943, \quad T_{max} = 30.820594$

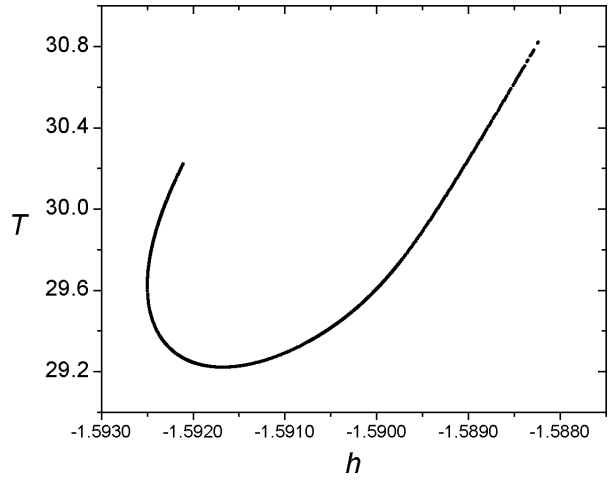
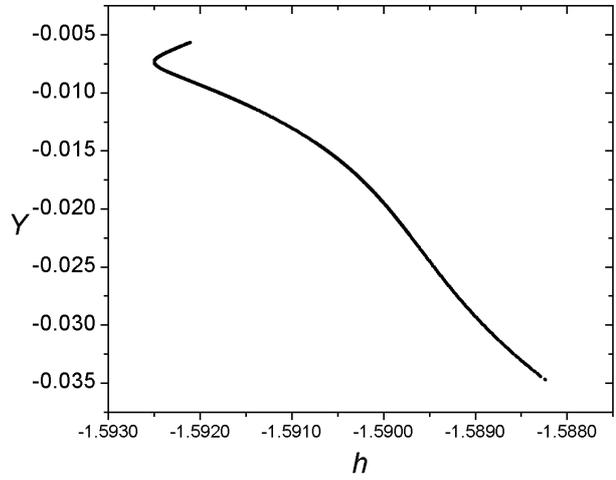

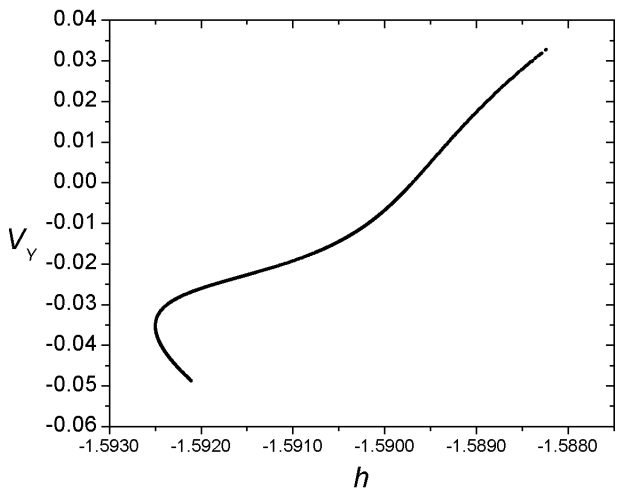
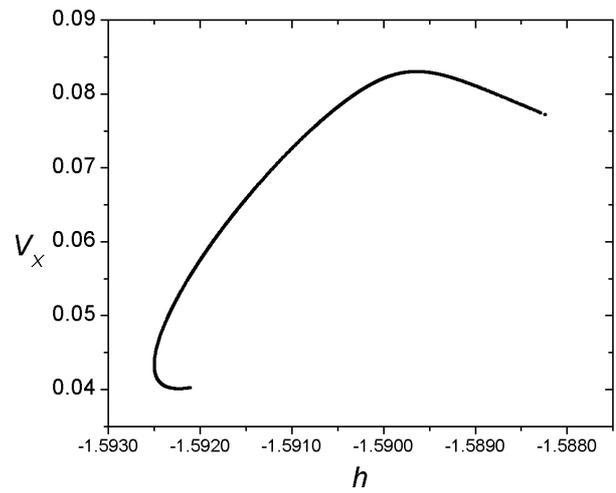

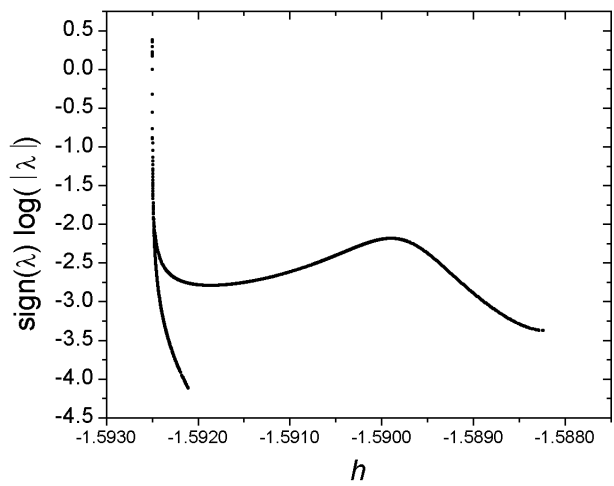
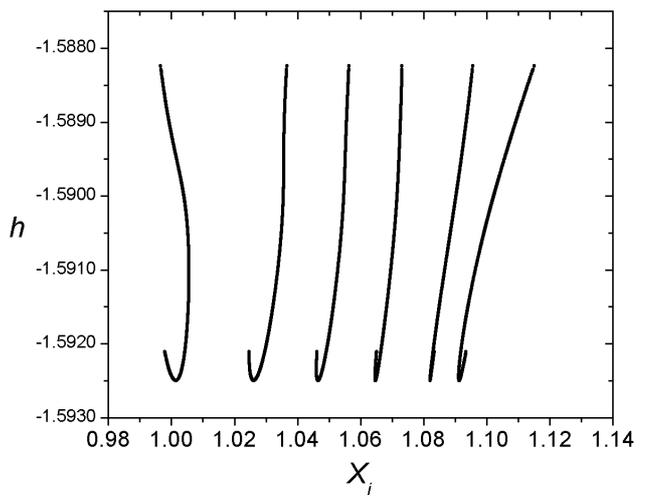

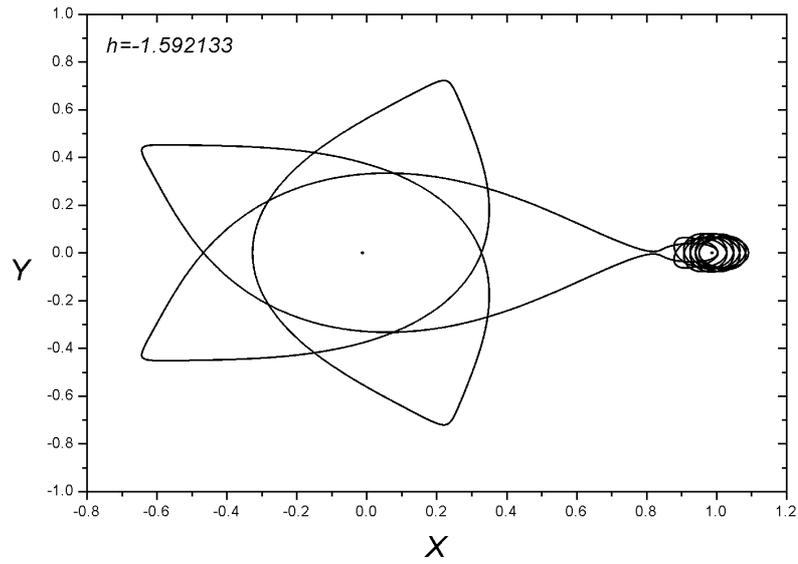

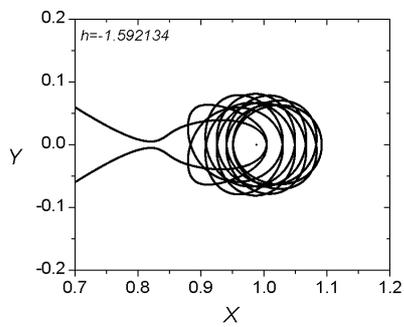
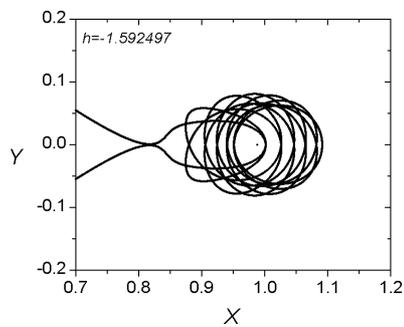
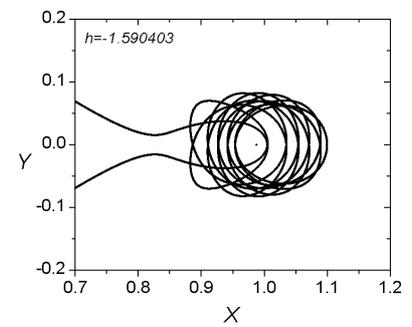

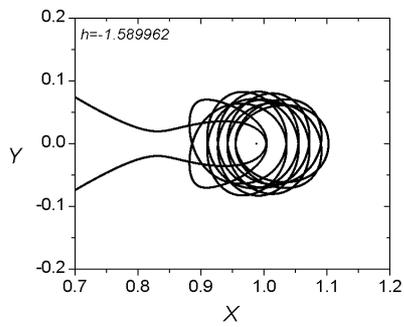
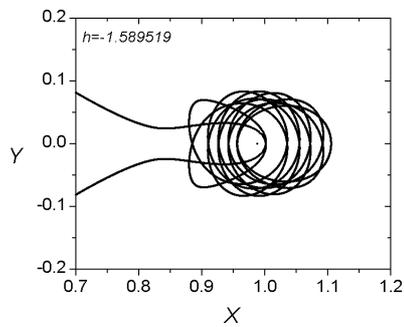
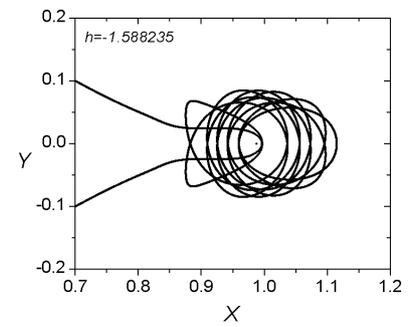

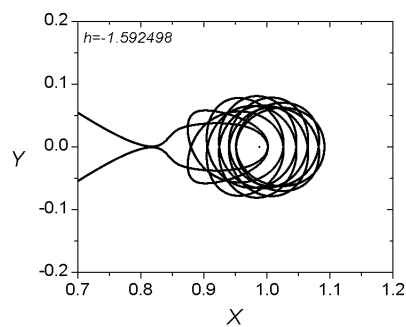
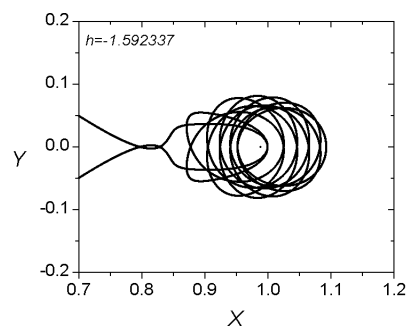
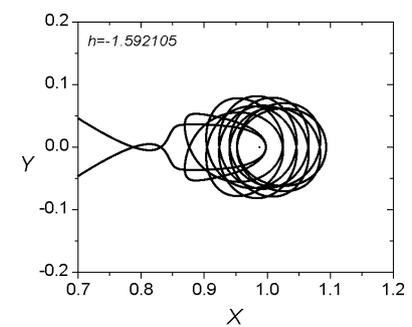

## *Family 209 - Asymmetric family of asymmetric POs*

$h_{min}$ = -1.592584,  $h_{max}$ =  -1.591028,  $T_{min}$ = 30.572004,  $T_{max}$ =  31.977814

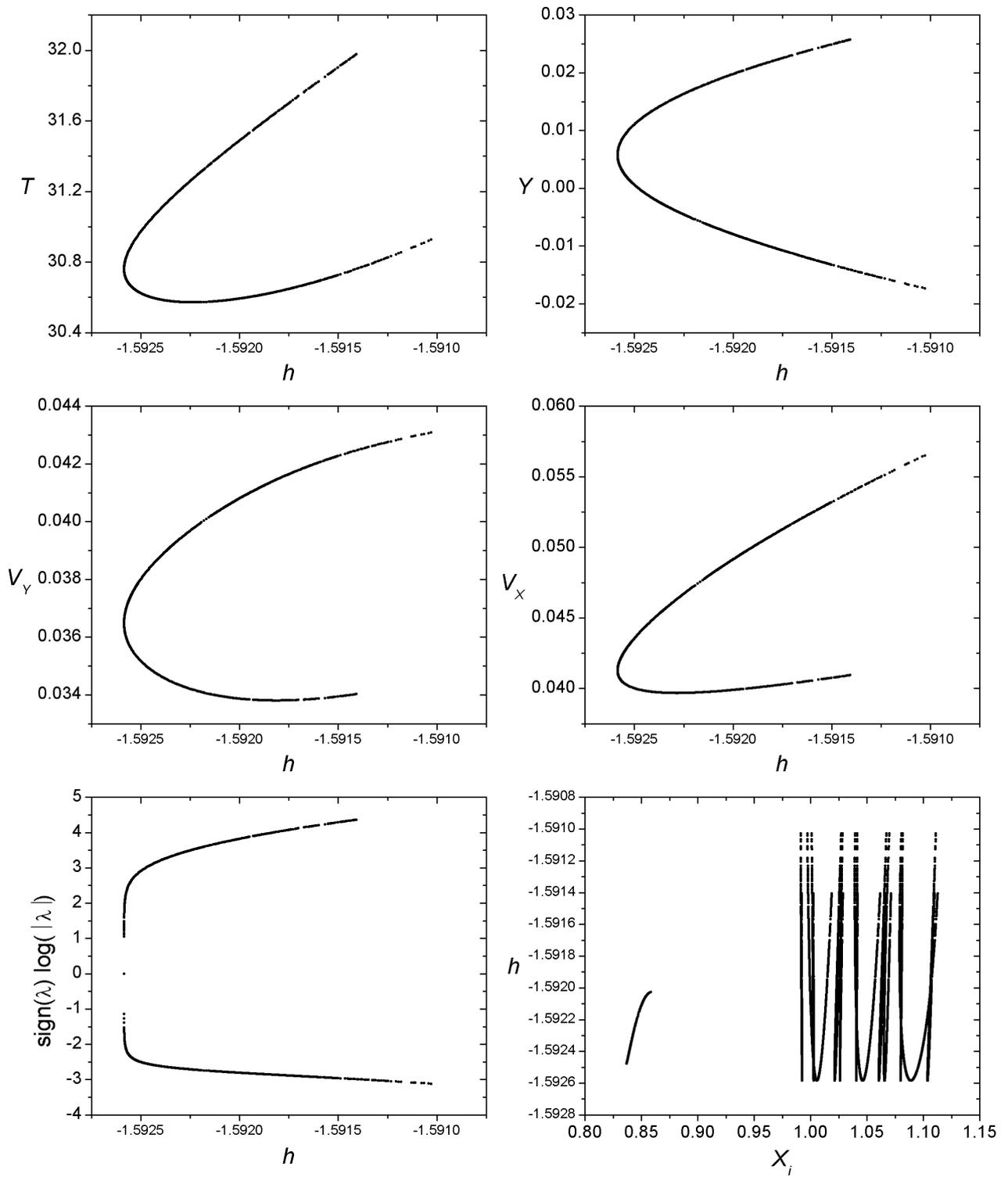

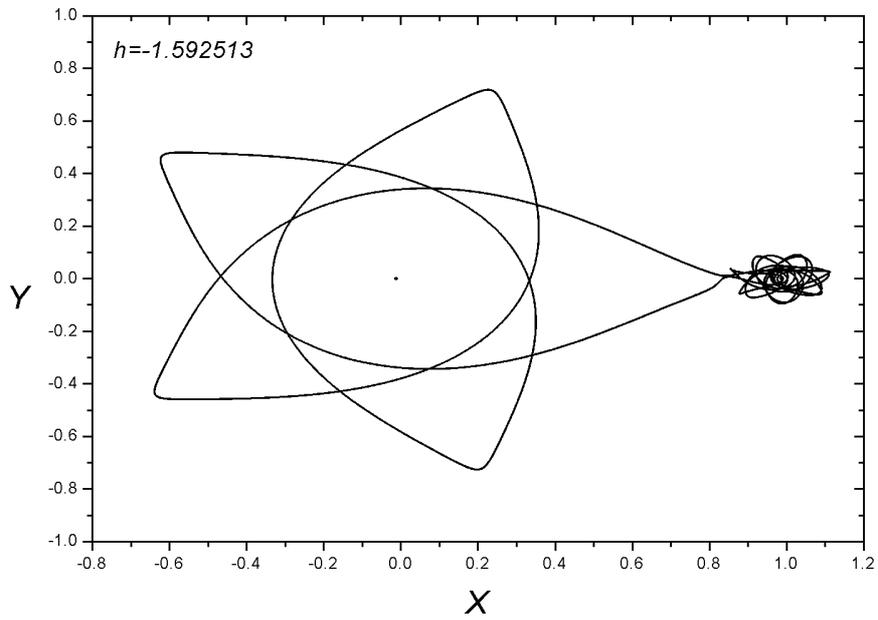

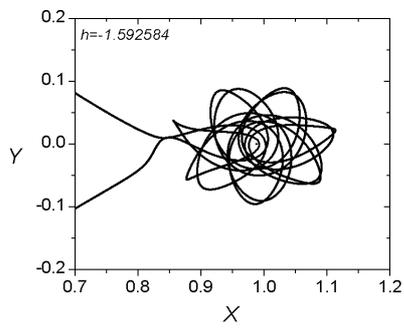
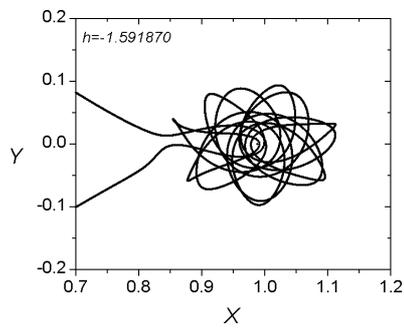
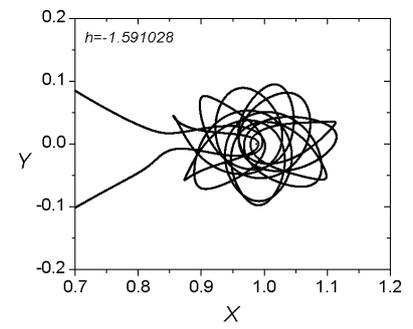

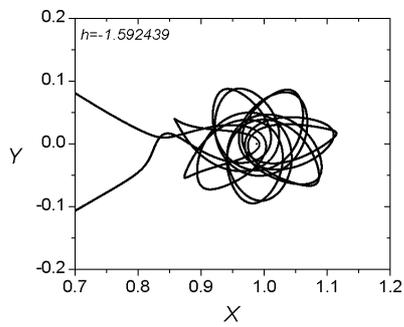
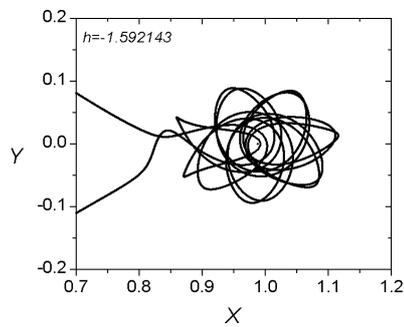
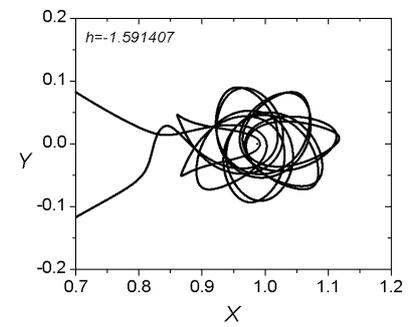

## Family 232 - Asymmetric family of asymmetric POs

$h_{min} = -1.592584$,  $h_{max} = -1.588964$,  $T_{min} = 30.572004$,  $T_{max} = 32.373619$

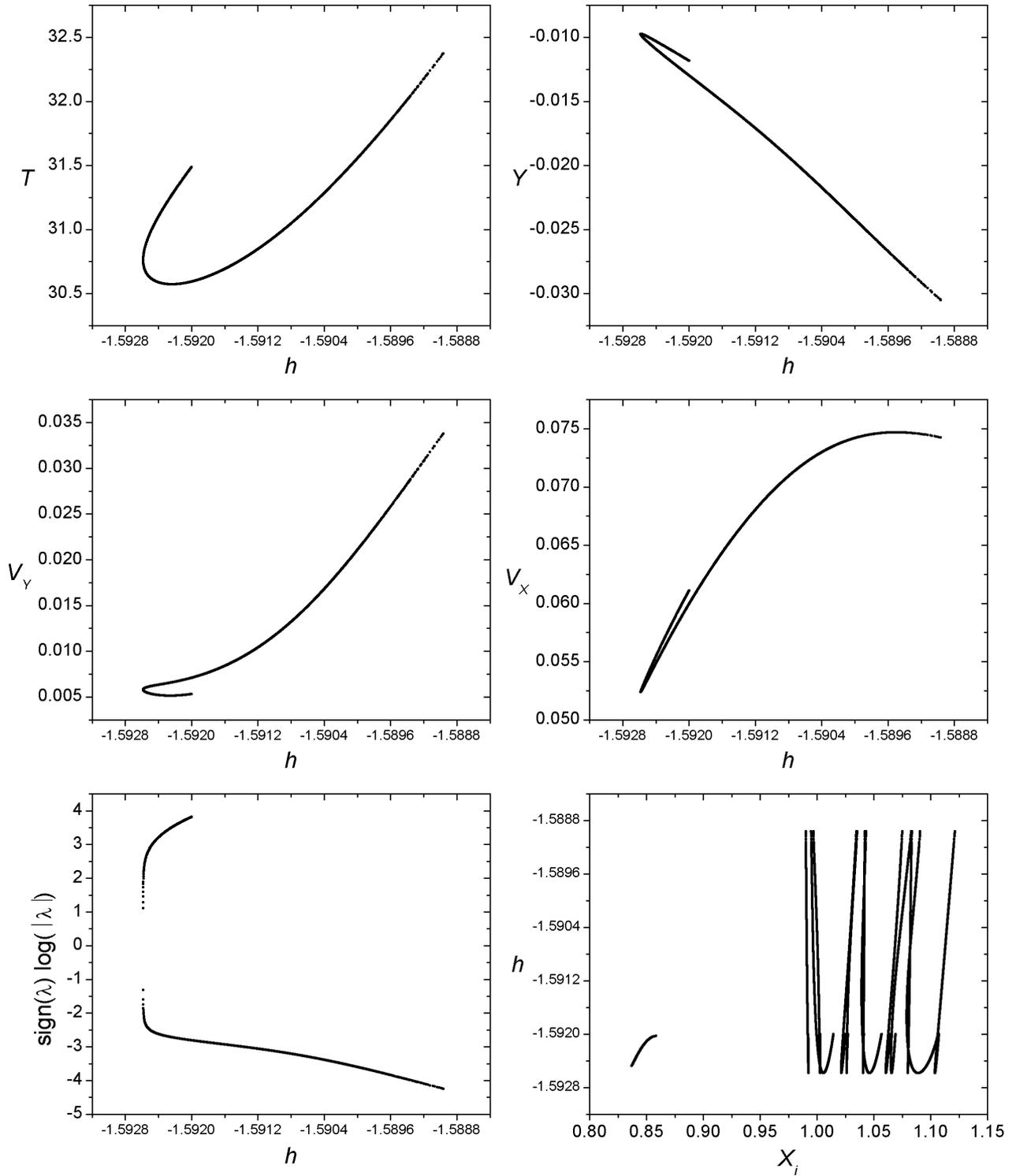

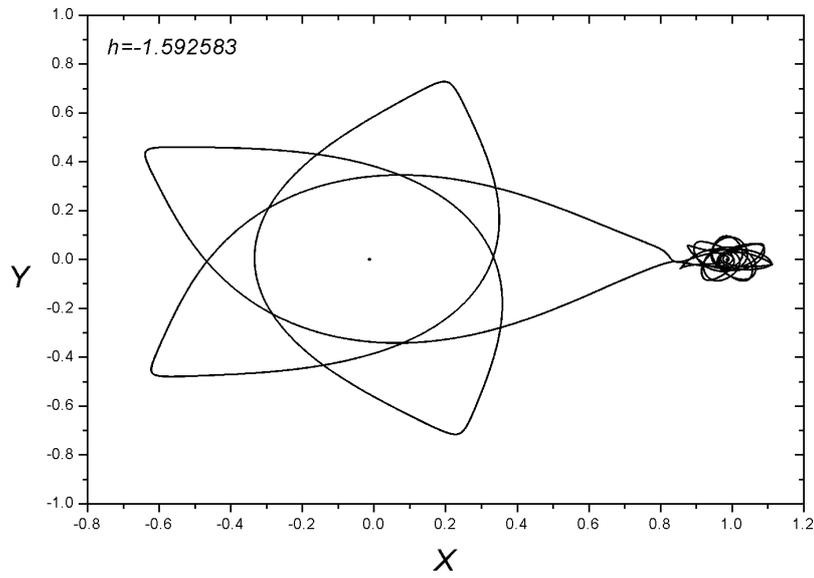

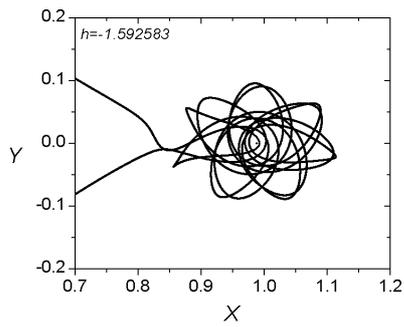
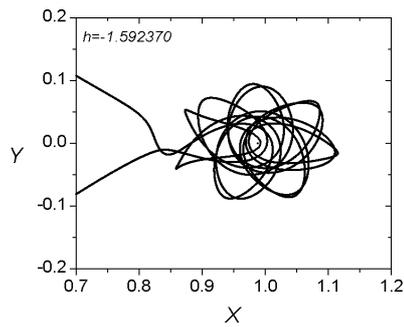
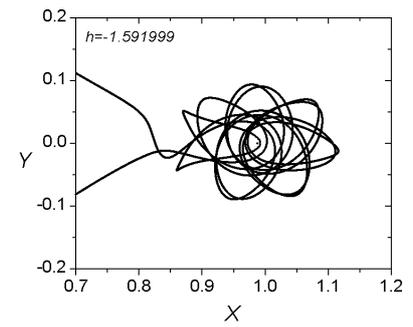

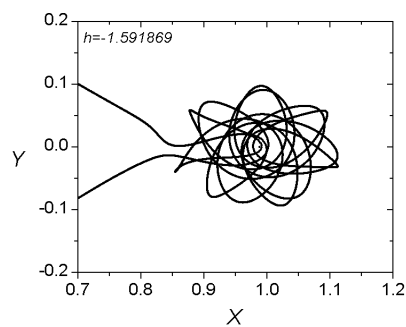
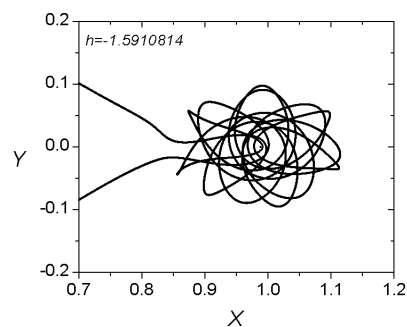
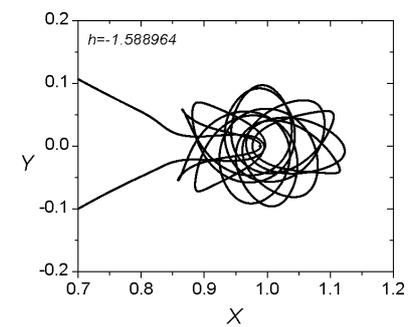

## Families 058 A - 058 B

*Bifurcation Point*

|  | *h* | *T* | *y* | *v_y* | *v_x* |
|---|---|---|---|---|---|
| *P₁* | −1.587665 | 31.348078 | 0.035164 | -0.030898 | 0.084255 |

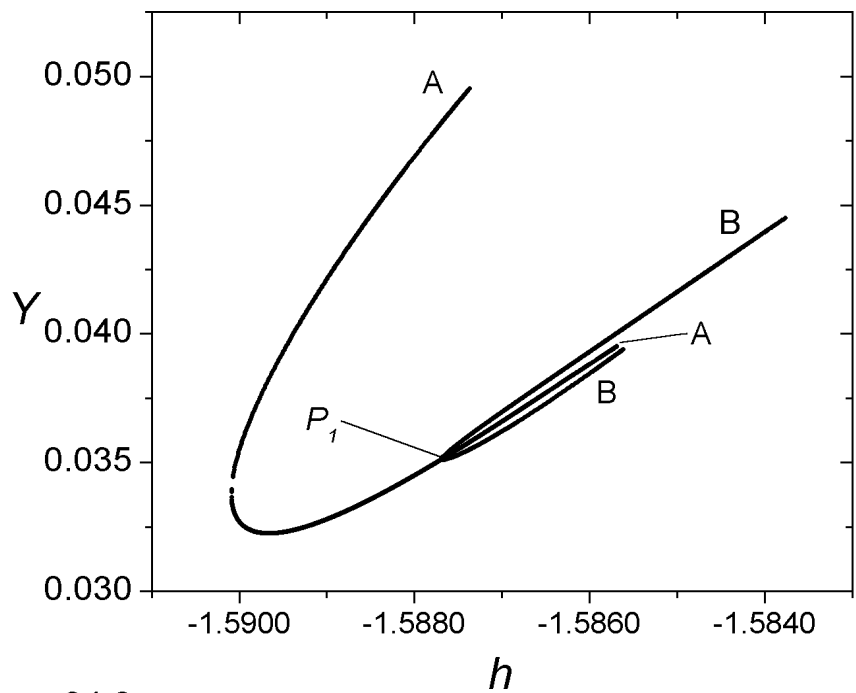

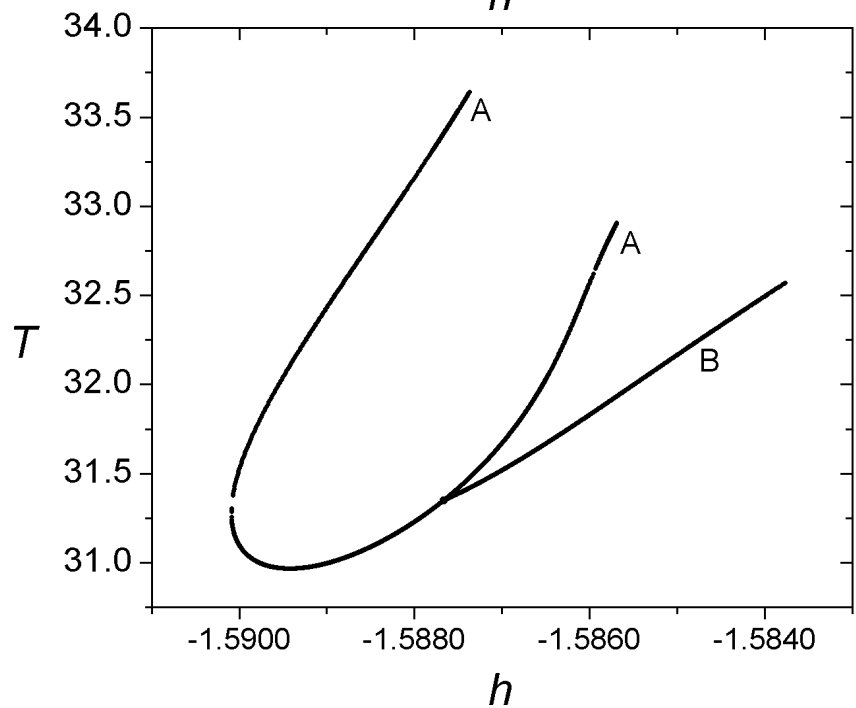

## Family 058 A - Symmetric family of symmetric POs

$h_{min} = -1.590087$, $h_{max} = -1.585685$, $T_{min} = 30.966292$, $T_{max} = 33.641214$

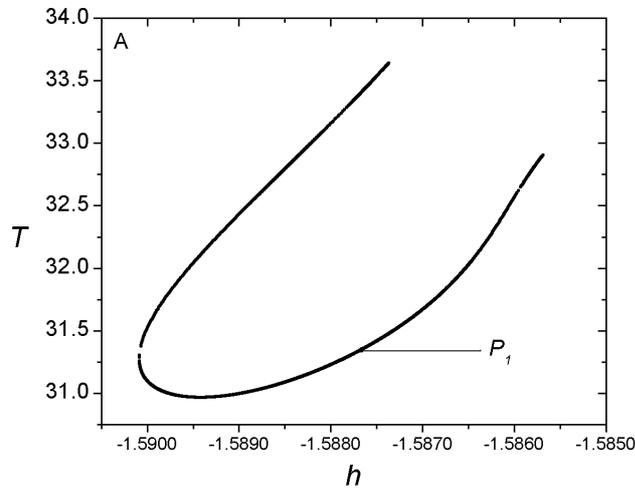
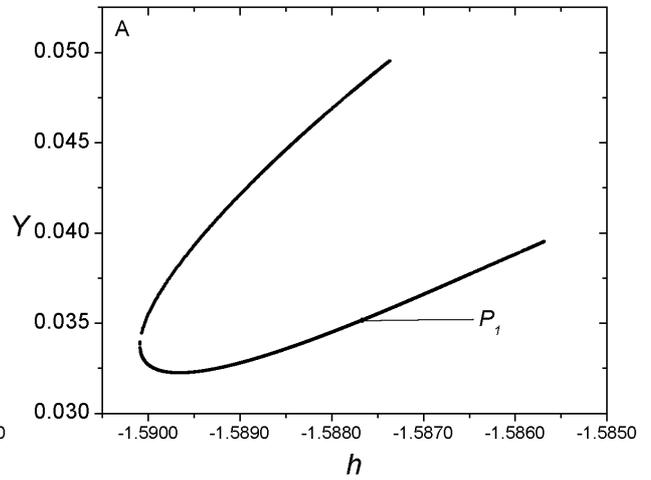
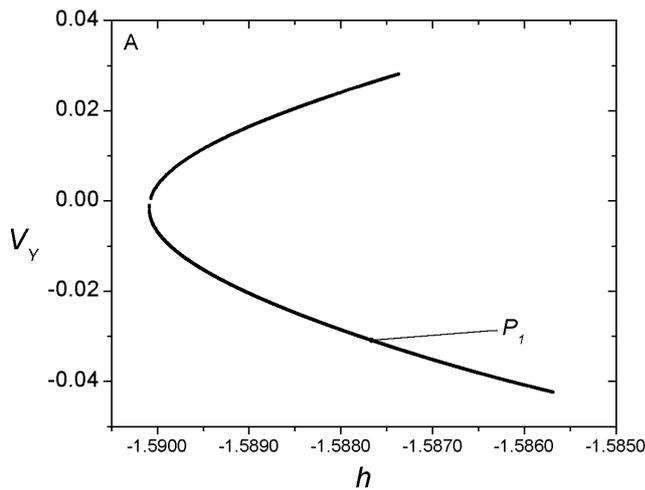
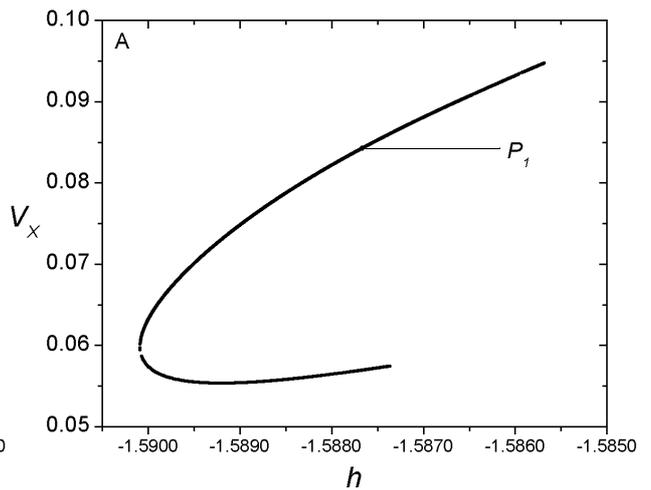
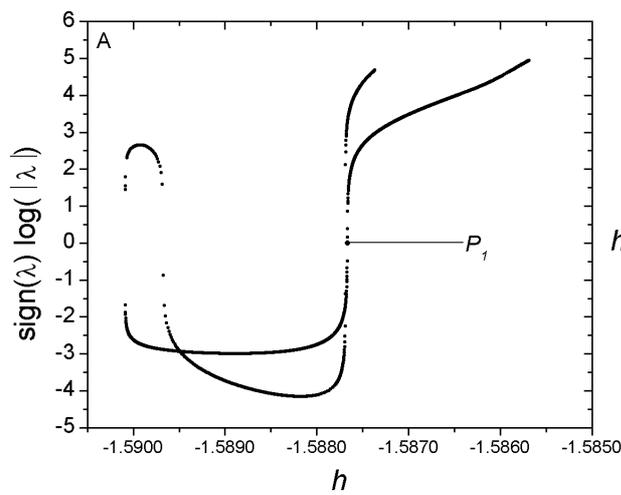
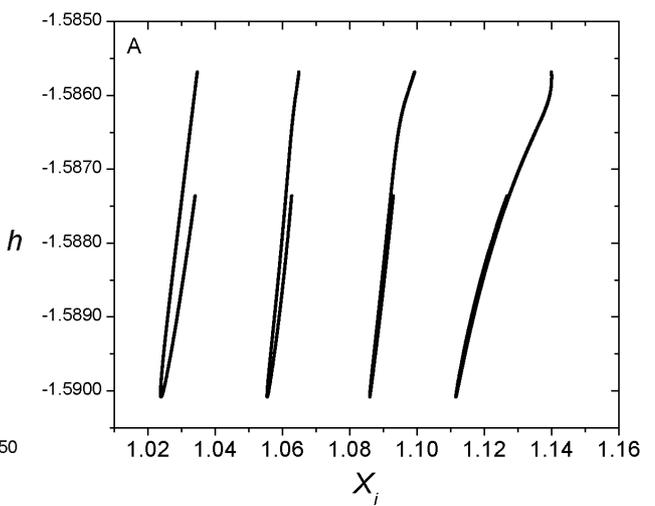

## *Family 058 B - Symmetric family of asymmetric POs*

$h_{min} = -1.587667$, $h_{max} = -1.583764$, $T_{min} = 31.347506$, $T_{max} = 32.568419$

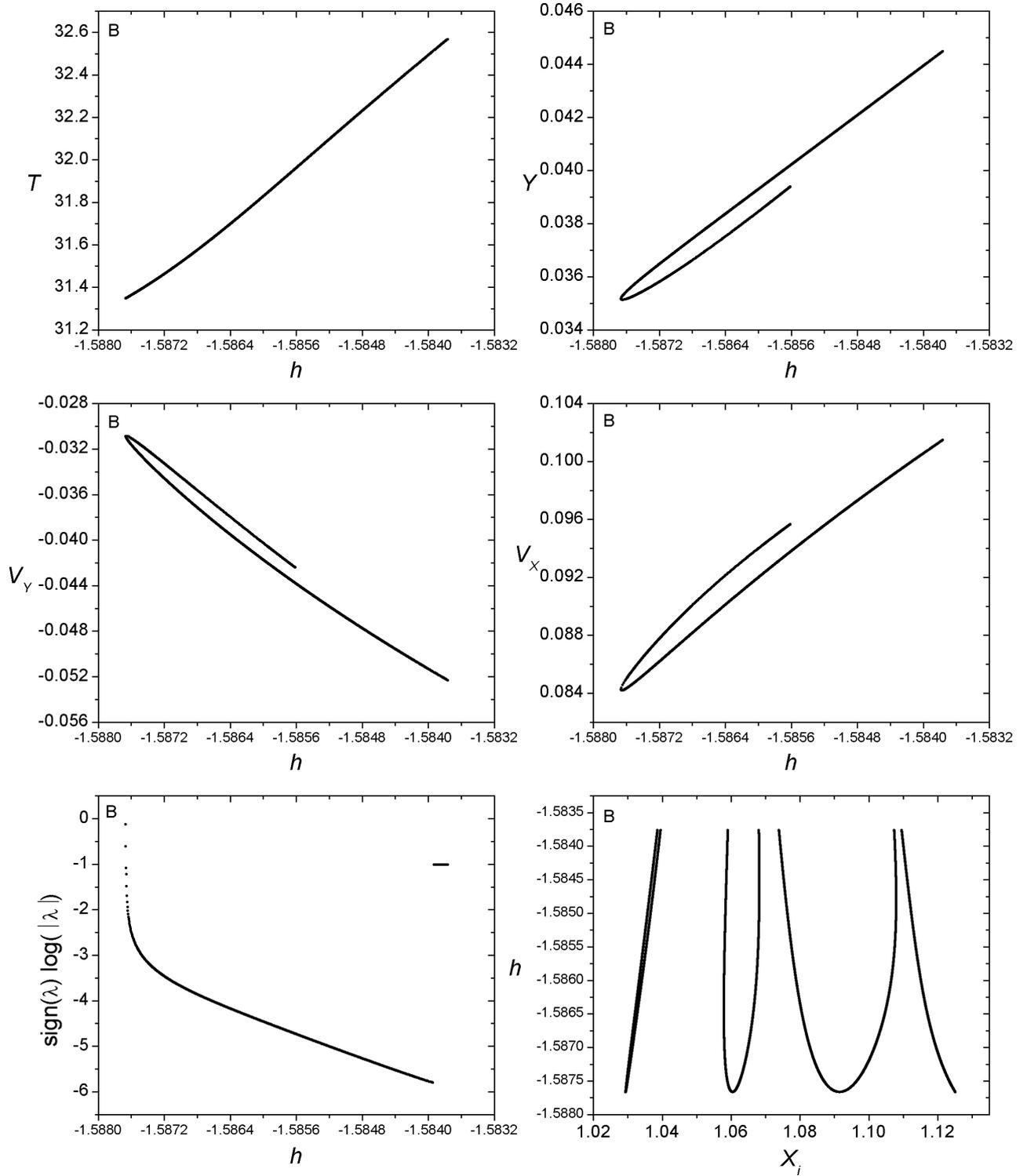

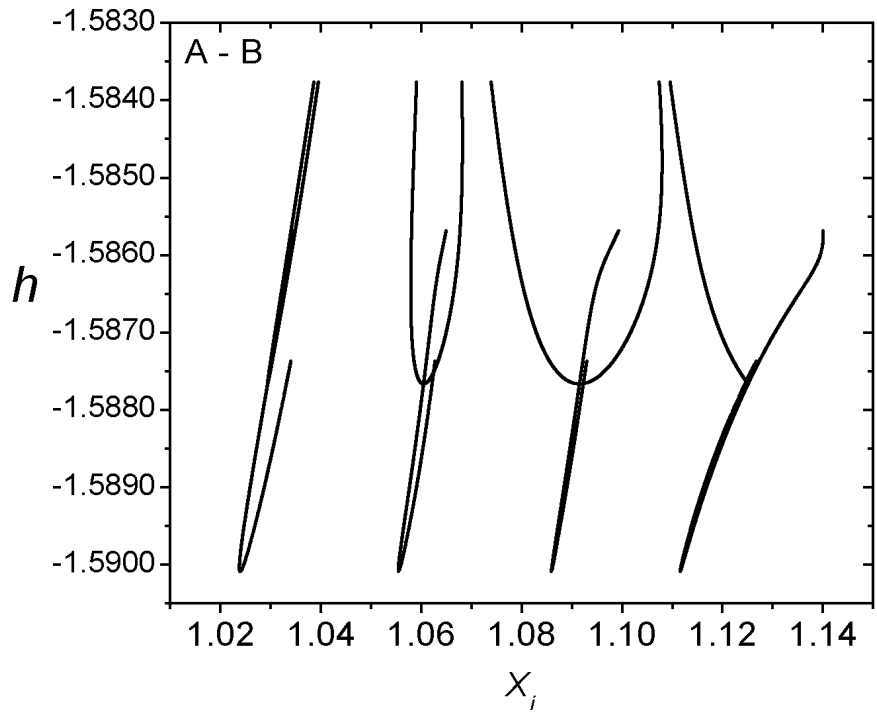

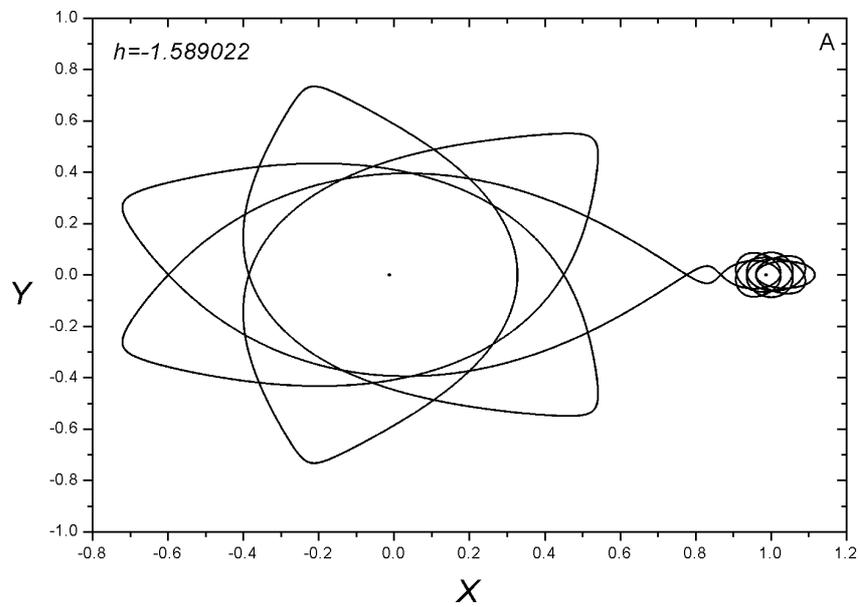

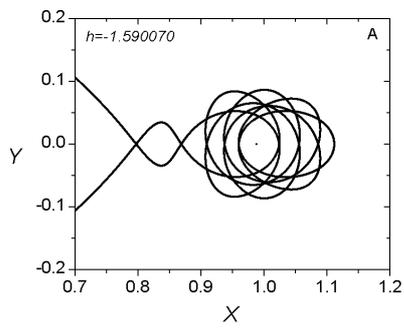
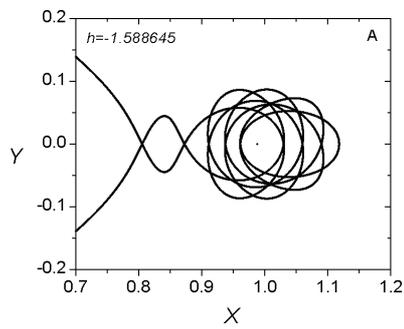
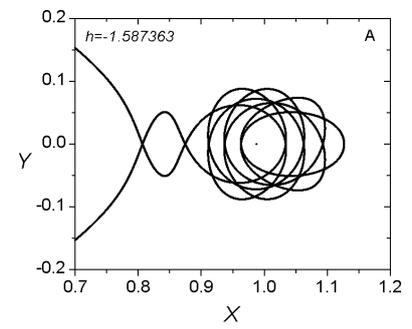

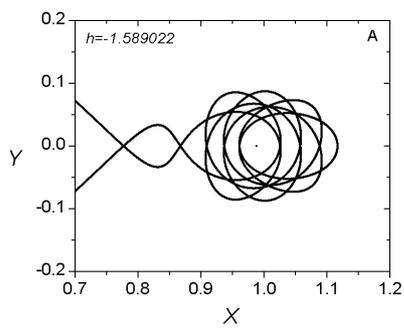
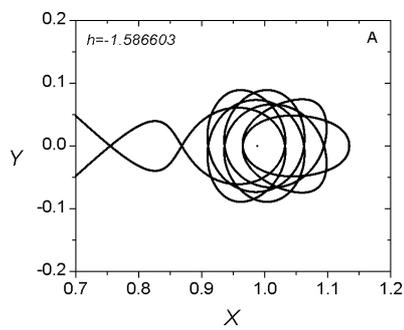
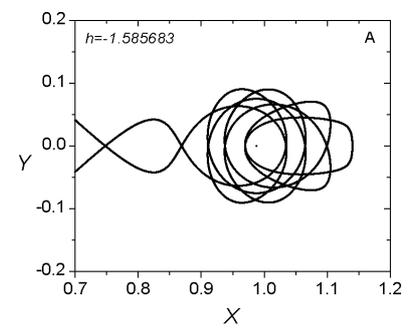

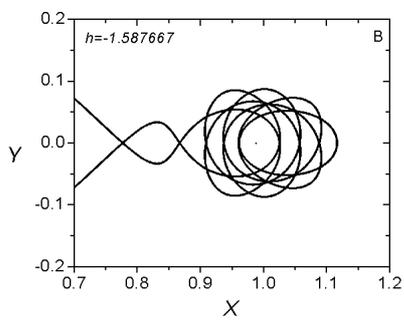
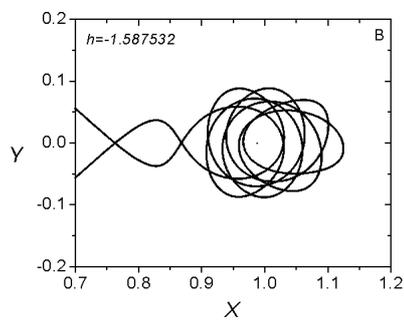
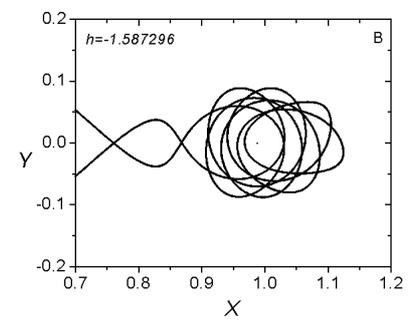

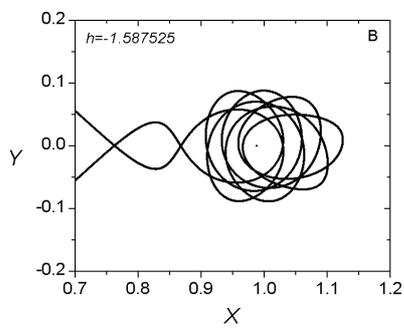
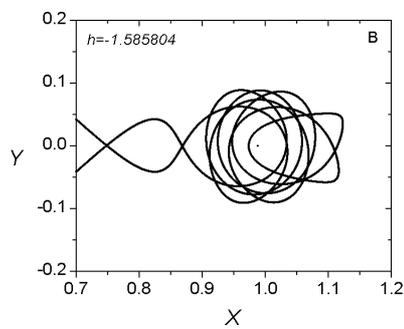
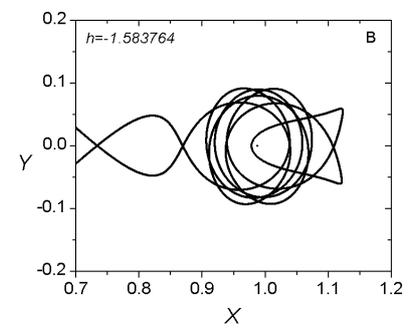

## Families 263 A - 263 B

*Bifurcation Point*

|       | $h$        | $T$       | $y$      | $v_y$    | $v_x$    |
|-------|------------|-----------|----------|----------|----------|
| $P_1$ | −1.593844  | 32.508829 | 0.000382 | 0.008559 | 0.024062 |

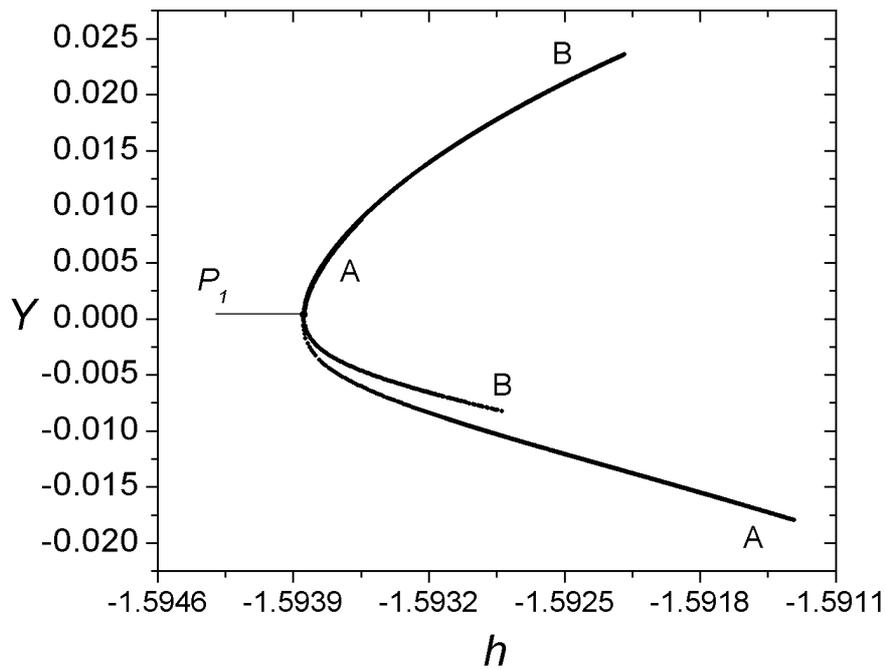

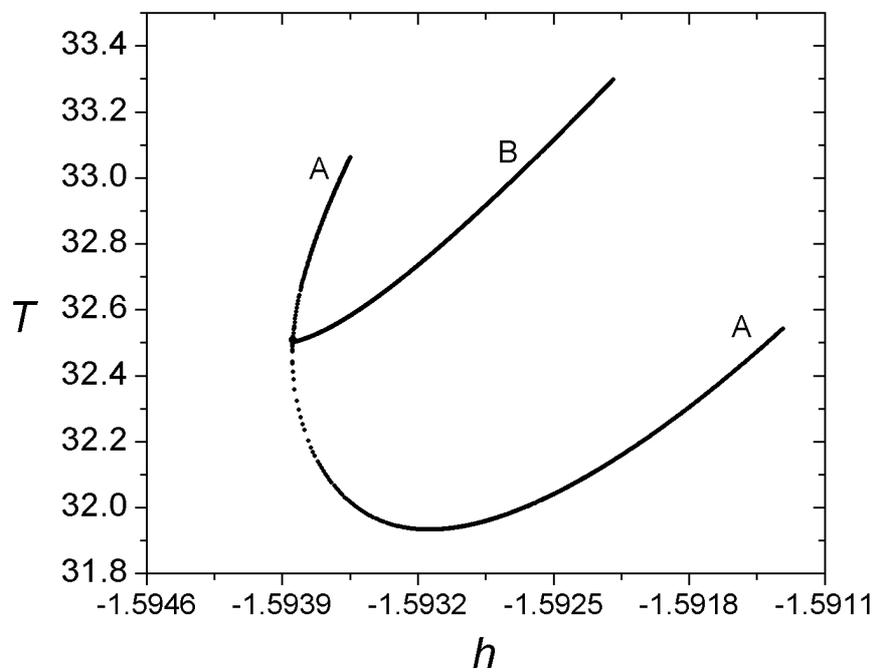

## *Family 263 A - Symmetric family of symmetric POs*

$h_{min} = -1.593847$,  $h_{max} = -1.591315$,  $T_{min} = 31.933083$,  $T_{max} = 33.061924$

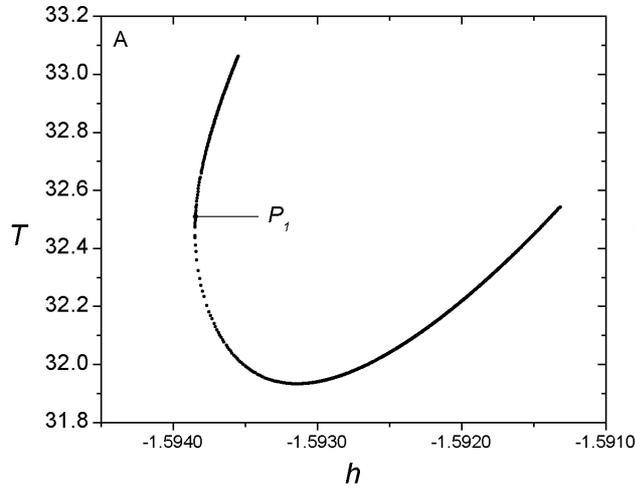
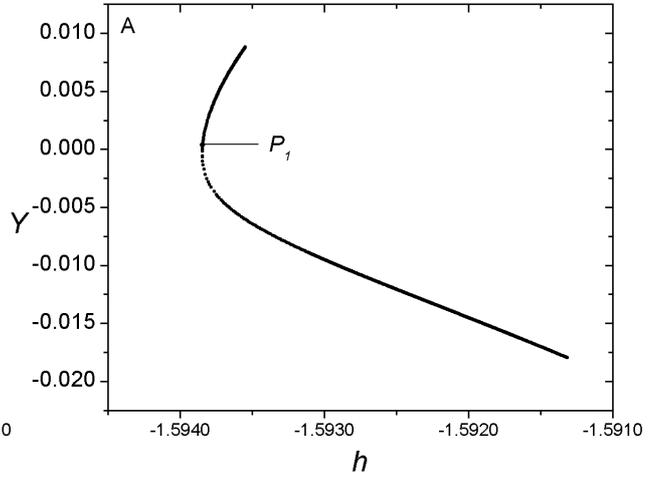

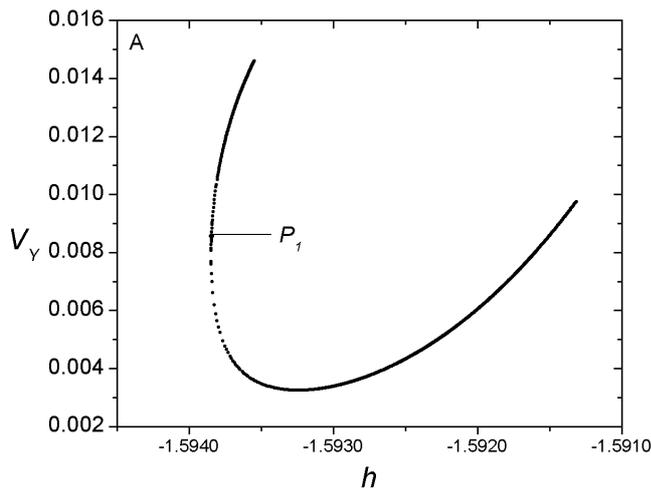
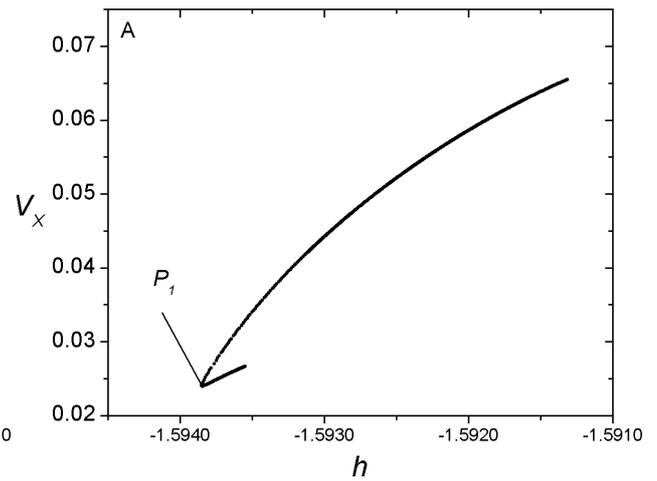

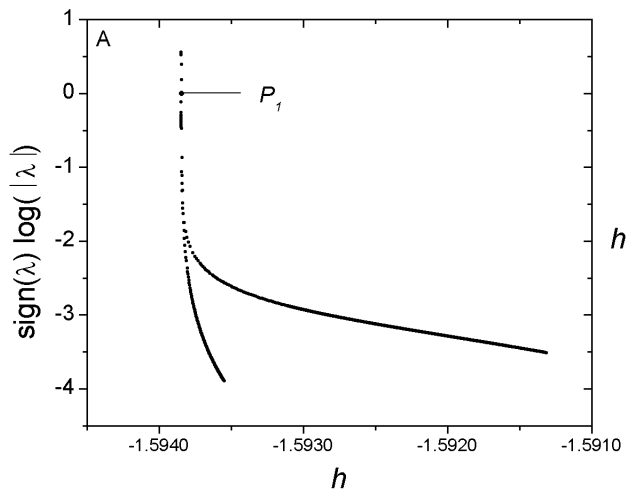
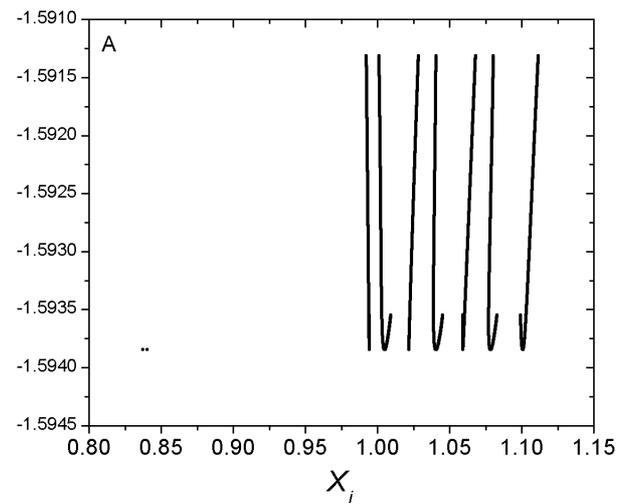

## Family 263 B - Symmetric family of asymmetric POs

$h_{min} = -1.593846$, $h_{max} = -1.592189$, $T_{min} = 32.489625$, $T_{max} = 33.298942$

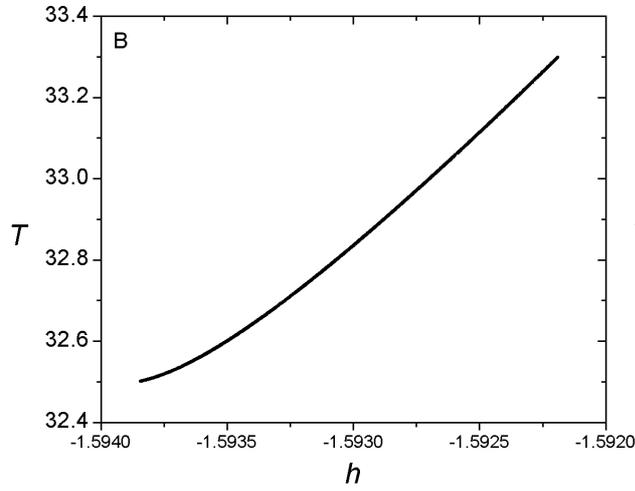

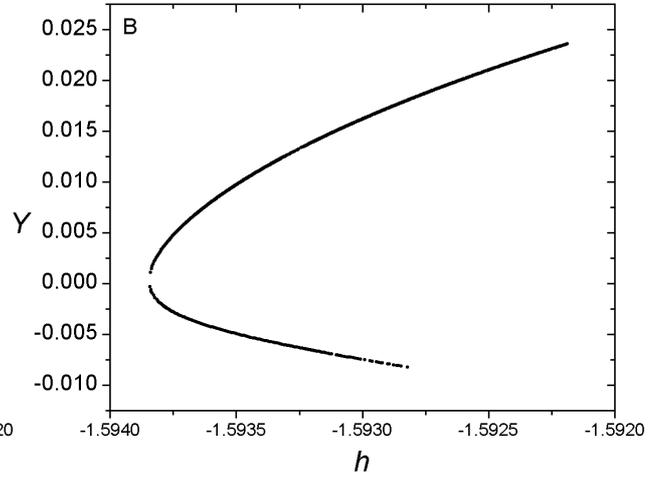

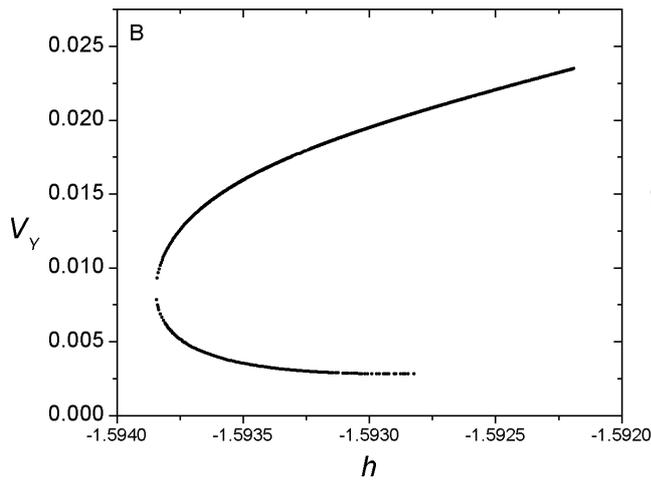

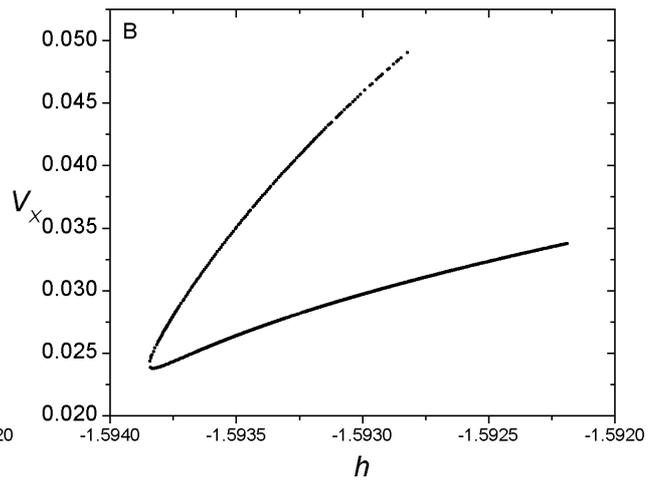

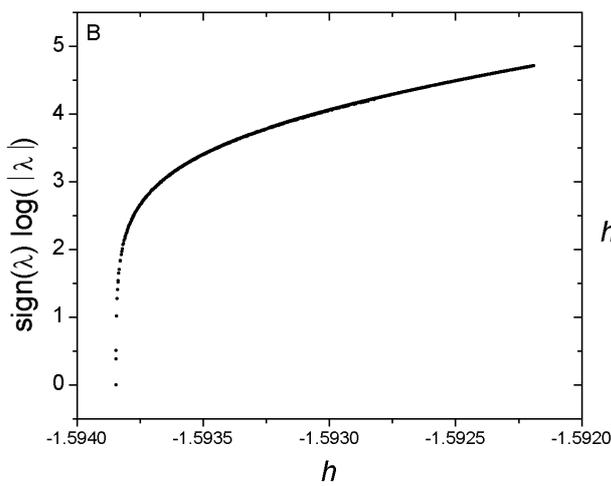

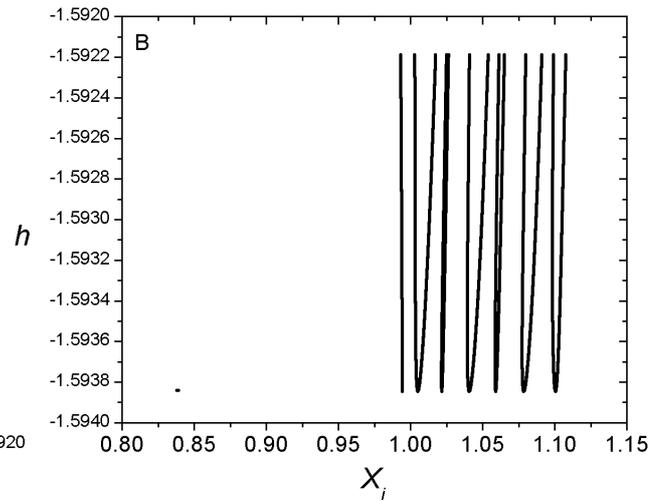

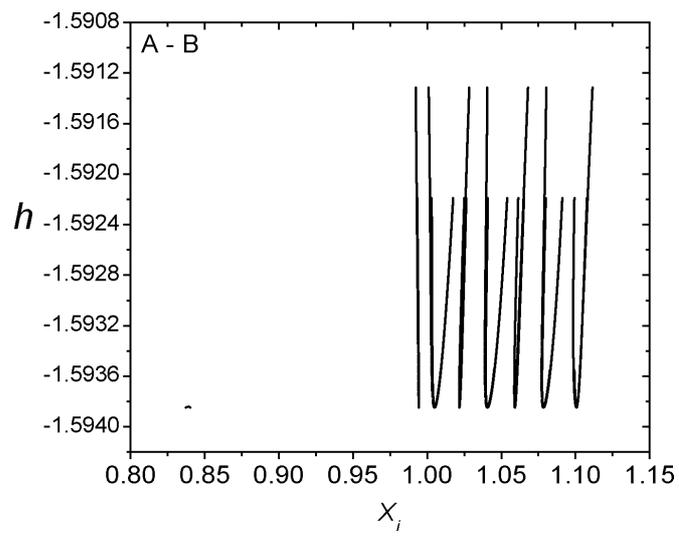

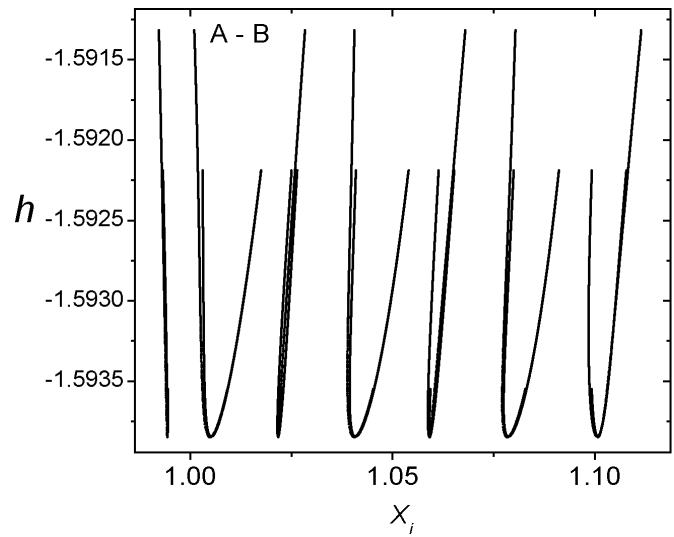

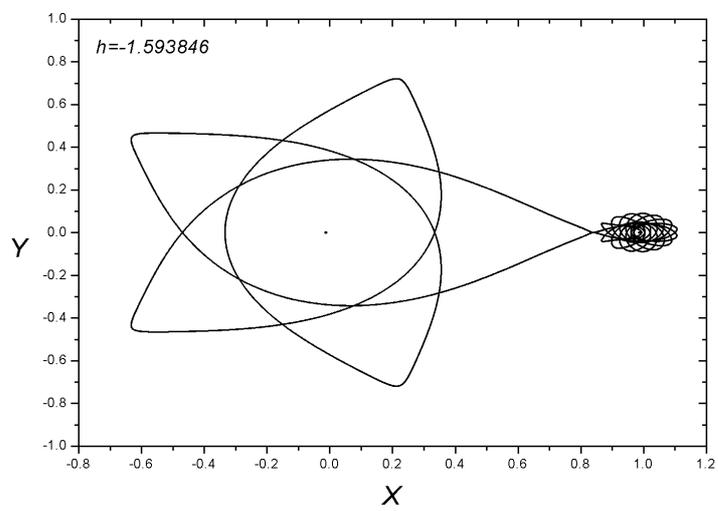

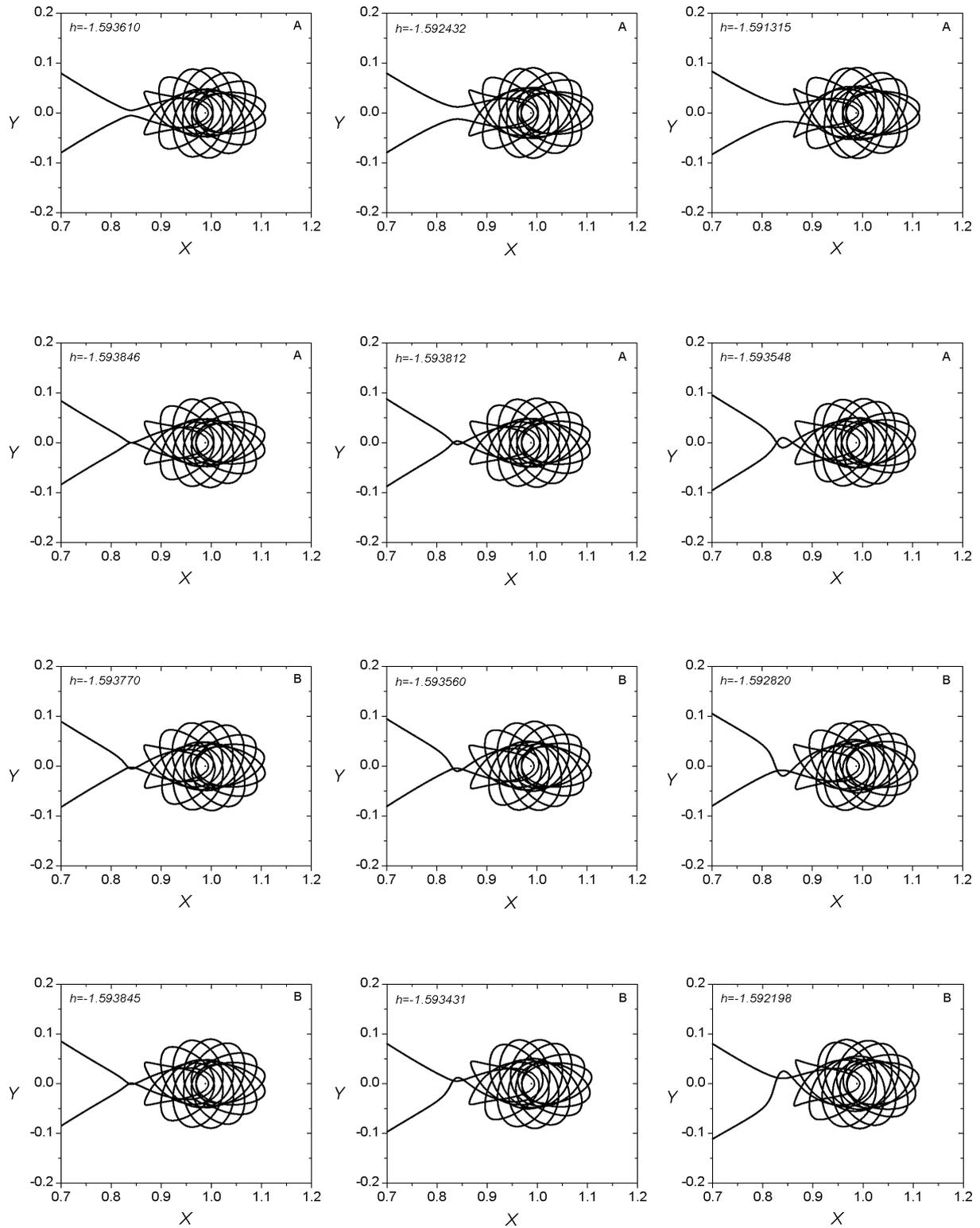

## Family 287 - Symmetric family of symmetric POs

$h_{min} = -1.594138, \ h_{max} = -1.593394, \ T_{min} = 32.562769, \ T_{max} = 34.433050$

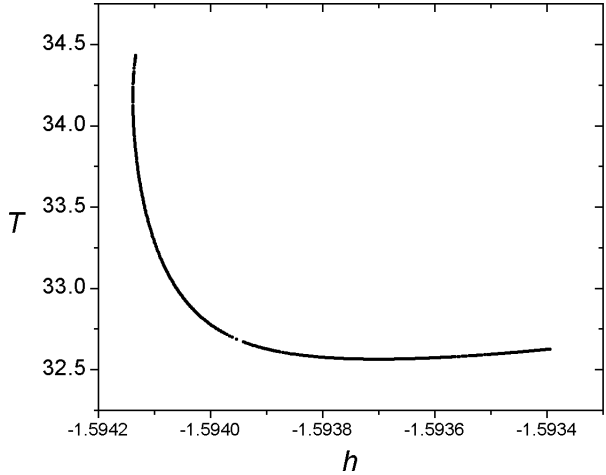
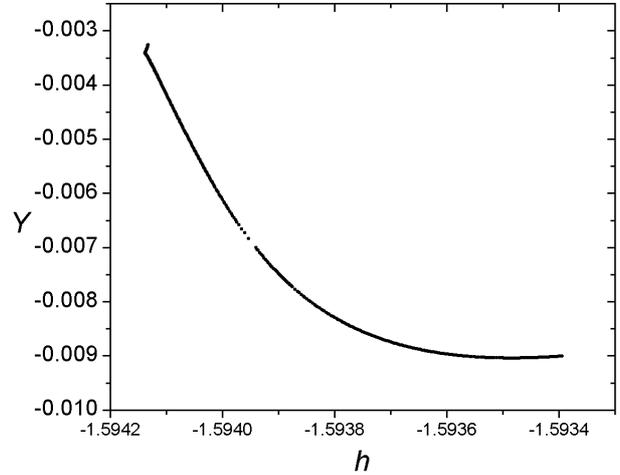

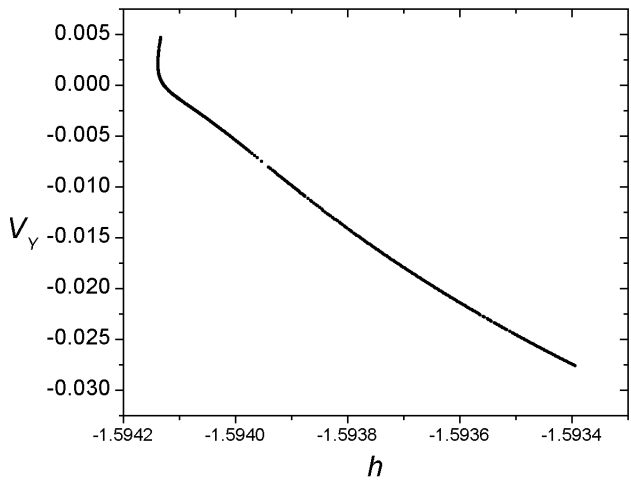
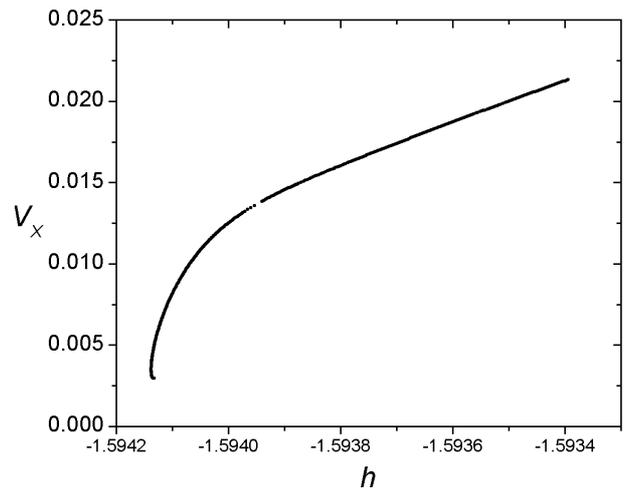

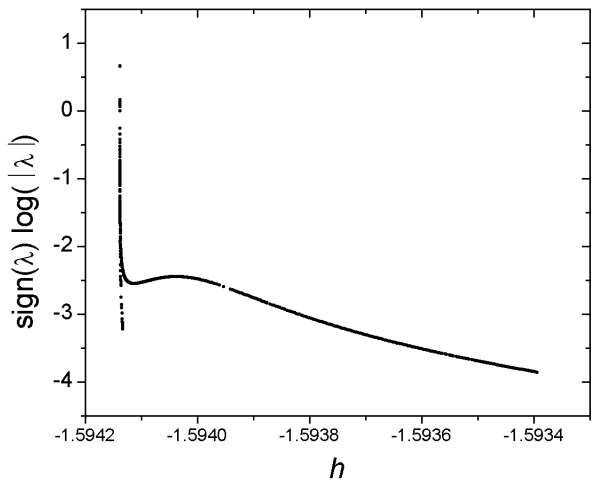
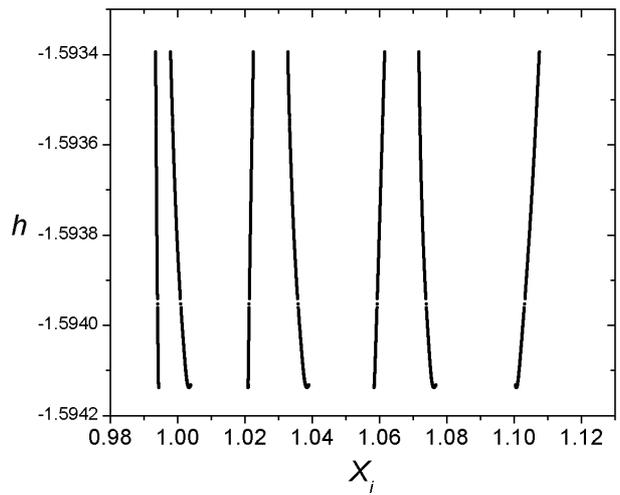

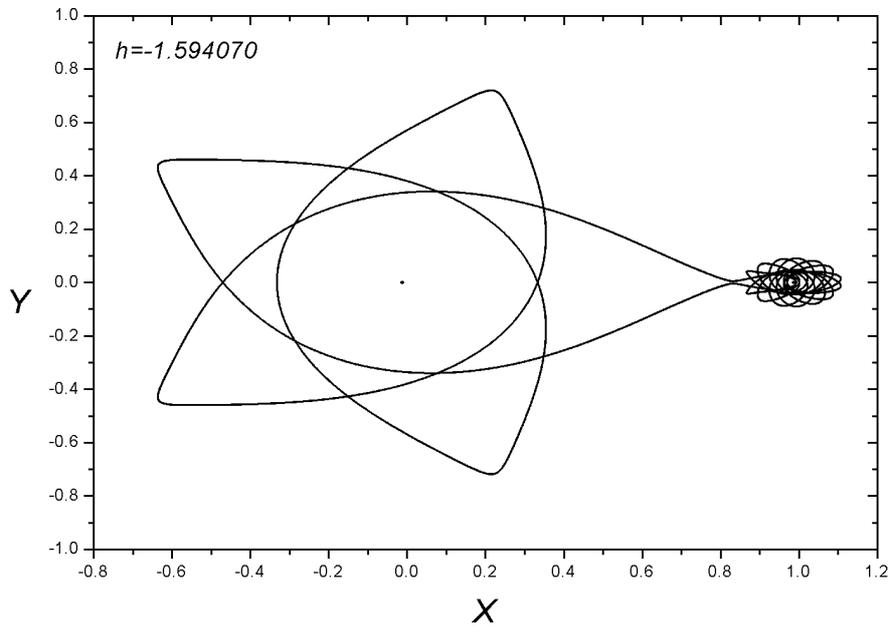

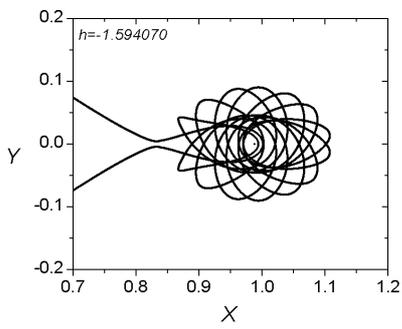
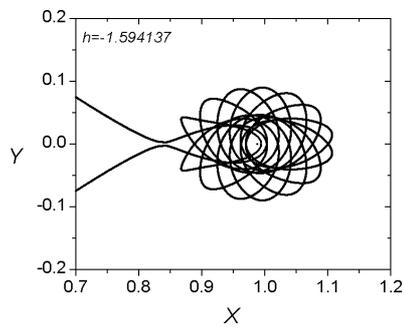
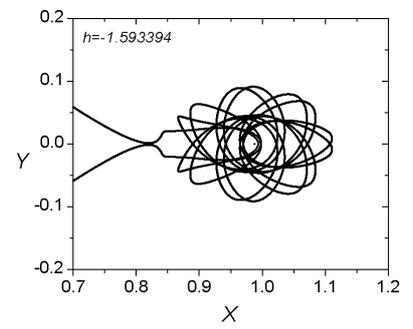

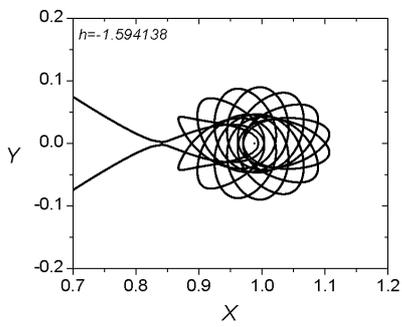
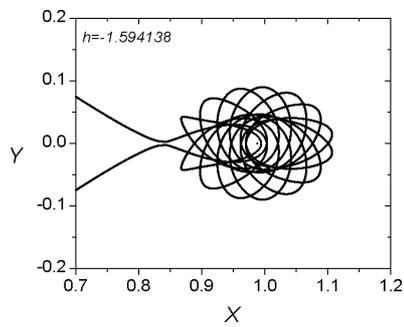
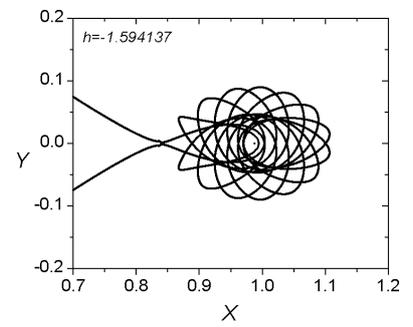

## *Families 032 A - 032 B*

*Bifurcation Point*

|       | $h$        | $T$       | $y$      | $v_y$     | $v_x$    |
|-------|------------|-----------|----------|-----------|----------|
| $P_1$ | −1.587120  | 33.892706 | 0.023970 | -0.001461 | 0.108411 |

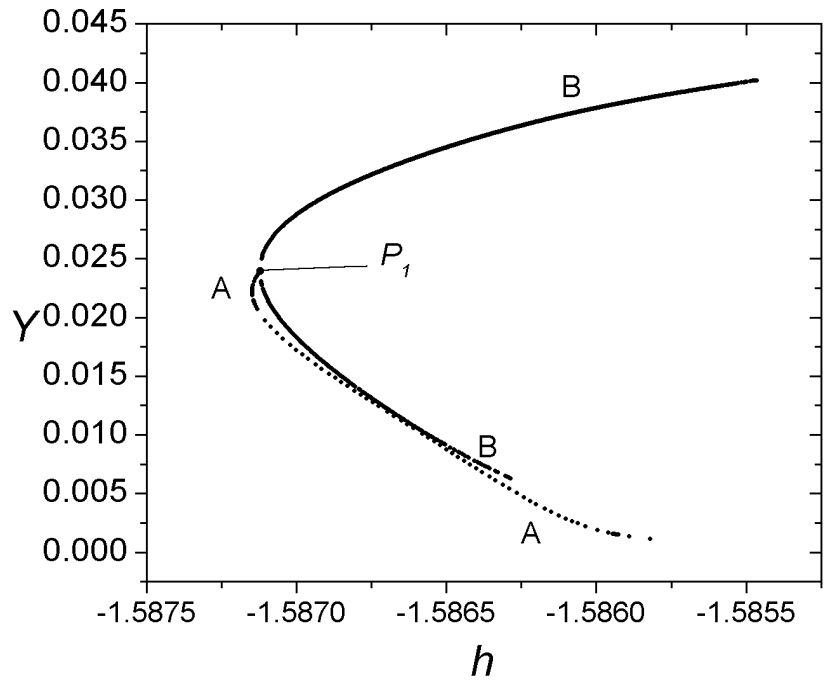

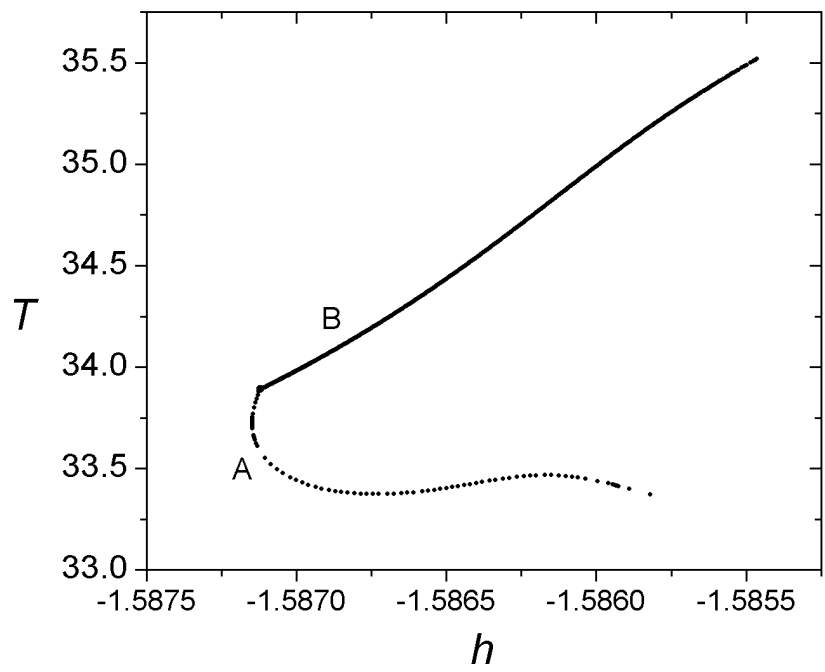

## Family 032 A - Symmetric family of symmetric POs

$h_{min} = -1.587147$,  $h_{max} = -1.585821$,  $T_{min} = 33.370152$,  $T_{max} = 33.892706$

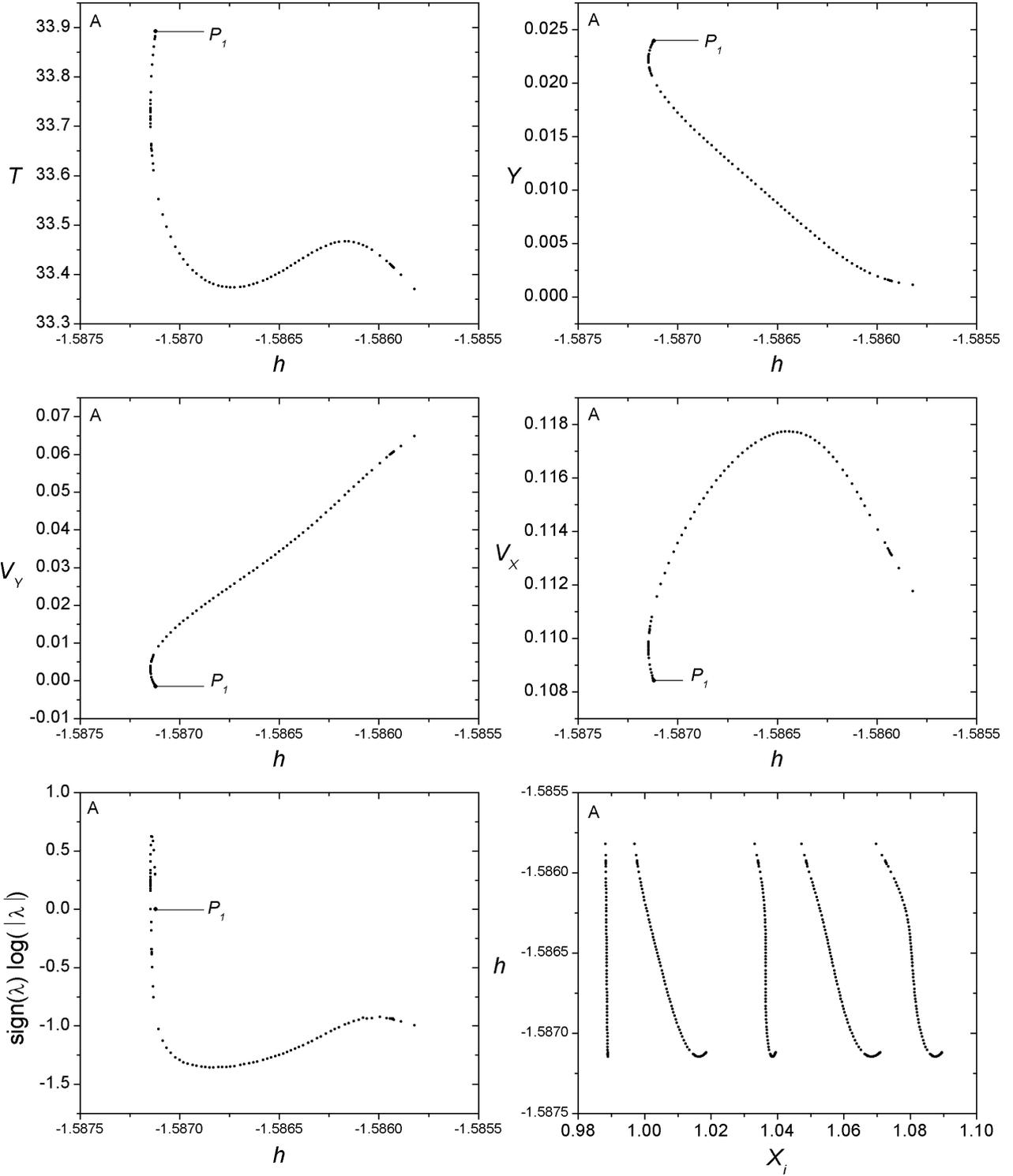

## Family 032 B - Symmetric family of asymmetric POs

$h_{min} = -1.587118$, $h_{max} = -1.585464$, $T_{min} = 33.891167$, $T_{max} = 33.518105$

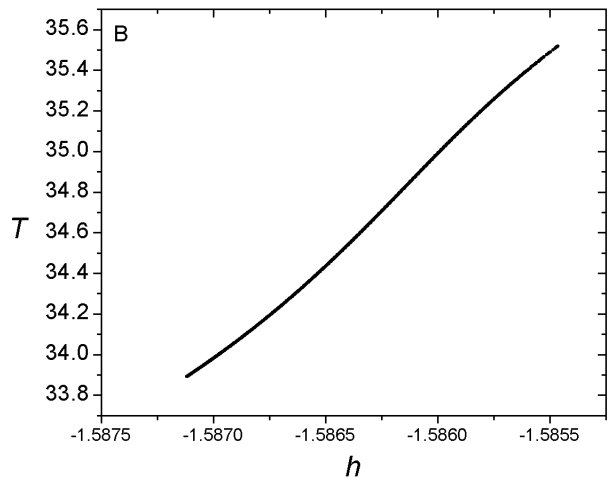
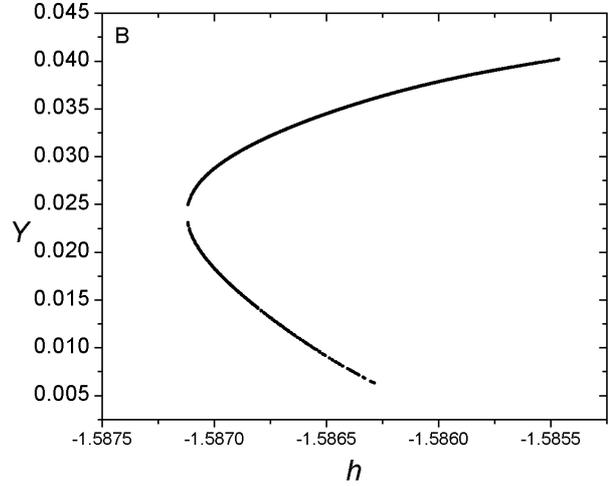

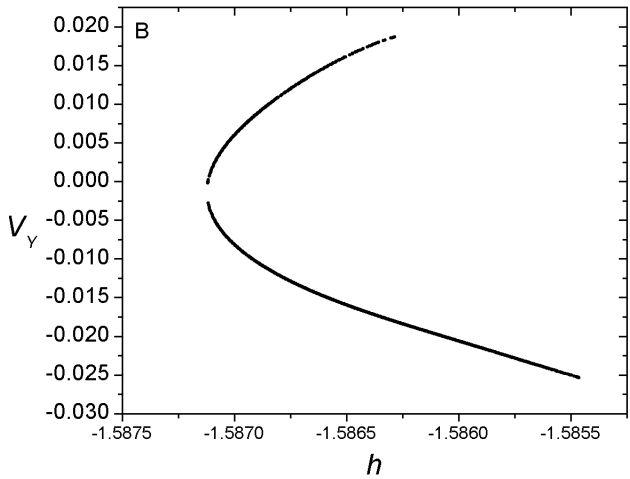
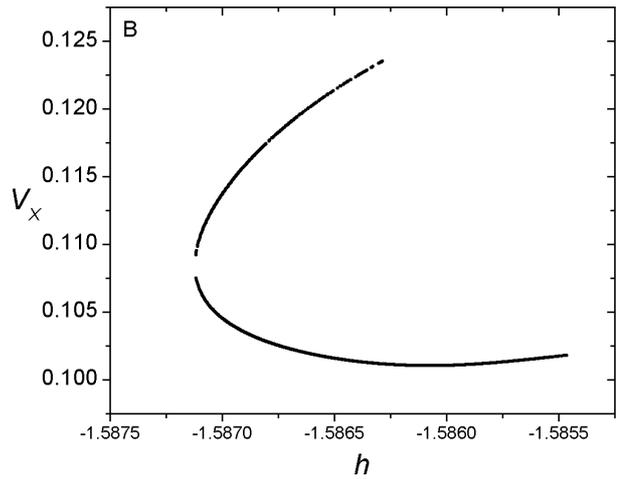

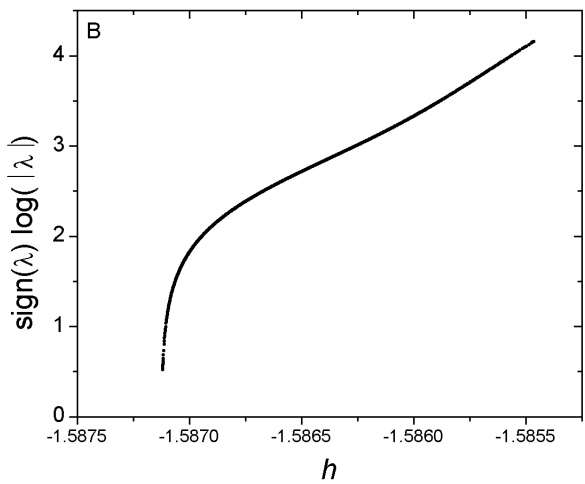
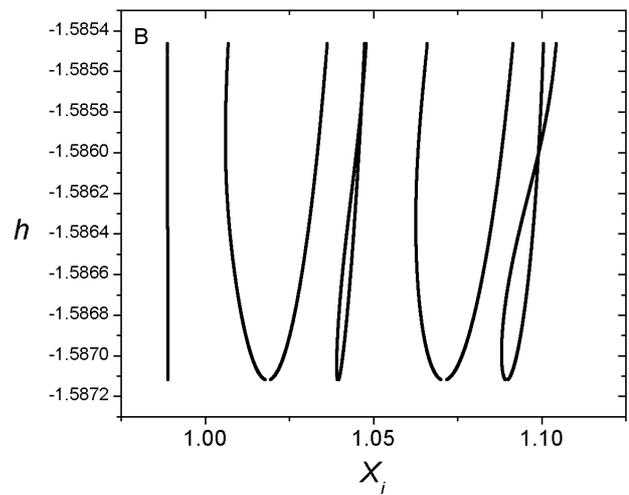

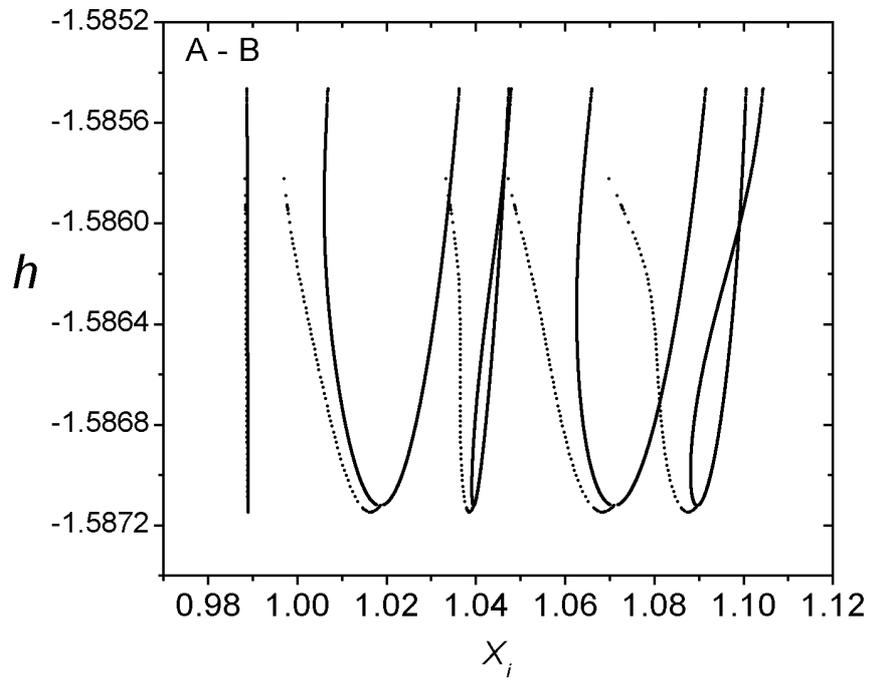

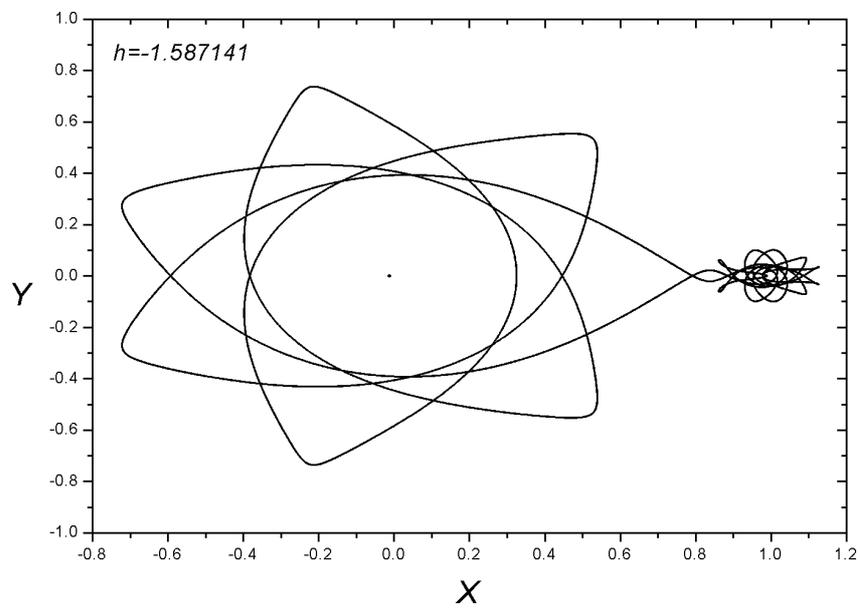

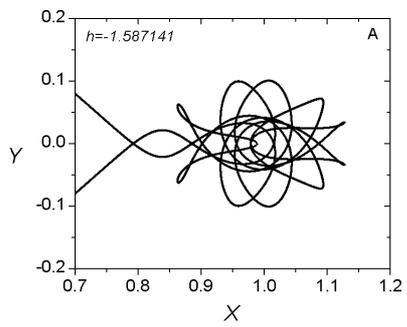
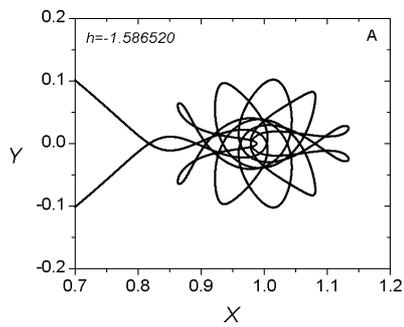
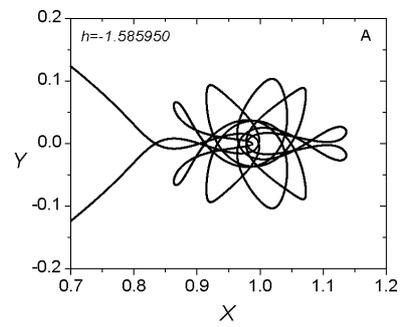

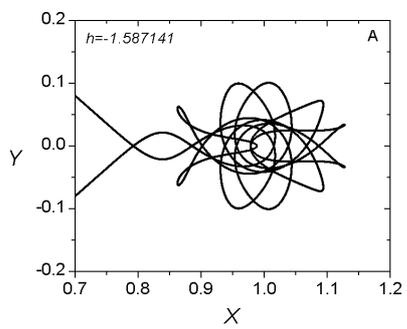
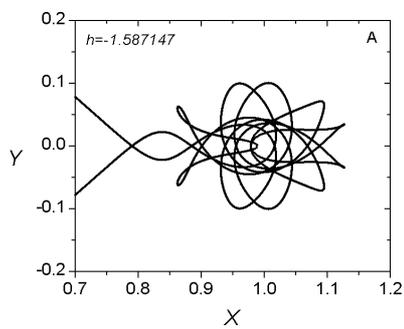
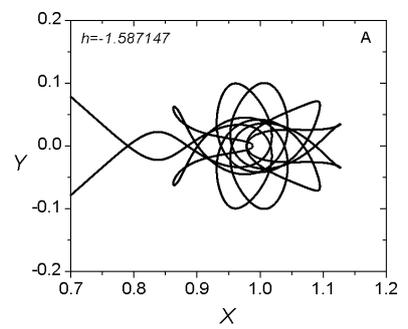

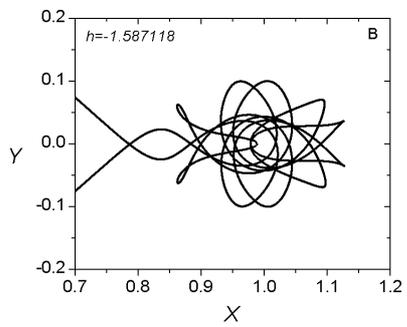
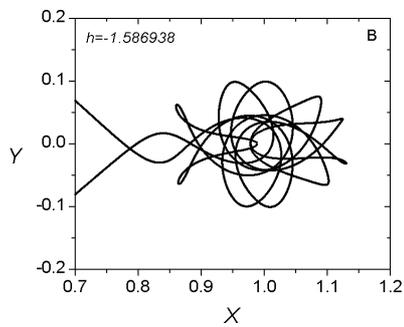
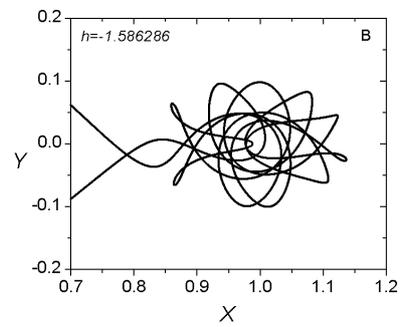

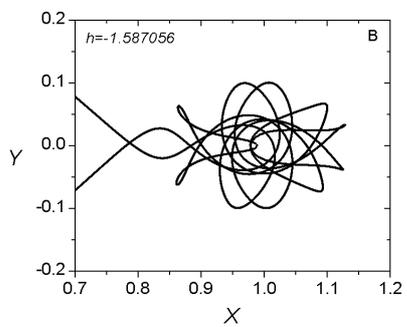
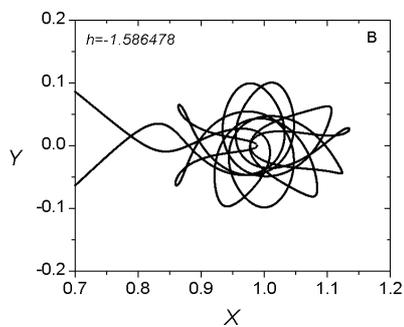
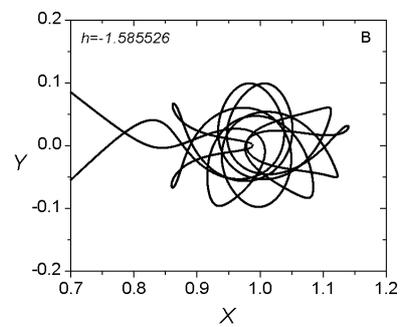

## Family 254  - *Asymmetric family of asymmetric POs*

$h_{min} = -1.593180$,  $h_{max} = -1.592432$,  $T_{min} = 33.984921$,  $T_{max} = 34.194128$

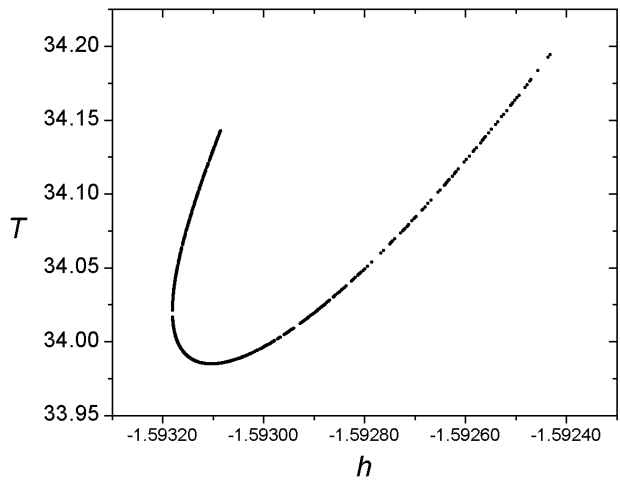
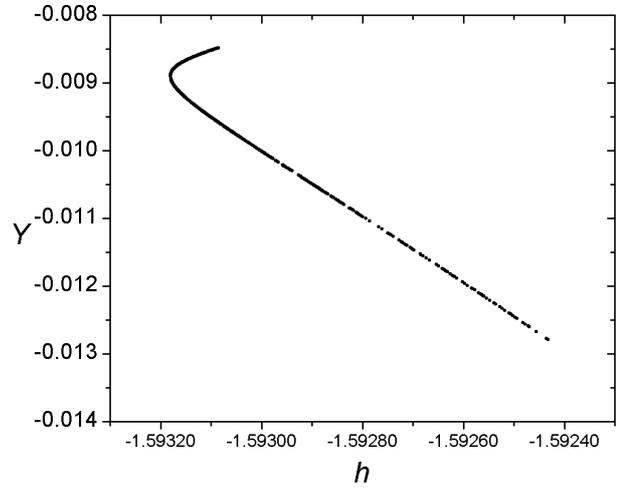

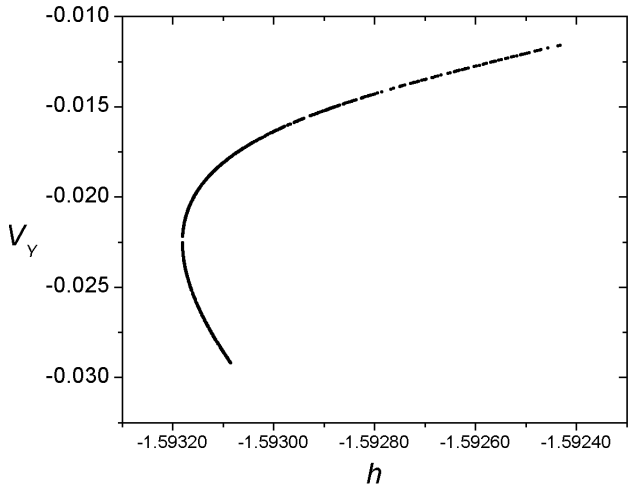
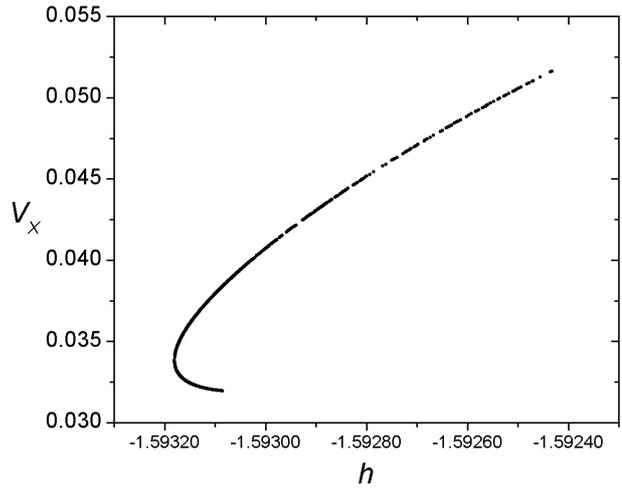

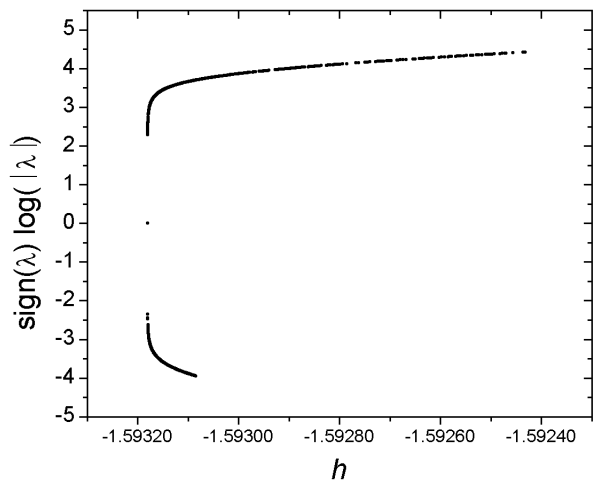
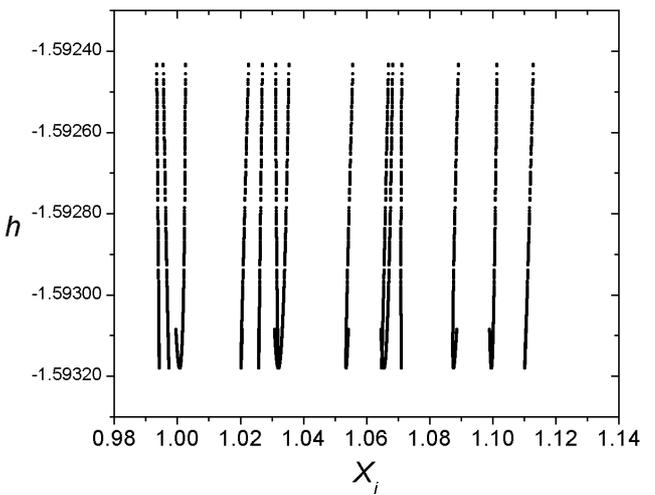

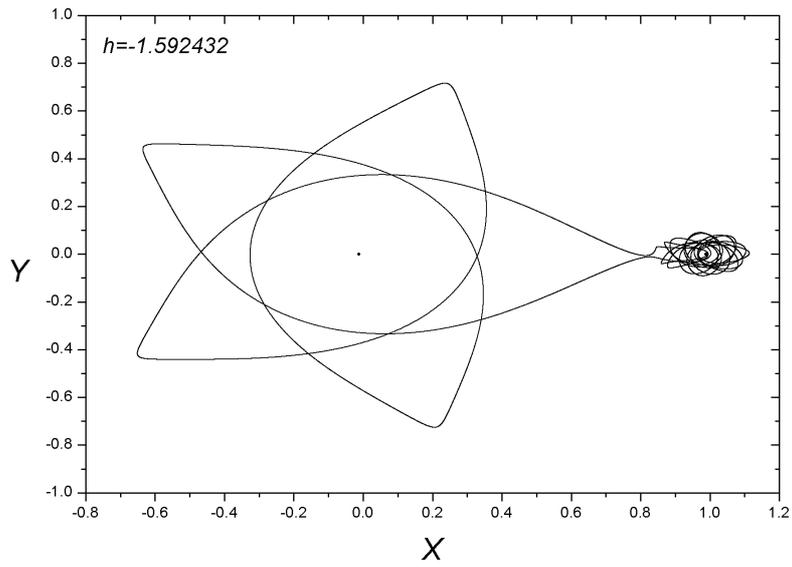

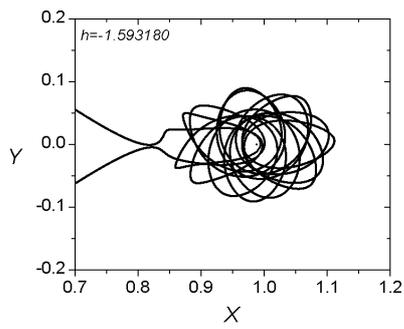
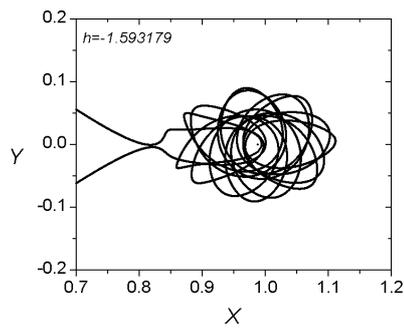
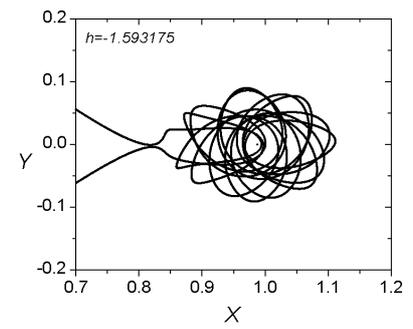

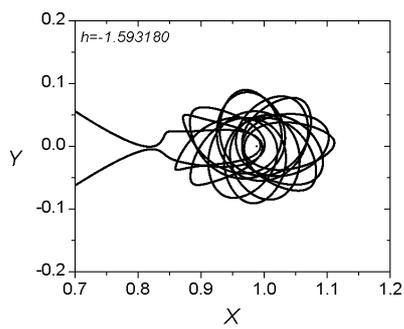
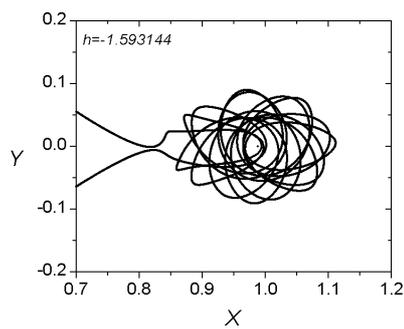
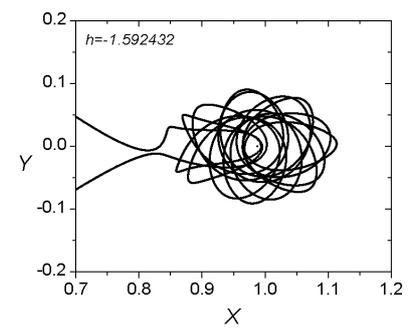

## Family 255 - Asymmetric family of asymmetric POs

$h_{min} = -1.593180$, $h_{max} = -1.592951$, $T_{min} = 33.984922$, $T_{max} = 34.110649$

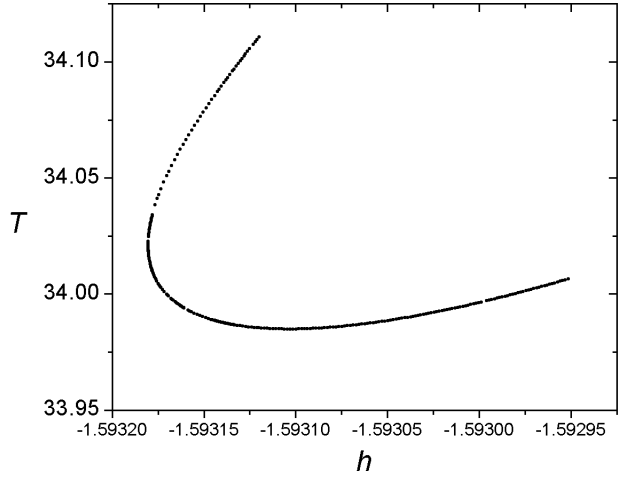
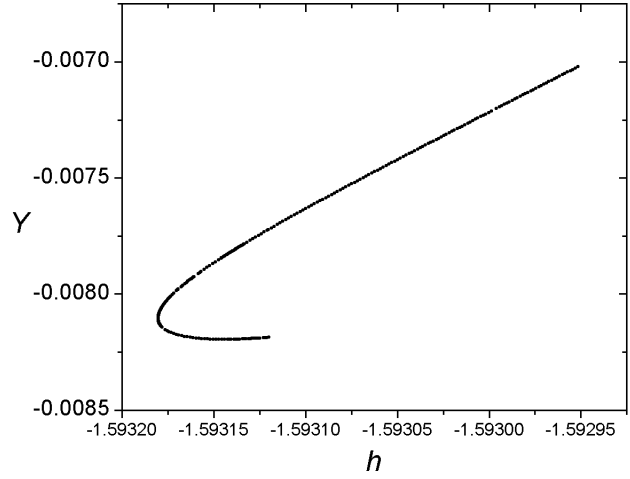

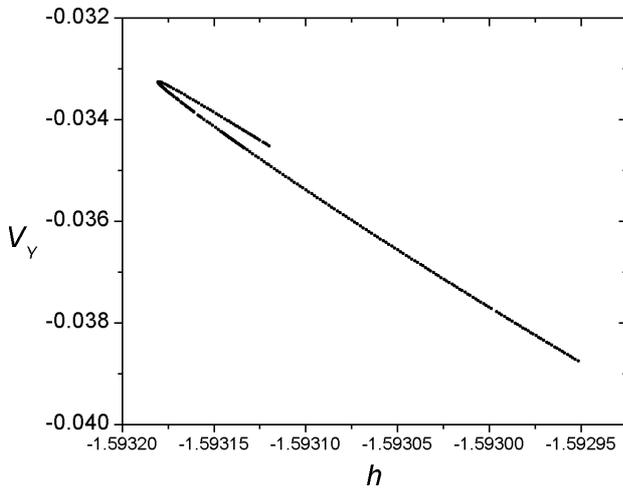
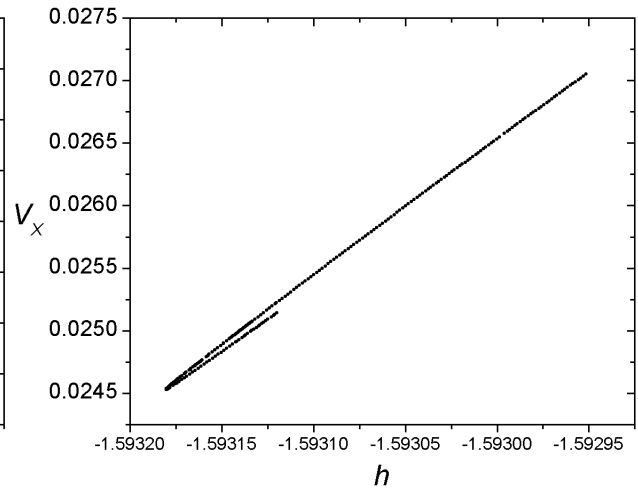

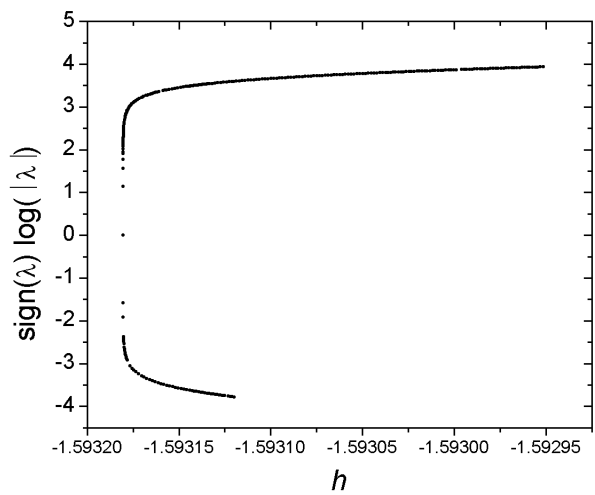
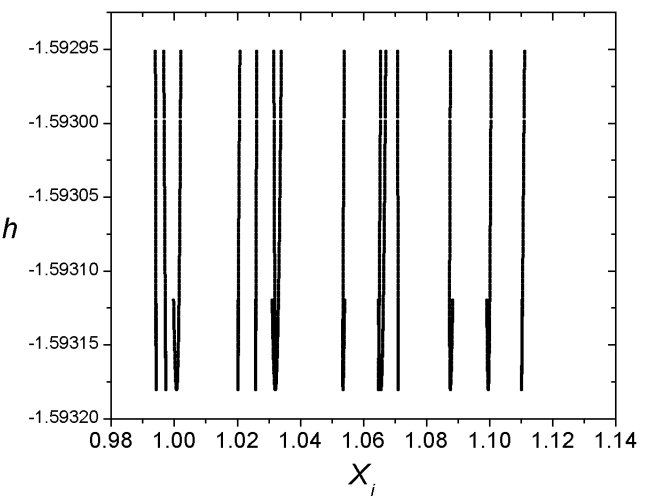

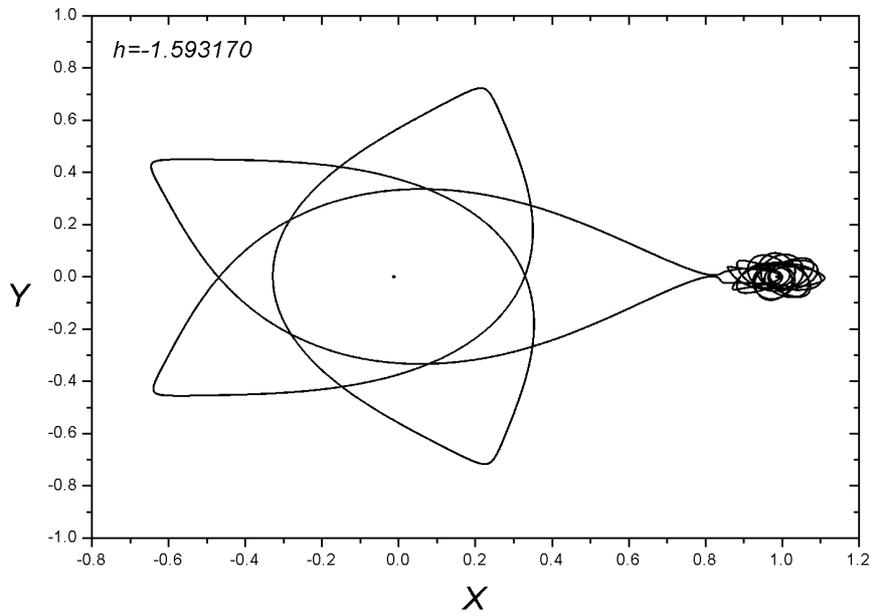

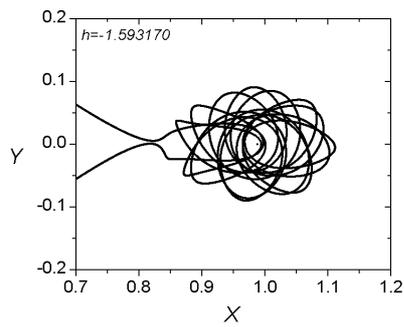
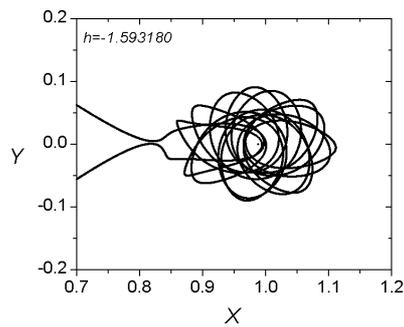
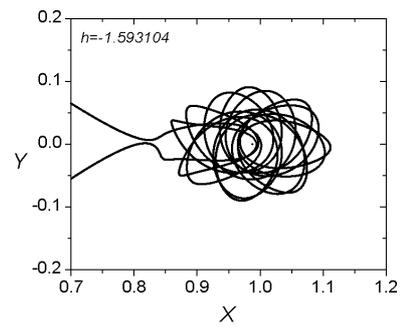

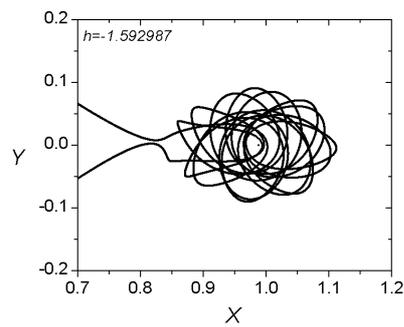
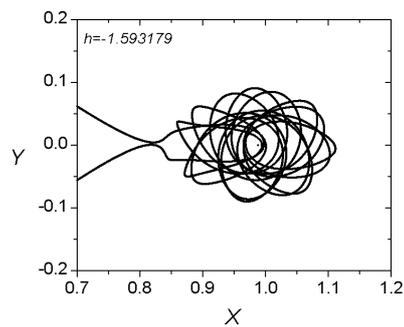
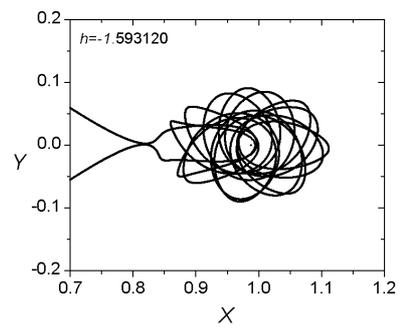

### *Family 026  - Asymmetric family of asymmetric POs*

$h_{min} = -1.587204$,  $h_{max} = -1.561246$,  $T_{min} = 34.017748$,  $T_{max} = 34.834543$

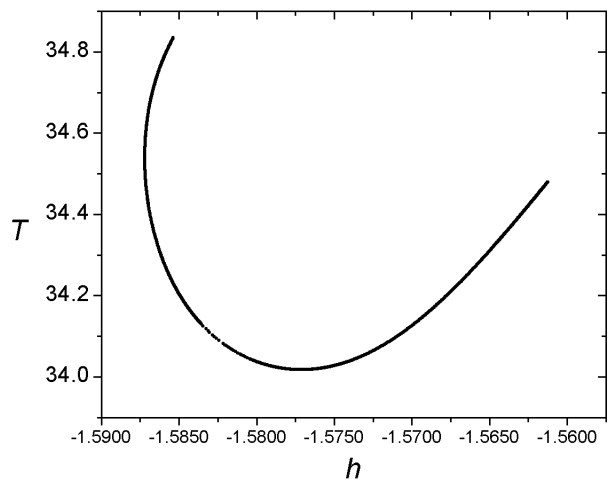

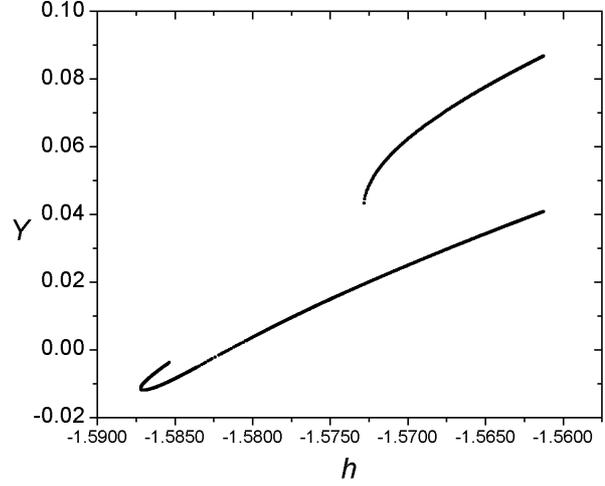

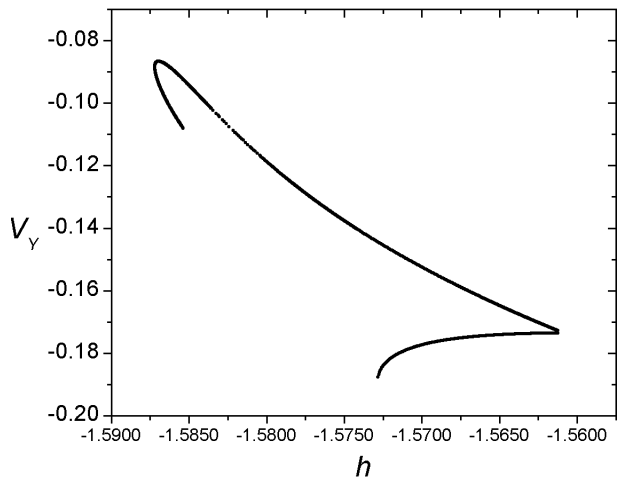

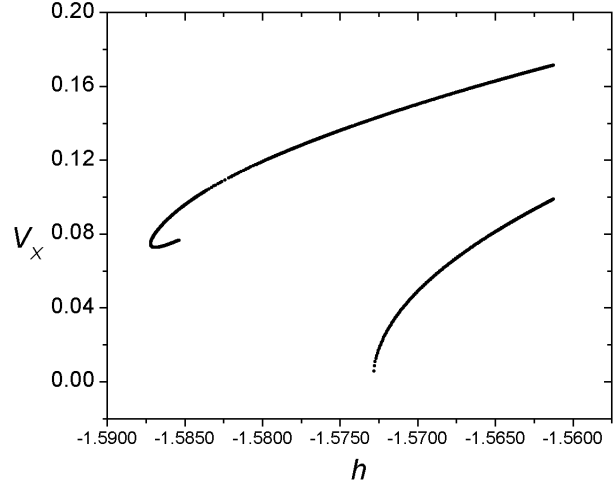

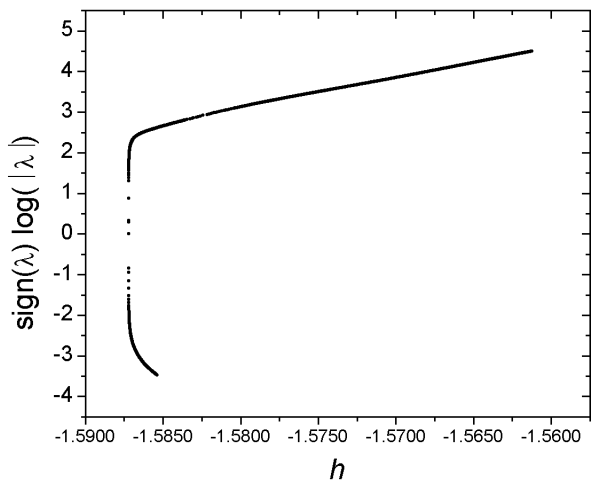

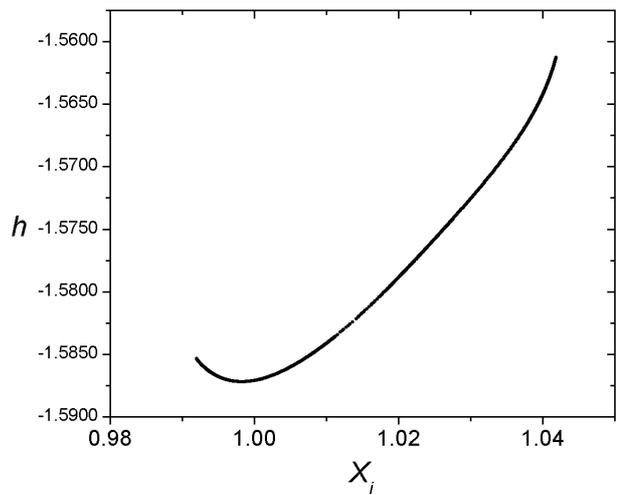

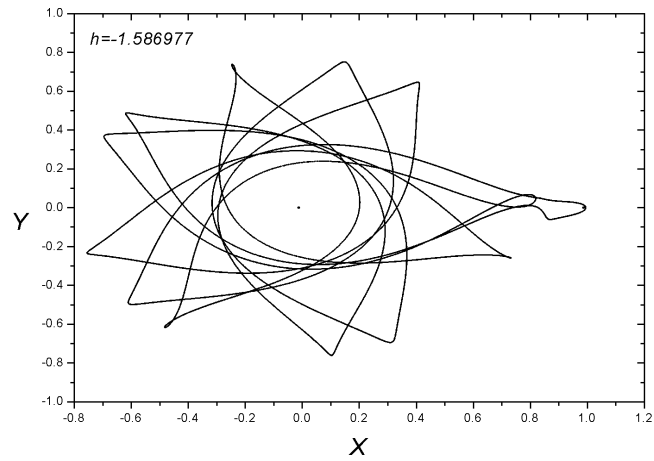

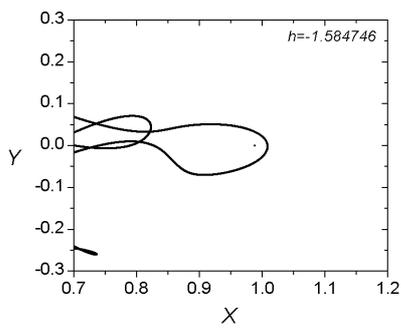
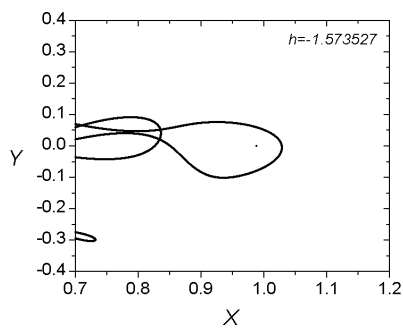
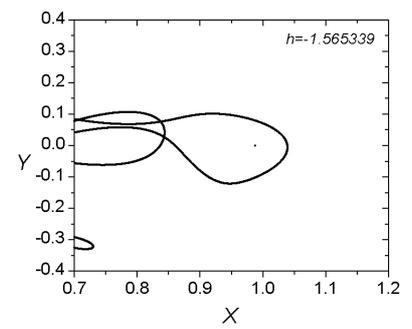

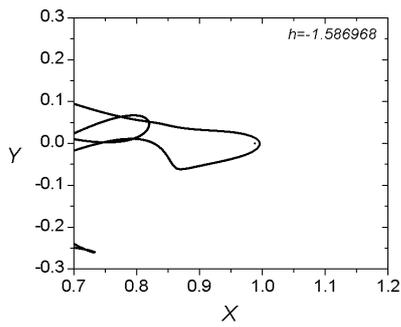
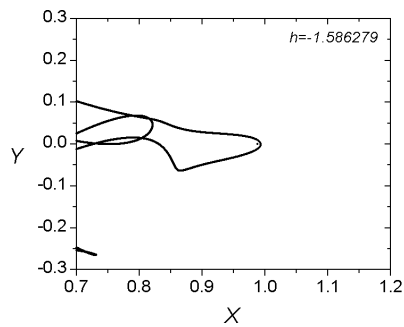
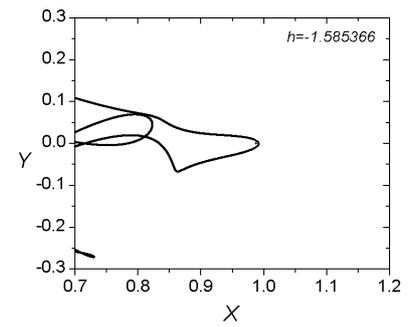

## Family 027 - *Asymmetric family of asymmetric POs*

$h_{min} = -1.587204$, $h_{max} = -1.565404$, $T_{min} = 34.017748$, $T_{max} = 34.790317$

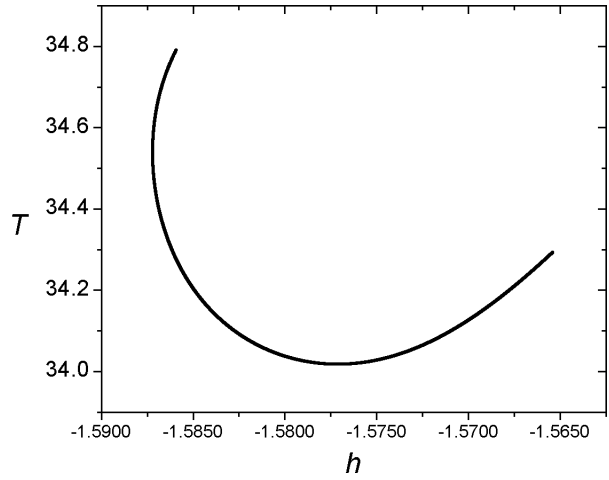
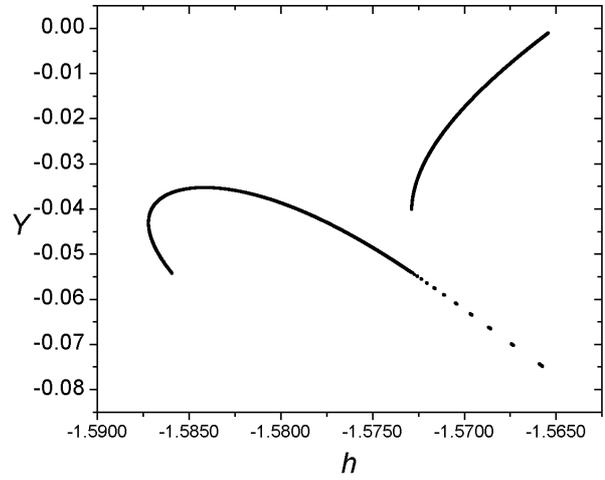
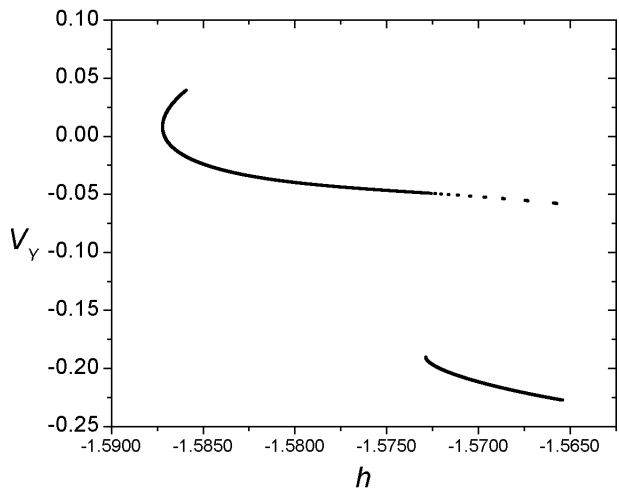
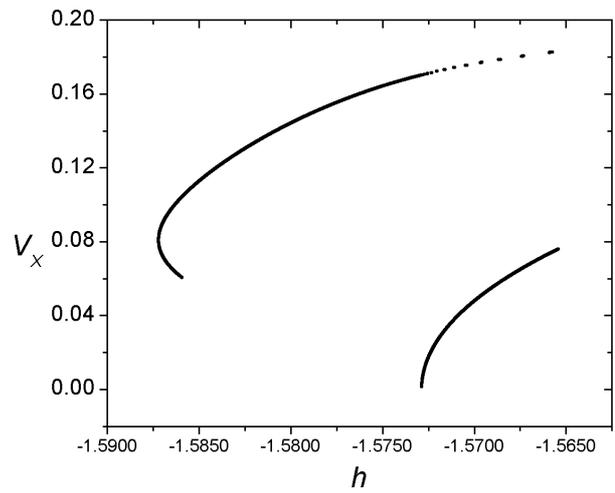
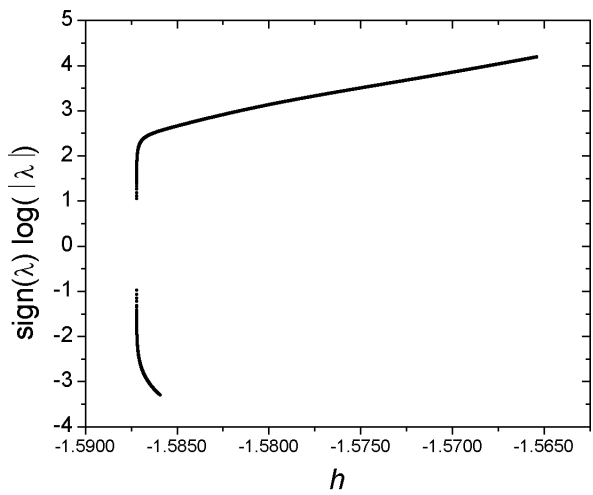
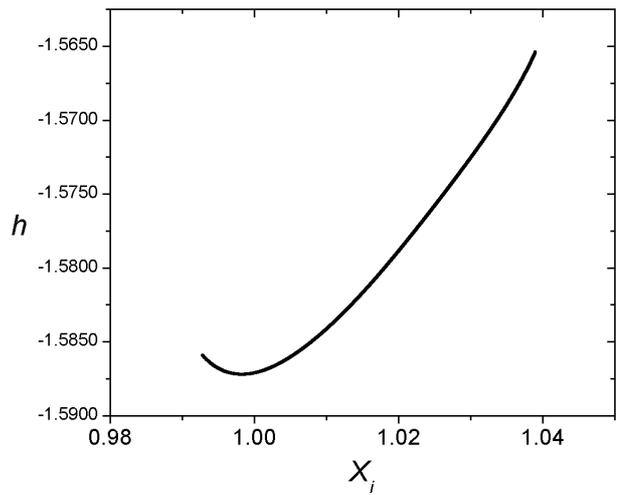

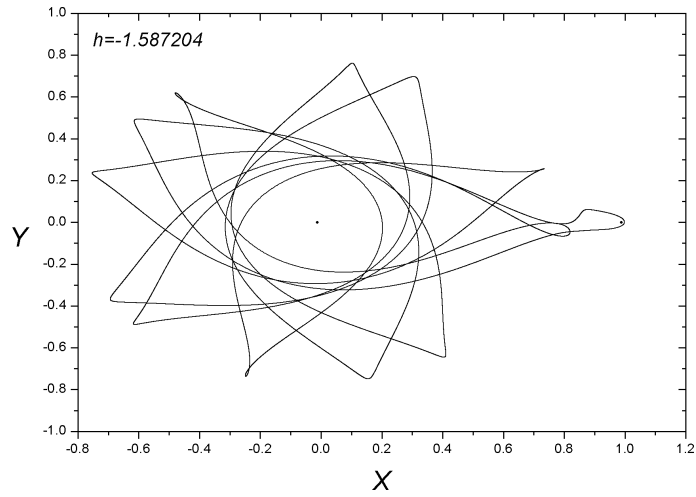

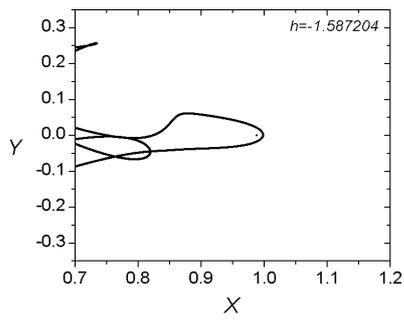
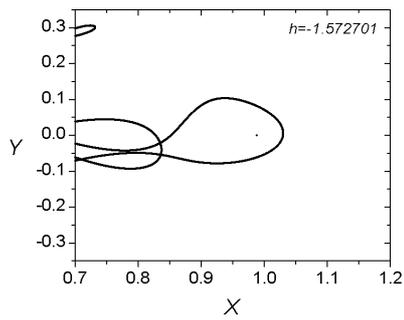
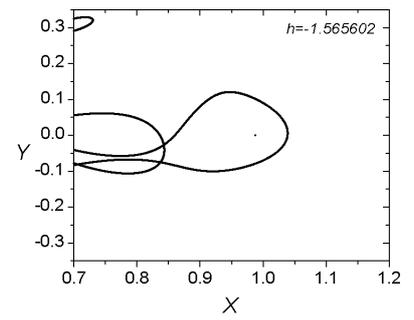

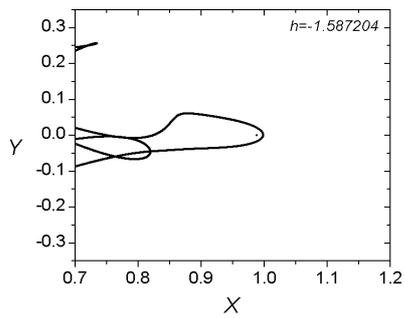
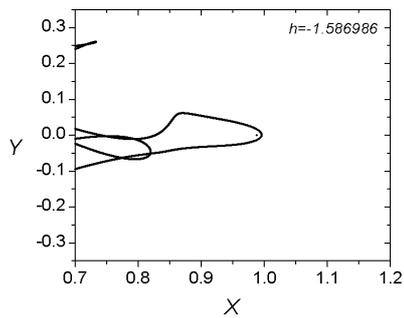
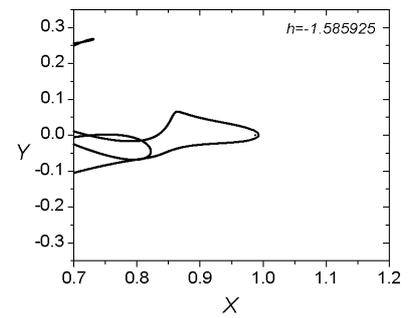

## *Family 044 - Asymmetric family of asymmetric POs*

$h_{min} = -1.587818,\ \ h_{max} = -1.581369,\ \ T_{min} = 34.285575,\ T_{max} = 34.988350$

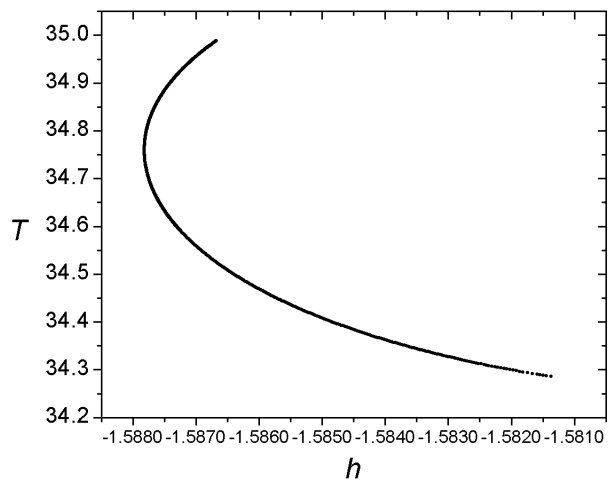
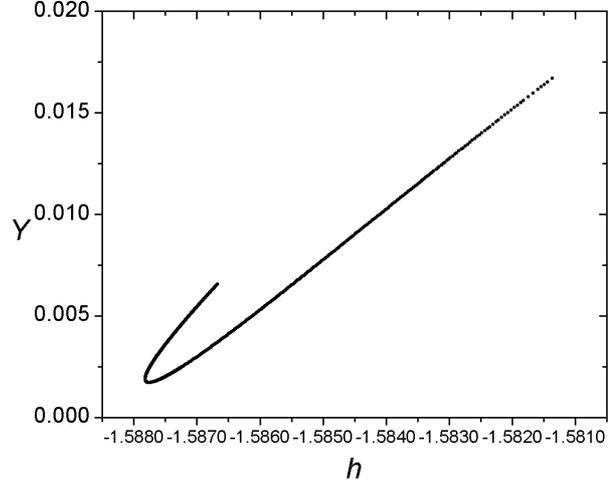
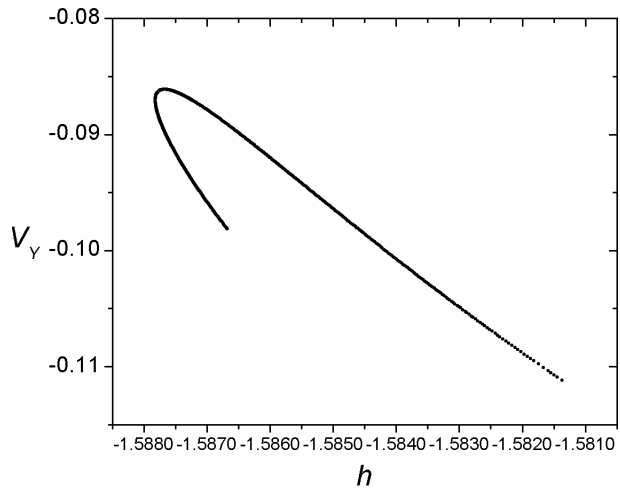
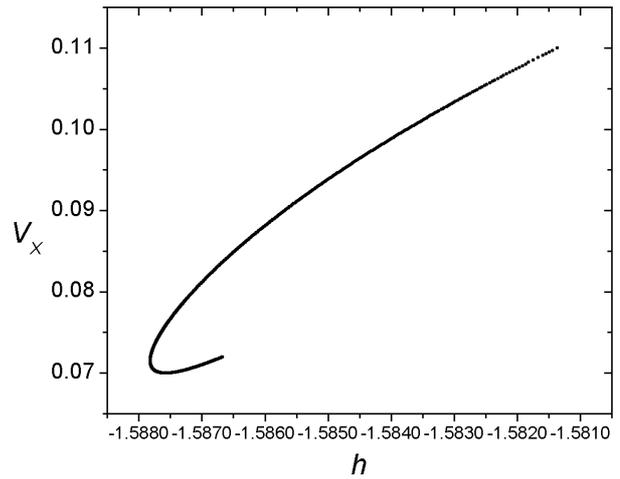
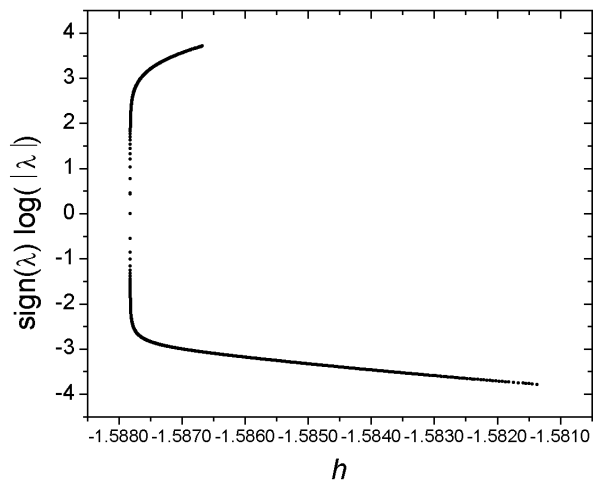
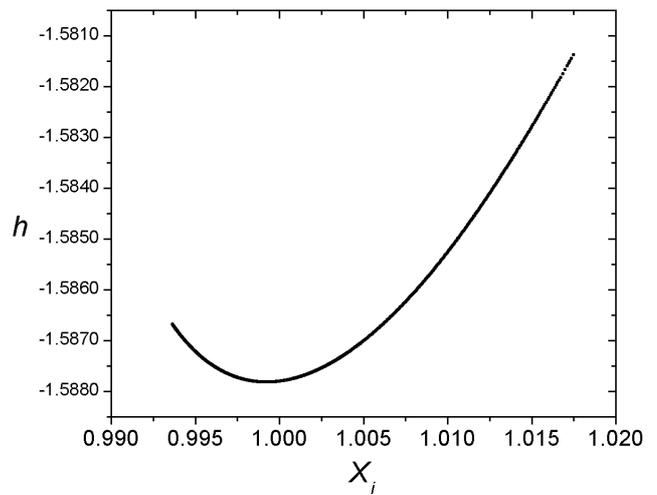

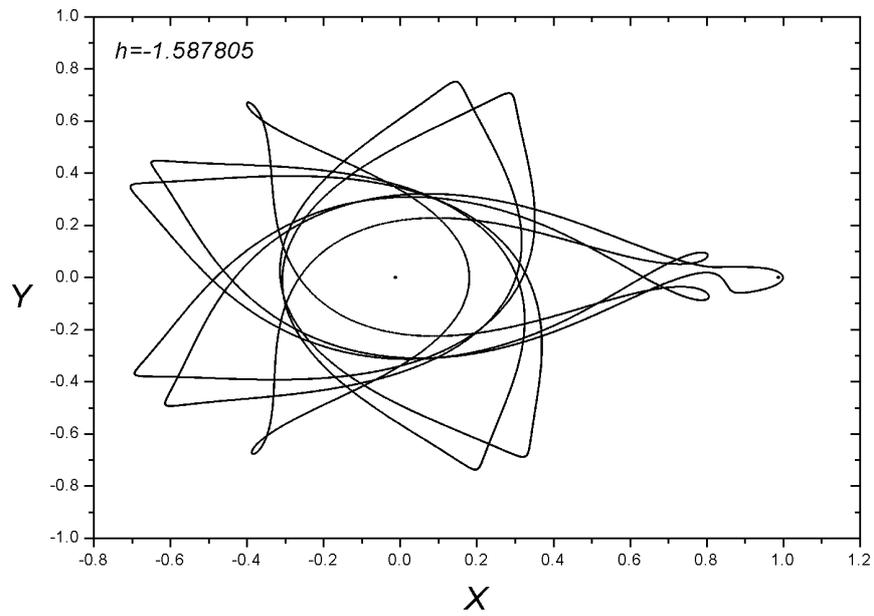

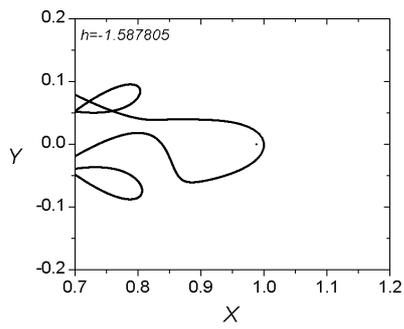
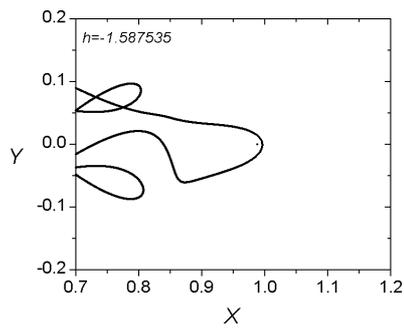
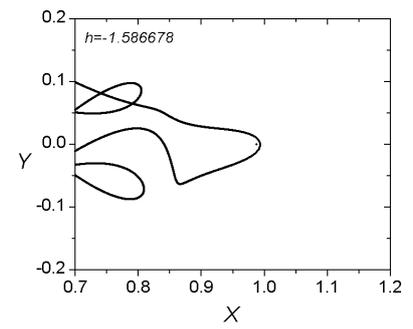

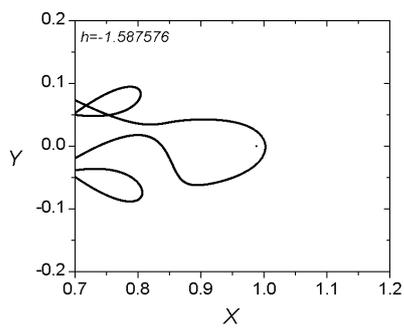
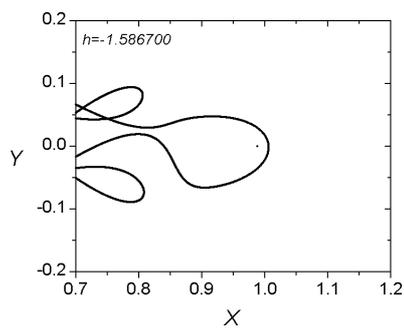
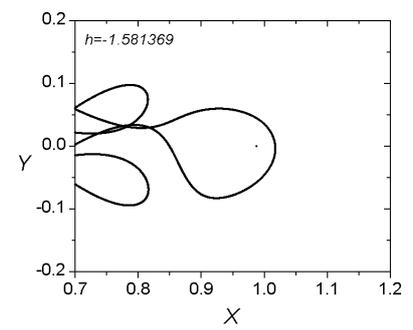

# Family 045  - Asymmetric family of asymmetric POs

$h_{min} = -1.587818, \ h_{max} = -1.581568, \ T_{min} = 34.289797, \ T_{max} = 34.936683$

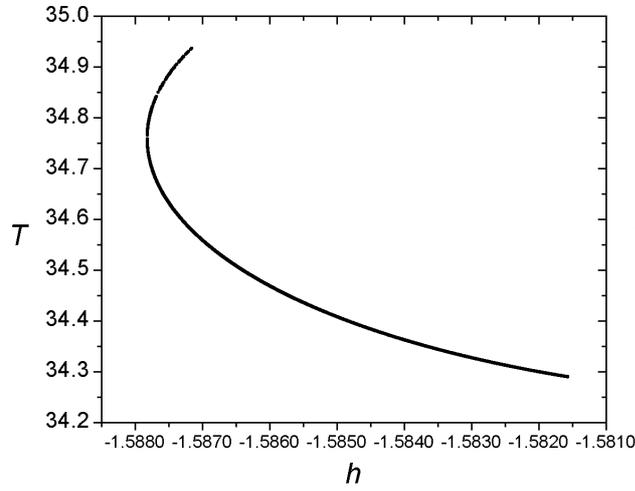
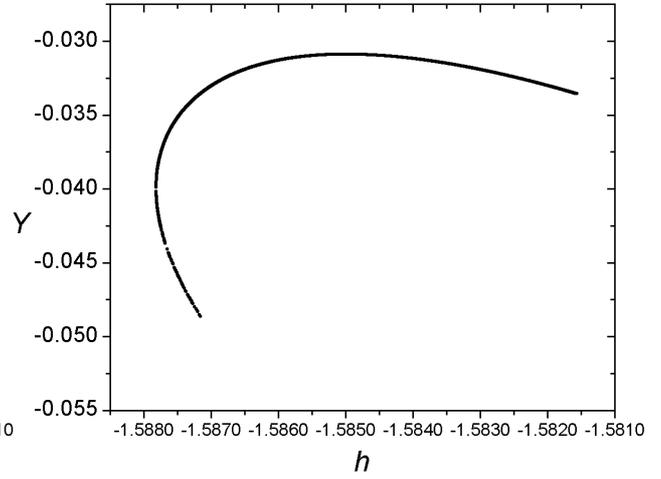
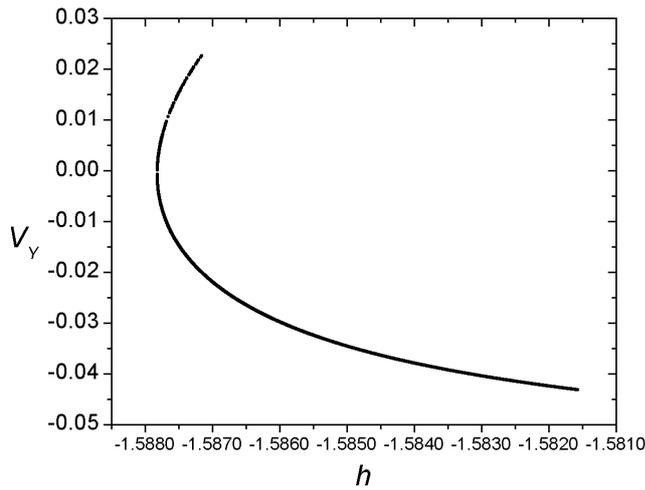
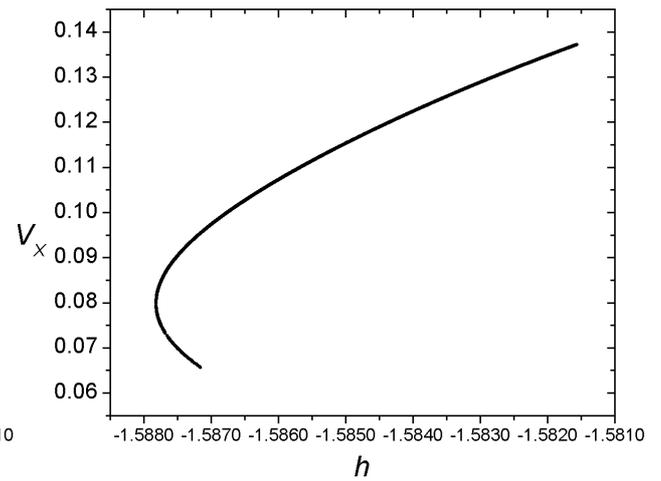
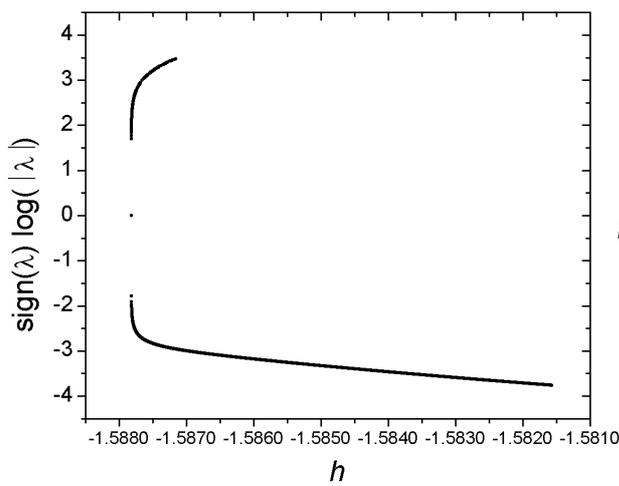
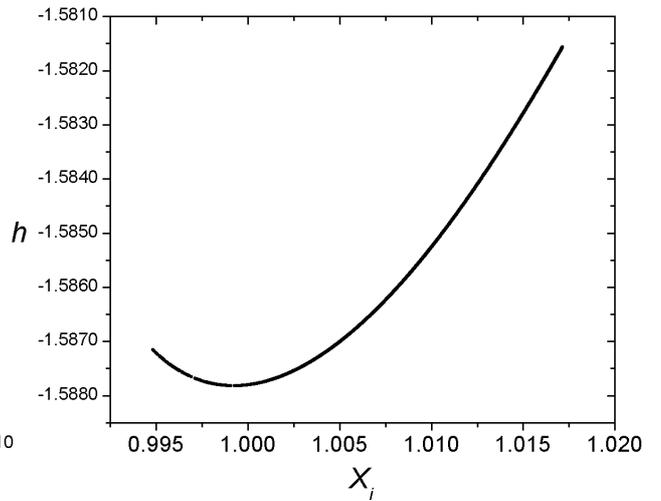

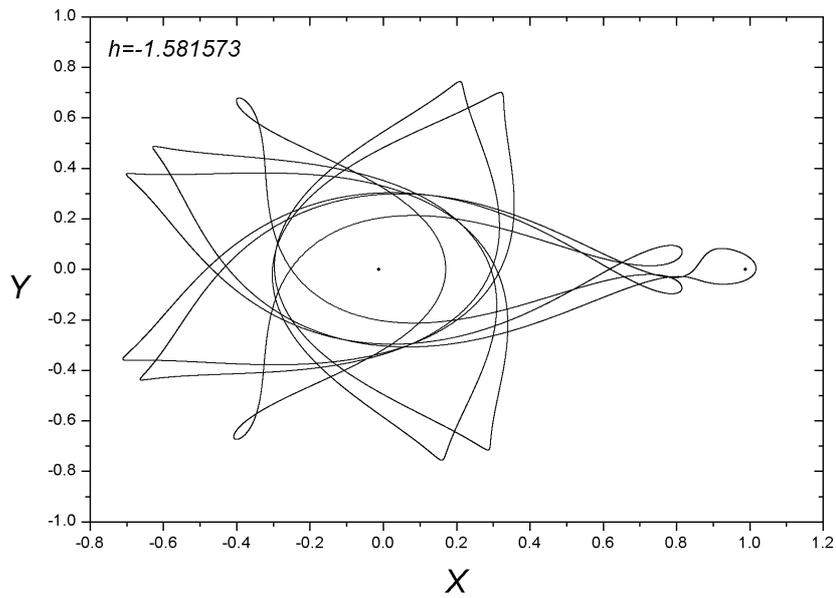

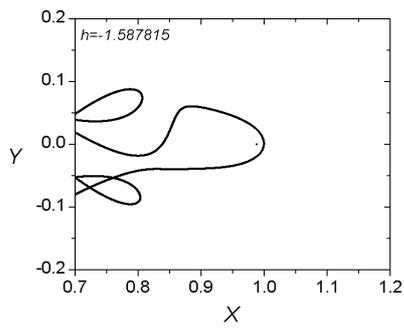
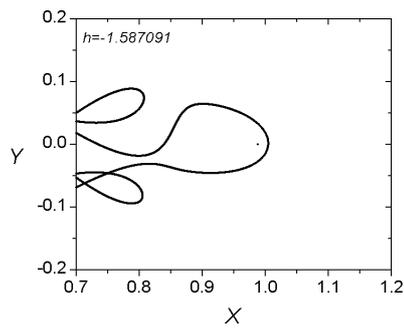
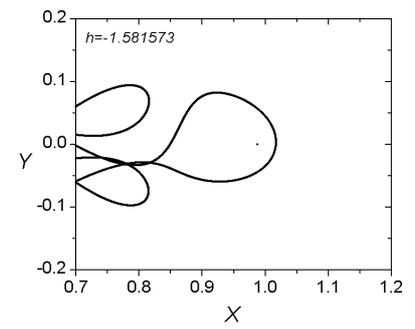

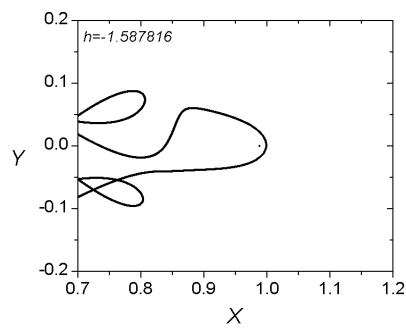
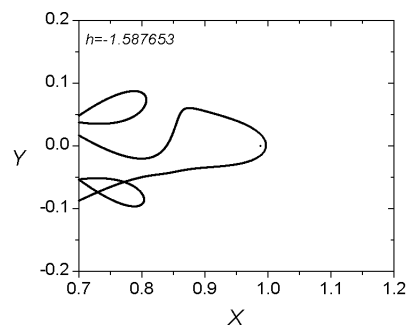
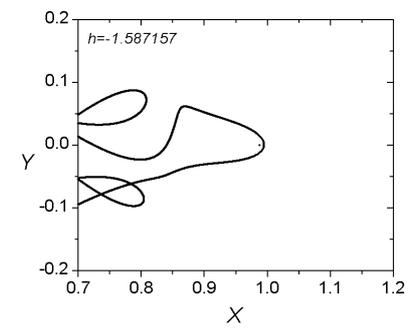

### Family 132 - Asymmetric family of asymmetric POs

$h_{min} = -1.590049, \ h_{max} = -1.586482, \ T_{min} = 34.758144, \ T_{max} = 35.307247$

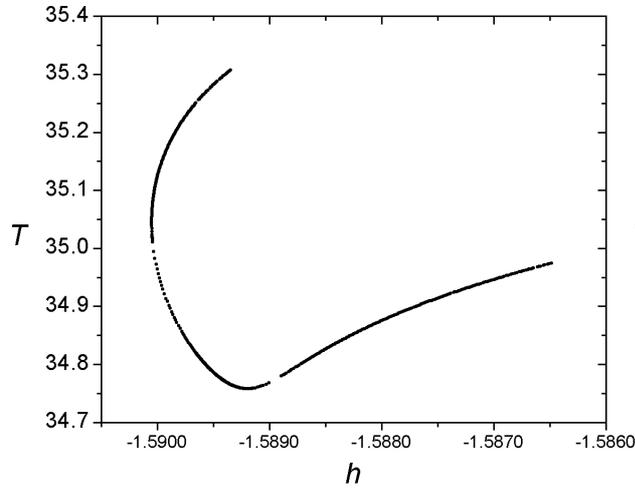
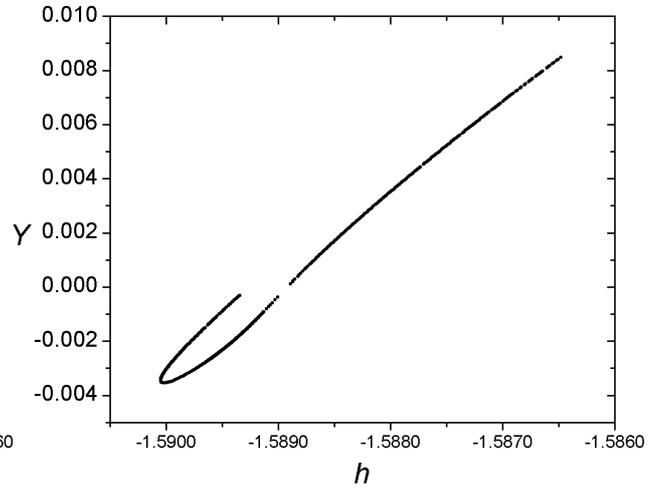

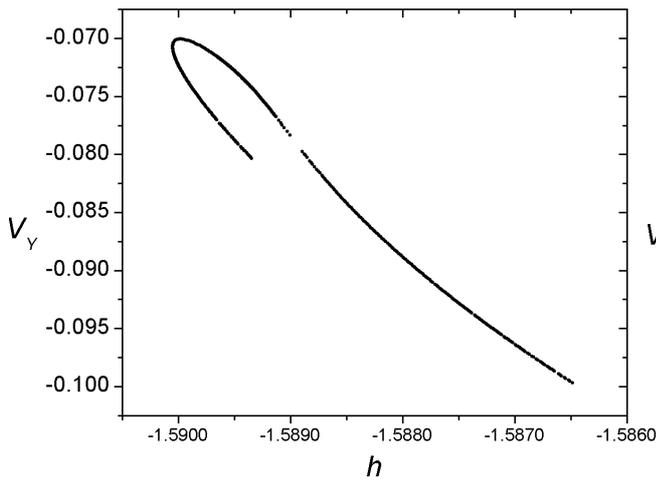
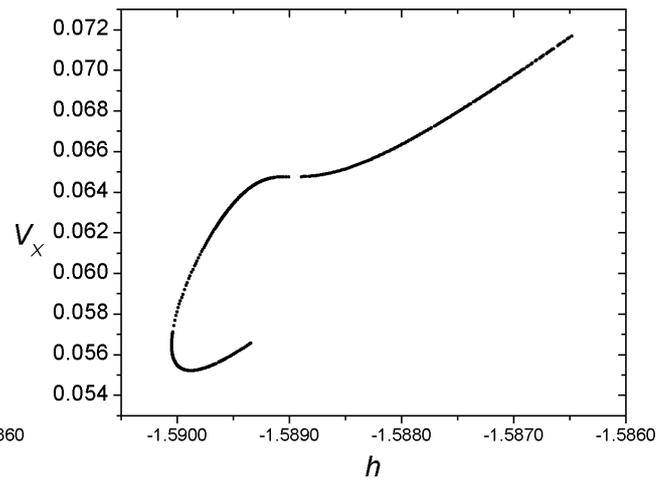

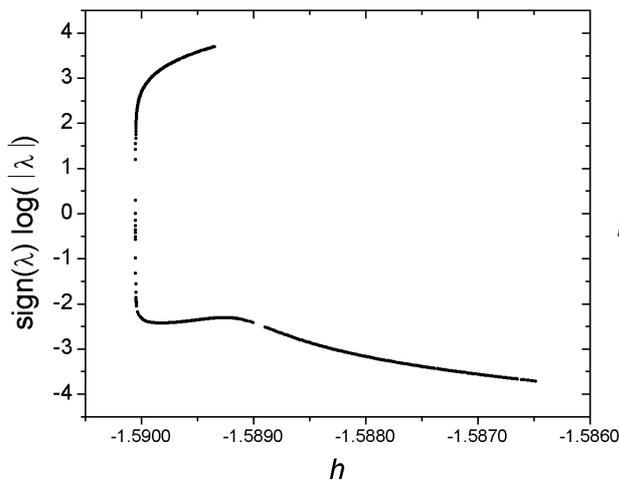
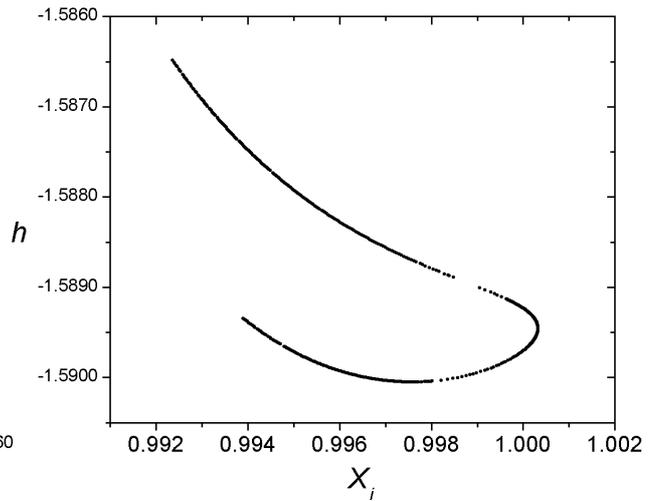

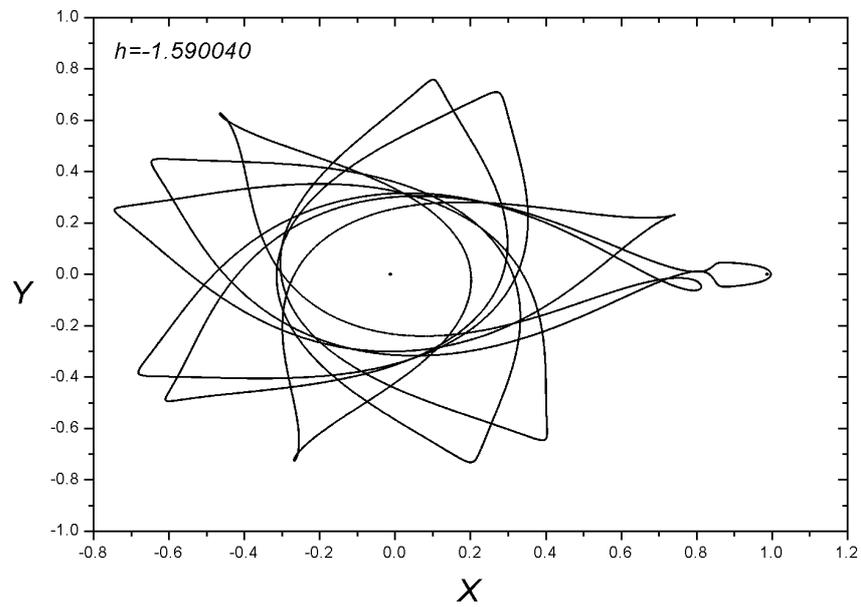

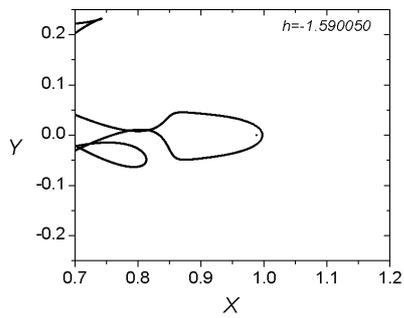

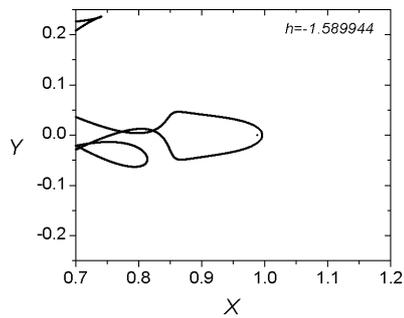

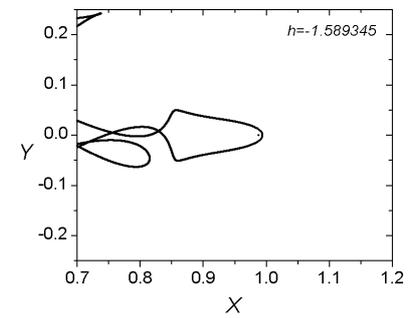

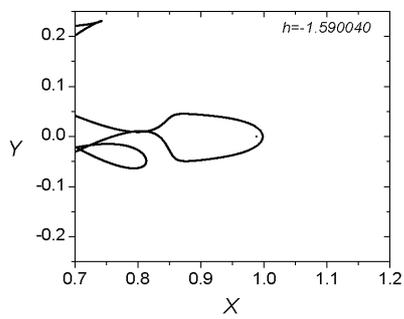

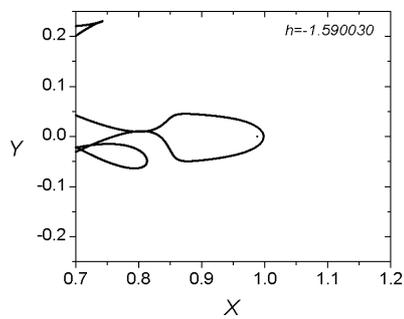

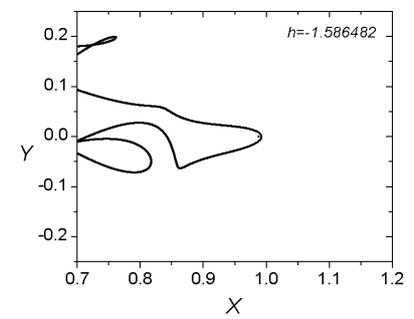

## Family 133  - *Asymmetric family of asymmetric POs*

$h_{min} = -1.590049, \ h_{max} = -1.587059, \ T_{min} = 34.758144, \ T_{max} = 35.299056$

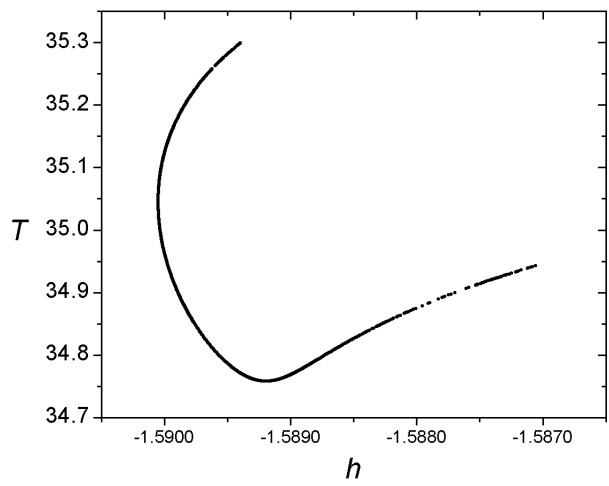
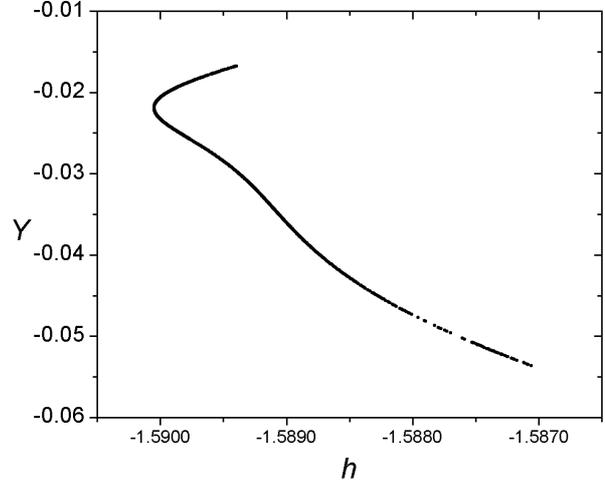

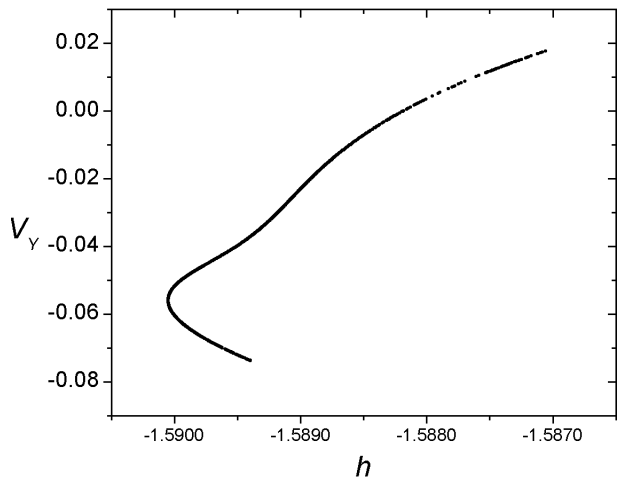
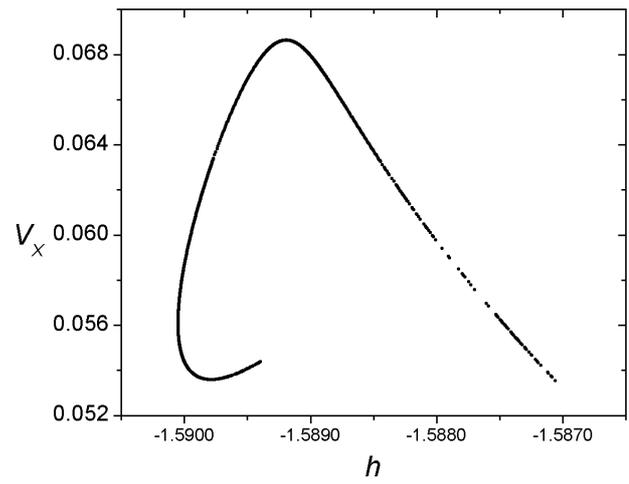

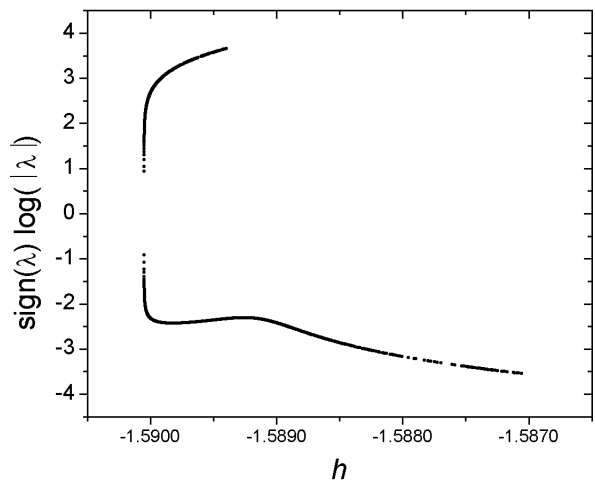
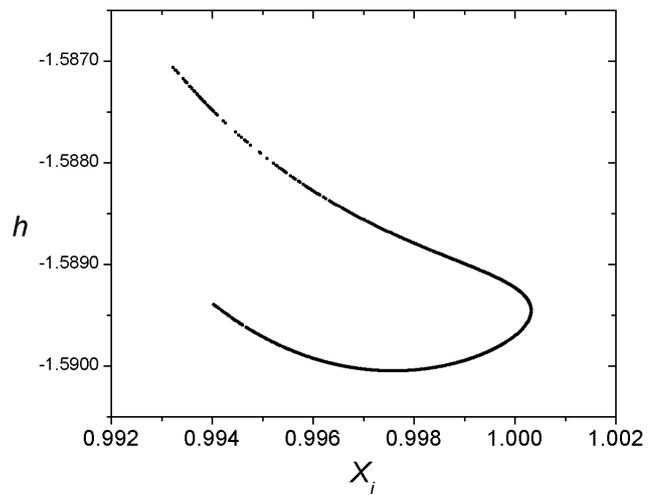

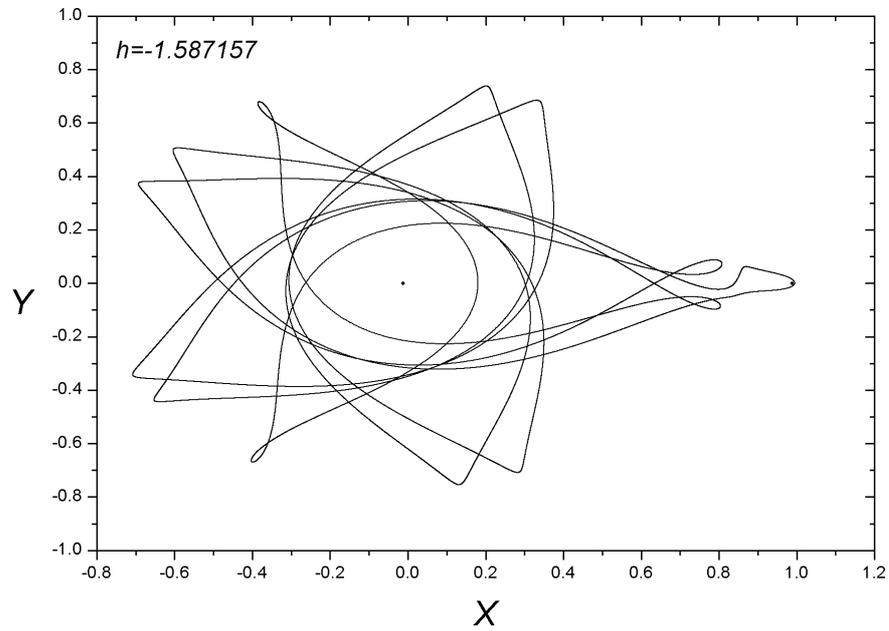

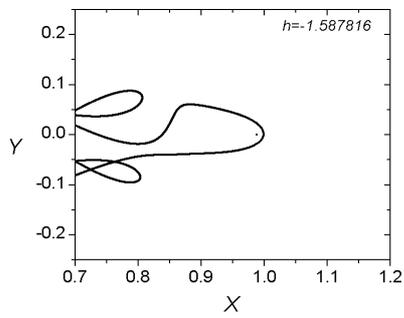 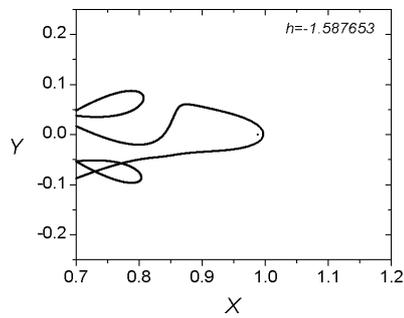 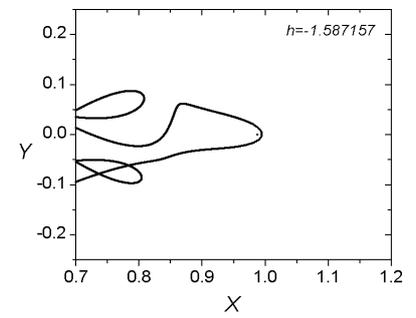

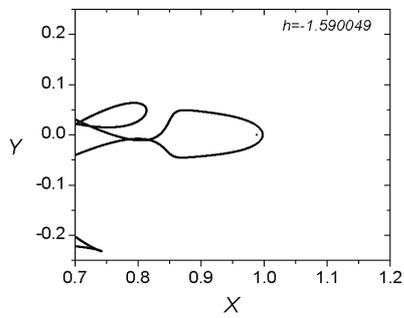 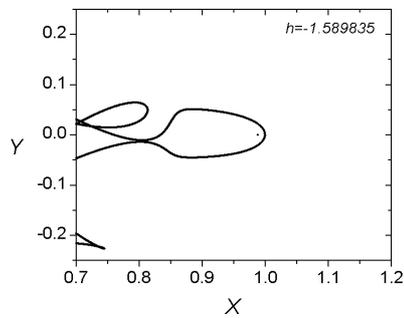 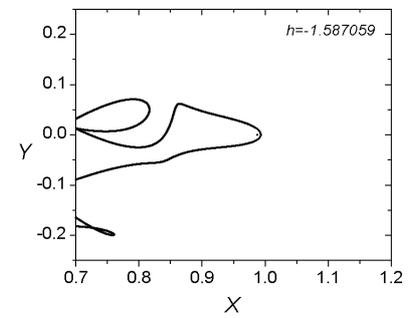

## Families  256 A - 256 B

*Bifurcation Point*

|  | $h$ | $T$ | $y$ | $v_y$ | $v_x$ |
|---|---|---|---|---|---|
| $P_1$ | -1.593178 | 35.263386 | -0.008313 | -0.023497 | 0.033857 |

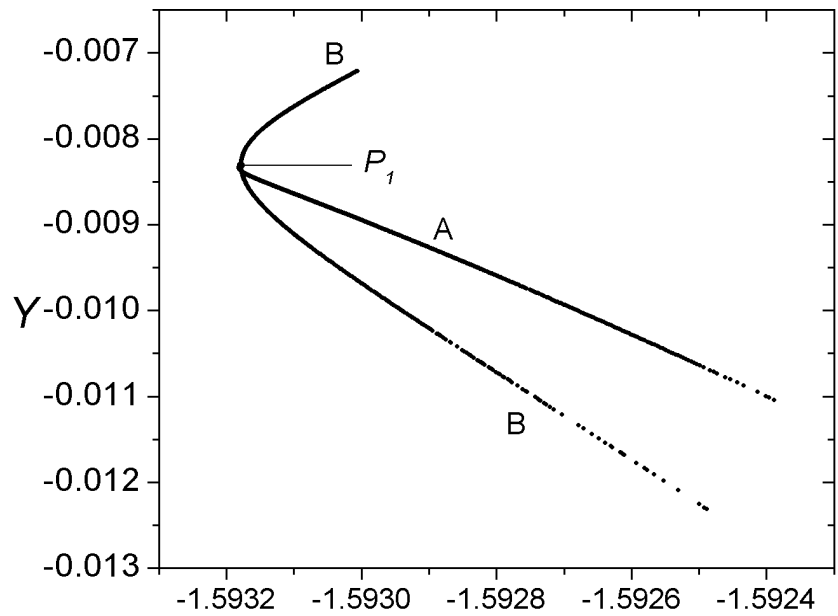

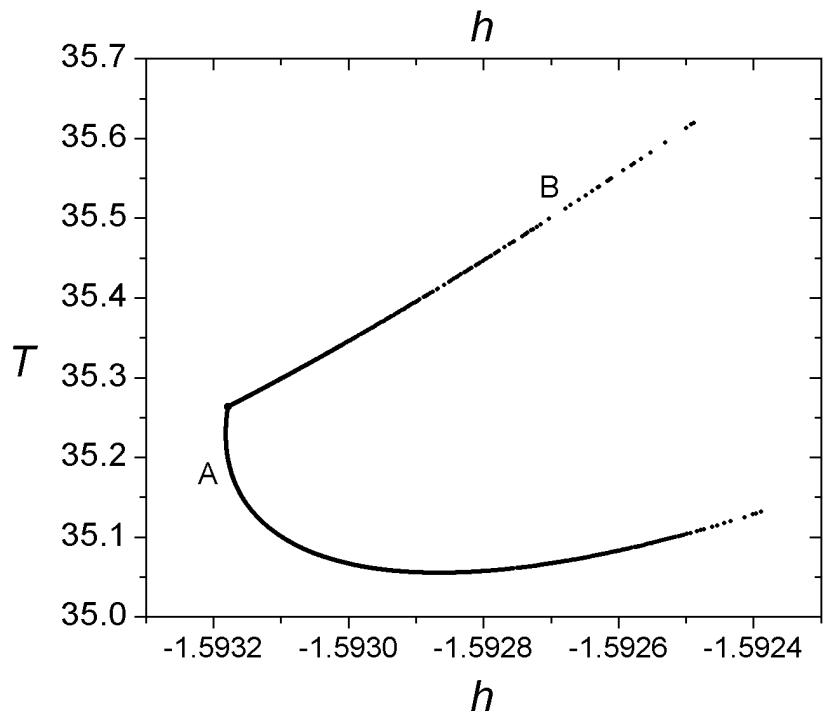

## Family 256 A - Symmetric family of symmetric POs

$h_{min} = -1.593181, \ h_{max} = -1.592388, \ T_{min} = 35.055198, \ T_{max} = 35.263456$

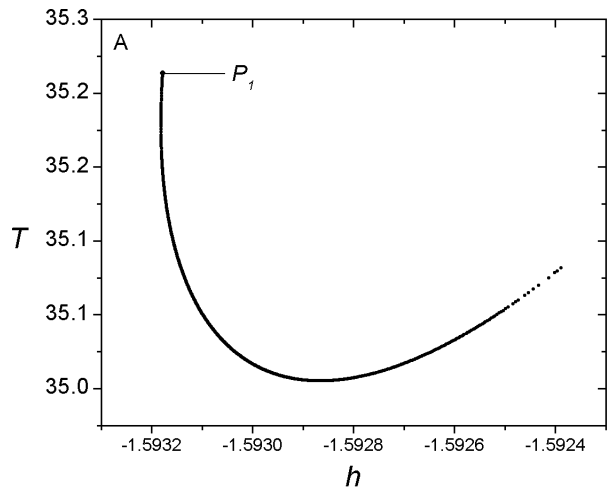
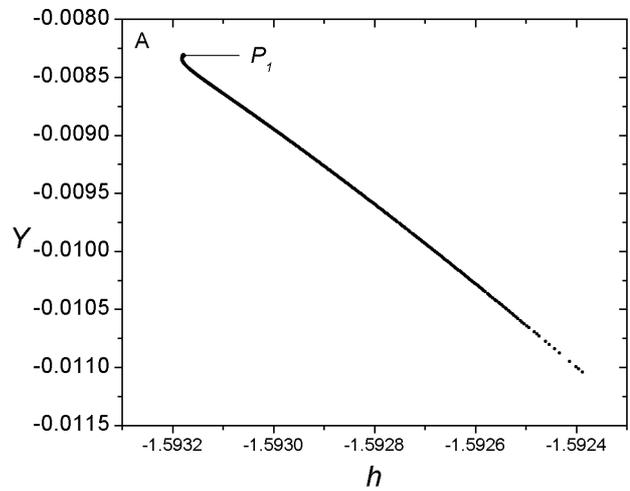

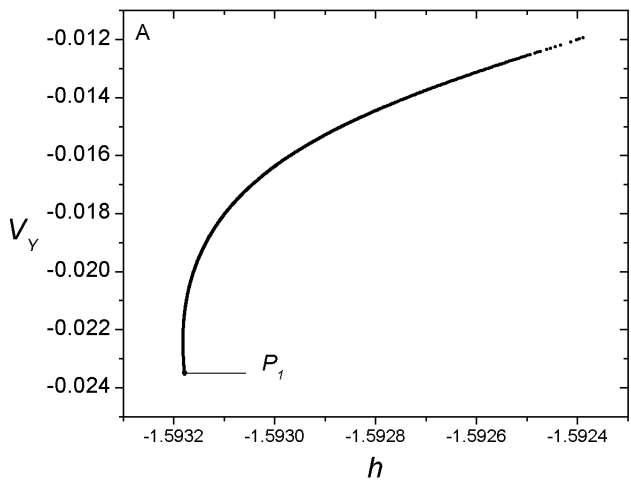
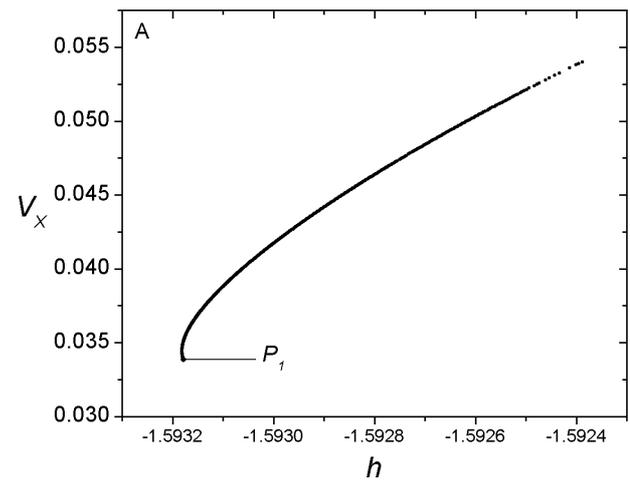

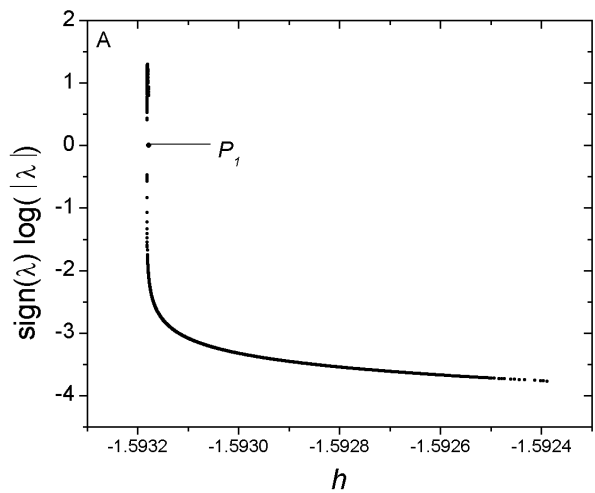
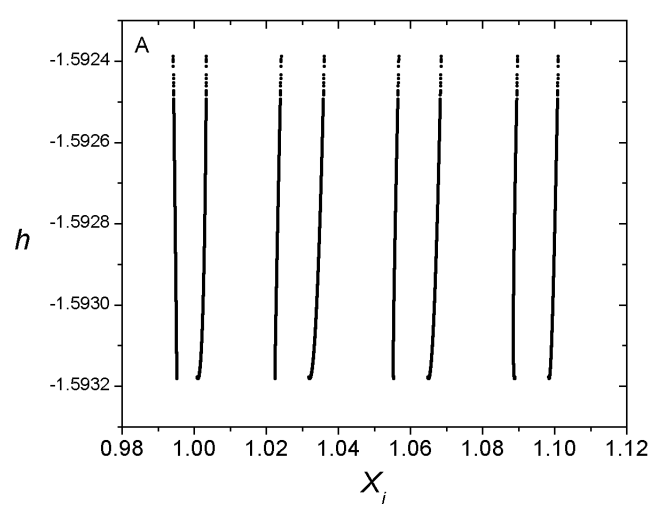

## Family 256 B - Symmetric family of asymmetric POs

$h_{min} = -1.593178$, $h_{max} = -1.592489$, $T_{min} = 35.263374$, $T_{max} = 35.619251$

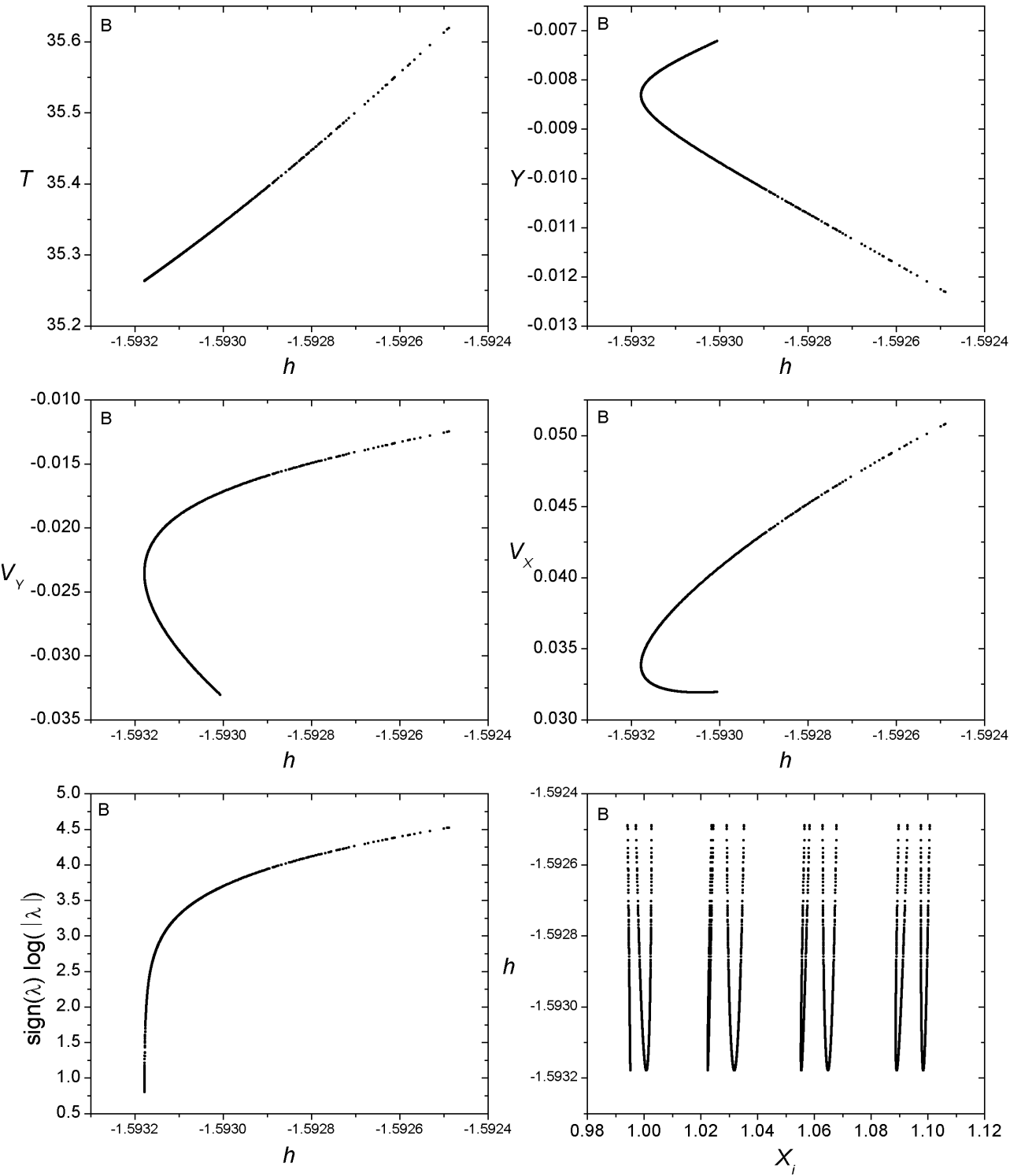

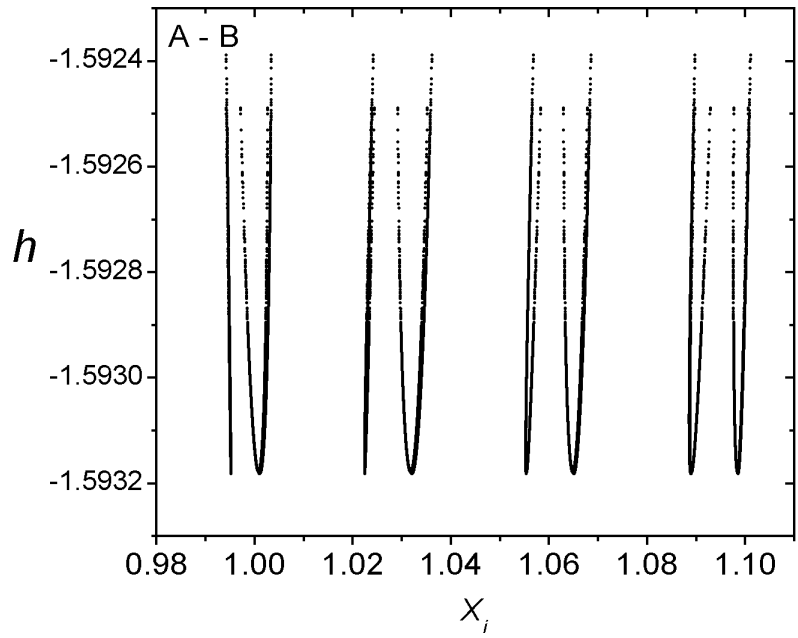

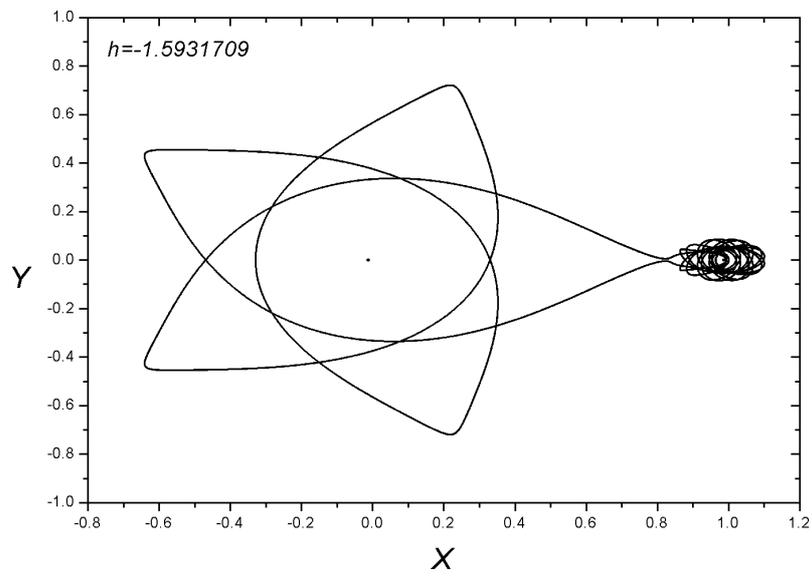

## Families  300 A - 300 B - 300 C - 300 D

*Bifurcation Points*

|       | $h$       | $T$       | $y$       | $v_y$     | $v_x$     |
|-------|-----------|-----------|-----------|-----------|-----------|
| $P_1$ | -1.588863 | 36.260497 | -0.013748 | -0.029820 | 0.094579  |
| $P_2$ | -1.588614 | 36.286616 | -0.018195 | -0.020671 | 0.096566  |
| $P_3$ | -1.588600 | 36.416067 | -0.021787 | -0.012071 | 0.095145  |
| $P_4$ | -1.588572 | 36.625223 | -0.025731 | -0.001808 | 0.092185  |
| $P_5$ | -1.588950 | 36.510494 | -0.026499 | -0.002950 | 0.087044  |
| $P_6$ | -1.588950 | 36.510494 | -0.006758 | -0.040001 | 0.093015  |

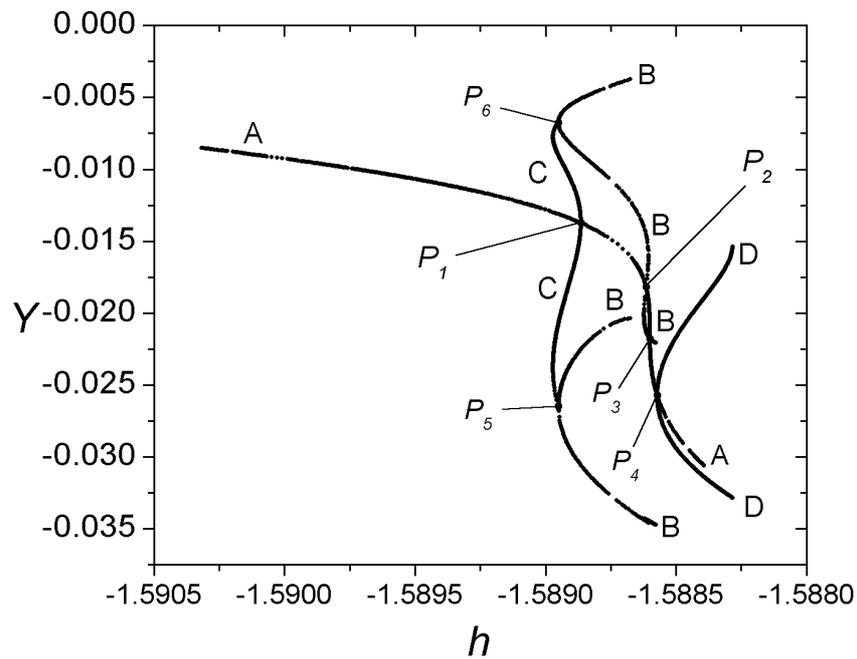

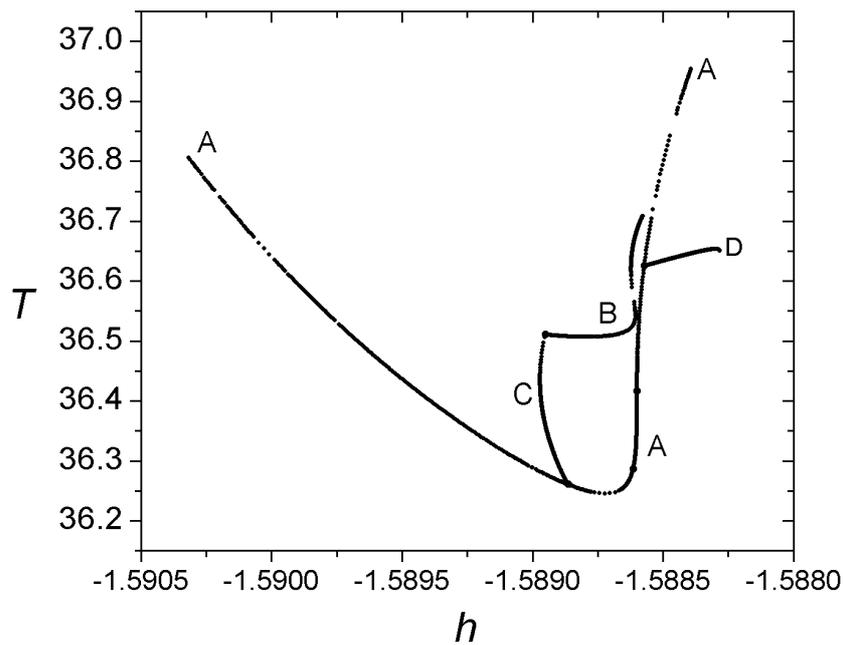

### Family 300 A - *Symmetric family of symmetric POs*

$h_{min} = -1.590317, \quad h_{max} = -1.588392, \quad T_{min} = 36.245398, \quad T_{max} = 36.954144$

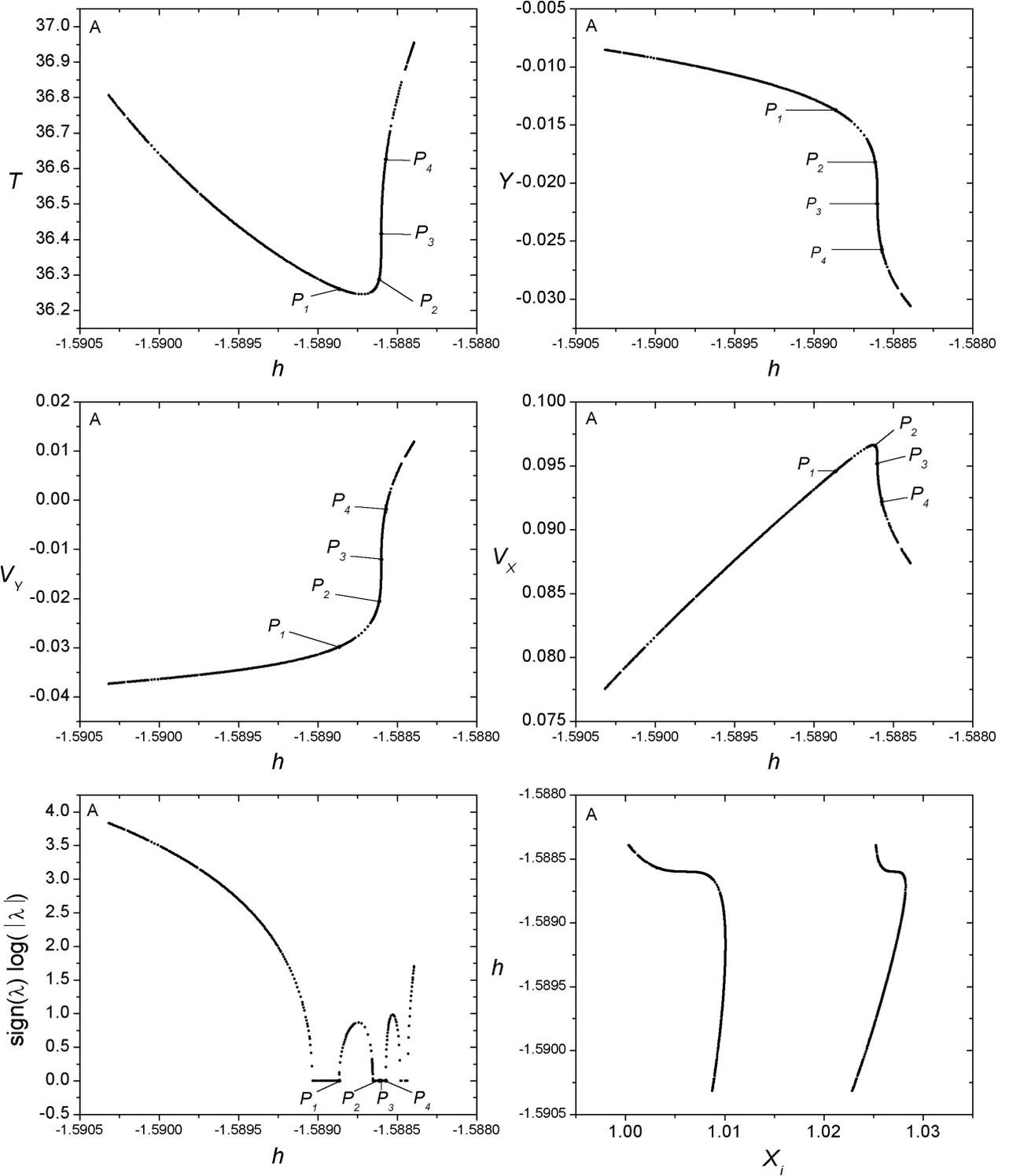

## Family 300 B - Symmetric family of asymmetric POs

$h_{min} = -1.588950, \quad h_{max} = -1.588577, \quad T_{min} = 36.506826, \quad T_{max} = 36.708906$

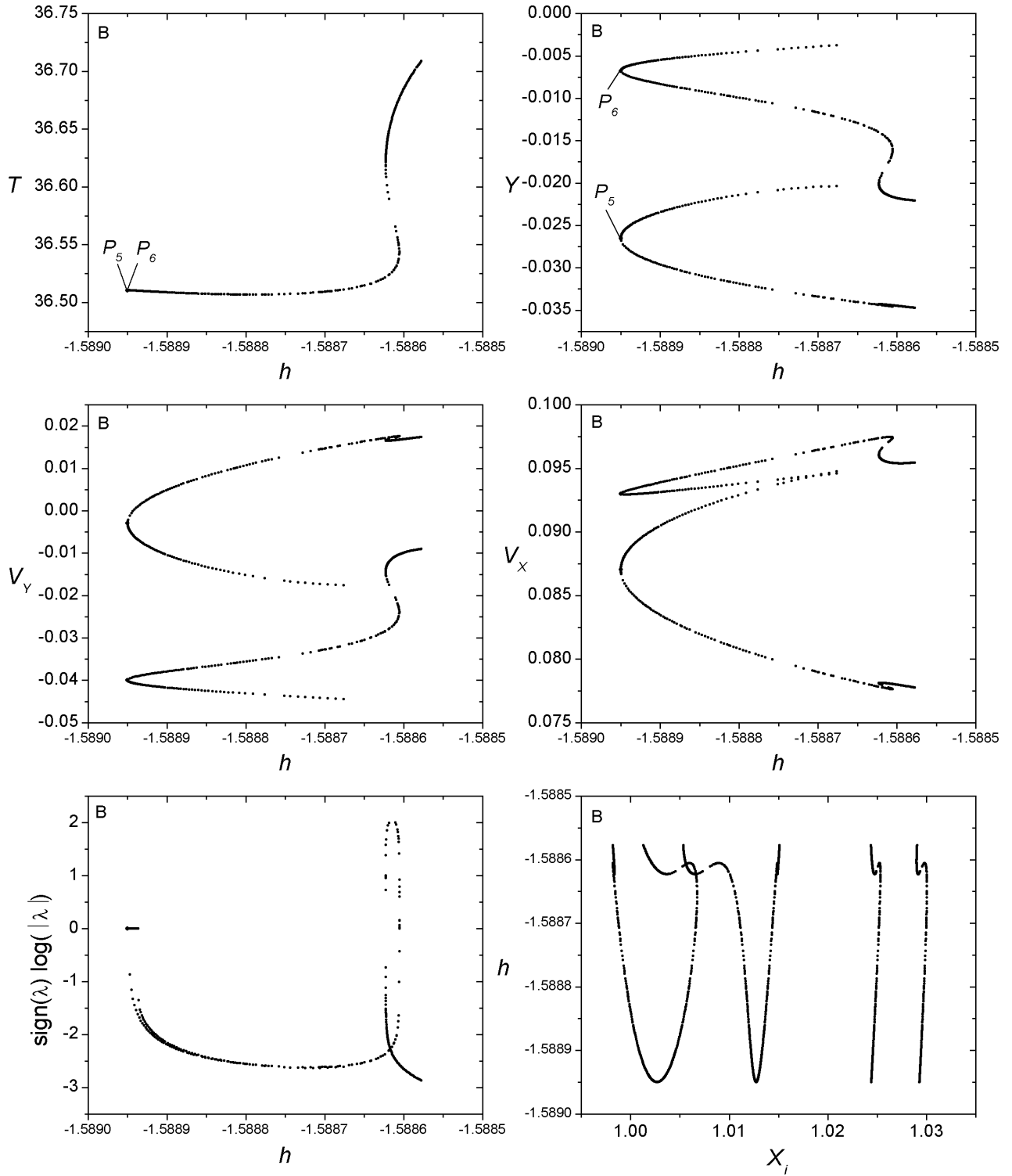

# Family 300 C - Symmetric family of symmetric POs

$h_{min} = -1.588971, \quad h_{max} = -1.588864, \quad T_{min} = 36.260612, \quad T_{max} = 36.510494$

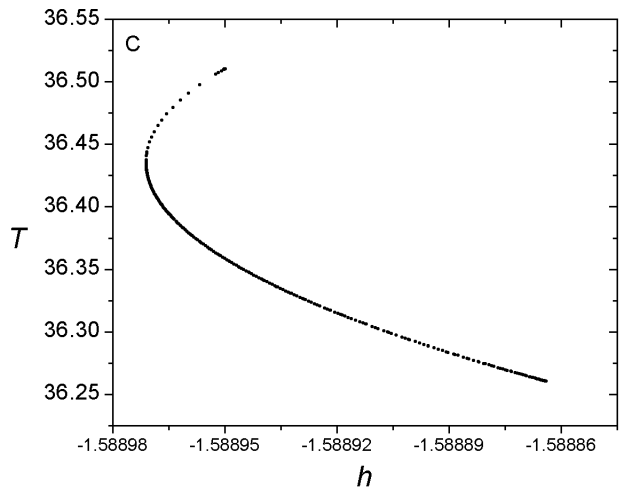
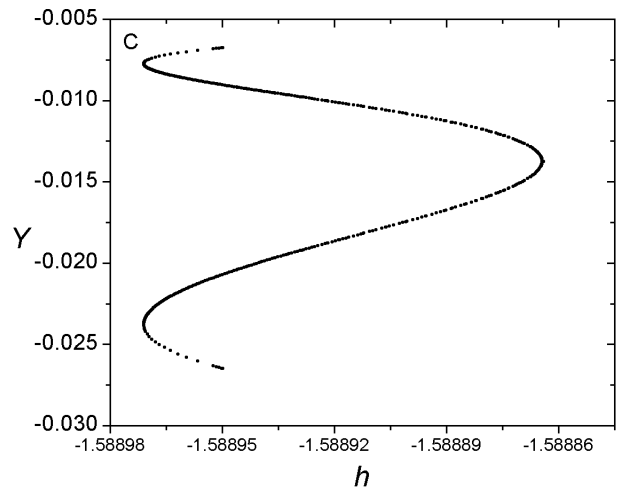
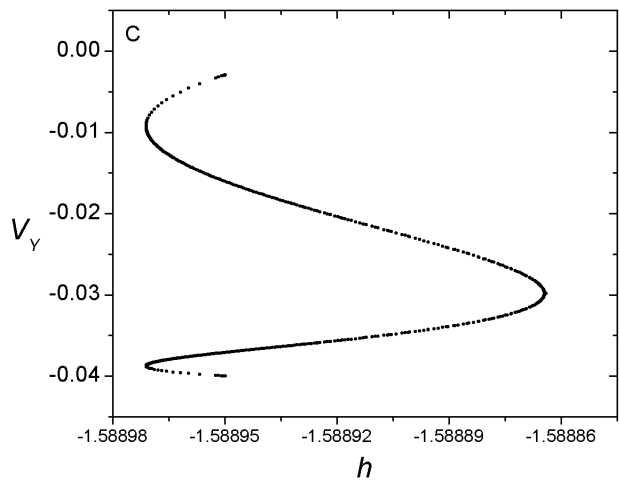
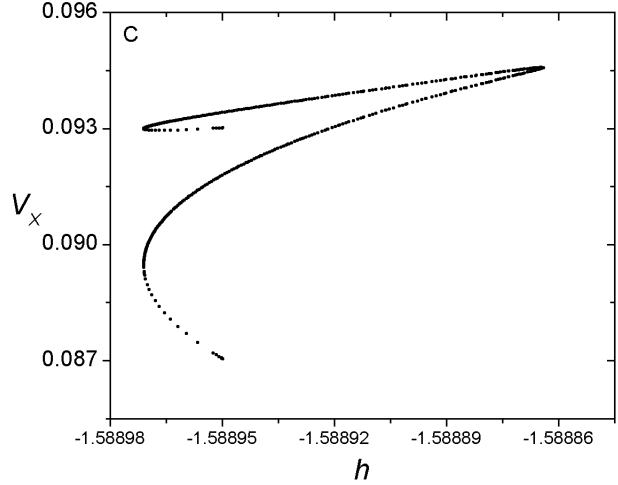
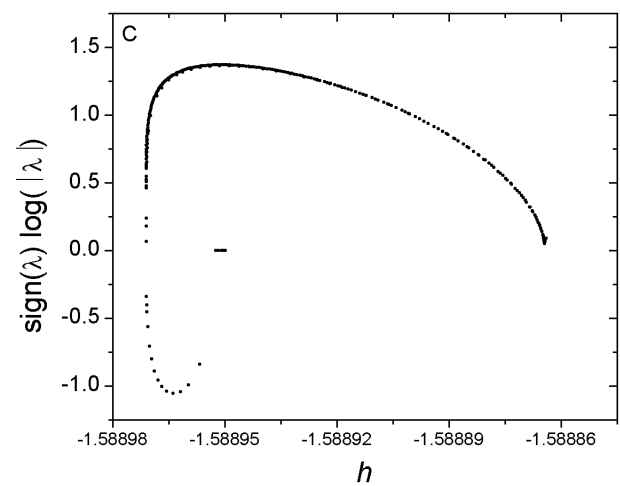
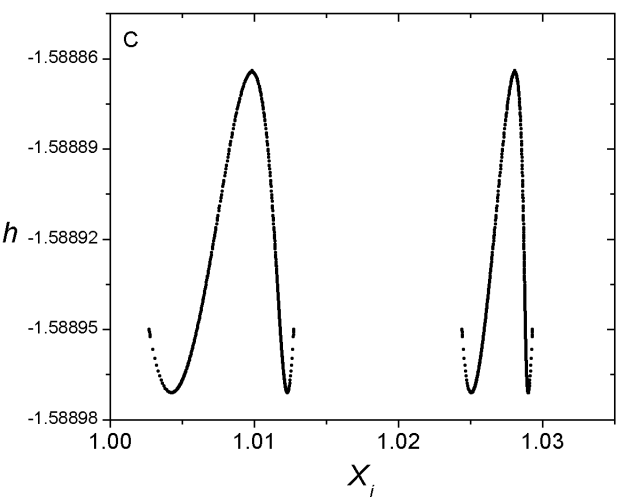

## Family 300 D - Symmetric family of symmetric POs

$h_{min} = -1.588572,\ \ h_{max} = -1.588284,\ \ T_{min} = 36.625264,\ \ T_{max} = 36.653674$

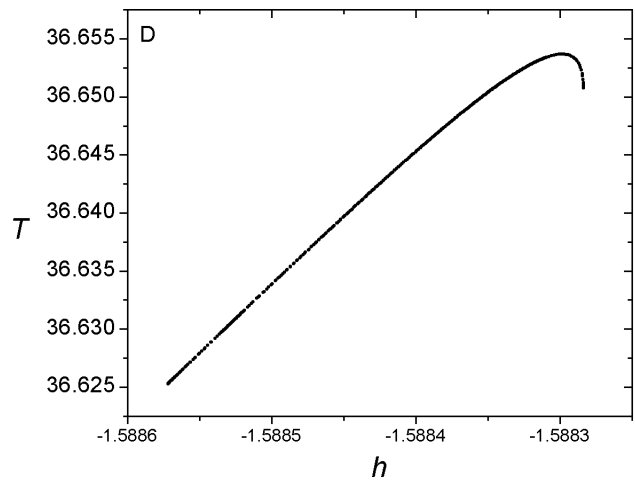
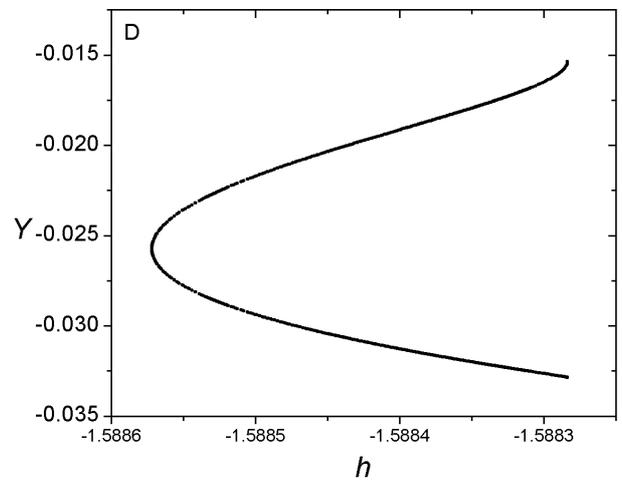

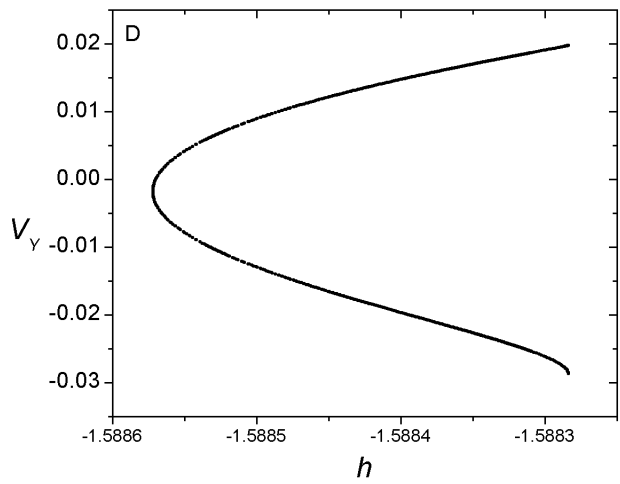
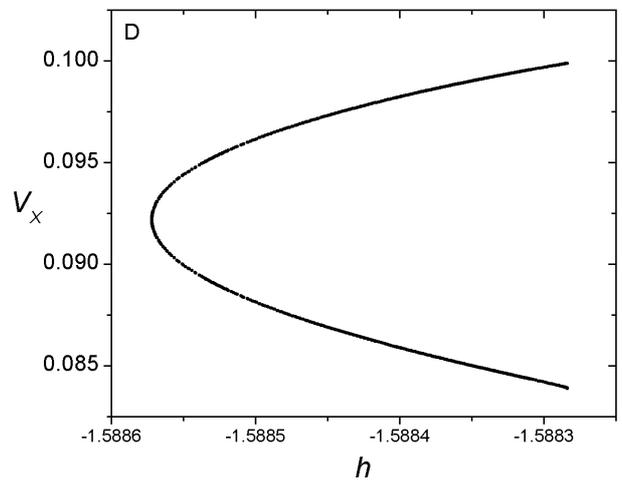

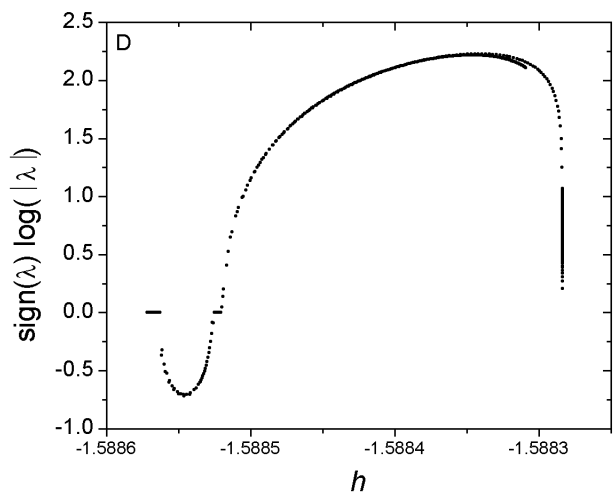
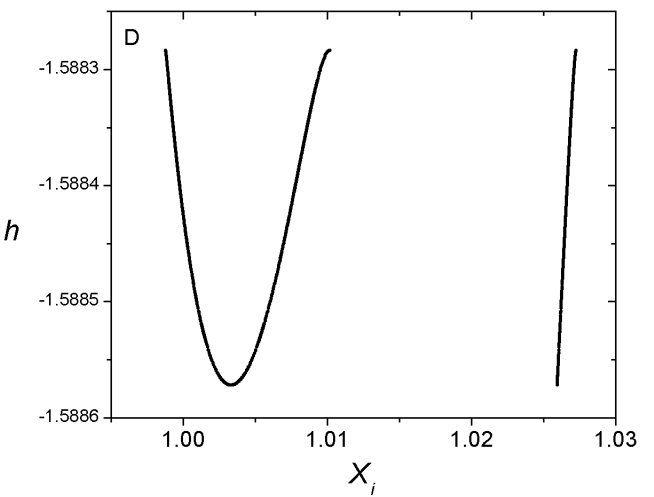

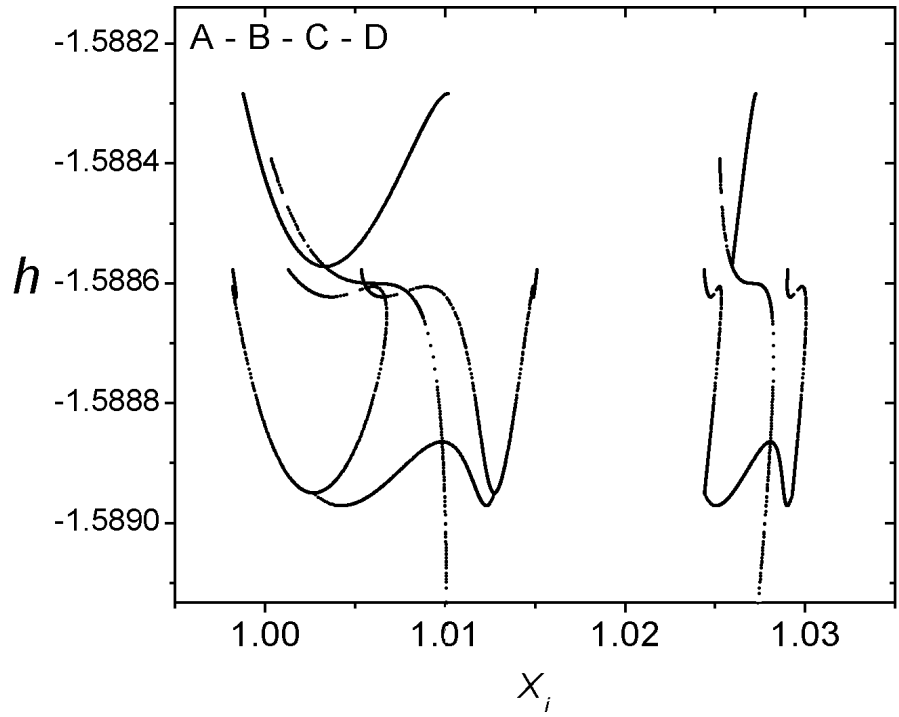

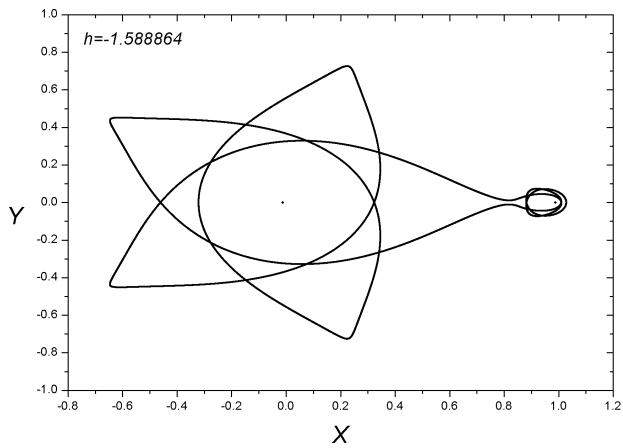

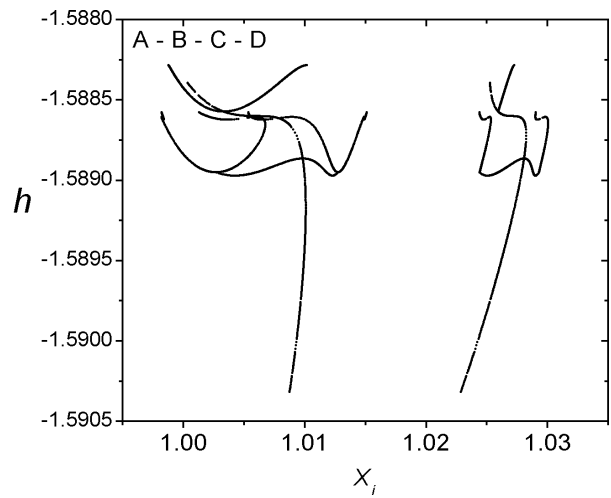

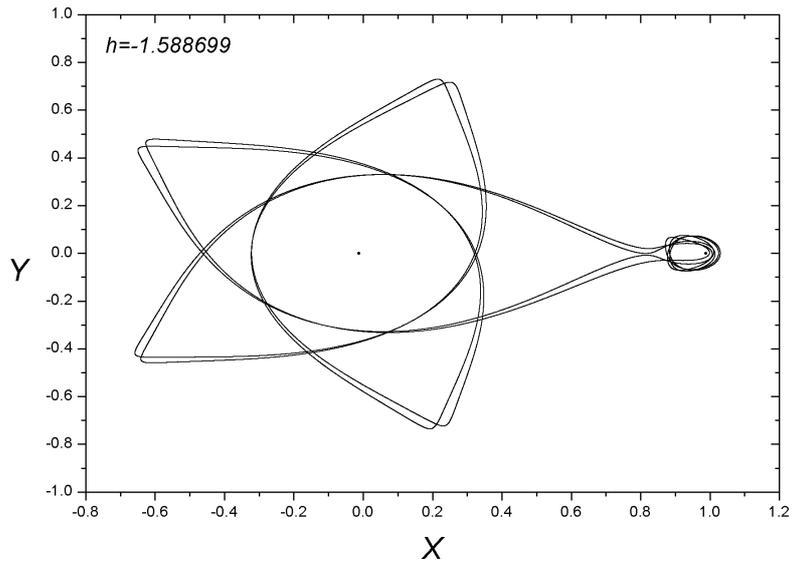

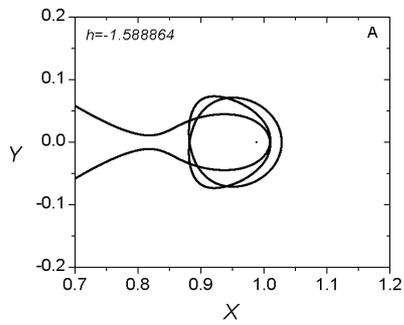
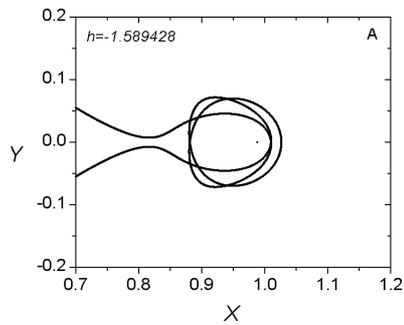
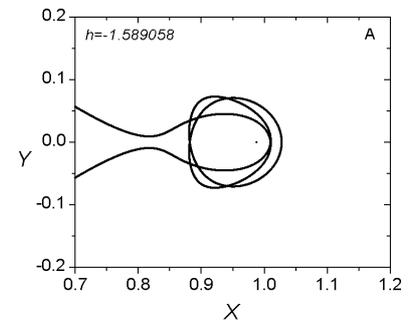

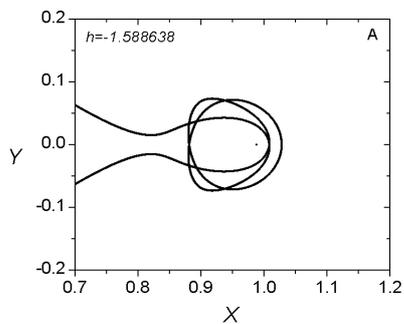
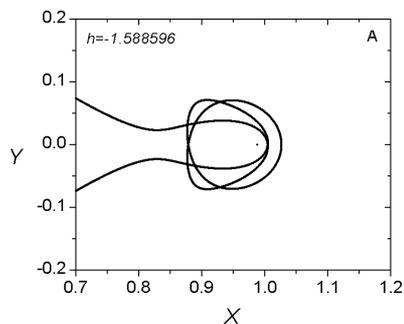
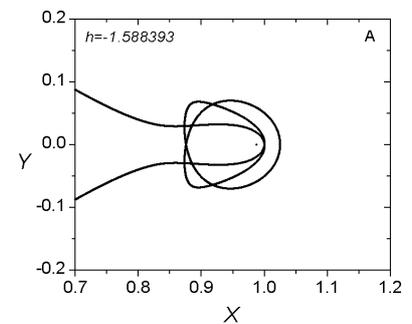

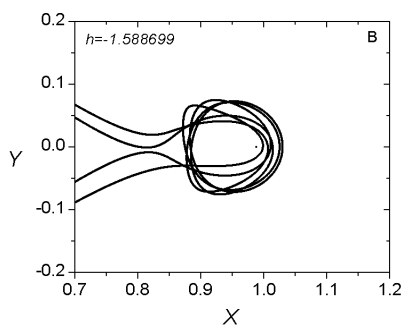
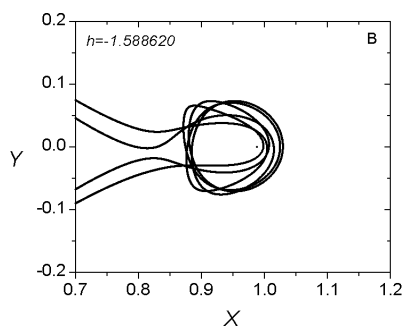
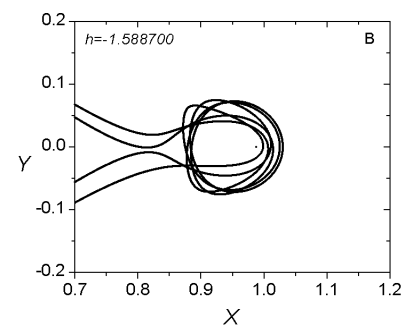

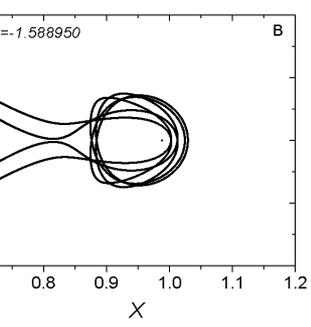
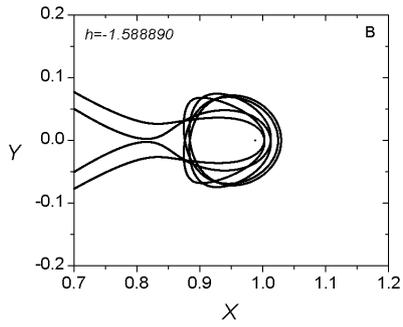
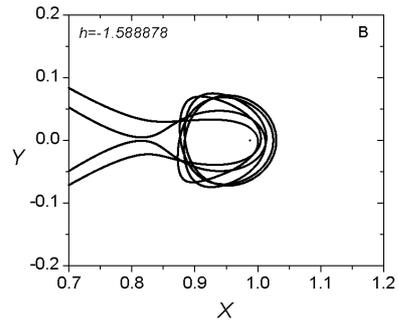

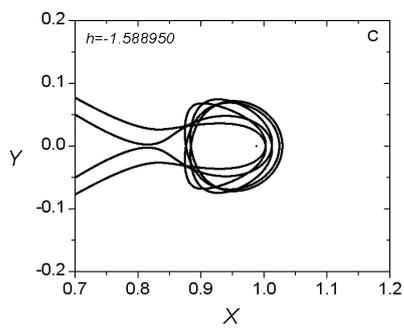
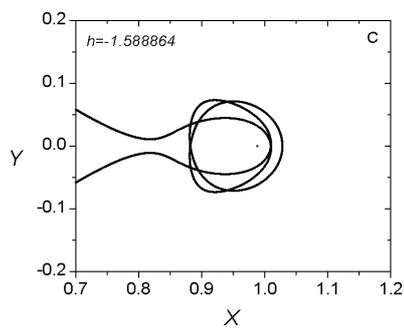
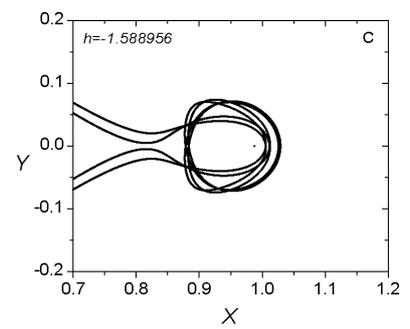

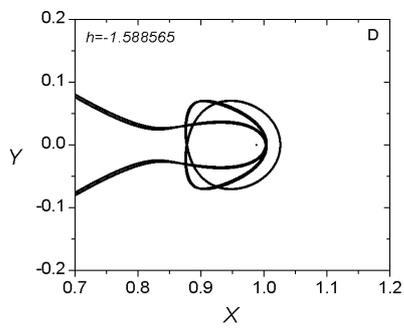
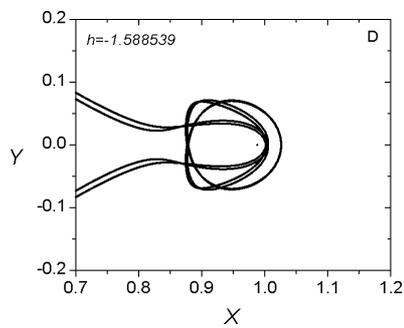
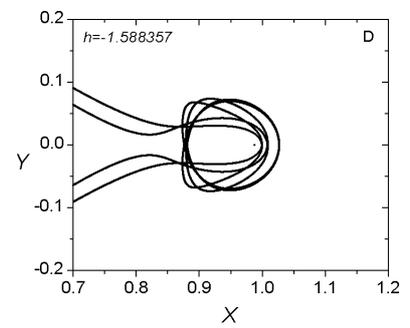

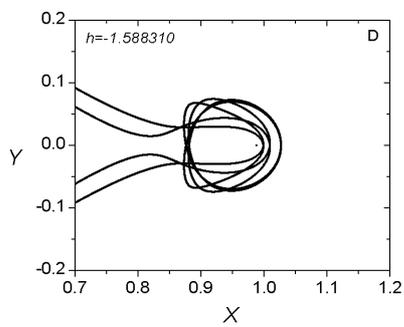
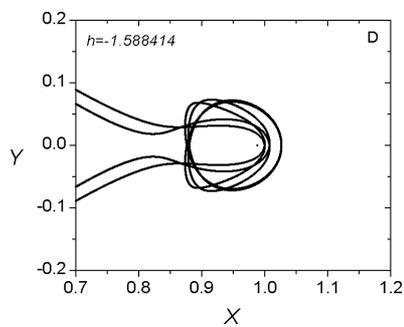
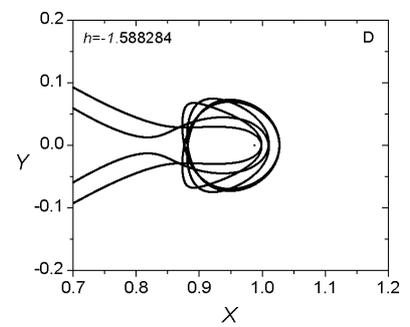

# Family 238  - *Symmetric family of symmetric POs*

$h_{min} = -1.593350, \ h_{max} = -1.590782, \ T_{min} = 36.403708, \ T_{max} = 37.626047$

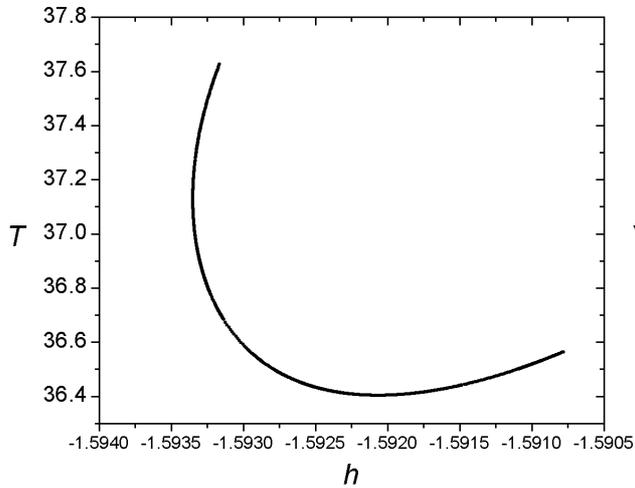

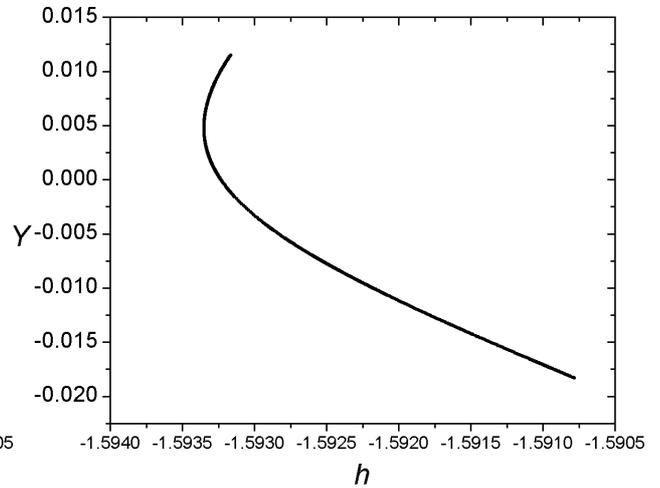

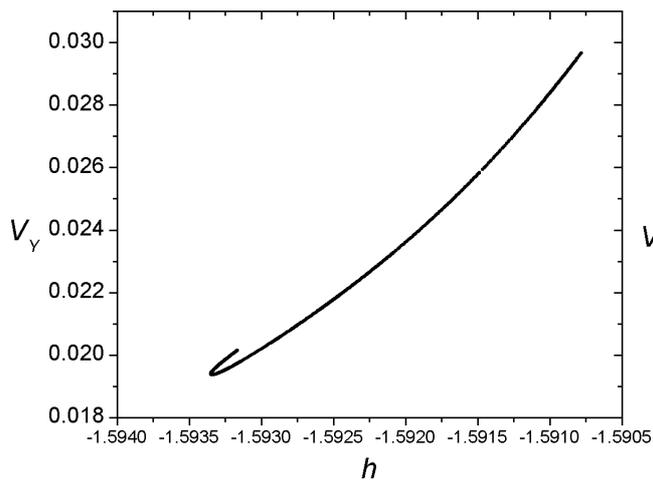

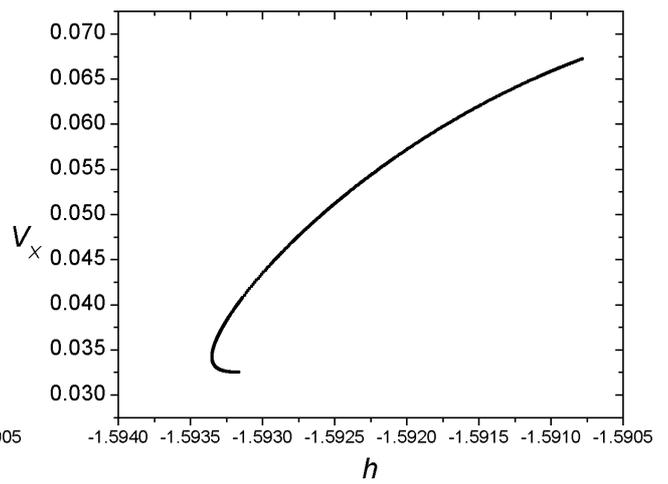

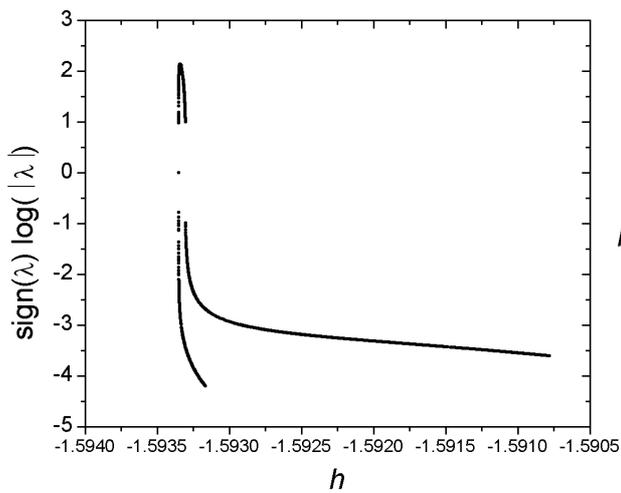

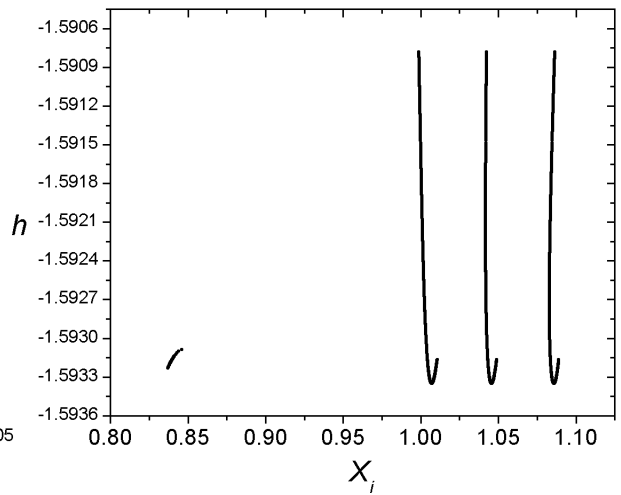

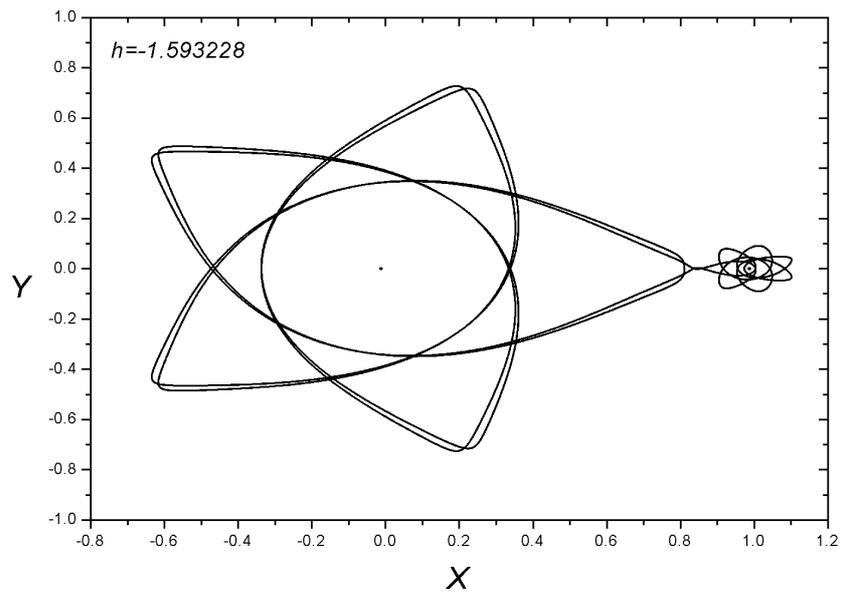

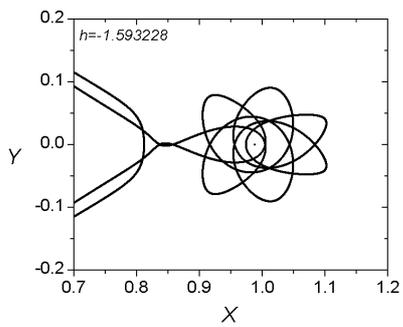
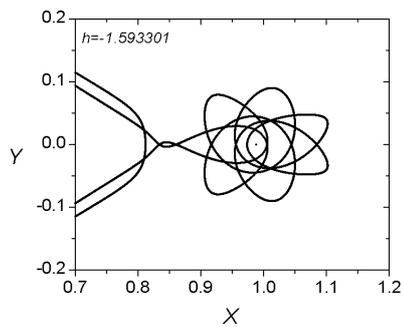
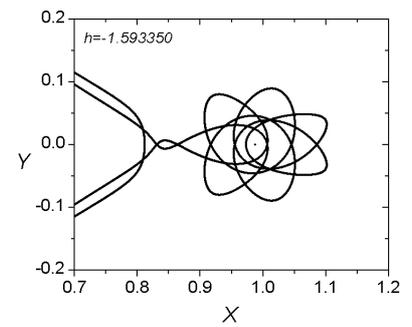

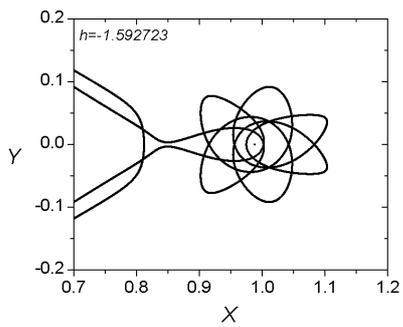
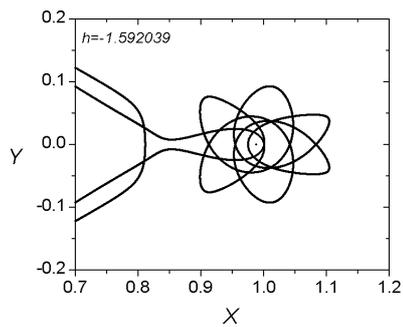
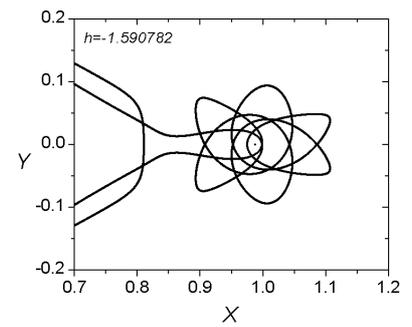

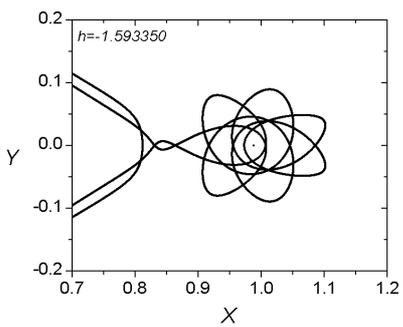
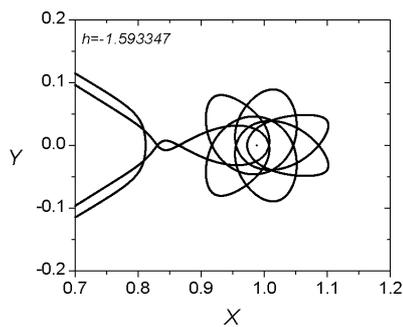
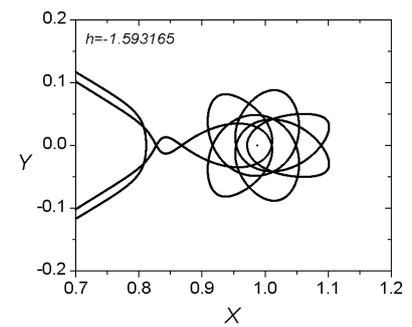

## Families 262 A - 262 B - 262 C

### Bifurcation Points

|       | h          | T          | y         | $v_y$     | $v_x$     |
|-------|------------|------------|-----------|-----------|-----------|
| $P_1$ | −1.593579  | 39.357940  | 0.002564  | 0.016132  | 0.029908  |
| $P_2$ | −1.593392  | 38.815243  | 0.005048  | 0.011887  | 0.036181  |

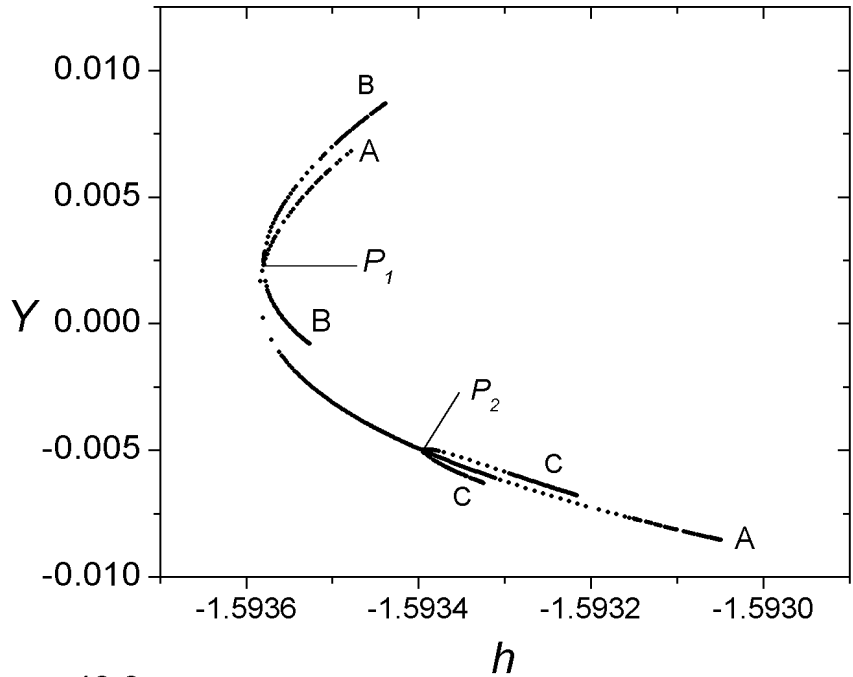

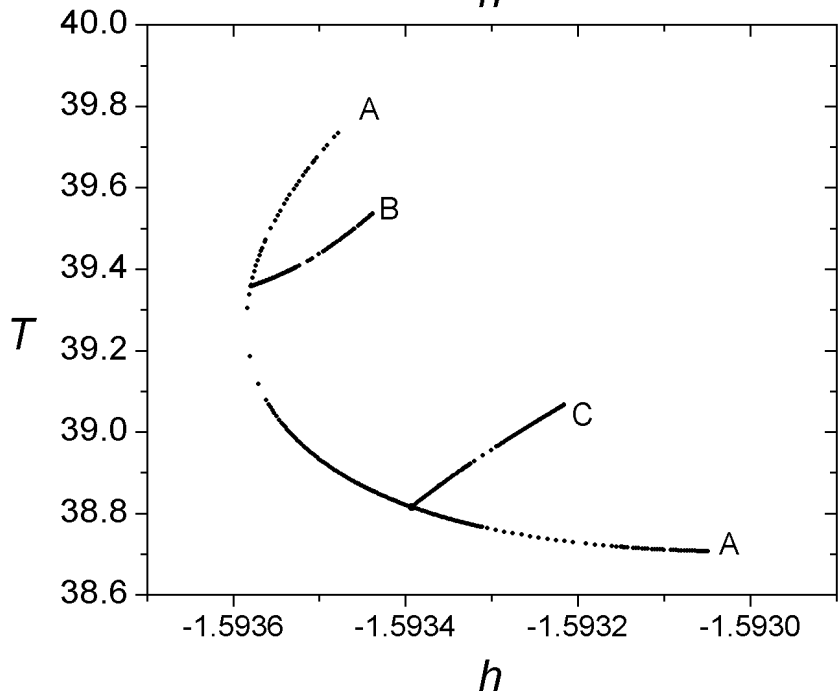

## Family 262 A  - Symmetric family of symmetric POs

$h_{min} = -1.593583, \quad h_{max} = -1.593049, \quad T_{min} = 38.707003, \quad T_{max} = 39.733886$

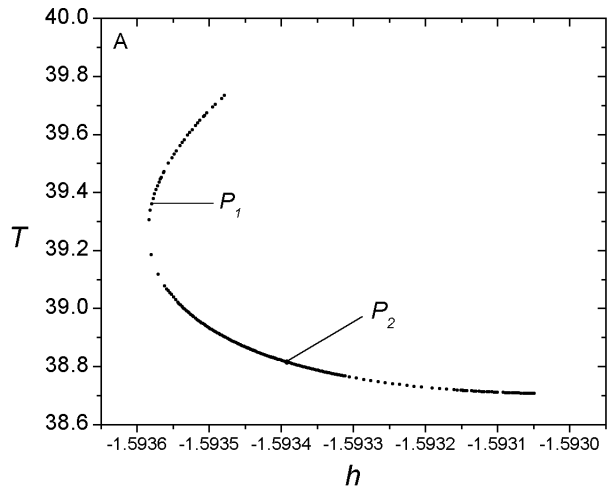

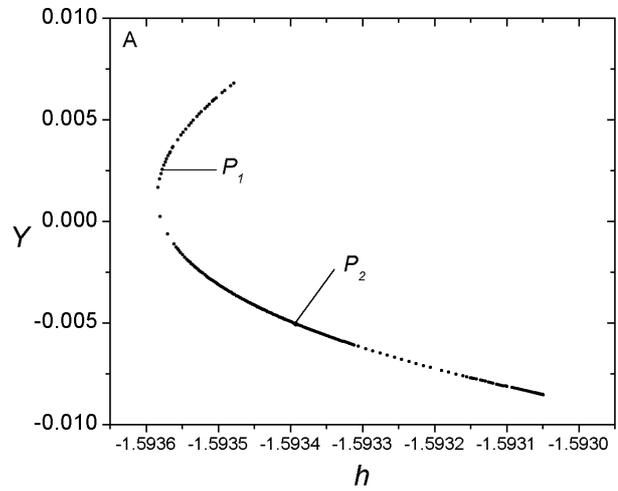

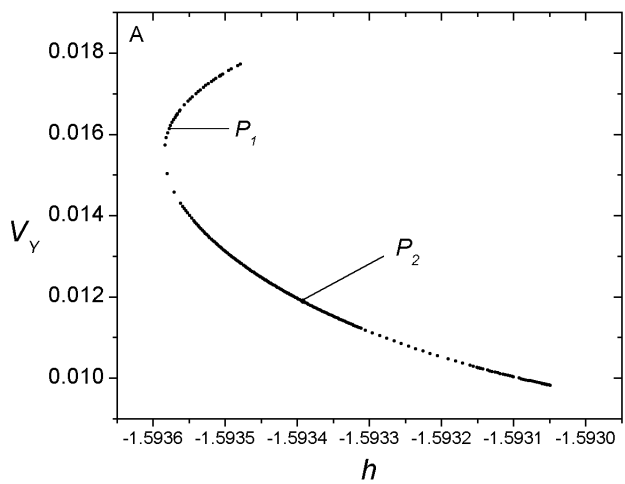

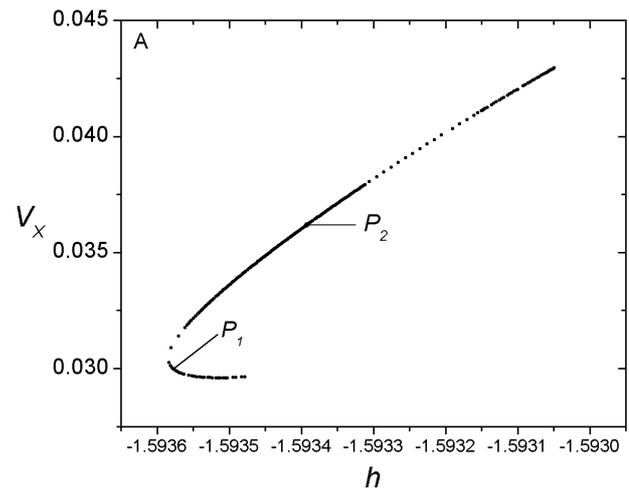

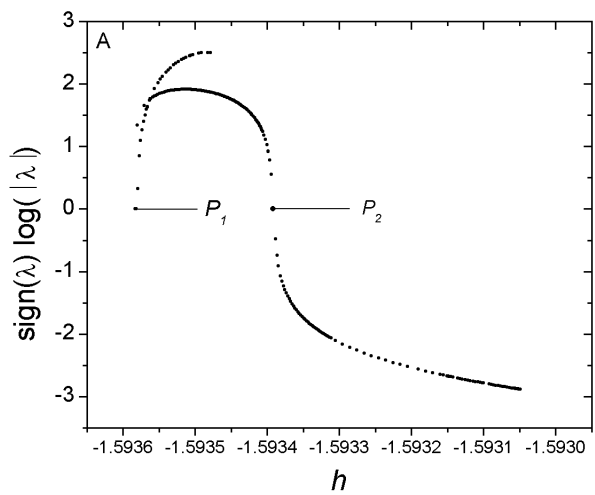

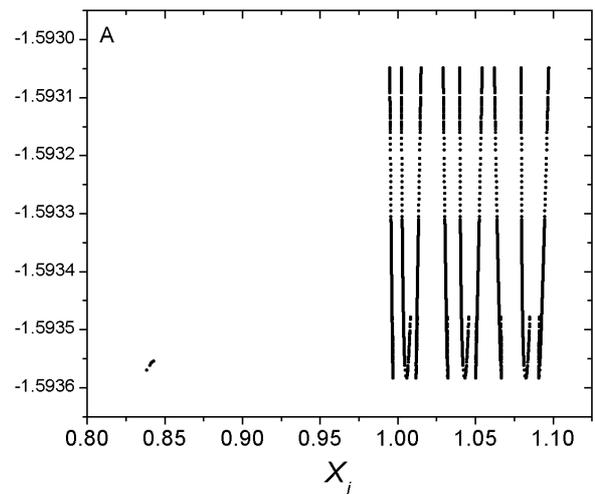

## Family 262 B  - Symmetric family of asymmetric POs

$h_{min} = -1.593580, \quad h_{max} = -1.593438, \quad T_{min} = 39.357787, \quad T_{max} = 39.536105$

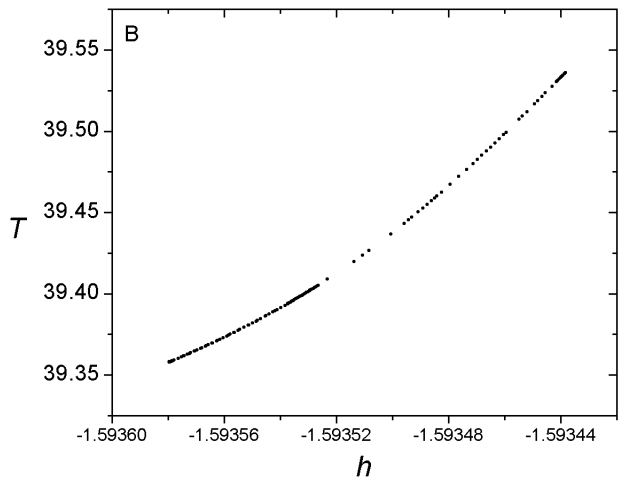

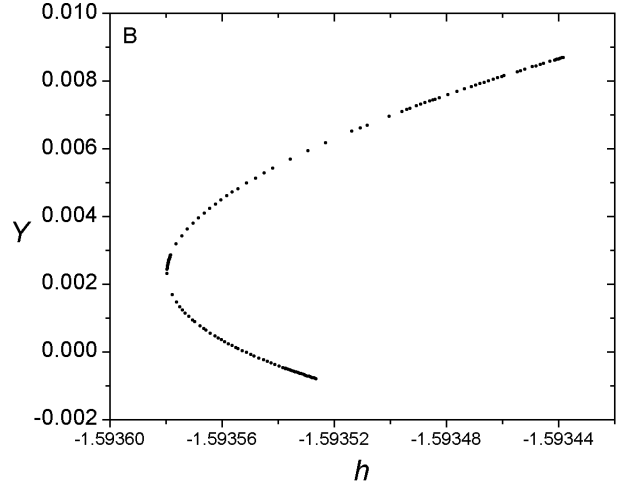

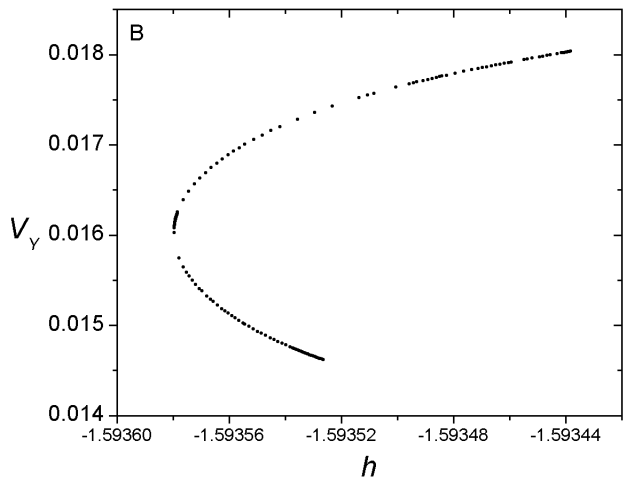

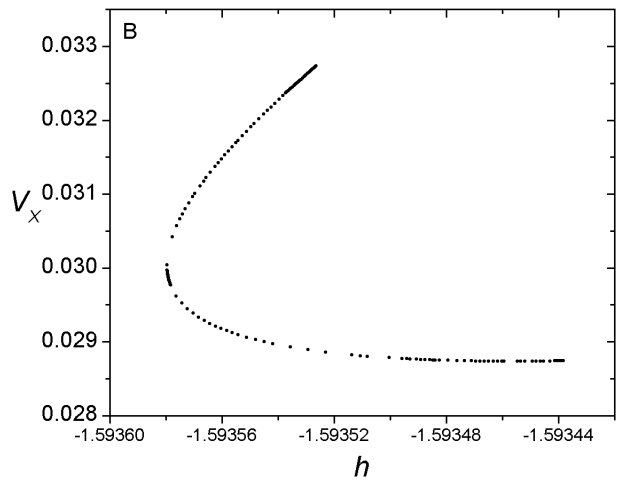

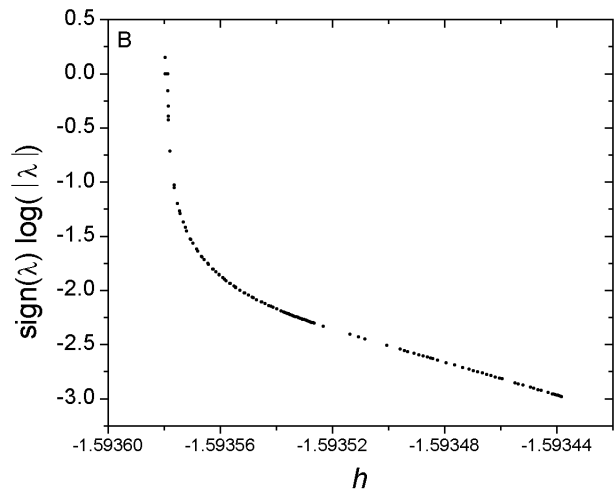

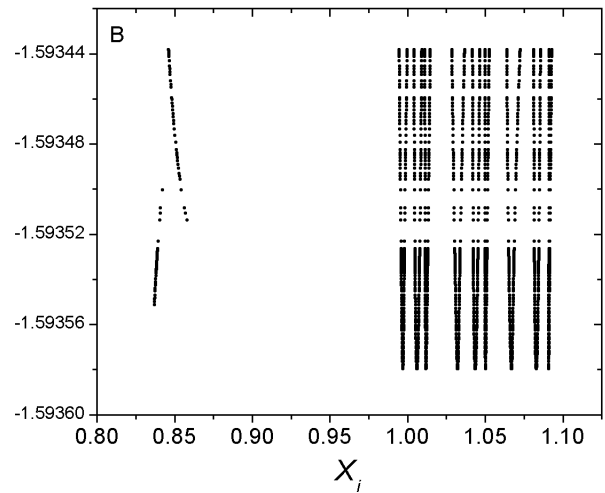

## *Family 262 C - Symmetric family of asymmetric POs*

$h_{min} = -1.593393, \quad h_{max} = -1.593216, \quad T_{min} = 38.815998, \quad T_{max} = 39.066561$

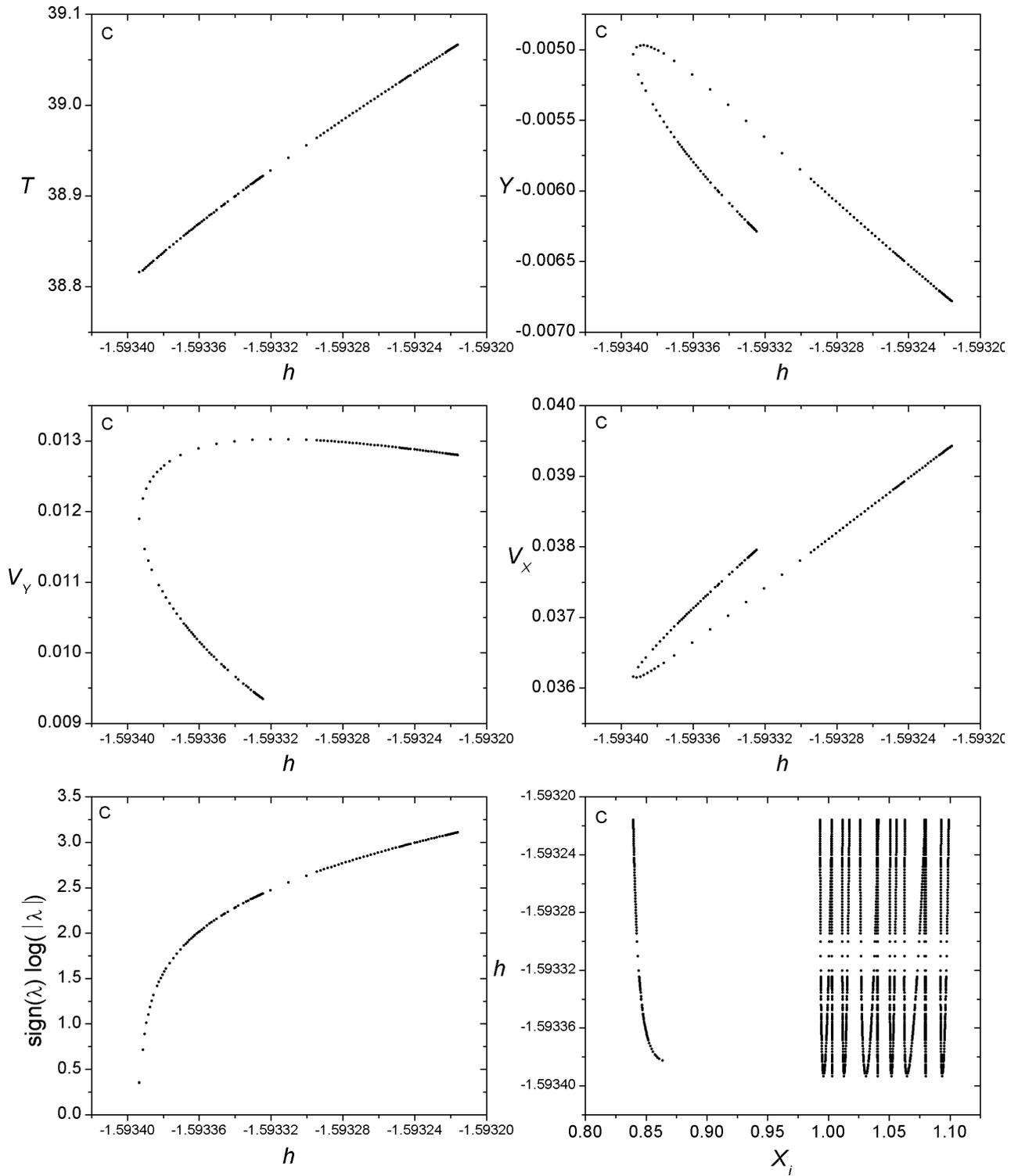

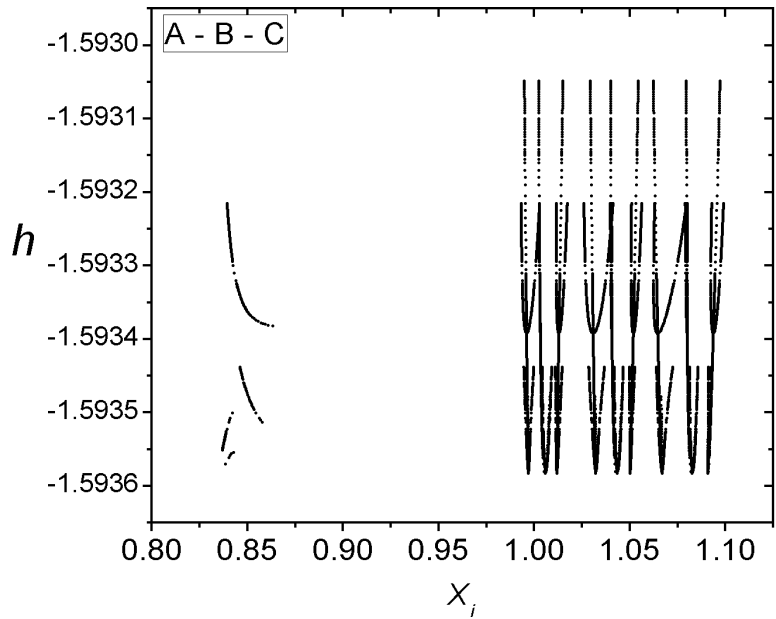

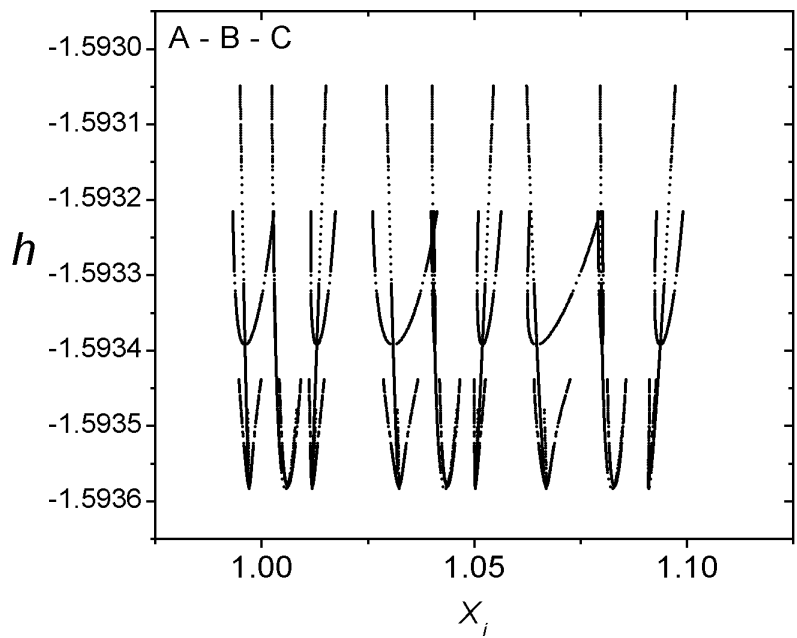

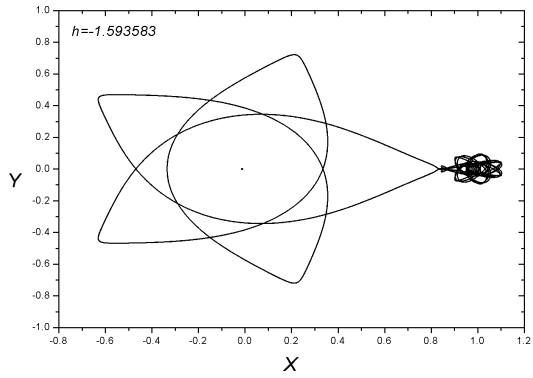

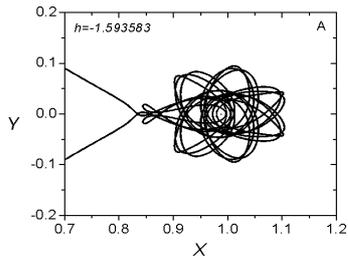
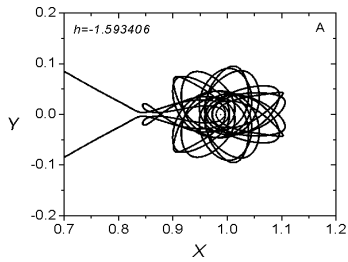
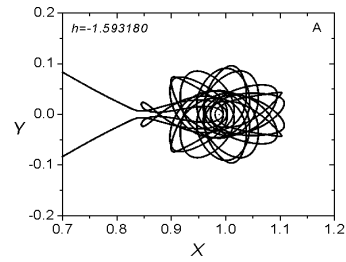

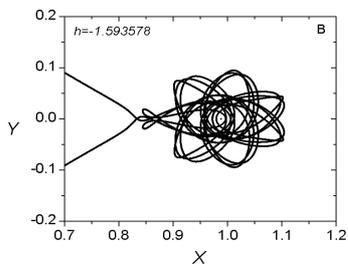
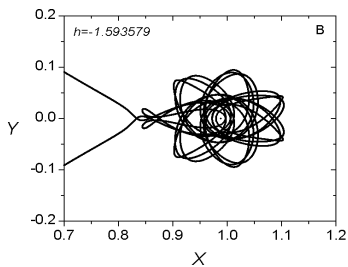
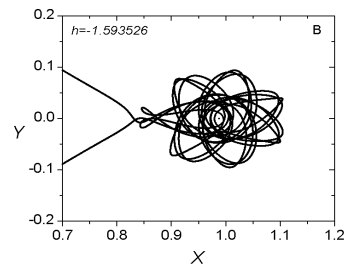

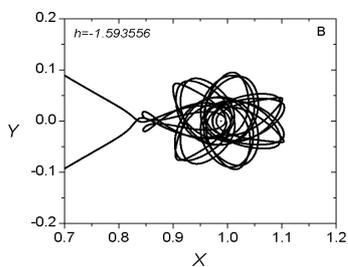
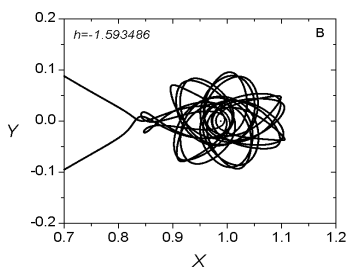
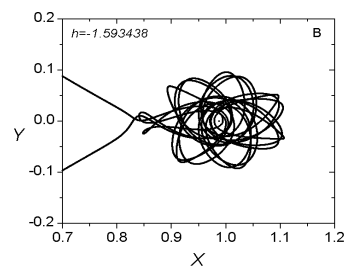

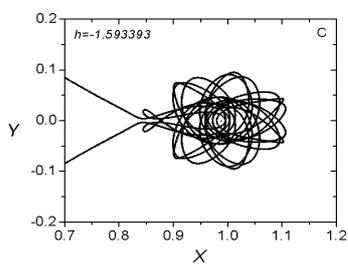
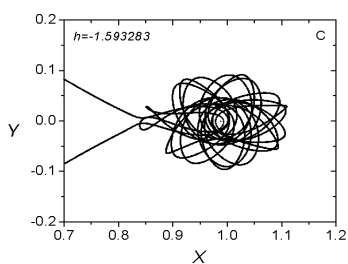
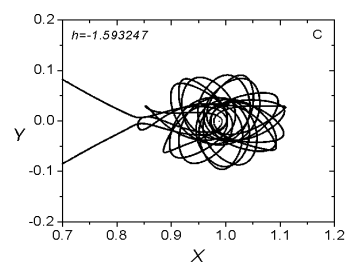

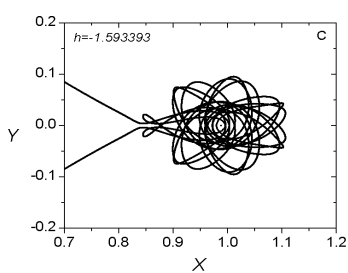
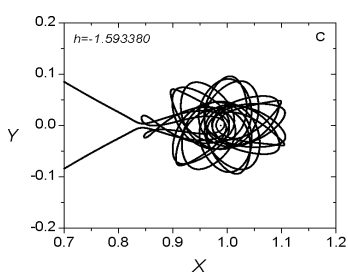
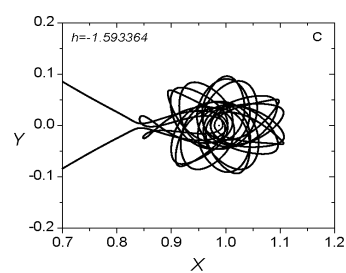

### *Family 008  - Symmetric family of symmetric POs*

$h_{min} = -1.587113,\ \ h_{max} = \ -1.584601,\ \ T_{min} = 39.423500,\ \ T_{max} = \ 39.999187$

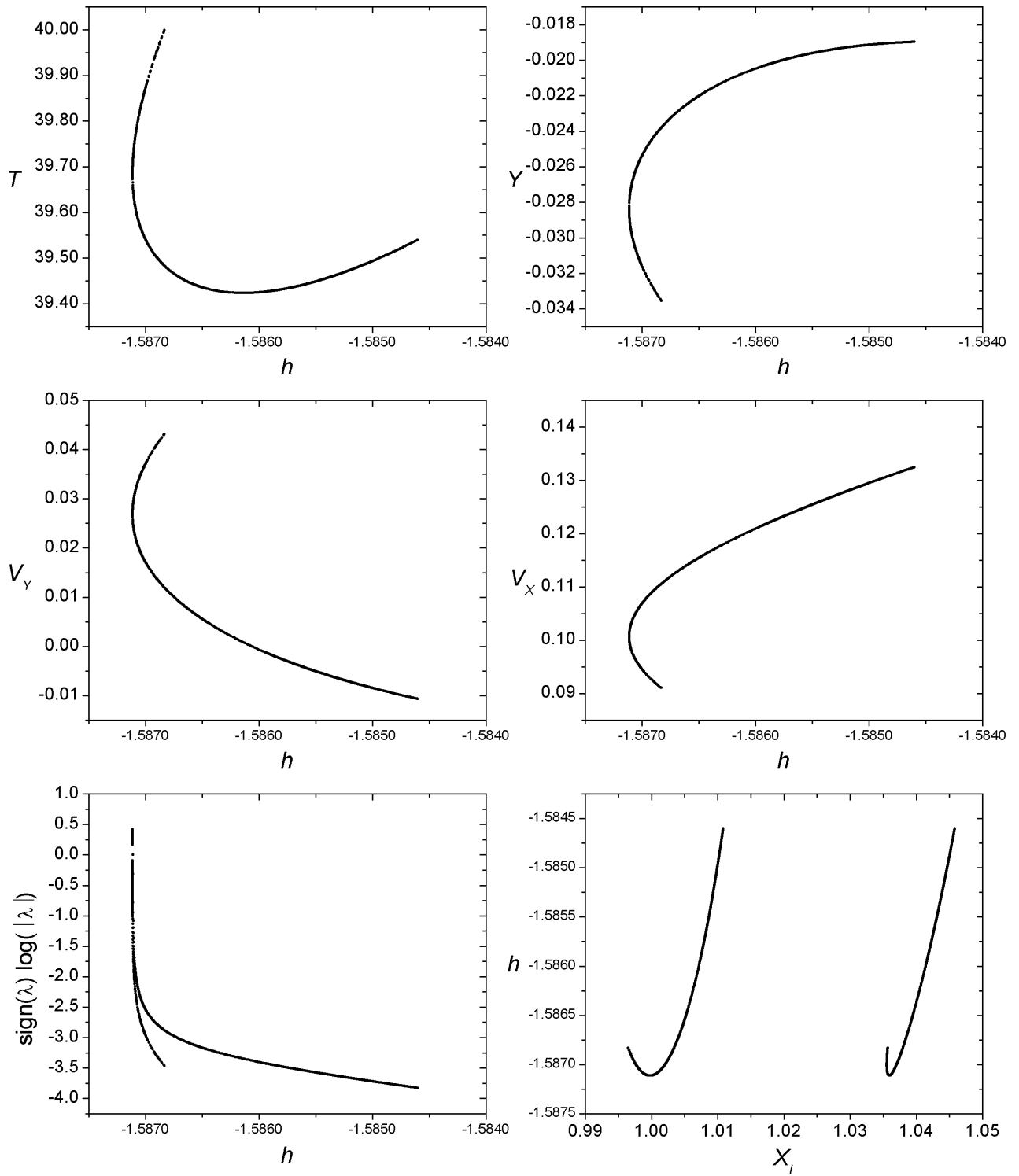

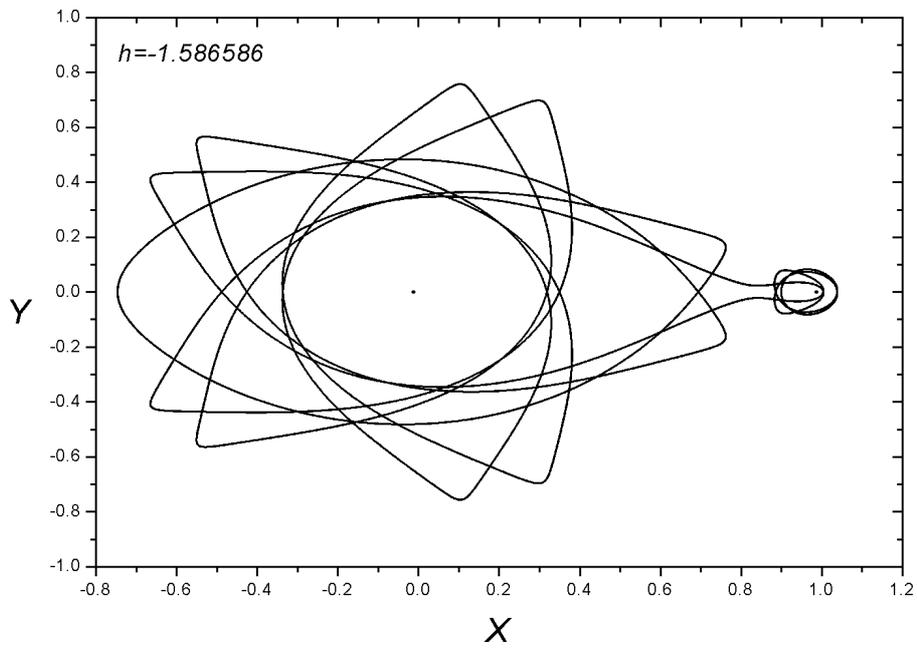

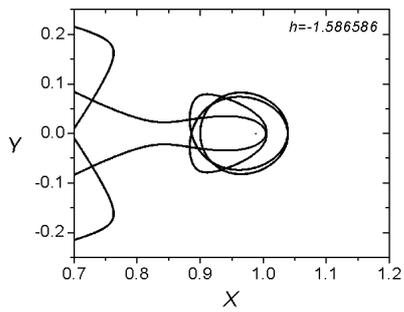
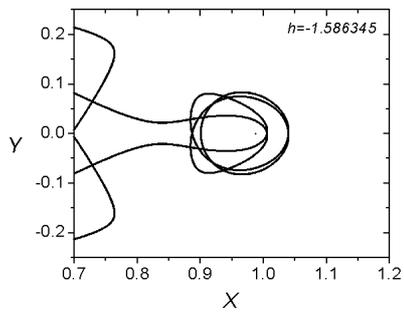
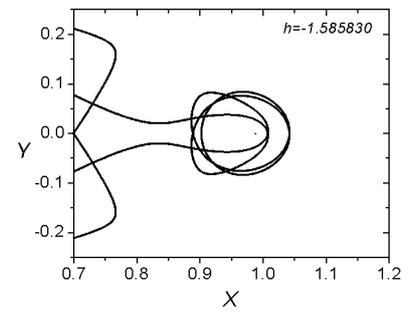

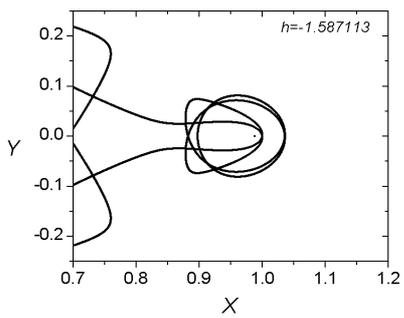
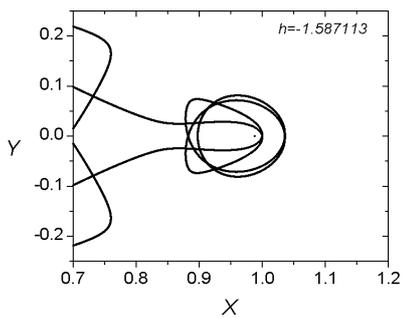
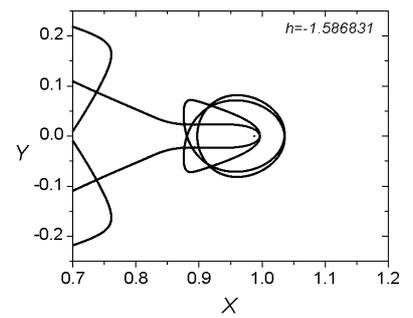

# 6. Discussion

## 6.1 Completitude

The first point we want to address is the completitude status of the Atlas presented. After the initial search in Section 3.1, we performed several more similar searches for POs in the range $h_1 < h < h_2$, $T < 40$, with finer grid spacings in $y$ and $\dot{y}$; none of them turned up POs not belonging to the families shown here. We also constructed dynamical maps of all the orbits emanating from these initial condition grids on $\Sigma_1$ for some selected values of $h$, which showed that their return distances $\delta$ were small only in the neighbourhood of the already found fixed points. These results are not shown here, but will be presented in subsequent works.

This gives us a high confidence when saying that the Atlas presented here is essentially complete, for families of periodic transfer orbits in the range $h_1 < h < h_2$, $T < 40$ in the Earth-Moon CR3BP. Known or suspected incompletitudes are described below.

First, analytical continuation of the families (see Sec. 3.2) has been stopped either when the orbits collided with a primary (the Moon, in all cases presented here) or became too unstable for the Newton-Raphson algorithm to reconverge to a fixed point from the return point on $\Sigma_1$. In the first case we are confident that the finer numerical searches performed would have spotted the family branches after a collision if they were not have been on the Atlas. So in this respect the only incompletitude remaining would be the determination of which families in the Atlas are connected to which others (if any) thorough a collision. In the second case the only way to extend the analytical continuation would be to resort to higher precision arithmetics (here we have been working in double precision), which is much slower. It must be remembered that attempting continuation on a Poincaré section different from $\Sigma_1$ does not improve convergence, since the map eigenvalues are the same at any point of a given orbit (Parker and Chua, 1989). Also, we want to remark that though some families have been continuated for $h > h_2$, no systematic search or continuation has been performed in this region; correspondingly, we by no means claim completitude of the Atlas for $h > h_2$.

Second, symmetries of the CR3BP (see Sec. 3.3) have been used to find or complete some families or branches that were hard to continuate either from the POs found by the initial search of Sec. 3.1, or from the bifurcation points found. An example is family 027, an asymmetric family of asymmetric orbits symmetrical to family 026, from which it was completed. Others are the families of asymmetric orbits (*e.g.* families 146 B-C-D), where symmetry has been used to complete some segments. This procedure has been carried on thoroughly, for each and every family in the Atlas comprising asymmetric orbits, so in this respect the Atlas is complete.

Third, period doubling bifurcations (see Sec. 3.4) can be used to find new families from known ones, as has been done with family 300 A, obtained from family 197A. Since we are restricting ourselves to $T < 40$ the candidates to generate new families through this technique must have $T < 20$. These are families 357, 037, 043, 056, 053, 084, 077, 180 A-B, 197 B-C-D, and 146 A-B-C. The application of the aforementioned procedure to these families is yet to be carried out, and in this respect the Atlas can be as yet incomplete.

Fourth, continuation in $\mu$ (see Sec. 3.5) can be used both to find "hidden" relationships between families in the Atlas, as happened with families 037, 043, and 056, and to discover new families related to them, as was the case with family 357. Families 037 and 043 are both asymmetric families of asymmetric orbits, symmetrical to each other, while family 056 is a symmetric family of symmetric orbits; the three are similar in period, energy and morphology.

There are other groups of families in the Atlas showing similar realationships, *e.g.* families 077 and 084, and families 146 A and B. The application of this procedure to them has not yet been carried out, and in this respect the Atlas can be as yet incomplete. It is worth mentioning that a recent work by Bruno and Varin (2006) also addresses the continuation in μ of PO families in the CR3BP; however, it is restricted to symmetric orbits, while most of the families in the present Atlas comprise asymmetric orbits. The superposition with the present work is thus only partial, but it provides a welcome way of cross-checking the results of both.

The definitive completion of the Atlas by higher-precision continuation, period doubling bifurcations, and continuation in μ, is the subject of current research. The results found will be added to this work.

## 6.2 Applications

Although the study of periodic orbits in the CR3BP has a clear academic interest, the possibility of realizing some of these orbits in practice would make them useful in the design of long-duration space missions. However, in the case of the Earth--Moon system the CR3BP is a very poor approximation, since the gravitational pull of the Sun tends to quickly destroy periodic orbits (Szebehely 1967). This is particularly true for the kind of orbits in the present Atlas, which besides spanning distances of the same order of the size of the Earth-Moon system, pass very slowly near to the Lagrangian point L1, where the gravitational pull of Earth and Moon almost cancel and the influence of the Sun dominates. This makes control methods like the one devised by Briozzo and Leiva (2006a) inapplicable, and turns almost impossible to add the Sun's influence perturbatively. Instead, it is better to start from a more elaborate model for the Sun-Earth-Moon system, like the Quasi-Bicircular problem (QBCP) (Andreu 1998); but this being a periodic Hamiltonian system, POs are there isolated, and much harder to find. We recently showed (Leiva and Briozzo 2005) how under suitable conditions some UPOs can be extended from the Earth-Moon CR3BP to the Sun-Earth-Moon QBCP by analytical continuation in an adequate parameter. In work soon to be submitted for publication, we will present a systematic way to extend some POs from families in the Atlas to equivalent POs in the Sun-Earth-Moon QBCP, together with a suitable control algorithm. In this way, low-energy and short-period transfer orbits in the real Earth-Moon system will become much nearer to realization.

There are several works on the issue of low energy Earth--Moon transfer orbits, from the papers by Bollt and Meiss (1995) and Schroer and Ott (1997) to some recent works by Yagasaki (2004a, b) and Ross et al. (2003), but as far as we know all of them are aimed to find a single, one-time transfer orbit from the Earth to the Moon which minimizes fuel consumption. It must be noted that Ross et al. (2003) and Yagasaki (2004a) only address the Earth-Moon CR3BP, and though Yagasaki (2004b) introduces the Sun perturbation, it does it through the Bicircular problem, which is not a self-consistent model (Andreu 1998) and is less accurate than the QBCP.

It is worth to review the succession of orbits found by Bollt and Meiss (1995), Schroer and Ott (1997), and Ross et al. (2003), and see how as they become simpler (consisting of less joined arcs) and faster, they progressively resemble some of the POs in the present Atlas. In particular, the orbit in Fig. 2b of Ross et al. (2003) closely resembles some of the orbits in families 056 or 053. Also, the compromise between time of flight and fuel consumption found by Ross et al. (2003) has a clear correlate in the Atlas: a comparison between Fig. 2c there and the characteristic curves $T(h)$ for all branches of all families in the this Atlas shows that the lower-left envelope of our $T(h)$ curves resembles closely their graph for the time-of-flight vs. consumption tradeoff between several orbits. In all, this allows us to affirm with some confidence that this tradeoff is an unavoidable consequence of  the dynamics of the Earth-Moon CR3BP, and that transfer orbits with periods shorter than 14 will only be found at much higher energies.

## Acknowledgments

This work has been partially financed by Grant No.162/06 from the Secretaría de Ciencia y Tecnología de la Universidad Nacional de Córdoba.